# Models for nuclear fusion in the solid state

Peter L. Hagelstein[1,*], Florian Metzler[1,*], Matt K. Lilley[1,*], Jonah F. Messinger[1,2] and Nicola Galvanetto[1,3,*]

[1]*Massachusetts Institute of Technology, Cambridge, MA, USA*
[2]*Cavendish Laboratory of Physics, University of Cambridge, UK*
[3]*University of Zurich, Zurich, Switzerland*
[*]*plh@mit.edu, fmetzler@mit.edu, mklilley@mit.edu, galvanet@mit.edu*

This article presents a theoretical framework for enhancing nuclear fusion rates in solid-state environments under near-ambient conditions. Drawing on quantum tunneling, electron screening, and resonance energy transfer, the study proposes rate enhancements of more than 40 orders of magnitude for deuterium-deuterium (D-D) fusion in palladium lattices. A generalized nuclear Dicke model describes a fusion-fission process as a result of energy transfer between D-D and palladium mediated by lattice vibrations. Practical challenges such as decoherence, destructive interference, receiver decay, and achieving resonance between donor and receiver systems are addressed. Experimental strategies to validate the model are proposed along with its implications for the advancement of solid-state fusion as a potential pathway to sustainable energy technologies.

# Contents

# 1 Overview

Nuclear fusion has remained a prime topic of interest since its discovery during the early 20th century [1, 2]. It represents a comparably "clean" option to access nuclear binding energy with typically fewer reaction products than in nuclear fission, and it symbolizes the achievement of greater degrees of control over the transmutation of elements.

The known property of palladium to absorb hydrogen into its metal lattice led researchers — as early as the 1920s — to contemplate the effect of condensed matter on fusion processes [3, 4]. Subsequently, rapid progress in plasma science led to a shift of attention towards achieving fusion in plasmas via intense pressures, high temperatures, and long confinement times. This was a route that could be described theoretically through comparatively simple and well-established formalisms [5]. The hope was that the associated technical challenges could be overcome in due time. Much progress has been made in the study of nuclear fusion in the plasma state [6]. However, many challenges remain, especially in view of the scalability and economics of proposed concepts for commercial power production [7].

Whether the simple two-body formalisms used to calculate nuclear fusion rates in plasmas remain appropriate for modeling fusion under all circumstances has been called into question. In 1965, Terhune and Baldwin [8] pointed out that nuclei, when coupled through shared modes in a lattice, must not be treated in isolation but call for modeling via many-body formalisms prevalent in condensed matter physics and quantum optics. The events of March 1989 [9] added much fuel to such considerations, when electrochemists Martin Fleischmann and Stanley Pons announced experimental observations of heat beyond chemical levels emanating from electrochemically treated palladium-deuterium samples [10].

Fleischmann and Pons received criticism for the dearth of documentation that surrounded their experiments, at least in the early days after their initial announcement [11]. However, some leading experts in nuclear physics and quantum optics, including Edward Teller [12], Julian Schwinger [13], and Giuliano Preparata [14], recognized — irrespective of the experimental details — that the results reported by Fleischmann and Pons were at least in principle conceivable, when considering more complicated many-body formalisms to model nuclear fusion processes, beyond the well-established but simple two-body formalisms. This position is reflected in Julian Schwinger's assertion [15]:

> *"I have asked myself not whether Pons and Fleischmann are right — but whether a mechanism can be identified that will produce nuclear energy by manipulations at the atomic — the chemical — level. Of course, the acceptance of that interpretation of their data is needed as a working hypothesis, in order to have quantitative tests of proposed mechanisms."*

In our group, we have long advocated for and pursued a two-pronged approach [16]: experiments that claim accelerated fusion in solid state environments need to become better characterized and documented; and theoretical models aimed at explaining such processes need to become more and more quantitative and compatible with realistic assumptions.

The work here now represents — for the first time to our knowledge — a detailed and comprehensive argument on how nuclear fusion can be accelerated to observable rates in solid state samples at ambient conditions.

The framework draws on and combines known physical principles such as quantum tunneling, electron screening, and excitation transfer dynamics, underpinned by the application of a generalized nuclear Dicke model. This approach leverages the interactions between nuclei and lattice oscillations, such as phonons and plasmons, to facilitate energy transfer between donor nuclei (e.g., deuteron pairs) and receiver systems (e.g., groups of palladium nuclei). Palladium nuclei, with their high density of excited states, are identified as suitable receiver systems, enabling rapid energy transfer and corresponding fusion rates that extend into the observable range.

The study also addresses practical considerations such as suitable and realistic lattice structures, decoherence, energy loss, and lattice dynamics. The resulting models not only match reports of accelerated fusion rates but also imply a large range of different reported reaction products, including nuclear particles and excess heat — consistent with experimental reports.

By bridging theoretical predictions with experimental relevance, this work seeks to lay the foundation for advancing nuclear fusion in the solid state towards a mature field of study with impactful technological applications [17].

The structure is as follows:

- **Section 2** outlines the operational definitions, nomenclature, and historical context of low-energy nuclear fusion within solid-state physics.
- **Section 3** discusses the quantum dynamics involved in nuclear fusion, including collision models and spontaneous fusion processes.
- **Section 4** provides a simplified explanation of the complex quantum interactions between nuclei and lattice oscillations that facilitate energy transfer.
- **Section 5** — the core of this document — explores in-depth theoretical models and calculations that describe the excitation transfer dynamics essential for enhancing fusion rates. It also proposes experimental strategies to validate the theoretical models and explores the practical challenges and implications for advancing solid-state fusion technology.

An earlier version of this document first appeared as Supplementary Information for [18].

# 2 Introduction

## 2.1 Operative definition of readily observable fusion as used in this article

Here we derive estimates for deuterium-deuterium (D-D) fusion rates corresponding to experimental observables that are macroscopically and uncontroversially detectable. Such experimental observables can include the emission of neutrons and charged particles or the production of thermal energy that far exceeds the chemical potential energy available in the system.

In terms of neutron emission, Jones et al. 1989 [19] report an observed fusion rate of $10^{-23}$ $s^{-1}$ based on neutron detection of (4.1 ± 0.8) × $10^{-3}$ counts $s^{-1}$ (foreground counts minus background counts) at a neutron detection efficiency of (1.0 ± 0.3)% for a 3 g $TiD_2$ sample comprising about $4\times10^{22}$ Ti atoms and about the same number of D pairs. A somewhat higher fusion rate of $10^{-20}$ $s^{-1}$ would be readily detectable.

More generally speaking, we considered 100 mW to be a power level that is readily measurable. For instance, the release of 100 mW in a 3 g metal sample in ambient air at 25°C leads to a temperature increase of about 20°C. Translating to nanoscopic units: 100 mW = 100 mJ $s^{-1}$ = $6.2\times10^{11}$ MeV $s^{-1}$. For D-D fusion, 23.8 MeV of energy is released per reaction (considering ground state helium as a product), therefore the above corresponds to $2.6\times10^{10}$ fusion reactions per second. A 3 g $TiD_2$ sample comprising about 4 × $10^{22}$ Ti atoms and approximately the same number of D pairs, then 100 mW of observed power release corresponds to a D-D fusion rate of about $10^{-12}$ $s^{-1}$ per D pair.

The above considerations suggest that D-D fusion rates in the range between $10^{-23}$ $s^{-1}$ to $10^{-12}$ $s^{-1}$ and higher can be considered readily observable—subject to experimental details and specific reaction products. Comparing this range with the spontaneous fusion rate in $D_2$ gas (~$10^{-64}$ $s^{-1}$ per deuteron pair at the molecular distance of 74 pm, as estimated by Koonin and Nauenberg 1989 [20]) sets a target for fusion rate enhancement of 40-50 orders of magnitude.

## 2.2 Nomenclature when referring to deuterium-deuterium fusion

The nomenclature for deuterons differs across scientific communities and across different framings of the fusion problem.

In nuclear physics, it is customary to refer to a deuterium nucleus as *d* and to a deuteron-deuteron fusion reaction as *d*+*d*. Here, the occurrence of collisions with accelerated deuterons is typically implied. The expression *d*+*d* is also sometimes used to refer to a 2+2 cluster configuration of nucleons in an atomic nucleus (see different states of the $^4$He nucleus, as discussed in detail in section 5 and specifically in 6.2). Other cluster configurations include the 3+1 configuration of nucleons, which serves as a precursor to what we refer to as 3+1 fusion. The phrase 3+1 fusion points to deuteron-deuteron fusion that results in either a triton (3) and a proton (1), or a $^3$He nucleus (3) and a neutron (1). It does not include the variant of deuterium-deuterium fusion that results in a $^4$He nucleus. The latter is of considerable interest throughout this document. For an explicit discussion of branching ratios, see 5.15.

In chemistry and materials science, deuterium atoms are referred to as D, which form $D_2$ molecules in the gas phase.

In quantum dynamics, a deuteron pair such as in the case of a deuterium molecule can be viewed as a single quantum system with a specific energetic state associated with it (at the atomic and at the nuclear level). Such a system can be referred to as $|D_2\rangle$.

In this article, we use the *d* terminology, when referring to bare nuclei such as in a collision framing. We refer to D, when describing a deuterium atom and DD when referring to a pair of deuterium atoms (which can be in the form of a molecule $D_2$). We use $|D_2\rangle$ terminology, when a quantum dynamics framing is explicitly adopted and to be highlighted. In the most general sense, we refer to deuterium-deuterium fusion as D-D fusion, a phrase that encompasses all conceivable framings of the process.

### 2.3 Earlier reports of energetic particle emission from metal-hydrogen systems at low energies

A number of articles report observations of energetic particles emitted from metal-hydrogen systems stimulated at energies far below those typically associated with the initiation of nuclear reactions.

A selection of such articles and a brief summary of their reported observations are given below:

In Chambers et al. 1990 [21], the authors report the observation of charged particles with energies ~28 MeV from Pd foils that were bombarded with deuterium ions at energies <1.5 keV.

In Takahashi et al. 1990 [22], the authors report the observation of neutron emission in the 3-7 MeV range from electrochemically loaded PdD samples stimulated by electric current pulses.

In Menlove et al. 1990 [23] the authors observed repeated bursts of neutron emission using high-efficiency neutron detectors in various forms of Pd and Ti metal in pressurized $D_2$ gas cells and $D_2O$ electrolysis cells.

In Chambers et al. 1991 [24], the authors report the observation of ~5 MeV charged particles from the bombardment of Ti foils with deuterium ions at energies of 350-400 eV.

In Ziehm 2024 [25], the author reports the observation of 138 keV charged particles from the bombardment of Pd foils with deuterium ions at energies <500 eV.

# 3 Nuclear fusion rate calculations

Here we review the standard approach to fusion rate calculations and show how it can be adapted it for the case of the spontaneous fusion of a deuteron pair (*e.g.*, in the form of a dideuterium complex, see section 6.21). We will also highlight the importance of electron screening for fusion in solid state. Throughout in the text, we will adjust our fusion rates based on experimentally observed screening values.

We provide more details on the fusion rate calculations in 6.3. We also provide a GitHub-hosted Python Notebook to go the step-by-step through the calculations on `https://github.com/project-ida/nuclear-reactions`.

## 3.1 General

Fusion can be described as a two-step process:

1. A quantum tunneling event through a potential barrier, with the barrier defined by the interatomic potential between two nuclei.
2. A relaxation or decay of the highly excited nuclei into some ground state or decay products (with the concomitant release of energy)

Step 1 is concerned with solving the Schrödinger equation to calculate the probability, $P$, that the the nuclei have passed the Coulomb barrier and are within each other's nuclear volume. Step 2 is concerned with nuclear physics and proceeds at a rate $\gamma$ that's typically extremely fast (on the order of $10^{-20}$ s$^{-1}$).

The fusion rate per pair of nuclei can then be simply written as:

$$\Gamma = P\gamma \tag{1}$$

This two-step approach is the most accurate way to calculate the fusion rate and we will use it in section 6.3 (based on the numerical integration of wave functions introduced in section 6.1). For the purposes of including the effect of electron screening, we'll work with a simpler description, which is however commonly used - fusion as a collision process - and we'll adapt it to the spontaneous fusion of two deuterons.

## 3.2 Fusion upon collision

A variety of approaches for performing fusion rate calculations based on a two-body tunneling model are provided by multiple authors [26, 27, 28, 29, 30, 31, 32, 33, 34, 35, 36, 37, 38]. Central to all these approaches of fusion rate calculations is the tunneling factor *T* that represents the probability of two fusing nuclei reaching the range of strong force attraction (fm range). *T* is also referred to as the barrier penetration factor.

The tunneling factor *T(d,Ue)* depends on the fusion reactants' proximity *d* as well as the screening potential in their vicinity *Ue*. The defining difference between the received approaches is the choice of interatomic potential *V*, and to a lesser but still significant extent the choice of nuclear potential. The interatomic potential can be approached as a numerical potential [20, 39] or as a parameterized potential [40, 41] for numerical evaluation of the integral that determines *T*. Analytical expressions of the integral that serve as approximations are given by some authors [37, 40, 38].

In the collision picture, a fusion event occurs instantaneously when two nuclei collide with enough energy, $E$, to tunnel to within each others' nuclear cross section $\sigma(T)$. The fusion rate for the case of a single nucleus colliding at speed $v$ into a target with density $n$ is the particle flux multiplied by the fusion cross section:

$$\Gamma = n\sigma(T)v \tag{2}$$

## 3.3 Spontaneous fusion from tunneling

Although there are no collisions in the spontaneous fusion of a deuterons in close proximity, we can adapt the fusion rate expression in Eq. 2 through a helpful interpretation of $n\sigma v$:

- Without the tunneling factor, $\sigma v$ represents the "reactable nuclear volume" that's swept out by a colliding nucleus per unit time.
- $n\sigma v$ then counts how many nuclei are in this reactable volume to give us number of reactions per second.
- The tunneling factor effectively reduces the density of particles in the reactable volume due to the repulsion between the nuclei.

For the case of a dideuterium complex or a $D_2$ molecule:

- The reactable volume is just the volume of the nucleus $v_{nuc}$
- The rate at which particles in this volume react is given by $\gamma$
- The density of nuclei is calculated by the inverse the volume occupied by a single molecule $1/v_{mol}$.
- The effective density of nuclei in the reacting volume is $T/v_{mol}$ due to the repulsion of the nuclei.

This results in a fusion rate of:

$$\Gamma = \frac{v_{nuc}}{v_{mol}} T\gamma \tag{3}$$

## 3.4 Electron screening

In a first approximation, electron screening of nuclei can be expressed as a correction factor ($e^{-r/a}$) applied to the Coulomb potential (*V(r)*), where *a* is the screening length [42, 43, 44]. The screening length in turn consists of a constant multiplied by $n^{-1/6}$, where *n* is the free electron density of the metal. For palladium, *n* is approximately 3.15 Å$^{-3}$ [45]. The resulting screened Coulomb potential is now expressed as:

$$V(r) = \frac{Z_1 Z_2 e^2}{r}\, e^{\frac{-r}{a}} \tag{4}$$

This approach requires the assumption that the free electron density of the metal is treated as a Fermi gas [46]. Application of this approach to a hypothetical system consisting of a $D_2$ molecule within a palladium vacancy results in the contraction of interatomic distance from 74 pm to 57 pm [47]. The screening length, *a*, can be expressed as terms of a screening energy $U_e = Z_1 Z_2 e^2/a$.

For high beam energies $E \gg U_e$, the screened potential can be simplified to a subtraction by a constant *Ue*:

$$V(r) = \frac{Z_1 Z_2 q_e^2}{r} - U_e \tag{5}$$

The simplification is allowed because the classical tuning points for tunneling satisfy $r \ll a$ in the high energy case. For very low energies, the full exponential must be used.

Other considerations—such as the impact of positive ions on the screening potential as well as dynamic effects—can be introduced to the Coulomb potential as additional corrections [48]. For the purpose of this discussion, they will be considered negligible and fusion rates will be considered through the context of a constant screening potential *Ue*, as introduced above.

A more detailed expression for the screening potential is given in Czerski et al. 2016 [49]:

$$U_e = e^2 k_{TF} = \frac{2e^3}{\hbar \pi^{1/3}} (3\pi n)^{1/6} \sqrt{m^*} \tag{6}$$

where *kTF* is the Thomas-Fermi wave number, $n$ the electron density, and $m^*$ the effective electron mass. In first approximation, the effective electron mass is the electron rest mass *me*. However, in some solid-state environments, the effective electron rest mass is higher due to local electronic band structure changes that result from specific geometries [50, 51, 52, 53].

The two electrons of a gas phase hydrogen (or deuterium) molecule correspond to a screening potential *Ue* of ~25 eV [54]. Reported theoretical *Ue* values range from 50 to 150 eV for different metals, with Li at the lower range and Pd at the upper range (see Figure 1 and [43]). While theoretically predicted screening energies have been calculated for a variety of metals, it should be noted that screening energies derived from experimental data (based on observed fusion rates) are higher than predicted values [55]. Experiments to determine screening energies have been carried out by multiple groups [56, 57, 58] and typically involve low-energy deuteron bombardment of different metal targets with concurrent measurement of resulting nuclear byproducts (*e.g.*, neutrons and charged particles). For the same materials, where the theoretical screening energy range is given as 50 to 150 eV, the experimental screening energy range reported is 150 to 300 eV (Figure 1) [43, 59] and beyond [51]. In other words, the experimentally observed fusion rates are substantially higher than expected if one were to only consider proximity and screening within the Gamow model. The typical approach in the literature is to parameterize such discrepancies and include them in the phenomenological correction factor *A* without understanding all aspects of them causally.

Theoretical models for screening typically assume effects due to free electrons, which can begin to account for the large screening observed in low-energy ion beam experiments. The screening energies predicted from such models are substantial for relative deuteron kinetic energy in excess of 1 keV, but is much smaller at zero relative energy [58]. Note that the $(1s)^2$ electron orbitals, which are relevant to two deuterons in close proximity, are unlikely candidates for acting as free electrons.

It may be that an alternate mechanism is relevant in this context. A deuteron pair (*e.g.*, in the form of a dideuterium complex, see section 6.21) in proximity to a Pd atom can be imagined to undergo more complex dynamics, where both of the deuterons tunnel into the Pd electron orbitals before tunneling through the Coulomb barrier between them. In this kind of scheme, there is the possibility of screening due to bound electrons at much higher electron density (and therefore stronger screening). It would be possible to model this kind of screening with a special purpose quantum chemistry code that is constructed for or adapted to the problem. This would represent another mechanism for enhanced screening—however, this conjecture is yet to be explored in more detail.

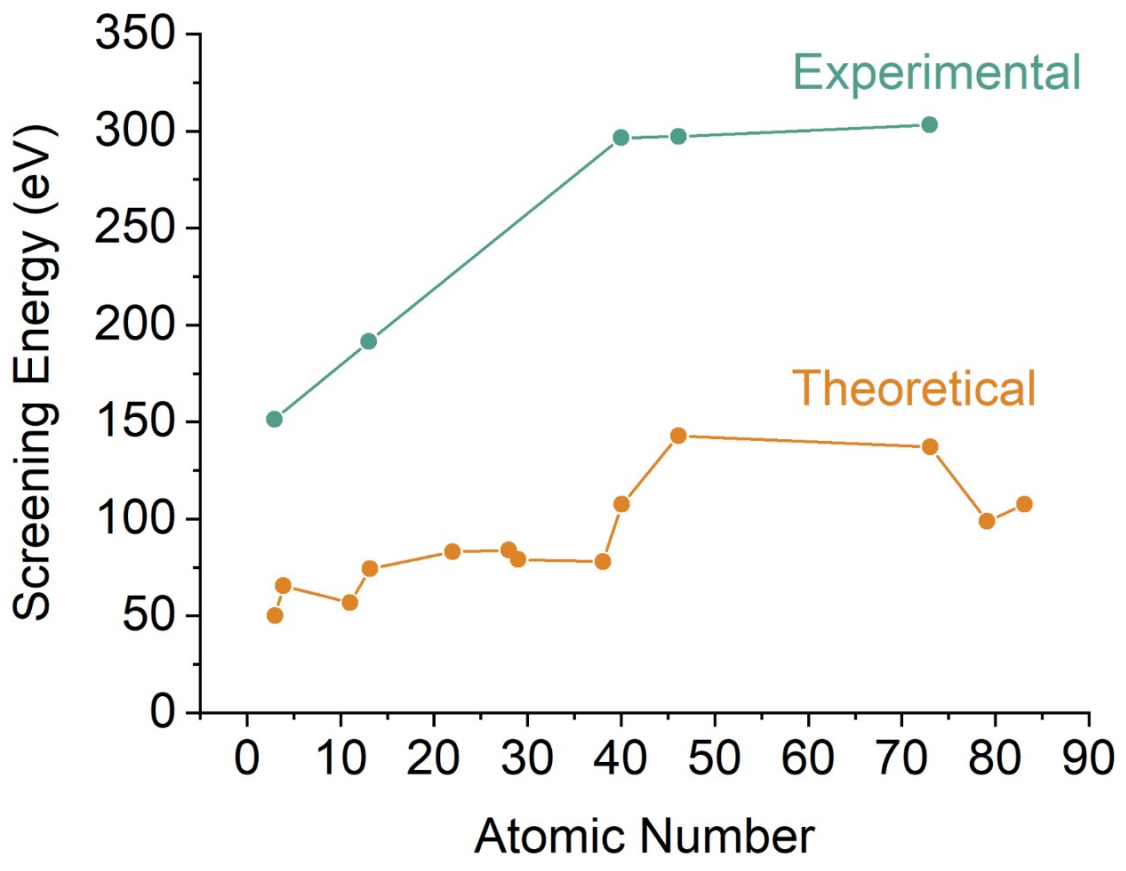

**Figure 1:** Reported experimental (orange) and theoretical (teal) screening energy values. Reproduced with permission from Huke et al. 2008 [43].

# 4 Quantum dynamics of nuclear fusion in condensed matter (overview)

Nuclear fusion can be modeled in first approximation as a two-level system (TLS) undergoing a spontaneous transition from a (highly) metastable excited state to the ground state at a rate corresponding to what we can refer to as the fusion rate. In other words, an excited TLS (*i.e.*, the fusion reactants) relaxes to its ground state (*i.e.*, the fusion products) [13].

If the number of final states is large enough to be represented as a continuum of states, the so-called Golden Rule can be applied to extract a rate, as shown below [60]. That is indeed the case in the conventional thermonuclear fusion picture, where the possible final states represent a continuum (fusion products as particles with momentum in all directions of space). The equivalence of modeling tunneling via the Wentzel–Kramers–Brillouin (WKB) approximation, *i.e.*, the Gamow factor approach (as discussed in section 3 and shown in more detail in section 6.3) and modeling spontaneous emission via the Golden Rule (as shown here) is explicated by Raju [61].

The simplest Hamiltonian representation of a TLS (*HTLS*) that relaxes from an initial excited state (*i*) to its final ground state (*f*) is

$$H_{TLS} = \begin{bmatrix} i & V \\ V^* & f \end{bmatrix} \tag{7}$$

where $V$ represents the coupling between the states *i* and *f*. A corresponding diagram is shown in Figure 2. The described treatment of the D-D system as a two-level quantum system is discussed in greater depth by Hagelstein [62, 63] and in section 5.1.

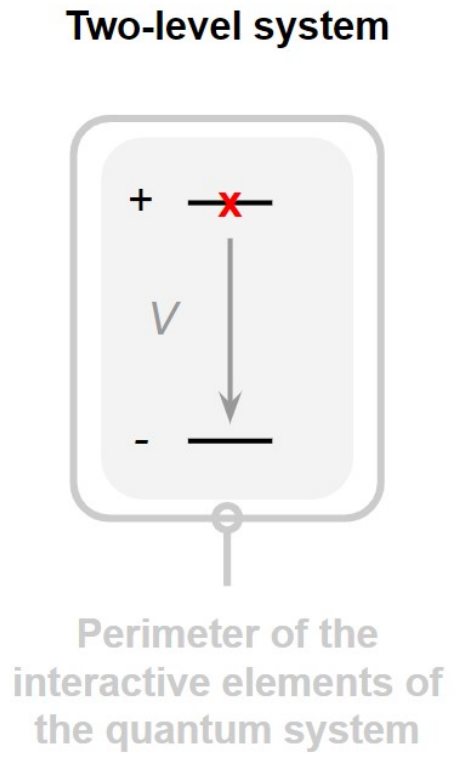

**Figure 2:** Essential features of a two-level system (TLS): An excited state (+) is coupled to a ground state (-) whereby the coupling is represented by *V*. A perimeter encompasses all interacting elements of the quantum system, which in this case is only the TLS itself.

Now consider the $^4$He + $\gamma$ channel of D-D fusion: the photon can be emitted in many angular configurations effectively making the number of final states infinite and leading to irreversible dynamics [64] (practically zero probability of reabsorption). In this case, the Hamiltonian matrix more accurately is

$$H_{GoldenRule} = \begin{bmatrix} i & V & V & V & V & V & \dots \\ V & f_1 & 0 & 0 & 0 & 0 & \dots \\ V & 0 & f_2 & 0 & 0 & 0 & \dots \\ V & 0 & 0 & f_3 & 0 & 0 & \dots \\ V & 0 & 0 & 0 & f_4 & 0 & \dots \\ V & 0 & 0 & 0 & 0 & f_5 & \dots \\ \vdots & \vdots & \vdots & \vdots & \vdots & \vdots & \ddots \end{bmatrix} \tag{8}$$

The transition rate ($\Gamma$), per the Golden Rule, from an initial state $i$ to a continuum of final states $f$ is [60]

$$\Gamma_{i\to f} = \frac{2\pi}{\hbar}|\langle f|H|i\rangle|^2 \tag{9}$$

With the latter term taking on the value of

$$\langle f|H|i\rangle = V \propto e^{-G} \tag{10}$$

where the transition matrix element $V$ between $i$ and $f$ is represented by the inverse exponential of the Gamow factor ($G$), which represents the penetration probability. The small value of $V$ represents the low probability of the nuclear transition. This treatment results in an equivalency between the transition rate and the fusion rate for a two-nucleus system with their characteristic exponential probability distribution (Figure 3)

$$\Gamma_{i\to f} \propto e^{-2G} \tag{11}$$

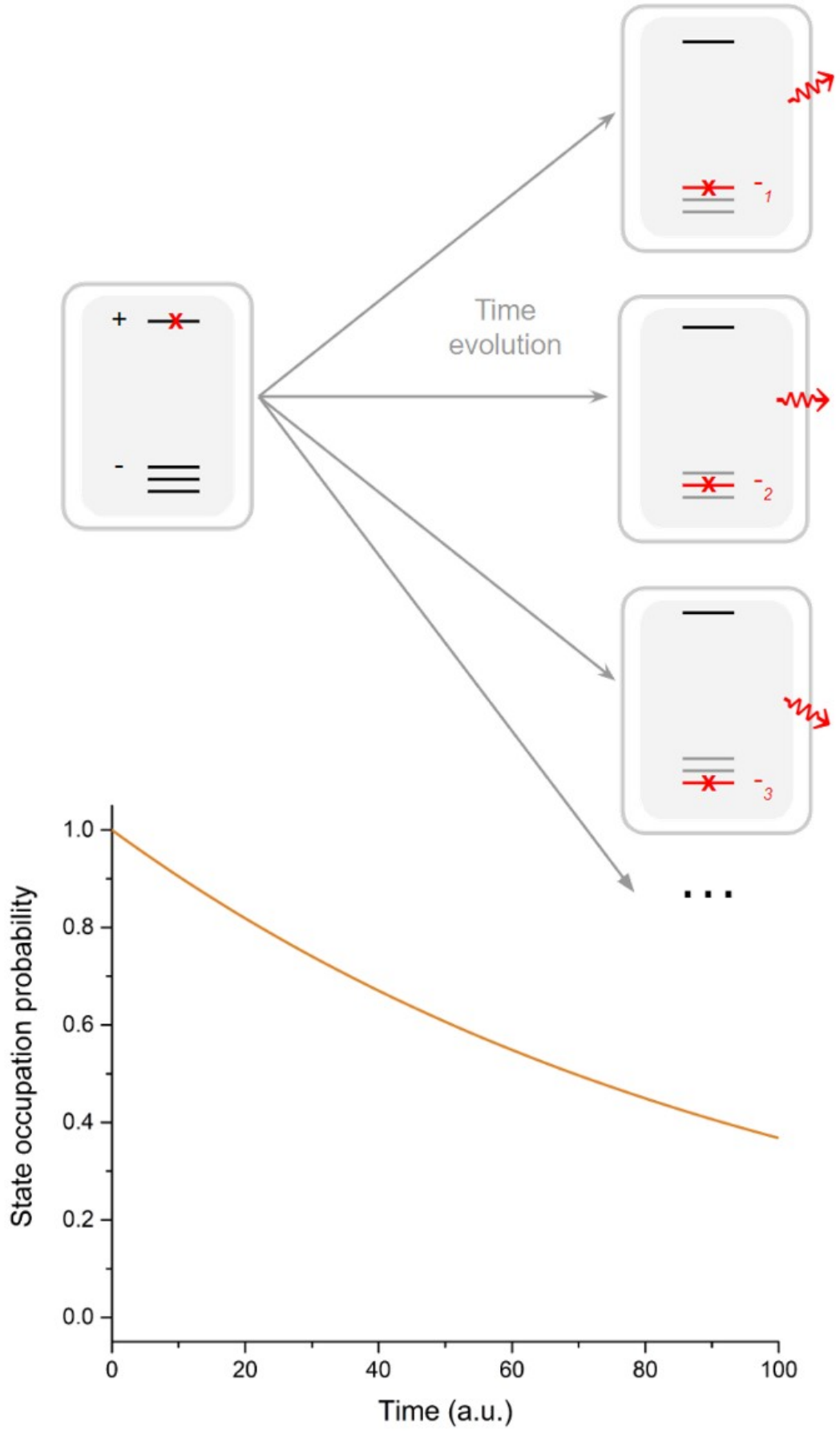

**Figure 3:** A two-level system (TLS) with a continuum of final states: when a TLS can relax into a near-infinite number of final states, the transition probability takes the form of a decaying exponential per the Golden Rule expression (Eq. 11). Here, the near-infinite number of final states are represented by small differences in final state energies and in differences in the angular orientation of resulting photon emission.

$\Gamma i \to f$ represents a decay channel that is always available at the given rate in a two-nucleus system.

However, it is not necessarily the only decay channel. If a coupling to a resonant or near-resonant quantum system exists, additional dynamics must be considered. Specifically, in the case of a resonance with a receiver system (sometimes referred to as an acceptor system), energy transfer can occur (see for instance the nuclear quantum dynamics described in [65]).

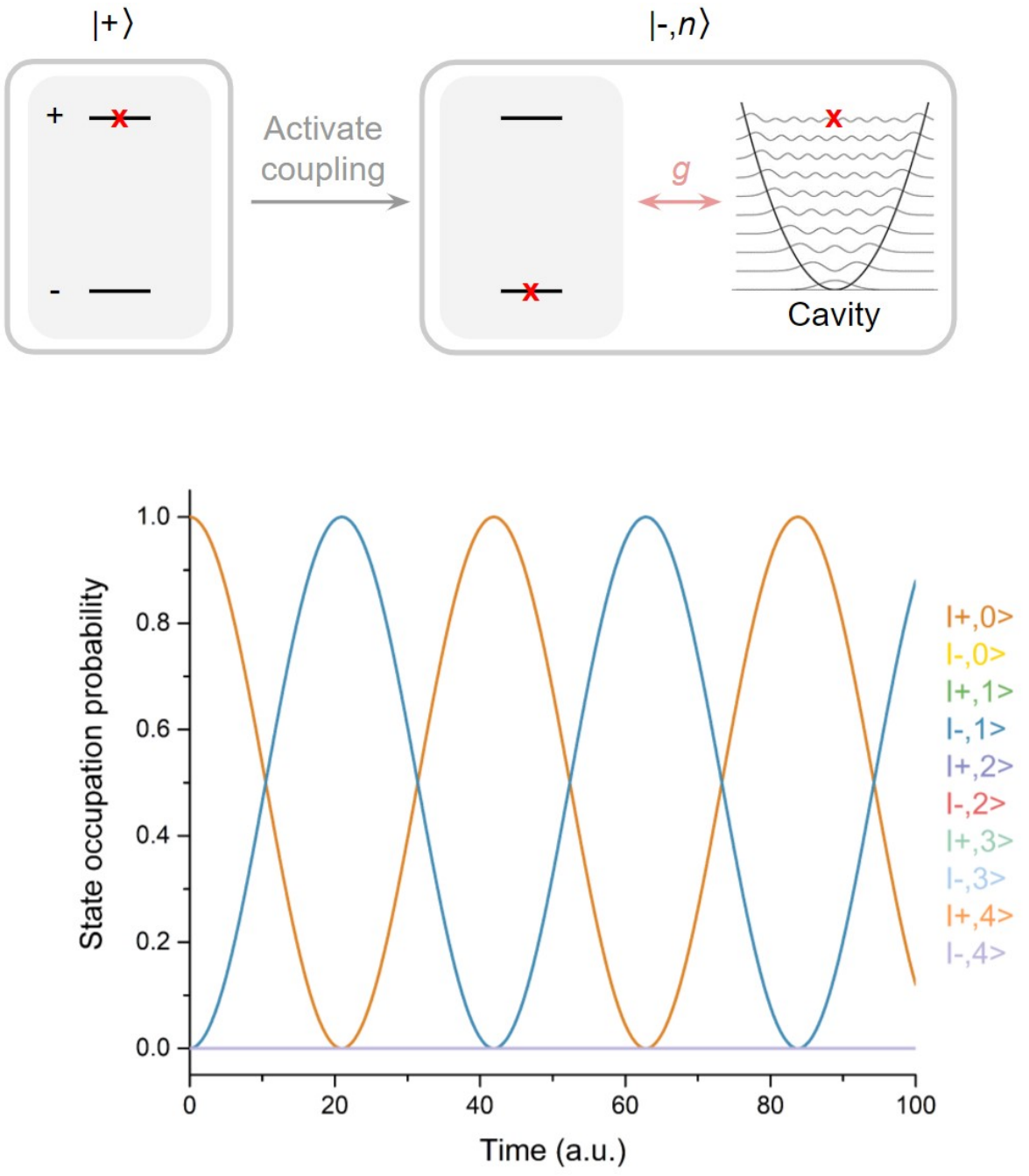

**Figure 4:** A two-level system (TLS) coupled to a resonant oscillator that allows for direct energy exchange between the two, resulting in Rabi oscillations between states $|+, 0\rangle$ and $|-, 1\rangle$. The simulation is based on the quantum dynamics Python library QuTiP and available at: https://github.com/project-ida/two-state-quantum-systems/

This can give rise to complicated dynamics, including linear and nonlinear Rabi oscillations, where occupation probability oscillates back and forth between donor and receiver systems. Generic examples of receiver systems in this context include oscillators and additional TLSs.

To first approximation, the Rabi frequency, which characterizes the oscillation rate, depends on the coupling strength and detuning between the two systems [66]. It can be expressed as:

$$\Omega_{\text{Rabi}} = \sqrt{|\Omega_0|^2 + \Delta^2} \tag{12}$$

Here, $\Omega_0$ represents the coupling strength, and $\Delta$ is the detuning parameter, defined as the difference in transition frequencies between the two systems:

$$\Delta = \omega_{\text{system 1}} - \omega_{\text{system 2}} \tag{13}$$

When the two systems are in resonance ($\Delta = 0$), the Rabi frequency reduces to the coupling strength $\Omega_0$, resulting in maximum energy transfer between the states.

Bringing this back to the nuclear fusion context, if a fusion transition can be coupled to one or multiple receiver nuclei (acceptor nuclei) with matching transitions, and if the transfer rate of that process is faster than the spontaneous decay rate, then there is an alternative fusion pathway.

We will later see that $\Omega$ is a function of the fusion hindrance factor $e^{-G}$ but also of other variables that can be used to increase that frequency.

For a TLS coupled to an oscillator, a so-called Rabi Hamiltonian can be used to describe the dynamics [67]:

$$H_{Rabi} = \overbrace{\hbar\omega_0\sigma_z}^{tls} + \overbrace{\hbar\omega a^\dagger a}^{osc} + \overbrace{V\left(a^\dagger + a\right)\sigma_x}^{interaction} \tag{14}$$

Here $\hbar\omega_0$ is the transition energy in the TLS, $\hbar\omega$ is the energy in the oscillator mode, and $V$ refers to the interaction between them. $\sigma x$, $\sigma z$ are Pauli matrices and $a\dagger$ and $a$ are creation and annihilation operators. A resonant oscillator would correspond to the case where $\omega_0 = \omega$.

The time-evolution of such a resonant system yields an occupation probability that Rabi-oscillates between the TLS and the oscillator (see Figure 4). While the above implementation illustrates what a coherent D-D fusion channel could look like, no oscillator can be practically implemented at this time that can readily absorb the 23.85 MeV energy that would result from the $|D_2\rangle \rightarrow |^4He\rangle$ reaction.

However, a modified version of such a system can be implemented, where another TLS with a resonant excited state (such as another nucleus) absorbs the released energy and where a non-resonant oscillator mediates the transfer.

The dynamics of the general case with many such resonant TLSs can be described with a so-called Dicke Hamiltonian:

$$H_{Dicke} = \hbar\omega_0 \sum_{j=1}^{N} \sigma_z^j + \hbar\omega a^\dagger a + V \sum_j \sigma_x^j (a + a^\dagger) \tag{15}$$

Here index $j$ counts over $N$ nuclei and $N$ corresponding interaction terms with the shared oscillator mode.

Time evolution for the specific case of two resonant TLSs yields an occupation probability that Rabi oscillates, but this time between donor TLS and receiver TLS where the oscillator merely mediates the transfer rather than fully participating in it (Figure 5). In such a case of non-radiative transfer, the mediating oscillator can facilitate the transfer of a large quantum of excitation (in our case 23.8 MeV) without ever having to hold the entirety of that energy itself [68, 69, 70]. This is a critical point, since many readers may intuitively assume that in such a context, the oscillator needs to be able to fully absorb the energy quanta that are to be transferred. That is, however, not the case, as can be shown simply from time-evolving corresponding versions of the well-known Dicke model (as demonstrated in the Python Notebooks hosted at https://github.com/project-ida/two-state-quantum-systems/).

The coupling between nuclei in that case is indirect, *i.e.*, via the mediating oscillator modes, whereas the coupling from each nucleus to the oscillator mode is direct. Examples of oscillators relevant in this case are phonons and plasmons.

Note that a key feature of such a Dicke model related transfer scheme is the acceleration of the transfer rate as a function of the number of participating systems (see section 5.2).

When considering $|D_2\rangle$ as a donor system in such excitation transfer dynamics (a TLS in the excited state, capable of undergoing a $|D_2\rangle \rightarrow |^4He\rangle$ transition with its 23.85 MeV transition energy), then a resonant receiver system is needed as well as a shared oscillator mode that both systems are coupled to. A ground state $^4He$ nucleus offers a perfectly resonant excited state via the $|^4He\rangle \rightarrow |D_2\rangle$ transition. However, in this case, the donor side $|D_2\rangle \rightarrow |^4He\rangle$ transition and the receiver side $|^4He\rangle \rightarrow |D_2\rangle$ transition would remain within a closed system, *e.g.*, $|D_2\ {}^4He\rangle$ turns into $|^4He\ D_2\rangle$. Accordingly, the process would be difficult to be observed for lack of clear reaction products—and it would also likely still be too slow to reach the observable range (see section 5.3 for corresponding rate estimates based on concrete parameters).

A key reason for the comparatively slow rates, despite coherent dynamics, is the fact that the Coulomb barrier, as a key hindrance to the involved transitions, comes into the rate equation twice per transfer: once on the donor side $|D_2\rangle \rightarrow |^4He\rangle$ transition and once on the receiver side $|^4He\rangle \rightarrow |D_2\rangle$ transition.

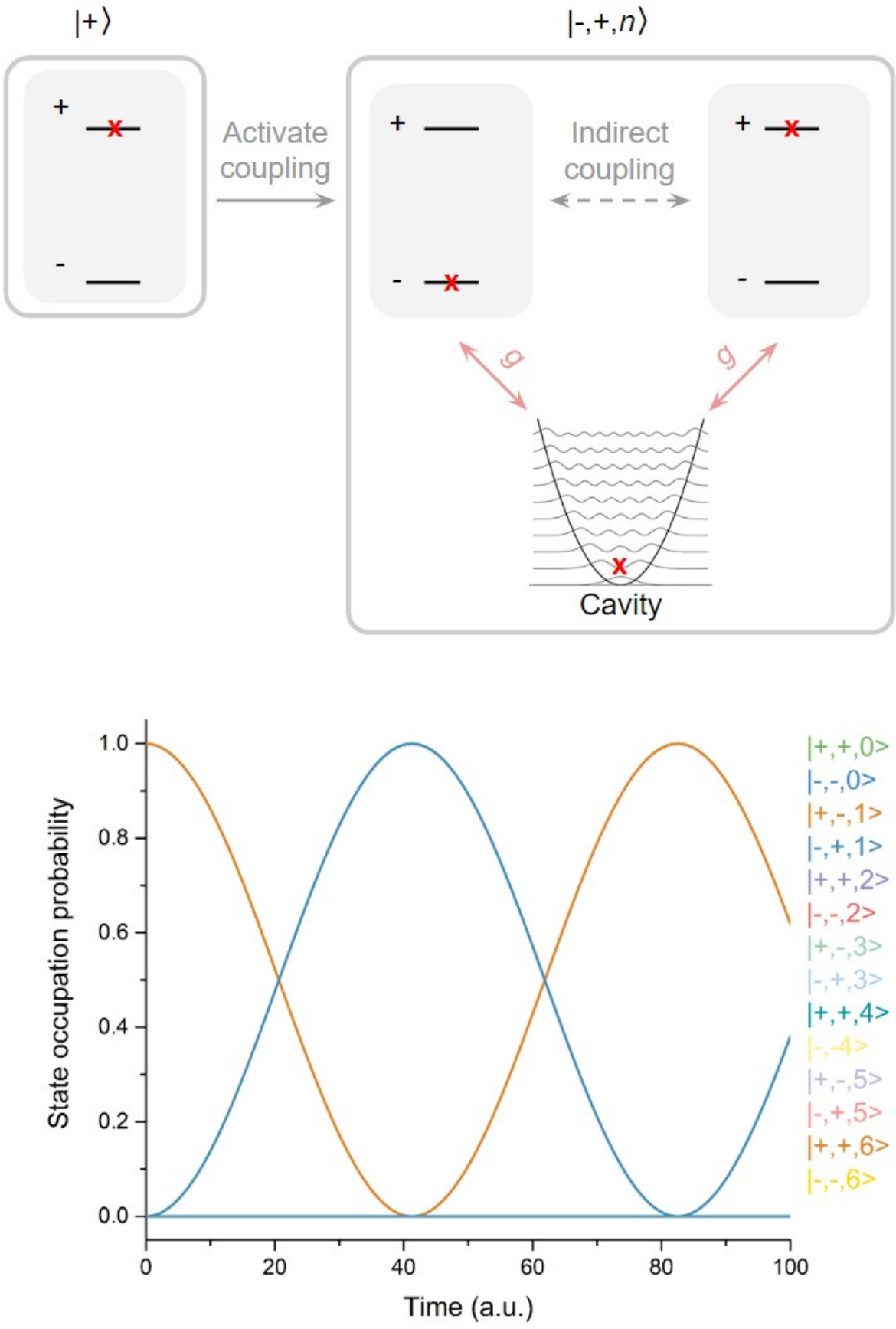

**Figure 5:** Two two-level systems (TLS) coupled to a shared oscillator mode with resonant TLS states: the coupled resonant TLS allows for indirect energy exchange between the two TLSs via the shared oscillator mode, resulting in Rabi oscillations between states $|-,+,1\rangle$ and $|+,-,1\rangle$. The simulation is based on the quantum dynamics Python library QuTiP and available at: https://github.com/project-ida/two-state-quantum-systems/

This is not the case with many alternative receiver systems. In section 5.4, we will consider Pd nuclei as candidate receiver systems. Transfer to Pd on the receiver side instead of $^4$He is expected to result in faster rates, since transitions from Pd ground states to excited states $|\mathrm{Pd}\rangle \rightarrow |\mathrm{Pd}^*\rangle$ are less hindered than the $|^4\mathrm{He}\rangle \rightarrow |\mathrm{D_2}\rangle$ transition. In that case, while there is an analogous transition hindrance factor that we will refer to as the $O$-value (section 5.4), this hindrance factor is many orders of magnitude smaller compared to the impact of the deuterium-deuterium Coulomb barrier on the receiver side (see section 5.5 and Figure 23).

Considering a different nuclear species as receiver systems raises the question of how to achieve resonance with the donor transition. While it is in principle conceivable that a Pd nucleus exhibits a suitable excited state precisely near the resonance condition of 23,848,109 eV, such a coincidence would be unlikely. However, as will be discussed in section 5.7 and subsequent sections, combinations of multiple lower-energy (and longer-lived) Pd excited states result in an extremely high density of states near 23.85 MeV and above, allowing for the resonance condition to be met.

In practice, other issues need to be considered such as the effect of high-power transients of the oscillatory modes of interest, which lead to a temporary increase in coupling strengths (due to the dependence of the coupling strengths on the dissipated power in the oscillator modes, see section 5.1). High-power transients can arise from shocks deliberately applied to the lattice, *e.g.*, through laser pulses, and also from internal lattice dynamics, *e.g.*, phase changes. The impact of such high-power transients on evolving excitation transfer dynamics will be considered in section 5.6.

As for reaction products, working with a different nuclear species than $^4$He as receiver systems offers other advantages with respect to observables. If the receiver nuclei were to disintegrate upon receipt of the transferred excitation—thus turning the closed quantum system into an open one—the process becomes observable. Many heavy nuclei can resonantly absorb energies near 23.85 MeV (and higher) and would promptly decay via alpha, proton, or neutron emission upon receipt of such large quanta of excitation [71]. This final step at the end of a rather complex process of excitation transfer could then account for the reported energetic particle emission from metal-hydrogen systems, such as those summarized in section 2.3.

An even more detailed treatment of the problem also recognizes that the density of states on the receiver side gets very high, when allowing for energy exchange between the nuclear states and the oscillator modes (section 5.10). In that case, there is not just a single discrete receiver state, as in the idealized example of transfer from $D_2$ donor systems to $^4$He receiver systems, but many possible states on the receiver side. This in turn requires a Golden Rule treatment when undertaking rate calculations and a corresponding model. We find that rates in such a model are particularly high and argue that such a model may be well suited to connected with the kinds of experimental reports mentioned above. Energy exchange from excited nuclear states with oscillator modes of the lattice—when there is a high density of states that nuclear systems can traverse as they emit energy into phonon and plasmon modes—provides an intrinsic explanation for so-called excess heat effects observed in metal-hydrogen systems. In such experiments, researchers have reported the accumulation of heat far beyond the applied input energy and the chemical potential energy in the system.

The discussions in this section sought to provide a broad overview of nuclear excitation transfer dynamics. A more detailed and comprehensive treatment with estimates for real systems is provided in section 5.

# 5 Quantum dynamics of nuclear fusion in condensed matter (detailed treatment)

Section 4 laid out a basic conceptual overview of relevant dynamics. This section here will provide a more detailed treatment that goes deeper into individual points raised. It will present arguments for how, in specific kinds of metal-hydrogen systems, known quantum dynamical energy transfer mechanisms—in combination with electron screening and molecule-like proximity of hydrogen isotopes—can result in greatly accelerated fusion rates.

Some of the quantitative examples in this section are idealized and primarily pedagogical in nature, seeking to help the reader appreciate different aspects of this complex problem. Other quantitative examples are based on realistic assumptions, and we argue that the resulting fusion rate estimates in the observable range align with reports of energetic particle production as well as excess heat production in metal-hydrogen experiments such as the ones briefly discussed in section 2.

In their development, these arguments have undergone many iterations—and we expect further refinement and expansion in the future. Nevertheless, we are confident that the mechanisms laid out here, and their combined effects in specific metal-hydrogen systems, represent a viable approach to engaging with experimental reports productively. We propose that this approach makes solid-state fusion actionable not only from an experimental perspective (such as with the ARPA-E program on low-energy nuclear reactions that commenced in 2023 [72]—see section 5.16) but also from a theoretical and modeling perspective.

As we make more progress in developing and refining these arguments, we will provide updated versions of this document (and subsections of it) via our group's GitHub site at `https://github.com/project-ida`. This includes links to evolving preprints as well as computational notebooks and a regularly updated list of related published materials.

### Nomenclature and visual language

This section will make use of figures that include schematic representations of atomic and nuclear structure. We therefore adopt different visual languages to represent systems such as $D_2$, $^4$He, $^A$Pd, $^A$Pd* (excited states of Pd isotopes, as discussed in 5.4).Figure 6 summarizes these visual representations at atomic and nuclear scales.

We will frequently refer to "transitions" in this document, and the usage of this word, along with related notations, requires some elaboration. Depending on the context, "transition" may be used in a static manner, referring to the energy level spacing in a two-level or multi-level system. For instance, the level spacing between a palladium excited state (Pd*) and a palladium ground state (Pd) represents a transition in that sense of the word.

In such cases, we use the notation Pd*/Pd to denote the upper and lower energetic state of the system. Since $D_2$ is viewed as an excited state of $^4$He in this document, we often refer to the $D_2$/$^4$He transition in this way (which is associated with the 23.85 MeV mass defect energy that defines the level spacing).

The other use of "transition" is a more dynamic one, referring to an actual state change rather than to the static configuration of energy levels. For instance, the Pd*/Pd transition can undergo a state change, when an occupied excited state decays to the ground state. In this case, we use the notation Pd* $\rightarrow$ Pd to represent this process, and we refer to it as a Pd* $\rightarrow$ Pd transition. The reverse of the latter is a Pd $\rightarrow$ Pd* transition, in the case of an excitation instead of a deexcitation.

In most instances, the meaning of "transition" will be clear from the context. However, where ambiguity might arise, we will explicitly refer to "two-level system" to indicate the static use of the term and "state change" to indicate the dynamic use.

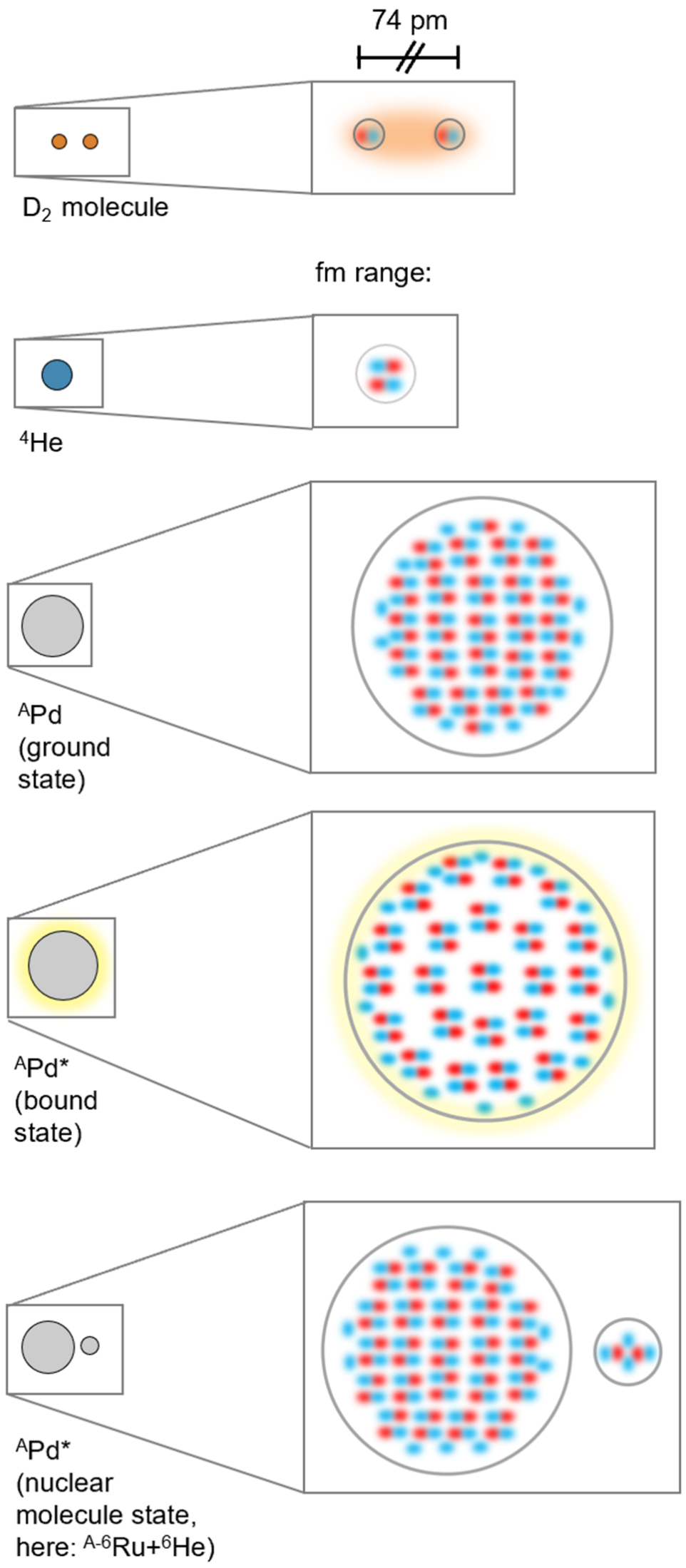

**Figure 6:** Different visual representations at atomic and nuclear scales used in this document. The A in $^A$Pd can be used to refer to a specific Pd isotope whereas the use of the expression "$^A$Pd" indicates that any of the (stable) Pd isotopes can be considered here.

## 5.1 From a comprehensive Hamiltonian describing nuclei in a lattice to a nuclear Dicke model

The starting point for this exploration of nuclear quantum dynamics is a comprehensive Hamiltonian that describes nuclei in a lattice as well as various oscillator modes that the nuclear states couple to (weakly).

This starting point is consistent with Terhune and Baldwin's [8] perspective who in a 1965 Physical Review Letter laid out a basic approach to the modeling of nuclear dynamics in a lattice with a single comprehensive Hamiltonian that accounts for nuclei, lattice modes, radiation, and interactions. The need for such a comprehensive Hamiltonian is motivated by them as follows:

> *"In a solid composed of N identical two-level nuclei in a perfect crystal lattice at a uniform and low temperature, correlations in the internal motions of the radiators are more probable than in the case of a, gas. [..] The usual assumption' that each nucleus radiates independently of the states of other nuclei in the system is incompatible with the coupling of the nuclei through the common electromagnetic and phonon fields."*

The most generic form of the comprehensive Hamiltonian that Terhune and Baldwin provide is:

$$\hat{H} \;=\; \hat{H}_{nuclei} + \hat{H}_{lattice} + \hat{H}_{radiation} + \hat{V}_{interaction} \tag{16}$$

A more specific form of the comprehensive Hamiltonian used here is:

$$\hat{H} \;=\; \hat{H}_{nuclei} + \hat{H}_{phonons} + \hat{H}_{plasmons} + \hat{H}_{magnons} + \hat{V}_{nuclei,phonons} + \hat{V}_{nuclei,plasmons} + \hat{V}_{nuclei,magnons} \tag{17}$$

In this equation, there is a Hamiltonian to account for ground states and excited states of the nuclei in the system; there are condensed matter Hamiltonians for phonons, plasmons and magnons; and there are interaction terms that describe coupling between the nuclear states and the phonons, plasmons and magnons in the system.

Since such a comprehensive Hamiltonian describes a multitude of dynamics and is essentially intractable, it will be gradually simplified in this section with the goal of identifying key dynamics that under specific circumstances yield observable effects. For the sake of simplicity, we will start by dropping the magnon terms. The focus for calculations in this document will be on phonons, but extensions for plasmons are discussed in section 6.27. The potential role of magnons will have to be explored in future work.

The nuclei can be represented as in Hagelstein 2024 [73]

$$\hat{H}_{nuclei} \;=\; \sum_{j,k} \Big( |\Phi_j\rangle M_j c^2 \langle\Phi_j| \Big)_k \tag{18}$$

where the mass energies $M_j c^2$ of the different nuclear states $j$ at the different sites $k$ in the lattice are written in a multi-level system notation (in contrast to two-level system notation). $\Phi_j$ represents nuclear states in terms of internal degrees of freedom.

The focus here will be on phonons and plasmons whose Hamiltonians are of the form

$$\hat{H}_{phonons} \;=\; \sum_{\mathbf{k},\nu} \hbar\omega_{\mathbf{k},\nu}^{(phonons)} \hat{a}_{\mathbf{k},\nu}^{\dagger} \hat{a}_{\mathbf{k},\nu} \tag{19}$$

$$\hat{H}_{plasmons} \;=\; \sum_{\mathbf{k},\nu} \hbar\omega_{\mathbf{k},\nu}^{(plasmons)} \hat{b}^{\dagger}_{\mathbf{k},\nu}\hat{b}_{\mathbf{k},\nu} \tag{20}$$

where **k** is the wave vector of the phonon and plasmon modes, $\nu$ is describes which mode is considered, and $\omega_{\mathbf{k},\nu}$ is the frequency of the respective mode.

When considering both electric and magnetic interactions associated with oscillations of the nuclei when vibrations occur, the interaction term can be expressed via a multi-level system formalism according to [74, 75]

$$\hat{V}_{nuclei,phonons} \;=\; -\sum_{j,j',k}\left(|\Phi_j\rangle\langle\Phi_j|\mathbf{d}\cdot\hat{\mathbf{E}}|\Phi_{j'}\rangle\langle\Phi_{j'}|\right)_k - \sum_{j,j',k}\left(|\Phi_j\rangle\langle\Phi_j|\boldsymbol{\mu}\cdot\hat{\mathbf{B}}|\Phi_{j'}\rangle\langle\Phi_{j'}|\right)_k \tag{21}$$

where $\hat{\mathbf{E}}$ is the electric field strength, $\hat{\mathbf{B}}$ is the magnetic field strength, **d** is the electric dipole operator, $\boldsymbol{\mu}$ is the magnetic dipole operator, and where $j$ and $j'$ are the nuclear states included, and $k$ indexes the sites in space where the nuclei are positioned.

For simplicity, in this section, we will restrict ourselves to the case of electric dipole interactions. However, the overall dynamics explored here are in principle agnostic to the type of coupling that is focused on and we describe dynamics and provide rate estimates based on different couplings and different nuclear transitions later in this text (section 5.3 and subsequent sections).

In terms of nomenclature, we use $V$ here to refer to a coupling generically and we use subscripts to refer to specific couplings, *e.g.*, $V_{electric}$. Therefore, we can write here:

$$\hat{V}_{nuclei,phonons} \;=\; \hat{V}_{electric} \;=\; -\sum_{j,j',k}\left(|\Phi_j\rangle\langle\Phi_j|\mathbf{d}\cdot\hat{\mathbf{E}}|\Phi_{j'}\rangle\langle\Phi_{j'}|\right)_k \tag{22}$$

Electric coupling in this context can be thought of as follows:

Acceleration of an individual nucleus occurs because of the local oscillatory field in a classical picture according to

$$M_j\frac{d^2}{dt^2}\mathbf{R}_j \;=\; Z_j e\mathbf{E}_k \tag{23}$$

where $\mathbf{R}_j$ is the center of mass position, $Z_j$ is the nuclear charge, and *e* is the unit charge.

In the quantum version of the model, the position operator is expressed in terms of the different phonon modes according to

$$\hat{\mathbf{R}}(j,l) \;=\; \mathbf{R}(j,l) + \sum_{\mathbf{k},\nu}\mathbf{e}(j,\mathbf{k},\nu)\sqrt{\frac{\hbar}{2NM_j\omega_{\mathbf{k},\nu}}}\left(e^{i\mathbf{k}\cdot\mathbf{R}(j,l)}\hat{a}_{\mathbf{k},\nu} + e^{-i\mathbf{k}\cdot\mathbf{R}(j,l)}\hat{a}^{\dagger}_{\mathbf{k},\nu}\right) \tag{24}$$

where $l$ denotes the unit cell and where $\mathbf{e}(j,\mathbf{k},\nu)$ is the phonon polarization vector ([76]). $j$ index the positions of the nuclei, **k** is the phonon wave vector and $\nu$ indexes the phonon branch (*e.g.*, acoustic, optical, compressional and transverse).

A corresponding electric field operator can then be defined at a given nucleus position as a function of phonon modes

$$\hat{\mathbf{E}}(j,l) \;=\; -\sum_{\mathbf{k},\nu}\mathbf{e}(j,\mathbf{k},\nu)\sqrt{\frac{M_j\hbar\omega^3_{\mathbf{k},\nu}}{2NZ_j^2e^2}}\left(e^{i\mathbf{k}\cdot\mathbf{R}(j,l)}\hat{a}_{\mathbf{k},\nu} + e^{-i\mathbf{k}\cdot\mathbf{R}(j,l)}\hat{a}^{\dagger}_{\mathbf{k},\nu}\right) \tag{25}$$

### Focus on spatially uniform interaction

A first glance at the overall Hamiltonian draws attention to several critical aspects: the coupling energies in the $-\mathbf{d}\cdot\mathbf{E}$ and $-\boldsymbol{\mu}\cdot\mathbf{B}$ interaction terms are comparatively small and the state transition energies in the nuclear terms are comparatively large. The small coupling energies have to do with the small sizes of nuclei, which limit their electric dipole moments.

Large transition energies in the MeV range are typical for nuclear state transitions and intrinsic to nuclear physics. Oscillator transitions in turn are near an eV and less. Consequently, first-order processes (emission or absorption of single phonons, plasmons, or magnons) are not expected to be important—which agrees with basic intuition and experience because, due to the large mismatch, energy cannot be conserved in a first-order exchange process. However, energy can be conserved in second-order processes and also in higher order processes. The focus below will be on second-order oscillator-mediated excitation transfer processes. Later on we will make heavy use of higher-order generalized excitation transfer processes (section 5.10).

Despite the relevance of excitation transfer and higher-order processes, dynamics can be expected to be slow due to the weak coupling strengths between nuclear state transitions and condensed matter degrees of freedom (*e.g.*, phonons, plasmons, and magnons). However, Dicke enhancement [77] can compensate for this and accelerate dynamics. Dicke enhancement is expected in many-body quantum systems interacting with a uniform common oscillator. Originally applied to a coherently radiating gas, its relevance for accelerating nuclear state transitions and nuclear reaction parameters was first described in 1965 [8] and 1990 [78], respectively. In the context of excitation transfer, Dicke enhancement has been referred to as supertransfer [79].

Since only Dicke-enhanced interactions can be expected to yield observable effects in the given systems, the modeling focus will be reduced to such interactions. This results in a restricted Hamiltonian of the form

$$\hat{H} = \sum_{j,k} \Big( |\Phi_j\rangle M_j c^2 \langle \Phi_j| \Big)_k + \hbar\omega_A \hat{a}_A^\dagger \hat{a}_A + \hbar\omega_O \hat{a}_O^\dagger \hat{a}_O + \hbar\omega_P \hat{a}_P^\dagger \hat{a}_P - \sum_{j,j',k} \Big( |\Phi_j\rangle\langle\Phi_j|\mathbf{d}\cdot\hat{\mathbf{E}}|\Phi_{j'}\rangle\langle\Phi_{j'}| \Big)_k \quad (26)$$

In this idealization, the different nuclear states interact with a single (uniform) acoustic phonon mode ($A$), with a single (uniform) optical phonon mode ($O$), and with a single (uniform) plasmon mode ($P$) via the electric interaction $\mathbf{d}\cdot\hat{\mathbf{E}}$. The electric field operator is now uniform across all nuclei in the domain, and simplifies to

$$\hat{\mathbf{E}} \rightarrow -\mathbf{e}_A\sqrt{\frac{M_j\hbar\omega_A^3}{2NZ_j^2e^2}}\left(\hat{a}_A + \hat{a}_A^\dagger\right) - \mathbf{e}_O\sqrt{\frac{M_j\hbar\omega_O^3}{2NZ_j^2e^2}}\left(\hat{a}_O + \hat{a}_O^\dagger\right) - \mathbf{e}_P\sqrt{\frac{m_e\hbar\omega_P^3}{2N_ee^2}}\left(\hat{a}_P + \hat{a}_P^\dagger\right) \quad (27)$$

where $\mathbf{e}$ unit vectors account for the polarization and amplitude that are involved for a specific nucleus.

This model describes many multi-level nuclear systems (with large state transition energies) collectively and uniformly interacting with three different condensed matter oscillators at low energy. As such, its state transition and energy transfer dynamics are subject to cooperative Dicke enhancement similar to the supertransfer collective excitation transfer dynamics described in [79].

A remaining difference is that here we have multi-level systems and the typical Dicke model works with two-level systems. We will next describe how the above Hamiltonian can be further simplified to correspond to the generalized Dicke model.

### Nuclei idealized as two-level systems

Although the model above is already much simpler compared to the starting point, a further idealization will be helpful in understanding the associated coherent dynamics. If we replace the multi-level systems that describe the nuclear states with simpler two-level systems, then we end up with an idealization that is more amenable to analysis.

Considering a deuteron pair (such as in molecular $D_2$), the different nuclear states are shown schematically in Figure 7 (and also in Figure 3 of [80]). The allowed states include $^1$S, $^5$S, $^3$P, $^1$D, $^5$D, ... states, all of which (including sub-levels) can be described in an multi-level model formalism. Note that half the states that one may expect are missing because the generalized Pauli principle makes the construction of anti-symmetric states impossible. A more detailed discussion of the relevant $D_2$ states and their wave functions as well as the relevant $^4$He states and their wave functions is presented in section 6.1. A brief review of known excited states of $^4$He in the literature is provided in section 6.2.

In a two-level system description, as used below, this collection of states is replaced by a single (generic) $D_2$ molecule state as the excited state (highly clustered four-nucleon configuration) and an unexcited $^4$He nucleus as the ground state (compact four-nucleon configuration) of a two-level system.

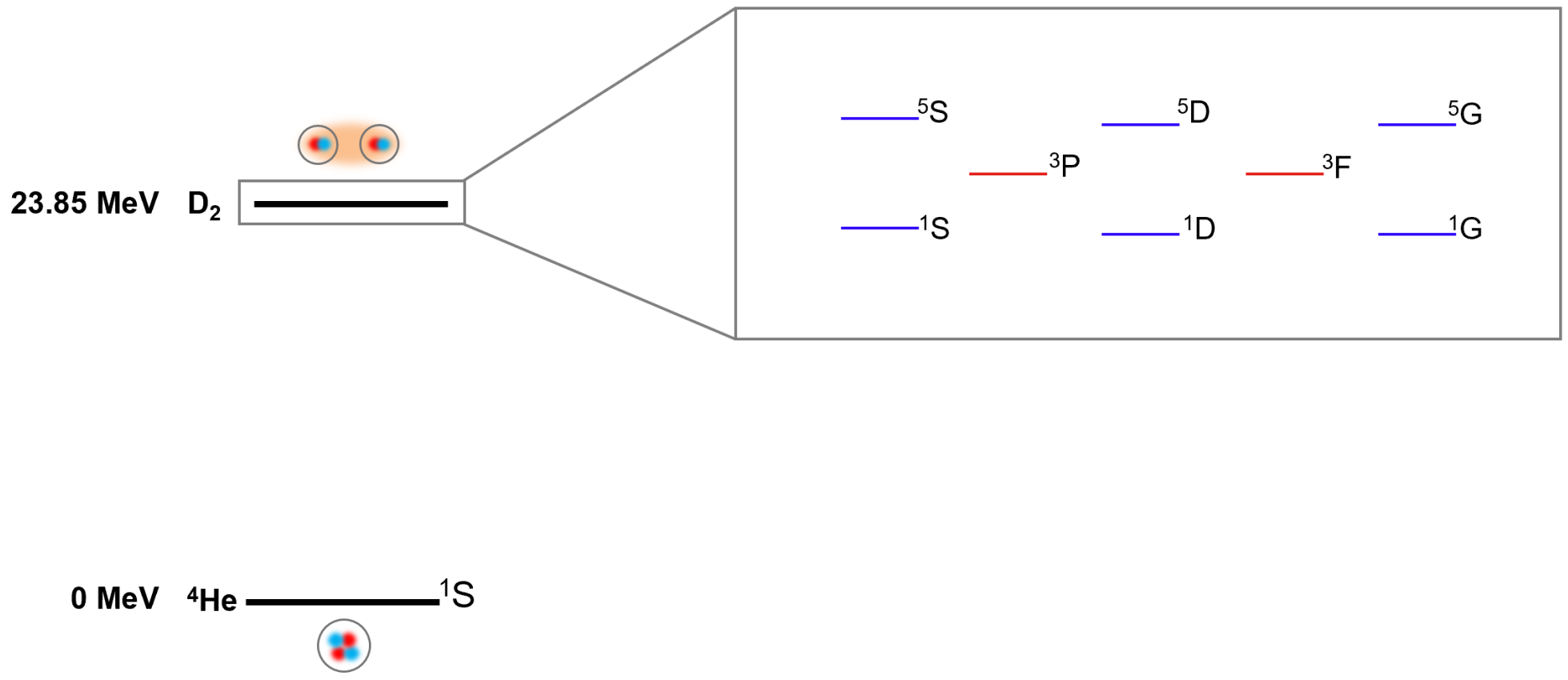

**Figure 7:** Molecular states and nuclear spin states of a $D_2$ molecule versus a $^4$He nucleus. The $D_2$ states are effectively degenerate with rotational energies on the order of meV. Even parity states are indicated in blue, and odd parity states are in red. Related $^4$He* excited states are shown in Figure 47 in section 6.2.

### Pseudo-spin formalism for ensembles of nuclei

Two-level systems are conveniently modeled making use of Pauli matrices. Here, the pseudo-spin formalism is employed so as to focus not just on a set number of two-level systems but on the collective behavior of an extensive system of such entities (this formalism is a common mathematical tool to describe collections of two-level systems and is named after spins even though in many cases there is no spin involved at all in the problem).

For a single two-level system, a formalism based on a single pseudo-spin $\hat{\mathbf{s}}$ operator can be used according to

$$\frac{\hbar}{2}\boldsymbol{\sigma} = \hat{\mathbf{s}} \tag{28}$$

where $\boldsymbol{\sigma}$ is a vector made up of Pauli matrices

$$\begin{aligned}\boldsymbol{\sigma} &= \hat{\mathbf{i}}_x\sigma_x + \hat{\mathbf{i}}_y\sigma_y + \hat{\mathbf{i}}_z\sigma_z \\ &= \hat{\mathbf{i}}_x\begin{pmatrix} 0 & 1 \\ 1 & 0 \end{pmatrix} + \hat{\mathbf{i}}_y\begin{pmatrix} 0 & -i \\ i & 0 \end{pmatrix} + \hat{\mathbf{i}}_z\begin{pmatrix} 1 & 0 \\ 0 & -1 \end{pmatrix}\end{aligned} \tag{29}$$

where $\hat{\mathbf{i}}$ are the Cartesian unit vectors. A single pseudo-spin operator is useful for modeling a single two-level system. If there are many two-level systems, then it is convenient to work with bigger pseudo-spin operators defined according to

$$\hat{\mathbf{S}} = \sum_j \hat{\mathbf{s}}_j = \frac{\hbar}{2}\sum_j \boldsymbol{\sigma}_j \tag{30}$$

This aids in the analysis of models where cooperative enhancements occur, since the associated many-pseudo-spin operators leads to cooperative factors naturally.

### Generalized nuclear Dicke model

We can write the two-level idealization of the model above making use of pseudo-spin operators according to

$$\hat{H} = \sum_j \Delta M_j c^2 \frac{S_z^{(j)}}{\hbar} + \hbar\omega_A \hat{a}_A^\dagger \hat{a}_A + \hbar\omega_O \hat{a}_O^\dagger \hat{a}_O + \hbar\omega_P \hat{a}_P^\dagger \hat{a}_P - \sum_j \langle \Phi_2 | \mathbf{d}\cdot\hat{\mathbf{E}} | \Phi_1 \rangle_j \left( \frac{S_+^{(j)}}{\hbar} + \frac{S_-^{(j)}}{\hbar} \right) \tag{31}$$

where $S_+^{(j)}$ is a pseudo-spin raising operator and $S_-^{(j)}$ is a pseudo-spin lowering operator and the nuclear dipole matrix elements are assumed to be real. The model is now a Dicke model.

In this derivation, the focus was placed on electric coupling. However, the Hamiltonian can be written in more general form and different interactions $V$ can be readily considered (as will be done in section 5.3 and in subsequent sections):

$$\begin{aligned}\hat{H} = & \sum_j \Delta M_j c^2 \frac{S_z^{(j)}}{\hbar} + \hbar\omega_A \hat{a}_A^\dagger \hat{a}_A + \hbar\omega_O \hat{a}_O^\dagger \hat{a}_O + \hbar\omega_P \hat{a}_P^\dagger \hat{a}_P \\ & + \sum_j V_j^{(A)} (\hat{a}_A^\dagger + \hat{a}_A) \left( \frac{S_+^{(j)}}{\hbar} + \frac{S_-^{(j)}}{\hbar} \right) + \sum_j V_j^{(O)} (\hat{a}_O^\dagger + \hat{a}_O) \left( \frac{S_+^{(j)}}{\hbar} + \frac{S_-^{(j)}}{\hbar} \right) + \sum_j V_j^{(P)} (\hat{a}_P^\dagger + \hat{a}_P) \left( \frac{S_+^{(j)}}{\hbar} + \frac{S_-^{(j)}}{\hbar} \right)\end{aligned} \tag{32}$$

Note that this Dicke model is similar to supertransfer models in the literature such as Eq. 1 in Lloyd and Mohseni [79]

$$\hat{H} = -\sum_j \hbar \frac{\omega}{2} \sigma_z^j + \hbar\omega a^\dagger a + \sum_j \hbar\gamma \sigma_x^j (a + a^\dagger) \tag{33}$$

## 5.2 Excitation transfer dynamics between $D_2$ and a resonant receiver nuclear state

The above-described Dicke model is still complex, containing a multitude of possible dynamics, and it can be applied to many different physical systems. Herein, we focus on resonant nuclear excitation transfer between $D_2$ molecules and $^4$He nuclei, which represent excited and ground states of an idealized two-level system per the discussions in sections 5.1 and 6.1 (where molecular $D_2$ is viewed as being an excited state of a multi-body system where $^4$He is the associated ground state). This physical setup implies that the receiver system ($^4$He) is perfectly resonant with the donor system ($D_2$).

To proceed this way, we consider two ensembles of nuclei $a$ and $b$ whereby ensemble $a$ denotes the group of $D_2$ nuclei in the coupled system (which can undergo the $D_2 \to {}^4$He transition) and $b$ denotes the group of $^4$He nuclei in the coupled system (which can undergo the $^4\text{He} \to D_2$ transition). Note that the collective transition of group $a$ occurs coherently together with the corresponding transition in group $b$—although intermediate (sometimes referred to as virtual) states are involved, as will be discussed later in this section.

Our starting point is the generalized nuclear Dicke model (Eq. 32). Recall that the corresponding Hamiltonian, based on first-order coupling between the nuclei and the oscillator modes, is

$$\hat{H} = \sum_j \Delta M_j c^2 \frac{S_z^{(j)}}{\hbar} + \hbar\omega_A \hat{a}_A^\dagger \hat{a}_A + \hbar\omega_O \hat{a}_O^\dagger \hat{a}_O + \hbar\omega_P \hat{a}_P^\dagger \hat{a}_P$$

$$+\sum_j V_j^{(A)}(\hat{a}_A^\dagger + \hat{a}_A)\left(\frac{S_+^{(j)}}{\hbar} + \frac{S_-^{(j)}}{\hbar}\right) + \sum_j V_j^{(O)}(\hat{a}_O^\dagger + \hat{a}_O)\left(\frac{S_+^{(j)}}{\hbar} + \frac{S_-^{(j)}}{\hbar}\right) + \sum_j V_j^{(P)}(\hat{a}_P^\dagger + \hat{a}_P)\left(\frac{S_+^{(j)}}{\hbar} + \frac{S_-^{(j)}}{\hbar}\right) \quad (34)$$

Again, for the sake of simplicity, and following 5.1, we focus here on electric coupling. This reduces the generic expression of Eq. 32 to Eq. 31:

$$\begin{aligned} \hat{H} &= \Delta M c^2 \frac{S_z^{(a)}}{\hbar} + \Delta M c^2 \frac{S_z^{(b)}}{\hbar} + \hbar\omega_A \hat{a}_A^\dagger \hat{a}_A + \hbar\omega_O \hat{a}_O^\dagger \hat{a}_O + \hbar\omega_P \hat{a}_P^\dagger \hat{a}_P \\ &\quad -\langle \Phi_{D_2} | \mathbf{d} \cdot \hat{\mathbf{E}} | \Phi_{^4\text{He}} \rangle \left(\frac{S_+^{(a)}}{\hbar} + \frac{S_-^{(a)}}{\hbar}\right) - \langle \Phi_{D_2} | \mathbf{d} \cdot \hat{\mathbf{E}} | \Phi_{^4\text{He}} \rangle \left(\frac{S_+^{(b)}}{\hbar} + \frac{S_-^{(b)}}{\hbar}\right) \end{aligned} \quad (35)$$

We can denote this as

$$\hat{H} = \hat{H}_0 + \hat{V} \quad (36)$$

where

$$\begin{aligned} \hat{H}_0 &= \Delta M c^2 \frac{S_z^{(a)}}{\hbar} + \Delta M c^2 \frac{S_z^{(b)}}{\hbar} + \hbar\omega_A \hat{a}_A^\dagger \hat{a}_A + \hbar\omega_O \hat{a}_O^\dagger \hat{a}_O + \hbar\omega_P \hat{a}_P^\dagger \hat{a}_P \\ \hat{V} &= -\langle \Phi_{D_2} | \mathbf{d} \cdot \hat{\mathbf{E}} | \Phi_{^4\text{He}} \rangle \left(\frac{S_+^{(a)}}{\hbar} + \frac{S_-^{(a)}}{\hbar}\right) - \langle \Phi_{D_2} | \mathbf{d} \cdot \hat{\mathbf{E}} | \Phi_{^4\text{He}} \rangle \left(\frac{S_+^{(b)}}{\hbar} + \frac{S_-^{(b)}}{\hbar}\right) \end{aligned} \quad (37)$$

In terms of nomenclature, we use $V$ here to describe a coupling generically, *i.e.*, when the emphasis is less on a specific transition and coupling, and more on the exploration of generalizable dynamics (although we will at times use concrete examples such as electric coupling for the $D_2/{}^4$He transition as in Eq. 35). Later in this text, we will use the letter $U$ specifically to refer to coupling for the $D_2/{}^4$He transition. This distinction in nomenclature is particularly relevant, when rate estimates are developed (section 5.3 and subsequent sections), where we need to make use of concrete matrix elements that are characteristic of specific transitions and couplings.

**Excitation transfer as the result of second-order interactions**

Since first-order processes cannot be resonant due to the energy mismatch between the nuclear system and the oscillators, we are interested in second-order excitation transfer, which can be resonant. The first-order interactions in the Hamiltonian can be replaced by second-order interactions (for a derivation of second-order interactions see section 6.4)

$$\hat{H} \ \rightarrow \ \hat{H}_0 + \hat{V} \frac{1}{E - \hat{H}_0} \hat{V} \tag{38}$$

The terms in this Hamiltonian couple to states on resonance, near resonance (with exchange of oscillator quanta), above resonance by 23.85 MeV, below resonance by 23.85 MeV as well as above resonance by 47.7 MeV and below resonance by 47.7 MeV. To see this more explicitly we can define:

$$\begin{aligned} \hat{V}_+ \ &= \ -\langle \Phi_{\mathrm{D}_2} | \mathbf{d} \cdot \hat{\mathbf{E}} | \Phi_{^4\mathrm{He}} \rangle \left( \frac{S_+^{(a)}}{\hbar} + \frac{S_+^{(b)}}{\hbar} \right) \\ \hat{V}_- \ &= \ -\langle \Phi_{\mathrm{D}_2} | \mathbf{d} \cdot \hat{\mathbf{E}} | \Phi_{^4\mathrm{He}} \rangle \left( \frac{S_-^{(a)}}{\hbar} + \frac{S_-^{(b)}}{\hbar} \right) \end{aligned} \tag{39}$$

which then allows us to write the second-order Hamiltonion as

$$\hat{H} \ \rightarrow \ \hat{H}_0 + \hat{V}_- \frac{1}{E - \hat{H}_0} \hat{V}_- + \hat{V}_+ \frac{1}{E - \hat{H}_0} \hat{V}_+ + \hat{V}_+ \frac{1}{E - \hat{H}_0} \hat{V}_- + \hat{V}_- \frac{1}{E - \hat{H}_0} \hat{V}_+ \tag{40}$$

This second-order (*i.e.*, excitation transfer) Hamiltonian [81] includes coupling to intermediate states that can be massively off of resonance. 'Off of resonance' in this context means that we consider the occupation of states that have more or less of the total energy available during the dynamical process, but since the process is fast, this is technically allowed by the formalism. For focus on the most likely dynamics, the coupling that leads to terms with double deexcitation or double excitation (the first two coupling terms in Eq. 40) are eliminated by restricting the terms kept in the Hamiltonian (to focus on second-order excitation transfer to off-resonant nuclear states with single excitation or deexcitation) according to

$$\hat{H} \ \rightarrow \ \hat{H}_0 + \hat{V}_+ \frac{1}{E - \hat{H}_0} \hat{V}_- + \hat{V}_- \frac{1}{E - \hat{H}_0} \hat{V}_+ \tag{41}$$

In other words, the scenario shown in Figure 8, while conceivable, is excluded from the modeling effort proposed here since it is considered less relevant than excitation transfer via single excitation and deexcitation. The latter dynamics are illustrated in Figure 9.

If the coupling is weak (as it is in our case), then the self-energy terms can be expected to be small. This suggests that the self-energy terms can be dispensed with and the focus is on the indirect transitions by working with

$$\hat{H} \ \rightarrow \ \hat{H}_0 + \hat{V}_+^{(b)} \frac{1}{E - \hat{H}_0} \hat{V}_-^{(a)} + \hat{V}_+^{(a)} \frac{1}{E - \hat{H}_0} \hat{V}_-^{(b)} + \hat{V}_-^{(b)} \frac{1}{E - \hat{H}_0} \hat{V}_+^{(a)} + \hat{V}_-^{(a)} \frac{1}{E - \hat{H}_0} \hat{V}_+^{(b)} \tag{42}$$

where

$$\begin{aligned} \hat{V}_\pm^{(a)} \ &= \ -\langle \Phi_{\mathrm{D}_2} | \mathbf{d} \cdot \hat{\mathbf{E}} | \Phi_{^4\mathrm{He}} \rangle \frac{S_\pm^{(a)}}{\hbar} \\ \hat{V}_\pm^{(b)} \ &= \ -\langle \Phi_{\mathrm{D}_2} | \mathbf{d} \cdot \hat{\mathbf{E}} | \Phi_{^4\mathrm{He}} \rangle \frac{S_\pm^{(b)}}{\hbar} \end{aligned} \tag{43}$$

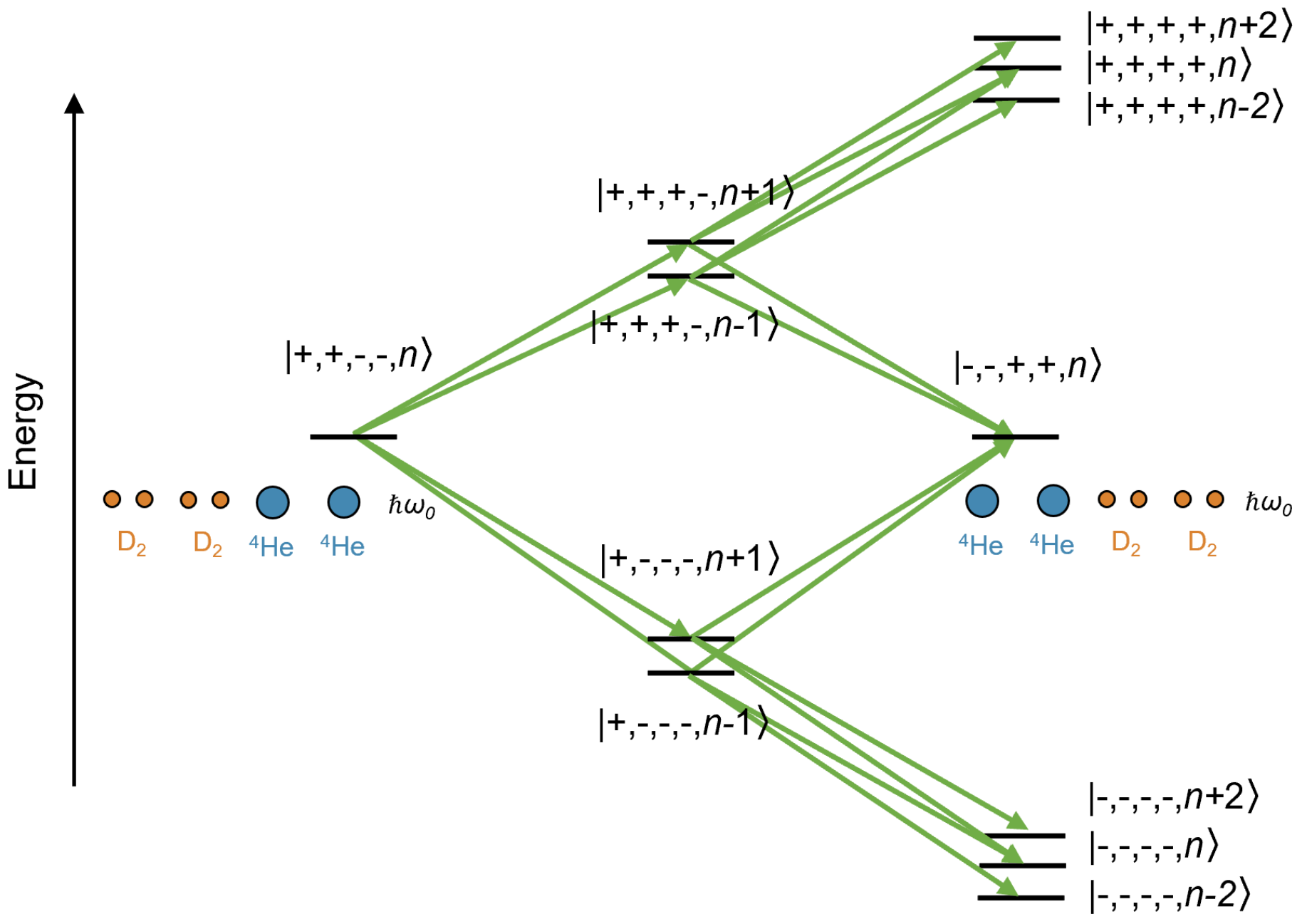

**Figure 8:** Excited $D_2$ and ground $^4$He nuclear states are coupled to a resonant nuclear state via a shared oscillator mode ($\hbar\omega_0$). Resonant nuclear energy transfer occurs through the temporary occupation of virtual states, in which either more (upper pathway) or fewer (lower pathway) of the nuclear systems are excited. When there is more than one donor and one receiver present, as is the case in the systems we consider with many donors and receivers (shown here as an example are two $D_2$ donors and two $^4$He receivers), then double excitation and double deexcitation can occur. For the sake of simplicity, we do however exclude scenarios with double excitation and deexcitation from the modeling in Eq. 41.

The resulting model is one that focuses only on terms responsible for nearly resonant excitation transfer (*i.e.*, the receiver system matches the donor system in energy) via second-order (indirect) coupling through the traversal of off-resonant intermediate states. Such a model will help move toward analyzing the coherent dynamics of interest.

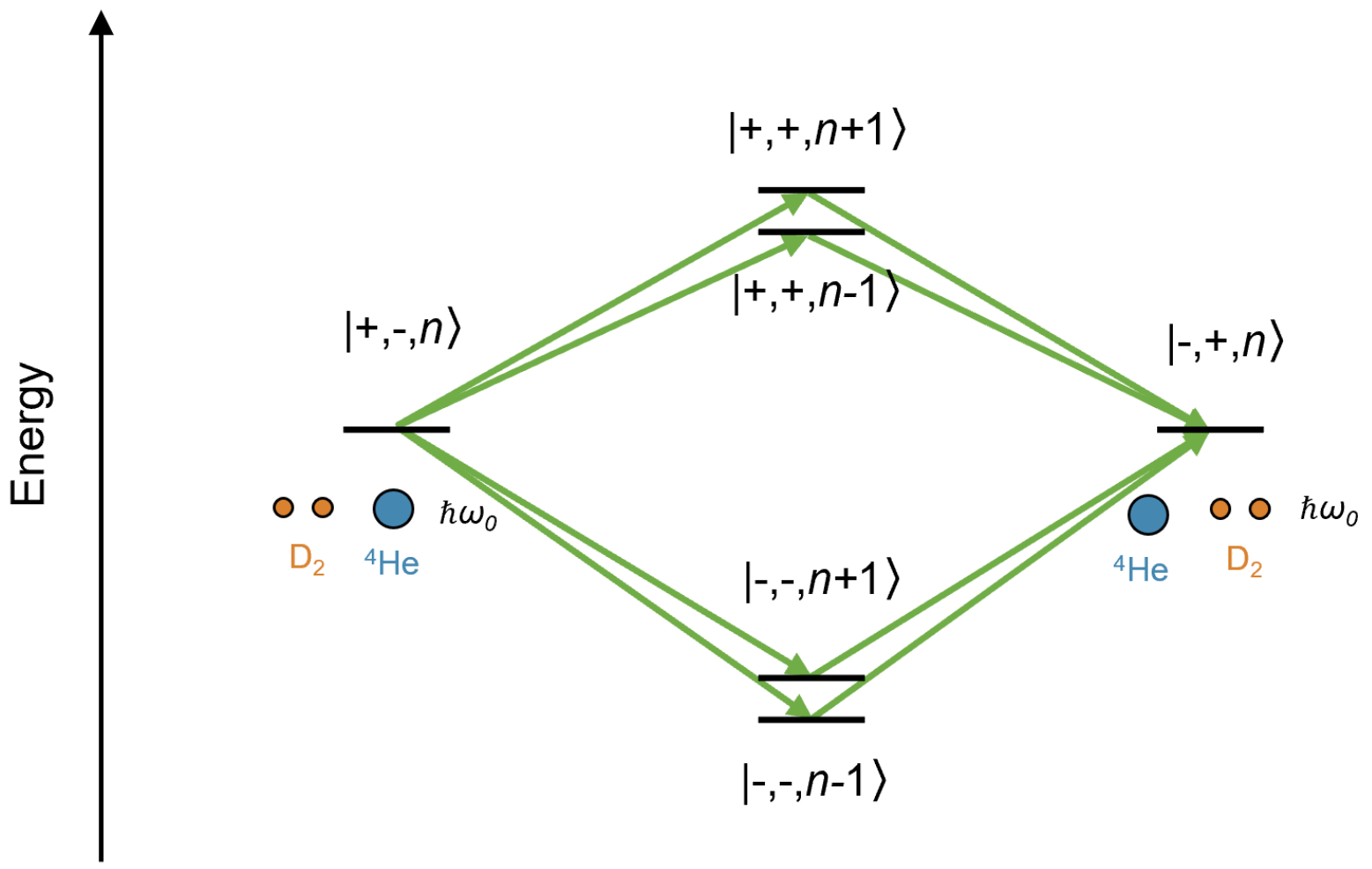

**Figure 9:** Excited $D_2$ and ground $^4$He nuclear states are coupled to a resonant nuclear state via a shared oscillator mode ($\hbar\omega_0$). Resonant nuclear energy transfer occurs through the temporary occupation of virtual states, in which either both (upper pathway) or neither (lower pathway) of the two nuclear systems are excited. Note that, due to the symmetry in the upper and lower pathways, destructive interference greatly hinders indirect nuclear excitation transfer.

The upper and lower parts in Figure 9 represent different off-resonant pathways between real states in the Hamiltonian via intermediate states, which are also sometimes described as virtual states. In a Hamiltonian matrix such intermediate states are represented through off-diagonal matrix elements.

**Destructive interference and its reduction through loss**

An approximate evaluation of the denominators of Eq. 42 leads to

$$\hat{H} \ \rightarrow \ \hat{H}_0 + \hat{V}_+^{(b)} \frac{1}{\Delta M c^2} \hat{V}_-^{(a)} + \hat{V}_+^{(a)} \frac{1}{\Delta M c^2} \hat{V}_-^{(b)} + \hat{V}_-^{(b)} \frac{1}{-\Delta M c^2} \hat{V}_+^{(a)} + \hat{V}_-^{(a)} \frac{1}{-\Delta M c^2} \hat{V}_+^{(b)} \tag{44}$$

$$\hat{H} \ \rightarrow \ \hat{H}_0 - \frac{1}{\Delta M c^2} \left( \hat{V}_-^{(b)} \hat{V}_+^{(a)} + \hat{V}_-^{(a)} \hat{V}_+^{(b)} - \hat{V}_+^{(b)} \hat{V}_-^{(a)} - \hat{V}_+^{(a)} \hat{V}_-^{(b)} \right) \tag{45}$$

where

$$\hat{V}_-^{(b)} \hat{V}_+^{(a)} + \hat{V}_-^{(a)} \hat{V}_+^{(b)} - \hat{V}_+^{(b)} \hat{V}_-^{(a)} - \hat{V}_+^{(a)} \hat{V}_-^{(b)} \sim 0 \tag{46}$$

and therefore

$$\hat{H} \ \rightarrow \ \hat{H}_0 \tag{47}$$

As can be seen from the diagram in Figure 9 and from the corresponding simplified Eq. 47, the symmetry of upper and lower pathways causes destructive interference, resulting in an interaction term of 0. In practice this represents a localization effect, where coupling would be dominated by nearest neighbor interactions, with the result of no or only very weak indirect coupling between the real states of more distant nuclei. In practice, this would mean no or unobservably slow transfer rates to distant nuclei.

This kind of destructive interference is reduced if the symmetry of the available pathways through intermediate states is broken and certain pathways are removed or weakened. This is indeed the case when loss selectively affects some pathways more than others. Related to this concept, the impact of dephasing and dissipation loss on coherent excitation transfer dynamics were considered by Plenio and Huelga [82], who found an enhancement of an excitation transfer rate due to interactions of a closed quantum system with a noisy environment.

Consider again the case where 23.85 MeV is transferred from a $D_2$ molecule (representing a highly metastable $^4$He excited state) to a $^4$He nucleus ($^4$He ground state) via indirect coupling per the above process. The intermediate states at the bottom of the diagram in Figure 9 are off of resonance by 23.85 MeV. Consider the situation where there are relevant loss channels available for the $|-,-\rangle$ states on the lower part of the diagram, as depicted in Figure 10, where the system can dissipate 23.85 MeV to bring itself back onto resonance; and where such loss channels are not available for the $|+,+\rangle$ intermediate states on the upper part of the diagram. This is indeed how loss can affect such a system selectively, because the basis state energy lies well above the energy eigenvalue. If the system is perturbed when one of the lower intermediate virtual states is occupied, the system can terminate at the $|-,-\rangle$ state resulting in a transfer of the off-resonance energy $\Delta E$ from the oscillator to a loss channel. This kind of mechanism was proposed in Hagelstein 2002 [83] in connection with the closely related problem of multi-phonon exchange, and modeled in subsequent papers (see, for example, [84]); in connection with excitation transfer this was discussed in [85]. The basic idea is that quantum systems behave differently on resonance and off of resonance, and decay channels can be very different in the intermediate states involved in excitation transfer due to their being off of resonance. In this picture, off-resonant loss was considered to be very fast, since the phonon mode in the model under consideration couples to all of the nuclei in the lattice, potentially providing access to many loss channels.

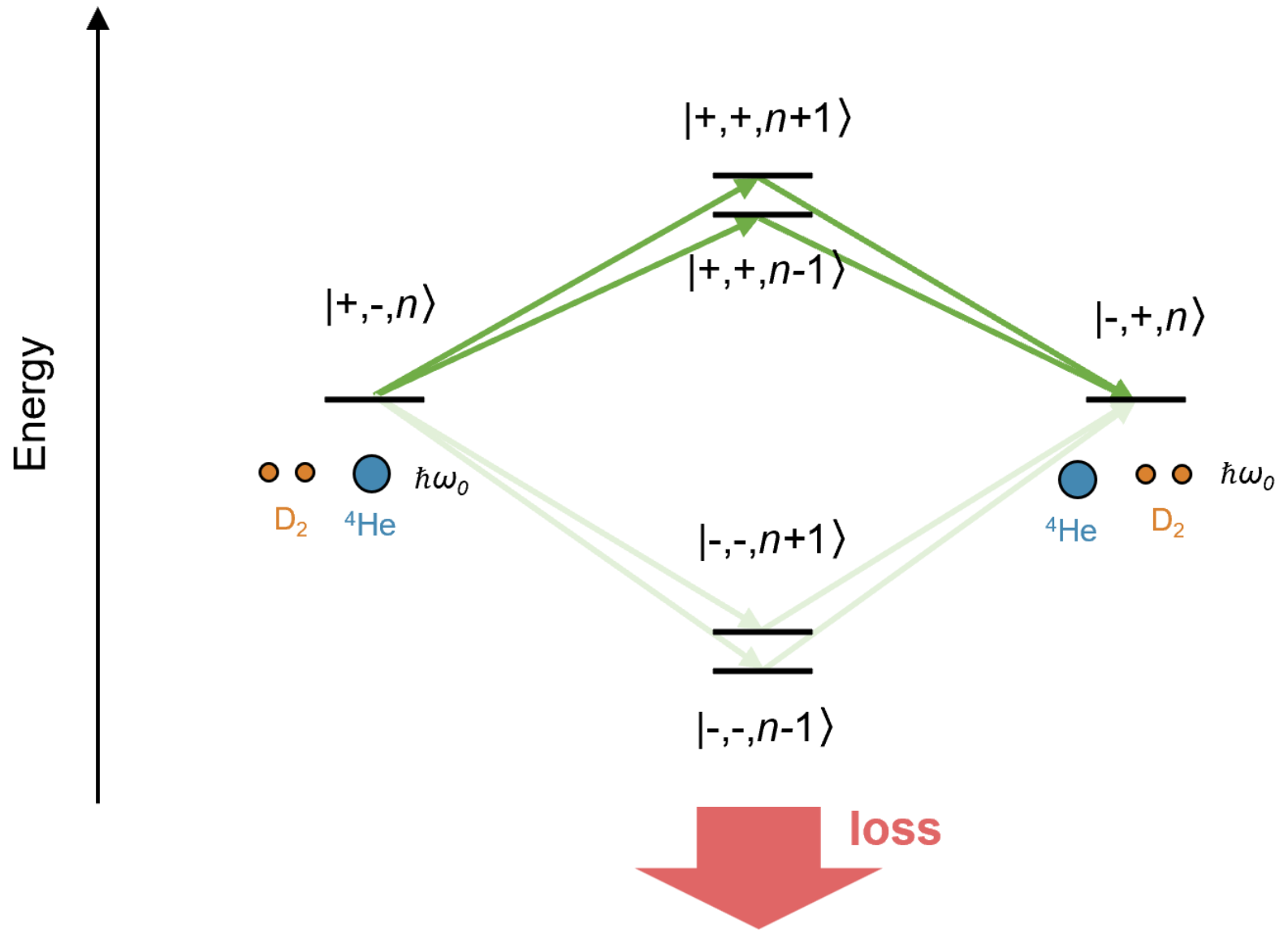

**Figure 10:** While the symmetry present in the scenario of Figure 9 causes destructive interference and greatly hinders excitation transfer, we can conceive of a scenario, where a form of idealized loss eliminates the bottom pathway and reduces destructive inference. In that case, excitation transfer is much less hindered.

In terms of physical mechanisms for causing this kind of selective loss in the MeV range, nuclear disintegration of lattice nuclei was considered to be dominant [86], leading to the complete removal of all lower pathways and therefore the complete elimination of destructive interference. This kind of loss can be considered to be "*ideal loss*" with regard to the acceleration of excitation transfer. However, it was subsequently understood that it is not possible for there to be contributions from all the nuclei in the lattice off of resonance and this form of loss to lattice nuclei comes with its own form of destructive interference.

In contrast to focusing on other coupled lattice nuclei, there are loss processes associated with the nuclei directly involved in the excitation transfer that do not suffer from this kind of destructive interference. Such a form of loss is described in more detail in section 6.5. The conclusion of that discussion is that it is possible to work with the conceptually simple picture of ideal loss that assumes a complete elimination of destructive interference (via removal of all lower pathways in the diagram), and then apply a *correction factor* $|1-\eta|$ to account for the imperfect elimination of destructive interference in actual physical systems (per section 6.5). The correction factor derives from the loss-based asymmetry between the upper pathway and the lower pathway (with coupling constants $U$ and coupling constants $U' = \eta U$ respectively in Figure 11). An estimate for this correction factor on the order of 0.1 is given in section 6.5.

The model simplifies considerably when the system enters a regime where loss rate associated with the overall system ($\Gamma$) is fast compared to the two-level system energy [85], *i.e.*, when

$$\frac{1}{2}\hbar\Gamma \gg \Delta M c^2 \tag{48}$$

Now suppose that we augment the second-order Hamiltonian of Eq. 42 by including off-resonant loss for the terms that involve an energy excess in the intermediate states, then we would obtain

$$\hat{H} \rightarrow \hat{H}_0 + \hat{V}_+^{(b)} \frac{1}{E - \hat{H}_0 + i\hbar\Gamma/2} \hat{V}_-^{(a)} + \hat{V}_+^{(a)} \frac{1}{E - \hat{H}_0 + i\hbar\Gamma/2} \hat{V}_-^{(b)} + \hat{V}_-^{(b)} \frac{1}{E - \hat{H}_0} \hat{V}_+^{(a)} + \hat{V}_-^{(a)} \frac{1}{E - \hat{H}_0} \hat{V}_+^{(b)} \tag{49}$$

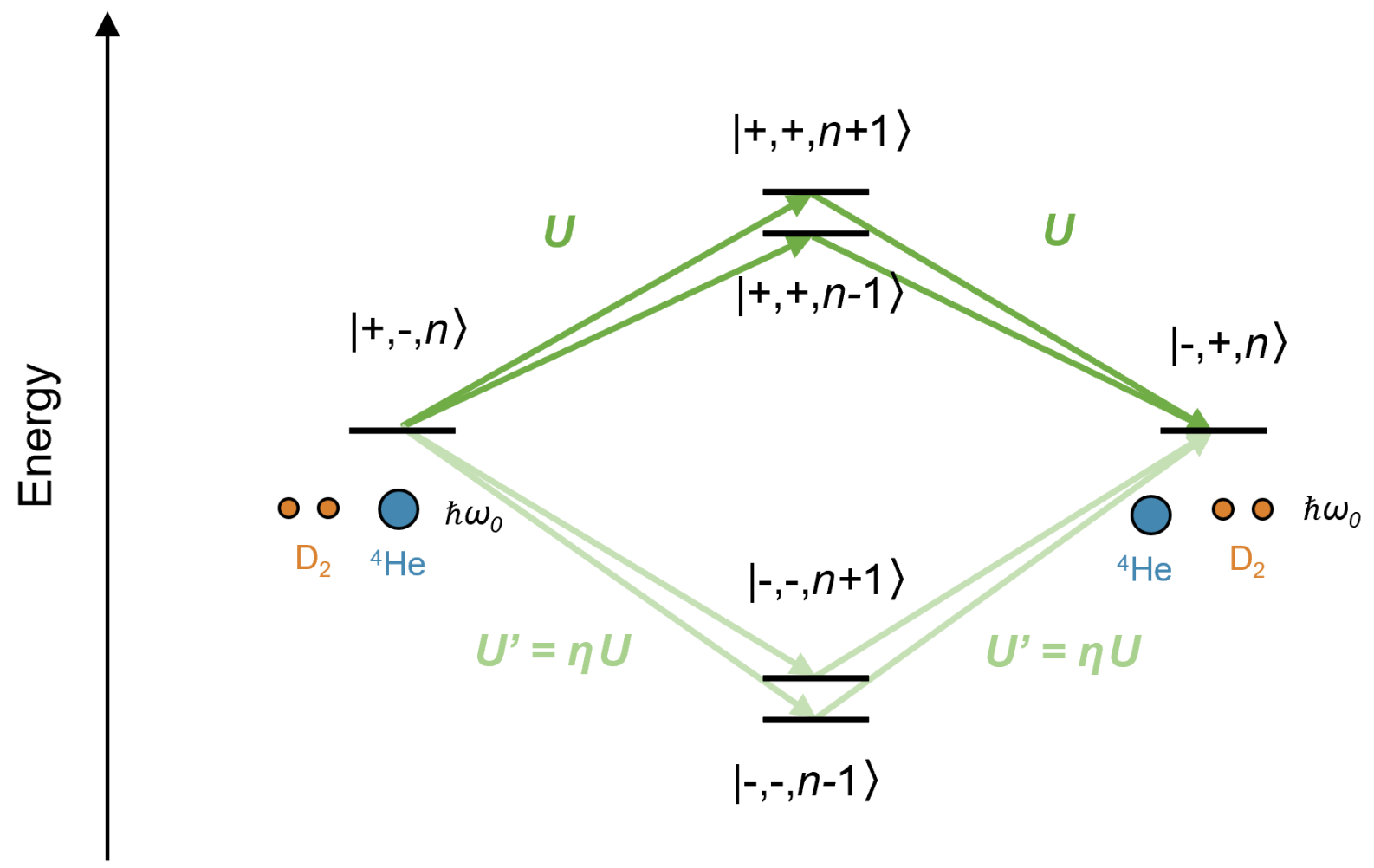

**Figure 11:** The complete elimination of lower pathways in Figure 10 depends on a form of loss that is idealized. In reality, while there are strong loss channels affecting the lower pathways (see section 6.5), the reduction of destructive interference is not complete. To account for the presence of loss in the lower pathways (and the partial reduction of destructive interference), we introduce $\eta$ to represent the difference in coupling between states in the lower pathways ($U'$) and states in the upper pathways ($U$). Note that we use here $U$ to refer to coupling in the $D_2$/$^4$He transition (instead of the generic $V$), consistent with section 5.3 and subsequent sections.

An approximate evaluation of the denominators leads to

$$\hat{H} \rightarrow \hat{H}_0 + \hat{V}_+^{(b)} \frac{1}{\Delta Mc^2 + i\hbar\Gamma/2} \hat{V}_-^{(a)} + \hat{V}_+^{(a)} \frac{1}{\Delta Mc^2 + i\hbar\Gamma/2} \hat{V}_-^{(b)} + \hat{V}_-^{(b)} \frac{1}{-\Delta Mc^2} \hat{V}_+^{(a)} + \hat{V}_-^{(a)} \frac{1}{-\Delta Mc^2} \hat{V}_+^{(b)} \quad (50)$$

In the case of large loss (per Eq. 48), destructive interference is eliminated, and the Hamiltonian reduces to

$$\hat{H} \rightarrow \hat{H}_0 - \frac{1}{\Delta Mc^2} \left( \hat{V}_-^{(b)} \hat{V}_+^{(a)} + \hat{V}_-^{(a)} \hat{V}_+^{(b)} \right) \quad (51)$$

where

$$\hat{V}_-^{(b)} \hat{V}_+^{(a)} + \hat{V}_-^{(a)} \hat{V}_+^{(b)} > 0 \quad (52)$$

In other words, in the presence of strong selective loss, there exists a nonzero indirect coupling between real states of different nuclei coupled to the same oscillator mode (including distant nuclei). This indirect coupling enables excitation transfer.

Finally, we recall that this indirect coupling is based on the assumption of ideal loss. To get a physically more accurate result, we apply the correction factor per section 6.5:

$$\hat{H} \rightarrow \hat{H}_0 - \frac{1}{\Delta Mc^2} \left( \hat{V}_-^{(b)} \hat{V}_+^{(a)} + \hat{V}_-^{(a)} \hat{V}_+^{(b)} \right) |1 - \eta| \quad (53)$$

For the sake of simplicity, we will consider the case of ideal loss for the remainder of this section and not carry the correction factor $|1 - \eta|$ through every expression. We point out, however, that it is necessary to apply it, when determining numbers intended to match experiments, as is the case in later sections of this text.

### Resonance condition: averaging over the oscillators

We seek to further simplify the main Hamiltonian (Eq. 53) so that we can analyze excitation transfer dynamics. In this subsection, we formally introduce the resonance condition, which assumes that the donor and receiver systems in groups *a* and *b* have energy levels that are matched. This is the case in our example with $D_2$ and $^4$He nuclei which represent the same physical system in different states.

We will need to keep in mind this condition later when considering excitation transfer across different species of nuclei, *e.g.*, between $D_2$ systems and $^A$Pd systems, as discussed in section 5.5 and onward.

The electric field operator creates or annihilates oscillator quanta, which means that the model under consideration couples to states that involve a different number of quanta. This can be seen by expanding out the Hamiltonian of Eq. 51 according to

$$\hat{H} \ \rightarrow \ \hat{H}_0 - \frac{1}{\Delta M c^2} \langle \Phi_{D_2} | \mathbf{d} | \Phi_{^4\text{He}} \rangle \cdot \hat{\mathbf{E}}\hat{\mathbf{E}} \cdot \langle \Phi_{D_2} | \mathbf{d} | \Phi_{^4\text{He}} \rangle \left( \frac{S_-^{(b)}}{\hbar} \frac{S_+^{(a)}}{\hbar} + \frac{S_-^{(a)}}{\hbar} \frac{S_+^{(b)}}{\hbar} \right) \tag{54}$$

This interaction involves the square of the electric field, which means that there is coupling between states that differ by 2, 0 and -2 oscillator quanta. Since we are here interested only in resonant processes, it is possible to eliminate the coupling to off-resonant (oscillator) final states by taking an expectation value over the oscillator degrees of freedom, yielding

$$\hat{H} \ \rightarrow \ \langle \hat{H}_0 \rangle_{osc} - \frac{1}{\Delta M c^2} \langle \Phi_{D_2} | \mathbf{d} | \Phi_{^4\text{He}} \rangle \cdot \langle \hat{\mathbf{E}}\hat{\mathbf{E}} \rangle_{osc} \cdot \langle \Phi_{D_2} | \mathbf{d} | \Phi_{^4\text{He}} \rangle \left( \frac{S_-^{(b)}}{\hbar} \frac{S_+^{(a)}}{\hbar} + \frac{S_-^{(a)}}{\hbar} \frac{S_+^{(b)}}{\hbar} \right) \tag{55}$$

Thus the model that results is maximally simplified to include only resonant excitation transfer processes. Note that since in resonant excitation transfer no net change in the two-level system energy results, the two-level system energy is constant. Note also that since there is no coupling to off-resonant terms of the oscillators, all of the basis state energies will be the same. This means that $\langle \hat{H}_0 \rangle_{osc}$ will not contribute to the associated coherent dynamics, and that we can work with a Hamiltonian of the form

$$\hat{H} \ = \ \frac{1}{2} \hbar \Omega_{ab} \left( \frac{S_-^{(b)}}{\hbar} \frac{S_+^{(a)}}{\hbar} + \frac{S_-^{(a)}}{\hbar} \frac{S_+^{(b)}}{\hbar} \right) \tag{56}$$

where $\Omega_{ab}$ is the generalized Rabi frequency which depends on the square of the interaction $V$

$$-\frac{1}{2} \hbar \Omega_{ab} \ = \ \frac{1}{\Delta M c^2} \langle \Phi_{D_2} | \mathbf{d} | \Phi_{^4\text{He}} \rangle \cdot \langle \hat{\mathbf{E}}\hat{\mathbf{E}} \rangle_{osc} \cdot \langle \Phi_{D_2} | \mathbf{d} | \Phi_{^4\text{He}} \rangle \ = \ \frac{V^2}{\Delta M c^2} \tag{57}$$

Eq. 56 is now the simplified Hamiltonian which describes resonant excitation transfer between sets of many two-level systems. Note that no self-energy terms are included as we assume weak coupling within groups *a* and *b*.

### Approximate classical dynamics from Ehrenfest's theorem

We can now proceed to explore the coherent dynamics associated with cooperative resonant excitation transfer between two degenerate levels, as described by the simplified Hamiltonian in Eq. 56.

As discussed by Hagelstein in [87], Ehrenfest's theorem can be applied

$$\frac{d}{dt} \langle \hat{Q} \rangle \ = \ \left\langle \frac{d\hat{Q}}{dt} \right\rangle + \frac{1}{i\hbar} \langle [\hat{Q}, \hat{H}] \rangle \tag{58}$$

and approximate classical evolution equations can be developed for this system. This yields

$$
\begin{aligned}
\frac{d}{dt}S_x^{(a)}(t) &= \Omega_{ab}\frac{S_y^{(b)}(t)}{\hbar}S_z^{(a)}(t) \\
\frac{d}{dt}S_y^{(a)}(t) &= -\Omega_{ab}\frac{S_x^{(b)}(t)}{\hbar}S_z^{(a)}(t) \\
\frac{d}{dt}S_z^{(a)}(t) &= \Omega_{ab}\left(\frac{S_x^{(b)}(t)}{\hbar}S_y^{(a)}(t) - \frac{S_y^{(b)}(t)}{\hbar}S_x^{(a)}(t)\right)
\end{aligned}
\tag{59}
$$

and

$$
\begin{aligned}
\frac{d}{dt}S_x^{(b)}(t) &= \Omega_{ab}\frac{S_y^{(a)}(t)}{\hbar}S_z^{(b)}(t) \\
\frac{d}{dt}S_y^{(b)}(t) &= -\Omega_{ab}\frac{S_x^{(a)}(t)}{\hbar}S_z^{(b)}(t) \\
\frac{d}{dt}S_z^{(b)}(t) &= \Omega_{ab}\left(\frac{S_x^{(a)}(t)}{\hbar}S_y^{(b)}(t) - \frac{S_y^{(a)}(t)}{\hbar}S_x^{(b)}(t)\right)
\end{aligned}
\tag{60}
$$

Since these dynamics are relatively simple, it is possible to eliminate the $x$ and $y$ components and obtain

$$
\begin{aligned}
\frac{1}{\Omega_{ab}^2}\frac{d^2}{dt^2}S_z^{(a)}(t) + \frac{|\mathbf{S}^{(b)}|^2 - (S_z^{(b)})^2(t)}{\hbar^2}S_z^{(a)}(t) &= \frac{|\mathbf{S}^{(a)}|^2 - (S_z^{(a)})^2(t)}{\hbar^2}S_z^{(b)}(t) \\
\frac{1}{\Omega_{ab}^2}\frac{d^2}{dt^2}S_z^{(b)}(t) + \frac{|\mathbf{S}^{(a)}|^2 - (S_z^{(a)})^2(t)}{\hbar^2}S_z^{(b)}(t) &= \frac{|\mathbf{S}^{(b)}|^2 - (S_z^{(b)})^2(t)}{\hbar^2}S_z^{(a)}(t)
\end{aligned}
\tag{61}
$$

### Nonlinear Rabi oscillations

Solutions to these equations are, in general, nonlinear Rabi oscillations [88, 89], in which occupation probability periodically oscillates from ensemble $a$ to ensemble $b$ and back in a pulsed manner with a delay between pulses.

It is possible to develop an analytic solution for a single excitation transfer pulse of this sort under conditions where the excitation is transferred from ensemble $a$ to ensemble $b$. When the number of two-level systems in the two sets ($a$ and $b$) differ, then we can write

$$
S_z^{(b)}(t) = -|\mathbf{S}^{(b)}| + 2|\mathbf{S}^{(b)}|\frac{|\mathbf{S}^{(a)}|\left[1 - \tanh^2\left(\sqrt{\frac{|\mathbf{S}^{(a)}||\mathbf{S}^{(b)}|}{\hbar^2}}\Omega_{ab}t\right)\right]}{|\mathbf{S}^{(a)}| - |\mathbf{S}^{(b)}|\tanh^2\left(\sqrt{\frac{|\mathbf{S}^{(a)}||\mathbf{S}^{(b)}|}{\hbar^2}}\Omega_{ab}t\right)}
\tag{62}
$$

This results in a pulse-like increase and decrease of Dicke enhancement, whose shape is depicted in Figure 12.

The pulse length ($\Delta\tau_{Dicke}$) when the populations are mismatched is

$$
|\Omega_{ab}|\Delta\tau_{Dicke} = \frac{\pi}{\sqrt{12}}\sqrt{\frac{\hbar}{|\mathbf{S}^{(a)}|}\frac{\hbar}{|\mathbf{S}^{(b)}|}}
\tag{63}
$$

where $\frac{|\mathbf{S}^{(a)}|}{\hbar}$ is the initial population of group $a$ (in our case $N_{\mathrm{D}_2}$ initially) divided by 2.

Note that for this highly idealized system that we have not taken into account de-coherence or loss processes.

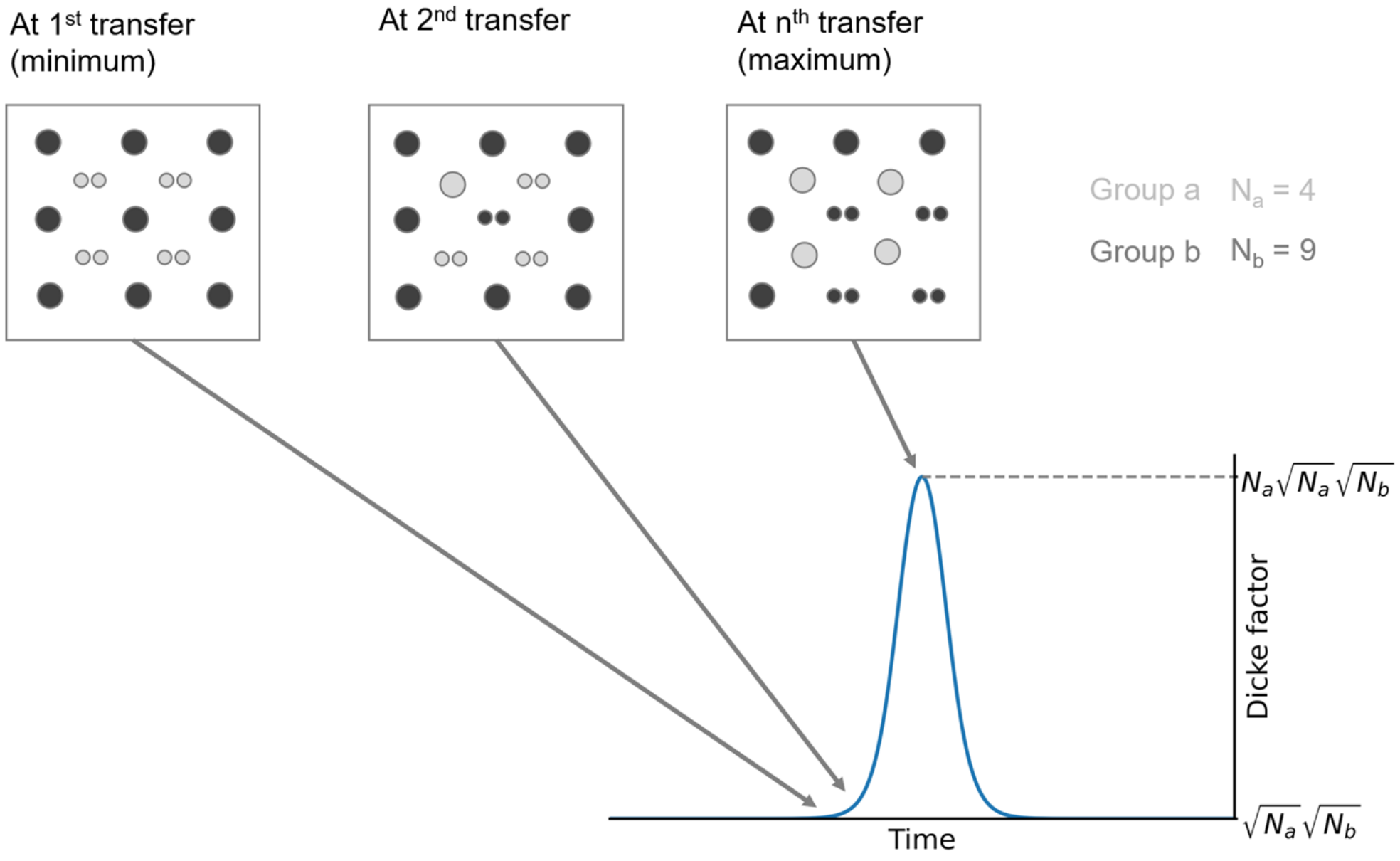

**Figure 12:** Illustration of the Dicke enhancement pulse in Eq. 62.

### Minimum and maximum Dicke factors

Based on the analytical solution in Eq. 62 and Eq. 63, we can estimate the maximum transfer rate $\Gamma_{max}$ (for a mismatched system) to be on the order of

$$\Gamma_{max} \sim \frac{1}{\Delta\tau}2\text{min}\left\{\frac{|\mathbf{S}^{(a)}|}{\hbar}, \frac{|\mathbf{S}^{(b)}|}{\hbar}\right\} \approx |\Omega_{ab}|\sqrt{\frac{|\mathbf{S}^{(a)}|}{\hbar}\frac{|\mathbf{S}^{(b)}|}{\hbar}}\text{min}\left\{\frac{|\mathbf{S}^{(a)}|}{\hbar}, \frac{|\mathbf{S}^{(b)}|}{\hbar}\right\} \tag{64}$$

Minimum and maximum Dicke enhancement factors can also be derived from the pseudo-spin Hamiltonian in Eq. 56 (see section 6.12 for the derivations), yielding

$$\hat{H} = \frac{1}{2}\hbar\Omega_{ab}\sqrt{S_a(S_a+1)-(M_a-1)M_a}\sqrt{S_b(S_b+1)-(M_b+1)M_b} \tag{65}$$

with the transfer rate

$$\Gamma = \Omega_{ab}\sqrt{S_a(S_a+1)-(M_a-1)M_a}\sqrt{S_b(S_b+1)-(M_b+1)M_b} \tag{66}$$

We can define the Dicke enhancement factor $\beta_{Dicke}$ as

$$\beta_{Dicke} = \sqrt{S_a(S_a+1)-(M_a-1)M_a}\sqrt{S_b(S_b+1)-(M_b+1)M_b} \tag{67}$$

The Dicke enhancement factor assumes its minimum value, at $t = 0$ when the dynamics just start up. It then applies to the rate for the first transition.

In this case we can take

$$M_a = S_a \qquad M_b = -S_b$$

and the expression simplifies to

$$\Gamma \;=\; \Omega_{ab}\sqrt{2S_a}\sqrt{2S_b} \tag{68}$$

We can connect with the number of $D_2$ molecules and with the number of ground state $^4$He atoms according to

$$2S_a \;=\; N_{\mathrm{D}_2} \qquad 2S_b \;=\; N_{^4\mathrm{He}}$$

We end up with

$$\Gamma_{min} \;=\; \Omega_{ab}\sqrt{N_{\mathrm{D}_2}}\sqrt{N_{^4\mathrm{He}}} \tag{69}$$

for the minimum rate.

For the second transfer, we can take

$$M_a \;=\; S_a - 1 \qquad M_b \;=\; -S_b + 1$$

which results in a rate

$$\Gamma_{min+1} \;=\; \Omega\, 2\sqrt{N_{\mathrm{D}_2}}\sqrt{N_{^4\mathrm{He}}} \tag{70}$$

For the *n*-th transfer, we can take

$$M_a \;=\; S_a - n \qquad M_b \;=\; -S_b + n$$

which results in a rate

$$\Gamma_{min+n} \;=\; \Omega\, n\sqrt{N_{\mathrm{D}_2}}\sqrt{N_{^4\mathrm{He}}} \tag{71}$$

The Dicke enhancement factor takes on its maximum value, when a large number of Dicke transitions have occurred. In that case there is an additional enhancement of the order of $N_{\mathrm{D}_2}$, resulting in

$$\Gamma_{max} \;=\; \Omega\frac{1}{4}N_{\mathrm{D}_2}\sqrt{N_{\mathrm{D}_2}}\sqrt{N_{^4\mathrm{He}}} \tag{72}$$

for the maximum rate.

## 5.3 Transfer rate estimates: $D_2$ to $^4$He nuclei

In the previous part, the dynamics of excitation transfer were analyzed in the nuclear Dicke model and it was found that excitation transfer in the form of nonlinear Rabi oscillations can occur between two ensembles of nuclei $a$ and $b$ that are coupled indirectly through shared oscillator modes. In other words, excitation (or rather the occupation probability of excitation) oscillates between groups $a$ and $b$. In this subsection, a first rate estimate is provided for this process, continuing with the earlier, conceptually simple example of $D_2$ donor systems comprising group $a$ and $^4$He receiver systems comprising group $b$. At the end of this section stands a transfer rate estimate for excitation transfer based on each of the three couplings between nuclei and vibrational modes to be considered in this context: magnetic dipole coupling, electric dipole coupling, and relativistic coupling. Because of this structure, this section also contains a quantitative comparison of these couplings, yielding the insight that relativistic coupling is substantially stronger than magnetic and electric dipole coupling.

An overview of this particular case of nuclear excitation transfer is given in Figure 13, which shows explicitly the hindrance represented by the Coulomb barrier part of the interatomic potential of the $D_2$ molecule, which is represented by the expression with the Gamow factor. Figure 14 illustrates the same process but emphasizes its time evolution, which includes traversal of off-resonant intermediate states per the discussion in section 5.2.

In the given case, the Gamow hindrance factor comes into the transfer rate expression twice: once, due to the $D_2$ molecule transition to the $^4$He state on the donor side, and once due to the $^4$He nucleus transition to the $D_2$ molecule state on the receiver side. This is not the case in the faster nuclear excitation transfer scenarios explored later in this text. For the scenario here, we will see that, while the overall transfer rate is a function of several parameters and is therefore amenable to change and acceleration more so than the spontaneous fusion rate, the predicted transfer rate is still too low to yield observable results. Nevertheless, the picture developed here will serve as an important stepping stone towards scenarios of greater interest—which are predicted to yield observable results—introduced and discussed in the latter sections.

When evaluating obtained rates for nuclear excitation transfer, it is important to compare them with the expected decoherence times. If the mean transfer rate in a given system is longer than the corresponding mean decoherence time, then that system is unlikely to yield observable results at scale.

Typical qubits used in quantum computers, where the decoherence time is from microseconds to seconds depending on the technology used, all operate at temperatures close to 0 K. The case of coherence between nuclei is quite different. On the one hand, nuclei interact much less with the environment, which has to do with the smaller size of their dipoles. However, since we consider nuclear systems across multiple atoms—such as in the case of nuclear wave functions of $D_2$ molecules—spatial degrees of freedom, which include molecular dissociation and angular momentum changes, has the potential to disrupt coherence.

As shown in section 6.22, we expect a mean decoherence time of about 1 ns. Another decoherence mechanism is spin relaxation, which is, however, much slower than molecular dissociation. A proxy for spin relaxation in the systems of interest is provided by nuclear magnetic resonance experiments. In the study of deuteron spin relaxation in metal deuterides at room temperature, there are two time parameters of interest: $T_1$, which is a measure of the (longer) lifetime associated with a spin correlation, and $T_2$, which is a measure of the (shorter) dephasing time. A typical value for spin dephasing of hydrogen in palladium is 10 ms at room temperature [90], so we can expect the same order of magnitude for deuterium or helium.

For now, we will adopt a decoherence time on the order of 1 ns. The later section 5.8 considers this kind of decoherence as an explicit constraint, which an excitation transfer process needs to exceed for it to yield observable results. In that section, basic experimental parameters will be given under which the predicted transfer rate does indeed exceed the decoherence time. Section 5.8 will build on section 5.5, which contains an extended variant of the model applied here.

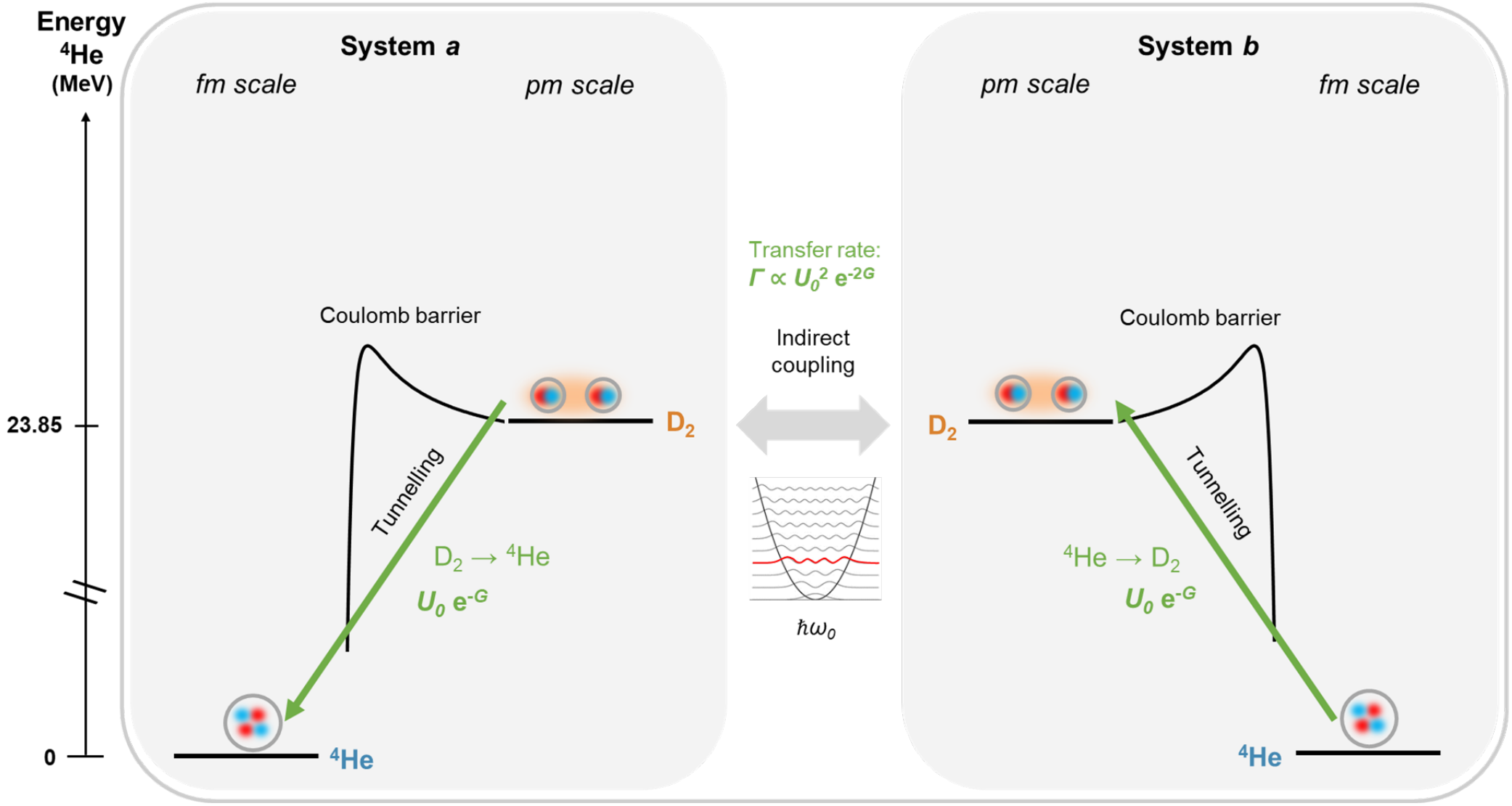

**Figure 13:** Resonant nuclear excitation transfer of 23.85 MeV from the $D_2 \to {}^4He$ fusion decay to the ${}^4He \to D_2$ excitation. We consider two ensembles of nuclei $a$ and $b$ whereby ensemble $a$ denotes the group of $D_2$ nuclei in the coupled system (which can undergo the $D_2 \to {}^4He$ transition) and $b$ denotes the group of ${}^4He$ nuclei in the coupled system (which can undergo the ${}^4He \to D_2$ transition). The collective transition of group $a$ occurs coherently together with the corresponding transition in group $b$. Note that in actuality, this kind of excitation transfer process involves pathways through different possible intermediate states, which exhibit an asymmetry in how they are affected by loss (represented by $\eta$ in Fig. 14).

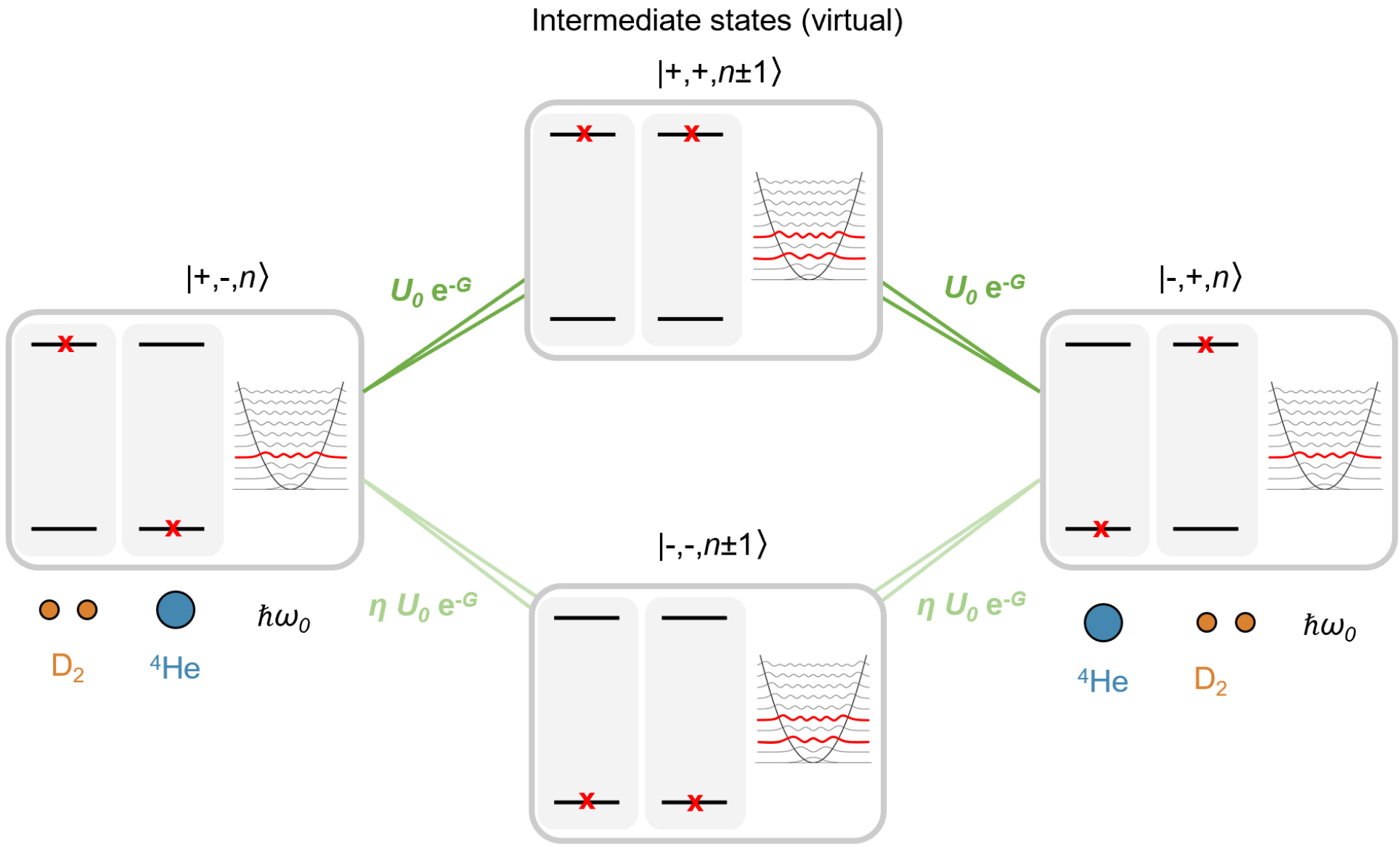

**Figure 14:** Resonant nuclear excitation transfer between a $D_2/{}^4He$ donor transition and a $D_2/{}^4He$ receiver transition by indirect coupling via a shared oscillator mode. Destructive interference between the upper and lower intermediate (*i.e.*, virtual) states is reduced by selective loss in the lower pathway, allowing for resonant nuclear excitation transfer to occur (here represented by $\eta$ per section 6.5).

## Recalling the expression for the excitation transfer rate

The derivation of Dicke factors in 6.12 yields expressions for the minimum and maximum rate of excitation transfer. Following Eq. 69 the minimum rate is

$$\Gamma_{min} \;=\; |\Omega_{ab}|\sqrt{N_{\mathrm{D}_2} N_{^4\mathrm{He}}} \tag{73}$$

where $\Omega_{ab}$ refers to the generalized Rabi frequency for oscillating between the transitions in group $a$ and $b$.

Following Eq. 72 the maximum transfer rate is (when a large number of transitions have occurred)

$$\Gamma_{max} \;=\; |\Omega_{ab}|\frac{1}{4}N_{\mathrm{D}_2}\sqrt{N_{\mathrm{D}_2} N_{^4\mathrm{He}}} \tag{74}$$

The generalized Rabi frequency is (Eq. 57)

$$\frac{1}{2}\hbar|\Omega_{ab}| = \frac{V_{gen}^2}{\Delta E} = \frac{V_{gen}^2}{\Delta M c^2} \tag{75}$$

where notationwise we added here a subscript to $V$ to highlight when a generic coupling is referred to rather than a specific one. For the transitions considered in this section, as shown in Figure 13, the generic coupling $V_{gen}$ is represented as

$$V_{gen} = U = U_0\, e^{-G} \tag{76}$$

since, as we will see shortly, the Gamow factor term $e^{-G}$ can be explicated as a key factor in all of the indirect coupling terms in this section (where $U_0$ differs based on the nature of the coupling, *i.e.*, electric, magnetic, and relativistic coupling).

Note that for rate calculations under realistic conditions, we need to work with realistic loss scenarios instead of ideal loss. Therefore, per Eq. 53 we bring in again the correction factor $|1-\eta|$ (in this case twice, because for excitation transfer from $D_2$ to $^4$He, nonideal loss occurs at the donor and the receiver side, see 6.5):

$$\frac{1}{2}\hbar|\Omega_{ab,loss}| = 2|1-\eta|\frac{V_{gen}^2}{\Delta M c^2} \tag{77}$$

resulting in

$$\Gamma_{min} \;=\; \frac{2}{\hbar}2|1-\eta|\frac{V_{gen}^2}{\Delta M c^2}\sqrt{N_{\mathrm{D}_2} N_{^4\mathrm{He}}} \tag{78}$$

For being able to use these simple versions of the minimum and maximum Dicke factors we need to assume that there is a greater number of receiver ($^4$He) systems than donor ($D_2$) systems.

Concretely, we consider a uniform oscillator mode in PdD spanning $10^{18}$ unit cells (corresponding to a volume defined by a mode frequency in the MHz range). In this volume, we then consider to be present $N_{\mathrm{D}_2} = 10^{15}$ donor systems (representing group $a$) and $N_{^4\mathrm{He}} = 10^{16}$ receiver systems (representing group $b$). This then results in ratios of $\frac{N_{\mathrm{D}_2}}{N}$ of 0.001 and $\frac{N_{^4\mathrm{He}}}{N}$ of 0.01. All nuclei in group $a$ and $b$ are uniformly coupled to the same oscillator mode(s) per the discussion in section 5.1.

Most physical metal-hydrogen systems will not exhibit such high helium content, although they can be prepared this way by, for example, loading the metal lattice with helium in addition to deuterium. Note also that there is a helium content on the order of 1 ppm in natural air and we can expect some helium to be present in all metal lattices. That being said, the assumption of a large number of $^4$He nuclei is made here for pedagogical purposes, as we will later (5.5 and onwards) consider a case where the receiver systems are Pd nuclei, which are indeed greater in number than deuteron pairs in palladium-deuterium systems.

Next, the Rabi frequency will be evaluated for the three couplings available: magnetic dipole coupling $U_{magnetic}$, electric dipole coupling $U_{electric}$, and relativistic coupling $U_{relativistic}$.

**Magnetic dipole coupling**

For magnetic dipole coupling, the indirect coupling strength between a $D_2/^4He$ donor transition (assumed to start out in the excited state, *i.e.*, as $D_2$) and a $D_2/^4He$ receiver transition (assumed to start out in the ground state, *i.e.*, as $^4He$) per the nuclear excitation transfer model discussed in section 5.1 and 5.2 is

$$U^2_{magnetic} = \langle \Phi_{D_2} | \boldsymbol{\mu} | \Phi_{^4He} \rangle \cdot \langle \Psi_{osc} | \hat{\mathbf{B}} \hat{\mathbf{B}} | \Psi_{osc} \rangle \cdot \langle \Psi_{D_2} | \boldsymbol{\mu} | \Psi_{^4He} \rangle \tag{79}$$

where $\Psi_{osc}$ describes the state of the spins which give rise to the magnetic fields at the sites of the $D_2$ donor systems and the $^4He$ receiver systems.

See section 6.13 for a derivation of the terms, which yields

$$U^2_{magnetic} \;=\; \left( 10 \sqrt{\frac{v_{nuc}}{v_{mol}}} e^{-G} \right)^2 (\mu_N B)^2 \tag{80}$$

where $v_{nuc}/v_{mol}$ is the ratio of the characteristic nuclear to molecule volume as discussed on section 6.3.

Note that magnetic coupling affects the $^5D$ state of the nuclear wave function for $D_2$ [91].

Recall that the minimum rate (starting rate) is (Eq. 78)

$$\Gamma_{min} \;=\; \frac{2}{\hbar} 2 |1-\eta| \frac{V^2_{gen}}{\Delta M c^2} \sqrt{N_{D_2} N_{^4He}} \tag{81}$$

Plugging in $U^2_{magnetic}$ for $V^2_{gen}$ yields

$$\Gamma_{min} \;=\; \frac{2}{\hbar} 2 |1-\eta| \left( 10 \sqrt{\frac{v_{nuc}}{v_{mol}}} e^{-G} \right)^2 \frac{(\mu_N B)^2}{\Delta M c^2} \sqrt{N_{D_2} N_{^4He}} \tag{82}$$

To evaluate this expression we need to consider the values of all relevant parameters. We take:

$$|1-\eta| \;=\; 0.1 \qquad \frac{v_{nuc}}{v_{mol}} \;=\; 6.64 \times 10^{-12} \qquad G = 94.8 \tag{83}$$

where the volume ratio and the Gamow factor are for the $^5D$ state and where $|1-\eta|$ is for the $D_2$/3+1/$^4He$ contribution to the $D_2/^4He$ transition per section 6.5. The volume ratio and the (unscreened) Gamow factor G take on the values determined in section 6.3 and are discussed in more detail there.

Moreover, we follow the scenario given in the introduction with

$$\frac{N_{D_2}}{N} \;=\; 0.001 \times 0.3 \times \frac{1}{25} \qquad \frac{N_{^4He}}{N} \;=\; 0.01 \tag{84}$$

where we assume a loaded lattice where every 1000$^{th}$ unit cell contains a $D_2$ molecule and where 30% of the $D_2$ molecules are in the $^5D$ state (see Figure 38 in section 6.1); with only one of the 25 possible $^5D$ states (with $J = 1$) being involved in the $-\boldsymbol{\mu} \cdot \mathbf{B}$ transition for a $z$-directed magnetic field; and where every 100$^{th}$ unit cell contains a $^4He$ nucleus.

The nuclear magneton has a value of

$$\mu_N \;=\; 3.15 \times 10^{-8}\,\frac{\mathrm{eV}}{\mathrm{T}} \tag{85}$$

and we assume a magnetic field strength of

$$B \;=\; 0.1\,\mathrm{T} \tag{86}$$

Note that this model would be most relevant under conditions where the nuclei that undergo transitions are in the vicinity of a strong magnet such that the interaction of the nuclei with any individual electronic spin is idealized to be uniform.

Plugging in all of the above values results in a minimum transfer rate (unscreened) of

$$\Gamma_{min,unscreened} \;=\; 2.64 \times 10^{-89}\,\mathrm{s}^{-1} \tag{87}$$

Since we used the unscreened Gamow factor $G$ above, we can consider the effects of screening by multiplying with (per section 6.3)

$$e^{2\Delta G_{scr}} \tag{88}$$

where

$$\Delta G_{scr} = 47.8 \tag{89}$$

accounts for electron screening of the $^5$D channel of the $\mathbf{D_2}$ nuclear wave function with 350 eV screening potential (see section 6.3). The screening factor per section 6.3 comes in twice here since screening affects the donor side transition as well as the receiver side transition.

This results in

$$\Gamma_{min,screened} \;=\; 8.73 \times 10^{-48}\,\mathrm{s}^{-1} \tag{90}$$

To obtain the maximum rate (peak rate of the Dicke enhancement pulse), we apply the additional Dicke enhancement, per section 5.2:

$$\Gamma_{max} \;=\; \Gamma_{min}\frac{1}{4}N_{\mathrm{D}_2} \tag{91}$$

For $N_{\mathrm{D}_2} = 10^{15}$ this yields

$$\Gamma_{max,screened} \;=\; 2.18 \times 10^{-33}\,\mathrm{s}^{-1} \tag{92}$$

Again, this is the population-level transfer rate, *i.e.*, the rate at which excitation from the ensemble in group *a* transfers to the ensemble in group *b*. To obtain a rate per deuteron pair that is comparable to the estimate given by Koonin and Nauenberg 1989 [20], we normalize the expression. Applying this normalization to the Dicke-enhanced expression we obtain

$$\frac{\Gamma_{max,screened}}{N_{\mathrm{D}_2}} = 2.18 \times 10^{-48} \text{ s}^{-1} \tag{93}$$

and

$$\frac{\Gamma_{min,screened}}{N_{\mathrm{D}_2}} = 8.73 \times 10^{-63} \text{ s}^{-1} \tag{94}$$

Note that this rate is not in the observable range (see 2.1). Also note that the same amount of screening (350 eV) applied to the base rate given by Koonin and Nauenberg 1989 [20] and reproduced in section 6.3 (see table 8) yields a rate per deuteron pair on the order of $10^{-20}$ s$^{-1}$. Therefore, inducing the $\mathbf{D_2} \rightarrow {}^4$He transition through nuclear excitation transfer based on magnetic coupling is an interesting scenario to study, but it does not appear to outcompete the spontaneous fusion rate.

**Electric dipole coupling**

Next we consider a situation that corresponds to the system discussed in 5.2 where electric dipole coupling results from oscillations of the nuclei (phonons and plasmons).

For electric dipole coupling, the indirect coupling strength between a $D_2/{}^4$He donor transition (assumed to start out in the excited state, *i.e.*, as $D_2$) and a $D_2/{}^4$He receiver transition (assumed to start out in the ground state, *i.e.*, as ${}^4$He) is

$$U^2_{electric} = \langle \Phi_{\mathrm{D}_2} | \mathbf{d} | \Phi_{^4\mathrm{He}} \rangle \cdot \langle \hat{\mathbf{E}}_{^4He} \hat{\mathbf{E}}_{^4He} \rangle_{osc} \cdot \langle \Phi_{\mathrm{D}_2} | \mathbf{d} | \Phi_{^4\mathrm{He}} \rangle \tag{95}$$

See section 6.14 for a derivation of the terms, which yields (here given for acoustic phonons)

$$U^2_{electric} = \frac{3}{4\alpha} \frac{\hbar \gamma_{rad}}{\Delta M c^2} \frac{(\hbar \omega_A)^2}{(\Delta M c^2)^2} \frac{\sqrt{M_{D_2} M_{^4He}} c^2}{N Z^2} (P_{diss} \tau_A) \tag{96}$$

where $\gamma_{rad}$ is the radiative DD fusion rate of the photon emitting channel, which is given in [92] and shown below. The fine structure constant is given by $\alpha$ and $\tau_A$ is the acoustic phonon mode lifetime as a function of the frequency of the mode, $f_A$.

Variables specific to this coupling are defined as follows [92]:

$$\alpha = \frac{e^2}{4\pi\epsilon_0 \hbar c} \tag{97}$$

$$\gamma_{rad} = 6 \times 10^{-8} \gamma_{DD} \tag{98}$$

$$\tau_A = 10^{-12} \left( 10^{-7} \frac{f_A}{1 \text{ MHz}} \right)^{-3/2} \text{sec} \tag{99}$$

where $\gamma_{DD}$ is the rate formula in terms of the (unscreened) DD rate given by Koonin and Nauenberg 1989 [20].

Recall that the minimum rate (starting rate) is (Eq. 78)

$$\Gamma_{min} = \frac{2}{\hbar} 2 |1 - \eta| \frac{V^2_{gen}}{\Delta M c^2} \sqrt{N_{\mathrm{D}_2} N_{^4\mathrm{He}}} \tag{100}$$

Plugging in $U^2_{electric}$ for $V^2_{gen}$ yields

$$\Gamma_{min} \;=\; |1-\eta| \frac{3}{\alpha Z^2} \frac{(\hbar\omega_A)^2}{(\Delta Mc^2)^2} \frac{\sqrt{M_{D_2} M_{^4He}} c^2}{\Delta Mc^2} \frac{P_{diss}\tau_A}{\Delta Mc^2} \frac{\gamma_{rad}}{\gamma_{DD}} \gamma_{DD} \frac{\sqrt{N_{\mathrm{D}_2} N_{^4\mathrm{He}}}}{N} \tag{101}$$

Here, we take advantage of knowing the ratio $\gamma_{rad}$ to $\gamma_{DD}$ so we can express this rate formula in terms of the (unscreened) DD rate, which allows for easy comparison. $\gamma_{DD}$ is given as (section 6.3):

$$\gamma_{DD} = 3 \times 10^{-64} s^{-1} \tag{102}$$

The above unscreened rate $\gamma_{DD}$ is consistent with the use of a Gamow factor of:

$$G = 90.4 \tag{103}$$

which is appropriate for the $^3$P states considered for electric dipole coupling (see section 6.3).

To evaluate $\Gamma_{min}$, we need to consider all the relevant paraemters.

We choose experimentally plausible values for vibrational power dissipation and frequency applied to the lattice as

$$P_{diss} \;=\; 1\,\mathrm{W} \qquad f_A \;=\; 5\,\mathrm{MHz} \tag{104}$$

which gives a phonon lifetime of $\tau_A \approx 0.003$sec.

We use the same deuterium and helium loading values as for magnetic dipole coupling (see Eq. 84)

$$\frac{N_{\mathrm{D}_2}}{N} \;=\; 0.001 \times 0.3 \times \frac{1}{25} \qquad \frac{N_{^4\mathrm{He}}}{N} \;=\; 0.01 \tag{105}$$

and the same loss correction factor $|1-\eta| = 0.1$ as for magnetic dipole coupling (see Eq. 83).

Evaluating the rate expression based on the above values (using $Z = 2$) results in a minimum transfer rate (unscreened) of

$$\Gamma_{min,unscreened} \;=\; 3.39 \times 10^{-94} s^{-1} \tag{106}$$

Since we used the unscreened Gamow factor G above, we can consider the effects of screening by multiplying with (per section 6.3)

$$e^{2\Delta G_{scr}} \tag{107}$$

where

$$\Delta G_{scr} = 49.2 \tag{108}$$

accounts for electron screening of the $^3$P channel of the $\mathbf{D_2}$ nuclear wave function with 350 eV screening potential (see section 6.3).

The screening factor per section 6.3 comes in twice here since screening affects the donor side transition as well as the receiver side transition.

This results in

$$\Gamma_{min,screened} = 1.84 \times 10^{-51}\ \mathrm{s}^{-1} \tag{109}$$

To obtain the maximum rate (peak rate of the Dicke enhancement pulse), we apply the additional Dicke enhancement per section 5.2:

$$\Gamma_{max} = \Gamma_{min}\frac{1}{4}N_{\mathrm{D}_2} \tag{110}$$

For $N_{\mathrm{D}_2} = 10^{15}$ this yields

$$\Gamma_{max,screened} = 4.6 \times 10^{-37}\ \mathrm{s}^{-1} \tag{111}$$

Again, this is the population-level transfer rate, *i.e.*, the rate at which the ensemble in group *a* transfer excitation to the ensemble in group *b*. To obtain a rate per deuteron pair that is comparable to the estimate given by Koonin and Nauenberg 1989 [20], we normalize the expression. Taking the Dicke-enhanced expression we obtain

$$\frac{\Gamma_{max,screened}}{N_{\mathrm{D}_2}} = 4.6 \times 10^{-52}\ \mathrm{s}^{-1} \tag{112}$$

and

$$\frac{\Gamma_{min,screened}}{N_{\mathrm{D}_2}} = 1.84 \times 10^{-66}\ \mathrm{s}^{-1} \tag{113}$$

Similar to the previous subsection, this obtained rate estimate is neither in the observable range nor does it appear to outcompete the spontaneous fusion rate.

**Relativistic coupling**

In the same physical system as considered in the previous subsections (nuclei oscillating), a different interaction is expected to apply as well. Relativistic coupling results from the fact that oscillating nuclei in the lattice undergo Lorentz transformations, which implies a coupling between center of mass motion and internal nuclear degrees of freedom (see section 6.7 and [93] for more details about this coupling mechanism).

For relativistic coupling, the indirect coupling strength between a $D_2$/$^4$He donor transition (assumed to start out in the excited state, *i.e.*, as $D_2$) and a $D_2$/$^4$He receiver transition (assumed to start out in the ground state, *i.e.*, as $^4$He) is

$$U^2_{relativistic} = \langle \Phi_{^4\mathrm{He}}|\mathbf{a}|\Phi_{^4\mathrm{He}}\rangle \cdot c^2 \langle \hat{\mathbf{P}}_{^4He}\hat{\mathbf{P}}_{^4He}\rangle \cdot \langle \Phi_{\mathrm{D}_2}|\mathbf{a}|\Phi_{^4\mathrm{He}}\rangle \tag{114}$$

See section 6.15 for a derivation of the terms, which yields (here given for acoustic phonons)

$$U^2_{relativistic} = \left(0.0362\sqrt{\frac{v_{nuc}}{v_{mol}}}e^{-G}\right)^2 c^2 \frac{\sqrt{M_{D_2}M_{^4He}}}{N} P_{diss}\tau_A \tag{115}$$

where the variables are the same as in prior subsections. Note that the masses $M_{D_2}$ and $M_{^4He}$ are essentially the same as far as the phonons are concerned.

Recall that the minimum rate (starting rate) is (Eq. 78)

$$\Gamma_{min} \;=\; \frac{2}{\hbar} 2|1-\eta| \frac{V_{gen}^2}{\Delta M c^2} \sqrt{N_{\mathrm{D}_2} N_{^4\mathrm{He}}} \tag{116}$$

Plugging in $U_{relativistic}^2$ for $V_{gen}^2$ yields

$$\Gamma_{min} \;=\; 4|1-\eta| \left( 0.0362 \sqrt{\frac{v_{nuc}}{v_{mol}}} e^{-G} \right)^2 \frac{\sqrt{M_{D_2} M_{^4He}} c^2}{\Delta M c^2} \frac{P_{diss} \tau_A}{\hbar} \frac{\sqrt{N_{\mathrm{D}_2} N_{^4\mathrm{He}}}}{N} \tag{117}$$

To evaluate this expression we need to consider the values of all relevant parameters. We take:

$$|1-\eta| \;=\; 0.1 \qquad \frac{v_{nuc}}{v_{mol}} \;=\; 6.65 \times 10^{-12} \qquad G = 90.35 \tag{118}$$

where the volume ratio and the Gamow factor are for the $^3$P state and where $|1-\eta|$ is for the $D_2$/3+1/$^4$He contribution to the $D_2$/$^4$He transition per section 6.5.

Moreover (per this section's introduction):

$$\frac{N_{\mathrm{D}_2}}{N} \;=\; 0.001 \times 0.25 \times \frac{1}{9} \qquad \frac{N_{^4\mathrm{He}}}{N} \;=\; 0.01 \tag{119}$$

where we assume a loaded lattice where every 1000th unit cell contains a $D_2$ molecule and where 25% of the $D_2$ molecules are in the $^3$P state (see Figure 38 in section 6.1); with only one of the nine possible $^3$P states (with $J = 1$) being involved in the $\mathbf{a} \cdot c\mathbf{P}$ transition (see section 6.9 and 6.8); and where every 100th unit cell contains a $^4$He nucleus.

We choose experimentally plausible values for vibrational power dissipation and frequency applied to the lattice as

$$P_{diss} \;=\; 1\,\mathrm{W} \qquad f_A \;=\; 5\,\mathrm{MHz} \tag{120}$$

which gives a phonon lifetime of:

$$\tau_A \;=\; 10^{-12} \left( 10^{-7} \frac{f_A}{1\,\mathrm{MHz}} \right)^{-3/2} \mathrm{sec}$$
$$\approx 0.003\ \mathrm{sec} \tag{121}$$

Evaluating the rate expression based on the above values results in a minimum transfer rate (unscreened) of

$$\Gamma_{min,unscreened} \;=\; 2.32 \times 10^{-63}\ \mathrm{s}^{-1} \tag{122}$$

Since we used the unscreened Gamow factor G above, we can consider the effects of screening by multiplying with (per section 6.3)

$$e^{2\Delta G_{scr}} \tag{123}$$

where

$$\Delta G_{scr} = 49.2 \tag{124}$$

accounts for electron screening of the $^3$P channel of the $D_2$ nuclear wave function with 350 eV screening potential (see section 6.3).

This results in

$$\Gamma_{min,screened} = 1.26 \times 10^{-20}\ \mathrm{s}^{-1} \tag{125}$$

To obtain the maximum rate (peak rate of the Dicke enhancement pulse), we apply the additional Dicke enhancement per section 5.2:

$$\Gamma_{max} = \Gamma_{min}\frac{1}{4}N_{\mathrm{D}_2} \tag{126}$$

For $N_{\mathrm{D}_2} = 10^{15}$ this yields

$$\Gamma_{max,screened} = 3.15 \times 10^{-6}\ \mathrm{s}^{-1} \tag{127}$$

Again, this is the population-level transfer rate, *i.e.*, the rate at which the ensemble in group $a$ transfer excitation to the ensemble in group $b$. To obtain a rate per deuteron pair that is comparable to the estimate given by Koonin and Nauenberg 1989 [20], we normalize the expression. Taking the Dicke-enhanced expression we obtain

$$\frac{\Gamma_{max,screened}}{N_{\mathrm{D}_2}} = 3.15 \times 10^{-21}\ \mathrm{s}^{-1} \tag{128}$$

Such a rate estimate means that this kind of excitation transfer would get into the observable range. Let us therefore also consider the minimum rate per $D_2$ to get a better sense of whether the process of time-evolving Dicke factors from the minimum to the maximum rate has a chance to get started:

$$\frac{\Gamma_{min,screened}}{N_{\mathrm{D}_2}} = 1.26 \times 10^{-35}\ \mathrm{s}^{-1} \tag{129}$$

Several points can be noted here. We can see that the estimated transfer rate based on relativistic coupling is substantially faster than the obtained rates for magnetic and electric dipole coupling. However, these rates are still not in the observable range. While the maximum rate per deuteron pair outcompetes the spontaneous rate from Koonin and Nauenberg 1989 [20] (see section 6.3), the minimum rate per deuteron pair does not.

Note also that the time evolution of Dicke factors has to take place within the constraint of the decoherence time of the system. We provide an estimate for the decoherence time in section 6.21, yielding a value of about 1 ns.

### Comparison of coupling strengths

The calculations in this section suggests that the maximum rate for cooperatively enhanced resonant excitation transfer based on the relativistic $\mathbf{a} \cdot c\mathbf{P}$ interaction is orders of magnitude larger than what would be expected from the $\mathbf{d} \cdot \mathbf{E}$ and $\boldsymbol{\mu} \cdot \mathbf{B}$ interactions above.

However, all rates estimated in this section are below the observable range. As discussed in the introduction and as can be seen from Eqs. 75 and 76 as well as Figure 13, a major hindrance factor is the transition through the Coulomb barrier that occurs twice per each transfer (once for the $D_2$ molecule tunneling inward to the $^4$He state and once for the $^4$He nucleus tunneling apart to the $D_2$ state). After this exercise, a faster transfer that we consider to be relevant to experimental reports of observable rates will be considered in the next section.

## 5.4 Pd nuclei as receiver systems

In addition to helium as receiver nuclei, as discussed in earlier sections, heavier lattice nuclei can be considered as candidate receiver systems for the transfer of excitation from $D_2$.

It is unlikely that such nuclei exhibit excited states precisely resonant with the $D_2/^4He$ transition at 23,848,109 eV. However, several lower-energy transitions can be resonant with the $D_2/^4He$ transition, as will be discussed in detail in section 5.7 and the subsequent sections. Comparatively small energy differences can be made up via energy exchange with oscillator modes when the system first starts up. After the system starts up, the receiver system will have more energy, so that states at higher energy (in the vicinity of multiple $D_2/^4He$ transition energies) can be accessed, where the density of nuclear molecule cluster states can be high.

As will be discussed in section 5.10 in more detail, significant energy exchange with optical phonons and plasmons looks to be possible under conditions where the density of states (including states from multiple transitions) is high, and when the phonon mode (or plasmon mode) that is coupled to and emitted into is highly excited.

This section provides an overview of nuclear excited states available in a Pd lattice, although most points raised apply to other elements as well. Natural Pd contains six stable isotopes—$^{102}$Pd, $^{104}$Pd, $^{105}$Pd, $^{106}$Pd, $^{108}$Pd, and $^{110}$Pd—so excited states from all these isotopes can be expected to be amply available in a Pd lattice. When seeking to identify states suitable for nuclear excitation transfer dynamics in mid-sized nuclei such as Pd, two categories of nuclear excited states are candidates that are discussed here. First, there are nuclear bound states; and second, there are nuclear cluster states.

We will begin with an overview of nuclear bound states in Pd isotopes and then proceed with an overview of nuclear cluster states, with a focus on nuclear molecule states, which represent a comparatively long-lived subset of nuclear cluster states. This section concludes with a discussion of key properties that make candidate excited states suitable as participants in nuclear excitation transfer dynamics.

### Pd bound states

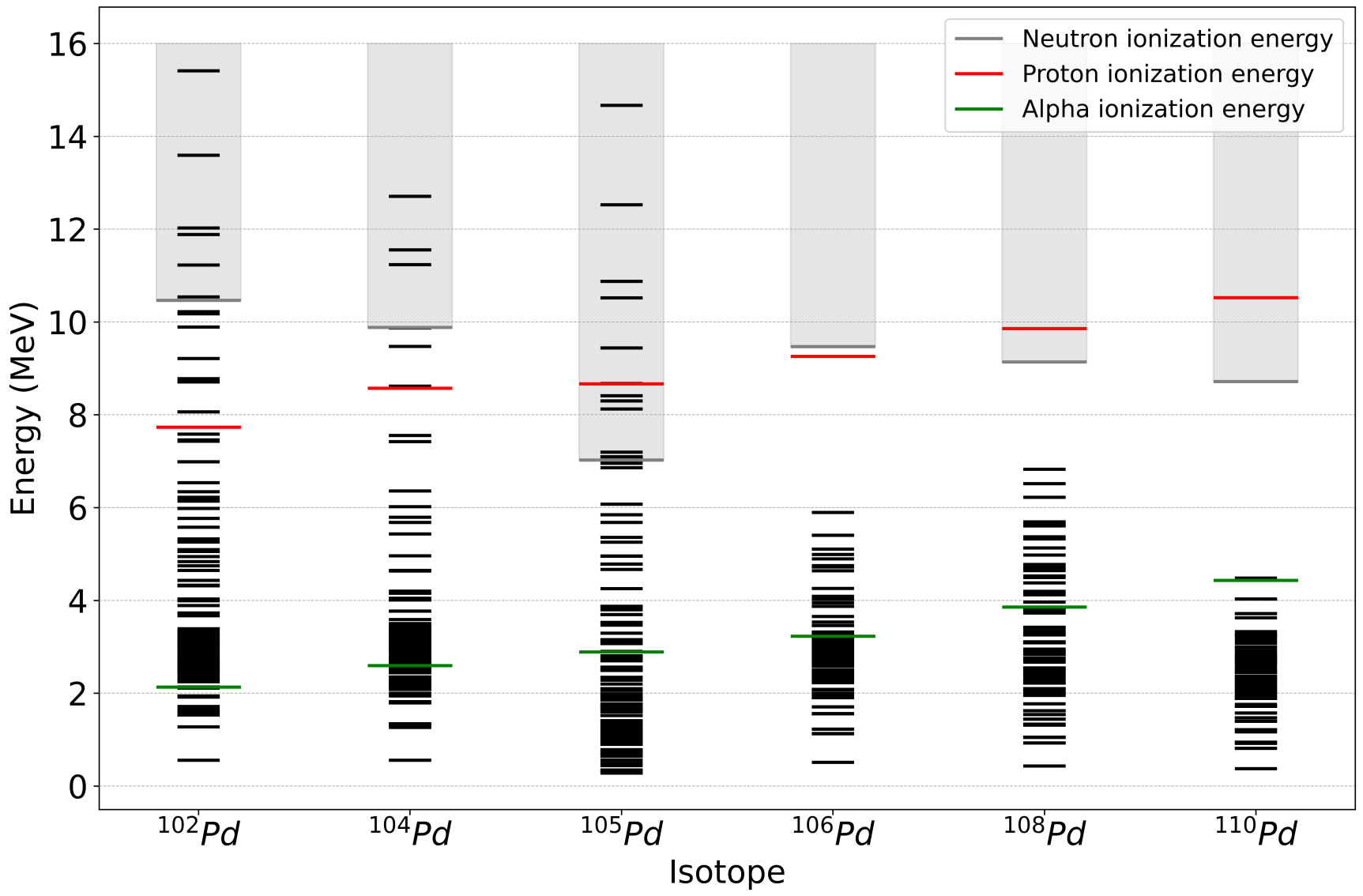

**Figure 15:** Known nuclear bound states for stable palladium (Pd) isotopes from the NuDat database.

Nuclear bound states have been widely studied for many isotopes, with aggregated data on corresponding energy levels, lifetimes, and transitions publicly available in nuclear databases such as NuDat. Known bound states of the stable Pd nuclei from NuDat are shown in Figure 15. Above the neutron ionization threshold (in the 7.0-10.5 MeV range, depending on the Pd isotope), the bound states are quite unstable against neutron decay. Above the proton ionization threshold (in the 7.7-10.5 MeV range, again depending on the isotope), the bound states become increasingly unstable against proton decay. Similarly, above the alpha ionization threshold (in the 2.1-4.4 MeV range, depending on the isotope) the bound states are unstable against alpha decay, again the more unstable the higher above the ionization energy. Proton, neutron and alpha removal energies are tabulated in section 6.17.

The relativistic $\mathbf{a} \cdot c\mathbf{P}$ interaction is a tensor operator with magnetic quadrupole selection rules (see section 6.8), and, when considering Pd nuclei as receiver systems for nuclear excitation transfer, we are interested in Dicke enhanced transitions from the ground states of the stable Pd isotopes. If the energies, lifetimes and multipolarities of transitions coupling to the ground state were available for all of the Pd bound states, then we would be able to simply make use of the databases to compile nuclear data with which we could analyze the excitation transfer schemes of interest. However, the datasets of nuclear bound states are not complete. The density of nuclear bound states increases with energy, but the density of energy levels in the canonical nuclear databases such as NuDat do not follow this expected relationship for levels above about 3 MeV. This is because at present techniques are lacking to study excited states efficiently and systematically. In addition, there have been no important applications to motivate such studies until now.

The nuclear bound states of some isotopes have been studied more comprehensively than is the case with the Pd isotopes to date. We draw attention to the bound states of $^{120}$Sn, as reported in [94]. An energy level diagram is seen in Figure 16. Here we can see that, the density of states increases with energy, as expected. This suggests that fewer states are missing from that dataset. Since the number of nucleons in $^{120}$Sn nuclei is similar to that in Pd isotopes, we can derive important lessons from the $^{120}$Sn dataset for Pd nuclei, as will be laid out below.

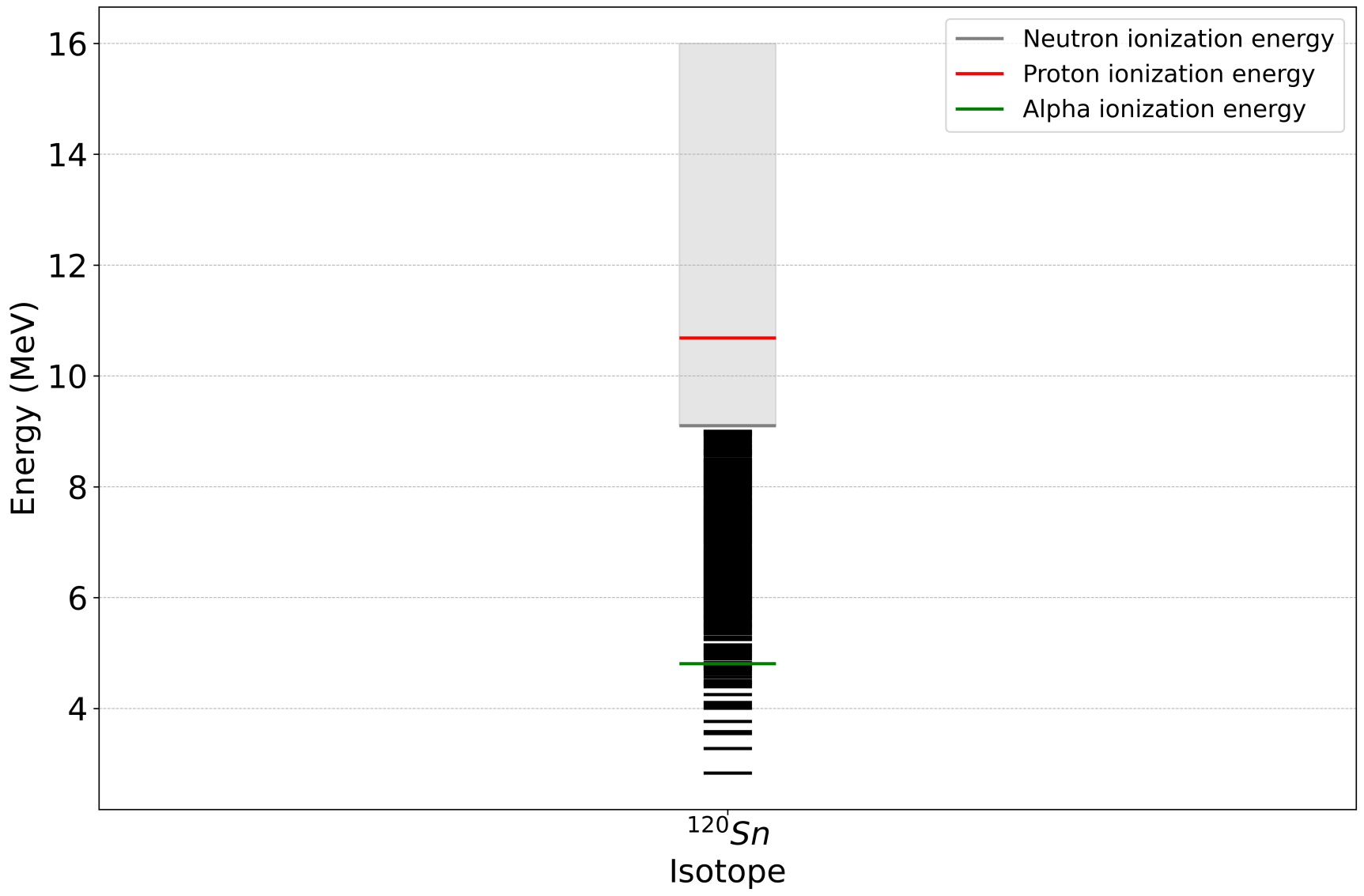

**Figure 16:** Known nuclear bound states for the $^{120}$Sn (tin) isotope.

### Pd bound states with M2 transitions to the ground state

Nuclear level diagrams have been developed from experiments involving radioactive decay from neighboring isotopes, and from experiments involving neutron, proton and alpha irradiation. Most of the transitions between the ground states and excited states at a few MeV are electric quadrupole (E2) transitions, which tend to yield large signals in accelerator-based measurement campaigns. There are fewer electric dipole (E1) transitions, but these are generally easy to detect as well. The transitions of interest to us involve weaker magnetic quadrupole (M2) coupling with the ground states, which tend to be excited less efficiently in a collision. Moreover, M2 gamma lines tend to be weaker because electromagnetic quadrupole coupling in radiative decay is a relatively weak higher-order effect. And as will become clear later on, we are interested in *weak* M2 transitions, which are in general even easier to miss in an experiment. Consequently, it is likely the case that the transitions that are of particular interest are severely undercounted in the databases.

**M2 radiative decay**

As will be discussed in more detailed in section 5.8, coherent nuclear dynamics are most efficient in the case of excited states that are reasonably stable. This is the case because quantum coherence is degraded or lost when an excited state decays. The coherent dynamics are fastest in the case of transitions that have a stronger coupling. However, there is a trade-off. The strength of coupling relevant to the coherent dynamics is proportional to the square root of the normalized transition strength $\sqrt{O^2}$, which can be expressed through a dimensionless coupling constant

$$V \sim O \tag{130}$$

whereas the spontaneous radiative decay rate is proportional to the normalized transition strength $O^2$:

$$\Gamma_{M2} \ \sim \ 2.2 \times 10^7 \ A^{2/3} \ \left(\frac{\epsilon}{1 \text{ MeV}}\right)^5 \ O^2_{\text{Pd}} \, \text{s}^{-1} \tag{131}$$

with $\Gamma_{M2}$ being the Weisskopf estimate scaled by the the transition strength which will discussed further below.

For more background on the normalized transition strength $O^2$, see section 5.8 and below.

Transitions that might be assumed to be particularly suitable for coherent dynamics in the model presented here—due to their strong coupling strength—are also the transitions which will suffer the fastest radiative loss. This competition between coherent and radiative dynamics, and its impact on nuclear excitation transfer at large, will be discussed in greater detail in section 5.8.

To estimate how fast this radiative loss might be, we can make use of the Weisskopf estimate for radiative decay. The Weisskopf estimate for radiative decay is grounded in the concept of single-particle transitions, assuming that the transition matrix elements are primarily determined by the motion of a single nucleon. The model incorporates basic assumptions about the nuclear structure, such as the spherical shape of the nucleus and the use of harmonic oscillator wave functions for the nucleons. It calculates transition probabilities by considering the multipolarity of the emitted radiation (*e.g.*, electric dipole, quadrupole) and derives standard transition rates for each type of multipole radiation.

Weisskopf estimates for the radiative decay rate for M2 transitions are shown in Figure 17 for the six stable Pd isotopes. Later on we will make use of this estimate when determining the minimum excitation transfer rates from the $D_2$/$^4$He transition needed to outpace decoherence processes and to get coherent dynamics into a stable regime. This is in addition to an estimate for the $D_2$ decoherence time, which is on the order of 1 ns. Inspection of Figure 17 indicates that even though M2 transitions are not strong, if fully allowed, the transitions are sufficiently fast so as to impact the coherent dynamics, when the transition energy exceeds 1 MeV. Given that there are very few M2 transitions at or below 1 MeV, this suggests that we are interested in weak M2 transitions at higher energies.

As the name indicates, the Weisskopf estimate of a state's lifetime is only an estimate. Actual lifetimes depend on the particular dynamics of nucleons in a transition, which can go far beyond the comparatively simple considerations underlying the Weisskopf estimate.

We capture the dynamics that the Weisskopf estimates fails to estimate in a normalized transition strength $O^2$ that satisfies

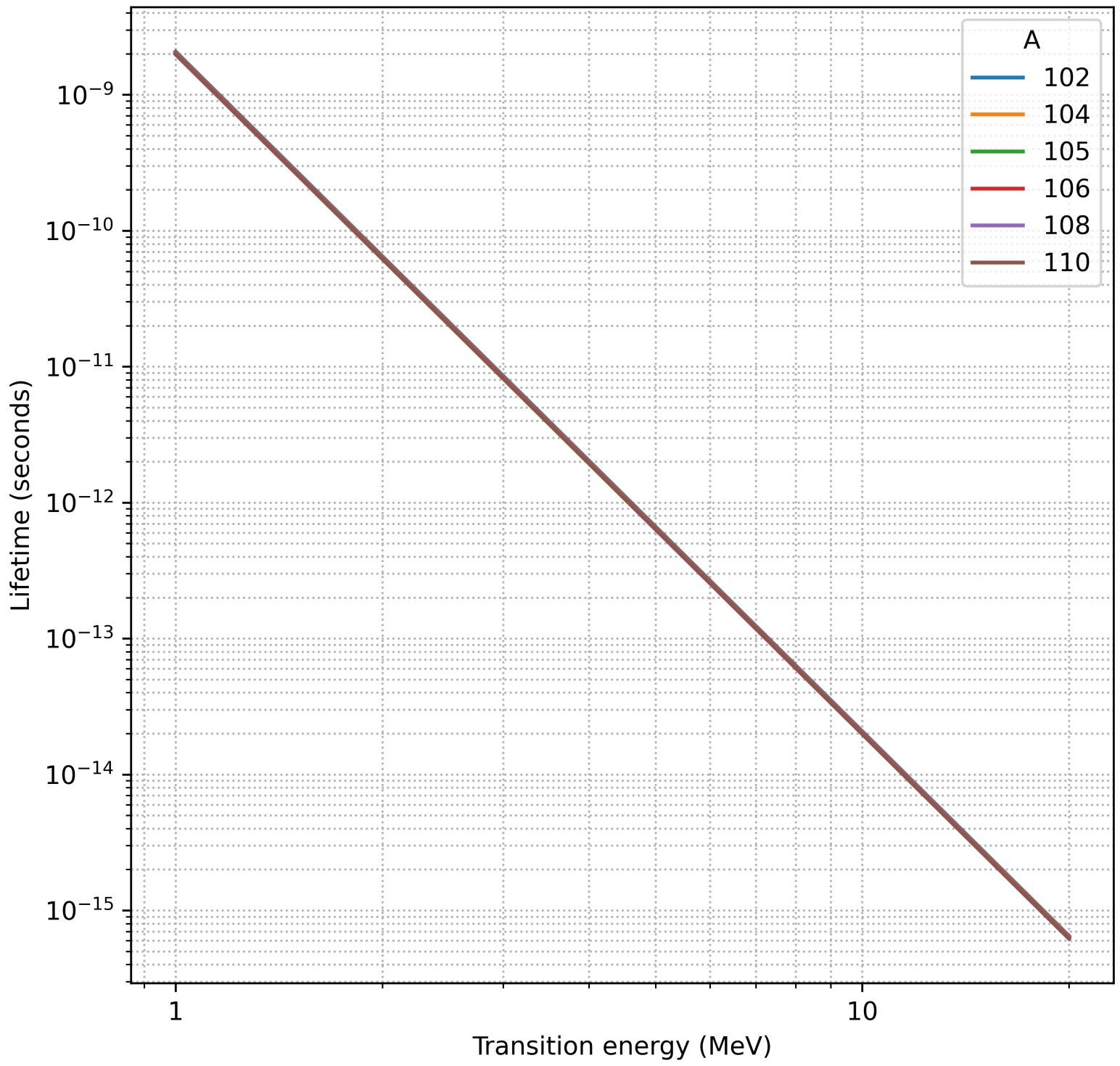

**Figure 17:** Estimated lifetimes for fully allowed M2 transitions in the stable Pd isotopes.

$$O^2 \approx \frac{\Gamma_{actual}}{\Gamma_{Weisskopf}} \tag{132}$$

With respect to relativistic coupling and associated transitions, the $O$-factor can be thought of as the square root of the normalized M2 transition strength for the transition. Since our knowledge of how big these transitions matrix elements are in the Pd isotopes is limited, we will express matrix elements and rates in terms of $O$-values, so as to explicitly distinguish between transition-specific differences. Consequently, if a transition is particularly weak—and it's lifetime is underestimated by the Weisskopf estimate—its associated $O$-value would be $\ll 1$.

### Data set for E1 transitions in $^{120}$Sn

In Müscher et al. (2022) [94], a list of 249 energy levels are given for $^{120}$Sn, along with estimates for the associated electric dipole E1 transition strength.

We might expect there to be a similar distribution of levels and transition strengths for magnetic quadrupole (M2) transitions in the stable Pd isotopes. We used the $^{120}$Sn data set in place of the stable Pd isotopes, where we estimated transition strengths according to

$$BE(E1) \uparrow \;=\; O^2 e^2 \text{fm}^2 \tag{133}$$

The resulting distribution of $O$-values and energies are shown in Figure 18.

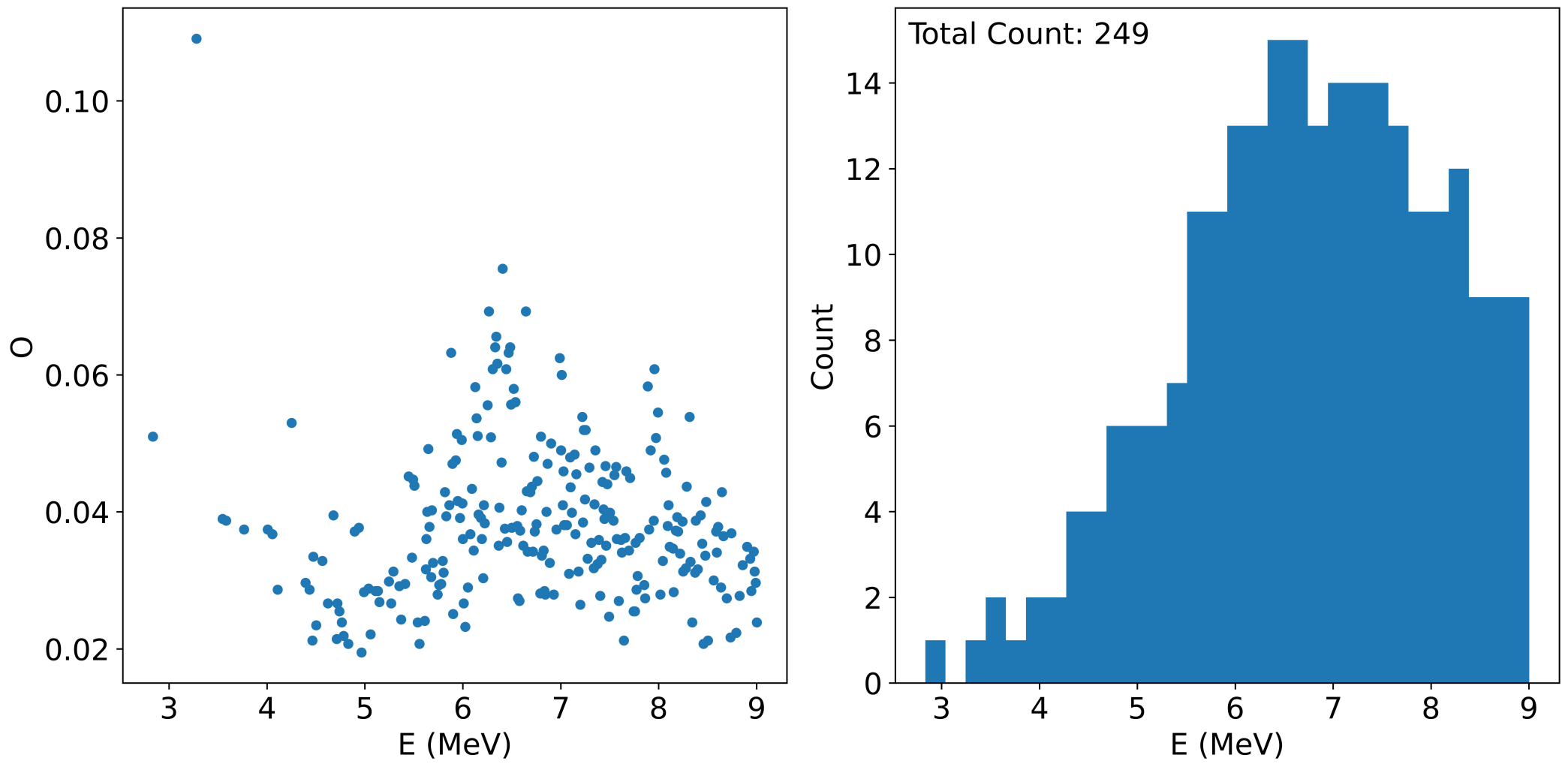

**Figure 18:** Transition strength $O$-values extracted from Müscher et al. (2020).

## Cluster-based nuclear molecule states

Earlier efforts to study the possibility of excitation transfer among excited nuclear states ran into difficulties due to the lack of reasonably stable states near the deuterium-deuterium fusion transition energy at 23.85 MeV. A possible solution to this problem lied in the recognition that reasonably stable high-energy states can form as so-called nuclear molecule states, a specific type of nuclear cluster states. In some metal-hydrogen experiments, observations that suggest fission-like transmutation of lattice nuclei has been reported. If coherent nuclear dynamics can result in the excitation of reasonably stable and highly-excited nuclear molecule states, and if these nuclear molecule states decay by tunneling, that would represent the basis of a possible explanation for such report.

This motivated the development of liquid drop model codes with which to calculate nuclear deformations, leading to the realization that the models used for deformed nuclei in existing fission and fusion literatures do not predict reasonably stable non-rotational nuclear molecules in the stable Pd isotopes. A solution to this problem involves the notion of nuclear molecules based on clusters, where two or more clusters remain nearly stationary at short range. In liquid drop models there is no significant potential barrier keeping nuclear clusters apart, so one daughter can move into the other daughter easily in a binary nuclear molecule in a liquid drop model calculation. However, in more sophisticated cluster models there is a barrier. This suggests the possibility of nuclear molecule cluster states in the Pd isotopes, where daughter parts of a nucleus are held close to the parent by strong force interactions, but where a potential barrier prevents them from coalescing.

In the literature, such nuclear cluster states have been considered in conjunction with $^{12}$C nuclei that can form molecule-like states of $^{24}$Mg. Calculations done for such states suggest the existence of reasonably stable nuclear molecule states that involve separated clusters, rather than extreme nuclear deformation. The idea is that the daughters remain localized as separate units with essentially no net relative kinetic energy. An illustration of a few cluster states in $^{24}$Mg from a recent paper of Adsley et al. (2022) [95] is shown in Figure 19. Work so far on these states has focused on low mass nuclei, where relevant experimental data exists. We conjecture that this kind of nuclear molecule cluster state also occurs in the Pd isotopes.

Nuclear cluster states are much less studied than nuclear bound states, and no comprehensive databases exist. We were interested in understanding at what energies relevant nuclear cluster states might occur in the Pd isotopes, and also how stable they might be against tunnel decay. For this, we developed a simple model for binary nuclear cluster states that consist of spherical daughters with surfaces separated by 0.5 fm, with a total energy that corresponds to known mass energies from isotope mass tables, and with folded Coulomb and strong force contributions from the finite range liquid drop model. This model is discussed further in section 6.20.

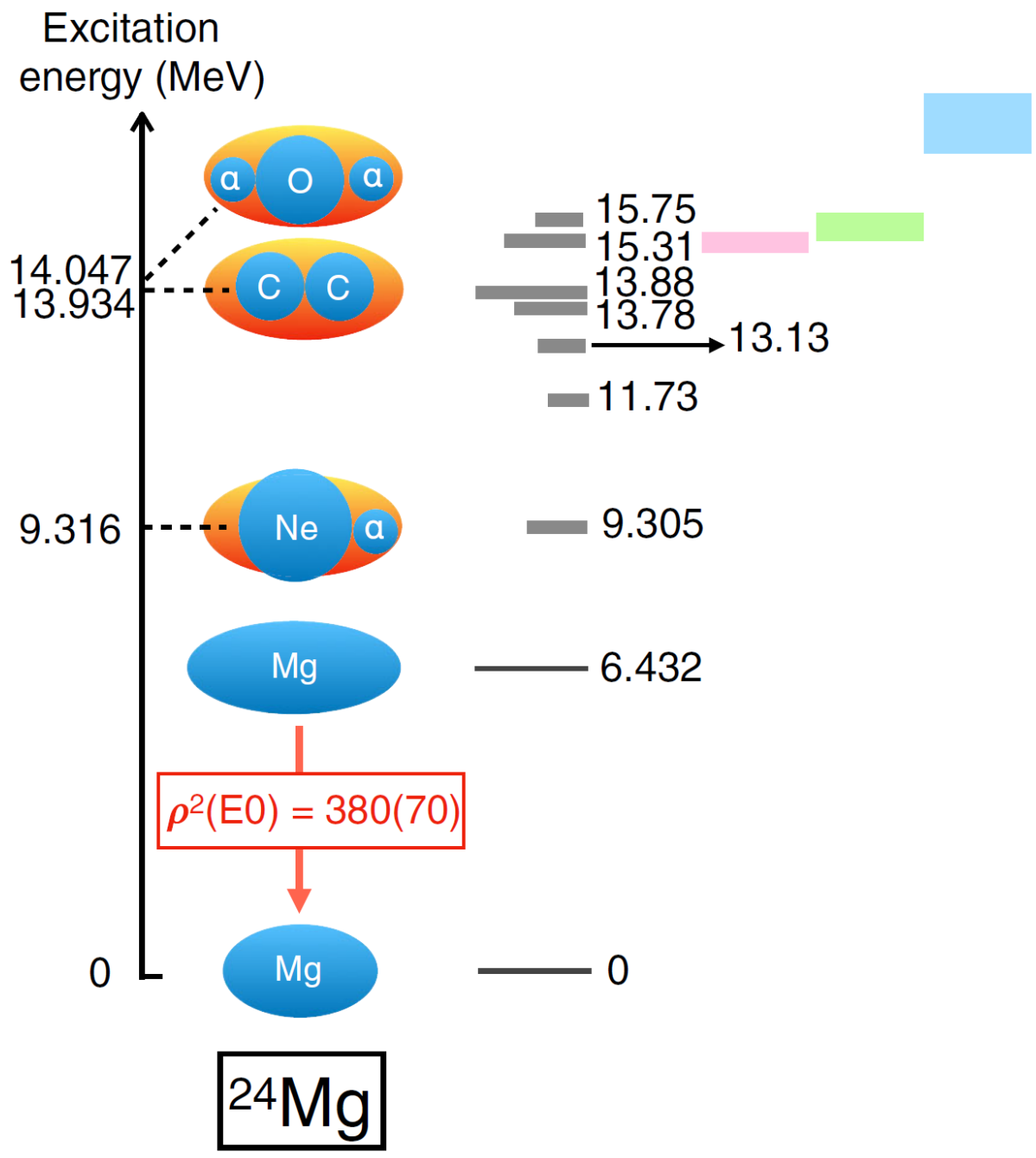

**Figure 19:** Level diagram for $^{12}$Mg depicting highly-excited (nuclear molecule) cluster states from Adsley et al. (2022) [95], reproduced with permission. Note that the indicated level positions for the nuclear cluster states only include the mass energies, with no inclusion of the binding energies which will be several MeV.

Excited state energies predicted from the model are illustrated in Figure 20. The more symmetric nuclear molecule cluster states generally appear at higher energy as a consequence of the large Coulomb repulsion, with a substantial spread in energy due to differences in the individual cluster mass energies (highly unstable ground state Pd isotopes can have significant extra mass energy). At lower energies there are more asymmetric binary clusters, including clusters of Rh + H isotopes, and Ru + He isotopes. For illustrative purposes, we will refer to Ru + He states as exemplary nuclear molecule states later in this text.

The model predicts more than 1000 non-rotational binary cluster states. The associated density of states as a function of tunnel plus beta decay lifetime is shown in Figure 21. Note that many of these binary nuclear molecule states are predicted to be quite stable against tunnel decay, with tunneling times well in excess of 1 second. In future work, we seek to refine such models and extend them so as to allow for the estimation of **a**-matrix elements connecting these states to their ground states.

Simple forms of resonant excitation transfer require a precise energy match between the initial state transition and a final state transition. More complicated generalized resonant excitation transfer requires a precise energy match between a possible combination of initial state transitions and a possible combination of final state transitions. Yet more complicated models of excitation transfer allow for energy exchange with the oscillator—in our case plasmons and phonons—to make up for energy mismatch.

The interest in states with energy levels close to the energy of the donor transition, *i.e.*, 23.85 MeV, motivates an interest in nuclear molecule cluster states more complicated than binary clusters. Such variations of nuclear molecule states offer many more configurations and therefore states in regions of interest. Most easily visualized are binary nuclear molecule cluster states with two daughters. However, the concept extends to more complicated ternary and tertiary nuclear molecule cluster states. A collection of different cluster configurations is illustrated in Figure 22.

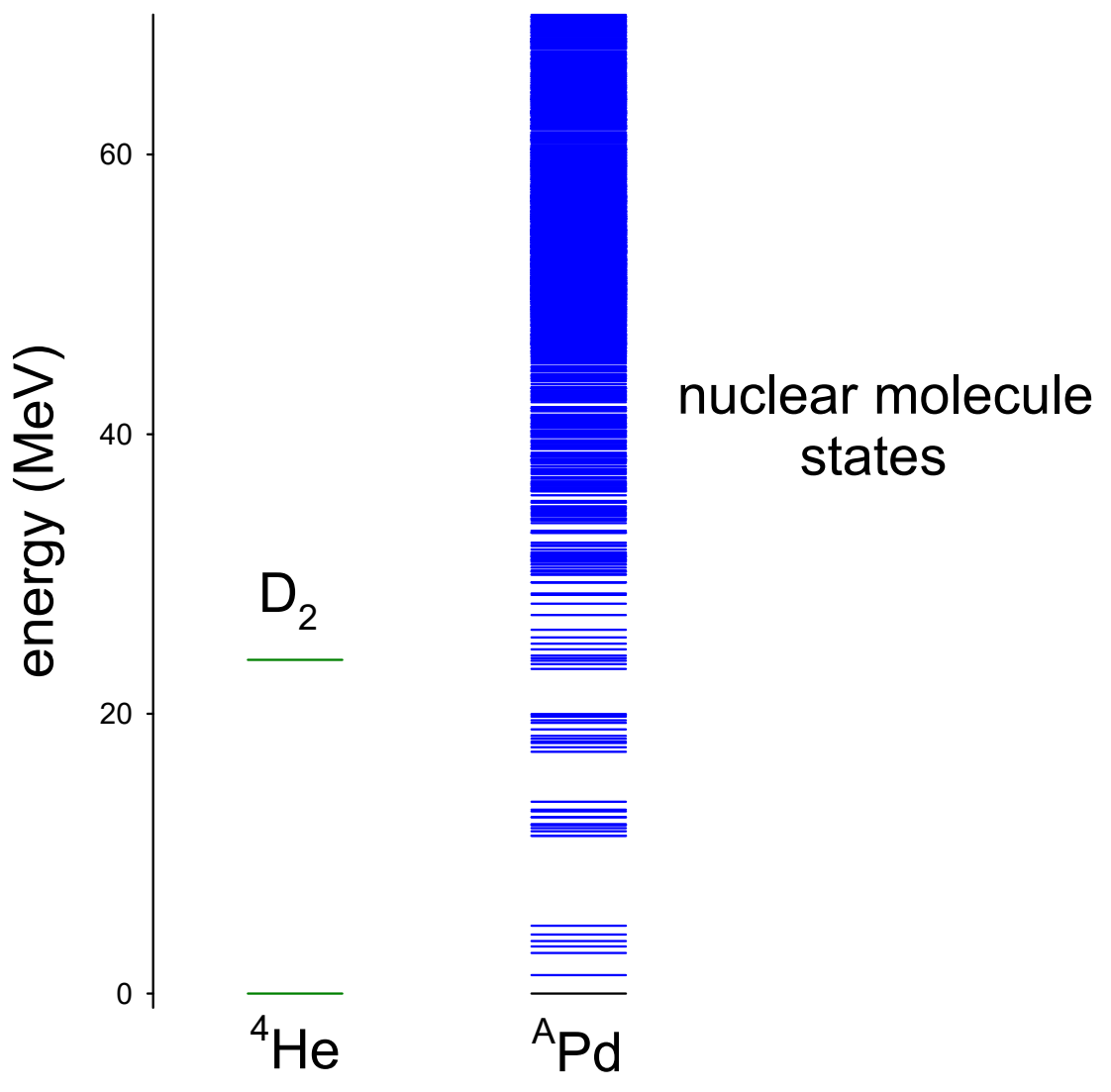

**Figure 20:** Molecular $D_2$ and ground-state $^4He$ levels along with (naive) non-rotational and non-vibrational binary nuclear cluster states up to 70 MeV of the stable Pd isotopes.

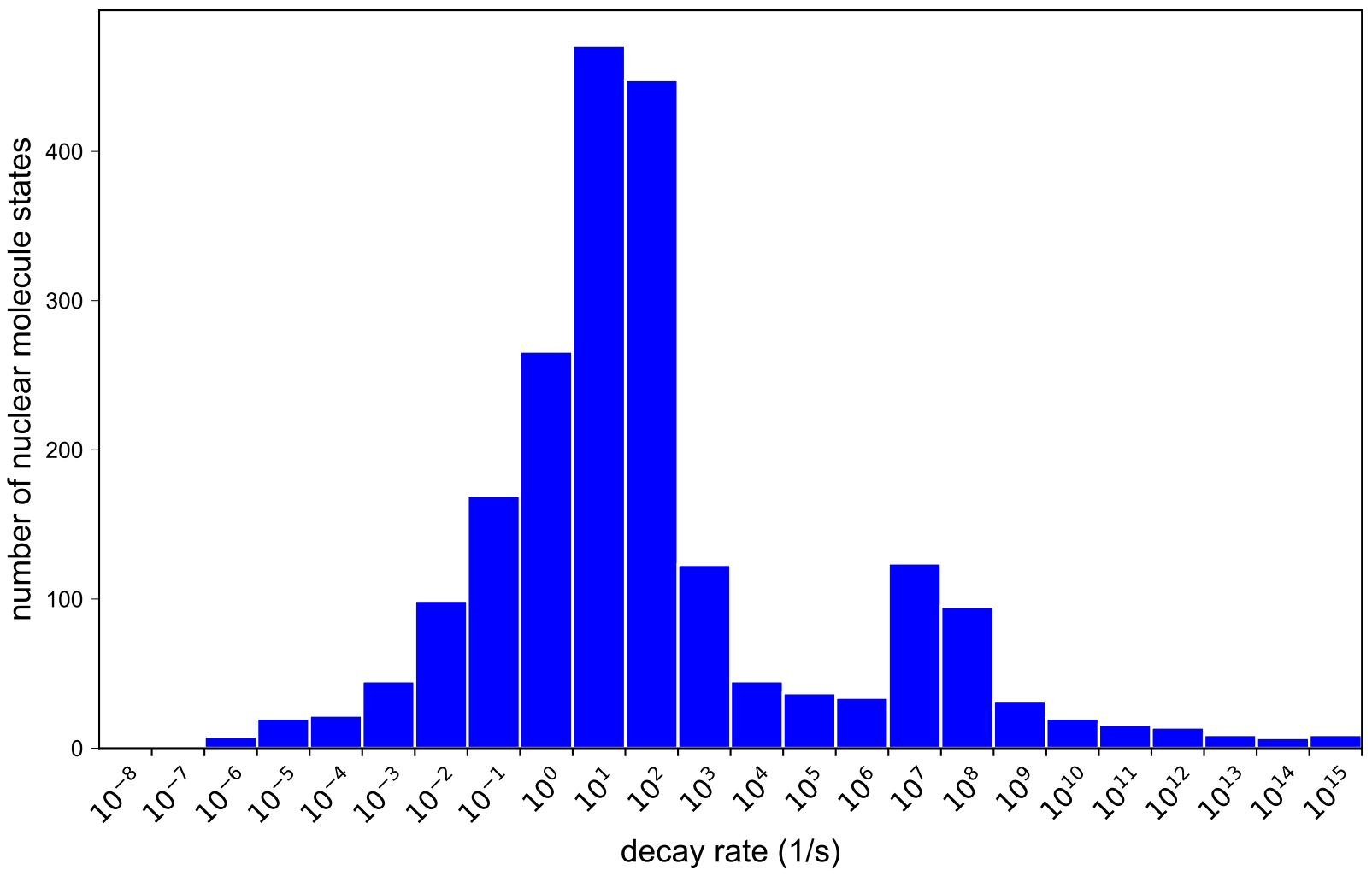

**Figure 21:** Distribution of nuclear molecule cluster state lifetimes by order of magnitude.

### Excited Pd* states and different nuclear excitation transfer schemes

The question is now what would be required to extend the notion of resonant nuclear excitation transfer–as developed in 5.1, 5.2, and 5.3–to include Pd nuclei as receiver systems. The situation would be conceptually simplest if all that was required was to find a reasonably stable highly-excited Pd* state that is precisely matched to the $D_2$/$^4$He fusion mass energy. In that hypothetical case, the previously introduced model of excitation transfer based on (simple) second-order excitation transfer could be employed, analogous to the transfer of excitation from $D_2$ to $^4$He, as considered in section 5.3.

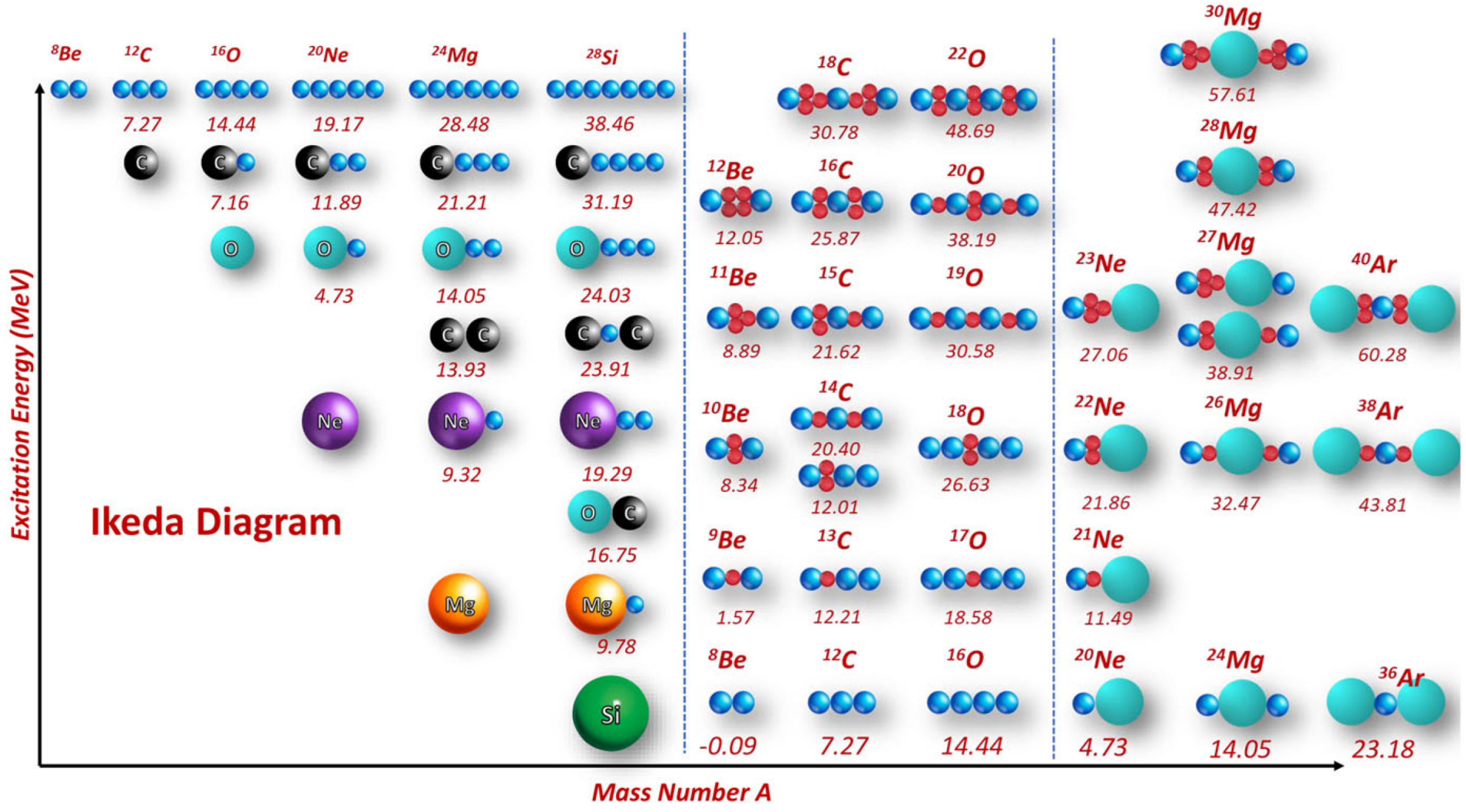

**Figure 22:** Ikeda diagram of nuclear cluster states of various isotopes from Lombardo and Aquila (2023) [96], reproduced with permission.

In reality, however, present knowledge suggests that it is unlikely for a state to exist that is nearly resonant with the donor transition of interest, the $D_2$/$^4$He transition with its 23.85 MeV energy difference. That remains to be the case, even when tertiary and other higher level types of nuclear molecule configurations are considered, which contribute to the density of states.

Consequently, a theory of nuclear excitation transfer, where the donor system is $D_2$, needs to consider scenarios that are more complicated. Here, we will, in anticipation of later sections, focus attention and aspects of nuclear excited states that are particularly relevant to the problem at hand.

In section 5.7 and onwards, we will describe a more complicated scheme in which excitation is transferred from a single transition to multiple state transitions. This is an attractive prospect, because lower receiver side state transition energies tend to correspond to slower M2 radiative decay. Moreover, with multiple lower-energy states comes the possibility of combining states and the number of available combinations—and therefore the implied density of states—grows rapidly with every state that can participate in such combinations.

For this kind of scheme, we care a lot about M2 transitions among excited Pd* bound states. What is currently impeding modeling efforts is the lack of comprehensive knowledge of bound state level energies, lifetimes, and M2 transition strengths, which are needed for more precise predictions of dynamics under this scheme. Consequently, the best option at present is to parameterize the transition strengths and work with estimates for the density of M2 accessible levels.

In section 5.9, we estimate rates for transfer to multiple transitions. To anticipate findings discussed in that section, the very high density of states that results from combinations of even a modest number of M2 transitions, suggests that the resonance condition can likely be met this way, especially when the system is in the strong coupling regime—as we find is possible under realistic circumstances. While this is a model closer to realistic conditions compared to previous idealizations, we find that unphysically extreme high power transients would be required for transfer rates predicted by this version of the model to be in the observable range.

We also consider an extension of the nuclear excitation transfer scheme presented above, where we include energy exchange with uniform highly-excited vibrational or plasmon modes. Suppose that there is a Pd* state sufficiently close that a precise resonance can be achieved with energy exchange. For example, according to the simple nuclear molecule cluster state model, the closest binary cluster states to 23.85 MeV for excitation from a stable Pd isotope ($^{A}$Pd) involves $^{A-6}$Ru+$^6$He binary cluster state. In this kind of scenario we can imagine a resonant excitation transfer scheme based on

$$M_{\mathrm{D}_2}c^2 - M_{^4\mathrm{He}}c^2 \;=\; M^*_{\mathrm{Pd}}c^2 - M_{\mathrm{Pd}}c^2 + \Delta n_P \hbar\omega_P + \Delta n_O \hbar\omega_O + \Delta n_A \hbar\omega_A \tag{134}$$

where $Pd^* = Pd^*_{Ru+6He}$.

In 5.10, we will explore how such an additional feature of energy exchange with phonon and plasmon modes affects excitation transfer rates, when considered in combination with the very high density of states associated with combinations of bound states. If the system is able to make a generalized excitation transfer from one transition to many transitions, starting from the $D_2$/$^4$He fusion transition, then it will be able to make generalized excitation transfer transitions starting from a set of many transitions, and going to a different set of many transitions (as long as all of the transitions are in the strong coupling regime, and as long as the energies match). Such a system has the potential to exchange energy with optical phonons and plasmons efficiently, as long as the magnitude of the relevant $\mathbf{a} \cdot c\mathbf{P}$ transition matrix elements for the optical phonon transitions or plasmon transitions are greater than the average energy separation between the associated multi-transition “states”.

This picture is not only conducive to high rates of nuclear excitation transfer that can outcompete decoherence rates, it also has important implications for the reaction products that become observable from such a process. Reaction products will be discussed in more detail in section 5.14, but we can anticipate already in this section that repeated excitation transfer from sets of multiple transitions to other sets multiple transitions can take place in a regime, where the density of states is so high (meV and smaller differences between states) that the exchange of small amounts of excitation energy with a coupled phonon or plasmon mode becomes possible along with every such excitation transfer. This is especially the case if the interaction is with a uniform optical phonon mode (or uniform plasmon mode) with substantial excitation (so that the interaction is in the strong coupling regime with respect to the fine-splitting between the multi-transition “states”).

This conjecture draws attention to the density of multi-transition “states” such that energy can be exchanged efficiently. In this regime, when the total energy of a coupled system gets into the high-density-of-states region, mass energy from additional $D_2$/$^4$He fusion transitions during the same coherence cycle has the potential to be converted to optical phonon and plasmon energy as part of the repeated generalized excitation transfer process.

Nuclear molecule states, with the exceptionally long lifetimes that many of them exhibit, also play an important role in this picture. They may serve as natural endpoints with comparably high stability, where a system arrives after repeated emission of energy into lattice modes. Over time, we conjecture this can lead to the build up of large populations of particular nuclear molecule states, whose eventual decay—through tunneling, *i.e.*, fission, can account for reports of low-Z element production in certain metal-hydrogen experiments.

Our interest in the Pd* bound excited states is ultimately focused on their contribution to the fine structure of the spectrum of multi-transition states. Our interest in the Pd* nuclear molecule cluster states is ultimately similarly focused on their contributions to the fine-structure of the multi-transition states, and to a lesser degree in connection with low-level nuclear emission and low-Z element production.

## 5.5 Transfer rate estimates: $D_2$ to Pd nuclei

Section 5.3 provided a first estimate for excitation transfer rates from $D_2$ donor systems to $^4$He receiver systems. It was found that under realistic experimental conditions, expected transfer rates would be unobservably low. Another key insight from that section was that relativistic coupling is significantly stronger—and therefore more relevant for the processes under consideration here—than magnetic dipole coupling and electric dipole coupling. In section 5.4 Pd nuclei were introduced as receiver systems and different kinds of Pd excited states discussed.

We expect indirect coupling between the two transitions—$D_2$/$^4$He and Pd*/Pd—to work similarly to the resonant excitation transfer model of section 5.2. Because of the comparatively high strength of relativistic coupling–as seen in 5.3– we focus on the $\mathbf{a} \cdot c\mathbf{P}$ interaction for both transitions. We know that the magnetic quadrupole (M2) of the $\mathbf{a} \cdot c\mathbf{P}$ interaction can provide for direct coupling between the $D_2$ $^3$P $J$=1,2 states and the $^4$He state, so that the use of a direct matrix element is appropriate. We do not expect there to be a significant direct coupling between the ground state of a Pd isotope and excited states (bound states or cluster-based nuclear molecule states, *e.g.*, a Ru+$^6$He state. The direct matrix element would be hindered, depending on what fractional admixture (superposition) of such clusters were already present as part of the Pd isotope ground state.

Before getting to the actual rate estimates, we start with a discussion of notation, as small variations in notation will become useful in later considerations.

**Recalling the expression for the excitation transfer rate**

The derivation of Dicke factors in 6.12 yields expressions for the minimum and maximum rate of excitation transfer. Following Eq. 69 the minimum rate is

$$\Gamma_{min} = |\Omega_{ab}|\sqrt{N_{\mathrm{D}_2} N_{^{\mathrm{A}}\mathrm{Pd}}} \tag{135}$$

where $\Omega_{ab}$ refers to the generalized Rabi frequency for oscillating between the transitions in group $a$ and $b$.

The generalized Rabi frequency is (Eq. 57)

$$\frac{1}{2}\hbar|\Omega_{ab}| = \frac{V_{gen}^2}{\Delta E} = \frac{V_{gen}^2}{\Delta Mc^2} \tag{136}$$

where again we add a subscript $V_{gen}$, as previously done in Eq. 75, to differentiate referral to a generic interaction from a concrete one.

The minimum rate with the corresponding Dicke factors is therefore:

$$\Gamma_{min} = \frac{2}{\hbar}\frac{V_{gen}^2}{\Delta Mc^2}\sqrt{N_{\mathrm{D}_2} N_{^{\mathrm{A}}\mathrm{Pd}}} \tag{137}$$

In the case of transfer from $D_2$ donors to $^4$He receivers, the transitions involved on the donor and receiver sides are the same and so for $V_{gen}^2$ we had $U^2$. For the case of transfer from $D_2$ donors to Pd receivers, the variable $U = U_0 e^{-G}$ denotes the coupling for the fusion transition and $V = V_0 O_{Pd}$ denotes the coupling for the Pd transition (per the diagrams in Figs. 23 and 24) and for $V_{gen}^2$ we get $UV$.

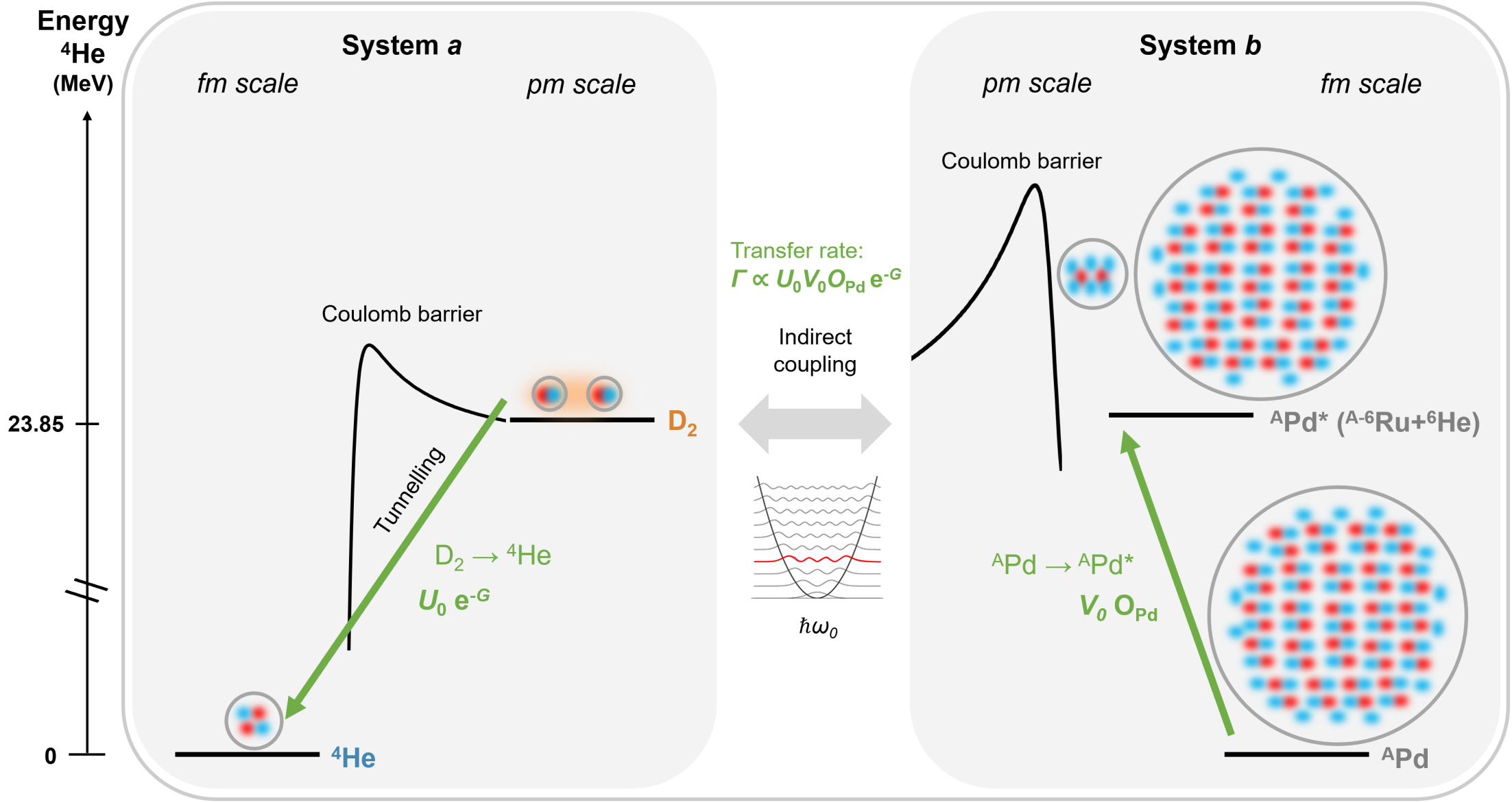

**Figure 23:** Resonant nuclear excitation transfer of 23.85 MeV from the $D_2 \rightarrow {}^4$He fusion decay to the $^A$Pd $\rightarrow$ $^A$Pd* excitation, where the latter state is an exemplary case of $^A$Pd*. Note that in actuality, this kind of excitation transfer process involves pathways through different possible intermediate states, which exhibit an asymmetry in how they are affected by loss (represented by $\eta$ in Fig. 24).

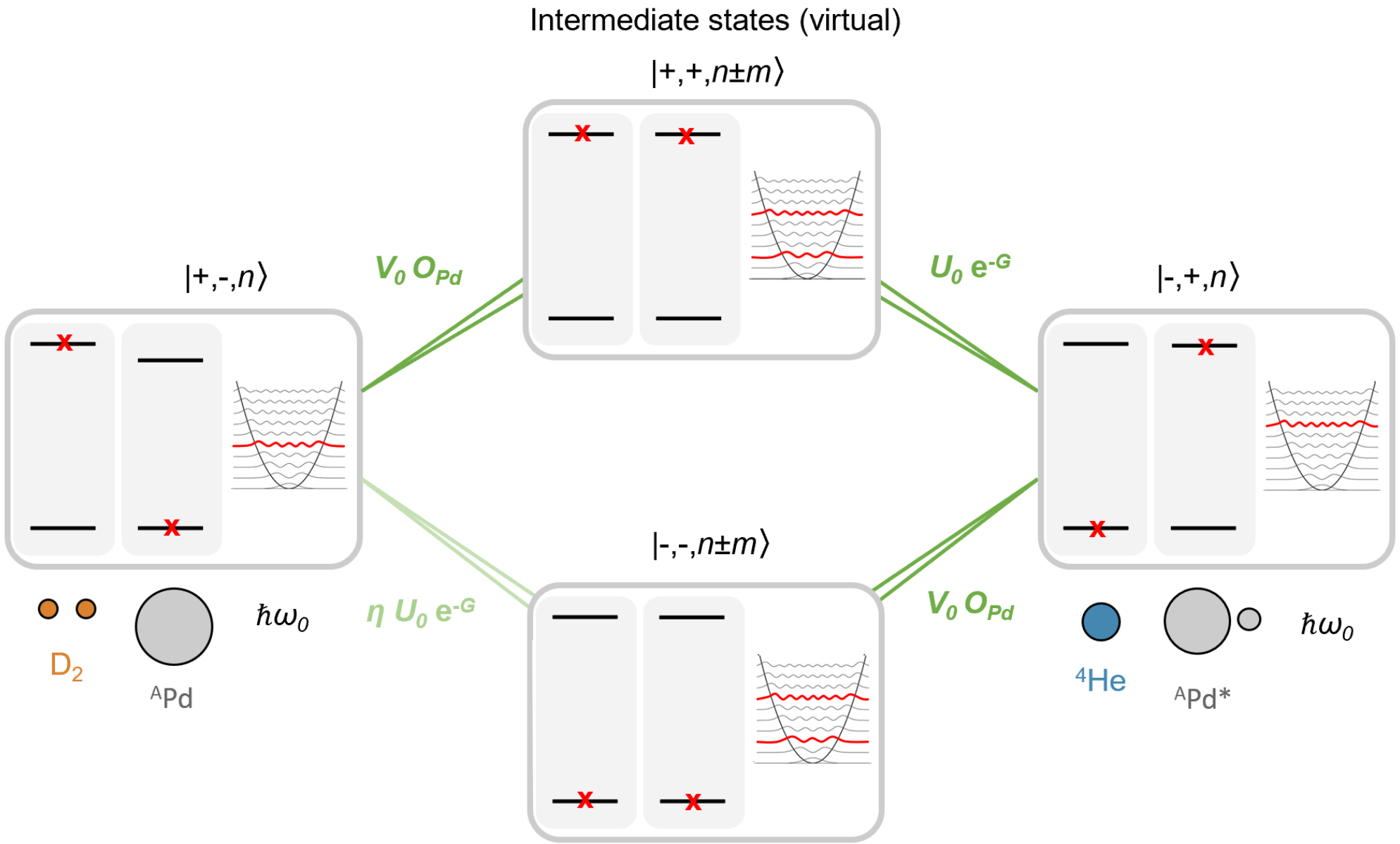

**Figure 24:** Resonant nuclear excitation transfer between a $D_2/{}^4$He decay transition and a (hypothetical) resonant Pd*/Pd excitation transition by indirect coupling via a shared oscillator mode. Destructive interference between the upper and lower intermediate (*i.e.*, virtual) states is reduced by selective loss in the lower pathway, allowing for resonant nuclear excitation transfer to occur (here represented by $\eta$ per sections 6.5 and 6.6).

Note that Eq. 137 refers to the ideal loss case (see section 5.2). To account for realistic loss the correction factor $|1-\eta|$ needs to be applied, resulting in:

$$\Gamma_{min} \;=\; \frac{2}{\hbar}|1-\eta|\frac{UV}{\Delta Mc^2}\sqrt{N_{\mathrm{D}_2}N_{^{\mathrm{A}}\mathrm{Pd}}} \tag{138}$$

In section 5.3 we found that among the three couplings considered—magnetic coupling, electric coupling, and relativistic coupling—relativistic coupling is the strongest, which is why we proceed focusing on relativistic coupling.

For relativistic coupling for the $D_2$/$^4$He transition (recall from Eq. 5.3):

$$U \;=\; |\langle \mathrm{D}_2|a_z|^4\mathrm{He}\rangle| c\sqrt{\langle P^2_{^4\mathrm{He}}\rangle} \tag{139}$$

Equivalently, for the Pd*/Pd transition we have:

$$V \;=\; |\langle ^{\mathrm{A}}\mathrm{Pd}^*|a_z|^{\mathrm{A}}\mathrm{Pd}\rangle| c\sqrt{\langle P^2_{^{\mathrm{A}}\mathrm{Pd}}\rangle} \tag{140}$$

**Useful variations in notation**

We can simplify the notation by pulling in the Dicke factors with the variables for the coupling strengths according to (here exemplary for the minimum rate case)

$$\mathcal{U} \;=\; U\sqrt{N_{\mathrm{D}_2}} \tag{141}$$

$$\mathcal{V} \;=\; V\sqrt{N_{^{\mathrm{A}}\mathrm{Pd}}} \tag{142}$$

resulting in

$$\Gamma_{transfer} \;=\; \frac{2}{\hbar}|1-\eta|\frac{\mathcal{U}\mathcal{V}}{\Delta Mc^2} \tag{143}$$

We can further simplify this notation by pulling in the correction factor with $U$, resulting in

$$\Delta U \;=\; |1-\eta| U \tag{144}$$

and

$$\Delta \mathcal{U} \;=\; |1-\eta| \mathcal{U} \tag{145}$$

The notion of $\Delta U$ refers to the origin of that term as the difference between the two pathways in the diagram in Figure 24, where one pathway is affected by loss and the other is not.

Finally, it will be useful to have normalized the coupling strength of a transition, such as $\mathcal{V}$, by the transition energy $\Delta E \equiv \Delta Mc^2$. This then results in dimensionless coupling parameter g:

$$g \;=\; \frac{|\mathcal{V}|}{\Delta Mc^2} \tag{146}$$

The result is a further simplification of the expression for the excitation transfer rate Eq. 137. It can be alternatively written as

$$\Gamma_{transfer} = \frac{2}{\hbar}|\Delta\mathcal{U}|g \tag{147}$$

We expect perturbation theory to give good results for the weak coupling limit where $g \ll 1$. For the strong coupling limit $g \geq 1$ we will need to implement a non-perturbative analysis to get reliable results. We will consider the transition from weak coupling to strong coupling in section 5.6.

We will make use of this alternative notation later in the text since it will simplify a number of important considerations. For now, we will return to the notation of Eq. 143 and proceed based on:

$$\Gamma_{transfer} = \frac{2}{\hbar}|1-\eta|\frac{\mathcal{U}\mathcal{V}}{\Delta Mc^2} \tag{148}$$

with the starting place being the version with minimum Dicke factors:

$$\Gamma_{min} = \frac{2}{\hbar}|1-\eta|\frac{UV}{\Delta Mc^2}\sqrt{N_{\mathrm{D}_2}N_{^A\mathrm{Pd}}} \tag{149}$$

### First rate estimate for excitation transfer from $D_2$ to Pd

The key difference to the rate estimates of section 5.3 is the interaction term, which is now $UV$ instead of $U^2$ due to the Pd*/Pd transition (coupling $V$) being different from the $D_2$/$^4$He transition (coupling $U$).

This gives an overall coupling strength for excitation transfer between a $D_2$/$^4$He transition and a Pd*/Pd transition of:

$$(UV)_{relativistic} = \langle\Phi_{D_2}|\mathbf{a}|\Phi_{^4\mathrm{He}}\rangle \cdot c^2\langle\hat{\mathbf{P}}_{^4He}\hat{\mathbf{P}}_{^APd}\rangle \cdot \langle\Phi_{^A\mathrm{Pd}*}|\mathbf{a}\cdot\hat{\mathbf{i}}_z|\Phi_{^A\mathrm{Pd}}\rangle \tag{150}$$

where (per section 6.15)

$$\begin{aligned}|\langle\Phi_{D2}|\mathbf{a}\cdot\hat{\mathbf{i}}_z|\Phi_{^4\mathrm{He}}\rangle| &\sim \langle\Psi[^4\mathrm{He}]|\mathbf{a}|\Psi[J=1,M_J=0]\rangle \\ &= 0.0362\sqrt{\frac{v_{nuc}}{v_{mol}}}e^{-G}\end{aligned} \tag{151}$$

and (per section 6.16)

$$\begin{aligned}&\langle\Phi_{^A\mathrm{Pd}*}|\mathbf{a}\cdot\hat{\mathbf{i}}_z|\Phi_{^A\mathrm{Pd}}\rangle \\ &= 2.6\times10^{-5}\,O\end{aligned} \tag{152}$$

The idea here is that for a fully-allowed single nucleon transition at 23.85 MeV, we would expect the magnitude of the **a**-matrix element to be roughly $2.6 \times 10^{-5}$. The $O$-factor can be thought of as the square root of the normalized M2 transition strength for the transition. Since our knowledge of how big these transition matrix elements are in the Pd isotopes is limited, we will express matrix elements and rates in terms of $O$-values, keeping in mind that if the transition is weak that we would expect $O \ll 1$.

The above expression results in

$$(UV)_{relativistic} = \left(0.0362\sqrt{\frac{v_{nuc}}{v_{mol}}}e^{-G}\right)\frac{\sqrt{M_{^4\mathrm{He}}M_{^A\mathrm{Pd}}}c^2}{N}P_{diss}\tau\left(2.6\times10^{-5}\ O\right) \tag{153}$$

Consequently, based on Eq. 149, the minimum transfer rate is:

$$\Gamma_{min} = 2|1-\eta|\left(0.0362\sqrt{\frac{v_{nuc}}{v_{mol}}}e^{-G}\right)\left(2.6\times10^{-5}\ O\right)\frac{\sqrt{M_{^4\mathrm{He}}M_{^A\mathrm{Pd}}}c^2}{\Delta Mc^2}\frac{P_{diss}\tau}{\hbar}\frac{\sqrt{N_{\mathrm{D}_2}N_{^A\mathrm{Pd}}}}{N} \tag{154}$$

To evaluate this expression we need to consider the values of all relevant parameters. We take:

$$|1-\eta| = 0.1 \qquad \frac{v_{nuc}}{v_{mol}} = 6.65\times10^{-12} \qquad G = 90.35 \tag{155}$$

where $|1-\eta|$ is for the $D_2$/3+1/$^4$He contribution to the $D_2$/$^4$He transition per section 6.5 and where the volume ratio and the Gamow factor $G$ take on the values determined in section 6.3.

For the $O$-value, we consider the following a realistic choice for the purpose of this section (although we will provide a more detailed discussion of this parameter later in section 5.8):

$$O = 0.001 \tag{156}$$

Moreover:

$$\frac{N_{\mathrm{D}_2}}{N} = 0.25\times\frac{1}{9} \qquad \frac{N_{^A\mathrm{Pd}}}{N} = 0.25 \tag{157}$$

where we assume a highly loaded lattice with monovacancies where 25% of the $D_2$ molecules are in the $^3$P state (see Figure 38 in section 6.1); with only one of the 9 possible $^3$P states (with $J = 1$) being involved in the $\mathbf{a}\cdot c\mathbf{P}$ transition (see section 6.9 and 6.8); and where the natural abundance of the "average" Pd isotope involved in the Pd*/Pd transition is 0.25 (since we do not yet know exactly which Pd isotopes exhibit best suited levels).

Note that these values for the number of donor side and receiver side systems per unit cell differ from those used in section 5.3. Since we now consider Pd nuclei as receiver systems (instead of $^4$He nuclei), the $\frac{N_{^A\mathrm{Pd}}}{N}$ ratio is comparably high and the assumption that underlies the derivation of Dicke factors in section 5.2 (namely that the number of receiver systems exceeds the number of donor systems) is met. On the donor side, the choice of $\frac{N_{\mathrm{D}_2}}{N}$ is motivated section in section 6.21.

For the mode characteristics, we choose experimentally plausible values for vibrational power dissipation and frequency applied to the lattice as

$$P_{diss} = 1\ \mathrm{W} \qquad f_A = 5\ \mathrm{MHz} \tag{158}$$

resulting in a $\tau$ (for acoustic phonons) of

$$\begin{aligned}\tau_A &= 10^{-12}\left(10^{-7}\frac{f_A}{1\ \mathrm{MHz}}\right)^{-3/2}\ \mathrm{sec}\\ &\approx 0.003\ \mathrm{sec}\end{aligned} \tag{159}$$

Plugging in all of the above values results in a minimum transfer rate (unscreened) of

$$\Gamma_{min,unscreened} \;=\; 4.79 \times 10^{-22}\ \mathrm{s}^{-1} \tag{160}$$

Since we used the unscreened Gamow factor G above, we can consider the effects of screening by multiplying with (per section 6.3)

$$e^{\Delta G_{scr}} \tag{161}$$

where (for 350 eV screening potential; see section 6.3)

$$\Delta G_{scr} = 49.2 \tag{162}$$

This results in

$$\Gamma_{min,screened} \;=\; 1.12\ \mathrm{s}^{-1} \tag{163}$$

To obtain the maximum rate (peak rate of the Dicke enhancement pulse), we apply the additional Dicke enhancement, as derived in section 5.2:

$$\Gamma_{max} \;=\; \Gamma_{min}\frac{1}{4}N_{\mathrm{D}_2} \tag{164}$$

For $N_{\mathrm{D}_2} = 10^{15}$ this yields

$$\Gamma_{max,screened} \;=\; 2.79 \times 10^{14}\ \mathrm{s}^{-1} \tag{165}$$

Again, this is the population-level transfer rate, *i.e.*, the rate at which the ensemble in group *a* transfer excitation to the ensemble in group *b*. To obtain a rate per deuteron pair that is comparable to the estimate given by Koonin and Nauenberg 1989 [20], we normalize the expression. Taking the Dicke-enhanced expression we obtain

$$\frac{\Gamma_{max,screened}}{N_{\mathrm{D}_2}} \;=\; 0.28\ \mathrm{s}^{-1} \tag{166}$$

Such a rate estimate means that this kind of excitation transfer would get into the observable range. Let us therefore also consider the minimum rate per $D_2$ to get a better sense of whether the process of time-evolving Dicke factors from the minimum to the maximum rate has a chance to get started:

$$\frac{\Gamma_{min,screened}}{N_{\mathrm{D}_2}} \;=\; 1.12 \times 10^{-15}\ \mathrm{s}^{-1} \tag{167}$$

**Discussion**

We find that in the given scenario, minimum (and maximum) rates are in the observable range. Note, however, that we worked with an idealized model, which did not include dephasing, and which made assumptions about the existence of a single resonant receiver states. A model that includes reasonable estimates for dephasing would show no observable coherent effects due to this mechanisms. In later sections, we will consider more realistic models.

Note that for the sake of simplicity, in this section we assumed comparatively low steady-state dissipated power of 1 W. However, under realistic conditions, power can be expected to fluctuate. This means that instead of having a fixed value of $P_{diss}$ we can expect $P_{diss}(t)$. We refer to the power maxima as high-power transients. In the following sections, we will consider the effect of high-power transients on transfer rate estimates and startup dynamics.

Note also that the time evolution of Dicke factors has to take place within the constraint of the decoherence time of the system. We provided an estimate for the decoherence time in section 6.21, yielding a value of about 1 ns.

Moreover, in the estimate above we assumed a value for the $O$-factor of 0.001. Since many transitions in Pd isotopes (and other mid-sized nuclei) have not been identified and characterized, the existence of transitions with a wide range of different $O$-values is conceivable, impacting rate estimates and startup dynamics. Therefore, in addition to high-power transients, both the decoherence time and the $O$-values are important constraints. Section 5.8 is dedicated to a discussion of such constraints.

Finally, we assumed in this section the existence of a hypothetical Pd state at 23.85 MeV that meets the resonance condition for the $D_2/^4He$ transition. However, as pointed out earlier, the existence of a single state like that is highly unlikely. Section 5.7 contains a conjecture how the resonance condition can be met alternatively given multi-state transitions to a combination of realistic Pd states.

### 5.6 The effect of high power transients in the lattice modes on excitation transfer rates

In the previous section we obtained a first preliminary rate estimate for excitation transfer from $D_2$ to Pd.

The resulting minimum rate for the ensemble was large compared to the rate of spontaneous fusion but still small given a short decoherence time of about 1 ns. The estimated maximum rate is significantly higher but with a minimum rate that is too slow, a system does not have enough time to evolve to the maximum rate.

At the same time, the assumption of 1 W of dissipated power in a phonon mode across the coherence domain with cycles of about 3 ms is conservative. In this section, we consider the effect of occasional transients, where for a short time there is high-power peak in vibrational power. However, to anticipate the results of this section, we will find that in the present picture, the high-power transient requirements for observable fusion rates are unphysically extreme.

Studying the effect of high power transients suggests evaluation of the rate expression $\Gamma_{min,screened}$ of Eq. 154 as a function of $P_{diss}$.

Another variable of considerable uncertainty is the nuclear transition factor *O*, which is characteristic of the Pd transitions involved.

To gain intuition about the dependence of the transfer rate on those two variables, the goal of this section is the creation of contour plots that illuminate these relationships.

#### Testing for strong coupling

Before evaluating the transfer rate across a large parameters space of P and O values, we need to remind ourselves that the previous calculations were based on perturbation theory, which assumes that the coupling strength driving a transition is small compared to the transition energy. This relationship can be expressed through a dimensionless coupling constant g.

If g is close to unity or larger, then a perturbation theory approach needs to be replaced by a strong coupling approach. Since there are two transitions involved in the transfer from $D_2$ to Pd, we ought to evaluate a dimensional coupling constant for each. However, because the coupling strength for the fusion transition is hindered by the Coulomb barrier, as represented by the $e^{-G}$ factor, the $g$ associated with the fusion transition can be readily identified as $\ll 1$ .

For the dimensionless coupling constant on the $D_2$ side we have

$$
\begin{aligned}
g_{\mathrm{D_2}/^4\mathrm{He}} &= \frac{\mathcal{U}}{\Delta E} \\
&= \frac{|\langle \mathrm{D_2}|a_z|^4\mathrm{He}\rangle| c\sqrt{\langle P^2_{^4\mathrm{He}}\rangle}\sqrt{N_{\mathrm{D_2}}}}{\Delta M c^2}
\end{aligned}
\tag{168}
$$

For the dimensionless coupling constant on the Pd side we have

$$
\begin{aligned}
g_{Pd*/Pd} &= \frac{\mathcal{V}}{\Delta E} \\
&= \frac{|\langle \mathrm{Pd}^*|a_z|\mathrm{Pd}\rangle| c\sqrt{\langle P^2_{\mathrm{Pd}}\rangle}\sqrt{N_{^A\mathrm{Pd}}}}{\Delta M c^2}
\end{aligned}
\tag{169}
$$

For simplicity we take

$$
g = g_{\mathrm{Pd}^*/\mathrm{Pd}} \tag{170}
$$

and, following Eq. 147,

$$\Gamma_{transfer} = \frac{2}{\hbar}|\Delta\mathcal{U}|g \tag{171}$$

We are now ready to plot $g$. Evaluating Eq. 170 yields (recall the derivation of the Pd matrix element in section 6.16):

$$\begin{aligned} g &= \frac{|\langle \mathrm{Pd}^*|a_z|\mathrm{Pd}\rangle|\sqrt{\frac{M_{\mathrm{^APd}}c^2 P_{diss}\tau_A}{N}}}{\Delta Mc^2}\sqrt{N_{\mathrm{^APd}}} \\ &= \frac{|\langle \mathrm{Pd}^*|a_z|\mathrm{Pd}\rangle|\sqrt{M_{\mathrm{^APd}}c^2(1\text{ J})}}{\Delta Mc^2}\sqrt{\frac{P_{diss}\tau_A}{1\text{ J}}}\sqrt{\frac{N_{\mathrm{^APd}}}{N}} \\ &= 2.6\times 10^{-5}\, O_{\mathrm{Pd}}\, \frac{\sqrt{M_{\mathrm{^APd}}c^2(1\text{ J})}}{\Delta Mc^2(1\text{ J})}\sqrt{\frac{P_{diss}\tau_A}{1\text{ J}}}\sqrt{\frac{N_{\mathrm{^APd}}}{N}} \\ &= 865\, O_{\mathrm{Pd}}\sqrt{\frac{P_{diss}\tau_A}{1\text{ J}}}\sqrt{\frac{N_{\mathrm{^APd}}}{N}} \end{aligned} \tag{172}$$

The result is shown in Figure 25.

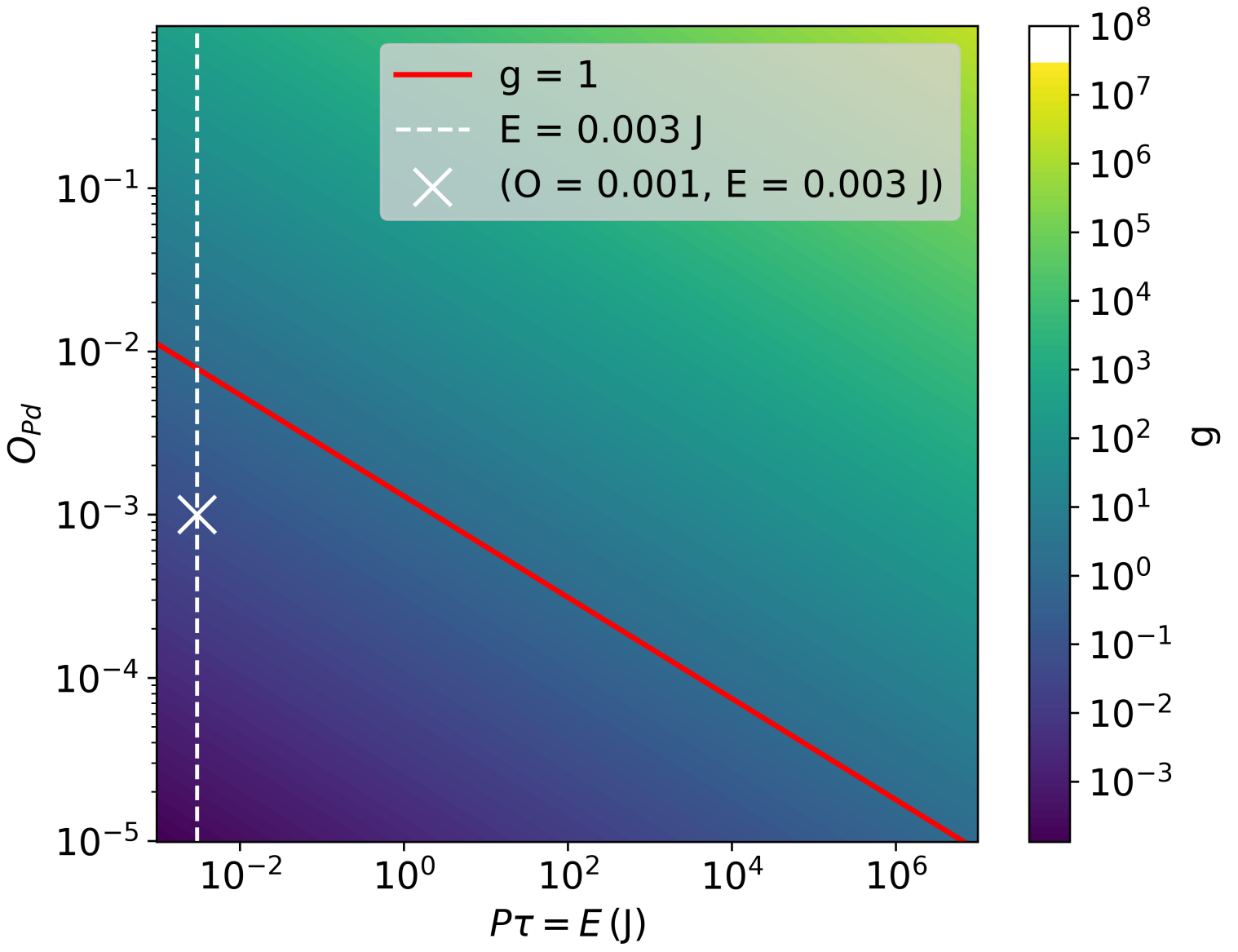

**Figure 25:** Plot of the dimensionless coupling constant, $g$, for the Pd nuclei as a function of the transient dissipated acoustic phonon energy (*i.e.*, the product of the dissipated transient power and pulse period in the acoustic mode) and the Pd $O$-value. The transition from a perturbative to a strongly coupled system is defined by g = 1 (red line) and the exemplary case of 0.003 J of dissipated energy from the acoustic phonon mode (dashed white line) and its intersection with a Pd*/Pd transition with an $O$-value of 0.001 (white X mark) are shown.

### Transfer rates with a crude extrapolation for the strong coupling regime

When $g$ is larger than unity, the relationship between $g$ and the transfer rate $\Gamma_{transfer}$ can no longer be considered linear. In the strong coupling limit the situation is more complicated as will be discussed in the next section. Based on what we know so far, the predicted excitation transfer rates are slower than our estimates for the decoherence and loss rates. What we will do here is a crude extrapolation based on

$$F(g) \sim \frac{g}{1+2g} \tag{173}$$

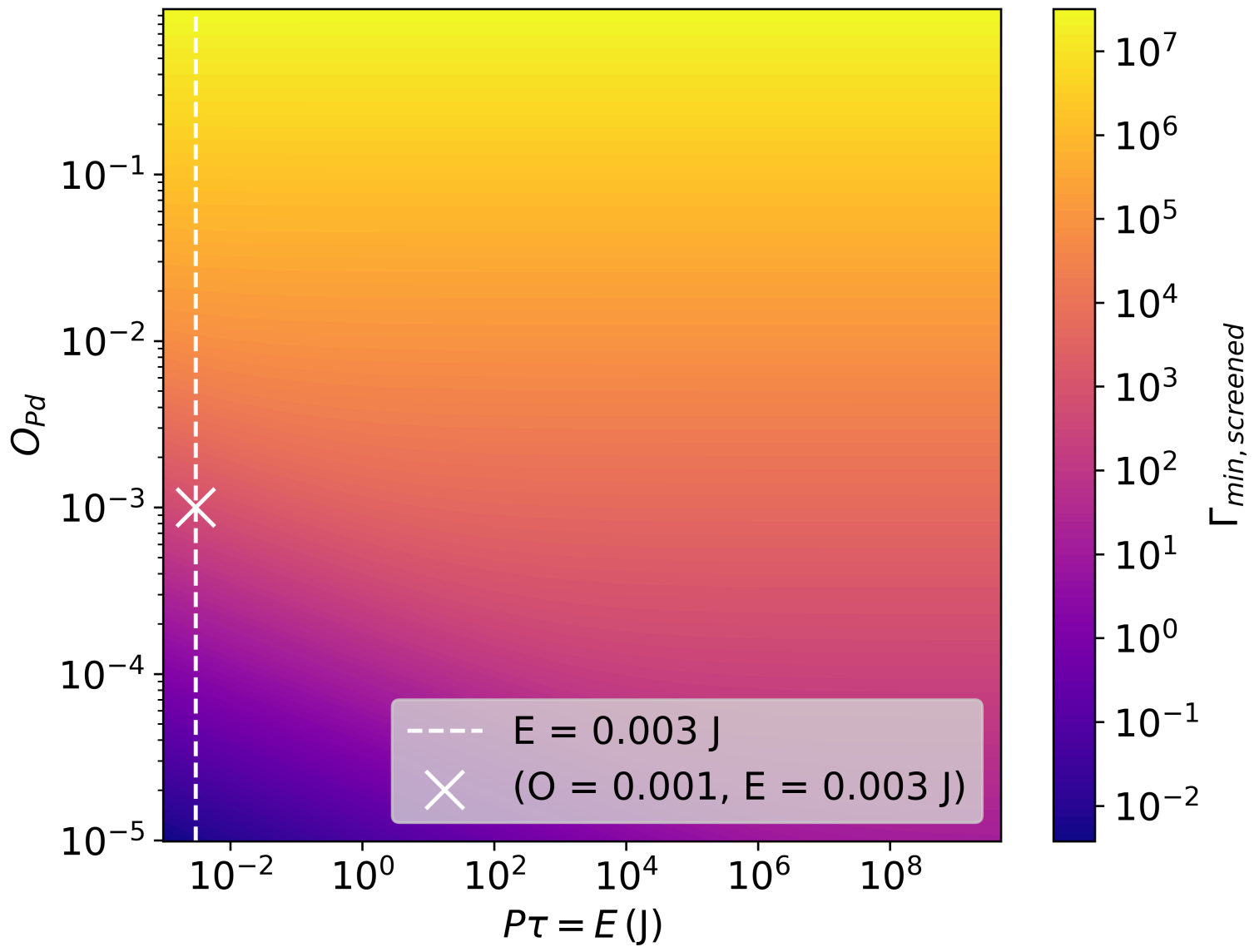

**Figure 26:** Plot of the minimum nuclear excitation transfer rate with 350 eV of electron screening ($\Gamma_{min,screened}$) from the $D_2$/$^4$He decay transition to the Pd*/Pd excitation transition as a function of the transient dissipated acoustic phonon energy (*i.e.*, the product of the dissipated transient power and pulse period in the acoustic mode) and the Pd $O$-value. The exemplary case of 0.003 J of dissipated energy from the acoustic phonon mode (dashed white line) and its intersection with a Pd*/Pd transition with an $O$-value of 0.001 (white X mark) are shown.

The idea is that at small values of $g$ this function will lead to results consistent with perturbation theory for weak coupling. We have some experience with the indirect coupling matrix element for excitation transfer, where one donor transition goes to two receiver transitions. In this model we see deviations from linearity that kick in around $g = 1/2$, which motivates in part the factor of 2g in the denominator. The numerical results for the indirect coupling matrix element were crudely oscillatory in the $g$ values as they increased above $g = 1$. So the form of $F(g)$ was chosen to establish a very rough limit for larger $g$. In the future there will need to be modeling done systematically for the indirect coupling matrix element in the strong coupling regime for excitation transfer to different numbers of Pd*/Pd transitions in order to develop accurate predictions.

The relationship in Eq. 173 is shown in Figure 27.

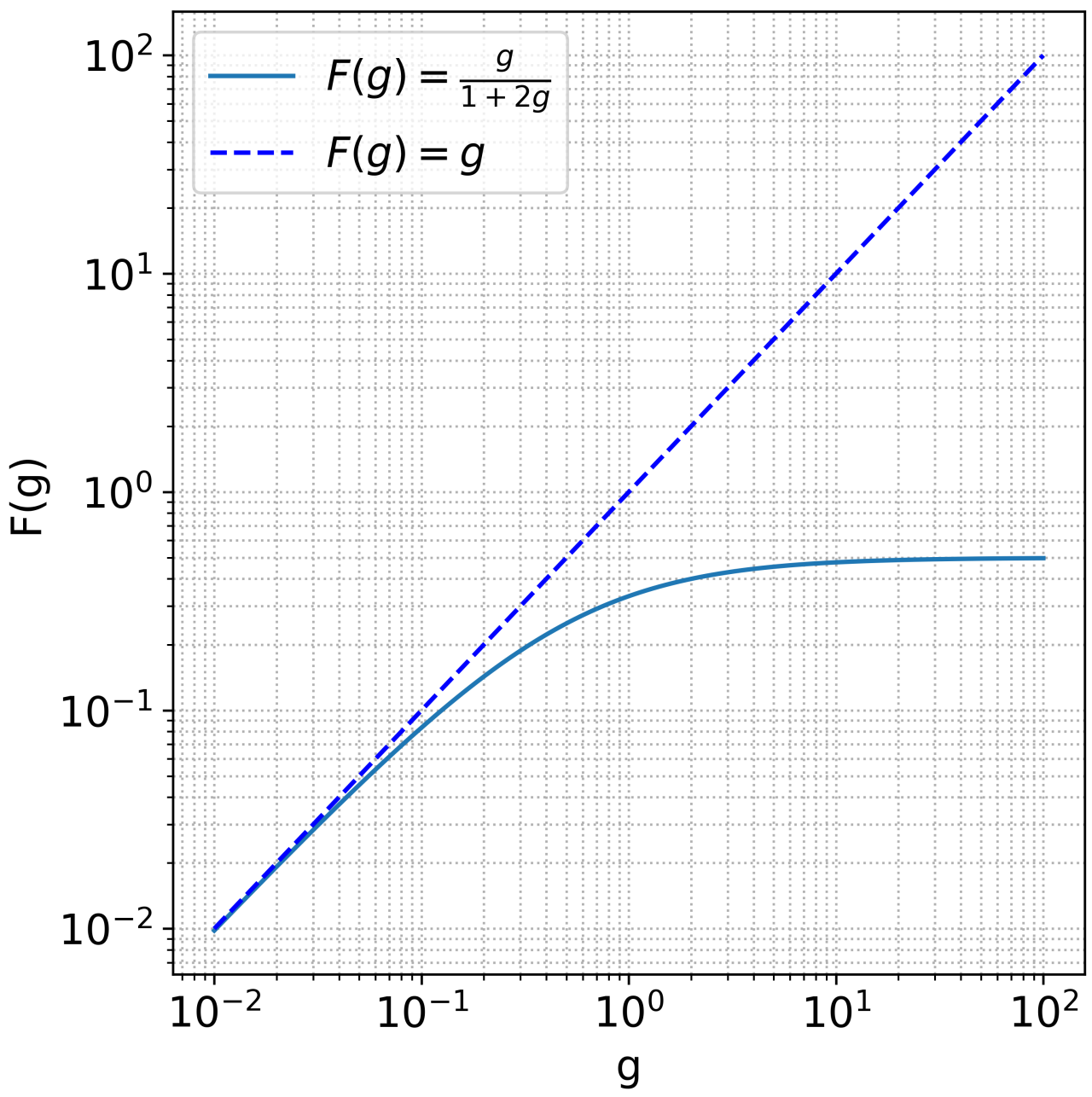

**Figure 27:** Comparison between linear scaling of normalised coupling constant ($g$) and strong coupling extrapolation ($F(g)$).

The excitation transfer rate with this extrapolation in the strong coupling regime is then:

$$\Gamma_{transfer} \rightarrow \frac{2}{\hbar}|\Delta\mathcal{U}|F(g) \tag{174}$$

After taking into account effects of entering the strong coupling regime this way, we can finally plot $\Gamma_{min,screened}$ rates at high power transients (Figure 26).

It can be seen from Figure 26 that unphysically extreme high power transients would be required for transfer rates to approach the observable range. In section 5.10, we will develop a more sophisticated version of the model that is even less idealized and exhibits higher rates.

First, however, we will address the other key issue raised by section 5.5: meeting the resonance condition beyond the hypothetical and unlikely Pd excited state at 23.85 MeV.

## 5.7 Transfer to multiple Pd transitions and meeting the resonance condition

In the sections above, we considered an idealized scenario in which the $D_2$/$^4$He transition of interest is matched in energy to a hypothetical Pd*/Pd transition where the Pd* state is stable enough to accommodate coherent dynamics.

The notion of excitation transfer from one transition to a second transition is conceptually simple and easy to analyze with perturbation theory when the coupling is weak. However, this simple model's presupposition of the existence of a single receiver state that is precisely resonant with the donor state can be readily challenged and needs to be seen as an idealization on the way to a more realistic model. While far from all nuclear excited states of common lattice nuclei are known, a single state that is so closely matched is unlikely to exist.

For the resonance condition we can write

$$E_{\mathrm{D}_2} - E_{^4\mathrm{He}} \; = \; \Delta M c^2 \; = \; \sum_j (E_{\mathrm{Pd}*} - E_{\mathrm{Pd}})_j \tag{175}$$

The existence of a resonance with a single transition at 23.85 MeV can be seen as very unlikely because candidate nuclear molecule states are separated by at least tens of keV. However, a receiver state can also comprise several lower energy palladium transitions whose sum meets the resonance condition as depicted in Figure 28. If combinations of available lower-energy states are considered, it is possible to get to within a fraction of an eV of the sought resonance. The specifics here depend on the available density of states. We provide an estimate based on related experimental data at the end of this section.

To make up residual energy mismatch, the system may need to exchange some phonons (preferably higher-energy phonons). We will not include this aspect in the analysis at this point, so the model under consideration will be an idealization that the resonance for excitation transfer is exact.

However, we will show in this section that excitation transfer can take place, in principle unmitigated to a combination of multiple receiver states as long as the sum of those states meets the resonance condition.

If we define dimensionless coupling coefficients according to

$$g_1 \; = \; \frac{|\mathcal{V}_1|}{\epsilon_1} \qquad g_2 \; = \; \frac{|\mathcal{V}_2|}{\epsilon_2} \qquad g_3 \; = \; \frac{|\mathcal{V}_3|}{\epsilon_3} \tag{176}$$

where $\epsilon_1$, $\epsilon_2$, and $\epsilon_3$ represent the energies associated with different transitions in Pd nuclei,

then we can write

$$H_{initial,final} \; = \; \Delta \mathcal{U} g_1 g_2 g_3 \tag{177}$$

and

$$\Gamma_{transfer} \; = \; \frac{2}{\hbar}|H_{initial,final}| \; = \; \frac{2}{\hbar}|\Delta \mathcal{U}| g_1 g_2 g_3 \tag{178}$$

See section 6.23 for a detailed derivation.

Once again we expect these perturbation theory based expressions to be accurate when all of the dimensionless coupling coefficients are much less than unity. A more extensive calculation is sought in the future to get accurate indirect coupling coefficients and rates in the strong coupling regime.

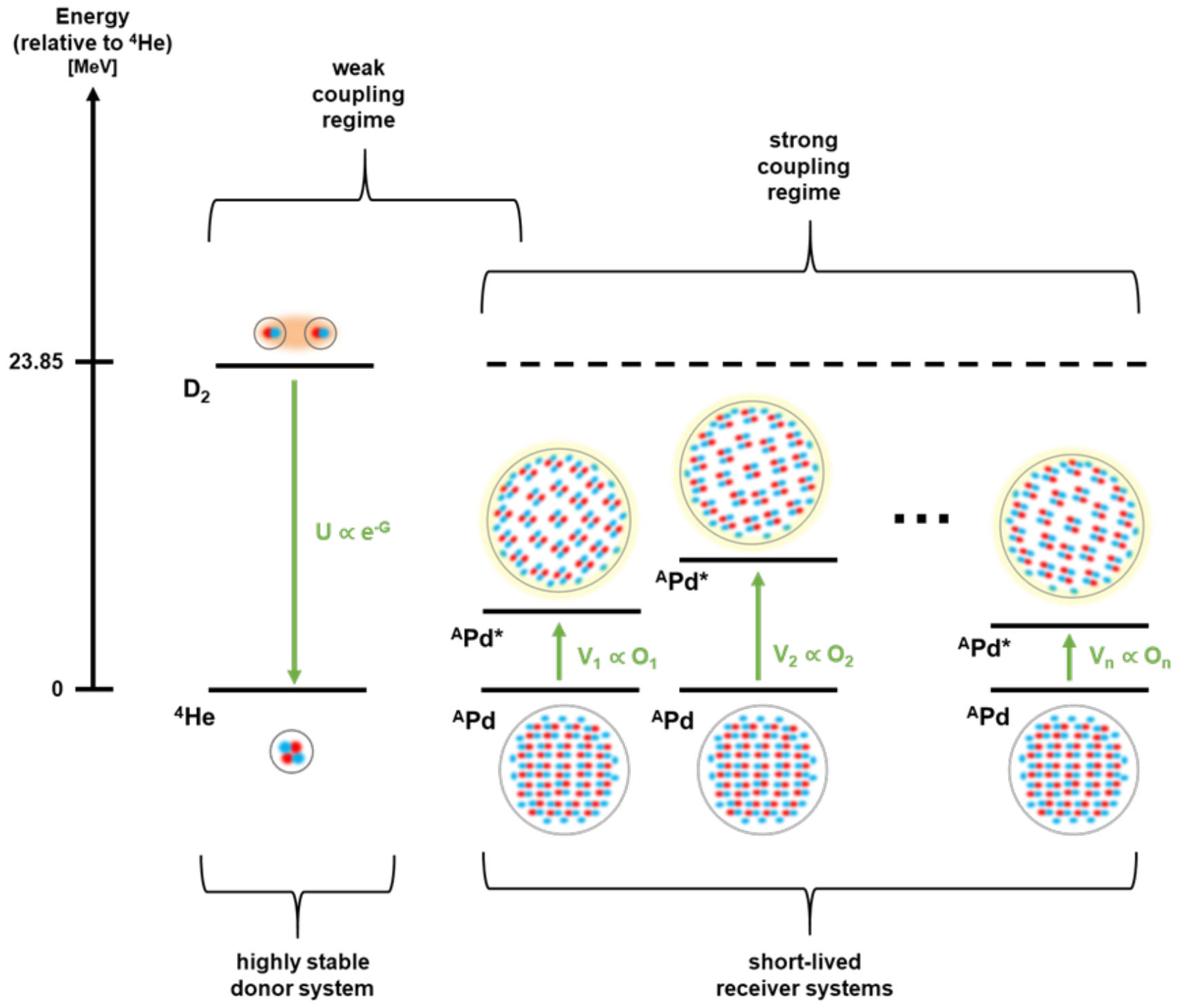

**Figure 28:** Weakly coupled nuclear excitation transfer from highly stable molecular $D_2$ donors via the $D_2 \to {}^4He$ transition to a strongly coupled system of ${}^{A}Pd^*$ bound state state acceptors via ${}^{A}Pd \to {}^{A}Pd^*$ transitions.

### Estimating the nuclear density of states based on related experimental data

We conclude this section with a brief discussion of the available density of states that results from combination of Pd nuclear bound states.

No tables are available in the literature yet with these kinds of data. However, a table close to what we need has been given for ${}^{120}$Sn by Müscher et al. (2022) [94], where results from a nuclear resonance fluorescence experiment on ${}^{120}$Sn are reported (see section 5.4).

An estimate for the density of states, based on the experimental data by Müscher et al, which can be expected to be similar for Pd isotopes, is developed in section 6.18. The key result is shown in Figure 29.

At the resonance condition of 23.85 MeV (more precisely: 23,848,109 eV), our estimated density of states is 171 states per eV. This implies that the closest nuclear state that results from a combination of lower energy Pd transitions is withing a fraction of an eV.

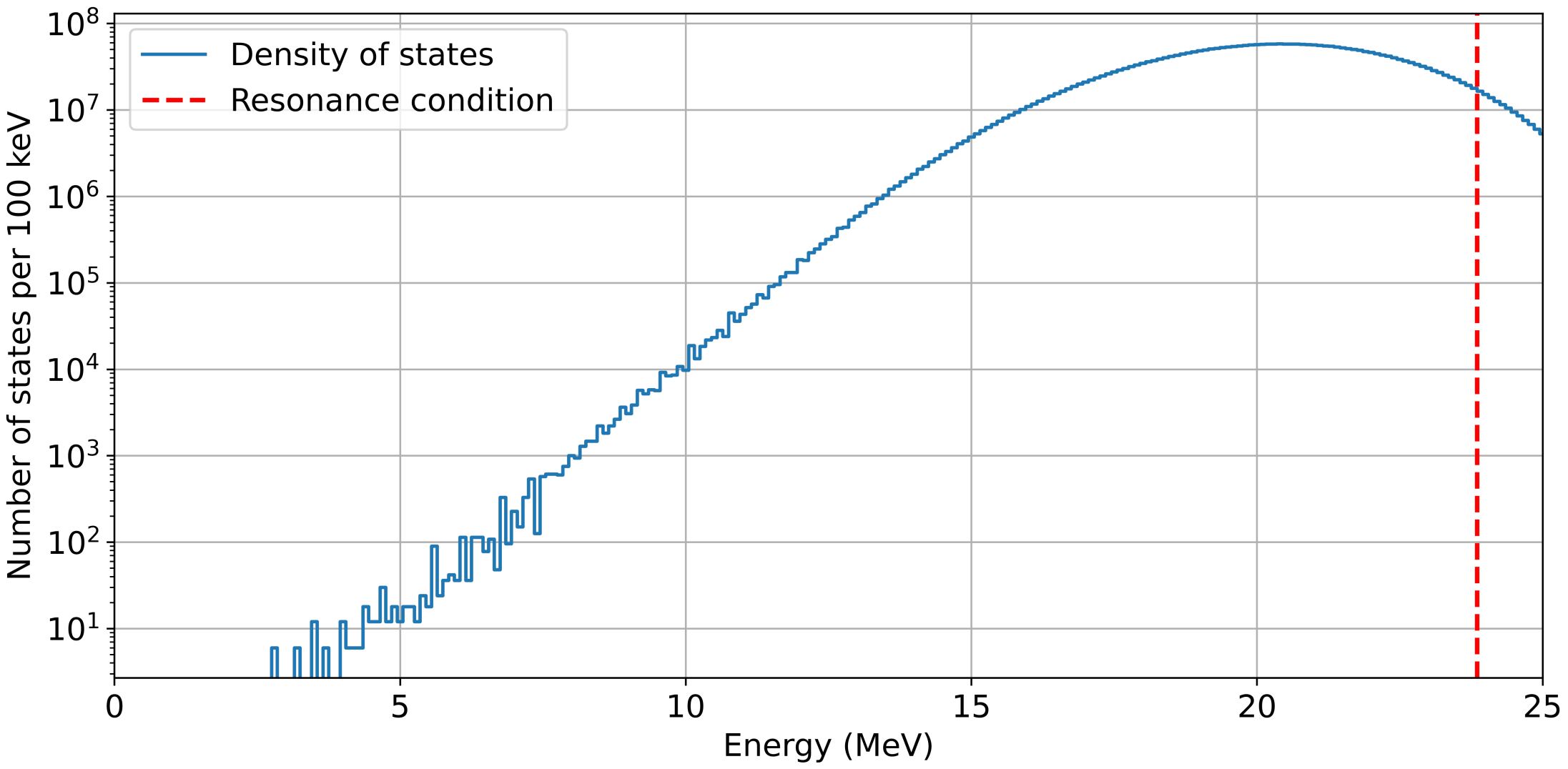

**Figure 29:** Result of the summing the density of states for combinations of one, two and three nuclear transitions, evaluated with the $^{120}$Sn data set (and assumed to be relevant for all stable Pd isotopes).

## 5.8 Constraints on coherent dynamics from decoherence and loss channels

A critical question that needs to be addressed is the question how coherent dynamics start up initially. Time evolution of the systems under consideration comprises two aspects: time evolution during a coherence cycle and time evolution across coherence cycles. We refer to these dynamics as startup dynamics and post-startup dynamics. This section is focused on startup dynamics. Post-startup dynamics will be discussed in subsequent sections 5.13 and 5.14.

Startup refers to the process of getting from the minimum excitation transfer rate to a large enough rate (via greater Dicke enhancement factors) to yield observable effects per the definition in section 2.1.

The discussion of startup dynamics is closely linked to the discussion of constraints. Besides needing enough initial power in the system to develop large enough minimum transfer rates (section 5.2), key constraints are the dephasing rates (*i.e.*, decoherence rates) of the two involved transitions (donor transition and receiver transition).

In this section, we develop a threshold condition based on the fundamental need for the excitation transfer rate to be greater than the total dephasing rate for the two transitions. Ultimately, this is the essential condition that needs to be met in order for the system to start up. The relationship between the achievable transfer rate and the dominant decoherence times is affected by the power provided to the system and by the nuclear transition factor of the Pd transitions involved. These variables are to be taken to span the parameter space for the threshold condition.

The issue is to determine how much energy needs to be in the highly-excited acoustic vibrational mode in order for the system to start up if it has an **a**-matrix element of a given strength.

### Constraints

Coherence on the $D_2$ molecule side is affected by spin-spin relaxation as well as the stability of deuteron pairs—whereby the latter, with a value of about 1 ns, is significantly shorter than the former (see section 6.22).

On the Pd side, the major competing process to coherent dynamics is radiative decay of the states involved. This is not the case on the donor side since a $D_2$ molecule can be viewed as a highly metastable excited state of $^4$He. However, on the receiver side, in the Pd receiver scenario, the nuclear states tend to be short-lived and the coherent dynamics need to outpace them, if they are to be dominant.

For Pd bound states, state lifetimes can be estimated by the Weisskopf estimate. We expect that a transition that has a strong $\mathbf{a} \cdot c\mathbf{P}$ coupling strength will also be able to radiatively decay. For this we will make use of the M2 Weisskopf estimate scaled by $O^2_{\text{Pd}}$ (also see the related discussion in section 5.4):

$$\Gamma_{M2} \sim 2.2 \times 10^7 \, A^{2/3} \left(\frac{\Delta M c^2}{1 \text{ MeV}}\right)^5 O^2_{\text{Pd}} \, \text{s}^{-1} \tag{179}$$

We evaluate it to obtain

$$\Gamma_{M2} \sim 3.8 \times 10^{15} \, O^2_{\text{Pd}} \, \text{s}^{-1} \tag{180}$$

Per the discussion in 6.22, we adopt a dissociation time for the $D_2$ molecule of

$$\tau_{\text{D}_2} \sim 1 \text{ ns} \tag{181}$$

In order for the coherent dynamics to get started, we require that the excitation transfer rate exceed the decoherence rate

$$\Gamma_{transfer} \geq \Gamma_{decoh} \rightarrow \Gamma_{M2} + \frac{1}{\tau_{\text{D}_2}} \tag{182}$$

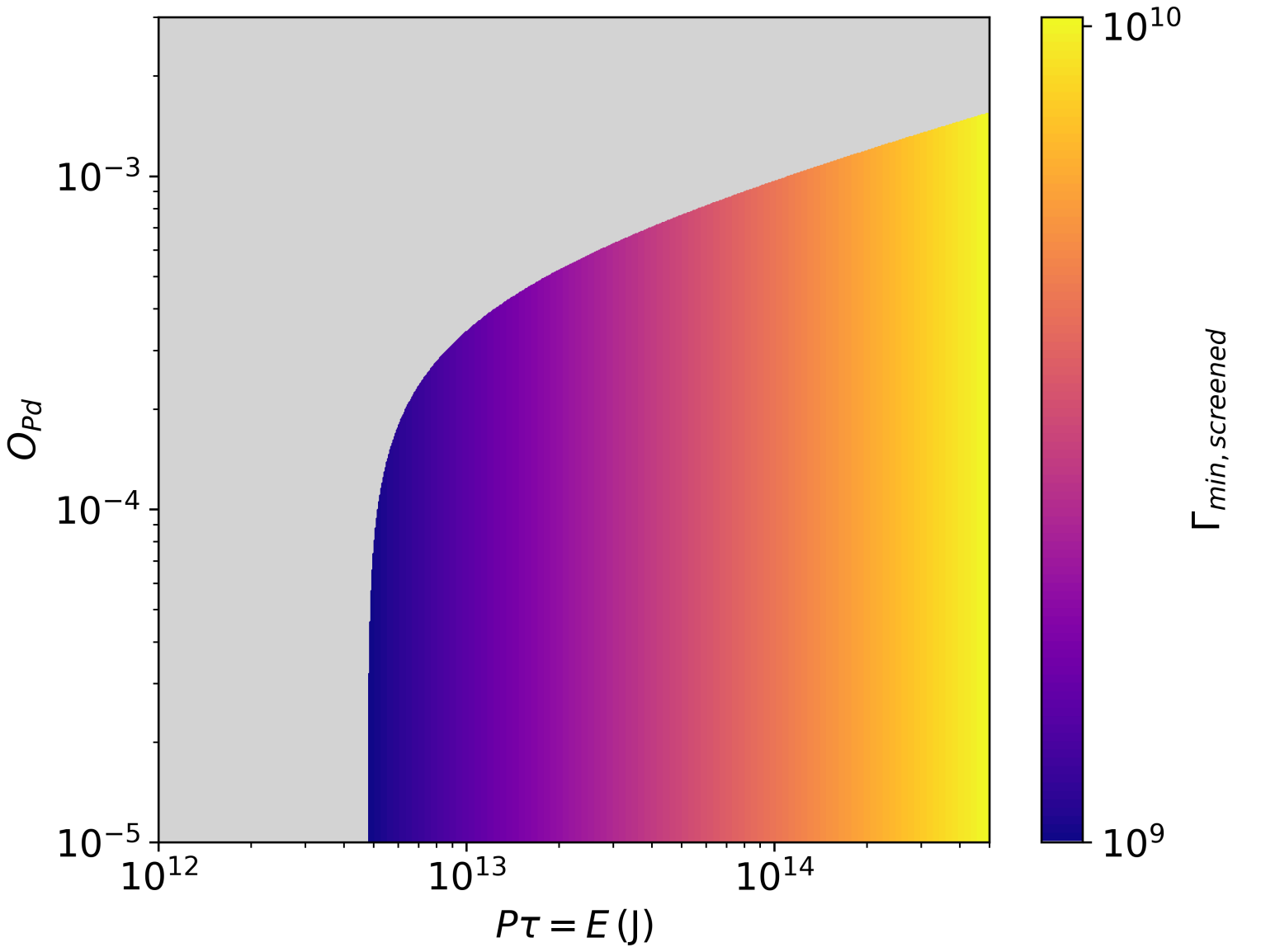

**Figure 30:** Plot of the minimum nuclear excitation transfer rate with 350 eV of electron screening ($\Gamma_{min,screened}$) from the $D_2$/$^4$He decay transition to the Pd*/Pd excitation transition as a function of the transient dissipated acoustic phonon energy (*i.e.*, the product of the dissipated transient power and pulse period in the acoustic mode) and the Pd $O$-value. The greyed out area represents rates that don't meet the minimum requirements determined by decoherence, *i.e.*, Eq. 182.

## Result and discussion

The results of this analysis are shown in Figure 30, where contours of the excitation transfer rate are shown as a function of acoustic energy $P_{diss}\tau_A$ and the Pd*/Pd transition $O_{\text{Pd}}$ value in the allowed part of the parameter regime where the excitation transfer rate exceeds the model decoherence rate. We've also used the same deuterium and palladium loading as in previous sections when calculating the rates, namely:

$$\frac{N_{\text{D}_2}}{N} = 0.25 \times \frac{1}{9} \qquad \frac{N_{^A\text{Pd}}}{N} = 0.25 \tag{183}$$

We see that it is possible within this idealized model for the coherent dynamics to start up; however, the amount of energy in the highly-excited acoustic mode is impractically large.

For this calculation we assumed the same model parameters as in the transfer rate calculation in section 5.5 based on Eq. 154.

A key driver of the high energy requirement in this scenario is the issue that the magnetic quadrupole (M2) radiative decay is very fast at 23.85 MeV if the transition is strong (meaning that it has a large $O_{\text{Pd}}$ value, which makes for a fast transfer rate but also fuels the competing decay). To reduce the radiative decay rate, the system needs to work with a "small" $O_{\text{Pd}}$ value so as to be affected by a more modest radiative decay rate. However, if the $O_{\text{Pd}}$ value is "small", the excitation transfer rate yielded from the expression in Eq. 154 is reduced. This in turn drives up the requirement for the acoustic energy needed in order to obtain an excitation transfer rate that exceeds the model decoherence rate (which is the sum of the Pd* radiative decay rate and molecular $D_2$ dissociation rate).

It is clear that this model does not connect well with reports of observable fusion in metal-hydrogen experiments with low-energy stimulation. In such experiments, orders of magnitude less energy than suggested in this section is reported to be involved, prompting the exploration of alternative schemes with reduced energy requirements. The basic issue that this model faces is that the magnetic quadrupole radiative decay rate scales as the fifth power of the transition energy, which means that the associated decay rate is very fast up at 23.85 MeV. This leads to a competition between the excitation transfer rate (which scales linearly in $O_{\mathrm{Pd}}$) and the radiative decay rate (which scales as the square of $O_{\mathrm{Pd}}$), which ends up driving the system to a low $O_{\mathrm{Pd}}$ value and consequently extremely large acoustic energy.

A solution is sketched out in the following section: transferring energy from the $D_2$ donor system to a combination of lower-energy transitions on the Pd side not only provides a more realistic scenario to meet the resonance condition, it also helps avoiding unstable high-energy states with very fast radiative decay.

## 5.9 Transfer rate estimates: $D_2$ to multiple Pd transitions

Here we present a possible solution to the challenges associated with the start-up problem, as discussed in section 5.8. We already introduced in section 5.7 the proposal to transfer excitation from the $D_2/^4He$ fusion transition to many transitions at lower energy (in our case in Pd nuclei). The approach has been motivated by its impact on the resonance condition, since combinations of transitions are much more likely to come close to the resonance conditions compared to a single transition.

There is another critical aspect to a model involving multiple receiver transitions that will be covered in this section: transferring to multiple transitions can improve startup dynamics through decreased competing loss channels. In the case of multiple lower-energy transitions on the receiver side of nuclear excitation transfer, the excited states produced tend to have a slower magnetic quadrupole (M2) decay rate (see the Weisskopf estimate for state lifetimes described in sections 5.4 and 5.8). We will explore this aspect and related aspects in some detail below.

In the notation of previous sections we start with

$$\Gamma_{transfer} = \frac{2}{\hbar}|\Delta \mathcal{U}| \prod_j g_j \tag{184}$$

as the result from perturbation theory for excitation transfer to one, two and three transitions (see section 6.23). As before, we expect this to work best, when the dimensionless coupling constants $g_j$ are less than unity. In the strong coupling regime, however, a more sophisticated analysis is needed to get accurate results. In order to develop estimates for the strong coupling regime, we assume that the excitation transfer rate gets larger according to

$$\Gamma_{transfer} \rightarrow \frac{2}{\hbar}|\Delta \mathcal{U}| \prod_j F(g_j) \tag{185}$$

again with

$$F(g) \sim \frac{g}{1+2g} \tag{186}$$

per section 5.6 (Eq. 173).

Nuclear data on excited states of Pd in the relevant energy levels at the MeV range is incomplete and many states have not been identified and characterized. Nevertheless, we can assume the existence of further M2 bound states, based on the confirmed existence of some such states, parameterize them and get some insight as to what the associated parameter space would look like.

### M2 states among Pd bound states

In the argument presented here we assume that $n$ reasonably strong and reasonably stable bound states are available with energies in the general vicinity of

$$\epsilon_j \rightarrow \epsilon = \frac{\Delta M c^2}{n} = \frac{23.85 \text{ MeV}}{n} = 2.98 \text{ MeV} \tag{187}$$

for the special case of

$$n = 8 \tag{188}$$

We will also assume that the different $g_j$ values are in the vicinity of a single average $g$ parameter

$$g_j \ \rightarrow \ g \tag{189}$$

The goal is the development of a first estimate for excitation of $n$ distinct transitions, but model all of the different transitions based on average transition parameters.

**Excitation transfer rate**

For the excitation transfer rate from $D_2$ donor transitions to multiple Pd receiver transitions we work with

$$\Gamma_{transfer} \ \rightarrow \ \frac{2}{\hbar}|\Delta \mathcal{U}| \bigg( F(g) \bigg)^n \tag{190}$$

with

$$|\Delta \mathcal{U}| \ = \ |1-\eta| |\langle \mathrm{D}_2 | a_z | ^4\mathrm{He} \rangle| c \sqrt{\langle P^2_{^4\mathrm{He}} \rangle} \sqrt{N_{\mathrm{D}_2}} \tag{191}$$

where the $\langle \mathrm{D}_2 | \mathbf{a} | ^4\mathrm{He} \rangle$ matrix element is the "ideal" no loss matrix element, and where $|1-\eta|$ takes into account the fraction of the "ideal" single path excitation transfer rate that we get due to the effect of the initial state fusion decay.

For the (average) dimensionless coupling constant we have

$$g \ = \ \frac{|\langle \mathrm{Pd}^* | a_z | \mathrm{Pd} \rangle| c \sqrt{\langle P^2_{\mathrm{Pd}} \rangle}}{\epsilon} \sqrt{N_{^A\mathrm{Pd}}} \tag{192}$$

where it should be noted that $|\langle \mathrm{Pd}^* | a_z | \mathrm{Pd} \rangle|$ depends on the energy of the transition $\epsilon$ (see Eq. 471).

We can develop numerical estimates for the quantities that appear in the excitation transfer rate. For the contribution of the fusion transition we can write

$$\begin{aligned} \frac{2}{\hbar}|\Delta \mathcal{U}| \ &= \ 2|1-\eta||\langle \mathrm{D}_2 | a_z | ^4\mathrm{He} \rangle| \frac{\sqrt{M_{^4He} c^2 (1 \ \mathrm{J})}}{\hbar} \sqrt{\frac{P_{diss}\tau_A}{1 \ \mathrm{J}}} \sqrt{\frac{N_{\mathrm{D}_2}}{N}} \\ &= \ 2|1-\eta| 0.0362 \sqrt{\frac{v_{nuc}}{v_{mol}}} e^{-G} \frac{\sqrt{M_{^4He} c^2 (1 \ \mathrm{J})}}{\hbar} \sqrt{\frac{P_{diss}\tau_A}{1 \ \mathrm{J}}} \sqrt{\frac{N_{\mathrm{D}_2}}{N}} \end{aligned} \tag{193}$$

To evaluate the expressions above, we use the same values for the parameters as we did in previous sections, namely:

$$|1-\eta| \ = \ 0.1 \qquad \frac{v_{nuc}}{v_{mol}} \ = \ 6.65 \times 10^{-12} \qquad G = 90.4 \tag{194}$$

The expression for $\Delta \mathcal{U}$ evaluates to

$$\frac{2}{\hbar}|\Delta \mathcal{U}| \ = \ 2.38 \times 10^{-18} \sqrt{\frac{P_{diss}\tau_A}{1 \ \mathrm{J}}} \sqrt{\frac{N_{\mathrm{D}_2}}{N}} \ \mathrm{s}^{-1} \tag{195}$$

Since we used the unscreened Gamow factor G above, we can consider the effects of screening by multiplying with (per section 6.3)

$$e^{\Delta G_{scr}} \tag{196}$$

where (for 350 eV screening potential; see section 6.3)

$$\Delta G_{scr} = 49.2 \tag{197}$$

This gives us

$$\frac{2}{\hbar}|\Delta\mathcal{U}| \; = \; 5536\sqrt{\frac{P_{diss}\tau_A}{1\text{ J}}}\sqrt{\frac{N_{\text{D}_2}}{N}}\text{ s}^{-1} \tag{198}$$

Because the “bare” excitation transfer rate is so “slow”, the system greatly benefits from being in the strong coupling regime, where it can obtain a sufficiently large rate increase needed for start up dynamics to kick off.

For the dimensionless coupling constant, because $|\langle \text{Pd}^*|a_z|\text{Pd}\rangle|$ is proportional to the energy of the transition (see Eq. 471) we can make use of the same expression from Eq. 172, namely

$$g \; = \; 865\, O_{\text{Pd}}\sqrt{\frac{P_{diss}\tau_A}{1\text{ J}}}\sqrt{\frac{N_{^{\text{A}}\text{Pd}}}{N}} \tag{199}$$

**Constraints**

*Pd* radiative decay*

We expect that a transition that has a strong $\mathbf{a}\cdot c\mathbf{P}$ coupling strength—as indicated by a large $O$-value— will also be able to radiatively decay comparatively fast. To estimate this decay we make use of the M2 Weisskopf estimate scaled by $O_{\text{Pd}}^2$. We estimate

$$\Gamma_{M2} \; \sim \; 2.2\times 10^7\, A^{2/3}\left(\frac{\epsilon}{1\text{ MeV}}\right)^5 O_{\text{Pd}}^2\text{ s}^{-1} \tag{200}$$

We evaluate it to obtain

$$\Gamma_{M2} \; \sim \; 1.16\times 10^{11}\, O_{\text{Pd}}^2\text{ s}^{-1} \tag{201}$$

*Molecular $D_2$ dissociation*

We again adopt for the molecular $D_2$ dissociation time

$$\tau_{\text{D}_2} \; \sim \; 1\text{ ns} \tag{202}$$

*Fundamental constraint*

As before we require that the excitation transfer rate exceed the decoherence rate

$$\Gamma_{transfer} \; \geq \; \Gamma_{decoh} \; \to \; \Gamma_{M2} + \frac{1}{\tau_{\text{D}_2}} \tag{203}$$

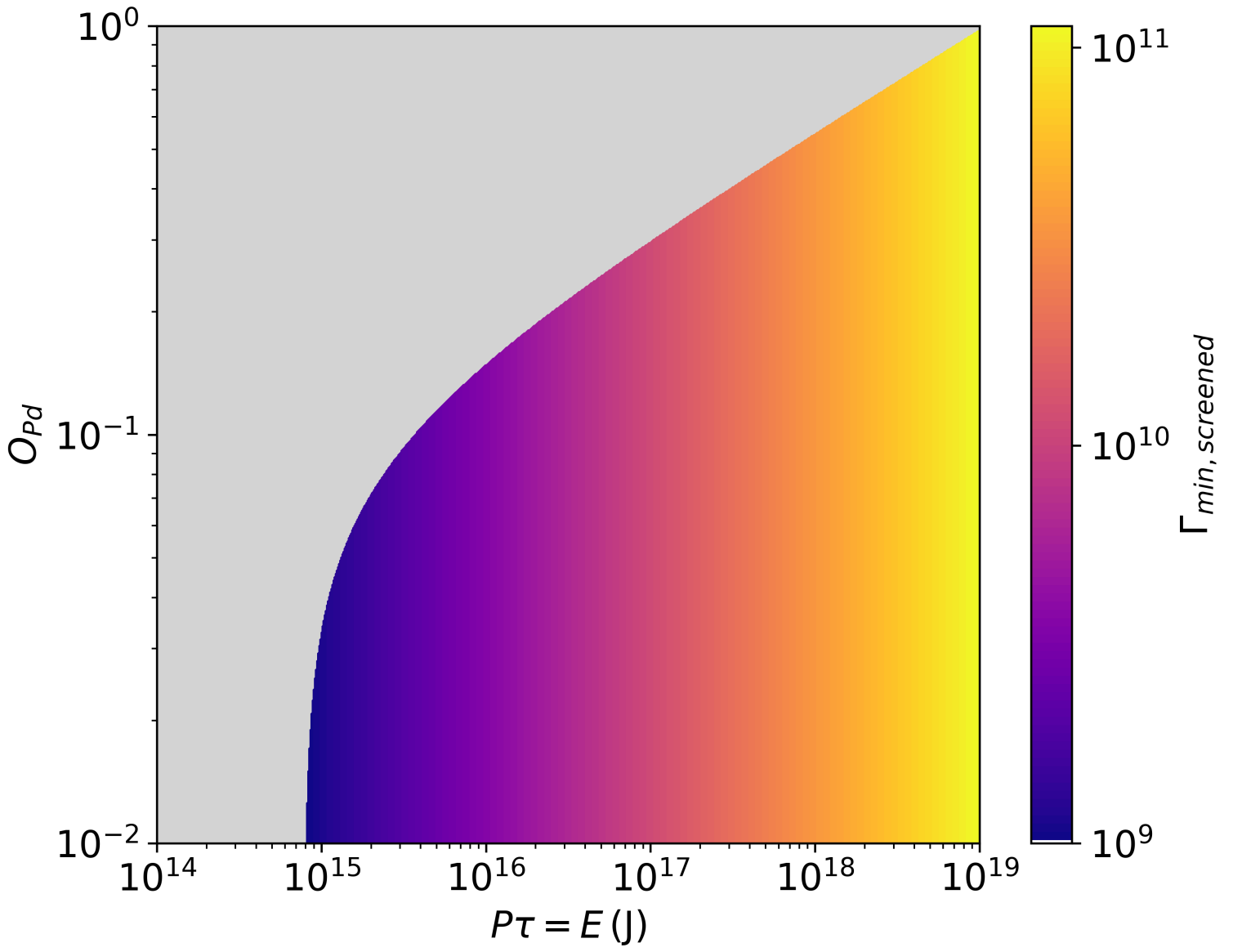

**Figure 31:** Plot of the minimum nuclear excitation transfer rate with 350 eV of electron screening ($\Gamma_{min,screened}$) from the $D_2$/$^4$He decay transition to the Pd*/Pd excitation transition as a function of the transient dissipated acoustic phonon energy (*i.e.*, the product of the dissipated transient power and pulse period in the acoustic mode) and the Pd $O$-value. The greyed out area represents rates that do not meet the minimum requirements determined by decoherence, *i.e.*, Eq. 203.

## Result and discussion

The results of this analysis are shown in Figure 31, where excitation transfer rates are shown as a function of acoustic energy $P_{diss}\tau_A$ and the $O$-value of the Pd*/Pd transition in the allowed part of the parameter regime where the excitation transfer rate exceeds the model decoherence rate. We've also used the same deuterium and palladium loading as in previous sections when calculating the rates, namely:

$$\frac{N_{\mathrm{D}_2}}{N} = 0.25 \times \frac{1}{9} \qquad \frac{N_{^{\mathrm{A}}\mathrm{Pd}}}{N} = 0.25 \tag{204}$$

We see that it is possible within this idealized model for the coherent dynamics to start up; however, the amount of energy in the highly-excited acoustic mode is impractically large.

For these rate calculations we assumed the same model parameters as in the transfer rate calculation in section 5.5 based on Eq. 154.

We point out that, after a considerable buildup with pedagogical but idealized examples in sections 5.3 and 5.5, we have arrived at a scheme in this section that comes closer to real physical systems expected to show observable effects. The approach explored in this section addresses issues such as the resonance condition and the constraints imposed by decoherence. The rates are, however, still not in the observable range, when considering real-world conditions.

We will explore another variant of the presented models in the following section, which is yet more complex but which yields results that suggest rates in the observable range under realistic assumptions.

## 5.10 Excitation transfer between $D_2$ and multiple receiver states

The structure of this document has been such that from 5.1 onwards, our goal has been to simplify the comprehensive and overbearingly rich lattice Hamiltonian in Eq. 17 so as to make things conceptually and computationally tractable.

This resulted in a first concrete example comprising excitation transfer from $D_2$ to $^4$He systems, a hypothetical scenario that involves a single perfectly resonant receiver state. While this example is straightforward to follow and yielded concrete rate estimates, it appears not directly relevant to experiments, since all predicted rates ended up in the unobservable range. Subsequently, we considered more complicated variations of the basic model: first, a different nuclear species as a receiver system in section 5.4 (focusing on Pd excited states), and then a combination of multiple lower-energy (but therefore also longer-lived) receiver transitions in section 5.9. In other words, we considered the possibility of multiple Pd transitions, including from multiple Pd nuclei, to combine.

Up to that point, we considered scenarios with a single receiver state (see Figure 32) – in the example with multiple Pd transitions, the single receiver state was the one closest to the resonance condition. However, they still suffered from certain unrealistic or incomplete assumptions and the need for extreme conditions (*e.g.*, regarding the required input energy).

Here, we will go one step further and consider the effects of multiple receiver states (see Figure 33). Multiple receiver states come into view when considering the high density of nuclear states that results from combinations of nuclear bound states (per the discussion in section 5.9). Note the difference between multiple receiver states (which can involve multiple transitions) and multiple transitions but a single receiver state (as was discussed in section 6.23).

Even more candidate receiver states become available when considering energy exchange with the lattice and resulting mixed nuclear-lattice states —which originally we deliberately excluded in section 5.1 to keep things simple at the beginning. We will bring energy exchange back in in the following section (see Figure 33 bottom for a preview of the implications).

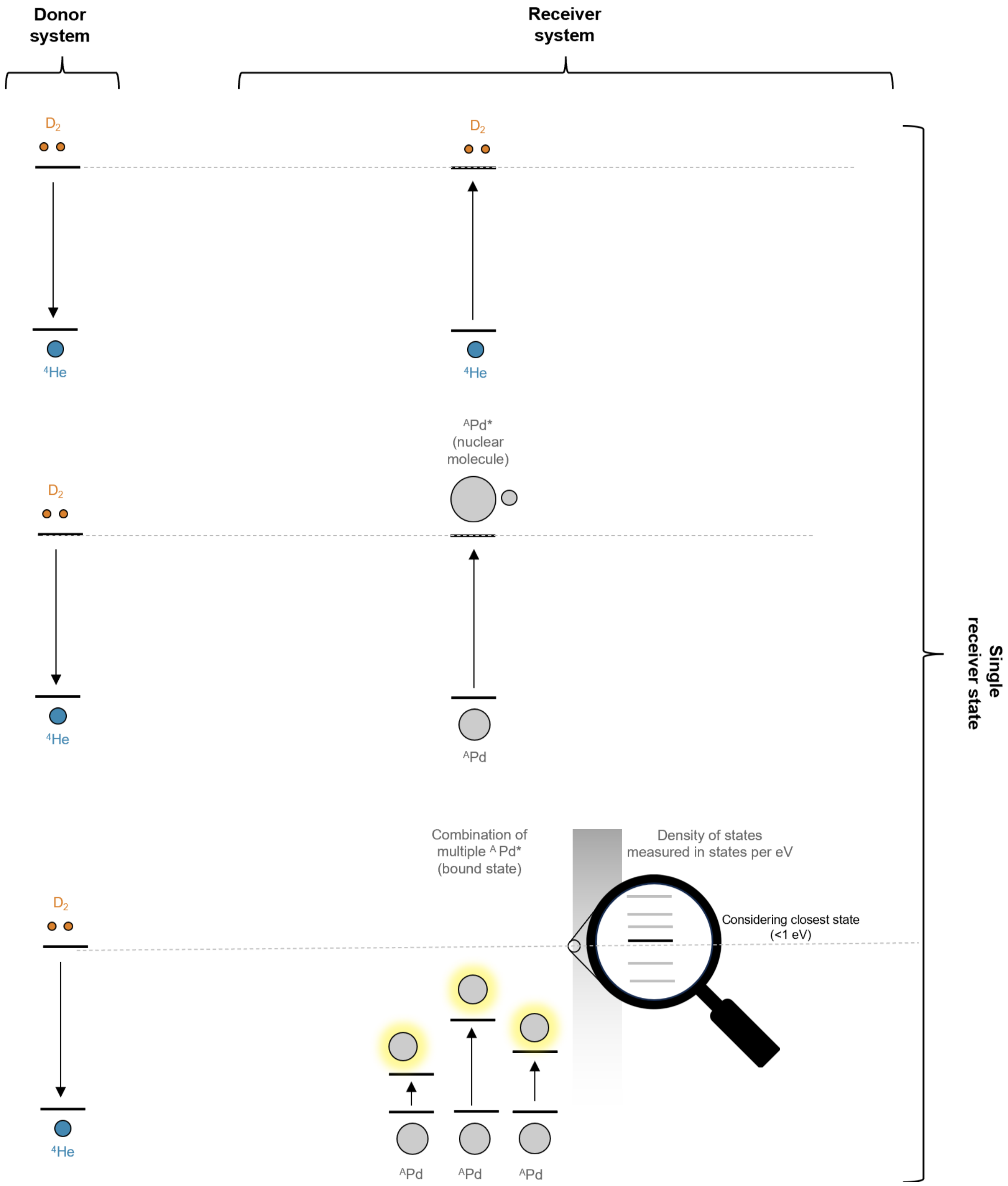

**Figure 32:** Overview of different single-receiver-state variants of the excitation transfer models discussed in this document: section 5.3 considered transfer from a $D_2$/$^4$He donor transition to a perfectly resonant $D_2$/$^4$He receiver transition but found that transfer rates were too low to yield observable results (top part); section 5.5 considered transfer to a single hypothetical Pd*/Pd receiver transition, where transfer rates are faster, but where perfect resonance was (unrealistically) assumed (middle part); section 5.9 considered transfer to multiple Pd*/Pd receiver transitions, where resonance can be realistically assumed to result from the high density of states due to combinations of nuclear bound states (bottom part). Note, however, that for single-receiver-state variants of the model, only the state closest to the resonance condition is considered as a receiver state.

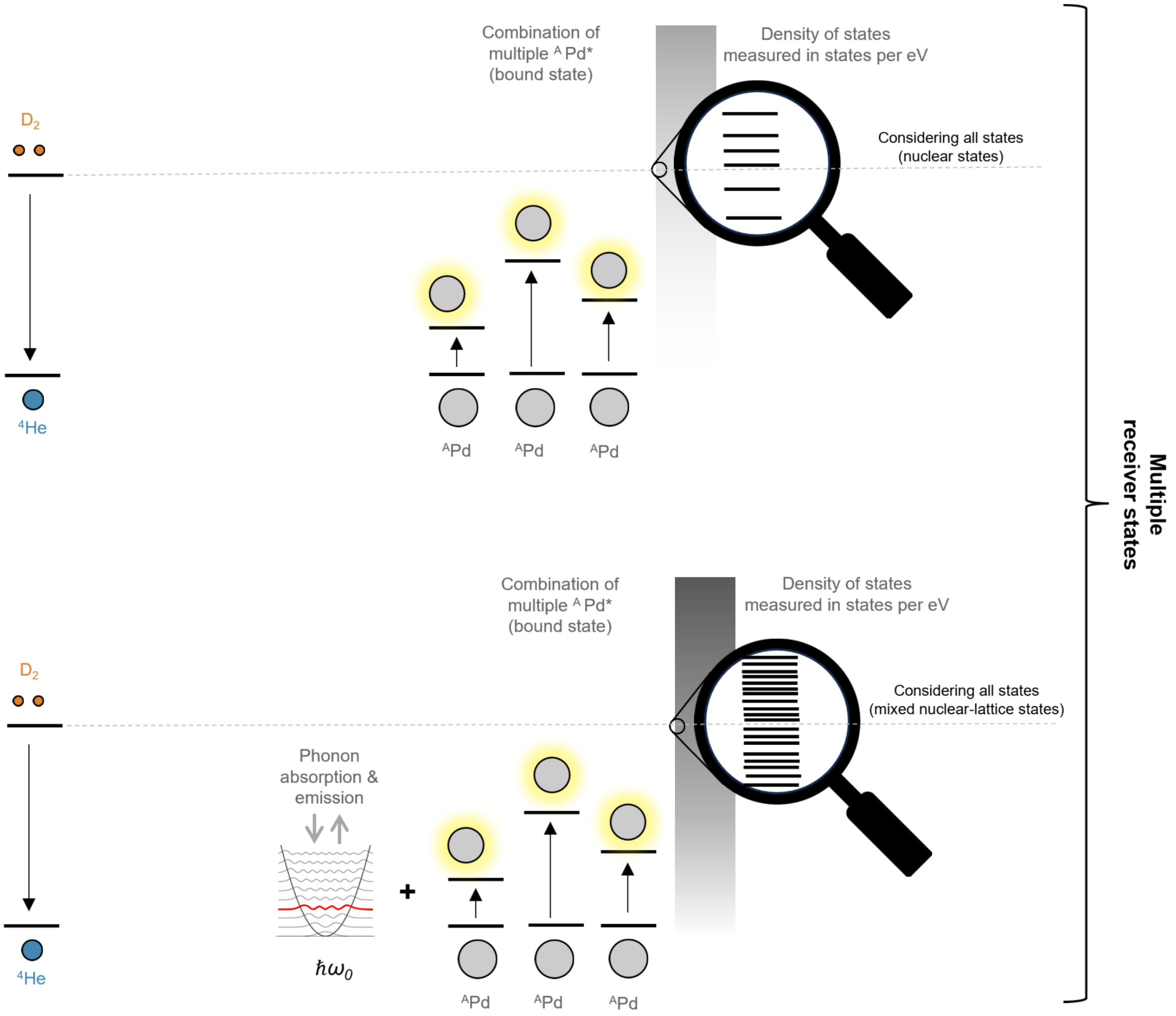

**Figure 33:** Overview of different multiple-receiver-state variants of the excitation transfer models discussed in this document: this section considers transfer from a $D_2$/$^4$He donor transition to multiple Pd*/Pd receiver transitions but—in contrast to section 5.9 (see the bottom part of Figure 32)—makes use of a formalism where the transfer rate calculation takes multiple possible receiver states into account (top part of the figure); the following section 5.11 goes one step further and includes phonon absorption and emission in the consideration, which effectively corresponds to a further increase in the density of receiver states (bottom part of the figure).

**Golden Rule formula for excitation transfer to multiple receiver states**

In the previous sections, we made use of a rate expression for transfer to a single receiver state. The general form of such a rate is

$$\Gamma_{transfer} \sim \frac{2}{\hbar}|V_{fi}| \tag{205}$$

When dealing with multiple receiver states, we need to make use of the so-called Golden Rule, which is often written as

$$\Gamma_{transfer} = \frac{2\pi}{\hbar}|V_{fi}|^2\rho \tag{206}$$

which we recognize as Dirac's formula [97]; where $V_{fi}$ is the interaction matrix element and where $\rho$ is the relevant density of states.

This formula is sometimes expressed as

$$\Gamma_{transfer} = \frac{2\pi}{\hbar}\sum_f |V_{fi}|^2\delta(E_i - E_f) \tag{207}$$

which indicates a summation/integration over all of the final states, where the $\delta$-function imposes energy conservation and gives rise to the density of states in Dirac's formula. In this way of thinking, we can identify $\sum_f \delta(E_i - E_f)$ with the density states $\rho$.

**The impact of multiple receiver states**

With the model of receiver states comprising multiple nuclear bound state transitions (in our case Pd transitions) from Eq. 185 the excitation transfer rate becomes

$$\Gamma_{transfer}(E) = \frac{2\pi}{\hbar}\sum_f |\Delta\mathcal{U}|^2\bigg(\prod F(g)\bigg)_f^2 \delta(E - E_f) \tag{208}$$

where the product considers the combinations of multiple Pd transitions and the sum considers the different receiver states.

Recall that $|\Delta\mathcal{U}|^2$ represents the fusion transition affected by selective loss (5.1). Since $|\Delta\mathcal{U}|^2$ is independent of the final state nuclear configuration, it can be pulled out from the summation and rewritten as

$$\Gamma_{transfer}(E) = \bigg(\frac{2\pi}{\hbar}|\Delta\mathcal{U}|^2\bigg)\bigg(\sum_f \bigg(\prod F(g)\bigg)_f^2 \delta(E - E_f)\bigg) \tag{209}$$

In this form, the Golden Rule rate is expressed as the product of the fusion transition rate $\frac{2\pi}{\hbar}|\Delta\mathcal{U}|^2$ and a term $\sum_f \bigg(\prod F(g)\bigg)_f^2 \delta(E - E_f)$. That term is essentially an effective density of states, which we call the *generalized density of states* $\tilde{\rho}_N(E)$. The generalized density of states includes all the possible transitions on the receiver side. We can define

$$\tilde{\rho}_N(E) \;=\; \sum_f \left( \prod F(g) \right)_f^2 \delta(E - E_f) \tag{210}$$

where we identify $\sum_f \delta(E - E_f)$ with the density of nuclear transitions $\rho_N(E)$.

Note that the generalized density of states depends on (via the $g$ terms) the energy in the phonon modes $E_A = P_{diss}\tau_A$ (see Eq. 199)

With this definition, the excitation transfer rate for transfer to multiple resonant receiver states can be expressed as

$$\Gamma_{transfer}(E) \;=\; \left( \frac{2\pi}{\hbar} |\Delta \mathcal{U}|^2 \right) \tilde{\rho}_N(E) \tag{211}$$

Since the energy transferred from the fusion transition is $E = \Delta Mc^2 = 23.85$ MeV, the excitation transfer rate that we are interested in is $\Gamma_{transfer}(\Delta Mc^2) = \Gamma_{transfer}(23.85 \text{ MeV})$.

**Estimating the generalized density of states based on related experimental data**

At this point we have an outline of a Golden Rule calculation than can be evaluated based on a list of energy levels for excited states of the Pd isotopes, along with M2 transition strength data for transitions to the ground state ($O$-values per section 5.4).

No such data sets are presently available for Pd, but in section 5.7 above, we have introduced an estimated nuclear density of states for combinations of Pd transitions based on data for $^{120}$Sn by Müscher et al. (2022) in [94], where results from a nuclear resonance fluorescence experiment on $^{120}$Sn are reported (see section 5.4 for more details on this data set). We have argued that such data is closely related to our problem and can stand in as a proxy for the time being.

Analogous to the nuclear density of states developed in section 6.18, an estimate for the generalized density of states is developed in section 6.19 based on the data from Müscher et al. (2022) [94]. Developing such an estimate is non-trivial. We found that the density of states—and the generalized density of states—can be approximated by a Gaussian and the resulting overall estimate is a sum of Gaussians for different numbers of combinations of states:

$$\tilde{\rho}_N(E) \;=\; \sum_m A_m \frac{1}{\sqrt{2\pi\sigma_m^2}} e^{-(E-\mu_m)^2/2\sigma_m^2} \tag{212}$$

Each successive Gaussian in the sum represents the contribution to the generalized density of states from one, two, three, etc. nuclear transitions.

Note that $\tilde{\rho}_N(E)$ has a dependence on $E_A$, *i.e.*, the larger the oscillator energy (acoustic phonon energy in the specific case considered here), the higher the effective density of states (see section 6.19 for an explanation of the $E_A$ dependence). So $\tilde{\rho}_N(E)$ is really $\tilde{\rho}_N(E, E_A)$.

This $E_A$ dependence can be seen in Figure 34 at the fixed value of E that is of most interest, namely the the deuterium fusion transition energy $E = \Delta Mc^2 = 23.85$ MeV.

**Excitation transfer rate to multiple receiver states with no significant oscillator energy exchange**

In this section, we are interested in the excitation transfer rate to multiple receiver states including only the nuclear contribution to receiver states, *i.e.*, in the absence of energy exchange with phonons and plasmons (which results in additional receiver states per Figure 33 and which will be discussed in the following section). Since this base rate plays such an important role in the model, we denote it as $\Gamma^{(0)}_{transfer}$, and write it as

$$\Gamma^{(0)}_{transfer} \;=\; \frac{2\pi}{\hbar}|\Delta\mathcal{U}|^2 \sum_f \left(\prod F(g)\right)^2_f \delta(\Delta Mc^2 - E_f) \;=\; \frac{2\pi}{\hbar}|\Delta\mathcal{U}|^2 \tilde{\rho}_N(\Delta Mc^2) \tag{213}$$

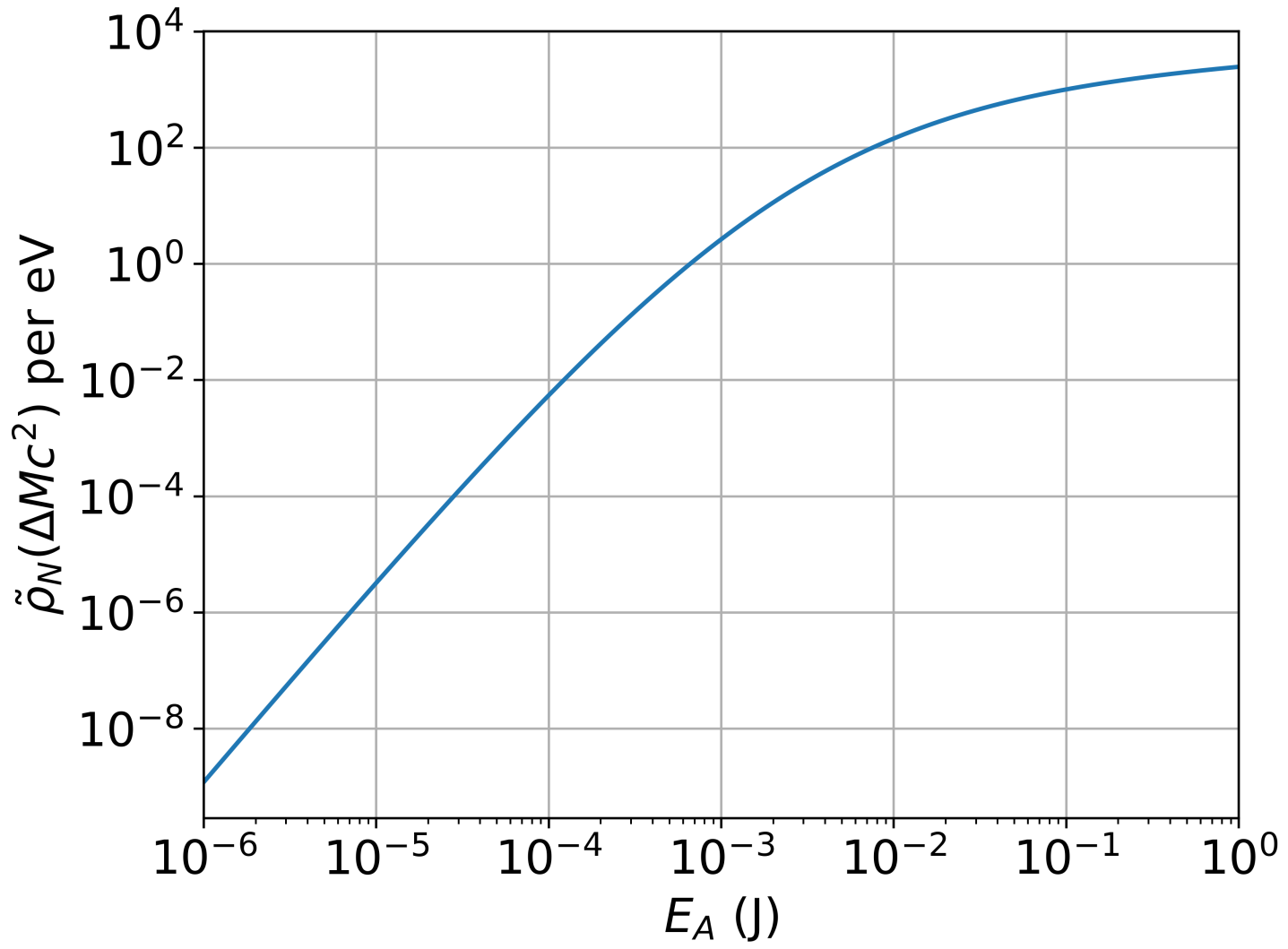

**Figure 34:** Generalized nuclear density of states $\tilde{\rho}_N(E_A)$ at $E = 23.85$ MeV.

It is useful to develop a numerical estimate for this rate. We recall from Eq. 198 that (for a screening energy of $U_e = 350$ eV)

$$\frac{2}{\hbar}|\Delta\mathcal{U}| \;=\; 5536\sqrt{\frac{E_A}{1\text{ J}}}\sqrt{\frac{N_{D2}}{N}}\ \text{s}^{-1} \tag{214}$$

We can use this to obtain

$$\Gamma^{(0)}_{transfer}(E) \;=\; 7.76\times 10^{-5}\left(\frac{E_A}{1\text{ J}}\right)\left(\frac{N_{D2}}{N}\right)\left(\frac{\tilde{\rho}_N(E)}{\tilde{\rho}^{(0)}_N}\right)\ \text{s}^{-1} \tag{215}$$

where

$$\tilde{\rho}^{(0)}_N \;=\; \tilde{\rho}_N(\Delta Mc^2)\Big|_{E_A=1\text{ J}} \;\to\; 2451\ \text{eV}^{-1} \tag{216}$$

Figure 35 shows Eq. 215 evaluated at different values of $E_A$ and $E$. The excitation transfer rate to multiple nuclear receiver states in the absence of energy exchange with phonons and plasmons is still lower than the estimated decoherence rate (see section 6.22).

However, note the exponentially increasing dependence on the transferred energy E when $E_A$ is above about 10 $\mu$J: if the energies involved in the transfer were larger, the rates could be faster. This is due to the much larger number of combinations of nuclear transitions available if more energy were available from the donor transition.

On first glance, this relationship does not appear to be of much help, since we work with a donor system at a fixed energy of 23.85 MeV, *i.e.*, the deuterium fusion transition energy. However, as we will see in the following section, it is possible to take advantage of this scaling relationship in other ways: if additional energy comes into the Pd system from elsewhere, it is possible to operate at a higher operating point. In the following section, we will show that energy can exchange rapidly between nuclear states and oscillator modes, resulting in a nonzero probability that the Pd nuclei are in a state of overall higher energy.

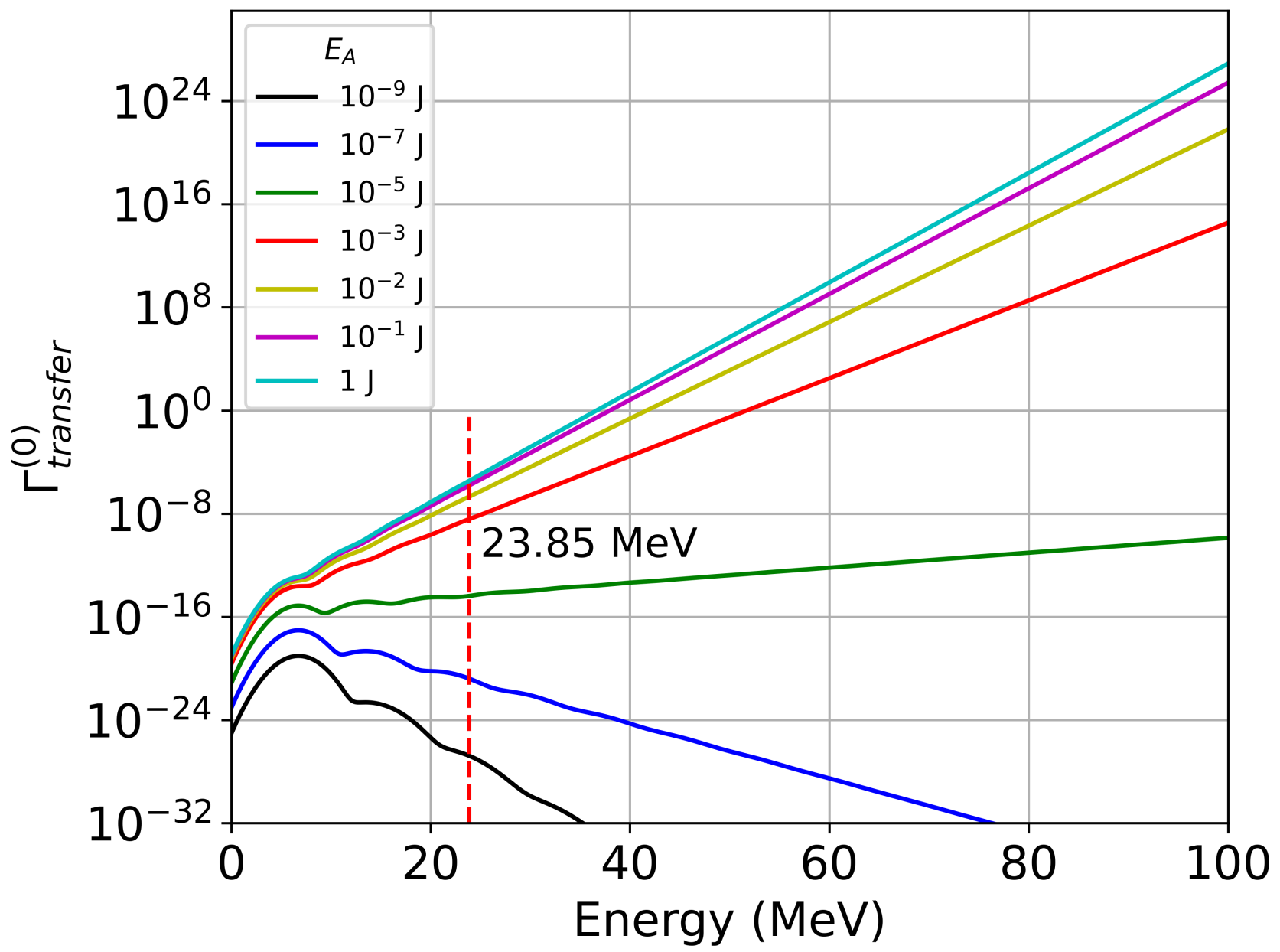

**Figure 35:** Excitation transfer rate to multiple receiver states from the $D_2/^4He$ fusion transition with no oscillator energy exchange for several acoustic phonon mode energies $E_A$.

## 5.11 Excitation transfer between $D_2$ and multiple receiver states with oscillator energy exchange

In the previous section, we introduced the notion of multiple receiver states and we developed a formalism, based on the Golden Rule, that resulted in a corresponding transfer rate equation (Eq. 213). A model that considers multiple receiver states is found to be more appropriate than a model that considers only single receiver states, given the high density of states near the resonance condition that results from the many possible combinations of nuclear bound states.

After having initially emphasized simplification of the comprehensive lattice Hamiltonian in section 5.1, we now allow for yet more complex (and at the same time more realistic) models in this section. Specifically, we now allow for energy exchange with the phonon modes or plasmon modes on top of the multiple receiver state model of the previous section. Note that in that context the transfer rate equation, Eq. 213 is the base rate against which rates that result from different extents of oscillator energy exchange can be compared.

Including energy exchange with oscillator modes makes the corresponding theoretical problem more difficult. A technical discussion of the model under consideration is given in section 6.24, where the basic approach and model is discussed. The excitation transfer rate from the fusion transition to many Pd*/Pd transitions with energy exchange with the oscillators is then given in terms of matrix elements of an approximation to the finite time transition matrix element.

In this section we step through the key issues and develop a simple approximate expression for the excitation transfer rate as a convolution of the generalized excitation transfer rate without oscillator exchange (Eq. 213) and the probability distributions for energy exchange with oscillators. The following section will then contain a quantitative evaluation of this rate expression that allows for ready comparison with rate estimates from earlier, simpler versions of the model.

### Hamiltonian for energy exchange with the lattice

As we are now concerned with energy exchange with the lattice, which will involve the net exchange of oscillator quanta, we need to make use of a Hamiltonian with nuclear degrees of freedom coupled to oscillator degrees of freedom.

Similar to what was done in section 5.1, we start out with a comprehensive Hamiltonian analogous to Eq. 17 (in this case not including magnons as we also did later in section 5.1 for simplicity):

$$\hat{H} = \hat{H}_{nuclei} + \hat{H}_{phonons} + \hat{H}_{plasmons} + \hat{V}_{nuclei,phonons} + \hat{V}_{nuclei,plasmons} \tag{217}$$

We include selective loss in the Hamiltonian for the nuclei, per section 5.2 and for the interaction part, we focus on the relativistic interaction as it has been shown to be the strongest among the relevant interactions (5.3).

This results in the Hamiltonian

$$\begin{aligned}\hat{H} = & \sum_j \sum_k \left\{ |\phi_j\rangle \left( M_j c^2 - i\frac{\hbar}{2}\gamma_j(E) \right) \langle\phi_j| \right\}_k + \hbar\omega_A \hat{a}_A^\dagger \hat{a}_A + \hbar\omega_O \hat{a}_O^\dagger \hat{a}_O + \hbar\omega_P \hat{a}_P^\dagger \hat{a}_P \\ & + \sum_{j,j'} \sum_k \left\{ |\phi_{j'}\rangle\langle\phi_{j'}| \mathbf{a} \cdot c\hat{\mathbf{P}}_j |\phi_j\rangle\langle\phi_j| \right\}_k \end{aligned} \tag{218}$$

The equation contains terms for the mass energy and decay rates for the excited nuclear states; we also include a uniform acoustic phonon mode, a uniform optical phonon mode, and a uniform plasmon mode; and we include $\mathbf{a} \cdot c\mathbf{P}$ mediated transitions between ground states and excited states, where the center of mass momentum operator depends on the nuclear mass and is uniform over the lattice, and includes contributions from the three different oscillators.

We recall from section 5.1 that the position operator can be written in terms of the phonon mode operators according to

$$\hat{\mathbf{R}}(j,l) \;=\; \mathbf{R}(j,l) + \sum_{\mathbf{k},\nu} \mathbf{e}(j,\mathbf{k},\nu) \sqrt{\frac{\hbar}{2NM_j\omega_{\mathbf{k},\nu}}} \left( e^{i\mathbf{k}\cdot\mathbf{R}(j,l)} \hat{a}_{\mathbf{k},\nu} + e^{-i\mathbf{k}\cdot\mathbf{R}(j,l)} \hat{a}^\dagger_{\mathbf{k},\nu} \right) \tag{219}$$

where all the parameters involved are described in section 5.1.

The corresponding expression for the center of mass momentum operator is

$$\hat{\mathbf{P}}(j,l) \;=\; \sum_{\mathbf{k},\nu} \mathbf{e}(j,\mathbf{k},\nu) \sqrt{\frac{\hbar M_j \omega_{\mathbf{k},\nu}}{2N}} \left( \frac{e^{i\mathbf{k}\cdot\mathbf{R}(j,l)} \hat{a}_{\mathbf{k},\nu} - e^{-i\mathbf{k}\cdot\mathbf{R}(j,l)} \hat{a}^\dagger_{\mathbf{k},\nu}}{i} \right) \tag{220}$$

If we consider contributions only from one (uniform) acoustic mode and one (uniform) optical mode, then this reduces to

$$\hat{\mathbf{P}}_j \;\to\; \mathbf{e}_j^{(A)} \sqrt{\frac{\hbar M_j \omega_A}{2N}} \left( \frac{\hat{a}_A - \hat{a}_A^\dagger}{i} \right) + \mathbf{e}_j^{(O)} \sqrt{\frac{\hbar M_j \omega_O}{2N}} \left( \frac{\hat{a}_O - \hat{a}_O^\dagger}{i} \right) \tag{221}$$

We can extend this to include coupling with a uniform plasmon mode by writing

$$\hat{\mathbf{P}}_j \;\to\; \mathbf{e}_j^{(A)} \sqrt{\frac{\hbar M_j \omega_A}{2N}} \left( \frac{\hat{a}_A - \hat{a}_A^\dagger}{i} \right) + \mathbf{e}_j^{(O)} \sqrt{\frac{\hbar M_j \omega_O}{2N}} \left( \frac{\hat{a}_O - \hat{a}_O^\dagger}{i} \right) + \mathbf{e}_j^{(P)} \sqrt{\frac{Z_j^2 \hbar M_j \omega_P}{2N_e}} \left( \frac{\hat{a}_P - \hat{a}_P^\dagger}{i} \right) \tag{222}$$

With this model we can describe $\mathbf{a} \cdot c\mathbf{P}$ interactions between nuclear states and acoustic phonons, optical phonons and plasmons conveniently within the basic formalism.

**Need for a strong coupling model**

Since we assume that the nuclei interact with a uniform acoustic phonon mode, a uniform optical phonon mode and a uniform plasmon mode, all of the nuclear transitions couple uniformly, which means that transitions are cooperatively enhanced. This cooperative enhancement will be reflected through the presence of Dicke factors below.

Perturbation theory runs into trouble in the strong coupling regime, so it is necessary to first evaluate how strong the coupling is. So far in the text, we have been assessing coupling strength by measuring how large the cooperatively enhanced coupling constant is compared to the nuclear transition energies, i.e.,

$$(g_j)_{nuc} \;=\; \sqrt{N_{Pd_j}} \frac{|\langle \mathrm{Pd}_j^* | \mathbf{a} \cdot c\mathbf{P} | \mathrm{Pd} \rangle|}{\epsilon_j} \tag{223}$$

Recall from 5.7 that $\epsilon_j$ is the transition energy.

For the conditions of interest to us, we have seen that this dimensionless coupling strength is greater than 1 for many of the Pd*/Pd transitions. Now that we consider energy exchange with the oscillators, we must also compare the cooperatively enhanced coupling constant to the energy quantum associated with the oscillator (here: an acoustic phonon mode), namely:

$$(g_j)_A \;=\; \sqrt{N_{Pd_j}} \frac{|\langle \mathrm{Pd}_j^* | \mathbf{a} \cdot c\mathbf{P} | \mathrm{Pd} \rangle|}{\hbar\omega_A} \;=\; \frac{\epsilon_j}{\hbar\omega_A} (g_j)_{nuc} \tag{224}$$

This extra comparison suggests that the dimensionless coupling strength for energy exchange between nuclear states and acoustic phonon modes is many orders of magnitude larger than the coupling strength for the nuclear transitions. A corresponding model therefore needs to operate in the extremely strong coupling regime and cannot rely on perturbation theory.

A relevant picture for the extremely strong coupling regime is one in which the transitions that are present during the excitation transfer exchange oscillator quanta freely with the lattice (at a rate $\Gamma_A$), sometimes creating and sometimes annihilating, during the whole time up until the excitation transfer event is complete ($\tau = 1/\Gamma_{transfer}$).

This is a probabilistic way of considering phonon exchange and it results in a probability distribution that describes how many phonons have been created or annihilated as a result of the interaction with the nuclear states. We will see shortly that such dynamics can lead to a significant enhancement of the excitation transfer rate. We will start by developing an expression for the probability distribution associated with this kind of phonon exchange ($f$). The probability distribution will depend on the rate of acoustic phonon exchange ($\Gamma_A$), which we will determine in a second step.

**Probability distribution for acoustic phonon exchange**

Each time an oscillator quantum is exchanged, either a quantum is created ($\delta n_A > 0$) or a quantum is destroyed ($\delta n_A < 0$) . If this were a classical Bernoulli process, then we would expect the spread in the number of phonon states ($\sigma_{n_A}$) to be proportional to the square root of the number of oscillator quantum exchanges (as with a random walk). However, quantum diffusion is much faster, and in the associated quantum diffusion model we would expect the number of oscillator states eventually occupied to be on the order of the number of oscillator quanta exchanged.

The probability distribution associated with phonon exchange that results after a time $\tau$ can be written as

$$f_{n_A}(\delta n_A, \tau) \;=\; \frac{1}{\sqrt{2\pi\sigma^2_{n_A}(\tau)}} \exp\left\{-\frac{\delta n_A^2}{2\sigma^2_{n_A}(\tau)}\right\} \tag{225}$$

with

$$\sigma_{n_A}(\tau) \;=\; \frac{1}{2}\sum_j \Gamma_A(j)\tau \;\rightarrow\; \frac{1}{2}\overline{n_j}\,\Gamma_A\tau \tag{226}$$

where $\Gamma_A$ is the average rate for acoustic phonon exchange from a single transition, and where $\overline{n_j}$ is the average number of transitions that contribute to phonon or plasmon exchange. We assume that the time available for phonon exchange $\tau$ is

$$\tau \;\sim\; \frac{1}{\Gamma_{transfer}} \tag{227}$$

where $\Gamma_{transfer}$ is the excitation transfer rate evaluated at $\Delta Mc^2$.

The associated energy distribution is

$$f_{\epsilon_A}(\epsilon_A, \tau) \;=\; \frac{1}{\sqrt{2\pi\sigma^2_{\epsilon_A}(\tau)}} \exp\left\{-\frac{\epsilon_A^2}{2\sigma^2_{\epsilon_A}(\tau)}\right\} \tag{228}$$

with

$$\epsilon_A = \delta n_A \hbar\omega_A \tag{229}$$

and

$$\sigma_{\epsilon_A} = \hbar\omega_A \sigma_{n_A} \tag{230}$$

Note that this distribution depends explicitly on the excitation transfer rate (through its inverse $\tau$). We are going to make use of this distribution to evaluate the excitation transfer rate, which means that a self-consistent solution will be required.

**Rate for acoustic phonon exchange with a single transition**

The rate of acoustic phonon exchange ($\Gamma_A$) for a single transition $j$ is

$$\begin{aligned} \Gamma_A(j) &= \frac{2}{\hbar}|\langle \text{Pd}^*(j)|\mathbf{a}\cdot c\mathbf{P}(j)|\text{Pd}\rangle_A| \\ &\approx \frac{2}{\hbar}|\mathbf{e}_j^{(A)}||\langle \text{Pd}^*(j)|a_z|\text{Pd}\rangle|\sqrt{M_j c^2 E_A}\sqrt{\frac{N_{Pdj}}{N}} \end{aligned} \tag{231}$$

Given that a great many individual transitions are involved, it is useful to work with an average according to

$$\begin{aligned} \Gamma_A &= \frac{2}{\hbar}\overline{|\langle \text{Pd}^*(j)|\mathbf{a}\cdot c\mathbf{P}(j)|\text{Pd}\rangle_A|} \\ &\approx \frac{2}{\hbar}\overline{|\mathbf{e}_j^{(A)}||\langle \text{Pd}^*(j)|a_z|\text{Pd}\rangle|\sqrt{M_j c^2 E_A}\sqrt{\frac{N_{Pdj}}{N}}} \end{aligned} \tag{232}$$

A numeric evaluation leads to

$$\Gamma_A \approx 6.21 \times 10^{25} \sqrt{\frac{E_A}{1\text{ J}}}\, |\mathbf{e}^{(A)}|\overline{O_j\sqrt{\frac{N_{Pdj}}{N}}} \text{ s}^{-1} \tag{233}$$

The rate of acoustic phonon exchange depends on the $O$-values of the transitions involved. If we average over the $^{120}$Sn levels and $O$-values that we have been using as an approximation for the Pd states (see sections 5.4 and following), and average over the natural abundance of the Pd isotopes, then we obtain

$$\overline{O_j\sqrt{\frac{N_{Pdj}}{N}}} = 0.0146 \tag{234}$$

This allows us to evaluate Eq. (233) and we find that the rate for coherent energy exchange between acoustic phonons and Pd*/Pd transitions is extremely fast.

The fastness of this rate implies that we should have included acoustic phonon exchange in the dynamicsof models discussed earlier. It impacts how we think about and model the strongly-coupled Pd and acoustic phonon systems. Ideally, we would like to work with the eigenfunctions of these two coupled systems in connection with modeling excitation transfer from the $D_2$/$^4$He fusion transition. However, at present we do not have available a suitable diagonalization. As a remedy, since acoustic phonon exchange is much faster than the excitation transfer rate, we model acoustic phonon exchange as a “free energy exchange” for the duration of an individual excitation transfer. This approach is discussed in detail in section 6.24.

### Excitation transfer rate with acoustic phonon energy exchange

As we have already seen, the excitation transfer rate, as determined by the Golden Rule, depends on the strength of the interactions and also the density of states at the relevant transition energy (to ensure energy conservation for the transition). For the nuclear excitation transfer scenario that we are interested in, the relevant transition energy in the absence of phonon exchange is the $D_2/^4He$ fusion energy $\Delta Mc^2 = 23.85$ MeV. The generalized density of states around that energy $\tilde{\rho}_N(\Delta Mc^2)$ is therefore what concerns us most. However, when energy exchange with the lattice is allowed, the density of states across a spectrum of energies becomes important.

Instead of considering energy conservation between the $D_2/^4He$ system and the Pd*/Pd system, we must consider the oscillator and Pd together as Pd,$n_A$/Pd*,$n_A + \delta n_A$. For example, if the coupled system of Pd nuclei has absorbed $5 \times 10^{14}$ quanta of phonon energy ($\epsilon_A = 5 \times 10^{14}\hbar\omega_A$) from the lattice during the nuclear excitation transfer period $\tau$ then the relevant density of states is going to be shifted up in energy $\tilde{\rho}_N(\Delta Mc^2 + \epsilon_A)$ to account for the already excited Pd.

We do not know deterministically ahead of time what the energy shift $\epsilon_A$ will be, because the energy exchange with the lattice is probabilistic. We can, however, create an average by calculating the base transfer rate $\Gamma^{(0)}_{transfer}$ for each energy shift (see Eq. 213 in the previous section), weighting each scenario according to how likely it is (taken from the probability distribution $f_{\epsilon_A}$) and then summing the results to obtain the resulting transfer rate estimate. This is represented visually in Fig 36.

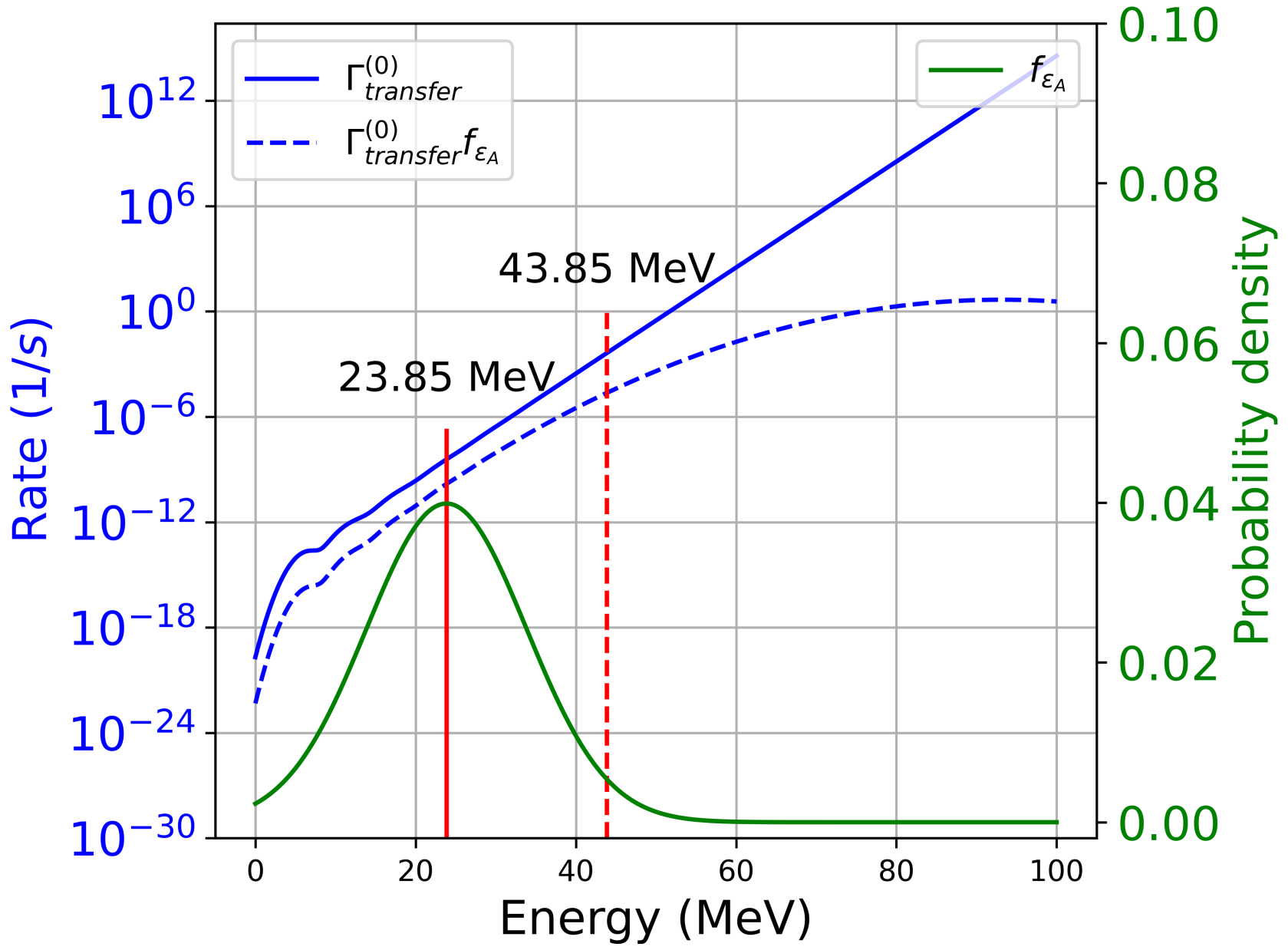

**Figure 36:** Calculating the excitation transfer rate including energy exchange with the lattice begins by looking at the base excitation transfer rate $\Gamma^{(0)}_{transfer}$ at 23.85MeV (intersection of red and blue solid lines) where energy exchange with the lattice is still excluded. In actuality, there is a non-zero probability (shown by the green line) that during the excitation transfer time $\tau$, the Pd will exchange energy with the lattice phonons, thus shifting the energy of the Pd that are due to receive the 23.85 MeV from the fusion transition. An exemplary shift of 20 MeV is shown by the red dashed line (with its likelihood indicated by the intersection of green line and the red dashed line). Such shifts require that we calculate the excitation transfer rate by evaluating $\Gamma^{(0)}_{transfer}$ at the shifted energy instead of at 23.85 MeV. To arrive at an overall rate estimate, each possible shift must be considered, weighted by how likely this particular shift is to occur (indicated by the blue dashed line). The end result for the excitation transfer rate comes from summing up all the weighted possibilities, *i.e.*, taking an integral of what is shown here as the blue dashed line.

When the above procedure is applied to different values of $E$ we get a similar plot to Figure 36 but with $f_{\epsilon_A}$ centered at $E$ instead of 23.85 MeV. This is described mathematically by the convolution below.

$$\begin{aligned}\Gamma_{transfer}(E) &= \left(\Gamma^{(0)}_{transfer} * f_{\epsilon_A}\right)(E) \\ &= \frac{2\pi}{\hbar}|\Delta\mathcal{U}|^2\left(\tilde{\rho}_N * f_{\epsilon_A}\right)(E) \\ &= \frac{2\pi}{\hbar}|\Delta\mathcal{U}|^2\int_{-\infty}^{\infty}\tilde{\rho}_N(E-\epsilon_A)f_{\epsilon_A}(\epsilon_A,\tau)d\epsilon \end{aligned} \qquad (235)$$

To develop an excitation transfer rate from this model we need to solve this constraint self-consistently at the fusion transition energy

$$\Gamma_{transfer} = \Gamma_{transfer}(\Delta Mc^2) = \frac{2\pi}{\hbar}|\Delta\mathcal{U}|^2\int_{-\infty}^{\infty}\tilde{\rho}_N(\Delta Mc^2-\epsilon_A)f_{\epsilon_A}(\epsilon_A,\tau)d\epsilon \qquad (236)$$

This convolution can be done exactly in the event that the generalized nuclear density of states is approximated by an exponential. An analytic model based on this kind of approximation is given in section 6.26.

A similar approach can be used to model optical phonon and plasmon exchange - see section 6.27 for details.

The following section will provide a quantitative estimate for expected transfer rates that result—under realistic experimental conditions—from the discussions in this section. These rate calculations will draw specifically on Eq. 236.

## 5.12 Transfer rate estimates: $D_2$ to multiple resonant receiver states with oscillator energy exchange

In the discussion above we have described a model for excitation transfer that involves multiple receiver states which results in an expression for the associated rate which must be solved for self-consistently. Here we focus on calculating the excitation transfer rate and other model parameters as a function of the energy in a highly excited acoustic mode ($E_A$).

To evaluate $\Gamma_{transfer}$ from Eq. 236 we first need to consider the values of all relevant parameters. We take values from section 5.5, specifically:

$$|1-\eta| \ = \ 0.1 \qquad \frac{v_{nuc}}{v_{mol}} \ = \ 6.65 \times 10^{-12} \qquad G = 90.35 \tag{237}$$

where $|1-\eta|$ is for the $D_2$/3+1/$^4$He contribution to the $D_2$/$^4$He transition per section 6.5 and where the volume ratio and the Gamow factor $G$ take on the values determined in section 6.3.

All parameters above feed into the expression for $\Delta\mathcal{U}$, which we remind the reader is (see section 5.9):

$$\begin{aligned}
\frac{2}{\hbar}\Delta\mathcal{U} \ &= \ \frac{2}{\hbar}|1-\eta| U \sqrt{N_{\mathrm{D}_2}} \\
&= \ 2|1-\eta| 0.0362 \sqrt{\frac{v_{nuc}}{v_{mol}}} e^{-G} \frac{\sqrt{M_{^4He} c^2 (1\ \mathrm{J})}}{\hbar} \sqrt{\frac{P_{diss}\tau_A}{1\ \mathrm{J}}} \sqrt{\frac{N_{\mathrm{D}_2}}{N}} \\
&= \ 2.38 \times 10^{-18} \sqrt{\frac{P_{diss}\tau_A}{1\ \mathrm{J}}} \sqrt{\frac{N_{\mathrm{D}_2}}{N}} \ \mathrm{s}^{-1}
\end{aligned} \tag{238}$$

As in previous sections, we can consider the effects of screening by multiplying $\Delta\mathcal{U}$ by (as per section 5.9)

$$e^{\Delta G_{scr}} \tag{239}$$

where (for 350 eV screening potential; see section 6.3)

$$\Delta G_{scr} = 49.2 \tag{240}$$

This gives us

$$\frac{2}{\hbar}|\Delta\mathcal{U}| \ = \ 5536 \sqrt{\frac{E_A}{1\ \mathrm{J}}} \sqrt{\frac{N_{D2}}{N}} \ \mathrm{s}^{-1} \tag{241}$$

The mode characteristics—$P_{diss}\tau_A = E_A$—feed into both $\Delta\mathcal{U}$ and the expression for $\Gamma_A$, which affects the generalized density of states for palladium $\tilde{\rho}_N$ and the phonon probability distribution $f$. For convenience, we reproduce $\Gamma_A$ from section 5.11:

$$\Gamma_A \ \approx \ 9.06 \times 10^{23} \sqrt{\frac{E_A}{1\ \mathrm{J}}} \ \mathrm{s}^{-1} \tag{242}$$

where, as in previous sections, we use a specific deuterium loading of

$$\frac{N_{\mathrm{D}_2}}{N} = 0.25 \times \frac{1}{9} \tag{243}$$

Putting this all together, the self-consistent solution for the excitation transfer rate from Eq. 236 including the effects of screening is shown in Figure 37 where we see an exponential relationship between $\Gamma_{transfer}$ and $P_{diss}\tau_A = E_A$.

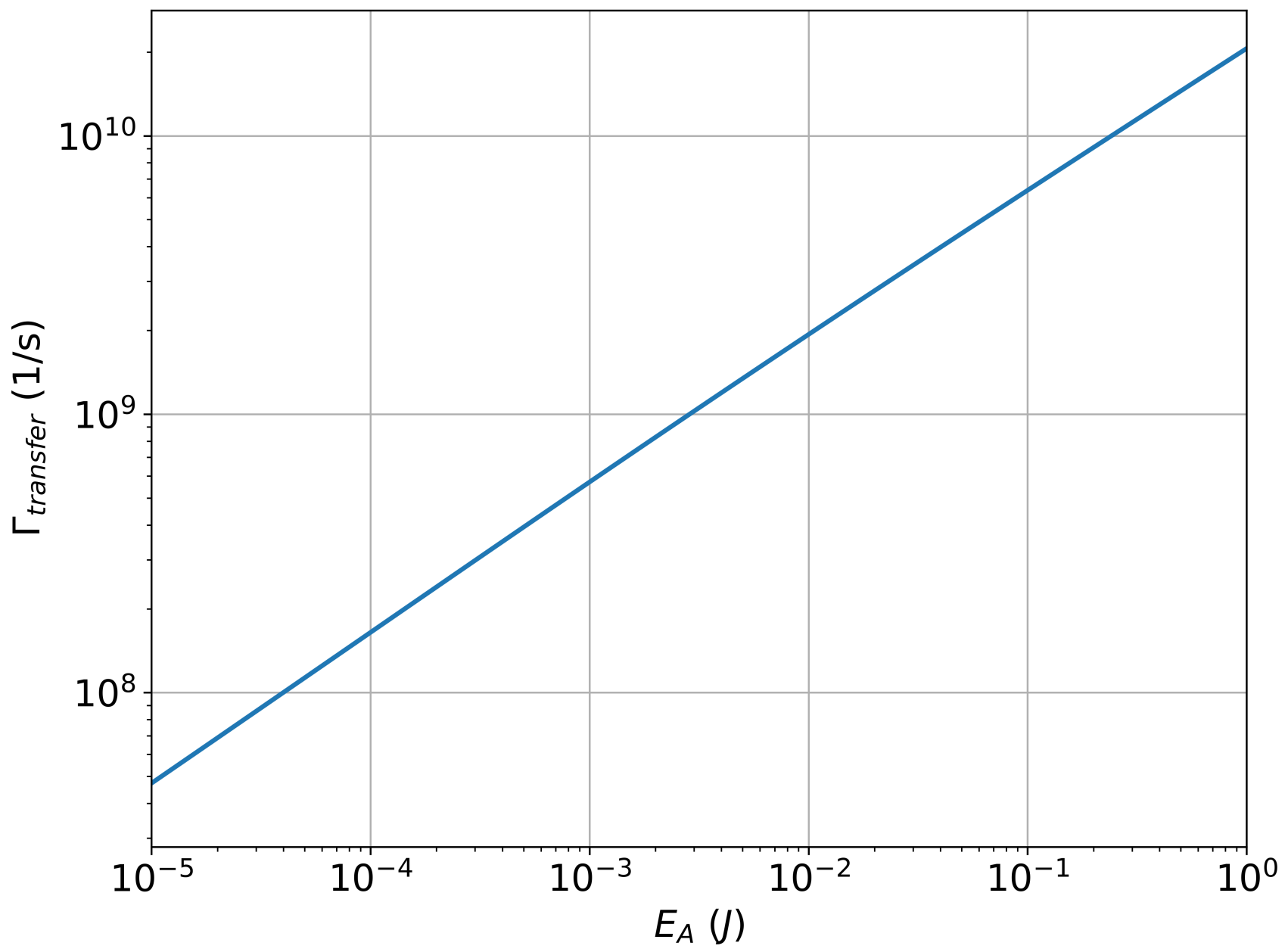

**Figure 37:** Excitation transfer rate as a function of $E_A$.

The self-consistent solution also allows us to study the associated average phonon energy exchanged and the spread of energies. See section 6.28 for more details.

We can extract an experimentally plausible value of $\Gamma_{transfer}$ from the plot by selecting values for vibrational power dissipation and frequency that we used in previous sections, namely:

$$P_{diss} = 1 \text{ W} \qquad f_A = 5 \text{ MHz} \tag{244}$$

resulting in a $\tau_A$ (for acoustic phonons) of

$$\begin{aligned} \tau_A &= 10^{-12}\left(10^{-7}\frac{f_A}{1\text{MHz}}\right)^{-3/2} \text{ sec} \\ &\approx 0.003 \text{ sec} \end{aligned} \tag{245}$$

We then arrive at an excitation transfer rate of:

$$\Gamma_{transfer} = 1 \times 10^9 \text{ s}^{-1} \tag{246}$$

This is the population-level transfer rate, *i.e.*, the rate at which excitation from the ensemble in group $a$ transfers to the ensemble in group $b$. To obtain a rate per deuteron pair that is comparable to the estimate given by Koonin and Nauenberg 1989 [20], we normalize the expression by $N_{\mathrm{D}_2} = 10^{15}$, which gives

$$\frac{\Gamma_{transfer}}{N_{\mathrm{D}_2}} = 1 \times 10^{-6}\ \mathrm{s}^{-1} \tag{247}$$

This is significantly greater than the Koonin and Nauenberg value of $10^{-64}\ \mathrm{s}^{-1}$ and also far beyond the observability threshold of $10^{-23}\ \mathrm{s}^{-1}$.

The population-level rate is on the order of the estimated $D_2$ decoherence rate and so we can expect Dicke factors to begin building up to create even higher rates over time. The path to building up these higher Dicke factors is discussed in more detail in section 6.29.

## 5.13 Pd* excited state stabilization

The sections above presented different versions of excitation transfer models with $D_2$ donor systems and Pd receiver systems at increasing degrees of complexity. The version of the model discussed in sections 5.11 and 5.12 yielded rates that not only exceed the observability threshold but that can also exceed the estimated $D_2$ decoherence time of 1 ns (given large enough phonon energies, *e.g.*, $E_A$ in the case of acoustic phonons, see Figure 37).

The expected result of such excitation transfer dynamics is the creation of a large number of Pd excited states. In other words, the principles laid out in this document imply the possibility of creating devices for the large-scale production of Pd excited states (or excited states of other receiver nuclei, when translating these findings to other materials).

This section will consider what the consequences are of such kind of excited state production, especially with respect to observable results. If the Pd nuclei in question were isolated, the issue would be trivial: each excited state would decay per the known (or estimated) state lifetimes. For the M2 states that we focused on because of the nature of $\mathbf{a} \cdot c\mathbf{P}$ coupling (see section 6.11), this decay is governed by magnetic quadrupole coupling with the electromagnetic field. However, the fact that the nuclei participate in a coupled quantum system—which causes their excitation in the first place—makes the more complex considerations in this section necessary.

We will focus here on the lifetime of the Pd excited states, while the next section will focus on the various loss, *i.e.*, decay channels, which determine the observable products. We will see that rapid phonon and plasmon exchange stabilizes radiative decay processes of unstable Pd* states. In essence, oscillator quanta exchange transitions happen so fast and the nuclear excited states are occupied for such short periods of time that their radiative decay is interrupted. These outcomes are related to the quantum Zeno effect discussed in the literature [98]. Another consequence of these dynamics, where energy exchange is fast, is that nuclear energy is expected to be converted efficiently into optical phonon energy and plasmon energy.

Note that changes to the lifetimes of excited states—beyond what was discussed about state lifetimes in section 5.4—also impact the excitation transfer dynamics. That is because the excited state lifetimes represent a constraint on those dynamics (per section 5.8). Stabilization and longer state lifetimes represent a loosening of that constraint.

### Brief review of Dirac's model

We will start the discussion with a brief review of radiative decay models. Exponential decay from quantum dynamics involving a continuum is described by Dirac's model [97], and we can work with a modified version of it here to address the issue of stabilization. Exponential decay can be derived from a finite basis model, where one state (labelled by 0) is coupled to a continuum of states (labelled by $j$), under conditions where the density of states is uniform and the coupling is the same to all states. This model can be written as

$$\begin{aligned} i\hbar \frac{d}{dt} c_0(t) &= E_0 c_0(t) + \sum_j V c_j(t) \\ i\hbar \frac{d}{dt} c_j(t) &= E_j c_j(t) + V c_0(t) \end{aligned} \tag{248}$$

where $c_j(t)$ ($j = 0, 1, 2, 3...$) is the probability amplitude for finding the system in basis state $j$ at some time $t$, $E_j$ ($j = 0, 1, 2, 3...$) is the energy of basis state $j$ in the absence of any coupling between the states and $V$ is the coupling between the special basis state with $j = 0$ and the other states.

The model can be solved (in the continuum limit) as an initial value problem (with $c_0(0) = 1$ and $c_j(0) = 0$) which results in exponential decay according to

$$c_0(t) = e^{-iE_0 t/\hbar} e^{-\gamma t/2} \tag{249}$$

with

$$\gamma = \frac{2\pi}{\hbar}|V|^2\rho \tag{250}$$

What is remarkable about this model is that the exponential decay begins immediately, which is a consequence of there being an infinite number of states arbitrarily far away in energy that contributes to the decay.

### A more realistic generalization of Dirac's model

It is possible to modify Dirac's model to account for variations in the coupling between the initial state and states in the continuum. In the case of radiative decay, the coupling matrix elements at long photon wavelength have a magnitude which is "large" compared to the "small" matrix elements expected when the photon wavelength is much smaller than the nucleus. We can modify Dirac's model to take this effect into account by writing

$$\begin{aligned} i\hbar\frac{d}{dt}c_0(t) &= E_0c_0(t) + \sum_j V_jc_j(t) \\ i\hbar\frac{d}{dt}c_j(t) &= E_jc_j(t) + V_jc_0(t) \end{aligned} \tag{251}$$

In this more realistic version of the model there are deviations from exponential decay, both at early times where the dynamics deviate from exponential decay, and at late times where the decay is algebraic (due to the absence of photon states with negative energy). At early times there is a delay before exponential decay sets in, since the relatively "strong" coupling near resonance does not extend far off of resonance.

### Decay in the case of transient occupation

Suppose that we are interested in the decay of a state which occurs for only a short time, and where the dynamics can be modeled approximately by

$$c_0(t) \approx e^{-iE_0t/\hbar}e^{-t^2/2\tau^2} \tag{252}$$

The idea here is that the state is occupied only temporarily due to the fast dynamics associated with phonon and plasmon exchange. We can develop an estimate for the continuum state occupation from perturbation theory according to

$$c_j(t) \approx \frac{V_j}{i\hbar}\int_{-\infty}^{t} e^{-iE_j(t-t')/\hbar}e^{-iE_0t'/\hbar}e^{-(t')^2/2\tau^2}dt' \tag{253}$$

We can use this to estimate the total probability in the continuum states according to

$$\sum_j |c_j(\infty)|^2 \approx \sum_j \frac{|V_j|^2}{\hbar^2}2\pi\tau^2e^{-(E_j-E_0)^2\tau^2/\hbar^2} \tag{254}$$

In the continuum limit this becomes

$$\sum_j |c_j(\infty)|^2 \rightarrow \frac{2\pi\tau^2}{\hbar^2}\int_{-\infty}^{\infty} V^2(\epsilon)e^{-\epsilon^2\tau^2/\hbar^2}\rho(\epsilon)d\epsilon \tag{255}$$

We can associate an effective decay rate in this case according to

$$\gamma \ = \ \frac{1}{\sqrt{\pi}\tau} \sum_j |c_j(\infty)|^2 \ = \ \frac{2\sqrt{\pi}\tau}{\hbar^2} \int_{-\infty}^{\infty} V^2(\epsilon) e^{-\epsilon^2\tau^2/\hbar^2} \rho(\epsilon) d\epsilon \tag{256}$$

We can perform a sense check on the effective decay rate solution above by reverting to a uniform density of states and constant coupling, i.e.

$$V(\epsilon) \ \rightarrow \ V \qquad \rho(\epsilon) \ \rightarrow \ \rho \tag{257}$$

In this case, the effective decay rate reduces to Eq. 250:

$$\gamma \ \rightarrow \ \frac{2\sqrt{\pi}\tau V^2 \rho}{\hbar^2} \int_{-\infty}^{\infty} e^{-\epsilon^2\tau^2/\hbar^2} d\epsilon \ = \ \frac{2\pi}{\hbar} |V|^2 \rho \tag{258}$$

**Stabilization of excited nuclear states**

The argument developed here is that this effect has the potential to arrest decay and therefore stabilize excited Pd$^*$ states under conditions where phonon and plasmon exchange is very fast. If $\tau$ is very small, then the decay rate approaches

$$\gamma \ \rightarrow \ \frac{2\sqrt{\pi}\tau}{\hbar^2} \int_{-\infty}^{\infty} V^2(\epsilon) \rho(\epsilon) d\epsilon \tag{259}$$

which becomes slower the faster the phonon and plasmon exchange is. How much stabilization occurs depends on the details of the excited nuclear state and the decay mechanism.

**Relation of this effect to the quantum Zeno effect**

The well-known quantum Zeno effect involves the disruption of spontaneous decay at early time, prior to when exponential decay sets in [98]. In the discussion here, the hindrance of spontaneous decay is due to fast coherent transitions in and out of an unstable state. We consider this effect to be closely related to the quantum Zeno effect, even though the underlying mechanism is different.

## 5.14 Pd* excited state decay

Earlier sections of this document described excitation transfer dynamics that can result in the production of a larger number of nuclear excited states (in our case, Pd* excited states).

In the previous section we presented an argument on how these kinds of Pd* excited states can be stabilized through rapid energy exchange with lattice oscillatory modes such as phonons. This stabilization relaxes the constraints on the nuclear excitation transfer model and widens the scope of its application.

The decay process can, however, not be arrested indefinitely. And indeed it is the decay process that ultimately determines how the energy that got released from the fusion transition results in observable reaction products. This section is dedicated to this issue.

#### Conventional gamma, neutron and charged particle emission from unstable Pd* states

When looking at conventional decay channels of excited Pd* bound states (as discussed in section 5.4), we can expect some low-level gamma emission in the 4-9 MeV part of the spectrum, alpha emission from states above the alpha ionization threshold, proton emission from states above the proton ionization threshold, and neutrons from states above the neutron ionization threshold (see section 6.17 for these thresholds).

In the case of a highly excited acoustic phonon mode, we expect rapid energy exchange between the nuclear states and the lattice (per the discussion in section 5.11) and a partial suppression of these decay channels due to the short occupation times of individual states (per section 5.13). Nonetheless, we expect such suppression not to be complete, which implies the prediction of some low-level nuclear particle emission from the conventional decay of excited Pd* per the previous paragraph.

#### Emission of small quanta from Pd* excited states to lattice modes

We noted in the previous section that we would conventionally expect the excited (M2) Pd* states to radiatively decay via magnetic quadrupole coupling with the electromagnetic field. The associated lifetimes are short (see section 5.4), which places severe limits on excitation transfer from the fusion transition to such receiver states. However, if the radiative decay of these states is slowed down per the arguments presented in the previous section, then this leaves not only more time for excitation transfer dynamics; it also implies that alternative pathways may be faster than the (suppressed) conventional loss channels.

Of particular interest is the conjectured transfer of excitation from nuclear states to oscillatory modes of the lattice (primarily optical phonons and plasmons). Relevant quanta here are in the meV and eV range, which is substantially smaller than the MeV level energies in the nuclear states. However, as we have seen in section 5.7, the density of nuclear states can be on the order of hundreds of states per eV and higher. Consequently, it is conceivable that energy from nuclear states gets emitted into lattice modes, which would macroscopically manifest as heat, as excitation transfers between nuclear states.

Ideally we would like to evaluate the rate for excitation transfer from the fusion transition in connection with excess heat. Such an approach goes beyond the scope of this paper, but we will provide a brief outlook here.

It is possible to consider a model for excess heat production based on Eq. 621 which we reproduce below:

$$\Gamma_{transfer} = \sum_{\{M'_j\}} \sum_{n'_A} \sum_{n'_O} \sum_{n'_P} \left| \left\langle |S, M-1\rangle_{fus} \left( \prod_j |S_j, M'_j\rangle \right)_{Pd} \left| \hat{T}'_{fus,Pd}(\tau) \right| |S, M\rangle_{fus} \left( \prod_j |S_j, M_j\rangle \right)_{Pd} \right\rangle \right|^2$$

$$\left| \left\langle n'_A, n'_O, n'_P \left| e^{-i\hat{H}_{free}\tau/\hbar} \right| n_A, n_O, n_P \right\rangle_{osc} \right|^2$$

$$\delta\left( \Delta Mc^2 - \sum_j \epsilon_j (M'_j - M_j) - (n'_A - n_A)\hbar\omega_A - (n'_O - n_O)\hbar\omega_O - (n'_P - n_P)\hbar\omega_P \right) \quad (260)$$

where a version of $\hat{H}_{free}$ in this expression would include all of the oscillators (instead of just a contribution from the acoustic phonons, as focused on earlier). The idea here is that after start-up there may be some excited Pd* nuclei present, and there may also be significant optical phonon excitation and/or plasmon excitation. Under these conditions there may be significant occupation of final states, where the fusion energy has gone nearly completely into the oscillator degrees of freedom. We would expect excitation transfer rates to be similar to what was estimated above under conditions of start-up, if not faster due to the accumulation of cooperative enhancement (Dicke) factors.

We do not yet have quantitative predictions from this model, but it provides a formal estimate for the rate of excitation transfer from the fusion transition in connection with excess heat production. The development of detailed models to describe this process is a major goal of our future research.

Note that even if the proposed process to convert nuclear energy into lattice modes is dominant, we would not expect that it outcompetes radiative decay in every instance. The implication is that we would still expect some nuclear particle emission per the previous subsection under many kinds of experimental conditions.

### Decay of Pd* nuclear molecule cluster states

The model under discussion also suggests the existence of an operating regime, where excitation transfers between nuclear states is at such a fast rate that there is not enough time neither for radiative decay nor for the exchange of quanta with lattice modes. In such a scenario, excitation may transfer rapidly between many nuclear states and eventually accumulate in naturally long-lived nuclear molecule cluster states (see section 5.4). Suppose that an excitation transfer receiver state has one unit of angular momentum, and radiatively decays quickly to a non-rotational state that is a long-lived nuclear molecule state. Such a state is essentially a dead end in connection with the nuclear dynamics considered here, since there is no $\mathbf{a} \cdot c\mathbf{P}$ interaction. Such states would remain occupied until ultimately undergoing tunnel decay, leading to (asymmetric and symmetric) fission products of the excited nuclei.

Fast acoustic phonon exchange is associated with large transition strengths (or large $O$-values), particularly in bound Pd*. However, transitions to highly-excited nuclear molecule cluster states are expected to have much smaller transition strengths, meaning that their spontaneous decay channels would not be suppressed.

If the excitation transfer from the fusion transition is slow, resulting in the receipt of single fusion transition quanta on the side of the receiver nuclei, then the decay of these states would be dominated by the tunneling of low-mass daughters (protons, alphas, neutrons). This mechanism could account for the low-level alpha emission at high energy reported in [99]. Conversely, if the excitation transfer from the fusion transition is faster, resulting in the receipt of multiple fusion transition quanta, then the system would be expected to reach more symmetric nuclear molecule cluster states at higher energies, with even lower transition strengths. Some of these transitions may involve one or more units of angular momentum and could radiatively decay to states without $\mathbf{a} \cdot c\mathbf{P}$ coupling. These non-rotating states, being long-lived, can undergo tunnel decay, potentially explaining the fission-type transmutation effects observed in certain experiments.

## 5.15 Other reaction products

The previous section discussed different types of reaction products that we expect to result from excitation transfer from $D_2$ donors to Pd receivers with a focus on reaction products resulting from the decay of Pd excited states caused by the transfer. While we expect these products to be the dominant ones in most operating regimes, they are not the only ones expected. This section discusses other reaction products not directly associated with Pd excited state decay.

Specifically, we expect as further reaction products a low level of 3+1 fusion products from the loss channels discussed in sections 5.2 (and in more detail in sections 6.5 and 6.6). These sections showed how loss channels—all associated with the highly unstable 3+1 state of $^4$He—play a key role in breaking destructive interference (with substantially larger transfer rates as a result of this). What these mentioned sections on loss did not emphasize were the reaction products that can result from such loss channels. This is the subject of the present section.

### Model to estimate 3+1 fusion reaction products

We can create a simple model to estimate the amount of 3+1 fusion reaction products by considering the excitation transfer event as involving an extended period where no 3+1 fusion occurs; a short period in which the 3+1 state is occupied, and during which 3+1 fusion is possible; and an extended period during which the $^4$He state is occupied and no 3+1 fusion occurs.

Based on this, we can develop an estimate for the 3+1 fusion rate based on

$$\Gamma_{3+1} = \frac{1}{2}\Gamma_{transfer} f_{3+1} P_{tunnel} \tag{261}$$

where $\Gamma_{transfer}$ is the excitation transfer rate derived in section 5.11, $f_{3+1}$ is the fraction of time that the 3+1 state is occupied for and $P_{tunnel}$ is the probability that 3+1 fusion occurs while the 3+1 state is occupied.

The time associated with an excitation transfer event is

$$\tau_{transfer} = \frac{1}{\Gamma_{transfer}} \tag{262}$$

For the time associated with 3+1 occupation we take the

$$\tau_{3+1} = \frac{\hbar}{\sqrt{|\mathcal{V}|^2 + \hbar^2\gamma_{tunnel}^2}} \tag{263}$$

where $\mathcal{V}$ is the Dicke enhanced version of $V$ from section 6.5

$$\mathcal{V} = \langle (3+1)|\mathbf{a}\cdot c\mathbf{P}|^4\text{He}\rangle \sqrt{S_a(S_a+1)-(M_a-1)M_a} \tag{264}$$

and $\hbar\gamma_{tunnel} = 5.95$ MeV is the line with associated with the 3+1 tunnel decay (see section 6.2).

This leads to

$$\begin{aligned} f_{3+1} &= \frac{\tau_{3+1}}{\tau_{transfer}} \\ &= \frac{\hbar\Gamma_{transfer}}{\sqrt{|\mathcal{V}|^2 + \hbar^2\gamma_{tunnel}^2}} \end{aligned} \tag{265}$$

The probability that 3+1 fusion occurs while the 3+1 state is occupied is

$$P_{tunnel} = \frac{\hbar\gamma_{tunnel}}{\sqrt{|\mathcal{V}|^2 + \hbar^2\gamma_{tunnel}^2}} \tag{266}$$

Putting this all together gives us a 3+1 fusion rate of

$$\begin{aligned} \Gamma_{3+1} &= \frac{1}{2}\Gamma_{transfer}\frac{\hbar\Gamma_{transfer}}{\sqrt{|\mathcal{V}|^2 + \hbar^2\gamma_{tunnel}^2}}\frac{\hbar\gamma_{tunnel}}{\sqrt{|\mathcal{V}|^2 + \hbar^2\gamma_{tunnel}^2}} \\ &= \frac{1}{2}\Gamma_{transfer}\frac{(\hbar\Gamma_{transfer})(\hbar\gamma_{tunnel})}{|\mathcal{V}|^2 + \hbar^2\gamma_{tunnel}^2} \end{aligned} \tag{267}$$

We see from this that the relative yield of 3+1 fusion reactions to excitation transfer transitions is

$$\frac{\Gamma_{3+1}}{\Gamma_{transfer}} = \frac{1}{2}\frac{(\hbar\Gamma_{transfer})(\hbar\gamma_{tunnel})}{|\mathcal{V}|^2 + \hbar^2\gamma_{tunnel}^2} \tag{268}$$

Our evaluation suggests that this is much less than unity. In other words, the expected rate of 3+1 fusion products via loss channels would be low. This is consistent with experimental reports such as by Jones and coworkers in 1989 [19] who reported low levels of neutron emission from a PdD experiment.

**Branching ratios**

The ratio in Eq. 268 effectively represents a kind of branching ratio between 3+1 fusion on the one hand and fusion from $D_2$ donors undergoing excitation transfer on the other hand, with only a small portion of the energy released going into 3+1 fusion products.

The absence of commensurable neutron products has been a primary concern surrounding claims of fusion in solid-state metal hydrides. The proposed model provides some insight as to why the observed branching ratio could be so different.

There is another branching ratio, namely the typical branching ratio of the two 3+1 fusion channels, which can be described as

$$R = \frac{\Gamma(n + ^3\text{He})}{\Gamma(p + t)} \tag{269}$$

From [100] we see that the $^4$He$^*$ d+d $T = 0$ $^5$S $J^\pi = 2^+$ state decays with an n/p branching ratio near 0.92, and the $^4$He$^*$ d+d $T = 0$ $^3$P $J^\pi = 1^-$ decays with an n/p branching ratio near 1.14. This indicates that the n/p branching ratio is channel dependent, and that measurements of the neutron-to-proton (n/p) ratio has the potential to provide information about what channels or states are involved.

This is particularly striking in muon-catalyzed fusion, where the R-matrix calculations of Hale [101] the expected ratio for S ($L = 0$) channel in muon-catalyzed dd fusion is 0.886, and for the P ($L = 1$) channel is 1.43. Experiments with muon-catalyzed dd fusion show a ratio for the $L = 1$ channel close to this predicted value [102].

In ion beam experiments the n/p ratio for d+d fusion reactions is near 1.0 [103], [104]. If reported low-level neutron emission in LENR experiments were caused by excessive levels of screening, this would be due to S-channel contribution, and would come with a n/p branching ratio near 1.0 as observed in the ion beam experiments.

Recall (section 5.4 and 6.8) that the kind of relativistic coupling discussed in this document involves transitions to $^3$P J=1 states, which would result in low-level dd-fusion when undergoing 3+1 decay. The n/p branching ratio for the $^4$He$^*$ (3+1) $T = 0$ $^3$P $J^\pi = 1^-$ state is near 0.93. This suggests that it might be possible to distinguish between an explanation for reported LENR effects based on an enhanced screening mechanism (which would lead to an n/p ratio near 1.0), and one based on nuclear excited transfer as discussed in this document (which would lead to an n/p ratio near 0.93) [105].

**Discussion**

In a study of nuclear-level excess heat from metal-hydrogen samples, in which neutron emission was monitored at the same time that excess heat was measured, there is little evidence of any low-level neutron emission correlated with excess heat production. An upper limit on the order of one neutron per 100 J of excess energy produced was estimated based on experimental reports [106]. In other words, experimental reports are consistent with the prediction of a very low branching ratio of 3+1 fusion to excitation transfer (Eq. 268)—with the latter including the possibility of excess heat production (per 5.15).

The view that we adopted over the years is one which suggests that excess heat production is associated with a "clean" reaction process that produces essentially no energetic nuclear particles when running efficiently. In terms of the models under discussion here, we consider this to correspond to conditions conducive to fast excitation transfer: a large ratio of $N_{D_2}/N$, large acoustic phonon mode excitation, large optical phonon mode excitation, and a large coherence domain; with excitation transfer rates from the fusion transition to Pd$^*$/Pd transitions, stabilization of corresponding Pd$^*$ states due to rapid acoustic phonon exchange, and fast exchange of energy from Pd$^*$ states to optical phonons.

Observable levels of 3+1 fusion (and potentially products from other loss processes) would be associated with a picture, in which the (primary) excess heat process is frustrated, *e.g.*, due to a lack of $D_2$ molecules present or insufficient acoustic phonon excitation.

## 5.16 Outlook

The notion that nuclear fusion and transmutation reactions might be modified when taking place in a metallic lattice goes back to the 1920s [3, 107] and received much attention again in 1989, when Fleischmann and Pons described an excess heat effect that they claimed to have observed in experiments with deuterium electrochemically loaded into Pd cathodes [10]. Fleischmann and Pons speculated that deuterons were fusing in some novel way to release large amounts of energy (as measured calorimetrically) and $^4$He (which had been reported from later experiments of the same kind [108, 109]). This was a truly remarkable claim, which, if true, would imply that nuclear fusion energy could be released near room temperature, in a small-scale, tabletop setup.

At the time, it was recognized by essentially all observers that the implications of such a hypothesized process would be enormous. This significance is reflected in a recent statement by the US Department of Energy's innovation agency ARPA-E that was part of a program announcement on low-energy nuclear reactions (LENR): "Based on its claimed characteristics, LENR may be an ideal form of nuclear energy with potentially low capital cost, high specific power and energy, and little-to-no radioactive byproducts. If LENR can be irrefutably demonstrated and scaled, it could potentially become a disruptive technology with myriad energy, defense, transportation, and space applications"

Note that ARPA-E refers to "low-energy nuclear reaction" to describe a hypothesized class of novel nuclear reactions associated with reported anomalies from metal-hydrogen samples such as those listed in section 2.3. In this document, we have referred to this hypothesized class of nuclear reactions as solid-state fusion reactions. ARPA-E's choice of words is rooted in the desire to be agnostic as to the cause of the reported anomalies and to include in the nomenclature the fact that *low-energy stimuli* are used in corresponding experiments.

A major contributing factor to the lack of research activity following the claims by Fleischmann and Pons in 1989 was the belief that the proposed effect was physically impossible. The basics of deuteron-deuteron fusion had been believed to be understood for many years, and the process was considered conceptually and theoretically straightforward. Nothing in the relevant theoretical or experimental scientific literature lent support to the possibility that Fleischmann and Pons might have observed a real effect.

In the many pages above and below, we describe a model that we suggest can lead to a sufficient understanding of the basic effect underlying the claims by Fleischmann and Pons as well as the subsequent (and some of the preceding) claims of many experimentalists in the LENR field.

In the physics community more generally, previous theoretical and experimental work on nuclear fusion has focused on it as a quantum mechanically spontaneous process, involving two, or in some cases three, nuclei at a time. In this frame, fusion (deuterium fusion and fusion with other reactants) is indeed conceptually simple: in essence, two nuclei collide with sufficient energy to tunnel through the Coulomb barrier. When two deuterons get close, the constituent nucleons can rearrange to a lower energy 3+1 configuration, and the fusion products leave at high relative energy as a result of the mass energy released. In particular, Koonin and Nauenberg's 1989 letter to Nature [20] and Leggett and Baym's 1989 article in Phys. Rev. Lett. [110] estimated upper bounds on electron-screened fusion reaction rates many orders of magnitude too slow to be observable. And in the case of the latter, the possibility of "an exotic mechanism relying on coherence between fusion processes involving different deuteron pairs" was deemed "extraordinarily implausible." Under discussion in this document is the possibility of fusion as an induced process, for which the underlying physical processes are very different.

We suggest that the ideas and models laid out in this document can contribute to the formation of a new field at the intersection of nuclear engineering and solid state physics. Towards that end, and towards the development of corresponding technology, a number of issues need attention in future research. These are briefly laid out below.

### Better characterization of LENR experiments

While there exists a substantial number of experimental reports that claim the observation of energetic particle emission, purported fission products and excess heat production consistent with the models presented here (see section 2.3 for a selection), more experimental work is needed to better understand what kinds of material systems can yield what kinds of results under what kinds of circumstances. In 2023, ARPA-E [72] launched a research program to improve the quality of LENR experiments and to develop and bring to bear best practices for materials characterization, elemental and isotopic analysis, gas analysis, and nuclear particle detection.

Such experimental efforts could also involve the development of new diagnostics for LENR effects, informed by the ideas in this document. For instance, we predict that under some circumstances nuclear energy can get emitted into lattice modes. Differences in the lattice strain resulting from large amplitude vibrations can in turn result in changes in the electron distribution at the surface due to the strain contribution to the electron energy. The expected result would be radio frequency (RF) generation, for which dedicated diagnostics could be developed and deployed in LENR experiments.

**Better understanding and control of sample preparation and stimulation**

There is also a need to better understand and achieve more control over the nanostructures and compositions of samples as well as samples' dynamic response to different kinds of stimuli. The former includes such aspects as the formation of vacancies and vacancy-hydrogen clusters in different materials and the preparation of isotopically pure and isotopically doped samples. The latter includes predicting and studying the effects of different kinds of stimulation mechanisms on samples such as laser pulses, electric pulses, acoustic/ultrasound stimulation, mechanical shocks, phase changes, etc. This pertains particularly to the phonon and plasmon modes that get excited as a result.

Specifically, techniques are needed to exert great control over the excitation of specific phonon and plasmon modes in samples that are of particular interest (as predicted by matching simulations). A better understanding is also needed regarding different roles that acoustic phonon modes, optical phonon modes, plasmon modes (and potentially other oscillatory modes such as magnon modes) play. We expect acoustic phonon modes to be more dominant than optical phonon modes (weaker coupling due to proportionality of the relativistic coupling to the square root of the energy 5.3) in mediating nuclear excitation transfer. We do expect, however, for optical phonons to play a significant role when it comes to energy exchange, and the same is the case for plasmons.

**Development of integrated and comprehensive codes linking sample and stimulation characteristics to LENR outcomes**

An ultimate goal of our effort is the deliberate design of LENR experiments (and later LENR technology) through rational design. This requires the development of codes that describe all relevant features of the system, from the nuclear states of the nuclei in the sample, to the structure and dynamics of the lattice, to the complex quantum dynamical processes that result from the above. We have already embarked on this kind of project, as is reflected in this document, but much more work is needed. This task may involve the creation or adaptation of different sets of code and data that then need to be integrated through well-defined interfaces.

**Towards more comprehensive nuclear data sets**

A critical input for the above-mentioned codes are nuclear data sets that include as comprehensively as possible the existing nuclear states for the types of nuclei to be used in the sample. Many such nuclear states are recorded in databases like NuDat. However, as we have discussed in section 5.4, many states are missing in existing databases and need to be identified through future experimental efforts or through simulations.

Specific focus should be on (bound) excited states that have M2 coupling with the ground state (associated energy levels and lifetimes as well as the radiative decay rates or M2 transition strengths). We expect large-scale nuclear shell model calculations to be highly effective for this.

#### Identification and characterization of nuclear molecule cluster states of mid-sized nuclei

Similarly, nuclear molecule cluster states play an important role in the dynamics described in this document (see 5.4 and 5.14). Numerous such states exist for different nuclei, but very few have been rigorously identified and characterized. Especially important are nuclei that form the lattices of samples or that are present in impurities or dopants.

Here, too, we care about energy levels, lifetimes, and estimates for the **a**-matrix elements for transitions from the ground state with M2 multipolarity.

#### Dedicated basic science experiments to further investigate individual parts of the presented argument

The argument presented in this document is complex and, while resting entirely on known physics, involves predicted effects that may be unexpected and perhaps surprising to many observers. This includes, for instance, the relativistic coupling between lattice modes and M2 nuclear states as a consequence of Lorentz invariance (6.7). While the predicted effects can be studied through LENR experiments, alternative platforms can be devised that allow for the dedicated investigation of such effects in isolation. This also applies to other effects such as breaking destructive interference through loss and nuclear supertransfer.

#### Consideration of alternative materials with different donor systems and receiver systems

Finally, as has been hinted at above, the ideas presented here can readily be extended and transferred to other materials systems beyond PdD, on which the focus has been laid in this document. On the donor side, other fusion transitions can be considered as a source of nuclear binding energy that drives excitation transfer dynamics. We direct particular attention toward the HD to $^3$He transition with its 5.49 MeV transition energy. Section 6.30 contains a dedicated discussion of that donor system and section 6.10 contains a preliminary calculation of the HD to $^3$He matrix element for the relativistic coupling (the coupling that we worked with in the majority of the document, see 5.3).

On the receiver side, many nuclei—and combinations thereof—can be considered. Critical factors are the proximity of the donor system nuclei that are enabled by the lattice (*e.g.*, deuterium molecule formation in vacancy-hydrogen clusters in Pd and equivalents in other materials), the amount of screening provided, the occupied oscillatory modes under relevant types of stimulation, and the nuclear states that determine resonances and transfer channels as well as transfer rates, stabilization and decay dynamics. Some reports from past LENR experiments suggest that materials such as nickel and titanium may represent viable alternatives to palladium.

#### Final discussion

The emergence of the LENR field exhibits some remarkable parallels to the emergence of the field of semiconductor engineering and modern electronics. Amplification effects in early semiconductor materials were reported as early as the 1920s [111], but it was not until the 1948 Physical Review Letter [112] on transistor action that experiments could be staged that yielded the sought amplification effects with some degree of reliability and with some degree of understanding. And still, many more questions needed to be answered after 1948, for instance with respect to the scalability, fabrication, economics of transistors and other electronic components.

This document seeks to make a contribution to the LENR field, helping it to move beyond sporadic reports of anomalies from metal-hydrogen samples towards the rational design of LENR experiments (and, later, LENR technology). In the case of the transistor, fundamental insights related to the control of electronic quantum states led not only to solid-state amplification devices but to the entire new field of electronics. Similarly, fundamental insights related to the control of nuclear quantum states may lead not only to LENR devices but to the new field of *nucleonics*.

The hope (and our expectation) is that this new field can follow the trajectory of the semiconductor field from the late 1940s onward and eventually produce devices that enhance and revolutionize—in this case in the field of energy instead of information—the lives of people across the planet.

# 6 Auxiliary sections for detailed quantum dynamics calculations

## 6.1 Relevant $D_2$ and $^4He$ nuclear states and their wave functions

### $D_2$ molecular wave functions

Just as electrons can occupy various configurations within an atom or molecule, nucleons (protons and neutrons) can also arrange themselves in different configurations within an individual nucleus and across nuclei that make up a molecule.

A $D_2$ molecule can be viewed from different perspectives, each of which can be justified as a starting point for model development. From a bottom-up perspective, we are looking at a system of four nucleons, where each nucleon has one close (< 5 fm) neighbor and two far (∼74 pm) neighbors. A more modular perspective recognizes pairs of protons and neutrons as deuterons, considers each deuteron by itself first, and then integrates the two deuterons into a $D_2$ molecule. A top-down perspective views the $D_2$ molecule as a single nuclear system that is highly clustered.

When one avoids simplifications, each perspective can result in the same mathematical description: a single wave function that describes the possible nuclear configurations across the expanse of a molecule such as $D_2$. Such a wave function must include the following information:

- Spatial (R) - How far apart are the nucleons that make up an individual nucleus and the nuclei that make up the molecule?
- Spin angular momentum (quantum number S) - What is the total spin of all the nuclei in the molecule?
- Orbital angular momentum (quantum number L) - How much are the nuclei rotating around each other?
- Isospin (quantum number T) - The total isospin of the nucleons, which helps describe how nucleons are differentiated (as protons or neutrons) and how these differences affect the nuclear interactions within the molecule.

In addition, total angular momentum $J$ can be important when spin-orbit (LS) coupling cannot be neglected—as is the case when matrix elements are calculated for coupling-mediated nuclear state transitions (see section 5.3 and subsequent sections).

Angular momentum plays a particularly critical role because it not only influences the potentials present in the Schrödinger equation, which shape the spatial wave function, but its values forbid certain transitions through specific selection rules (see section 6.8) that depend on the details of the interactions that mediate the transitions. Certain compact notations have therefore been developed to make it easier to navigate the angular momentum space. In particular "Term Symbols" represent the states as $^{2S+1}L_J$ where L is the total orbital quantum number in spectroscopic notation and $2S + 1$ indicates the total number of spin states which are enumerated by the spin magnetic quantum number $M_S = -S, -S + 1, ..., S - 1, S$. People therefore often refer to $S = 0, 1, 2$ as the singlet, triplet, quintet states. Sometimes the $J$ is not included in the notation if LS coupling is not relevant in a particular use case.

| $L$ | 0 | 1 | 2 | 3 | 4 | 5 | 6 | 7 | 8 | 9 | 10 | 11 | 12 | 13 | 14 | 15 | 16 |
|---|---|---|---|---|---|---|---|---|---|---|---|---|---|---|---|---|---|
| Symbol | S | P | D | F | G | H | I | K | L | M | N | O | Q | R | T | U | V |

**Table 1:** Correspondence between Orbital Angular Momentum Quantum Number ($L$) and Spectroscopic Symbols

For the specific case of a $D_2$ molecule, there are certain constraints based on our knowledge of an individual deuteron. Specifically, a deuteron has an isospin of $T = 0$ and a spin value of $S = +1$. A $D_2$ molecule therefore has isospin of $T = 0$ and spin $S = 0, 1, 2$, based on the rules of angular momentum addition.

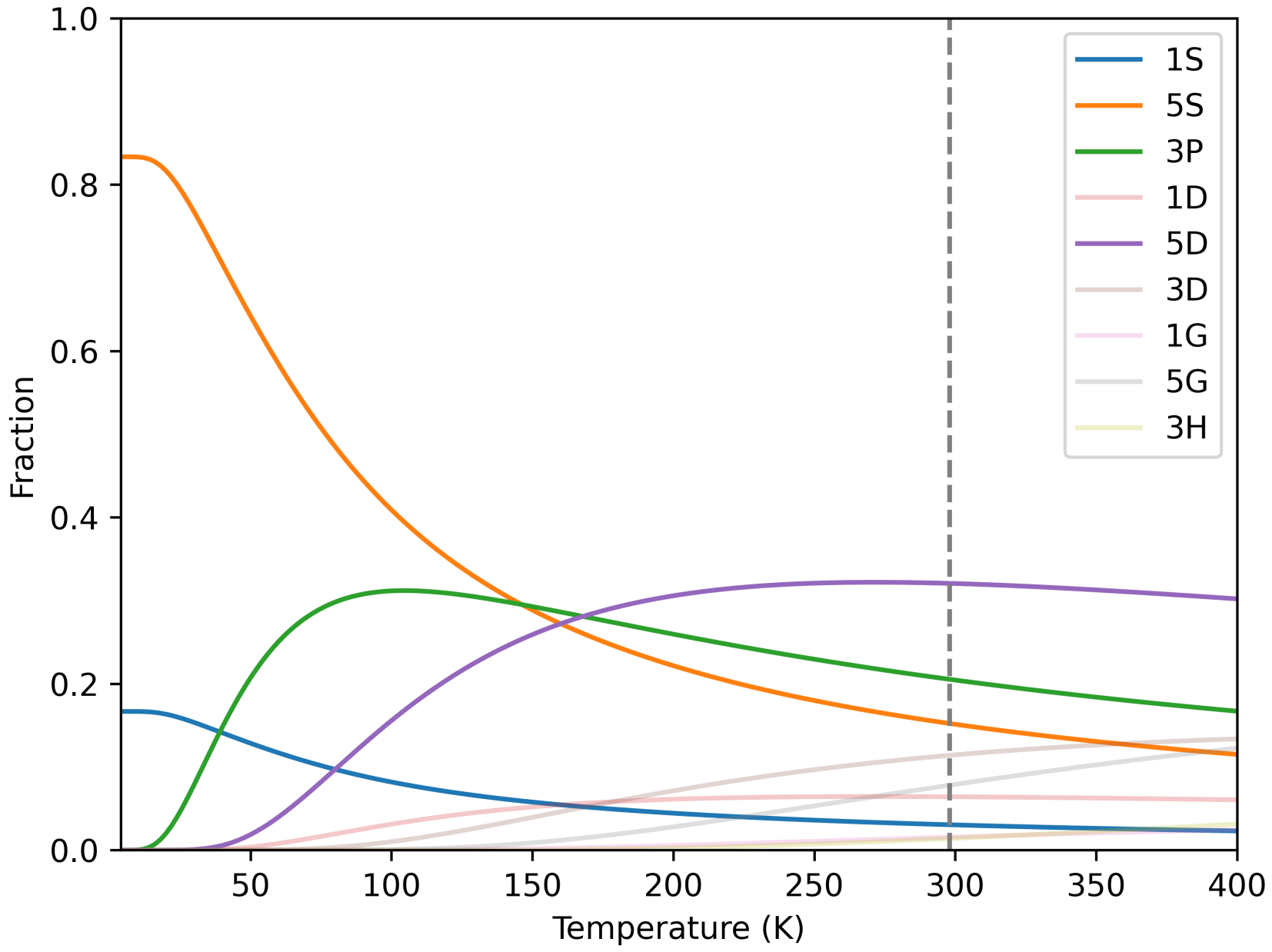

**Figure 38:** Temperature dependence of the distribution of angular momentum states for a $D_2$ molecule. Blue (1S), orange (5S), green (3P) and purple (5D) lines are bolded because they are the most significant states for our purposes.

To get a sense as to which angular momentum states are important for a $D_2$ molecule in practice, Figure 38 shows the relative proportion of each state in thermodynamic equilibrium across different temperatures.

The approach to constructing the nuclear wave functions for molecular $D_2$ involves first constructing a suitable unsymmetrized wave function that has appropriate space, spin, and isospin components. An unsymmetrized wavefunction is a preliminary wavefunction that is constructed without considering the requirement for the wave function to be antisymmetric under the exchange of identical particles (in this case, nucleons).

Then we can make use of an antisymmetrization operator $\mathcal{A}$ to produce a fully antisymmetric wavefunction, *i.e.*, a wave function that satisfies the Pauli exclusion principle. In general, we may write this as

$$\Psi = \mathcal{A}\left\{ \Phi_R \Phi_S \Phi_T \right\} \tag{270}$$

where $\Phi_R$, $\Phi_S$, and $\Phi_T$ denote space, spin, and isospin components, respectively.

We will not describe the antisymmeterisation procedure here (details can be found in Hagelstein 2013 [113]) and instead focus on how to construct the different parts of the wave function, beginning with the spatial part.

### $D_2$ molecular spatial wave function

To describe the positions of the nucleons in the $D_2$ molecule, instead of using the nucleon coordinates $\mathbf{r}_1, \mathbf{r}_2, \mathbf{r}_3, \mathbf{r}_4$, we choose a convenient coordinate system that uses the internal relative coordinates between the two nucleons in each deuteron, as well as the separation between the two center-of-mass positions of the deuterons. This is represented mathematically as:

$$\begin{aligned} \mathbf{R} &= \frac{\mathbf{r}_1 + \mathbf{r}_2 + \mathbf{r}_3 + \mathbf{r}_4}{4} \\ \mathbf{r}_a = \mathbf{r}_2 - \mathbf{r}_1 \qquad \mathbf{r}_b &= \mathbf{r}_4 - \mathbf{r}_3 \\ \mathbf{r} &= \frac{\mathbf{r}_3 + \mathbf{r}_4}{2} - \frac{\mathbf{r}_1 + \mathbf{r}_2}{2} \end{aligned} \tag{271}$$

and visually in Figure 39

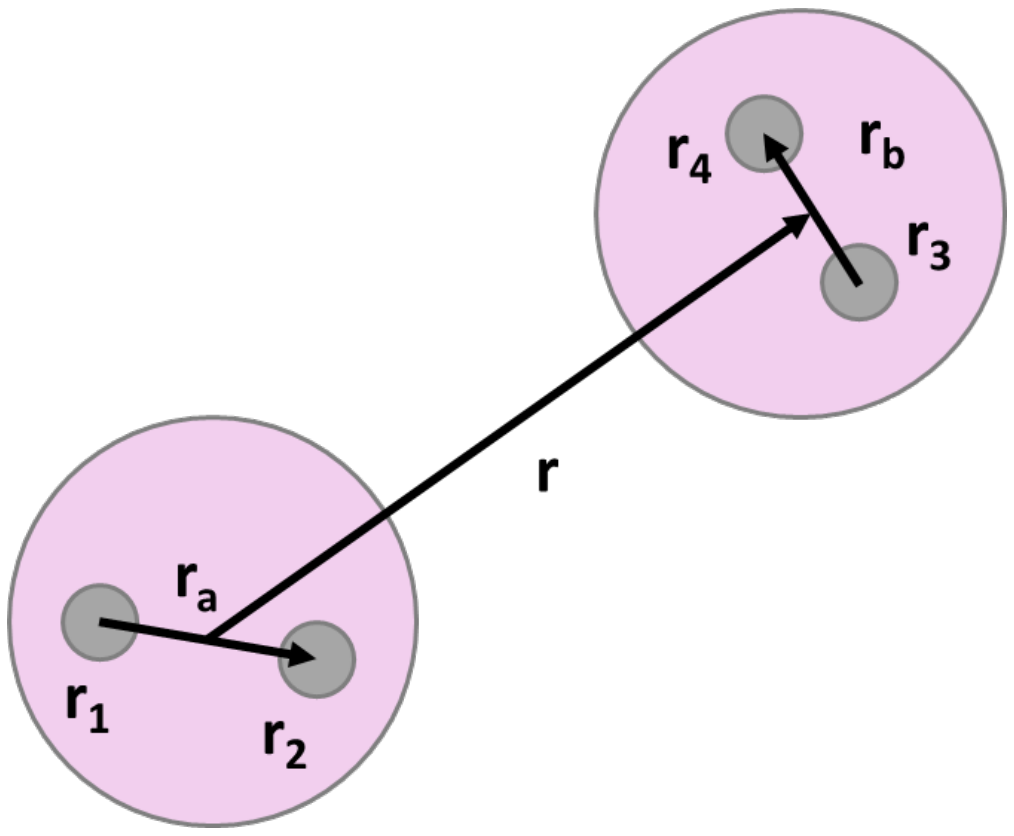

**Figure 39:** Internal coordinates for the $D_2$ molecule in the 12;34 permutation.

For the overall molecular $D_2$ spatial wave function we follow [113] and write

$$\Phi_R = \phi_d(|\mathbf{r}_a|)\phi_d(|\mathbf{r}_b|)R_{DD}(|\mathbf{r}|)Y_{LM} \tag{272}$$

where $\phi_d$ is deuteron relative wave function of each deuteron (assumed to carry no orbital angular momentum) and where $R_{DD}$ and $Y_{LM}$ are the radial wave function and relevant spherical harmonic associated with the deuteron-deuteron separation (**r**).

For the $D_2$/$^4$He **a**-matrix element calculation (6.9), we work with the $^3$S component of the deuteron relativ wave function $\phi_d$ from the chiral effective field theory NNNLO model of Epelbaum et al. (2005) [114] (see Figure 40).

Later in the text, we consider the HD/$^3$He transition as a donor system for nuclear excitation transfer and estimate the corresponding matrix element (6.10). In that calculation we used the following expression for the $^3$S component of the deuteron relative wave function [113]:

$$\phi_d(r) = \begin{cases} 0 & \text{for } r < r_0 \\ N_d \dfrac{\tanh[\gamma(r - r_0)]\, \mathrm{e}^{-\beta r}}{r} & \text{for } r_0 < r \end{cases} \tag{273}$$

For the radial wave function $R_{DD}$ we transform to $P_{DD} = rR_{DD}$ and solve the radial Schrödinger equation:

$$E\, P_{DD}(r) = \left( -\frac{\hbar^2}{2\mu}\frac{d^2}{dr^2} + \frac{\hbar^2 L(L+1)}{2\mu r^2} + V_{mol}(r) + V_N^{S,L}(r) \right) P_{DD}(r) \tag{274}$$

where $\mu$ is the reduced mass of the $D_2$ molecule, $\frac{\hbar^2 L(L+1)}{2\mu r^2}$ is the centripetal potential, $V_{mol}(r)$ is the molecular potential and $V_N^{S,L}(r)$ is the nuclear potential.

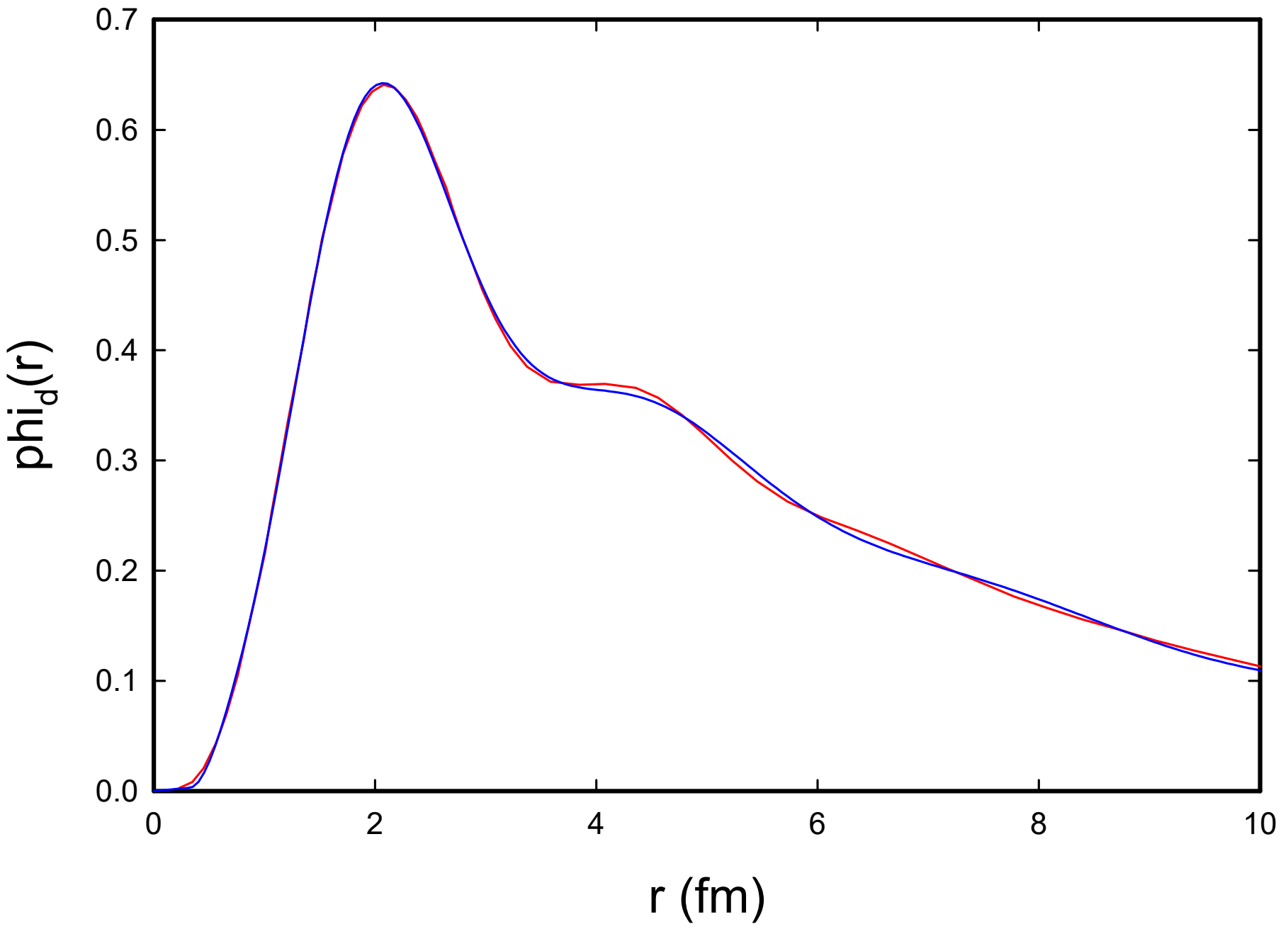

**Figure 40:** $^3$S component of the deuteron wavefunction based on the chiral effective field theory NNNLO interaction from Epelbaum et al. (2005) (red) [114]; fitted wave function used in the model for the $D_2$/$^4$He **a**-matrix element calculation (blue).

We use the nuclear potential (in MeV) from Weller (as described in [115]):

$$V_N^{S,L}(r) \ = \ \frac{V_0}{1 + e^{(r-r_S)/a_S}} \tag{275}$$

whose parameters depend on $S$ and $L$ as can be seen in table 2.

| State | $V_0$ (MeV) | $r_s$ (fm) | $a_s$ (fm) |
|---|---|---|---|
| $^1S$ | -74.0 | 1.70 | 0.90 |
| $^5S$ | -15.5 | 3.59 | 0.81 |
| $^3P$ | -13.5 | 5.04 | 0.79 |
| $^5D$ | -15.5 | 3.59 | 0.81 |

**Table 2:** Nuclear potential parameters

For a realistic molecular Coulomb potential, we draw on Kolos 1986 [116]. We parameterized the numerical Kolos potential as:

$$V_{mol}(r) = \frac{2}{r}(1 - b_1 r - b_2 r^2) e^{-\alpha r^s} \tag{276}$$

with $r$ in units of the Bohr radius ($a_0$), $V_{mol}$ is in Rydbergs and with

$$\begin{aligned} \alpha = 0.6255121237003474, \quad & b_1 = 1.4752572117720800, \\ b_2 = -0.2369829512108492, \quad & s = 1.0659864120418940 \end{aligned} \tag{277}$$

(where the fit only includes points out to $r = 4.8a_0$).

Putting this all together results in a combined potential $\frac{\hbar^2 L(L+1)}{2\mu r^2} + V_{mol}(r) + V_N^{S,L}(r)$ as seen in Figure 41. This enables a numerical solution to the radial Schrödinger equation using a 3 point scheme on a non-uniform grid based on a shooting method. For more details on this scheme, see `https://github.com/project-ida/nuclear-reactions`. Figure 42 shows the results for several values of L and S using a log scale to allow us to see the wave function across the molecular and nuclear length scales. Figure 43 and 44 shows the same results but focusing on the nuclear scale.

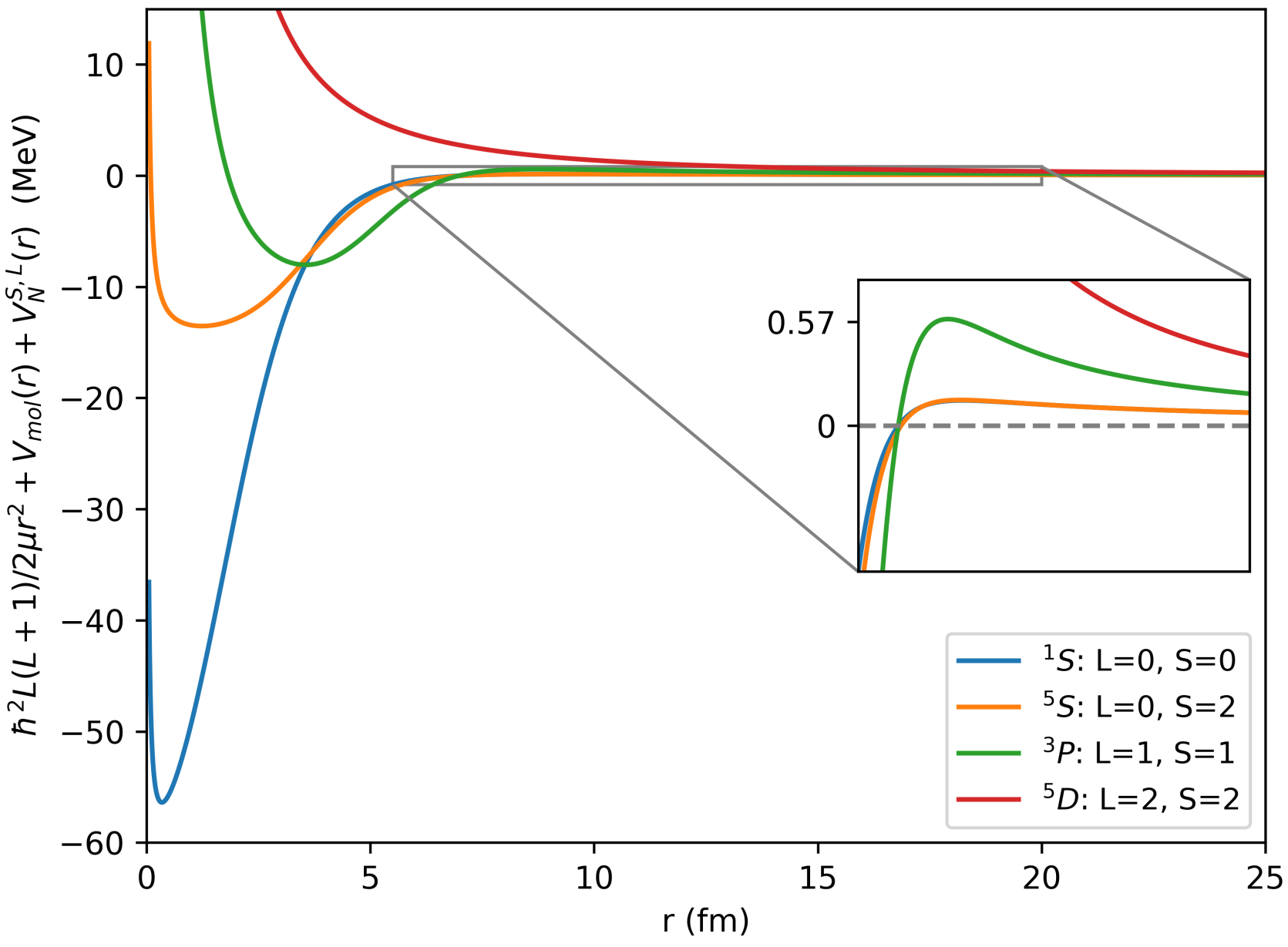

**Figure 41:** Total potential (centripetal + molecular + nuclear) that determines the radial wavefunction $R_{DD}$.

### Spin and isospin part of the $D_2$ molecular wave function

A deuteron is only stable with isospin $T = 0$ and spin $S = 1$. A $D_2$ molecule therefore has isospin $T = 0$ and 3 possible spin states of $S = 0, 1, 2$.

The isospin part of the $D_2$ wave function $\Phi_T$ is the simplest part to construct because it involves writing a singlet state

$$\begin{aligned} \Phi_T &= \left| T = 0, M_T = 0 \right\rangle_{12;34} \\ &= \left| T = 0, M_T = 0 \right\rangle_{12} \left| T = 0, M_T = 0 \right\rangle_{34} \\ &= \left( \frac{p_1 n_2 - n_1 p_2}{\sqrt{2}} \right) \left( \frac{p_3 n_4 - n_3 p_4}{\sqrt{2}} \right) \\ &= \frac{1}{2} \left( p_1 n_2 p_3 n_4 - p_1 n_2 n_3 p_4 - n_1 p_2 p_3 n_4 + n_1 p_2 n_3 p_4 \right) \end{aligned} \tag{278}$$

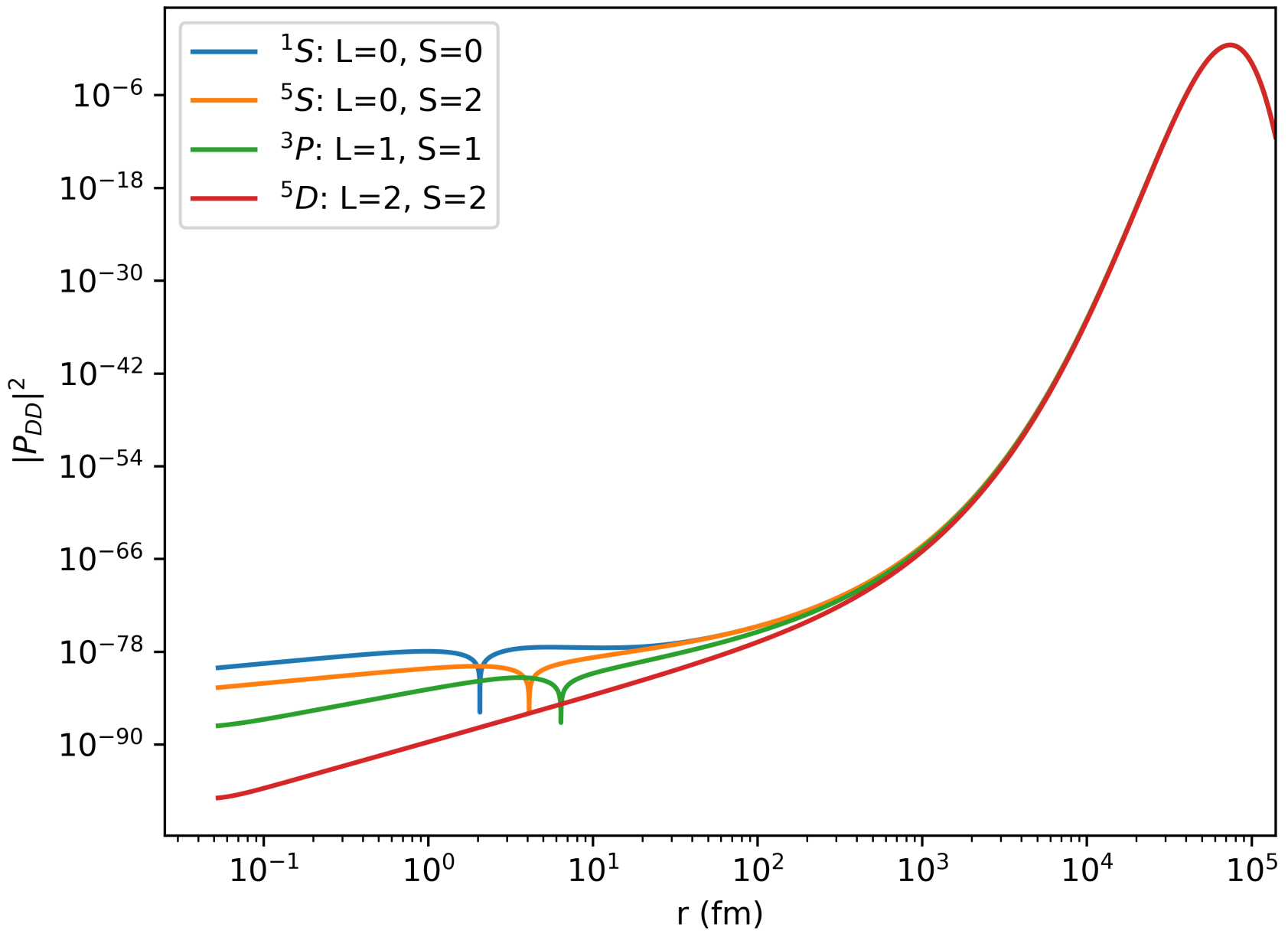

**Figure 42:** Normalised probability density associated with the solutions of the radial wave function equation 274.

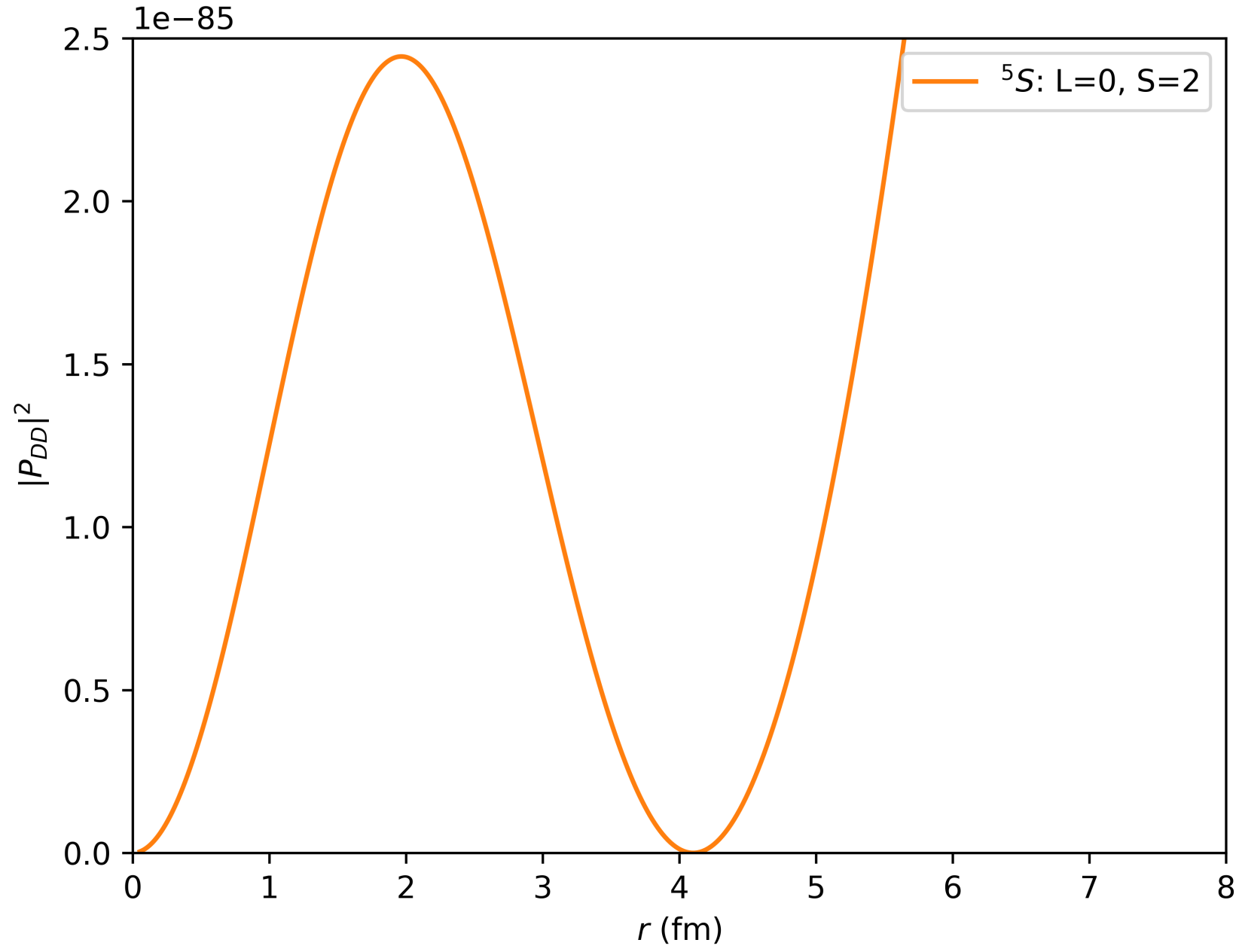

**Figure 43:** Normalised probability density associated with the solutions of the radial wave function equation 274.

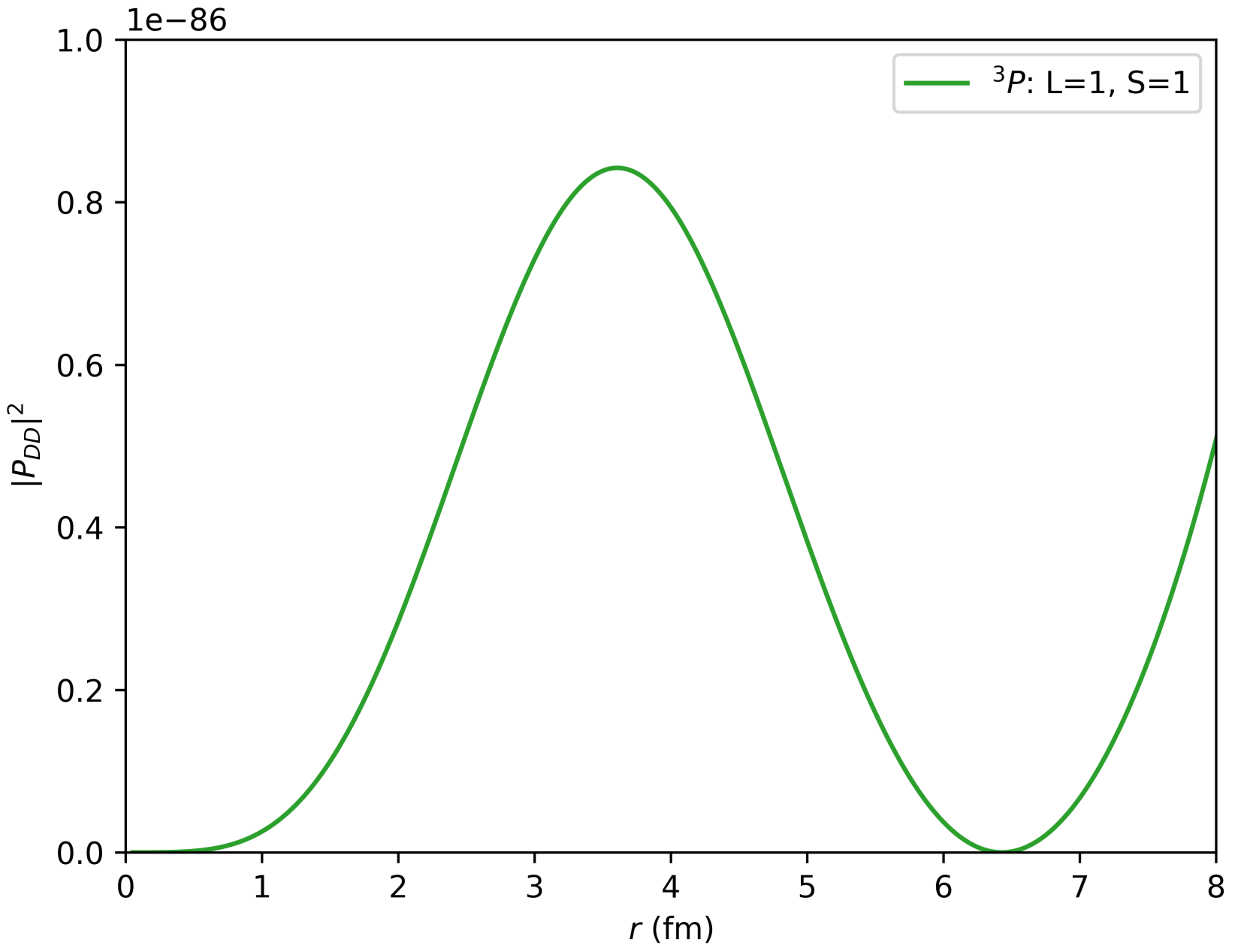

**Figure 44:** Normalised probability density associated with the solutions of the radial wave function equation 274.

For the spin part, we construct the wave function by enumerating all the possible combinations in which the total nuclear spin gives a desired spin component value $M_S$. The coefficients (including sign) for each combination then determine the overall spin value $S$. These coefficients are most conveniently found using Clebsch–Gordan coefficient lookup tables. We provide the $S = 1, M_S = 1$ state as an example:

$$\begin{aligned} \Phi_S &= \left| S = 1, M_S = 1 \right\rangle_{12;34} \\ &= \frac{1}{2} \left( \uparrow_1\uparrow_2\uparrow_3\downarrow_4 + \uparrow_1\uparrow_2\downarrow_3\uparrow_4 - \uparrow_1\downarrow_2\uparrow_3\uparrow_4 - \downarrow_1\uparrow_2\uparrow_3\uparrow_4 \right) \end{aligned} \tag{279}$$

Here we have 4 combinations in which 3 of the 4 nucleons have spin up and one has spin down - this results in $M_S = 1$.

From here, the different parts of the wave function are combined and the antisymmetrization procedure performed. This involves swapping nucleon labels to create different permutations and combining them together to create the allowed anti-symmetry.

### $^4$He ground state wave function

In a more complete description of the ground state of the alpha particle we might want to use the different $^1$S, $^3$P and $^5$D state components that result from group theory [117], [118]. For modeling the **a**-matrix element, we will make use of a simpler approximate wave function, focusing on the dominant $^1$S component with a fully symmetric spatial part.

We will follow the same procedure as for the $D_2$ case above, separating the wave function into spatial, spin and isospin parts:

$$\Psi = \Phi_R \Phi_S \Phi_T \tag{280}$$

The following subsections discuss those parts.

### $^4$He spatial wave function

The spatial part of the $^4$He ground state wave function depends on the four nucleon coordinates $\mathbf{r}_1$, $\mathbf{r}_2$, $\mathbf{r}_3$ and $\mathbf{r}_4$. We can visualize the associated physical system in Figure 45.

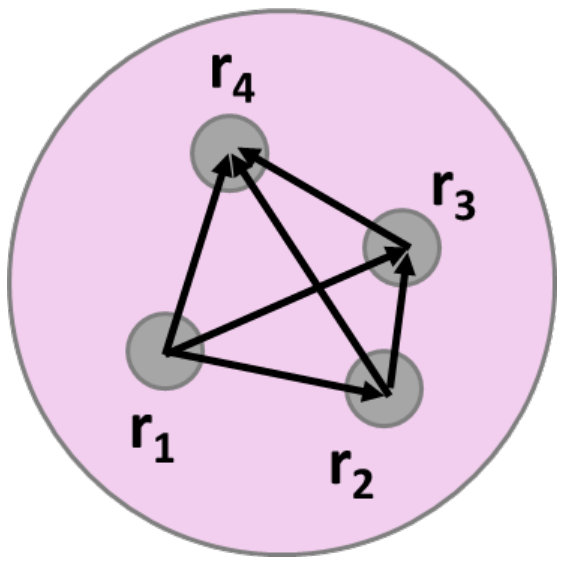

**Figure 45:** Comparison of the pair distribution from the model compared with the pair density from Kamada et al. (2001) [119].

We are working with the $^1S$ component and so the helium nucleus is taken to have no orbital angular momentum. $\Phi_R$ can be therefore be represented as:

$$\Phi_R \ = \ N_S \, u(|\mathbf{r}_2 - \mathbf{r}_1|)u(|\mathbf{r}_3 - \mathbf{r}_1|)u(|\mathbf{r}_4 - \mathbf{r}_1|)u(|\mathbf{r}_3 - \mathbf{r}_2|)u(|\mathbf{r}_4 - \mathbf{r}_2|)u(|\mathbf{r}_4 - \mathbf{r}_3|) \tag{281}$$

where N$_S$ is the normalization factor. The function $u$ is chosen to model what might be expected from a nucleon-nucleon potential with a “hard core”

$$u(r) \ = \ (r - r_0)^t e^{-\beta(r-r_0)^s} \tag{282}$$

whose coefficients are determined by integrating the wave function and ensuring its fit to the pair distribution according do Kamada et al. [119]. This results in the parameters:

$$\beta \ = \ 7.111527099157 \text{ fm}^{-1}, \ \ s \ = \ 0.335, \ \ t \ = \ 2.168033173083, \ \ r_0 \ = \ 0.3223838138079 \text{ fm} \tag{283}$$

The (radial) pair distribution function is defined according to

$$C(s) \ = \ \langle \Phi_R | \delta(s - |\mathbf{r}_{21}|) | \Phi_R \rangle \tag{284}$$

In Figure 46 is shown a comparison of the pair potential for the model wave function with the accurate 4-nucleon ground state $\alpha$ wave function of Kamada et al. (2001) [119]. Note that in this optimization we have fit a simple single configuration wave function so that it has roughly the same pair distribution as the much more accurate and complicated multi-configuration wave function of Kamada et al.

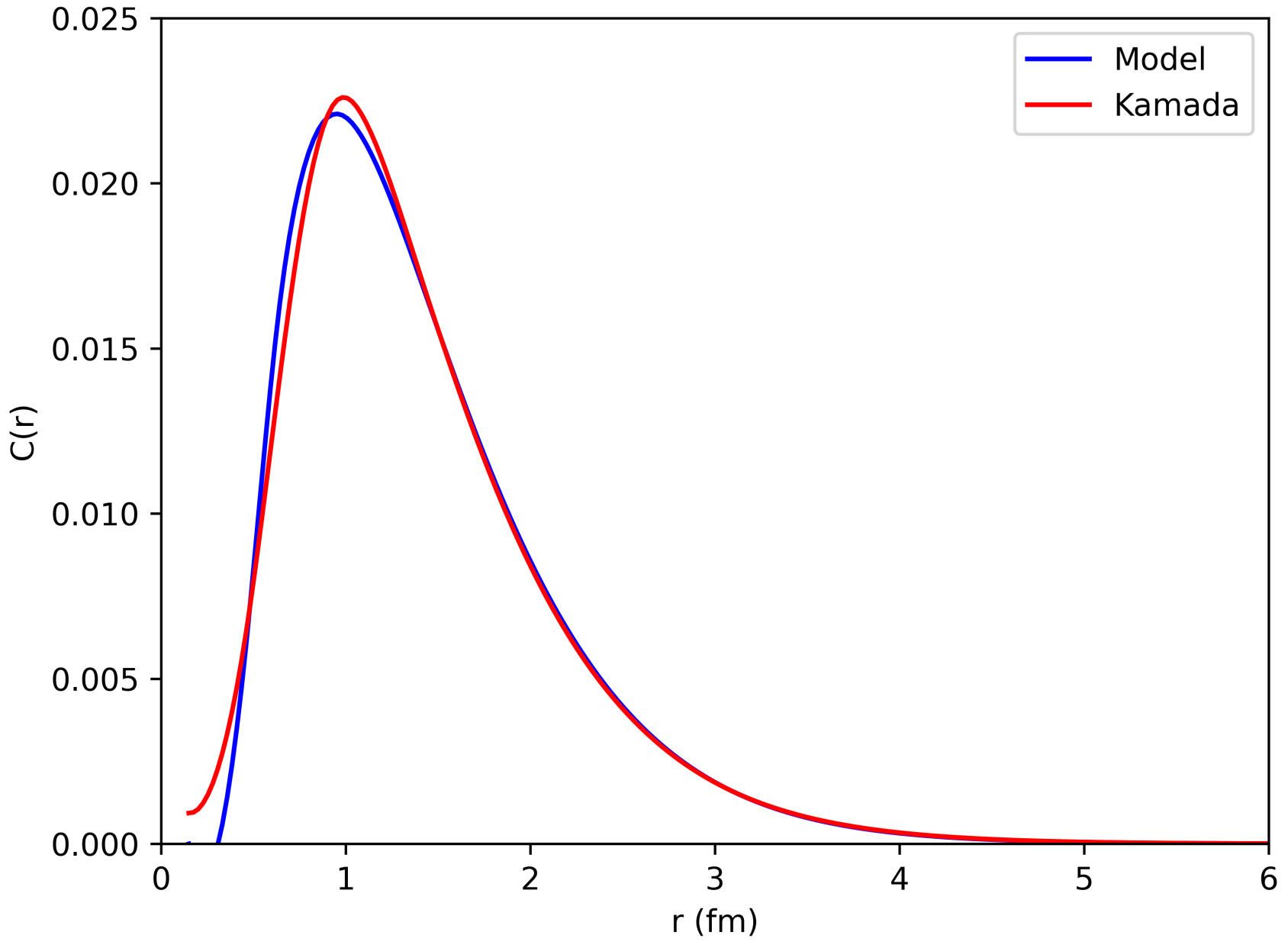

**Figure 46:** Model pair distribution function (blue) compared with the accurate pair distribution of Kamada et al. (2001) [119].

### Spin and isospin part of the $^4$He molecular wave function

Just like $D_2$ system, $^4$He system has some isospin and spin constraints. The total isospin has value $T = 0$ and for spin we take the $^1S$ state so that $S = 0$.

We can construct the spin and isospin part of the $^4$He nucleus by considering it to be composed of two deuterons whose individual spin is 1 but whose combined spin is 0.

We can therefore build on the result from the $D_2$ molecule for the isospin part, namely:

$$\begin{aligned} \Phi_T &= \Big|T = 0, M_T = 0\Big\rangle_{12;34} \\ &= \Big|T = 0, M_T = 0\Big\rangle_{12}\Big|T = 0, M_T = 0\Big\rangle_{34} \\ &= \left(\frac{p_1 n_2 - n_1 p_2}{\sqrt{2}}\right)\left(\frac{p_3 n_4 - n_3 p_4}{\sqrt{2}}\right) \\ &= \frac{1}{2}\left(p_1 n_2 p_3 n_4 - p_1 n_2 n_3 p_4 - n_1 p_2 p_3 n_4 + n_1 p_2 n_3 p_4\right) \end{aligned} \tag{285}$$

For the spin part we proceed similarly:

$$
\begin{aligned}
\Phi_S &= \Big|S=0, M_S=0\Big\rangle_{12;34} \\
&= \frac{1}{\sqrt{3}}\Big|S=1, M_S=1\Big\rangle_{12}\Big|S=1, M_S=-1\Big\rangle_{34} \\
&\quad - \frac{1}{\sqrt{3}}\Big|S=1, M_S=0\Big\rangle_{12}\Big|S=1, M_S=0\Big\rangle_{34} \\
&\quad + \frac{1}{\sqrt{3}}\Big|S=1, M_S=-1\Big\rangle_{12}\Big|S=1, M_S=1\Big\rangle_{34} \\
&= \frac{1}{\sqrt{3}}\uparrow_1\uparrow_2\downarrow_3\downarrow_4 \\
&\quad - \frac{1}{\sqrt{12}}\left(\uparrow_1\downarrow_2\uparrow_3\downarrow_4 + \uparrow_1\downarrow_2\downarrow_3\uparrow_4 + \downarrow_1\uparrow_2\uparrow_3\downarrow_4 + \downarrow_1\uparrow_2\downarrow_3\uparrow_4\right) \\
&\quad + \frac{1}{\sqrt{3}}\downarrow_1\downarrow_2\uparrow_3\uparrow_4
\end{aligned}
$$

The same antisymmetrization procedure that is used for the $D_2$ molecule can also be used for the $^4$He case.

### Construction of J states

For the cases where it is important to consider the total angular momentum $J$, we can construct the appropriate states using Clebsch-Gordan coefficients according to

$$
|JM_J\rangle \;=\; \sum_{M}\sum_{M_S}|L,M\rangle|S,M_S\rangle\Big\langle L,M;S,M_s\Big|J,M_J\Big\rangle \tag{286}
$$

For further details on constructing the $D_2$ and $^4$He wave functions, see [113].

## 6.2 Known $^4$He$^*$ excited states

Our discussion of excitation transfer and related processes will necessarily involve a consideration of the excited $^4$He$^*$ states, which motivates a discussion of these states in this section. Much theoretical and experimental work by different groups has led to a clarification of some of the states, with a detailed review having been given by Tilley and coworkers [100]. To given an overview, we start with the level diagram in Figure 47 taken from this reference.

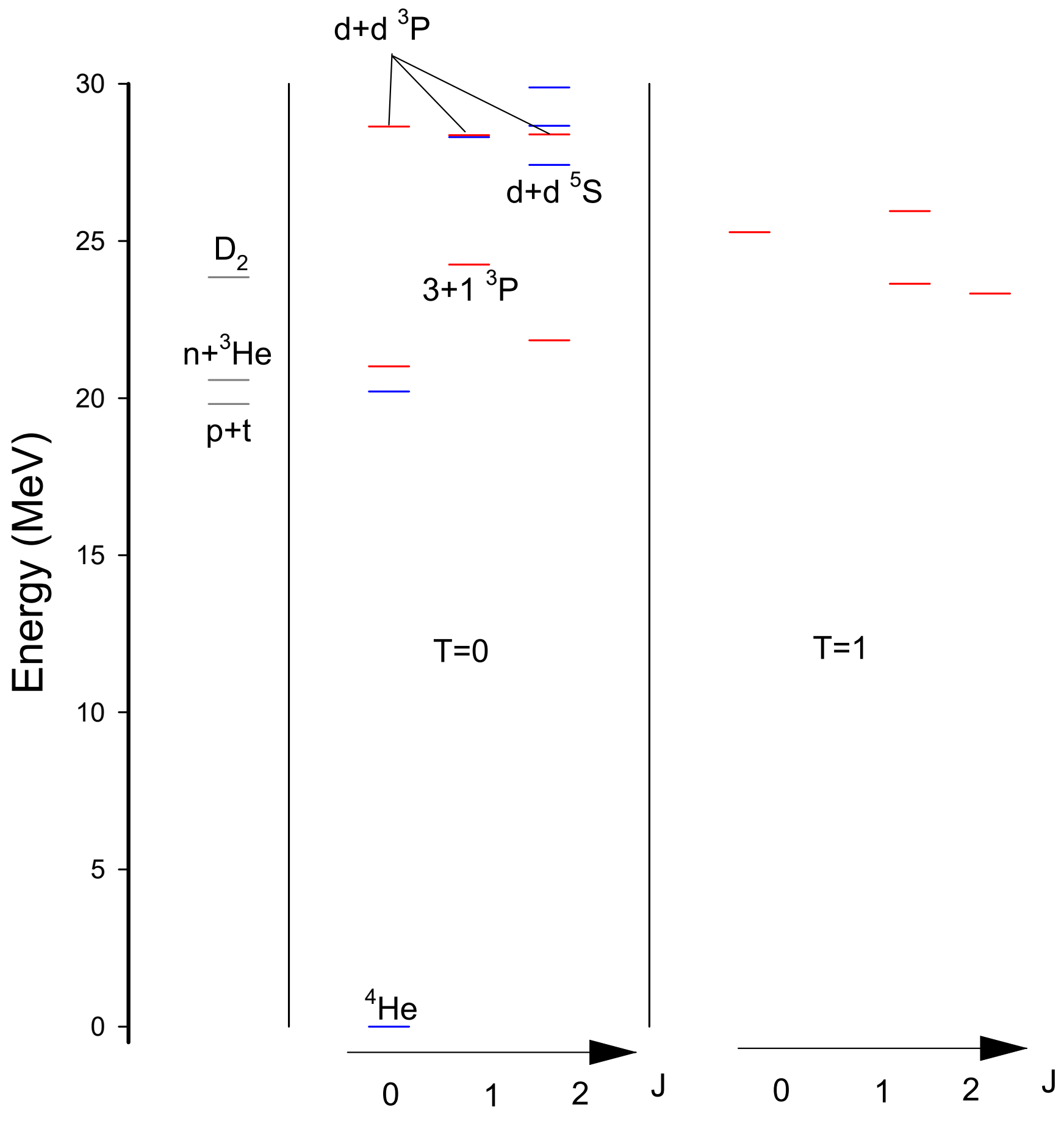

**Figure 47:** $^4$He ground state and $^4$He$^*$ excited states from [100]. The energies of free p+t, n+$^3$He and molecular $D_2$ are indicated on the left. The levels in the middle have isospin $T = 0$, and the levels on the right have isospin $T = 1$. Even parity states are indicated in blue, and odd parity states are in red. Note the different variations of the $D_2$ state, as discussed in sections 5.1 and 6.1 and shown in Figure 7, which are however very close in energy.

We note at the outset that all of the excited $^4$He states are very unstable, with lifetimes shorter than $10^{-20}$ seconds. Note that ultimately there is a lot of mixing between different contributing configurations, which results in a complicated level structure.

### A few states with relatively simple interpretations

Since the deuteron has isospin zero, excited $^4$H$^*$ states which are mostly d+d (*i.e.*, 2+2) must have isospin zero. The level at 27.42 MeV has been identified as mostly d+d $T = 0$ $^5$S $J^\pi = 2^+$. This state is different than the molecular $D_2$

$^5$S $J^\pi = 2^+$ state near 23.85 MeV, but it shares spin, isospin, total angular momentum, and parity quantum numbers.

There are three odd parity isospin zero states up near 28.5 MeV which have been identified as being primarily d+d $T = 0$ $^3$P. These states are potentially interesting to us in part because they share spin, isospin, total angular momentum, and parity quantum numbers with the molecular $D_2$ $^3$P $J^\pi = 0, 1, 2$ states (see section 6.1), and because interactions that couple to the molecular $D_2$ $^3$P $J^\pi = 0, 1, 2$ states also couples to these $^4$He$^*$ excited states.

Another state which draws our attention is the state at 24.25 MeV which is mostly (3+1) $T = 0$ $^3$P $J^\pi = 1^-$. It may seem odd that this state is labeled as 3+1, since we usually think of p+t and n+$^3$He for configurations made up of mass 3 and mass 1. The issue here is that if we try to construct fully antisymmetric states with quantum numbers $T = 0$ $^3$P $J^\pi = 1^-$, we find that it is impossible to make clean p+t or n+$^3$He states. But it is possible to construct a state based on a superposition of p+t and n+$^3$He states. Because of this, we use the notation 3+1. Since the constituent p+t and n+$^3$He particles can tunnel apart quickly, this state is very unstable, with a lifetime of about $10^{-21}$ seconds. This state is of interest to us since the d+d $T = 0$ $^3$P $J^\pi = 1^-$ state can mix with it, which opens up a decay channel where the canonical 3+1 fusion products result from the decay (see also sections 6.5 and 6.6 as well as section 5.15).

### Model for the unstable d+d $T = 0$ $^3$P $J^\pi = 1^-$ state at 28.37 MeV

We consider a simple model for the unstable d+d $T = 0$ $^3$P $J^\pi = 1^-$ state at 28.37 MeV based on two deuterons that interact through an attractive nuclear potential and repulsive Coulomb potential. In this calculation we focus on the tunnel decay which is dominant, and neglect the coupling to the 3+1 $T = 0$ $^3$P $J^\pi = 1^-$ state at 24.25, which is responsible for the decay by 3+1 fusion of this d+d state.

To model the unstable state we solve for the radial wave function that describes the separation between the two deuterons. We can write

$$E\,P(r) = \left( -\frac{\hbar^2}{2\mu}\frac{d^2}{dr^2} + \frac{\hbar^2 L(L+1)}{2\mu r^2} + V_{mol}(r) + V_N^{S,L}(r) \right) P(r) \tag{287}$$

subject to the boundary conditions

$$\begin{aligned} P(0) &= 0 \\ P(r) &\to Ae^{ikr} \end{aligned} \tag{288}$$

where $E$ and the wave vector $k$ are both complex.

For the potentials, we use the same as in section 6.1. Specifically, we use the Kolos molecular Coulomb potential (Eq. 276) and the Weller nuclear potential (Eq. 275) for the $^3P$ channel. Note that at short range, the molecular potential follows the simple Coulomb scaling. The combined potential is illustrated by the green line in Figure 41.

We developed a simulation code for this problem, utilizing a 3 point scheme on a non-uniform grid based on a shooting method (see `https://github.com/project-ida/nuclear-reactions` for details). The resulting complex energy eigenvalue are

$$\begin{aligned} E_r &= 4.61 \text{ MeV} \\ E_i &= -1.97 \text{ MeV} \end{aligned} \tag{289}$$

The energy relative to the ground state is

$$E = 4.6168 + 23.8465 = 28.463 \text{ MeV} \tag{290}$$

The d+d state energy value calculated this way matches reasonably well with the value reported in [100] of 28.37 MeV, giving us confidence in this method. For the line width we obtain

$$\Delta E \;=\; -2E_i \;=\; 3.94\,\text{MeV} \tag{291}$$

which also is in good agreement with the value (reported for tunneling) in [100] of 3.92 MeV.

The radial probability density for this unstable state is shown in Figure 48.

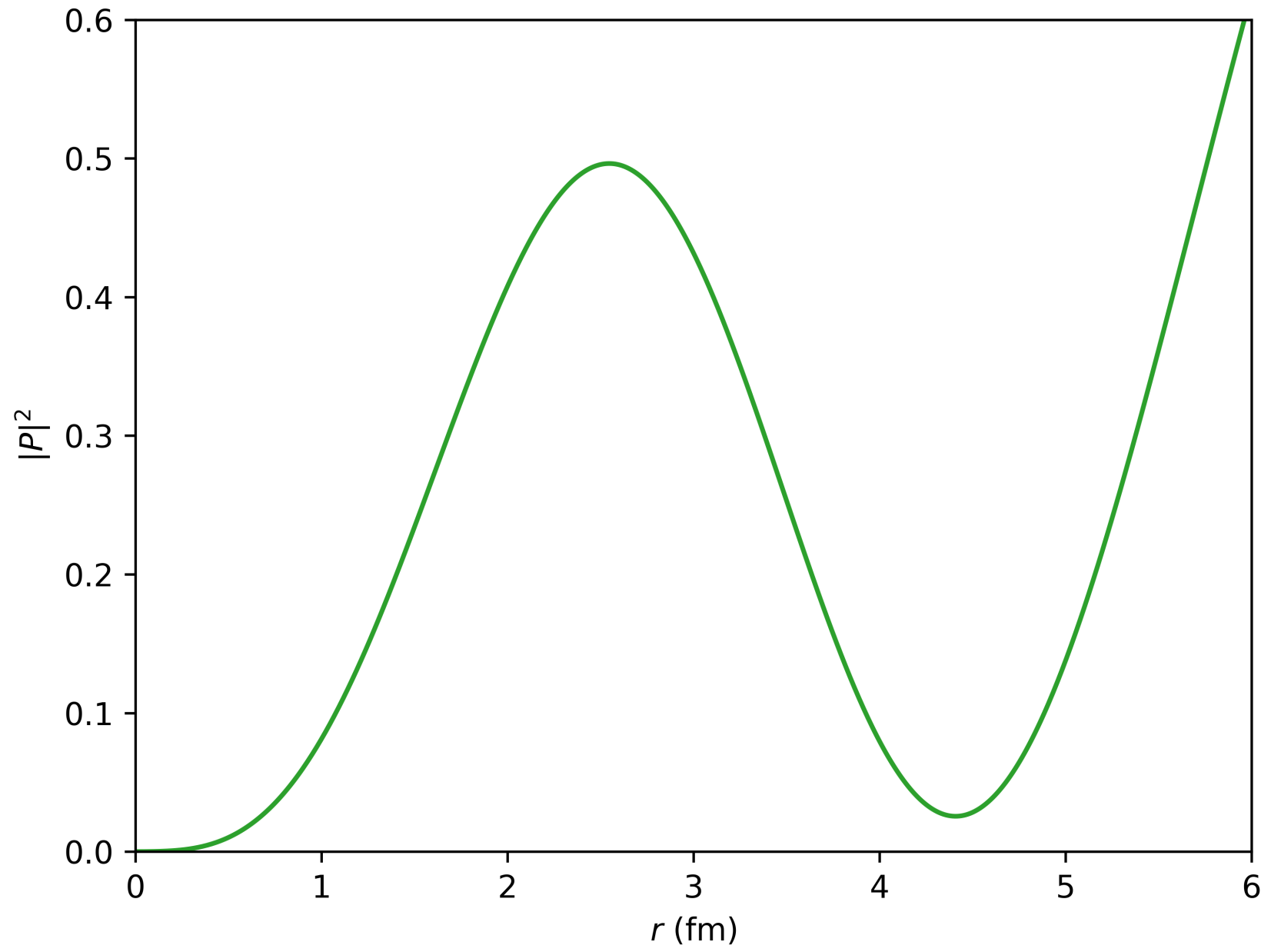

**Figure 48:** Radial probability density for the d+d $T = 0$ $^3$P $J^\pi = 1^-$ state at 28.37 as a function of $r$. The probability has been noramlized by integrating out to 4.4 fm and setting the result to equal 1.

If we compare Figure 48 above to the molecular $D_2$ $^3$P state calculated in section 6.1 (see Figure 44), we see that the molecular state can be seen as an admixture with partial occupation of this unstable $^4$He$^*$ excited state.

**Model for the unstable (3+1) $T = 0$ $^3$P $J^\pi = 1^-$ state at 24.25 MeV**

We consider a similar model for the unstable (3+1) $T = 0$ $^3$P $J^\pi = 1^-$ state at 24.25 MeV. As above, we work with the radial equation

$$E\,P(r) \;=\; \left(-\frac{\hbar^2}{2\mu}\frac{d^2}{dr^2} + \frac{\hbar^2 L(L+1)}{2\mu r^2} + V_{mol}(r) + V_N^{S,L}(r)\right)P(r) \tag{292}$$

subject to the boundary conditions

$$\begin{aligned} P(0) \;&=\; 0 \\ P(r) \;&\rightarrow\; Ae^{ikr} \end{aligned} \tag{293}$$

In this case, for the 3+1 nuclear potential model, we use

$$V_N^{S,L}(r) \;=\; V_0 e^{-\alpha r^2} \tag{294}$$

whose parameters can be seen in Table 3 [120] :

| State | $V_0$ (MeV) | $\alpha$ (fm$^{-2}$) |
|---|---|---|
| $^1P$ | -8.0 | 0.03 |

**Table 3:** Nuclear potential parameters

where we note that in the absence of a p+t $^3$P potential, we use the potential for $^1$P.

The 3+1 Coulomb potential is taken to be half of the p+t Coulomb potential because there is no Coulomb potential for the n+$^3$He component.

The energy eigenvalue that results is

$$\begin{aligned} E_r &= 4.15\,\text{MeV} \\ E_i &= -2.31\,\text{MeV} \end{aligned} \tag{295}$$

The energy relative to the ground state is

$$E \;=\; 4.1538 + 20.19574 \;=\; 24.350\,\text{MeV} \tag{296}$$

where 20.19574 MeV is half the combined mass energy of t+p and n+$^3$He.

The 3+1 state energy value calculated this way matches reasonably well with the value reported in [100] of 24.25 MeV. For the line width we obtain

$$\Delta E \;=\; -2E_i \;=\; 4.61\,\text{MeV} \tag{297}$$

which is lower than the value reported for tunneling in [100] of 5.95 MeV.

The radial probability density for this unstable state is shown in Figure 48.

We will come back to such $^4$He$^*$ excited states in sections 6.5 and 6.6 as well as section 5.15.

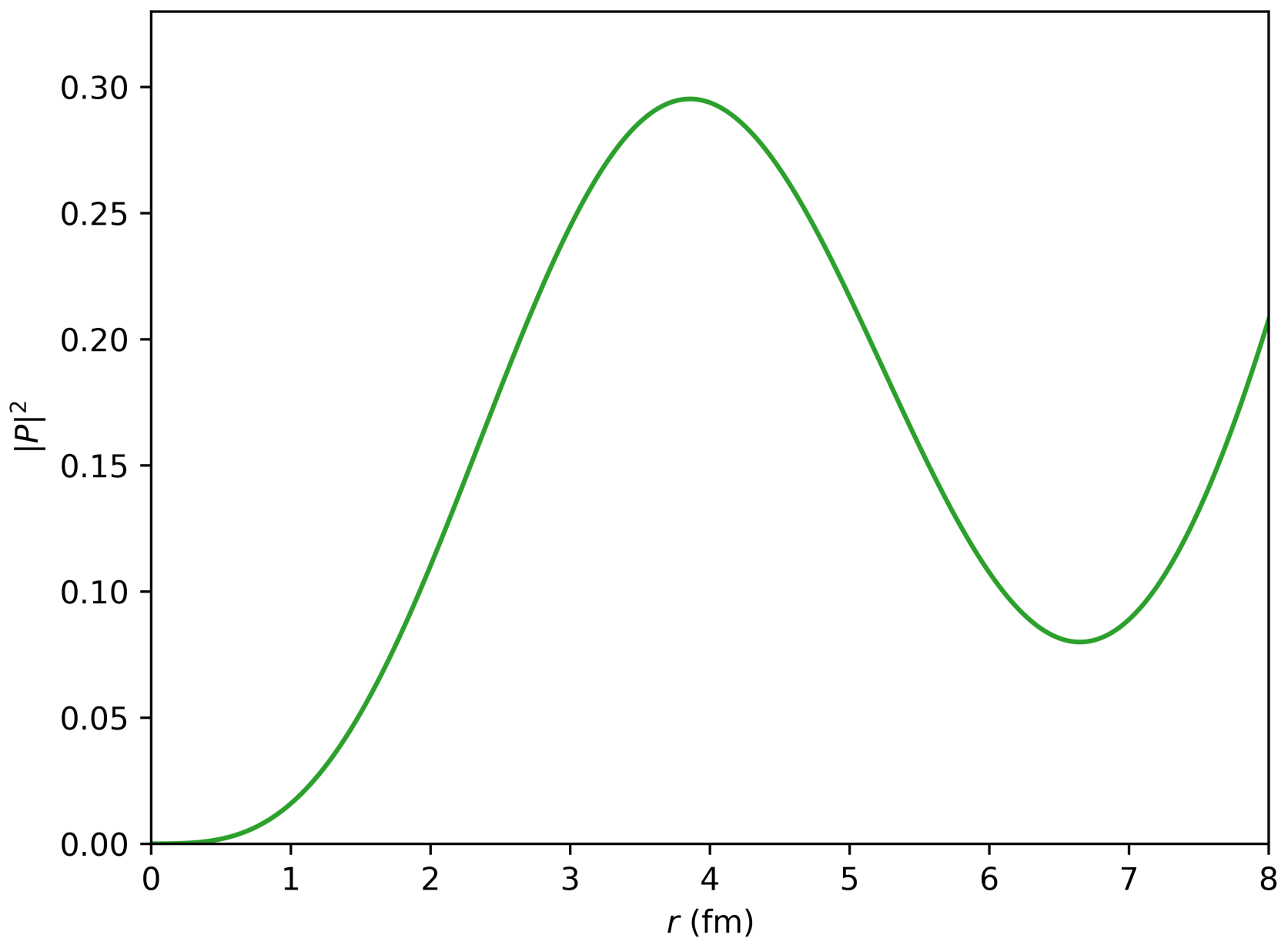

**Figure 49:** Radial probability density for the 3+1 $T = 0$ $^3$P $J^\pi = 1^-$ state at 24.25 as a function of $r$. The probability has been noramlized by integrating out to 6.65 fm and setting the result to equal 1.

## 6.3 Gamow factor and fusion rate calculations

In section 6.1, the $D_2$ nuclear wave functions were presented, at the center of which is the $D_2$ molecular wave function that governs the spatial orientation of the two deuteron pairs relative to each other (and therefore determines the mean distance between them).

Eq. 274 represents the radial wave equation as a form of the Schrödinger equation for spherically symmetric systems which we repeat here for convenience

$$E\,P_{DD}(r) \;=\; \left(-\frac{\hbar^2}{2\mu}\frac{d^2}{dr^2} + \frac{\hbar^2 L(L+1)}{2\mu r^2} + V_{mol}(r) + V_N^{S,L}(r)\right)P_{DD}(r) \tag{298}$$

This equation forms the basis for fusion rate estimates from deuterons tunneling through the potential barrier.

This section will discuss two different approaches, each of which has its own sets of merits: a numerical approach and the so-called WKB approximation.

Naturally, the numerical approach can be expected to be more accurate. However, as we will show at the end of this section, the WKB approximation performs well too, except for a comparatively small error at short radii. Nevertheless, there is an important advantage in using the WKB approximation, which is the reason why it has been used in such fusion rate calculations before: it allows for readily incorporating screening effects, which are caused by changes to the molecular potential due to free electrons (as represented by the value of the screening potential $U_e$).

Obtaining fusion rates based on the WKB approximation at different values of the screening potential $U_e$ is shown later in the section. In the end, we opt for a combined approach that draws on the strengths of both approaches: working with an unscreened base rate obtained from numerical integration, which is then adjusted to account for different screening potential values by a sort of "screening correction factor" that can be readily obtained from the WKB approach, as shown.

#### The 1989 Koonin and Nauenberg fusion rate estimate as a reference point

A relevant reference point is a short 1989 article by Koonin and Nauenberg, in which the authors present an estimate for the spontaneous deuterium fusion rate at room temperature, based on a numerical approach.

The rate reported by Koonin and Nauenberg 1989 [20] is

$$\gamma_{DD} = 3 \times 10^{-64}\ \mathrm{s}^{-1} \tag{299}$$

Note, however, that parts of the potential in Eq. 298 are state-dependent (see 6.1 for an overview of different states), which Koonin and Nauenberg neglect to take into account. Their estimate draws on the molecular part of the barrier only—an issue that we later remedy in our version of these calculations below.

#### Numerical approach to fusion rate estimation following Koonin and Nauenberg

From the Koonin and Nauenberg perspective, the fusion rate is the probability of being at small radius in the probability density of the wave function. Koonin and Nauenberg used 10 fm as the target radius. Therefore

$$\gamma_{DD} = A|\psi(10\ \mathrm{fm})|^2 \tag{300}$$

where A is a scaling factor that Koonin and Nauenberg give as

$$A = 1.5 \times 10^{-16} \frac{\text{cm}^3}{\text{sec}} \tag{301}$$

For the case of $L = 0$ and $M = 0$, we can take

$$\psi(r) = \frac{P_{DD}(r)}{r} Y_{LM}(\theta, \phi) = \frac{1}{\sqrt{4\pi}} \frac{P_{DD}(r)}{r} \tag{302}$$

Evaluating $P_{DD}$ at 10 fm (by solving Eq. 298, leaving out the nuclear potential as Koonin and Nauenberg did) yields

$$P_{DD}(10 \text{ fm}) = 6.9 \times 10^{-36} \text{ cm}^{-1/2} \tag{303}$$

which results in

$$\psi^2(10 \text{ fm}) = 3.8 \times 10^{-48} \text{ cm}^{-3} \tag{304}$$

and

$$\gamma_{DD} = 5.7 \times 10^{-64} \text{ s}^{-1} \tag{305}$$

**A state-specific numerical approach: fusion rate estimation for the molecular $D_2$ $^5$S state and the localized $^4$He$^*$ $^5$S state**

We expect the $^1$S and $^5$S states of molecular $D_2$ to decay the fastest, since there is no centripetal potential contribution to the overall barrier. We begin with the $^5$S state.

To evaluate the spontaneous fusion rate for the molecular $^5$S state, we make use of a numerical solution for the radial molecular wave function (orange line in Figure 42) to determine the fraction of the wave function that's present in the (off-resonant) localized $^4$He$^*$ state (Figure 43 at $r < 4.1$ fm). This localized excited state is an admixture with the highly unstable $d + d, ^5S\ T = 0\ J^\pi = 2^+$ state (see section 6.2), and so will rapidly decay.

The spontaneous fusion rate for the molecular $^5$S state is then given by:

$$\gamma_{DD,^5S} \ = \ P_{d+d,^5S} \gamma_{d+d,^5S} \tag{306}$$

where $P_{d+d,^5S}$ is the probability of occupation of the highly unstable $^4$He$^*$ state, and $\gamma_{d+d,^5S}$ is the associated decay rate.

The probability of occupation is obtained by integrating the wave function seen in Figure 43 between $0 < r < 4.1$ fm. The result is

$$P_{d+d,^5S} = 4.9 \times 10^{-85} \tag{307}$$

The line width of the localized $d + d, ^5S\ J^\pi = 2^+$ state on resonance at 27.42 MeV [100] is

$$\hbar\gamma_{d+d,^5S} \ = \ 0.25 + 0.23 \text{ MeV} \ = \ 0.48 \text{ MeV} \tag{308}$$

The associated decay rate and lifetime is

$$\gamma_{d+d,^5S} = \frac{0.48\ \text{MeV}}{\hbar} = 7.3 \times 10^{20}\ \text{s}^{-1}$$

$$\tau_{d+d,^5S} = \frac{\hbar}{0.48\ \text{MeV}} = 1.4 \times 10^{-21}\ \text{sec} \tag{309}$$

This leads to an overall fusion decay rate for the $D_2$ molecular $^5$S state of

$$\gamma_{DD,^5S} = 4.9 \times 10^{-85}\ \frac{0.48\ \text{MeV}}{\hbar} = 3.6 \times 10^{-64}\ \text{s}^{-1} \tag{310}$$

This state-specific rate is compatible with the Koonin and Nauenberg rate of $3 \times 10^{-64}$ s$^{-1}$ for $\gamma_{DD}$ [20].

**A state-specific numerical approach: fusion rate estimation for the molecular $D_2$ $^3$P state and the localized $^4$He$^*$ $^3$P state**

Since the molecular $D_2$ $^3$P states play such an important role in this paper, it is natural to be interested in the associated spontaneous fusion decay rate. We expect the rate to be slower than for the $D_2$ S states since there is one unit of angular momentum and an associated centripetal potential barrier. We can use an approach similar to what we did above for the $^5$S state.

For the $^3$P state we make use of the green line in Figure 42 to determine the fraction of the wave function that's present in the (off-resonant) localized $^4$He$^*$ state (Figure 44 at $r < 6.4$ fm). This localized excited state is an admixture with the highly unstable $d + d,^3 P\ J^\pi = 1^-$ state (see Figure 48 in section 6.2), and so will rapidly decay.

The spontaneous fusion rate for the molecular $^3$P state is then given by:

$$\gamma_{DD,^3P} = P_{d+d,^3P}\gamma_{d+d,^3P} \tag{311}$$

where $P_{d+d,^3P}$ is the probability of occupation of the highly unstable $^4$He$^*$ state, and $\gamma_{d+d,^3P}$ is the associated decay rate.

The probability of occupation is obtained by integrating the wave function seen in Figure 44 between $0 < r < 6.4$ fm. The result is

$$P_{d+d,^3P} = 2.3 \times 10^{-86} \tag{312}$$

The line width of the localized $^4$He$^*$ $^3$P $J^\pi = 1^-$ state on resonance at 28.37 MeV [100] is

$$\hbar\gamma_{d+d,^3P} = 0.07 + 0.08\ \text{MeV} = 0.15\ \text{MeV} \tag{313}$$

The associated decay rate and lifetime is

$$\gamma_{d+d,^3P} = \frac{0.15\ \text{MeV}}{\hbar} = 2.3 \times 10^{20}\ \text{s}^{-1}$$

$$\tau_{d+d,^3P} = \frac{\hbar}{0.15\ \text{MeV}} = 4.4 \times 10^{-21}\ \text{sec} \tag{314}$$

This leads to an overall fusion decay rate for the $D_2$ molecular $^3$P $J^\pi = 1^-$ state of

$$\gamma_{DD,^3P} = 2.3 \times 10^{-86} \frac{0.15\,\text{MeV}}{\hbar} = 5.2 \times 10^{-66}\,\text{s}^{-1} \tag{315}$$

We see that the centripetal potential (along with the difference in nuclear potential) leads to a spontaneous fusion decay rate that's substantially smaller than for the $^5$S state as expected.

### WKB approximation approach to fusion rate estimation

An alternative to the numerical approach of solving the radial wave equation (Eq. 298) is to draw on the WKB approximation.

Here, the probability of two deuterons being at close proximity, as previously obtained by numerical integration, ($P_{d+d,^5S}$ in the case above) is replaced by the analytical expression for the so-called tunneling probability:

$$T \approx e^{-2G} \tag{316}$$

where

$$G = \int_{r_1}^{r_2} \sqrt{\frac{2m}{\hbar^2}[V(r) - E]}\,dr \tag{317}$$

G is known as the Gamow factor.

To obtain an estimate for the fusion rate $\gamma_{DD}$ from the tunneling probability $T$, we need to multiply by a volume ratio factor $v_{nuc}/v_{mol}$ (as we saw in Eq. 3) and by the decay rate of the unstable $^4$He$^*$:

$$\gamma_{DD} \approx T \frac{v_{nuc}}{v_{mol}} \gamma_{^4\text{He}^*} \tag{318}$$

where $^4$He$^*$ is a specific $^4$He excited state such as $d + d,^5 S$.

Note that $\gamma_{DD}$ is state-specific (since all factors are state-specific).

### Volume ratio

We begin by defining the equilibrium bond lengths for different electronic states of a $D_2$ molecule in terms of the Bohr radius ($a_0$):

- For the S state: $R_{0S} = 1.401080 \times a_0$
- For the P state: $R_{0P} = R_{0S} + 0.001 \times 10^{-10}$ m
- For the D state: $R_{0D} = R_{0S} + 0.002 \times 10^{-10}$ m

Next, we calculate the de Broglie wavelength of the relative motion ($\Delta R$) using the formula:

$$\Delta R = a_0 \sqrt{\frac{I_H}{\hbar\omega_{0DD}}} \sqrt{\frac{2m_e c^2}{M_D c^2}} \tag{319}$$

where $I_H$ is the ionization potential of hydrogen in eV, $\hbar\omega_{0DD}$ is the zero-point energy (0.3862729 eV), $m_e c^2$ is the electron mass in MeV, and $M_D c^2$ is the deuteron mass in MeV.

The molecular volumes ($v_{mol}$) for the S, P, and D states are computed as follows:

- S: $v_{mol} = 2\pi R_{0S}\left(\pi \Delta R^2\right)$
- P: $v_{mol} = 2\pi R_{0P}\left(\pi \Delta R^2\right)$
- D: $v_{mol} = 2\pi R_{0D}\left(\pi \Delta R^2\right)$

The nuclear volume ($v_{nuc}$) of the deuterium nucleus is calculated assuming a spherical shape with a radius of 5 fm:

$$v_{nuc} = \frac{4}{3}\pi\left(5 \times 10^{-15}\ \text{m}\right)^3 \tag{320}$$

Finally, the volume ratios of the nuclear volume to the molecular volume for the S, P, and D states are calculated as:

- S: $\frac{v_{nuc}}{v_{mol}} = 6.66 \times 10^{-12}$
- P: $\frac{v_{nuc}}{v_{mol}} = 6.65 \times 10^{-12}$
- D: $\frac{v_{nuc}}{v_{mol}} = 6.64 \times 10^{-12}$

These ratios provide insight into the comparative scale of nuclear and electronic contributions to the physical structure of the molecule.

**Unscreened fusion rate estimates based on the WKB approximation**

Combining the above and plugging into Eq. 318 we obtain:

$$\gamma_{DD,{}^5S} \ \sim \ e^{-2G}\left(\frac{v_{nuc}}{v_{mol}}\right)\gamma_{d+d,{}^5S} \ = \ 1.6 \times 10^{-66}\ \text{s}^{-1} \tag{321}$$

This is low by more than two orders of magnitude, compared to the numerical result.

In order to do better with the WKB approximation, we would need to take account of the scaling of the $1/\sqrt{\alpha}$ factor (see Eq. 6.3) in the WKB transformation, and we would want to correct for the error associated with the WKB approximation itself (both of which are discussed briefly at the end of this section).

This suggests that numerical calculations are more accurate when evaluating wave functions and matrix elements compared to the WKB approximation based approach. However, we would like to make use of the WKB approximation in order to get estimates for the effect of different screening potentials, as is discussed next.

**Applying screening to unscreened fusion rate estimates**

Screening can be readily included in the WKB approximation based rate calculation via the screening length $\lambda$ [43]:

$$\lambda \ = \ \frac{1}{4\pi\epsilon_0}\frac{e^2}{U_e} \tag{322}$$

This results in a screened molecular potential:

$$V_{\text{mol,scr}} = V_{\text{mol}} \cdot e^{\frac{-r}{\lambda}} \tag{323}$$

### Screened fusion rate estimates based on the WKB approximation

Now fusion rates can be determined—for different screening energies—based on Eq. 318. Obtained values for the $^1$S configuration ($L = 0$, $S = 0$), the $^5$S configuration ($L = 0$, $S = 2$), $^3$P configuration ($L = 1$, $S = 1$), and the $^5$D configuration ($L = 2$, $S = 2$) are listed in Table 4, 5, 6, 7 respectively.

Values for the $^1$S configuration will be helpful for comparing with the result of Koonin and Nauenberg 1989 [20]; for the $^3$P configuration for calculations involving relativistic and electric dipole coupling; and for the $^5$D configuration for calculations involving magnetic dipole coupling.

The last column of each table shows the effect of screening on the tunneling probability via the expression $e^{\Delta G_{scr}}$ where $\Delta G_{scr} = G_{unscreened} - G_{screened}$.

**Table 4:** Gamow factors, fusion rates and screening parameters for the $^1$S state ($L = 0$, $S = 0$) of $D_2$.

| $U_e(eV)$ | $G$ | exp($-G$) | exp($-2G$) | Fusion rate $\gamma_{DD}$ (s$^{-1}$) | $\Delta G_{scr}$ | exp($\Delta G_{scr}$) |
|---|---|---|---|---|---|---|
| 0 | 88.0 | $5.78 \times 10^{-39}$ | $3.34 \times 10^{-77}$ | $1.62 \times 10^{-67}$ | 0 | 1 |
| 50 | 66.7 | $1.08 \times 10^{-29}$ | $1.16 \times 10^{-58}$ | $5.63 \times 10^{-49}$ | -21.3 | $1.86 \times 10^{9}$ |
| 100 | 57.7 | $8.66 \times 10^{-26}$ | $7.5 \times 10^{-51}$ | $3.64 \times 10^{-41}$ | -30.3 | $1.5 \times 10^{13}$ |
| 150 | 51.6 | $4.09 \times 10^{-23}$ | $1.67 \times 10^{-45}$ | $8.11 \times 10^{-36}$ | -36.5 | $7.08 \times 10^{15}$ |
| 200 | 46.9 | $4.49 \times 10^{-21}$ | $2.01 \times 10^{-41}$ | $9.78 \times 10^{-32}$ | -41.2 | $7.77 \times 10^{17}$ |
| 250 | 43.1 | $1.87 \times 10^{-19}$ | $3.48 \times 10^{-38}$ | $1.69 \times 10^{-28}$ | -44.9 | $3.23 \times 10^{19}$ |
| 300 | 40.1 | $3.87 \times 10^{-18}$ | $1.5 \times 10^{-35}$ | $7.28 \times 10^{-26}$ | -48.0 | $6.7 \times 10^{20}$ |
| 350 | 37.6 | $4.79 \times 10^{-17}$ | $2.3 \times 10^{-33}$ | $1.12 \times 10^{-23}$ | -50.5 | $8.30 \times 10^{21}$ |

**Table 5:** Gamow factors, fusion rates and screening parameters for the $^5$S state ($L = 0$, $S = 2$) of $D_2$.

| $U_e(eV)$ | $G$ | exp($-G$) | exp($-2G$) | Fusion rate $\gamma_{DD}$ (s$^{-1}$) | $\Delta G_{scr}$ | exp($\Delta G_{scr}$) |
|---|---|---|---|---|---|---|
| 0 | 88.0 | $5.8 \times 10^{-39}$ | $3.36 \times 10^{-77}$ | $1.63 \times 10^{-67}$ | 0 | 1 |
| 50 | 66.7 | $1.08 \times 10^{-29}$ | $1.17 \times 10^{-58}$ | $5.67 \times 10^{-49}$ | -21.3 | $1.86 \times 10^{9}$ |
| 100 | 57.7 | $8.69 \times 10^{-26}$ | $7.55 \times 10^{-51}$ | $3.67 \times 10^{-41}$ | -30.3 | $1.5 \times 10^{13}$ |
| 150 | 51.5 | $4.10 \times 10^{-23}$ | $1.68 \times 10^{-45}$ | $8.17 \times 10^{-36}$ | -36.5 | $7.08 \times 10^{15}$ |
| 200 | 46.8 | $4.50 \times 10^{-21}$ | $2.03 \times 10^{-41}$ | $9.84 \times 10^{-32}$ | -41.2 | $7.77 \times 10^{17}$ |
| 250 | 43.1 | $1.87 \times 10^{-19}$ | $3.51 \times 10^{-38}$ | $1.70 \times 10^{-28}$ | -44.9 | $3.23 \times 10^{19}$ |
| 300 | 40.1 | $3.88 \times 10^{-18}$ | $1.51 \times 10^{-35}$ | $7.32 \times 10^{-26}$ | -48.0 | $6.7 \times 10^{20}$ |
| 350 | 37.6 | $4.81 \times 10^{-17}$ | $2.31 \times 10^{-33}$ | $1.12 \times 10^{-23}$ | -50.5 | $8.30 \times 10^{21}$ |

**Table 6:** Gamow factors, fusion rates and screening parameters for the $^3$P state ($L = 1$, $S = 1$) of $D_2$.

| $U_e(eV)$ | $G$ | exp($-G$) | exp($-2G$) | Fusion rate $\gamma_{DD}$ (s$^{-1}$) | $\Delta G_{scr}$ | exp($\Delta G_{scr}$) |
|---|---|---|---|---|---|---|
| 0 | 90.4 | $5.75 \times 10^{-40}$ | $3.31 \times 10^{-79}$ | $1.61 \times 10^{-69}$ | 0 | 1 |
| 50 | 69.1 | $1.01 \times 10^{-30}$ | $1.03 \times 10^{-60}$ | $5.00 \times 10^{-51}$ | 21.3 | $1.76 \times 10^{9}$ |
| 100 | 60.1 | $7.56 \times 10^{-27}$ | $5.71 \times 10^{-53}$ | $2.77 \times 10^{-43}$ | 30.2 | $1.31 \times 10^{13}$ |
| 150 | 54.1 | $3.09 \times 10^{-24}$ | $9.53 \times 10^{-48}$ | $4.63 \times 10^{-38}$ | 36.2 | $5.37 \times 10^{15}$ |
| 200 | 49.9 | $2.03 \times 10^{-22}$ | $4.14 \times 10^{-44}$ | $2.01 \times 10^{-34}$ | 40.4 | $3.54 \times 10^{17}$ |
| 250 | 46.4 | $6.94 \times 10^{-21}$ | $4.82 \times 10^{-41}$ | $2.34 \times 10^{-31}$ | 43.9 | $1.21 \times 10^{19}$ |
| 300 | 43.6 | $1.22 \times 10^{-19}$ | $1.49 \times 10^{-38}$ | $7.21 \times 10^{-29}$ | 46.8 | $2.12 \times 10^{20}$ |
| 350 | 41.2 | $1.28 \times 10^{-18}$ | $1.65 \times 10^{-36}$ | $8.00 \times 10^{-27}$ | 49.2 | $2.23 \times 10^{21}$ |

**Table 7:** Gamow factors, fusion rates and screening parameters for the $^5$D state ($L = 2$, $S = 2$) of $D_2$.

| $U_e(eV)$ | $G$ | exp$(-G)$ | exp$(-2G)$ | Fusion rate $\gamma_{DD}$ (s$^{-1}$) | $\Delta G_{scr}$ | exp$(\Delta G_{scr})$ |
|---|---|---|---|---|---|---|
| 0 | 94.8 | $6.58 \times 10^{-42}$ | $4.33 \times 10^{-83}$ | $2.10 \times 10^{-73}$ | 0 | 1 |
| 50 | 73.6 | $1.04 \times 10^{-32}$ | $1.08 \times 10^{-64}$ | $5.24 \times 10^{-55}$ | 21.2 | $1.58 \times 10^{9}$ |
| 100 | 64.9 | $6.59 \times 10^{-29}$ | $4.35 \times 10^{-57}$ | $2.11 \times 10^{-47}$ | 29.9 | $1.00 \times 10^{13}$ |
| 150 | 59.6 | $1.37 \times 10^{-26}$ | $1.88 \times 10^{-52}$ | $9.14 \times 10^{-43}$ | 35.3 | $2.09 \times 10^{15}$ |
| 200 | 55.3 | $9.30 \times 10^{-25}$ | $8.66 \times 10^{-49}$ | $4.20 \times 10^{-39}$ | 39.5 | $1.41 \times 10^{17}$ |
| 250 | 52.0 | $2.74 \times 10^{-23}$ | $7.53 \times 10^{-46}$ | $3.66 \times 10^{-36}$ | 42.9 | $4.17 \times 10^{18}$ |
| 300 | 49.2 | $4.12 \times 10^{-22}$ | $1.70 \times 10^{-43}$ | $8.25 \times 10^{-34}$ | 45.6 | $6.26 \times 10^{19}$ |
| 350 | 47.0 | $3.75 \times 10^{-21}$ | $1.41 \times 10^{-41}$ | $6.83 \times 10^{-32}$ | 47.8 | $5.70 \times 10^{20}$ |

**Combining the numerical approach with screening factors obtained from the WKB approximation approach**

We found that the numerical integration approach to the radial wave equation Eq. 298 yields more accurate results. At the same time, we appreciate how screening effects can be determined via the WKB approximation approach. Here, we combine results from both approaches, by starting out with the (unscreened) Koonin and Nauenberg fusion rate and applying screening factors obtained from the WKB approximation appraoch for different screening potentials. This is represented by the expression:

$$\gamma_{DD,scr} = \gamma_{DD} e^{2\Delta G_{scr}} \tag{324}$$

The results are shown in Table 8 (screened Koonin and Nauenberg rates):

**Table 8:** Screening applied to the DD fusion base rate obtained from numerical integration (drawing on $^1$S screening values, consistent with Koonin and Nauenberg 1989).

| $U_e$ (eV) | $\Delta G_{scr}$ | Fusion rate (s$^{-1}$) |
|---|---|---|
| 0 | 0.0 | $3.00 \times 10^{-64}$ |
| 50 | 21.3 | $9.51 \times 10^{-46}$ |
| 100 | 30.3 | $6.24 \times 10^{-38}$ |
| 150 | 36.5 | $1.52 \times 10^{-32}$ |
| 200 | 41.2 | $1.83 \times 10^{-28}$ |
| 250 | 44.9 | $3.00 \times 10^{-25}$ |
| 300 | 48.0 | $1.48 \times 10^{-22}$ |
| 350 | 50.5 | $2.19 \times 10^{-20}$ |

**Characterizing the validity of the WKB approximation**

Since we have numerical solutions for the $D_2$ radial wave functions, we can check to see how good the WKB approximation is for molecular $D_2$ nuclear states. In the WKB approximation, we can relate the wave function to a spatially-dependent Gamow factor according to

$$P_{DD}(r) \approx \frac{C}{\sqrt{\alpha(r)}} e^{-G(r)} \tag{325}$$

with

$$\alpha(r) = \sqrt{\frac{2\mu(V(r)-E)}{\hbar^2}}$$

$$G(r) = \int_r^{r_{max}} \alpha(r')dr' \tag{326}$$

in the forbidden region. Here $C$ is a normalization constant. Results are shown in Figure 50. We see that the WKB approximation performs well away from the fm scale.

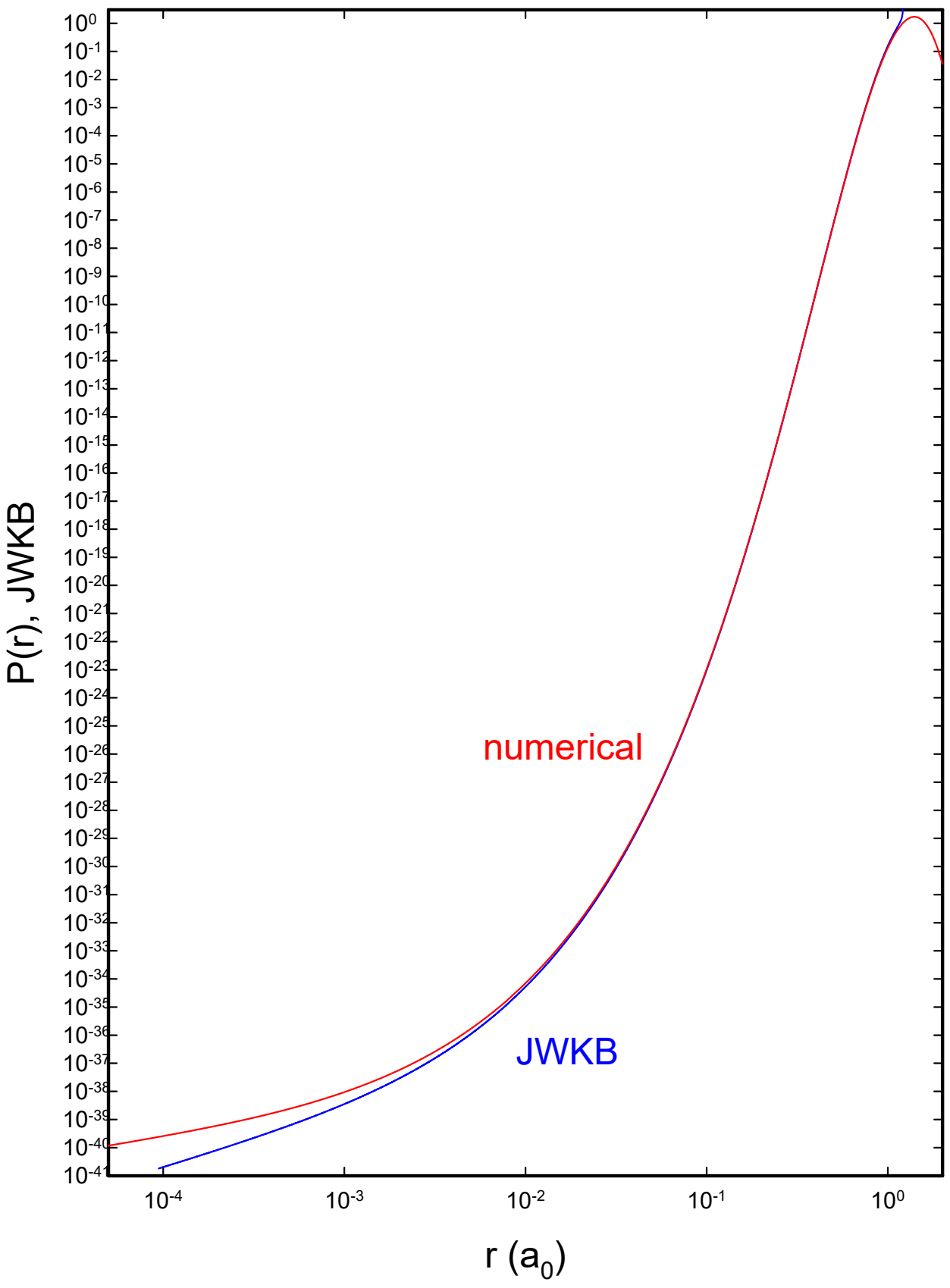

**Figure 50:** Numerical radial $D_2$ wave function (S-state), calculated with no nuclear potential (red) as a function of radius in $a_0$; WKB approximation for the same model (blue).

## 6.4 Derivation of indirect coupling coefficients in Dicke model excitation transfer

The notion of indirect coupling is simplest in the case of resonant second-order excitation transfer. Consider a simple excitation transfer problem with two identical two-level systems (TLSs) coupled to a common oscillator described by the spin-boson Hamiltonian

$$\hat{H} = \Delta E \frac{s_z^{(1)}}{\hbar} + \Delta E \frac{s_z^{(2)}}{\hbar} + \hbar\omega_0 \hat{a}^\dagger \hat{a} + V_0(\hat{a}+\hat{a}^\dagger)\frac{s_x^{(1)}}{\hbar} + V_0(\hat{a}+\hat{a}^\dagger)\frac{s_x^{(2)}}{\hbar} \tag{327}$$

It is possible to develop an approximate solution for the eigenvalue problem

$$E\Psi = \hat{H}\Psi \tag{328}$$

based on a finite basis approximation according to

$$\Psi = c_1\Phi_1 + \cdots + c_6\Phi_6 \tag{329}$$

where

$$\begin{aligned}
\Phi_1 &= \Big|\uparrow_1\downarrow_2, n\Big\rangle \\
\Phi_2 &= \Big|\downarrow_1\downarrow_2, n-1\Big\rangle \\
\Phi_3 &= \Big|\downarrow_1\downarrow_2, n+1\Big\rangle \\
\Phi_4 &= \Big|\uparrow_1\uparrow_2, n-1\Big\rangle \\
\Phi_5 &= \Big|\uparrow_1\uparrow_2, n+1\Big\rangle \\
\Phi_6 &= \Big|\downarrow_1\uparrow_2, n\Big\rangle
\end{aligned} \tag{330}$$

Resonant excitation transfer for this model would involve a transition from $\Phi_1$ to $\Phi_6$.

The expansion coefficients satisfy

$$\begin{aligned}
Ec_1 &= H_{11}c_1 + H_{12}c_2 + H_{13}c_3 + H_{14}c_4 + H_{15}c_5 \\
Ec_2 &= H_{21}c_1 + H_{22}c_2 + H_{26}c_6 \\
Ec_3 &= H_{31}c_1 + H_{33}c_3 + H_{36}c_6 \\
Ec_4 &= H_{41}c_1 + H_{44}c_4 + H_{46}c_6 \\
Ec_5 &= H_{51}c_1 + H_{55}c_5 + H_{56}c_6 \\
Ec_6 &= H_{62}c_2 + H_{63}c_3 + H_{64}c_4 + H_{65}c_5 + H_{66}c_6
\end{aligned} \tag{331}$$

where

$$
\begin{aligned}
H_{11} &= n\hbar\omega_0 \\
H_{22} &= -\Delta E + (n-1)\hbar\omega_0 \\
H_{33} &= -\Delta E + (n+1)\hbar\omega_0 \\
H_{44} &= \Delta E + (n-1)\hbar\omega_0 \\
H_{55} &= \Delta E + (n+1)\hbar\omega_0 \\
H_{66} &= n\hbar\omega_0 \\
H_{12} = H_{14} = H_{21} = H_{26} = H_{41} = H_{46} = H_{62} = H_{64} &= \sqrt{n}V_0 \\
H_{13} = H_{15} = H_{31} = H_{36} = H_{51} = H_{56} = H_{63} = H_{65} &= \sqrt{n+1}V_0
\end{aligned}
\tag{332}
$$

So far this is straightforward, and follows what one might find in a textbook. Suppose now that the excitation energy of the two-level systems is very much greater than the oscillator energy:

$$
\Delta E \gg \hbar\omega_0 \tag{333}
$$

The basis state energies of $\Phi_1$ and $\Phi_6$ are both $n\hbar\omega_0$, which is expected to be nearly resonant with the energy eigenvalue relevant for resonant excitation transfer if the coupling is weak

$$
\frac{\sqrt{n}V_0}{\Delta E} \ll 1 \tag{334}
$$

However, the other basis states are off of resonance by roughly $\Delta E$, so we might consider them to be far off of resonance. This means that the occupation of these states will be small under conditions where excitation transfer occurs (which is when the eigenvalue $E$ is near $n\hbar\omega_0$). We can solve for the (small) expansion coefficients $c_2, \cdots, c_5$ in terms of the (big) expansion coefficients $c_1$ and $c_6$ according to

$$
\begin{aligned}
c_2 &= \frac{H_{21}}{E-H_{22}}c_1 + \frac{H_{26}}{E-H_{22}}c_6 \\
c_3 &= \frac{H_{31}}{E-H_{33}}c_1 + \frac{H_{36}}{E-H_{33}}c_6 \\
c_4 &= \frac{H_{41}}{E-H_{44}}c_1 + \frac{H_{46}}{E-H_{44}}c_6 \\
c_5 &= \frac{H_{51}}{E-H_{55}}c_1 + \frac{H_{56}}{E-H_{55}}c_6
\end{aligned}
\tag{335}
$$

We can use these relations to eliminate the small expansion coefficients, which results in a contracted version of the problem that can be summarized as

$$
\begin{aligned}
Ec_1 &= (H_{11} + \Sigma_{11}(E))c_1 + H_{16}(E)c_6 \\
Ec_6 &= H_{61}(E)c_1 + (H_{66} + \Sigma_{66}(E))c_6
\end{aligned}
\tag{336}
$$

where the self-energies are

$$
\begin{aligned}
\Sigma_{11}(E) &= \frac{H_{12}H_{21}}{E-H_{22}} + \frac{H_{13}H_{31}}{E-H_{33}} + \frac{H_{14}H_{41}}{E-H_{44}} + \frac{H_{15}H_{51}}{E-H_{55}} \\
\Sigma_{66}(E) &= \frac{H_{62}H_{26}}{E-H_{22}} + \frac{H_{63}H_{36}}{E-H_{33}} + \frac{H_{64}H_{46}}{E-H_{44}} + \frac{H_{65}H_{56}}{E-H_{55}}
\end{aligned}
\tag{337}
$$

and where the indirect coupling coefficients are

$$H_{16}(E) \;=\; H_{61} \;=\; \frac{H_{12}H_{26}}{E-H_{22}} + \frac{H_{13}H_{36}}{E-H_{33}} + \frac{H_{14}H_{46}}{E-H_{44}} + \frac{H_{15}H_{56}}{E-H_{55}} \tag{338}$$

In this problem there is no direct coupling between $\Phi_1$ and $\Phi_6$, but from this discussion we see that there is indirect coupling that is accounted for by the indirect coupling coefficients $H_{16}(E)$ and $H_{61}(E)$.

In the literature one sometimes comes across a Hamiltonian of the form

$$\hat{H} \;=\; \hat{H}_0 + \hat{V} \tag{339}$$

where $\hat{V}$ accounts for a small perturbation. In the event that the intermediate states are off of resonance, as in the example above, the second-order coupling is sometimes described making use of [81]

$$\hat{H} \;\rightarrow\; \hat{H}_0 + \hat{V}\frac{1}{E-\hat{H}_0}\hat{V} \tag{340}$$

to indicate this kind of indirect coupling. This is the expression that we work with in section 5.1.

## 6.5 Loss and reduction of destructive interference for transfer from $D_2$ donors to $^4$He receivers

In the discussion in section 5.2 we are interested in the transfers from from $D_2$ donors to $^4$He receivers, where the energy from a $D_2$/$^4$He transition is used to promote a $^4$He nucleus from the ground state to the $D_2$ state. To evaluate excitation transfer rates for transfer from $D_2$ donors to $^4$He receivers, we need an estimate for the indirect coupling matrix element between the initial and final states.

The simplest approach to this is to consider only direct $D_2$/$^4$He transitions, ignoring pathways to states where energy can leave the system irreversibly (loss). In this oversimplified picture, the indirect coupling matrix element evaluated at second order using this approach is strongly hindered due to destructive interference. Taking into account loss, however, results in pathway dependent differences in contributions to the indirect coupling matrix elements. This removes some of the destructive interference and leads to a substantial indirect coupling effect.

In order for there to be a $D_2$/$^4$He transition in molecular $D_2$ in the kinds of excitation transfer schemes under discussion, the two deuterons must tunnel to get close at the few fm scale, and then there needs to be a transition to the ground state mediated by an oscillator. However, when the two deuterons are close at the fermi scale, there is a possibility that they fuse via the 3+1 state of $^4$He (see section 6.2), resulting in energetic p+t or n+$^3$He products. In such a case, the fusion energy is lost from the system because the energetic products escape irreversibly.

In some of the pathways that contribute to the indirect coupling matrix element, the intermediate state energy can be high, in which case the fusion tunnelling channel is closed. Because fusion loss depends on the pathway (in some of the pathways the loss channel is open, and in others the loss channel is closed), the presence of fusion loss can remove some of the destructive interference.

This section develops a first model for the indirect coupling matrix element that includes this effect. The scheme described here can be modeled in a non-Hermitian formalism. We can use imaginary potentials to take into account fusion loss in the pathways where fusion loss occurs. We find that the resulting indirect coupling matrix element that results is many orders of magnitude larger than what we get if loss were not included.

For reasons that will become clear in section 5.3, where we compare electric, magnetic and relativistic coupling strengths, we focus here on the $^3$P $J^\pi = 1^-$ channel as an example. This is because we find relativistic coupling to be the strongest among the available couplings and the relativistic interaction only couples from the ground state to the $^3$P $J^\pi = 1^-$ channels.

### Excitation transfer scheme for transfer from $D_2$ donors to $^4$He receivers

As noted in section 5.2, excitation transfer from $D_2$ donors to $^4$He receivers involves a downward transition from $D_2$ to $^4$He at one site and a matched upward $^4$He to $D_2$ transition at another site. To develop estimates for excitation transfer we need to evaluate the indirect coupling matrix between the initial state and final state.

To evaluate the indirect coupling matrix element between the initial state and final state, we need to take into account all of the states, couplings, and loss processes. The scheme that results is illustrated in Figure 51. We see that the fusion loss channels (denoted by red downward arrows) are only open for the 3+1 intermediate states near 0.4 MeV, and not for the 3+1 intermediate states at high energy above 20 MeV.

The fusion loss in this model occurs because there is a coupling matrix element that couples the localized d+d state to a localized 3+1 state, which tunnel decays (resulting in the canonical deuterium fusion products). We can think of the 3+1 state as a linear combination of p+t and n+$^3$He states, which have to combine into a 3+1 $T = 0$ $^3$P $J^\pi = 1^-$ channel, since the requirement that the overall 4-body wave function need to be anti-symmetric when any to particles are exchanged. It is not possible to construct individual p+t and n+$^3$He fully anti-symmetric states in this channel, but it possible to construct a 3+1 state.

This means that in general we can have direct $D_2$/$^4$He transitions as well as indirect $D_2$/(3+1)/$^4$He transitions for all excitation transfer processes involving the fusion transition. For excitation transfer from $D_2$ donors to $^4$He receivers in particular, we can have low-order contributions from two direct transitions, two indirect transitions, as well as combinations of direct and indirect transitions. The fusion loss channel will be open for some of the associated pathways, and closed for other pathways.

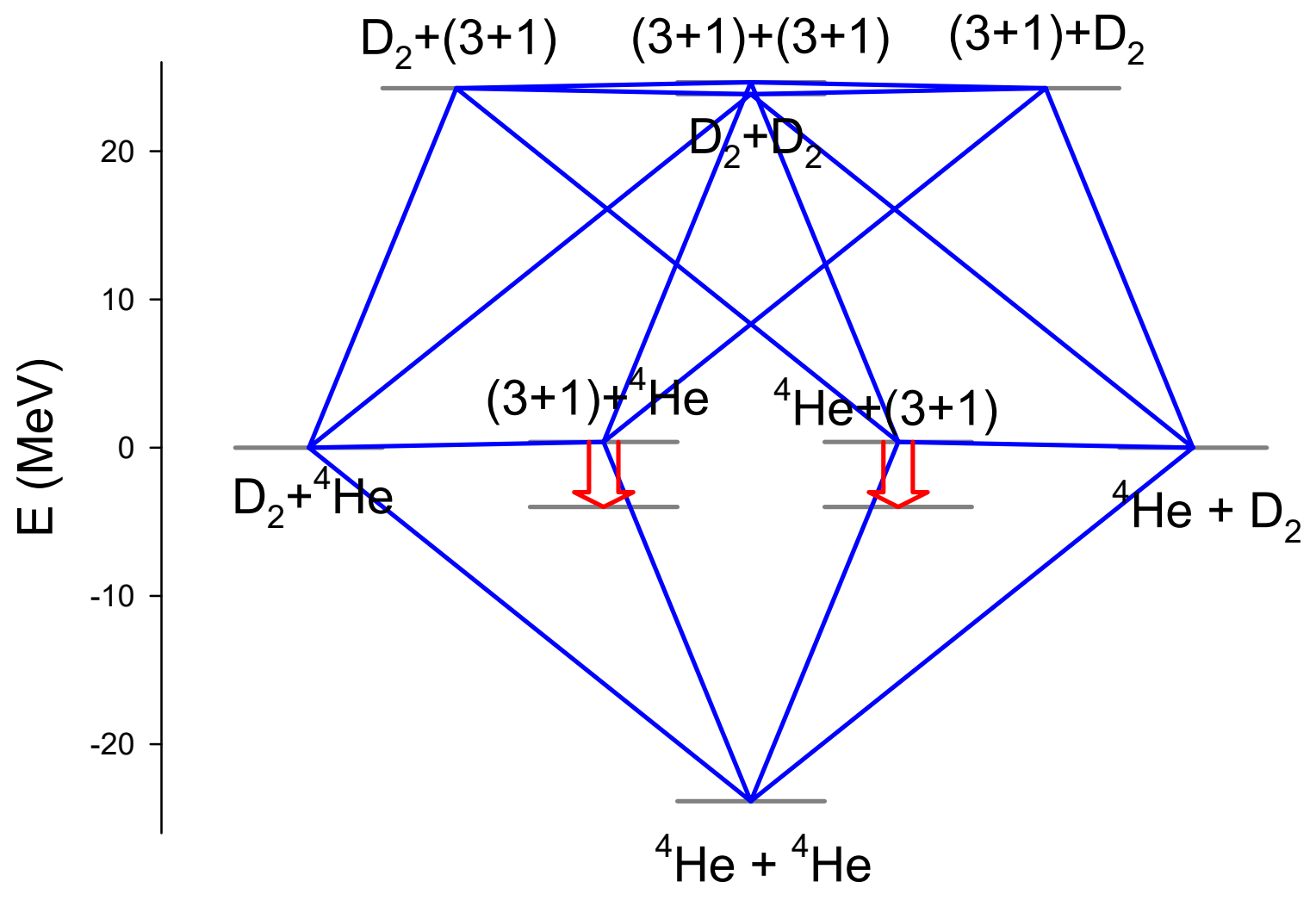

**Figure 51:** Excitation transfer scheme for transfer from $D_2$ donors to $^4$He receivers including direct and indirect transitions, as well as loss. The levels involved are indicated in gray, transitions in blue, and loss in red. State energies are indicated relative to the initial state.

### Analysis based on lossy channel wave functions

The approach pursued here uses a model in which loss is included in the calculation of the wave functions in the different channels. For example, in the $D_2$ $T = 0$ $^3$P $J^\pi = 1^-$ channel of interest, fusion decay due to coupling to the 3+1 state can be modeled through the use of a (non-Hermitian) complex potential (as discussed in 6.9). And for modeling tunnel decay in the unstable (3+1) $T = 0$ $^3$P $J^\pi = 1^-$ channel we could use boundary conditions at large $r$ appropriate for an outgoing wave.

This kind of approach leads to a version of the same scheme illustrated in Figure 52. For simplicity, we consider the simple case of excitation transfer for a single $D_2$ molecule and a single $^4$He nucleus both for the figure and for the analysis that follows, which ignores the associated Dicke factors. It is straightforward to include the Dicke factors at the end for the more general case.

### Contributing direct matrix elements

The $U'$ matrix element for the direct $D_2$/$^4$He transition is evaluated with an imaginary potential to take into account fusion loss. More specifically, the loss arises from self-energy terms resulting from coupling to the 3+1 state. The $U$ version is evaluated without an imaginary potential, since this loss pathway is energetically forbidden. Again, the excitation transfer mediating coupling considered here is relativistic coupling (as discussed in section 6.7), resulting in the $\mathbf{a} \cdot c\mathbf{P}$ matrix element.

In the absence of Dicke factors, $U$ and $U'$ are

$$
\begin{aligned}
U &= \langle \mathrm{D}_2 | \mathbf{a} \cdot c\mathbf{P} | ^4\mathrm{He} \rangle, \qquad \text{no loss} \\
U' &= \langle \mathrm{D}_2 | \mathbf{a} \cdot c\mathbf{P} | ^4\mathrm{He} \rangle, \qquad \text{with loss}
\end{aligned}
\tag{341}
$$

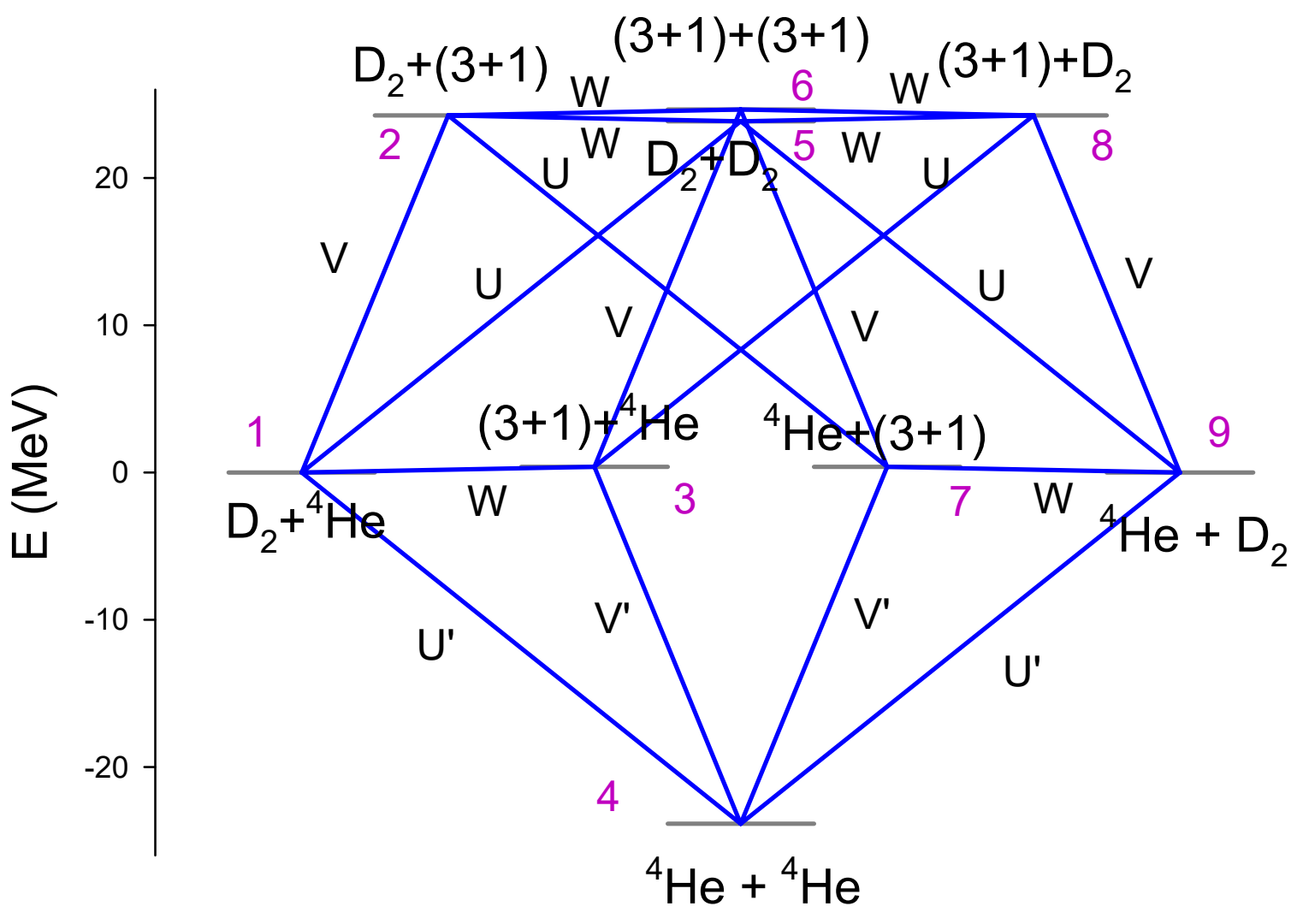

**Figure 52:** Excitation transfer scheme for transfer from $D_2$ donors to $^4$He receivers including direct and indirect transitions, where loss and tunnelling is included in the calculation of the transition matrix elements.

The momentum operator **P** creates and destroys oscillator quanta; however, our focus here is on the nuclear part of the problem.

Similarly, the $V'$ matrix element is evaluated using complex outgoing waves to take into account tunnel decay of the 3+1 state, and the $V$ matrix element is evaluated with no tunnel decay. Once again, in the absence of Dicke factors we can write

$$
\begin{aligned}
V &= \langle (3+1) | \mathbf{a} \cdot c\mathbf{P} | ^4\mathrm{He} \rangle, \qquad \text{no tunneling} \\
V' &= \langle (3+1) | \mathbf{a} \cdot c\mathbf{P} | ^4\mathrm{He} \rangle, \qquad \text{with tunneling}
\end{aligned}
\tag{342}
$$

Note that the reduction of destructive interference in this scheme comes from the difference between $U$ and $U'$, and from $V$ and $V'$. Because loss is present in some channels and not in others, terms that contribute at lowest-order in perturbation theory do not cancel.

Finally, for the $W$ matrix element we can write

$$
W = \langle \mathrm{D}_2 | V_{strong} + V_{Coulomb} | (3+1) \rangle \tag{343}
$$

where $V_{strong} + V_{Coulomb}$ represents the combined potential that includes contributions from the strong force and Coulomb force. In a more sophisticated treatment we would develop different $W$ matrix elements based on wave functions at the nuclear scale evaluated with loss or without loss. However, for simplicity here, we will model $W$ to be the same for all couplings in the model.

Now that we have defined all the parameters, we are in principle in a position to calculate the different pathways in Figure 52 and develop a sense of how much destructive interference is reduced by the 3+1 fusion loss.

**Estimating asymmetries between pathways**

The most important parameters involved in measuring the amount of destructive interference are $\eta_U = U'/U$ and $\eta_V = V'/V$. These parameters give us a sense of the asymmetries between upper and lower pathways in Figure 52. In this subsection, we will calculate those ratios.

For $U'/U$, we can use the numerical calculations of the **a**-matrix element for the $D_2$/$^4$He transition from 6.9. $U$ is calculated in the usual way yielding (here for the state $^3$P $J = 1$ $M_J = 0$)

$$U \propto \langle \Psi[^4\mathrm{He}]|\mathbf{a}|\Psi[J=1, M_J=0]\rangle = \hat{\mathbf{i}}_z \, 0.0362 \sqrt{\frac{v_{nuc}}{v_{mol}}} e^{-G} \tag{344}$$

and $U'$ is calculated by using imaginary potentials to represent loss, yielding (again for state $^3$P $J = 1$ $M_J = 0$)

$$U' \propto \langle \Psi[^4\mathrm{He}]|\mathbf{a}|\Psi[J=1, M_J=0]\rangle = \hat{\mathbf{i}}_z \, (0.0362 + i0.000477) \sqrt{\frac{v_{nuc}}{v_{mol}}} e^{-G} \tag{345}$$

We can then write for $\eta_U$

$$\eta_U = \frac{U'}{U} = \frac{0.0362 + i0.000477}{0.0362} = 1.000 + i0.013 \tag{346}$$

and

$$|1 - \eta_U| = 0.013 \tag{347}$$

For $V'/V$ we we need the **a**-matrix element for the (3+1)/$^4$He transition. We have developed numerical calculations for this transition based on a construction similar to that described in 6.9 using an approximate 3+1 state wave function for the no-loss case. The result for $V$, *i.e.*, the no loss case, is (for state $^3$P $J = 1$ $M_J = 0$)

$$V \propto \langle \Psi[^4\mathrm{He}]|\mathbf{a}|(3+1), \Psi[J=1, M_J=0]\rangle = -\hat{\mathbf{i}}_z \, 0.0030 \tag{348}$$

and $V'$ is calculated by using imaginary potentials to represent loss, yielding (again for the $^3$P $J = 1$ $M_J = 0$ state)

$$V' \propto \langle \Psi[^4\mathrm{He}]|\mathbf{a}|(3+1), \Psi[J=1, M_J=0]\rangle = -\hat{\mathbf{i}}_z \, (0.00035 + i0.00017) \tag{349}$$

We can then write for $\eta_V$

$$\eta_V = \frac{V'}{V} = \frac{0.00035 + i0.00017}{0.00030} = 1.16 - i0.55 \tag{350}$$

and

$$|1 - \eta_V| = 0.57 \tag{351}$$

### Discussion

Recall that $|1-\eta|$ is defined such that in the case of complete destructive interference $|1-\eta|=0$ and in the case of ideal loss (and complete breakage of destructive interference) $|1-\eta|=1$. We saw above that many pathways need to be considered and their relative contributions. However, in order to build some intuition, we estimated the asymmetry ratios for U and V based pathways. The result of this exercise was $|1-\eta_U| = 0.013$ and $|1-\eta_V| = 0.57$. This prompted us to estimate, in first approximation, an initial overall $|1-\eta|$ of 0.1. This is the value that we used in rate estimates in section 5.3 (Eq. 83).

A more detailed approach, which we undertook subsequently, is briefly laid out below and is intended to be pursued in future work. Our first rough estimation, based on this approach, suggests that the reduction of destructive interference is substantially higher than the initially assumed $|1-\eta|$ of 0.1. This suggests that the estimated rates may in fact be higher and that our initial estimate was a conservative one.

### Outlook towards more detailed estimates

To develop more detailed estimates for excitation transfer rates we need to evaluate the indirect coupling matrix between the initial state and final state. This involves taking into account all of the states, couplings, and loss processes. There are some technical issues associated with this, since this system cannot be described with two-level systems. The leading order contribution to the indirect coupling matrix element for this model is

$$\begin{aligned} H_{19} \;\rightarrow\; & \frac{|U'|^2-|U|^2}{\Delta Mc^2} + \frac{2UV-(U')^*V'-U'(V')^*}{\Delta Mc^2}\left(\frac{W}{\delta Mc^2}\right) + \frac{|V'|^2-|V|^2}{\Delta Mc^2}\left(\frac{W}{\delta Mc^2}\right)^2 \\ = \; & (|\eta_U|^2-1)\frac{U^2}{\Delta Mc^2} + (2-\eta_U^*\eta_V-\eta_U\eta_V^*)\frac{UV}{\Delta Mc^2}\left(\frac{W}{\delta Mc^2}\right) + (|\eta_V|^2-1)\frac{V^2}{\Delta Mc^2}\left(\frac{W}{\delta Mc^2}\right)^2 \end{aligned} \qquad (352)$$

The first term in Eq. 352 is due to direct $D_2$/$^4$He transitions on both the donor and receiver side—this is the only term considered in section 5.3. Note that for our case, where $|\eta_U|\approx 1$, the term $(|\eta_U|^2-1)$ approximates to $2(\mathrm{Re}(\eta_U)-1)$, as used in Eq. 77.

The last term in in Eq. 352 is due to indirect $D_2$/(3+1)/$^4$He transitions on both the donor and receiver side; and the second term is due to combinations of direct and indirect transitions.

The energy differences are

$$\begin{aligned} \Delta Mc^2 &= 2M_dc^2 - M_{^4He}c^2 = 23.85\ \text{MeV} \\ \delta Mc^2 &= M_{3+1}c^2 - 2M_dc^2 = 0.40\ \text{MeV} \end{aligned} \qquad (353)$$

We are familiar with $\Delta Mc^2$ being the fusion transition energy. To estimate the smaller energy difference $\delta Mc^2$ we make use of the $^4$He$^*$ 3+1 $T=0$ $J^\pi=1^-$ state at 24.25 MeV listed in [100].

As to the remaining variables, estimates for $U$ and $U'$ as well as $V$ and $V'$ are given in terms of the **a**-matrix elements above and involves averaging over the oscillator part of the wavefunction as is done in section 6.15. What remains is to calculate $W$. We have not yet developed a calculation for this, however, we can develop an estimate for the magnitude of the matrix element based on the calculations in [100].

The idea here is that the fusion loss rate for the microscopic d+d $^4$He$^*$ $T=0$ $^3$P $J^\pi=1^-$ state at 28.37 MeV is

$$\hbar\gamma_{d+d,^3P} = 0.15\ \text{MeV} \qquad (354)$$

This is a result of coupling with the (3+1) $^4$He$^*$ $T = 0$ $^3$P $J^\pi = 1^-$ state at 24.25 MeV, for which the tunneling rate is

$$\hbar\gamma_{tunnel} \ = \ 5.95 \text{ MeV} \tag{355}$$

We can use these numbers to extract a magnitude for the coupling matrix element

$$|\langle d+d|V_{strong} + V_{Coulomb}|(3+1)\rangle| \ = \ 1.15 \text{ MeV} \tag{356}$$

If we assume that this (local) matrix element is essentially the same as the molecular version $\langle \mathrm{D}_2|V_{strong}+V_{Coulomb}|(3+1)\rangle$ corrected for the (off-resonant) occupation of the local d+d state, we can write

$$\begin{aligned} |W| \ &= \ \sqrt{p_{d+d,{}^3P}}|\langle d+d|V_{strong} + V_{Coulomb}|(3+1)\rangle| \\ &= \ \left(102\sqrt{\frac{v_{nuc}}{v_{mol}}}e^{-G}\right)\left(1.15 \text{ MeV}\right) \end{aligned} \tag{357}$$

We now plug the estimates into $H_{19}$ and compare with the our initial approximation of $(|\eta|^2 - 1)\frac{U^2}{\Delta M c^2}$ (with $\eta = 0.9$). We find that the rates derived from our initial approximation are about 300 times smaller compared to rates based on the full $H_{19}$. This makes our initial estimate a conservative one.

## 6.6 Loss and reduction of destructive interference for transfer from $D_2$ donors to Pd receivers

In the discussion in section 5.4 we are interested in excitation transfer from the $D_2/^4He$ fusion transition to a Pd*/Pd transition, where the energy from a $D_2/^4He$ transition is used to promote a Pd nucleus from the ground state to a Pd* state. To evaluate excitation transfer rates (as done in section 5.5) we need an estimate for the indirect coupling matrix element between the initial and final states.

Based on the discussion in subsection 5.2, we expect that destructive interference will hinder excitation transfer if we do not include path-dependent loss. We also know that for the fusion transition that there is a contribution from the direct $D_2/^4He$ transition (that has path-dependent loss due to fusion decay), and also a contribution from the indirect $D_2/(3+1)/^4He$ transition (that has path-dependent loss due to tunnel decay). We would like to take into account both transitions in our analysis.

### Excitation transfer scheme

To evaluate the indirect coupling matrix element between the initial state and final state, we need to take into account all of the states, couplings, and loss processes. The scheme that results is illustrated in Figure 53. We see that the fusion loss channels (denoted by red downward arrows) are only open for the 3+1 intermediate states near 0.4 MeV, and not for the 3+1 intermediate states at high energy above 20 MeV.

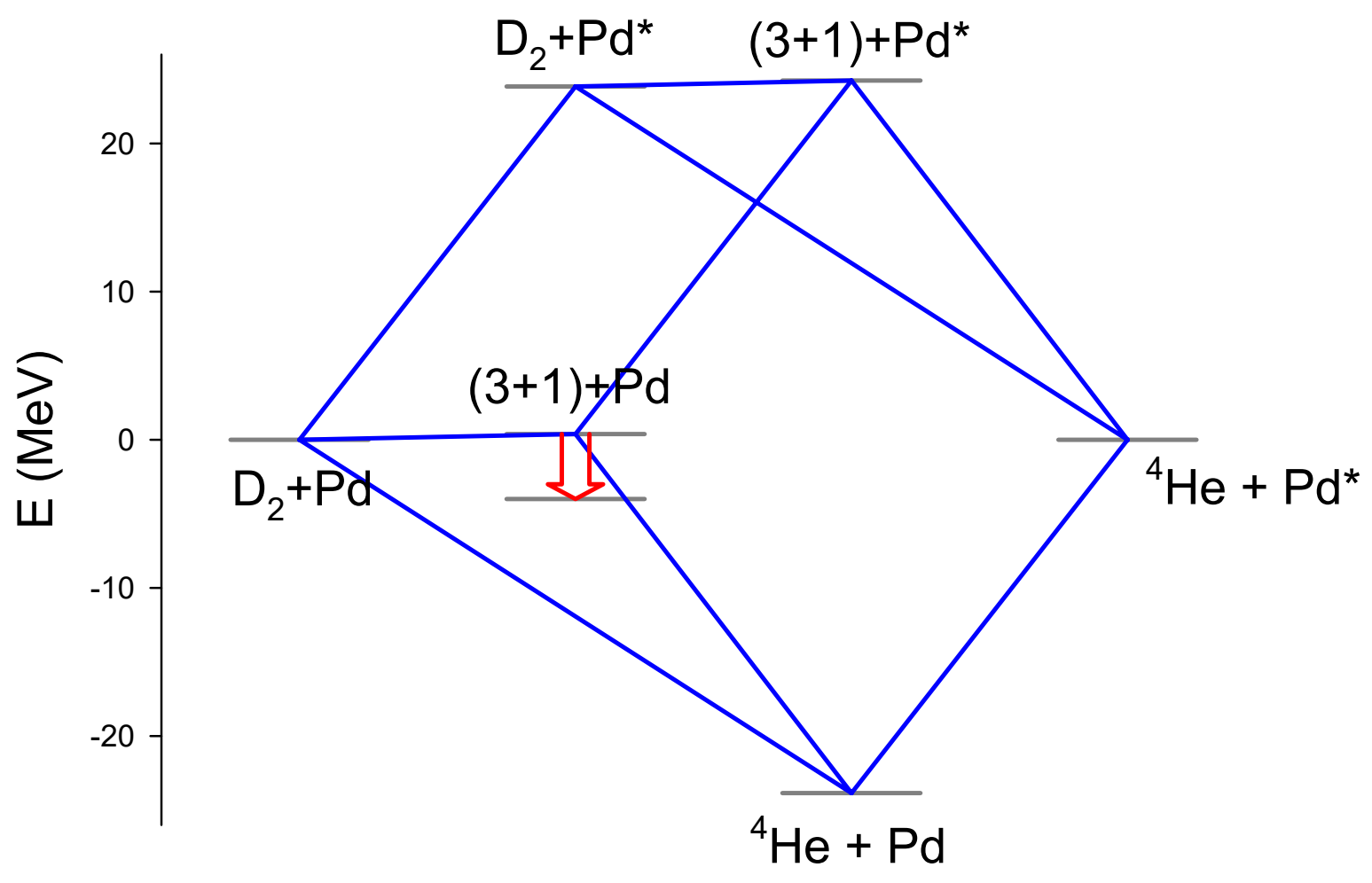

**Figure 53:** Excitation transfer scheme for transfer from $D_2$ donors to Pd receivers including direct and indirect transitions, as well as loss. The levels involved are indicated in gray, transitions in blue, and loss in red. State energies are indicated relative to the initial state.

### Analysis based on lossy channel wave functions

In the discussion of the indirect coupling matrix element for the transfer from $D_2$ donors to $^4He$ receivers, an approach was described in which fusion loss associated with the direct $D_2/^4He$ transitions was taken into account by making use of an imaginary potential in the calculation of the relative wave function, and where tunnel decay in the 3+1 channel was taken into account through the use of boundary conditions corresponding to an outgoing wave at large separation.

This kind of approach leads to a version of the same scheme illustrated in Figure 54. For simplicity, we consider the simple case of excitation transfer for a single $D_2$ molecule and a single Pd nucleus both for the figure and for the analysis that follows, which ignores the associated Dicke factors and makes things simpler. It is straightforward to include the Dicke factors at the end for the more general case.

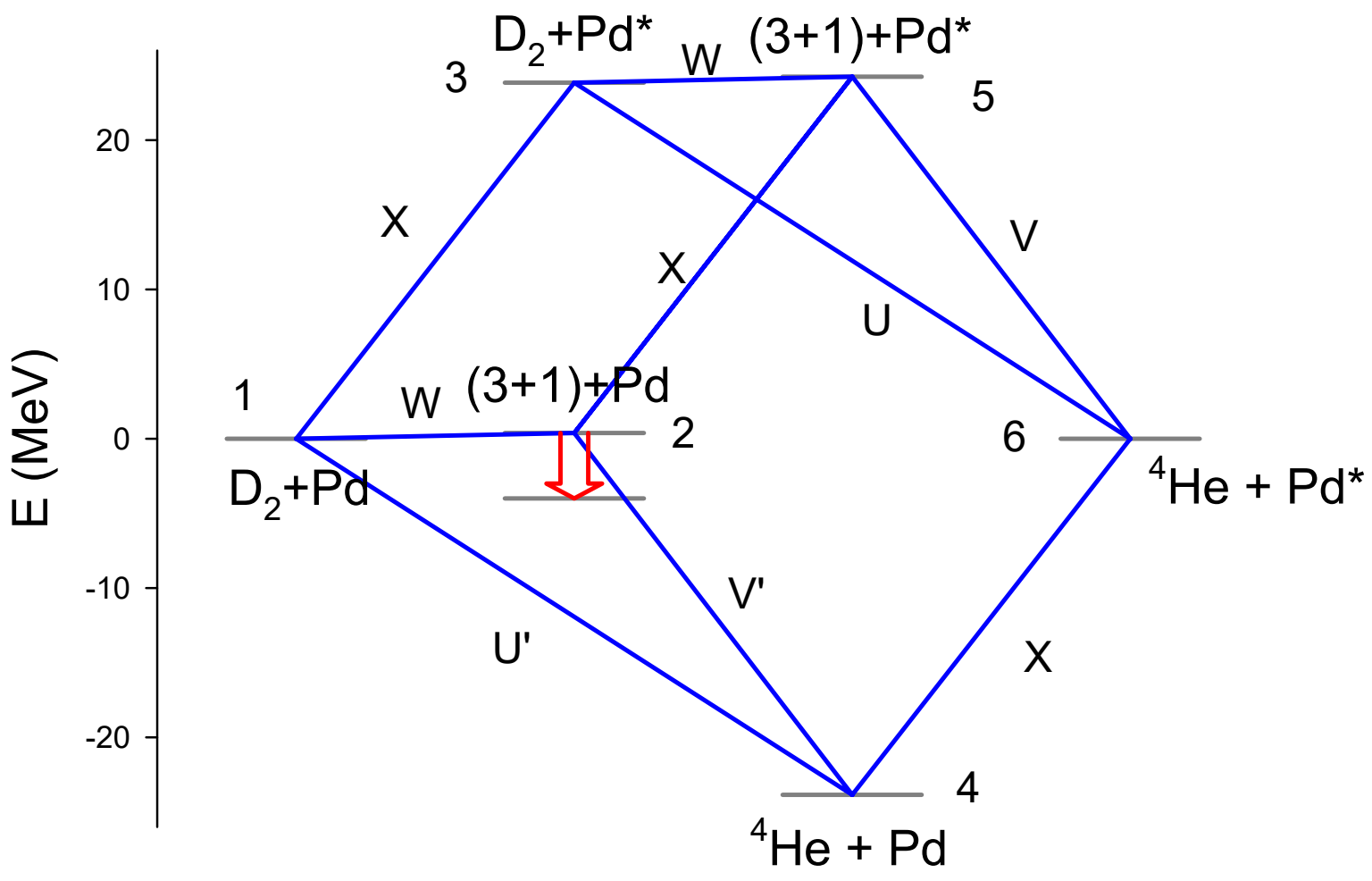

**Figure 54:** Excitation transfer from the fusion transition to the Pd*/Pd transition including direct and indirect transitions, as well as loss. State numbering and matrix elements are indicated.

## Contributing direct matrix elements

Using this approach the $U'$ matrix element for the direct $D_2$/$^4$He transition is evaluated with an imaginary potential to take into account fusion loss, and the $U$ version is evaluated without an imaginary potential. In the absence of Dicke factors this leads to

$$\begin{aligned} U &= \langle D_2 | \mathbf{a} \cdot c\mathbf{P} | ^4\text{He} \rangle, \quad \text{no loss} \\ U' &= \langle D_2 | \mathbf{a} \cdot c\mathbf{P} | ^4\text{He} \rangle, \quad \text{with loss} \end{aligned} \tag{358}$$

The momentum operator **P** creates and destroys oscillator quanta; however, our focus here is on the nuclear part of the problem

Similarly, the $V'$ matrix element is evaluated using complex outgoing waves to take into account tunnel decay of the 3+1 state, and the $V$ matrix element is evaluated with no tunnel decay. Once again, in the absence of Dicke factors we can write

$$\begin{aligned} V &= \langle (3+1) | \mathbf{a} \cdot c\mathbf{P} | ^4\text{He} \rangle, \quad \text{no tunneling} \\ V' &= \langle (3+1) | \mathbf{a} \cdot c\mathbf{P} | ^4\text{He} \rangle, \quad \text{with tunneling} \end{aligned} \tag{359}$$

Note that the reduction of destructive interference in this scheme comes from the difference between $U$ and $U'$, and from $V$ and $V'$. Because loss is present in some channels and not in others, terms that contribute at lowest-order in perturbation theory do not cancel.

For the $W$ matrix element we can write

$$W = \langle \text{D}_2 | V_{strong} + V_{Coulomb} | (3+1) \rangle \tag{360}$$

In a more sophisticated treatment we would develop different $W$ matrix elements based on wave functions at the nuclear scale evaluated with loss or without loss. However, for simplicity here we will model $W$ to be the same for all couplings in the model.

For the $X$ matrix element we can write

$$X = \langle \text{Pd} | \mathbf{a} \cdot c\mathbf{P} | \text{Pd}^* \rangle \tag{361}$$

Near 23.85 MeV there are unstable $Pd^*$ excited states that can tunnel decay. We will not include this possibility in the model under discussion.

Now that we have defined all the parameters, we're in principle a position to calculate the different pathways in Figure 54 and develop a sense of how much destructive interference is reduced by the fusion loss.

### Estimating asymmetries between pathways

Inspecting Figure 54, we can readily identify a transfer path involving $X$ and $V$ on the top and $X$ and $V'$ on the bottom as well as a path involving $X$ and $U$ on the top and $X$ and $U'$ on the bottom. Since $X$ is not affected by the 3+1 fusion loss, and therefore remains the same value on the top and the bottom, asymmetry again results from $\eta_U = U'/U \neq 1$ and $\eta_V = V'/V \neq 1$.

The results for $\eta_U$ and $\eta_V$ from the last section 6.5 therefore still apply in this scenario. Specifically,

$$|1 - \eta_U| = 0.013 \tag{362}$$

and

$$|1 - \eta_V| = 0.57 \tag{363}$$

### Discussion

The asymmetries in the different pathways are driven by the same ratios—namely $\eta_U$ and $\eta_V$—as in case where we have $D_2$ donors to $^4He$ receivers. We therefore continue to work with the same estimate $|1 - \eta = 0.1|$ as the correction factor representing the deviation from ideal loss (complete elimination of destructive interference) - see section 5.5 and following sections with rate estimates.

Again, as in the previous section 6.5, a more detailed analysis involves the evaluation of the complete indirect matrix element considering all pathways, in their relative contributions, from the initial to the final states. How this is done is outlined in the following subsection and will be implemented in future work.

As in the previous section, a first attempt at this suggests that the estimated rates may in fact be higher and that our initial estimate was a conservative one.

**Outlook towards more detailed estimates**

To develop more detailed estimates for excitation transfer rates we need to evaluate the indirect coupling matrix between the initial state and final state. This involves taking into account all of the states, couplings, and loss processes. There are some technical issues associated with this, since this system cannot be described with two-level systems. The leading order contribution to the indirect coupling matrix element for this model is

$$\begin{aligned} H_{16} \;\rightarrow\; & \frac{(U'-U)X}{\Delta Mc^2} - \frac{(V'-V)X}{\Delta Mc^2}\left(\frac{W}{\delta Mc^2}\right) \\ = (\eta_U - 1)U\left(\frac{X}{\Delta Mc^2}\right) & - (\eta_V - 1)V\left(\frac{X}{\Delta Mc^2}\right)\left(\frac{W}{\delta Mc^2}\right) \end{aligned} \tag{364}$$

The first term in Eq. 364 is due to a direct $D_2$/$^4$He transition - this is the only term considered in section 5.5. The second term in in Eq. 364 the second term is due to an indirect $D_2$/(3+1)/$^4$He transition.

The energy differences are

$$\begin{aligned} \Delta Mc^2 \;&=\; 2M_d c^2 - M_{^4He}c^2 \;=\; 23.85 \text{ MeV} \\ \delta Mc^2 \;&=\; M_{3+1}c^2 - 2M_d c^2 \;=\; 0.40 \text{ MeV} \end{aligned} \tag{365}$$

We are familiar with $\Delta Mc^2$ being the fusion transition energy. To estimate the smaller energy difference $\delta Mc^2$ we make use of the $^4$He$^*$ 3+1 $T = 0$ $J^\pi = 1^-$ state at 24.25 MeV listed in [100].

As to the remaining variables, estimates for $U$ and $U'$ as well as $V$ and $V'$ and $W$ given in the previous section 6.5

We now plug the estimates into $H_{16}$ and compare with the our initial approximation of $(\eta - 1)\frac{UX}{\Delta Mc^2}$ (with $\eta = 0.9$). We find that the rates derived from our initial approximation are about 10 times smaller compared to rates based on the full $H_{16}$. This makes our initial estimate a conservative one.

## 6.7 Relativistic phonon-nuclear coupling

We can make use of a Lorentz transformation to boost the at-rest wave function of a nucleus in order to develop a description in a relativistic moving frame (which would include contraction and spin rearrangement). Nuclear motion in a lattice is not close to being relativistic; however, the same basic idea applies. We expect the wave function of a nucleus that is vibrating to change depending on which direction it moves, and how fast it moves. When we expand the internal nuclear wave function in terms of the rest frame states, we find that an admixture of rest frame states develops, and it depends on the specifics of the velocity and direction. This argument implies the existence of a relativistic coupling between vibrations and internal nuclear states as a consequence of Lorentz invariance (see [121, 122, 105, 93], for more detailed versions of this argument).

Although nucleons are not Dirac particles, the Dirac equation is sometimes used to model them in order to account for relativistic effects. It is possible to extract the interaction Hamiltonian from a many-particle Dirac Hamiltonian directly. The lowest-order contribution to the interaction can be written in the form

$$\hat{H}_{int} \ = \ \mathbf{a} \cdot c\mathbf{P} \tag{366}$$

where **a** can be written in terms of Dirac $\boldsymbol{\alpha}$ and $\beta$ matrices according to (see Eq. 9 in Hagelstein 2023 [93])

$$\mathbf{a} \ = \ \frac{1}{Mc} \sum_j \beta_j \hat{\boldsymbol{\pi}}_j + \frac{1}{2Mc^2} \sum_{j<k} \left[ \beta_j \boldsymbol{\alpha}_j + \beta_k \boldsymbol{\alpha}_k, \hat{V}_{jk} \right] \tag{367}$$

One can find discussions in the literature, where researchers were interested in the elimination of this kind of interaction for uniform motion in free space (*e.g.*, [123]); however, only recently has it been proposed that this coupling might be important for coherent nuclear dynamics mediated by phonon exchange [124, 93]. As a result, most physicists are not familiar with this interaction and more research is needed to investigate it systematically.

In connection with the effects under discussion in this document, we work with this interaction due to its comparatively high strength compared to other lattice-nuclear interactions (see the discussion in section 5.3).

Matrix element calculations for different transitions of interest here are given in sections 6.9 ($D_2$/$^4$He transition), 6.10 (HD/$^3$He transition) and 6.11 (Pd*/Pd transition).

## 6.8 Selection rules

Selection rules play an important role in the discussion in this paper, which motivates us to consider selection rules and how they come into the various problems that we are interested in.

**Example: Selection rules associated with angular momentum**

At the most basic level, a selection rule is connected to the issue of whether a transition matrix element is finite or zero. The most straightforward illustration of this for us might be in the case of angular momentum. Suppose that we are considering electric dipole transitions between electrons of atomic hydrogen in a nonrelativistic approximation. In this case we might have two electronic states that we are interested in with wave functions given by

$$\begin{aligned} \psi_i &= \psi_{nlm} = R_{nl}(r)Y_{l,m}(\theta,\phi) \\ \psi_f &= \psi_{n'l'm'} = R_{n'l'}(r)Y_{l',m'}(\theta,\phi) \end{aligned} \tag{368}$$

If we consider the interaction with an electric field that is uniform over the atom, the interaction matrix element of interest would be

$$\langle\psi_f| - \mathbf{d}\cdot\mathbf{E}|\psi_i\rangle = \langle R_{n'l'}(r)Y_{l',m'}(\theta,\phi)|e\mathbf{r}\cdot\mathbf{E}|R_{nl}(r)Y_{l,m}(\theta,\phi)\rangle \tag{369}$$

For a transition to be driven by the electric field, the matrix element must be finite. Although simply stated, this is not such an easy problem. Since the electronic wave functions are known in terms of radial and angular pieces, it is most straightforward to express the dipole operator in terms of radial and spherical coordinates. This leads to

$$\begin{aligned} \mathbf{r} &= \hat{\mathbf{i}}_x x + \hat{\mathbf{i}}_y y + \hat{\mathbf{i}}_z z \\ &= \hat{\mathbf{i}}_x\sqrt{\frac{2\pi}{3}}r\Big(-Y_{1,1}(\theta,\phi)+Y_{1,-1}(\theta,\phi)\Big) + \hat{\mathbf{i}}_y\sqrt{\frac{2\pi}{3}}ir\Big(Y_{1,1}(\theta,\phi)+Y_{1,-1}(\theta,\phi)\Big) + \hat{\mathbf{i}}_z\sqrt{\frac{4\pi}{3}}rY_{1,0}(\theta,\phi) \end{aligned} \tag{370}$$

We can use this to write the matrix element as

$$\begin{aligned} \langle\psi_f| - \mathbf{d}\cdot\mathbf{E}|\psi_i\rangle &= e\sqrt{\frac{4\pi}{3}}\langle R_{n'l'}|r|R_{nl}\rangle\Bigg(E_x\frac{\langle Y_{l',m'}|-Y_{1,1}+Y_{1,-1}|Y_{l,m}\rangle}{\sqrt{2}} \\ &\quad + iE_y\frac{\langle Y_{l',m'}|Y_{1,1}+Y_{1,-1}|Y_{l,m}\rangle}{\sqrt{2}} + E_z\langle Y_{l',m'}|Y_{1,0}|Y_{l,m}\rangle\Bigg) \end{aligned} \tag{371}$$

Whether or not the transition matrix element is zero or not depends on the angular integral

$$\langle Y_{l_1,m_1}|Y_{l_2,m_2}|Y_{l_3,m_3}\rangle = (-1)^{m_1}\sqrt{\frac{(2l_1+1)(2l_2+1)(2l_3+1)}{4\pi}}\begin{pmatrix} l_1 & l_2 & l_3 \\ 0 & 0 & 0 \end{pmatrix}\begin{pmatrix} l_1 & l_2 & l_3 \\ -m_1 & m_2 & m_3 \end{pmatrix} \tag{372}$$

where the angular matrix element is expressed in terms of Wigner 3-j symbols.

Selection rules ultimately comes from the calculation of the transition matrix element. If we evaluate the angular integral, we will get zero unless

$$-m_1 + m_2 + m_3 \;=\; 0 \tag{373}$$

which we conclude is a consequence of conservation of $z$-directed angular momentum. This is a selection rule. When we evaluate the integral we will find that we also get zero unless

$$(-1)^{l_1+l_2+l_3} \;=\; 0 \tag{374}$$

which we conclude is a consequence of conservation of parity. We will also get zero unless the triangle selection rules

$$\begin{aligned} |l_1 - l_2| \;&\leq\; l_3 \leq l_1 + l_2 \\ |l_1 - l_3| \;&\leq\; l_2 \leq l_1 + l_3 \\ |l_2 - l_3| \;&\leq\; l_1 \leq l_2 + l_3 \end{aligned} \tag{375}$$

which we conclude is a consequence of total angular momentum conservation.

A similar approach works in the case of magnetic transitions.

### Selection rules for $^4$He in LS coupling

We might consider $^4$He to be in LS coupling in connection with radiative capture $d(d,\gamma)\alpha$ reaction as a first approximation (the situation is more complicated than is described by pure LS coupling). In this kind of a picture we would not expect radiative capture to occur through electric dipole (E1) interactions or magnetic dipole (M1) interactions. An electric dipole interaction would couple the ground state $^4$He $^1$S configuration to excited states (including continuum d+d channels) in $^1$P channels. But it is not possible to construct wave functions that are fully anti-symmetric for $^1$P states with two deuterons. Similarly, a magnetic dipole interaction would couple the ground state $^1$S to excited states (including continuum d+d channels) in $^3$S channels. Once again, it is impossible to construct anti-symmetric wave functions for two deuterons in a $^3$S state.

The conclusion from this argument is that electric and magnetic dipole transitions are forbidden in the radiative capture $d(d,\gamma)\alpha$ reaction. There is no low-order coupling with d+d $L = 0$ channels in LS coupling, which would lead to an expectation that the astrophysical S-factor (a measure of the cross section) should vanish at zero relative energy for these gamma capture reactions. However, we know from experiment [91] that both electric and magnetic dipole interactions are observed in such experiments. A conclusion that can be drawn is that simple LS coupling is insufficient to describe $d(d,\gamma)\alpha$ reactions. This is because the strong force does not respect LS coupling.

### Selection rules and tensor operators

Relativistic interactions and the nuclear potential both lead to a mixing between orbital angular momentum **L** and spin **S**, so that neither orbital angular momentum nor spin are good quantum numbers. However, total angular momentum **J**

$$\mathbf{J} \;=\; \mathbf{L} + \mathbf{S} \tag{376}$$

can be a good quantum number.

In light of this, we might expect to need to construct proper antisymmetric wave functions in LSJ coupling, and then the selection rules would come from the brute force evaluation of the resulting matrix elements. This situation would be discouraging if detailed calculations were needed for every problem that we wished to consider. What we would like for this situation is a generalization of the dipole calculation above that could work for more complicated LSJ wave functions.

This problem was considered early on by Wigner and others, with the conclusion that it is possible to develop a suitable generalization. The idea focuses on the abstract notion of a tensor operator, which is an operator that transforms like a tensor under a rotation

$$T_q^{(k)} = \sum_{q'} T_{q'}^{(k)} D_{q,q'}^{(k)}(\alpha, \beta, \gamma) \tag{377}$$

where $D_{q,q'}^{(k)}(\alpha, \beta, \gamma)$ is a rotation matrix and where $\alpha, \beta, \gamma$ here are Euler angles associated with a rotation [125]. This is a requirement for operators in physical systems, as we would like the underlying physics not to change if we work in a coordinate system that has been rotated. A test for whether an operator is a tensor operator is if an the operator satisfies

$$[J_\pm, T_q^{(k)}] = \hbar\sqrt{(k \mp q)(k \pm q + 1)} T_{q\pm 1}^{(k)} \tag{378}$$

where $J_\pm$ are raising and lowering operators for total angular momentum

$$\begin{aligned} J_+ &= J_x + iJ_y \\ J_- &= J_x - iJ_y \end{aligned} \tag{379}$$

The spherical harmonic $Y_{l,m}$ functions are tensor operators $T_m^{(l)}$ in this sense.

Matrix elements of tensor operators satisfy the Wigner-Eckart theorem

$$\langle jm|T_q^{(k)}|j'm'\rangle = \langle j||T^{(k)}||j'\rangle\langle j'm'; kq|jm\rangle \tag{380}$$

where $\langle j||T^{(k)}||j'\rangle$ is a reduced matrix element, and where $\langle j'm'; kq|jm\rangle$ is a Clebsch-Gordan coefficient. This can be used to determine selection rules now for more complicated atomic and nuclear systems described by LSJ coupling. If we make use of this formula to calculate matrix elements, we find that unless

$$-m' + q + m = 0 \tag{381}$$

is satisfied, the matrix elements are zero. We interpret this in terms of conservation of total momentum in the $z$-direction. Matrix elements vanish unless

$$(-1)^{\pi + \pi_q + \pi'} = 1 \tag{382}$$

where $\pi$ in this equation indicates parity. We can think of this as a statement that the spatial integrals will be zero if the integrand is odd in one of the spatial dimensions. The matrix elements are zero unless the triangle inequalities hold

$$\begin{aligned} |J - k| &\leq J' \leq J + k \\ |J - J'| &\leq k \leq J + J' \\ |k - J'| &\leq J \leq k + J' \end{aligned} \tag{383}$$

This comes from the requirement that overall total angular momentum must be conserved for there to be a transition.

### Selection rules for $^4$He in LSJ coupling

As mentioned above, observations of gamma multipolarity in the $d(d,\gamma)\alpha$ are not consistent with selection rules in LS coupling assuming a pure $^1$S $^4$He ground state. The strong force mixes different LS configurations to produce a $J^\pi = 0^+$ ground state, and we can make use of electric dipole or magnetic dipole interactions as tensor operators to understand the selection rules. The electric dipole operator **d** is a rank one tensor operator with negative parity, and the magnetic dipole operator $\boldsymbol{\mu}$ is a rank one tensor operator with positive parity.

However, LS selection rules are easier to visualize. People often think of the $^4$He ground state as made up of combinations of $^1$S, $^3$P and $^5$D states, all coupled to produce $J = 0$ states. For gamma capture at low energy, the contribution from the d+d $^5$S channel dominates due to electric quadrupole interactions coupling to the $^5$D admixture of the $^4$He ground state, aided by the short wavelength of the 23.85 MeV gamma. At higher energy there appears magnetic quadrupole (M2) coupling between the d+d $^3$P channel and the $^1$S component of the ground state. Magnetic dipole coupling is possible from the d+d 3P $J^\pi = 1^-$ channel to the $^3$P admixture of the $^4$He ground state. Electric dipole interactions can provide coupling from p+t and n+$^3$He admixtures of the d+d $J^\pi = 1^-$ channel to the ground state $J^\pi = 0^+$ [91].

### Selection rules and the a $\cdot$ $c$P interaction

The **a** $\cdot$ $c$**P** operator is a rank one ($k = 1$) tensor operator with negative parity, and can be verified to satisfy Equation (378) [105]. This means that we have good associated selection rules that we can take advantage of. For transitions from the $^4$He ground state (which is a $J^\pi = 0^+$ state), this operator can couple to $J^\pi = 1^-$ states. Since there is no dependence on nucleon charge, total isospin is preserved. This means that transitions are possible only to states with $T = 0$ $J^\pi = 1^-$. These include the molecular $D_2$ $^3$P $J^\pi = 1^-$ states (which have $T = 0$ since deuterons are isospin zero), and $^4$He$^*$ excited states also with $T = 0$ and with $J^\pi = 1^-$, which include the 3+1 state at 24.25 MeV and the d+d state at 28.37 MeV (illustrated in Figure 55).

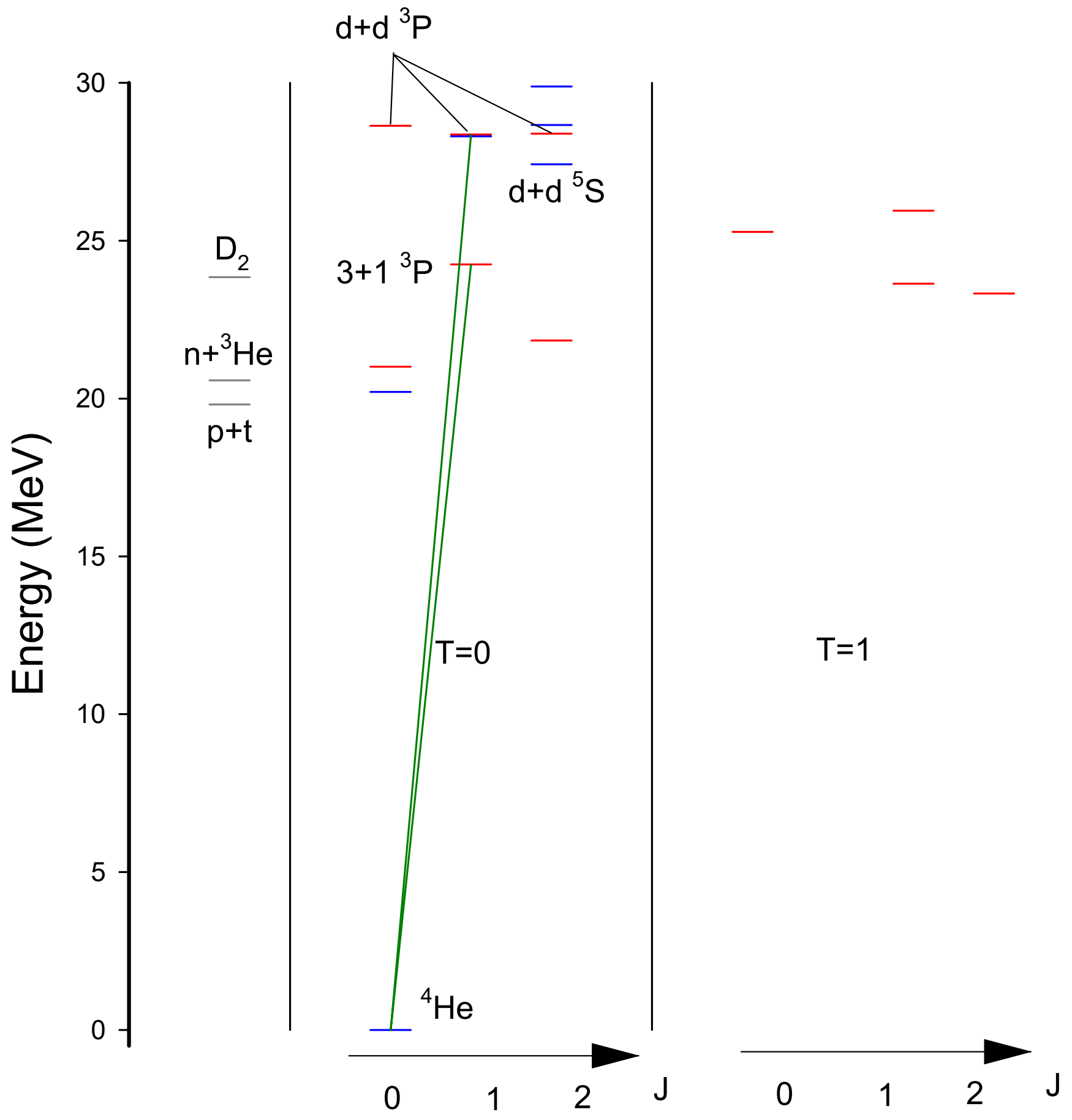

**Figure 55:** Allowed $\mathbf{a} \cdot c\mathbf{P}$ transitions from the $^4$He ground state to the $^4$He* excited levels with $J^\pi = 1^-$ are indicated in dark green.

## 6.9 Estimating the a-matrix element for the $D_2$/$^4$He transition

This section provides a brief summary of the calculation of the $\mathbf{a} \cdot c\mathbf{P}$ matrix element for the $D_2$/$^4$He transition (see section 6.7 for context). The two-body relativistic phonon-nuclear interaction satisfies an identity that allows matrix elements to be evaluated based on one-body operators according to (see Eq. 27 in Hagelstein 2023 [93]), which is reproduced here:

$$\langle \Psi_1 | \hat{H}_{int} | \Psi_2 \rangle \ \rightarrow \ -i \frac{(E_2 - E_1)}{2Mc^2} \left\langle \Psi_1 \left| \sum_j \boldsymbol{\sigma}_j \times \frac{\boldsymbol{\pi}_j}{mc} \right| \Psi_2 \right\rangle \cdot c\mathbf{P} \tag{384}$$

Here, $\Psi_1$ and $\Psi_2$ are the wave functions of the final and initial states, $E_2$ is the energy eigenvalue associated with state $\Psi_2$, $E_1$ is the energy eigenvalue associated with state $\Psi_1$, $M$ is the total mass of the nucleus, $m$ is the nucleon mass, $\boldsymbol{\sigma}_j$ is a vector consisting of Pauli spin matrices for nucleon j, $\boldsymbol{\pi}_j$ is relative momentum of nucleon $j$, and $\mathbf{P}$ is the center of mass momentum for the nucleus, which in a lattice will be a phonon operator.

For the $D_2$ to $^4$He transition, we have:

$$\begin{aligned} \Psi_1 &= \Psi[^4\text{He}] \\ \Psi_2 &= \Psi[\text{D}_2] \end{aligned} \tag{385}$$

$$\Psi[^4\text{He}] \ = \ \Phi_S \mathcal{A} \Big\{ p\uparrow p\downarrow n\uparrow n\downarrow \Big\} \tag{386}$$

where $\Phi_S$ is the spatial part of the singlet S ground state $^4$He wave function and

$$\Psi(L,S;J,M_J) \ = \ \sum_{M_L} \sum_{M_S} \mathcal{A} \Big\{ \Psi(L,M_L;S,M_S|12;34) \Big\} \langle L,M_L;S,M_S|J,M_J\rangle \tag{387}$$

We used Mathematica for a brute force reduction of the spin and isospin components of the matrix element. In the specific case of the $\boldsymbol{\sigma} \times \boldsymbol{\pi}$ matrix element for the $J=1, M_J=0$ molecular state we obtain

$$\begin{aligned} \left\langle \Psi[^4\text{He}] \left| \sum_j \boldsymbol{\sigma}_j \times \boldsymbol{\pi}_j \right| \Psi[J=1,M_J=0] \right\rangle \ &= \ \sqrt{\frac{v_{nuc}}{v_{mol}}} e^{-G} \sqrt{\frac{3}{4\pi}} \int d^3\mathbf{r} \int d^3\mathbf{r}_a \int d^3\mathbf{r}_b \\ & \Big\{ \Phi_S \phi_d(|\mathbf{r}_a|) \phi_d(|\mathbf{r}_b|) \Big[ -i\hat{\mathbf{i}}_z (p_x x + p_y y) \Big] F_{DD}(|\mathbf{r}|) \Big\} \end{aligned} \tag{388}$$

where $F_{DD}$ is defined for mathematical convenience via:

$$R_{DD}(r) \ = \ e^{-G} \sqrt{\frac{v_{nuc}}{v_{mol}}} F_{DD}(r) r \tag{389}$$

We remind the reader that $R_{DD}$ is the radial wave function that is often transformed into $P_{DD} = rR_{DD}$ as in Eq. 274

We see that the spin and isospin reduction leads to matrix elements in which the $\sigma \times \boldsymbol{\pi}$ operator results in spatial operators that depend on the relative separation of the two deuterons.

### Numerical calculation

After carrying out angular integrations, we end up with

$$\left\langle \Psi[^4\text{He}] \left| \sum_j \boldsymbol{\sigma}_j \times \boldsymbol{\pi}_j \right| \Psi[J=1, M_J=0] \right\rangle \ = \ -\hat{\mathbf{i}}_z \hbar \sqrt{\frac{v_{nuc}}{v_{mol}}} e^{-G} \sqrt{\frac{3}{4\pi}} \left( 2I + \frac{2}{3}K \right) \tag{390}$$

where $I$ and $K$ are the multidimensional integrals

$$\begin{aligned} I \ &= \ \int d^3\mathbf{r} \int d^3\mathbf{r}_a \int d^3\mathbf{r}_b \Big\{ \Phi_S(\mathbf{r}, \mathbf{r}_a, \mathbf{r}_b) \phi_d(|\mathbf{r}_a|) \phi_d(|\mathbf{r}_b|) F_{DD}(|\mathbf{r}|) \Big\} \\ K \ &= \ \int d^3\mathbf{r} \int d^3\mathbf{r}_a \int d^3\mathbf{r}_b \Big\{ \Phi_S(\mathbf{r}, \mathbf{r}_a, \mathbf{r}_b) \phi_d(|\mathbf{r}_a|) \phi_d(|\mathbf{r}_b|) r F'_{DD}(|\mathbf{r}|) \Big\} \end{aligned} \tag{391}$$

We have used Gaussian quadrature for the angular integrations and a second-order quadrature for the radial integrations with the result (in $\text{fm}^{-1}$)

$$I \ = \ -72.857 \qquad K \ = \ 53.330 \tag{392}$$

making use of the Gamow factor and the volume factor (see section 6.3)

$$G \ = \ 90.35 \qquad \frac{v_{nuc}}{v_{mol}} \ = \ 6.6498 \times 10^{-12} \tag{393}$$

### Results

The result of the calculation for the **a**-matrix elements for transitions from the three $^3$P $J^\pi = 1^-$ molecular $D_2$ states to the ground state $^4$He state are [105]

$$\langle \Psi[^4\text{He}] | \mathbf{a} | \Psi[J=1, M_J=+1] \rangle \ = \ -\left( \frac{\hat{\mathbf{i}}_x + i\hat{\mathbf{i}}_y}{\sqrt{2}} \right) 0.0362 \sqrt{\frac{v_{nuc}}{v_{mol}}} e^{-G} \tag{394}$$

$$\langle \Psi[^4\text{He}] | \mathbf{a} | \Psi[J=1, M_J=0] \rangle \ = \ \hat{\mathbf{i}}_z \ 0.0362 \sqrt{\frac{v_{nuc}}{v_{mol}}} e^{-G} \tag{395}$$

$$\langle \Psi[^4\text{He}] | \mathbf{a} | \Psi[J=1, M_J=-1] \rangle \ = \ \left( \frac{\hat{\mathbf{i}}_x - i\hat{\mathbf{i}}_y}{\sqrt{2}} \right) 0.0362 \sqrt{\frac{v_{nuc}}{v_{mol}}} e^{-G} \tag{396}$$

We expect errors on the order of a factor of 2 due to the use of approximate model nuclear wave functions.

### Including loss

In the calculation of the **a**-matrix elements above, we made use of a radial $D_2$ wave function calculated in the absence of loss due to deuteron-deuteron fusion to the 3+1 channels (see section 6.5 for a discussion of this kind of loss). This would be appropriate for an off-resonant process in which the basis energy is much higher than the energy eigenvalue such that 3+1 fusion is energetically forbidden. The situation is different on resonance where the 3+1 fusion process is allowed.

Consequently, we include an imaginary potential in the evaluation of the Schrödinger equation as discussed below. We then evaluate again the integrals $I$ and $K$ using complex wave functions calculated with the imaginary potential that is consistent with the known fusion rate from [100] for the d+d $J^\pi = 1^-$ state at 28.37 MeV. The $I$ and $K$ integrals are

$$I \;=\; -72.840 - i0.823 \qquad K \;=\; 53.322 + i0.290 \tag{397}$$

The **a**-matrix elements that result are

$$\langle \Psi[^4\text{He}]|\mathbf{a}|\Psi[J=1, M_J=+1]\rangle \;=\; -\left(\frac{\hat{\mathbf{i}}_x + i\hat{\mathbf{i}}_y}{\sqrt{2}}\right)(0.0362 + i0.000477)\sqrt{\frac{v_{nuc}}{v_{mol}}}e^{-G} \tag{398}$$

$$\langle \Psi[^4\text{He}]|\mathbf{a}|\Psi[J=1, M_J=0]\rangle \;=\; \hat{\mathbf{i}}_z\,(0.0362 + i0.000477)\sqrt{\frac{v_{nuc}}{v_{mol}}}e^{-G} \tag{399}$$

$$\langle \Psi[^4\text{He}]|\mathbf{a}|\Psi[J=1, M_J=-1]\rangle \;=\; \left(\frac{\hat{\mathbf{i}}_x - i\hat{\mathbf{i}}_y}{\sqrt{2}}\right)(0.0362 + i0.000477)\sqrt{\frac{v_{nuc}}{v_{mol}}}e^{-G} \tag{400}$$

The difference between these **a**-matrix elements and the matrix elements calculated in the absence of loss leads to a difference in the contributions from the different pathways associated with destructive interference. This can lead to a large acceleration of the excitation transfer rate.

### Imaginary potential

Loss due to dd-fusion (via spontaneous tunnel decay as described in section 6.3) impacts the relative deuteron-deuteron wave function, leading to observable effects in the elastic scattering channels as discussed by Chwieroth et al. (1972) [126]. We can include this in the model here by adding an imaginary potential to the Schrödinger equation (Eq. 274) according to

$$EP_{DD}(r) \;=\; \left(-\frac{\hbar^2}{2\mu}\frac{d^2}{dr^2} + \frac{\hbar^2 l(l+1)}{2\mu r^2} + V_{mol}(r) + V_{nuc}^{S,l}(r) + iW(r)\right)P_{DD}(r) \tag{401}$$

where the imaginary potential is of the form

$$W(r) \;=\; -W_0\left(\frac{1}{1+e^{(r-R_0)/a}} + \frac{4e^{(r-R_0)/a}}{(1+e^{(r-R_0)/a})^2}\right) \tag{402}$$

In the literature the parameters of the imaginary potential are obtained by fitting to scattering data. Unfortunately, this is problematic at very low relative energy, in the eV range. We used $R_0$ and $a$ values from Kurihara (1985) [127], and selected $W_0$ so that it would lead to a decay rate consistent with [100]. The model parameters are given by

$$
\begin{aligned}
W_0 &= 0.066 \text{ MeV} \\
R_0 &= 3.75 \text{ fm} \\
a &= 0.5 \text{ fm}
\end{aligned}
\tag{403}
$$

The imaginary potential is illustrated in Figure 56.

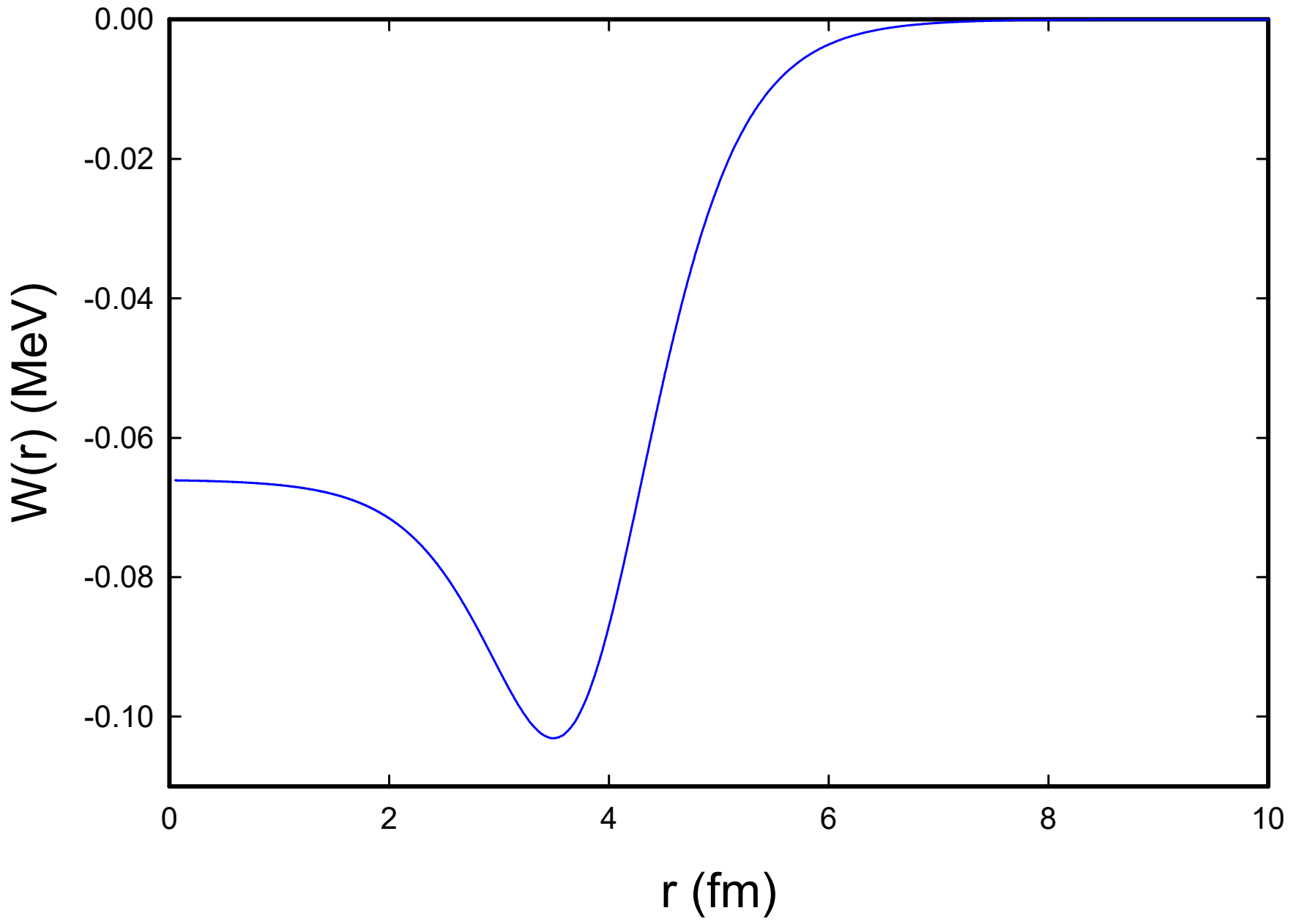

**Figure 56:** Imaginary potential $W(r)$ in the ${}^3P$ $D_2$ channel due to 3+1 fusion decay.

## 6.10 Estimating a-matrix elements for the HD/$^3$He transitions

This section provides a brief summary of the calculation of the $\mathbf{a} \cdot c\mathbf{P}$ matrix elements for the HD/$^3$He transitions. The model is similar to the one described above for the $D_2$/$^4$He transition.

### Ground state wave function

In a complete description of the $^3$He ground state we would make use of ten group theoretical basis states [128]; however, for the calculation in this subsection we will make use of a simpler approximate single configuration wave function. The overall single configuration wave function for the $^1$S configuration must be fully anti-symmetric, and one way to construct it is to make use of an anti-symmetrizer according to

$$\Psi[^3\text{He } ^2S, M_S = 1/2] \ = \ \mathcal{A}\Big\{(p\downarrow)_1(p\uparrow)_2(n\uparrow)_3\Phi_S(\mathbf{r}_1,\mathbf{r}_2,\mathbf{r}_3)\Big\} \tag{404}$$

where $\mathcal{A}$ is an anti-symmetrization operator; where $\Phi_S(\mathbf{r}_1,\mathbf{r}_2,\mathbf{r}_3)$ is the (totally symmetric) spatial wave function; and where $(p\downarrow)_1(p\uparrow)_2(n\uparrow)_3$ is the spin and isospin wave function for the $^1$S state prior to anti-symmetrization. We can expand this out and write

$$\Psi[^3\text{He } ^2S, M_S = 1/2] \ =$$
$$\frac{1}{\sqrt{6}}\bigg(-p\uparrow p\downarrow n\uparrow + p\downarrow p\uparrow n\uparrow + p\uparrow n\uparrow p\downarrow - p\downarrow n\uparrow p\uparrow - n\uparrow p\uparrow p\downarrow + n\uparrow p\downarrow p\uparrow\bigg)\Phi_S(\mathbf{r}_1,\mathbf{r}_2,\mathbf{r}_3) \tag{405}$$

where for simplicity we use a short cut in notation according to

$$p\uparrow p\downarrow n\uparrow \ = \ (p\uparrow)_1(p\downarrow)_2(n\uparrow)_3 \tag{406}$$

For the symmetric spatial wave function we use

$$\Phi_S(\mathbf{r}_1,\mathbf{r}_2,\mathbf{r}_3) \ = \ N_S u(|\mathbf{r}_2-\mathbf{r}_1|)u(|\mathbf{r}_3-\mathbf{r}_1|)u(|\mathbf{r}_3-\mathbf{r}_2|) \tag{407}$$

where $N_S$ is a normalization constant. The function $u$ is

$$u(r) \ = \ (r-r_0)e^{-\alpha r} \tag{408}$$

with

$$\alpha \ = \ 0.831 \text{ fm}^{-1} \qquad r_0 \ = \ 0.485 \text{ fm} \tag{409}$$

which results in

$$\sqrt{\langle|\mathbf{r}|^2\rangle} \ = \ 1.687 \text{ fm} \tag{410}$$

### Molecular state wave functions

For the molecular HD wave functions we use a similar approach. For simplicity we use a single-configuration $^3$S wave function for the deuteron. The overall wave function must be anti-symmetric, so we make use of an anti-symmetrization operator as above. Since the $\mathbf{a}\cdot c\mathbf{P}$ interaction preserves isospin, we can focus on molecular states with isospin $T = 1/2$ (and not be concerned with the $T = 3/2$ states). HD molecules can have total spin $S = 1/2$ and $S = 3/2$. Since the $\mathbf{a} \cdot c\mathbf{P}$ is a rank 1 tensor operator, we know that only the molecular $^2P$ and $^4$P states will given finite matrix elements. It is simplest to specify the states in LS coupling. For the different cases we can write

$$\Psi[\text{HD}\ ^4P, M_S = 3/2, M] \ = \ \mathcal{A}\left\{ \left( \frac{pn-np}{\sqrt{2}} \right)_{12} p_3 (\uparrow\uparrow)_{12} \uparrow_3 \phi_d(|\mathbf{r}_{21}|) R_{HD}(|\mathbf{r}_{12;3}|) Y_{1,M}(\Omega_{12;3}) \right\} \tag{411}$$

Prior to anti-symmetrization, the deuteron involves nucleons 1 and 2, and the proton is nucleon 3. The deuteron is in an isospin 0 state, leading to an overall isospin $T = 1/2$ state as required. All of the spins are up as appropriate for a $S = 3/2,\ M_S = 3/2$ state. The relative molecular separation is denoted as $\mathbf{r}_{12;3}$, indicating that the deuteron involves nucleons 1,2 and the proton involves nucleon 3. The spherical harmonic $Y_{L,M}$ has one unit of angular momentum $L = 1$ as appropriate for a molecular P state.

For the other $^4$P states we can write

$$\Psi[\text{HD}\ ^4P, M_S = 1/2, M] \ = \ \mathcal{A}\left\{ \left( \frac{pn-np}{\sqrt{2}} \right)_{12} p_3 \frac{1}{\sqrt{3}} \left( \ \uparrow\uparrow\downarrow + \uparrow\downarrow\uparrow + \downarrow\uparrow\uparrow \ \right) \phi_d(|\mathbf{r}_{21}|) R_{HD}(|\mathbf{r}_{12;3}|) Y_{1,M}(\Omega_{12;3}) \right\} \tag{412}$$

$$\Psi[\text{HD}\ ^4P, M_S = -1/2, M] \ = \ \mathcal{A}\left\{ \left( \frac{pn-np}{\sqrt{2}} \right)_{12} p_3 \frac{1}{\sqrt{3}} \left( \ \uparrow\downarrow\downarrow + \downarrow\uparrow\downarrow + \downarrow\downarrow\uparrow \ \right) \phi_d(|\mathbf{r}_{21}|) R_{HD}(|\mathbf{r}_{12;3}|) Y_{1,M}(\Omega_{12;3}) \right\} \tag{413}$$

$$\Psi[\text{HD}\ ^4P, M_S = -3/2, M] \ = \ \mathcal{A}\left\{ \left( \frac{pn-np}{\sqrt{2}} \right)_{12} p_3 (\downarrow\downarrow)_{12} \downarrow_3 \phi_d(|\mathbf{r}_{21}|) R_{HD}(|\mathbf{r}_{12;3}|) Y_{1,M}(\Omega_{12;3}) \right\} \tag{414}$$

For the $^2$P states we have

$$\Psi[\text{HD}\ ^2P, M_S = 1/2, M] \ = \ \mathcal{A}\left\{ \left( \frac{pn-np}{\sqrt{2}} \right)_{12} p_3 \frac{1}{\sqrt{6}} \left( 2 \uparrow\uparrow\downarrow - \uparrow\downarrow\uparrow - \downarrow\uparrow\uparrow \ \right) \phi_d(|\mathbf{r}_{21}|) R_{HD}(|\mathbf{r}_{12;3}|) Y_{1,M}(\Omega_{12;3}) \right\} \tag{415}$$

$$\Psi[\text{HD}\ ^2P, M_S = -1/2, M] \ = \ \mathcal{A}\left\{ \left( \frac{pn-np}{\sqrt{2}} \right)_{12} p_3 \frac{1}{\sqrt{6}} \left( \ \uparrow\downarrow\downarrow + \downarrow\uparrow\downarrow - 2 \downarrow\downarrow\uparrow \ \right) \phi_d(|\mathbf{r}_{21}|) R_{HD}(|\mathbf{r}_{12;3}|) Y_{1,M}(\Omega_{12;3}) \right\} \tag{416}$$

For the deuteron relative wave function for this matrix element calculation we used the following expression from [113]

$$\phi_d(r) \ = \ \begin{cases} 0 & \text{for } r < r_0 \\ N_d \dfrac{\tanh[\gamma(r-r_0)] \ e^{-\beta r}}{r} & \text{for } r_0 < r \end{cases} \tag{417}$$

which was fit to the $^3$S component of the deuteron wave function based on the old Hamada-Johnston potential model.

**Construction of LSJ wave functions**

As a rank 1 tensor operator, the $\mathbf{a} \cdot c\mathbf{P}$ operator satisfies the Wigner-Eckart theorem

$$\langle J'M'|T_q^{(k)}|JM\rangle = \langle J'M';kq|JM\rangle\langle J'||T^{(k)}||M\rangle \tag{418}$$

with $k = 1$. This provides motivation to work with states in LSJ coupling, which we can construct according to

$$|LSJ, M_J\rangle = \sum_{M,M_S} |L,M\rangle|S,M_S\rangle\langle L,M;S,M_S|J,M_J\rangle \tag{419}$$

where the $\langle L,M;S,M_S|J,M_J\rangle$ are Clebsch-Gordan coefficients. It will be useful to use LSJ states for the molecular HD states, since the associated selection is clean with these states (the interaction matrix elements are more complicated with LS-coupled states).

**Molecular coordinates**

It is convenient to define a set of coordinates that are specific to the HD molecule. These include the center of mass position $\mathbf{R}$, deuteron relative position $\mathbf{r}_a$, and HD molecular relative position $\mathbf{r}$. These coordinates can be related to the individual nucleon coordinates in the 12;3 permutation according to

$$\begin{aligned} \mathbf{R} &= \frac{\mathbf{r}_1 + \mathbf{r}_2 + \mathbf{r}_3}{3} \\ \mathbf{r}_a &= \mathbf{r}_2 - \mathbf{r}_1 \\ \mathbf{r} &= \mathbf{r}_3 - \frac{\mathbf{r}_1 + \mathbf{r}_2}{2} \end{aligned} \tag{420}$$

In matrix form this is

$$\begin{pmatrix} \mathbf{R} \\ \mathbf{r}_a \\ \mathbf{r} \end{pmatrix} = \mathbf{M} \begin{pmatrix} \mathbf{r}_1 \\ \mathbf{r}_2 \\ \mathbf{r}_3 \end{pmatrix} = \begin{pmatrix} \frac{1}{3} & \frac{1}{3} & \frac{1}{3} \\ -1 & 1 & 0 \\ -\frac{1}{2} & -\frac{1}{2} & 1 \end{pmatrix} \begin{pmatrix} \mathbf{r}_1 \\ \mathbf{r}_2 \\ \mathbf{r}_3 \end{pmatrix} \tag{421}$$

We can invert this and write

$$\begin{pmatrix} \mathbf{r}_1 \\ \mathbf{r}_2 \\ \mathbf{r}_3 \end{pmatrix} = \mathbf{M}^{-1} \begin{pmatrix} \mathbf{R} \\ \mathbf{r}_a \\ \mathbf{r} \end{pmatrix} = \begin{pmatrix} 1 & -\frac{1}{2} & -\frac{1}{3} \\ 1 & \frac{1}{2} & -\frac{1}{3} \\ 1 & 0 & \frac{2}{3} \end{pmatrix} \begin{pmatrix} \mathbf{R} \\ \mathbf{r}_a \\ \mathbf{r} \end{pmatrix} \tag{422}$$

or

$$\begin{aligned} \mathbf{r}_1 &= \mathbf{R} - \frac{1}{2}\mathbf{r}_a - \frac{1}{3}\mathbf{r} \\ \mathbf{r}_2 &= \mathbf{R} + \frac{1}{2}\mathbf{r}_a - \frac{1}{3}\mathbf{r} \\ \mathbf{r}_3 &= \mathbf{R} + \frac{2}{3}\mathbf{r} \end{aligned} \tag{423}$$

From these relations we can develop relations between the nucleon momenta and momenta associated with the molecular coordinates in the 12;3 permutation

$$\begin{pmatrix} \mathbf{P} \\ \mathbf{p}_a \\ \mathbf{p} \end{pmatrix} = (\mathbf{M}^T)^{-1} \begin{pmatrix} \mathbf{p}_1 \\ \mathbf{p}_2 \\ \mathbf{p}_3 \end{pmatrix} = \begin{pmatrix} 1 & 1 & 1 \\ -\frac{1}{2} & \frac{1}{2} & 0 \\ -\frac{1}{3} & -\frac{1}{3} & \frac{2}{3} \end{pmatrix} \begin{pmatrix} \mathbf{p}_1 \\ \mathbf{p}_2 \\ \mathbf{p}_3 \end{pmatrix} \tag{424}$$

and

$$\begin{pmatrix} \mathbf{p}_1 \\ \mathbf{p}_2 \\ \mathbf{p}_3 \end{pmatrix} = \mathbf{M}^T \begin{pmatrix} \mathbf{P} \\ \mathbf{p}_a \\ \mathbf{p} \end{pmatrix} = \begin{pmatrix} \frac{1}{3} & -1 & -\frac{1}{2} \\ \frac{1}{3} & 1 & -\frac{1}{2} \\ \frac{1}{3} & 0 & 1 \end{pmatrix} \begin{pmatrix} \mathbf{P} \\ \mathbf{p}_a \\ \mathbf{p} \end{pmatrix} \tag{425}$$

**Integrations**

For the integrations we have

$$\begin{aligned} \int d^3\mathbf{r}_1 \int d^3\mathbf{r}_2 \int d^3\mathbf{r}_3 \{\cdots\} &= \int d^3\mathbf{R} \int d^3\mathbf{r}_a \int d^3\mathbf{r} |\det\{\mathbf{M}\}^{-1}| \{\cdots\} \\ &= \int d^3\mathbf{R} \int d^3\mathbf{r}_a \int d^3\mathbf{r} \{\cdots\} \end{aligned} \tag{426}$$

since

$$|\det\{\mathbf{M}\}^{-1}| = 1 \tag{427}$$

**The $^3$He normalization integral**

We can make use of the molecular coordinate and integration formula to develop a normalization integral for the ground state wave function. We can write

$$\langle \Psi[^3\text{He}] | \Psi[^3\text{He}] \rangle = \int d^3\mathbf{r}_a \int d^3\mathbf{r} \Big\{ |\Phi_S|^2 \Big\} \tag{428}$$

where the spin and isospin components contribute a factor of unity. We can write the spatial wave function as

$$\Phi_S(\mathbf{r}_1, \mathbf{r}_2, \mathbf{r}_3) = N_S u(|\mathbf{r}_2 - \mathbf{r}_1|) u(|\mathbf{r}_3 - \mathbf{r}_1|) u(|\mathbf{r}_3 - \mathbf{r}_2|) \tag{429}$$

For the inter-particle separations we have

$$\begin{aligned} |\mathbf{r}_{21}|^2 &= |\mathbf{r}_a|^2 \\ |\mathbf{r}_{31}|^2 &= \left| \mathbf{r} + \frac{\mathbf{r}_a}{2} \right|^2 = |\mathbf{r}|^2 + \frac{1}{4} |\mathbf{r}_a|^2 + 2\mathbf{r} \cdot \mathbf{r}_a \\ |\mathbf{r}_{32}|^2 &= \left| \mathbf{r} - \frac{\mathbf{r}_a}{2} \right|^2 = |\mathbf{r}|^2 + \frac{1}{4} |\mathbf{r}_a|^2 - 2\mathbf{r} \cdot \mathbf{r}_a \end{aligned} \tag{430}$$

For the normalization integral we have

$$\langle\Psi[^3\text{He}]|\Psi[^3\text{He}]\rangle \;=\; \int d^3\mathbf{r}_a \int d^3\mathbf{r} \Big\{ N_S^2 u^2\Big(r_a\Big) u^2\Big(\sqrt{|\mathbf{r}|^2+\frac{1}{4}|\mathbf{r}_a|^2+2\mathbf{r}\cdot\mathbf{r}_a}\Big) u^2\Big(\sqrt{|\mathbf{r}|^2+\frac{1}{4}|\mathbf{r}_a|^2-2\mathbf{r}\cdot\mathbf{r}_a}\Big)\Big\} \tag{431}$$

We can expand this out according to

$$\begin{aligned}\langle\Psi[^3\text{He}]|\Psi[^3\text{He}]\rangle \;&=\; N_S^2\int_0^\infty r_a^2 dr_a \int_0^\pi \sin\theta_a d\theta_a \int_0^{2\pi} d\phi_a \int_0^\infty r^2 dr \int_0^\pi \sin\theta d\theta \int_0^{2\pi} d\phi \\ &\Big\{u^2\Big(r_a\Big)u^2\Big(\sqrt{r^2+\frac{1}{4}r_a^2+2rr_a\cos\theta_a}\Big)u^2\Big(\sqrt{r^2+\frac{1}{4}r_a^2-2rr_a\cos\theta_a}\Big)\Big\} \\ &=\; N_S^2 8\pi^2\int_0^\infty r_a^2 dr_a \int_0^\pi \sin\theta_a d\theta_a \int_0^\infty r^2 dr \\ &\Big\{u^2\Big(r_a\Big)u^2\Big(\sqrt{r^2+\frac{1}{4}r_a^2+2rr_a\cos\theta_a}\Big)u^2\Big(\sqrt{r^2+\frac{1}{4}r_a^2-2rr_a\cos\theta_a}\Big)\Big\} \\ &=\; 1\end{aligned} \tag{432}$$

**Reduction of the matrix elements**

We recall that the **a**-matrix element can be expressed in terms of $\boldsymbol{\sigma}\times\boldsymbol{\pi}$ operators according to

$$\langle\Phi_2|\mathbf{a}\cdot c\mathbf{P}|\Phi_1\rangle \;=\; i\frac{(E_2-E_1)}{2}\Big\langle\Phi_2\Big|\sum_j \boldsymbol{\sigma}_j\times\frac{\boldsymbol{\pi}_j}{mc}\Big|\Phi_1\Big\rangle\cdot\frac{\mathbf{P}}{Mc} \tag{433}$$

where $M$ is the total mass; where **P** is the center of mass momentum; where $\boldsymbol{\pi}_j$ is the relative momentum of nucleon $j$; where $\boldsymbol{\sigma}_j$ is the Pauli spin matrix for nucleon $j$; and where $E_1$ and $E_2$ are the energy eigenvalues for $\Phi_1$ and $\Phi_2$. We used Mathematica to evaluate the spin and isospin terms resulting in a reduction of the $\boldsymbol{\sigma}\times\boldsymbol{\pi}$ matrix elements into spatial integrals. The results are summarized according to:

$$\Big\langle\Psi[^3\text{He}]\Big|\sum_j\boldsymbol{\sigma}_j\times\boldsymbol{\pi}_j\Big|\Psi\Big[S=\frac{1}{2}:J=\frac{1}{2},M_J=\frac{1}{2}\Big]\Big\rangle \;=\; \hat{\mathbf{i}}_z\hbar\sqrt{\frac{3}{2\pi}}\sqrt{\frac{v_{nuc}}{v_{mol}}}e^{-G}\Big(I+\frac{1}{3}K\Big) \tag{434}$$

$$\Big\langle\Psi[^3\text{He}]\Big|\sum_j\boldsymbol{\sigma}_j\times\boldsymbol{\pi}_j\Big|\Psi\Big[S=\frac{1}{2}:J=\frac{1}{2},M_J=-\frac{1}{2}\Big]\Big\rangle \;=\; (\hat{\mathbf{i}}_x-i\hat{\mathbf{i}}_y)\hbar\sqrt{\frac{3}{2\pi}}\sqrt{\frac{v_{nuc}}{v_{mol}}}e^{-G}\Big(I+\frac{1}{3}K\Big) \tag{435}$$

$$\Big\langle\Psi[^3\text{He}]\Big|\sum_j\boldsymbol{\sigma}_j\times\boldsymbol{\pi}_j\Big|\Psi\Big[S=\frac{3}{2}:J=\frac{1}{2},M_J=\frac{1}{2}\Big]\Big\rangle \;=\; -\hat{\mathbf{i}}_z\hbar\sqrt{\frac{3}{4\pi}}\sqrt{\frac{v_{nuc}}{v_{mol}}}e^{-G}\Big(I+\frac{1}{3}K\Big) \tag{436}$$

$$\Big\langle\Psi[^3\text{He}]\Big|\sum_j\boldsymbol{\sigma}_j\times\boldsymbol{\pi}_j\Big|\Psi\Big[S=\frac{3}{2}:J=\frac{1}{2},M_J=-\frac{1}{2}\Big]\Big\rangle \;=\; -(\hat{\mathbf{i}}_x-i\hat{\mathbf{i}}_y)\hbar\sqrt{\frac{3}{4\pi}}\sqrt{\frac{v_{nuc}}{v_{mol}}}e^{-G}\Big(I+\frac{1}{3}K\Big) \tag{437}$$

$$\Big\langle\Psi[^3\text{He}]\Big|\sum_j\boldsymbol{\sigma}_j\times\boldsymbol{\pi}_j\Big|\Psi\Big[S=\frac{1}{2}:J=\frac{3}{2},M_J=\frac{3}{2}\Big]\Big\rangle \;=\; -(\hat{\mathbf{i}}_x+i\hat{\mathbf{i}}_y)\hbar\sqrt{\frac{9}{16\pi}}\sqrt{\frac{v_{nuc}}{v_{mol}}}e^{-G}\Big(I+\frac{1}{3}K\Big) \tag{438}$$

$$\Big\langle\Psi[^3\text{He}]\Big|\sum_j\boldsymbol{\sigma}_j\times\boldsymbol{\pi}_j\Big|\Psi\Big[S=\frac{1}{2}:J=\frac{3}{2},M_J=\frac{1}{2}\Big]\Big\rangle \;=\; \hat{\mathbf{i}}_z\hbar\sqrt{\frac{3}{4\pi}}\sqrt{\frac{v_{nuc}}{v_{mol}}}e^{-G}\Big(I+\frac{1}{3}K\Big) \tag{439}$$

$$\left\langle \Psi[^3\mathrm{He}] \left| \sum_j \boldsymbol{\sigma}_j \times \boldsymbol{\pi}_j \right| \Psi\left[ S = \frac{1}{2} : J = \frac{3}{2}, M_J = -\frac{1}{2} \right] \right\rangle = (\hat{\mathbf{i}}_x - i\hat{\mathbf{i}}_y)\hbar\sqrt{\frac{3}{16\pi}}\sqrt{\frac{v_{nuc}}{v_{mol}}}e^{-G}\left(I + \frac{1}{3}K\right) \quad (440)$$

$$\left\langle \Psi[^3\mathrm{He}] \left| \sum_j \boldsymbol{\sigma}_j \times \boldsymbol{\pi}_j \right| \Psi\left[ S = \frac{3}{2} : J = \frac{3}{2}, M_J = \frac{3}{2} \right] \right\rangle = -(\hat{\mathbf{i}}_x + i\hat{\mathbf{i}}_y)\hbar\sqrt{\frac{45}{16\pi}}\sqrt{\frac{v_{nuc}}{v_{mol}}}e^{-G}\left(I + \frac{1}{3}K\right) \quad (441)$$

$$\left\langle \Psi[^3\mathrm{He}] \left| \sum_j \boldsymbol{\sigma}_j \times \boldsymbol{\pi}_j \right| \Psi\left[ S = \frac{3}{2} : J = \frac{3}{2}, M_J = \frac{1}{2} \right] \right\rangle = \hat{\mathbf{i}}_z\hbar\sqrt{\frac{15}{16\pi}}\sqrt{\frac{v_{nuc}}{v_{mol}}}e^{-G}\left(I + \frac{1}{3}K\right) \quad (442)$$

$$\left\langle \Psi[^3\mathrm{He}] \left| \sum_j \boldsymbol{\sigma}_j \times \boldsymbol{\pi}_j \right| \Psi\left[ S = \frac{3}{2} : J = \frac{3}{2}, M_J = -\frac{1}{2} \right] \right\rangle = (\hat{\mathbf{i}}_x - i\hat{\mathbf{i}}_y)\hbar\sqrt{\frac{15}{16\pi}}\sqrt{\frac{v_{nuc}}{v_{mol}}}e^{-G}\left(I + \frac{1}{3}K\right) \quad (443)$$

where the spatial integrals are

$$I = \int\int \Phi_S \phi_d(r_a) F_{HD}(r) d^3\mathbf{r}_a d^3\mathbf{r} \quad (444)$$

$$K = \int d^3\mathbf{r} \int d^3\mathbf{r}_a \left\{ \Phi_S(\mathbf{r}, \mathbf{r}_a)\phi_d(|\mathbf{r}_a|) r F'_{HD}(|\mathbf{r}|) \right\} \quad (445)$$

In these formulas we defined $F_{HD}$ for mathematical convenience via:

$$R_{HD}(r) = \sqrt{\frac{v_{nuc}}{v_{mol}}} e^{-G} r F_{HD}(r) \quad (446)$$

**Numerical evaluation of the spatial integrals**

We have evaluated the spatial integrals and we obtained

$$\begin{aligned} I &= 77.774 \\ J &= -16.324 \end{aligned} \quad (447)$$

**Values for the a-matrix elements**

The **a**-matrix elements that result are

$$\left\langle \Psi[^3\mathrm{He}] \middle| \mathbf{a} \middle| \Psi\left[ S = \frac{1}{2} : J = \frac{1}{2}, M_J = \frac{1}{2} \right] \right\rangle = \hat{\mathbf{i}}_z\, 0.00931i \sqrt{\frac{v_{nuc}}{v_{mol}}} e^{-G} \quad (448)$$

$$\left\langle \Psi[^3\mathrm{He}] \middle| \mathbf{a} \middle| \Psi\left[ S = \frac{1}{2} : J = \frac{1}{2}, M_J = -\frac{1}{2} \right] \right\rangle = \frac{\hat{\mathbf{i}}_x - i\hat{\mathbf{i}}_y}{\sqrt{2}}\, 0.0132i \sqrt{\frac{v_{nuc}}{v_{mol}}} e^{-G} \quad (449)$$

$$\left\langle \Psi[^3\mathrm{He}] \middle| \mathbf{a} \middle| \Psi\left[ S = \frac{3}{2} : J = \frac{1}{2}, M_J = \frac{1}{2} \right] \right\rangle = -\hat{\mathbf{i}}_z\, 0.00659i \sqrt{\frac{v_{nuc}}{v_{mol}}} e^{-G} \quad (450)$$

$$\left\langle \Psi[^3\mathrm{He}] \middle| \mathbf{a} \middle| \Psi\left[ S = \frac{3}{2} : J = \frac{1}{2}, M_J = -\frac{1}{2} \right] \right\rangle = -\frac{\hat{\mathbf{i}}_x - i\hat{\mathbf{i}}_y}{\sqrt{2}}\, 0.00931i \sqrt{\frac{v_{nuc}}{v_{mol}}} e^{-G} \quad (451)$$

$$\left\langle \Psi[^3\mathrm{He}] \middle| \mathbf{a} \middle| \Psi\left[ S = \frac{1}{2} : J = \frac{3}{2}, M_J = \frac{3}{2} \right] \right\rangle = -\frac{\hat{\mathbf{i}}_x + i\hat{\mathbf{i}}_y}{\sqrt{2}}\, 0.00807i \sqrt{\frac{v_{nuc}}{v_{mol}}} e^{-G} \quad (452)$$

$$\left\langle \Psi[^3\text{He}] \middle| \mathbf{a} \middle| \Psi\left[S = \frac{1}{2} : J = \frac{3}{2}, M_J = \frac{1}{2}\right] \right\rangle = \hat{\mathbf{i}}_z\, 0.00659i \sqrt{\frac{v_{nuc}}{v_{mol}}} e^{-G} \tag{453}$$

$$\left\langle \Psi[^3\text{He}] \middle| \mathbf{a} \middle| \Psi\left[S = \frac{1}{2} : J = \frac{3}{2}, M_J = -\frac{1}{2}\right] \right\rangle = \frac{\hat{\mathbf{i}}_x - i\hat{\mathbf{i}}_y}{\sqrt{2}}\, 0.00466i \sqrt{\frac{v_{nuc}}{v_{mol}}} e^{-G} \tag{454}$$

$$\left\langle \Psi[^3\text{He}] \middle| \mathbf{a} \middle| \Psi\left[S = \frac{3}{2} : J = \frac{3}{2}, M_J = \frac{3}{2}\right] \right\rangle = -\frac{\hat{\mathbf{i}}_x + i\hat{\mathbf{i}}_y}{\sqrt{2}}\, 0.0180i \sqrt{\frac{v_{nuc}}{v_{mol}}} e^{-G} \tag{455}$$

$$\left\langle \Psi[^3\text{He}] \middle| \mathbf{a} \middle| \Psi\left[S = \frac{3}{2} : J = \frac{3}{2}, M_J = \frac{1}{2}\right] \right\rangle = \hat{\mathbf{i}}_z\, 0.00736i \sqrt{\frac{v_{nuc}}{v_{mol}}} e^{-G} \tag{456}$$

$$\left\langle \Psi[^3\text{He}] \middle| \mathbf{a} \middle| \Psi\left[S = \frac{3}{2} : J = \frac{3}{2}, M_J = -\frac{1}{2}\right] \right\rangle = \frac{\hat{\mathbf{i}}_x - i\hat{\mathbf{i}}_y}{\sqrt{2}}\, 0.0104i \sqrt{\frac{v_{nuc}}{v_{mol}}} e^{-G} \tag{457}$$

For these calculations we used

$$\begin{aligned} \sqrt{\frac{v_{nuc}}{v_{mol}}} &= 2.33 \times 10^{-6} \\ G &= 75.69 \end{aligned} \tag{458}$$

**Modeling loss on the matrix element**

Elsewhere we have commented on the fact that the **a**-matrix element for the $D_2$/$^4$He transition has one value if fusion loss is included, and another value if fusion loss is not included. We would expect the situation to be qualitative similar for the HD/$^3$He matrix element since there are loss mechanisms that would be expected to impact the **a**-matrix element. The main loss mechanisms include electric dipole (E1) radiative decay and internal conversion.

In the case of radiative decay, it would be possible to develop an imaginary potential by making use of

$$-i\frac{\hbar\gamma_{rad}}{2} = -i\frac{\hbar}{2}\frac{4}{3}\frac{1}{4\pi\epsilon_0}\frac{\omega^3}{\hbar c^3}|\langle \text{HD}|\mathbf{d}|^3\text{He}\rangle|^2 = i\langle \text{HD}|W(r)|\text{HD}\rangle \tag{459}$$

where $\gamma_{rad}$ is the radiative decay rate; where $\hbar\omega$ is the gamma energy; where **d** is the electric dipole operator

$$\mathbf{d} = \sum_j q_j \mathbf{r}_j \tag{460}$$

where $q_j$ is the nucleon effective charge and $\mathbf{r}_j$ is the (relative) position of nucleon $j$. $W(r)$ is the imaginary potential (operator) that can be used to model loss. It follows that the imaginary potential operator can be written as

$$W(r) = -\frac{2}{3}\frac{1}{4\pi\epsilon_0}\frac{\omega^3}{c^3}\mathbf{d}|^3\text{He}\rangle\langle^3\text{He}|\mathbf{d} \tag{461}$$

We have not yet set up such a calculation.

### Rough estimate for the impact of loss on the a-matrix element

We expect that the impact of loss on the **a**-matrix element will be proportional to the relevant decay rate. Based on this, we could get a rough estimate for the difference based on the ratio of the radiative decay rate for the HD/$^3$He transition (where we do not have a calculation) and the fusion loss rate for the $D_2$/$^4$He transition (where we do have an estimate). We would expect

$$|1-\eta|(\mathrm{HD}/{}^3\mathrm{He}) \ = \ \left|1-\eta(\mathrm{HD}/{}^3\mathrm{He})\right| \ \sim \ \left(\frac{\gamma_{rad}}{\gamma_{fusion}}\right)_{local} |1-\eta|(\mathrm{D}_2/{}^4\mathrm{He}) \tag{462}$$

where the relevant radiative and fusion rates are for nuclei localized on the fm scale; where $|1-\eta|$ is the ratio of the excitation rate including loss effects relative to the ideal; where $\eta$ is the (mostly) phase factor associated with loss in **a**-matrix element.

We have the estimate from Tilley et al. (1992) [100] for the fusion rate

$$\hbar\gamma_{fusion} \ = \ 0.15 \text{ MeV} \tag{463}$$

For the radiative decay rate we can make use of the Weisskopf estimate for an E1 transition

$$\gamma_{rad} \ \sim \ 1.0 \times 10^{14} A^{2/3}(\hbar\omega)^3 \ \to \ 3.4 \times 10^{16} \tag{464}$$

or

$$\hbar\gamma_{rad} \ \sim \ 23 \text{ eV} \tag{465}$$

This leads to the rough estimate

$$|1-\eta|(\mathrm{HD}/{}^3\mathrm{He}) \ \sim \ \left(\frac{23 \text{ eV}}{0.15 \text{ MeV}}\right) 0.013 \ = \ 2 \times 10^{-6} \tag{466}$$

### Discussion

It has long been argued that, because of the lower reduced mass and the smaller Gamow factor, HD/$^3$He fusion should be dominant in LENR processes. We can now make a quantitative version of this argument by comparing the ratio of the **a**-matrix elements (from this section and from section 6.9):

$$\frac{|\langle {}^4\mathrm{He}|a_z|\mathrm{D}_2\ {}^3P, J=0, M_J=0\rangle|}{|\langle {}^3\mathrm{He} J=1/2, M_J=1/2|a_z|\mathrm{HD}\ {}^3P, J=1/2, M_J=1/2\rangle|} \ = \ \frac{0.0362\left(\sqrt{\frac{v_{nuc}}{v_{mol}}}e^{-G}\right)_{\mathrm{D}_2}}{0.00931\left(\sqrt{\frac{v_{nuc}}{v_{mol}}}e^{-G}\right)_{HD}} \tag{467}$$

We can evaluate this ratio assuming $U_e = 350$ eV, with $G_{HD} = 35.87$ and $G_{\mathrm{D}_2} = 41.21$

$$\frac{|\langle {}^4\mathrm{He}|a_z|\mathrm{D}_2\ {}^3P, J=0, M_J=0\rangle|}{|\langle {}^3\mathrm{He} J=1/2, M_J=1/2|a_z|\mathrm{HD}\ {}^3P, J=1/2, M_J=1/2\rangle|} \ = \ 0.0053 \tag{468}$$

However, what is more important for excitation transfer is this ratio weighted by the appropriate $|1-\eta|$ values that represent selective loss and the suppression of destructive interference (see sections 6.5 and 6.6).

$$\frac{|1-\eta|(\mathrm{D}_2/{}^4\mathrm{He})|\langle {}^4\mathrm{He}|a_z|\mathrm{D}_2\ {}^3P, J=0, M_J=0\rangle|}{|1-\eta|(\mathrm{HD}/{}^3\mathrm{He})|\langle {}^3\mathrm{He} J=1/2, M_J=1/2|a_z|\mathrm{HD}\ {}^3P, J=1/2, M_J=1/2\rangle|} \sim \frac{0.1\times 0.0362\left(\sqrt{\frac{v_{nuc}}{v_{mol}}}e^{-G}\right)_{\mathrm{D}_2}}{2\times 10^{-6}\times 0.00931\left(\sqrt{\frac{v_{nuc}}{v_{mol}}}e^{-G}\right)_{HD}} = 270 \tag{469}$$

where we have used our rough estimate for the $D_2$/3+1/$^4$He contribution to the $D_2/^4$He transition for $|1-\eta|$. This suggests that even though tunneling is more difficult for the $D_2/^4$He transition, the faster loss likely results in a faster excitation transfer rate to receiver nuclei than for the HD/$^3$He transition.

## 6.11 Estimating a-matrix elements for Pd*/Pd transitions

Excitation transfer from the $D_2$/$^4$He fusion transition to Pd*/Pd transitions provides the foundation for nuclear effects in the $PdD_x$ system within the models under discussion in this document. In order to develop quantitative model predictions we need estimates for the excited state energy levels, for the lifetimes, multipolarities, and, if allowed by the selection rules, estimates for the **a**-matrix elements. At present, in general we lack much relevant data for the models (see the discussions in section 5.4 and section 5.16). It is nevertheless possible to develop a rough estimate for the magnitude of the **a**-matrix elements of Pd*/Pd transitions, as discussed in this section.

### M2 transitions from the ground state

Our focus in the models is on magnetic quadrupole (M2) transitions from the ground state, since the $\mathbf{a}\cdot c\mathbf{P}$ interaction has M2 multipolarity (see section 6.7), and since transitions from the ground state can develop cooperative enhancement (Dicke) factors. From the NuDat database there is only one such transition among the stable Pd isotopes listed, which is close to satisfying these requirements; this is an E1+M2 transition in $^{105}$Pd to a state at 644.7 keV with a lifetime listed as 126 ps. A gamma transition with M2 multipolarity can be relatively slow (since radiation with magnetic quadrupole multipolarity is a weak effect); however, an admixture with electric dipole (E1) multipolarity results in a much faster contribution to the radiative decay rate. That being said, we do not expect this to be the only M2 transition from the ground state in the Pd isotopes.

We can broaden our search by looking for levels with $J^\pi$ quantum numbers that are consistent with a transition from the ground state with M2 multipolarity. This is easiest among the even $A$ stable Pd isotopes, since we only need to concern ourselves with excited states with $J^\pi = 1^-$. These are listed in Table 9

**Table 9:** List of $J^\pi = 1^-$ excited states in the even mass stable Pd isotopes.

| Isotope | $E$ (keV) | $J^\pi$ | $T_{1/2}$ |
|---|---|---|---|
| Pd-106 | 2484.66 | (-1) | NA |
| Pd-106 | 2898.1 | (-1,-4) | NA |
| Pd-106 | 2908.7 | (-1) | NA |
| Pd-110 | 2125.3 | (-1) | NA |

The $(-1)$ notation in this table is taken from the NuDat database, which indicates uncertainty in the $J^\pi$ assignment. The lack of entries for the half-life reflects the fact that the states, which are potential candidates for us, are not well known. In $^{105}$Pd, there are states at 644.7 keV, 1088.2 keV, 1650.6 keV, and 2101.5 keV which are candidates, where the state at 644.7 keV was mentioned above.

### Prospects for modeling

It is possible to develop models from which **a**-matrix elements can be calculated. One such model is the interacting boson model, which has been applied to even-mass Pd nuclei (Scholten, 1980 [129]). A generalization of this model was used to model states and transitions in $^{105}$Pd (Meyer et al., 1996 [130]).

We expect that in the future large-scale shell model calculations can be done for the stable Pd isotopes (and other relevant nuclides), and used to develop approximate level energies, lifetimes and **a**-matrix elements. An example of large scale shell model calculations in a relevant mass region can be found in Sieja 2018 [131].

**Rough estimate for the a-matrix elements for allowed transitions**

We recall from section 6.7 that a matrix element of the relativistic phonon nuclear interaction can be written as

$$\langle \Phi_1 | \hat{H}_{int} | \Phi_2 \rangle \ \rightarrow \ -i \frac{(E_2 - E_1)}{2Mc^2} \left\langle \Phi_1 \middle| \sum_j \boldsymbol{\sigma}_j \times \frac{\boldsymbol{\pi}_j}{mc} \middle| \Phi_2 \right\rangle \cdot c\mathbf{P} \ = \ \langle \Phi_1 | \mathbf{a} | \Phi_2 \rangle \cdot c\mathbf{P} \tag{470}$$

We can develop a rough estimate for the **a**-matrix element in the case of an allowed transition at 23.85 MeV in a stable Pd isotope, in which a single nucleon makes a transition based on

$$\begin{aligned} |\langle \Phi_{Pd*} | \mathbf{a} \cdot \hat{\mathbf{i}}_z | \Phi_{Pd} \rangle| \ &\sim \ \frac{1}{2} \frac{23.85 \text{ MeV}}{106.41 \times 931.5 \text{ MeV}} \frac{\hbar \ (1 \text{ fm}^{-1})}{mc} \\ &= \ 2.6 \times 10^{-5} \end{aligned} \tag{471}$$

where the inverse reduced nucleon Compton wavelength is

$$\frac{mc}{\hbar} \ = \ \frac{mc^2}{\hbar c} \ = \ \frac{938 \text{ MeV}}{197.326 \text{ MeV fm}} \ = \ 4.76 \text{ fm}^{-1} \tag{472}$$

For this rough parametrization, the corresponding nucleon kinetic energy is approximately

$$\frac{p^2}{2m} \ \sim \ \frac{\hbar^2 (1 \text{ fm}^{-1})^2}{2m} \ = \ 20.7 \text{ MeV} \tag{473}$$

This is roughly the kinetic energy of a nucleon at the Fermi level.

**Normalized transition strength**

We would expect very few nuclear M2 transitions to be fully allowed corresponding to the model above. Instead, we expect the strong mixing between different nuclear configurations to lead to admixtures of states with some M2 strength, corresponding to a fractionation of transition strength of a fully allowed M2 transition over a great many individual states (especially at lower energy). To take this into account we scale the **a**-matrix element by the square root of the normalized transition strength according to

$$\begin{aligned} |\langle \Phi_{Pd*} | \mathbf{a} \cdot \hat{\mathbf{i}}_z | \Phi_{Pd} \rangle| \ &\sim \ \frac{1}{2} O \frac{23.85 \text{ MeV}}{106.41 \times 931.5 \text{ MeV}} \frac{\hbar \ (1 \text{fm}^{-1})}{mc} \\ &= \ 2.6 \times 10^{-5} \ O \end{aligned} \tag{474}$$

where $O^2$ is the normalized M2 transition strength. Unfortunately, we currently do not yet have reliable estimates for M2 $O$-values for precise **a**-matrix element determination. If we had estimates for the gamma decay rate for a pure M2 transition, we could develop an estimate for the normalized M2 transition strength according to

$$\gamma \ \approx \ O^2 \gamma^{(M2)}_{Weisskopf} \tag{475}$$

where the Weisskopf estimate for an M2 transition assumes a single nucleon transition that is fully allowed. Note that the M2 transition matrix element for the **a** operator is not the same as the M2 transition matrix element for radiative decay, so in a more precise model, we would keep track of $O$-values separately for the two different types of transitions. However, given the current lack of reliable estimates for either, making use of a general normalized M2 transition strength that we will use interchangeably for both will allow us to develop intuition about how we can expect **a**-matrix elements to work, given that we already have some intuition about how transition strengths work for the more familiar case of radiative decay.

### Transitions for nuclear molecule cluster states

The nuclear physics community has put much effort into the determination of the energy levels, $J^\pi$ quantum numbers, lifetimes, and gamma transition rates of numerous nuclear excited states over many decades. As a result, we can develop estimates in some cases of the associated transition strength based on experiment and theory. The situation in the case of the proposed highly-excited nuclear molecule cluster states of the stable Pd isotopes is different, since there is no widely developed literature.

We do have some intuition about how such states likely behave. A major issue is satisfying selection rules for a transition from the ground state to these states. Were we to insist on non-rotational binary nuclear molecule cluster states, then we immediately restrict the fraction of potentially allowable transitions. If we start from an even-even Pd ground state, then we have $J^\pi = 0^+$ in all cases. There would be no M2 transitions from these ground states to non-rotational nuclear molecule cluster states made up of $0^+$ daughters, since these states would be $0^+$ overall. This eliminates the majority of nuclear molecule cluster states as candidates for $\mathbf{a} \cdot c\mathbf{P}$ coupling with the ground states. Transitions to such states are also forbidden if we were to add binary nuclear molecule cluster states with one unit of rotational angular momentum (since a change of spin is required in addition to a change in angular momentum).

We also expect a hindrance factor associated with the splitting up for the initial parent nucleus into daughter clusters. This hindrance is expected to be minimized in the case of an alpha particle as a daughter, since there is evidence in the literature that nucleons in nuclei spend some of their time as alpha particles. We would expect the hindrance factor to be maximized in the case of nearly symmetric daughters, largely since the parent nuclei wouldn't be expected to have much admixture from such disparate configurations. Until such time as these states are modeled in detail (or addressed experimentally), we will not be able to have reliable estimates for the associated **a**-matrix elements from the grounds states. What we are able to do in the meantime is to make use of conceptual arguments that transitions to such states are weak, or extremely weak and estimates for specific $O$-values will be educated guesses.

## 6.12 Deriving minimum and maximum Dicke enhancement factors

In section 5.2 above, Eq. 56 was given as the Hamiltonian for the degenerate pseudo-spin version of the nuclear Dicke model, which is recalled to be

$$\hat{H} \ \rightarrow \ \frac{1}{2}\hbar\Omega_{ab}\bigg(\frac{S_-^{(b)}}{\hbar}\frac{S_+^{(a)}}{\hbar} + \frac{S_-^{(a)}}{\hbar}\frac{S_+^{(b)}}{\hbar}\bigg) \tag{476}$$

This Hamiltonian describes excitation transfer between two transitions.

### Transition matrix element

We can evaluate the transition matrix element directly

$$\begin{aligned} &\bigg\langle S_a, M_a\bigg|\bigg\langle S_b, M_b\bigg|\frac{1}{2}\hbar\Omega_{ab}\bigg(\frac{S_-^{(b)}}{\hbar}\frac{S_+^{(a)}}{\hbar} + \frac{S_-^{(a)}}{\hbar}\frac{S_+^{(b)}}{\hbar}\bigg)\bigg|S_a, M_a-1\bigg\rangle\bigg|S_b, M_b+1\bigg\rangle \\ &= \ \frac{1}{2}\hbar\Omega_{ab}\bigg\langle S_a, M_a\bigg|\bigg\langle S_b, M_b\bigg|\frac{S_-^{(b)}}{\hbar}\frac{S_+^{(a)}}{\hbar}\bigg|S_a, M_a-1\bigg\rangle\bigg|S_b, M_b+1\bigg\rangle \\ &= \ \frac{1}{2}\hbar\Omega_{ab}\bigg\langle S_a, M_a\bigg|\frac{S_+^{(a)}}{\hbar}\bigg|S_a, M_a-1\bigg\rangle\bigg\langle S_b, M_b\bigg|\frac{S_-^{(b)}}{\hbar}\bigg|S_b, M_b+1\bigg\rangle \end{aligned} \tag{477}$$

We recall that

$$\frac{S_\pm}{\hbar}|S, M_S\rangle \ = \ \sqrt{S(S+1) - M_S(M_S\pm 1)}|S, M_S\pm 1\rangle \tag{478}$$

which leads to

$$\begin{aligned} &\bigg\langle S_a, M_a\bigg|\bigg\langle S_b, M_b\bigg|\frac{1}{2}\hbar\Omega_{ab}\bigg(\frac{S_-^{(b)}}{\hbar}\frac{S_+^{(a)}}{\hbar} + \frac{S_-^{(a)}}{\hbar}\frac{S_+^{(b)}}{\hbar}\bigg)\bigg|S_a, M_a-1\bigg\rangle\bigg|S_b, M_b+1\bigg\rangle \\ &= \ \frac{1}{2}\hbar\Omega_{ab}\bigg\langle S_a, M_a\bigg|\sqrt{S_a(S_a+1)-(M_a-1)M_a}\bigg|S_a, M_a\bigg\rangle\bigg\langle S_b, M_b\bigg|\sqrt{S_b(S_b+1)-(M_b+1)M_b}\bigg|S_b, M_b\bigg\rangle \\ &= \ \frac{1}{2}\hbar\Omega_{ab}\sqrt{S_a(S_a+1)-(M_a-1)M_a}\sqrt{S_b(S_b+1)-(M_b+1)M_b} \end{aligned} \tag{479}$$

### Connection between rate and matrix element

To within better than an order of magnitude, the transition rate expected for a series of coherent transitions among degenerate states would scale like:

$$\Gamma \ \sim \ \frac{\langle S_a, M_a|\langle S_b, M_b|\hat{H}|S_a, M_a-1\rangle|S_b, M_b+1\rangle}{\hbar} \tag{480}$$

Indeed, a detailed calculation for the maximum rate was performed in [132] and was found to be:

$$\max\{\Gamma\} \;=\; 2\frac{|\langle S_a, M_a|\langle S_b, M_b|\hat{H}|S_a, M_a - 1\rangle|S_b, M_b + 1\rangle|}{\hbar} \tag{481}$$

In general one needs to solve the dynamics problem to determine the actual rate for a particular problem, which can be in the range

$$-2\frac{|\langle S_a, M_a|\langle S_b, M_b|\hat{H}|S_a, M_a - 1\rangle|S_b, M_b + 1\rangle|}{\hbar} \;\leq\; \Gamma \;\leq\; 2\frac{|\langle S_a, M_a|\langle S_b, M_b|\hat{H}|S_a, M_a - 1\rangle|S_b, M_b + 1\rangle|}{\hbar} \tag{482}$$

However, these systems "want" to evolve so that they proceed at their maximum rate [132]. Therefore, assuming that the dynamics proceeds at the maximum rate, we then can write

$$\Gamma \;=\; \Omega_{ab}\sqrt{S_a(S_a+1) - (M_a - 1)M_a}\sqrt{S_b(S_b+1) - (M_b+1)M_b} \tag{483}$$

## 6.13 Parameters for magnetic dipole coupling in excitation transfer from $D_2$ donors to $^4$He receivers

This section contains a derivation of the magnetic interaction strength

$$U^2_{magnetic} = \langle \Phi_{D_2} | \boldsymbol{\mu} | \Phi_{^4\text{He}} \rangle \cdot \langle \Psi_{osc} | \hat{\mathbf{B}}\hat{\mathbf{B}} | \Psi_{osc} \rangle \cdot \langle \Psi_{D_2} | \boldsymbol{\mu} | \Psi_{^4\text{He}} \rangle \tag{484}$$

In what follows in this section we develop a rough estimate for this expression.

The magnetic dipole transition involves coupling with the deuterons in the molecular $D_2$ $^5$D $J^\pi = 1^+$ $T = 0$ nuclear states [91], which can make spin-flip transitions to the $^5D$ admixture of the ground state $^4$He nucleus. To develop a rough approximation for the magnetic dipole matrix element we work with

$$|\langle \text{D}_2\ {}^5D\ J = 1 | \mu_z | {}^4\text{He}\ {}^1S \rangle|^2 \ \sim \ 0.15\, \mu_N^2 \int_0^{4 \text{ fm}} P^2_{DD}(r) dr \tag{485}$$

where we have taken the fraction of the ground state $^5$D admixture to be 15%. From a numerical integration, this evaluates to

$$|\langle \text{D}_2\ {}^5D\ J = 1 | \mu_z | {}^4\text{He}\ {}^1S \rangle|^2 \ \sim \ 3.8 \times 10^{-91}\ \mu_N^2 \tag{486}$$

In order to develop estimates under conditions where substantial screening occurs, we write this in terms of our estimate for the Gamow factor for the $^5$D channel evaluated with zero screening (see Table 7)

$$G \ = \ 94.8 \tag{487}$$

This leads to

$$|\langle \text{D}_2\ {}^5D\ J = 1 | \mu_z | {}^4\text{He}\ {}^1S \rangle|^2 \ \sim \ 100\ \mu_N^2 \frac{v_{nuc}}{v_{mol}} e^{-2G} \tag{488}$$

where (see 6.3)

$$\frac{v_{nuc}}{v_{mol}} \ = \ 6.64 \times 10^{-12} \tag{489}$$

We end up with the estimate

$$|\langle \text{D}_2\ {}^5D\ J = 1 | \mu_z | {}^4\text{He}\ {}^1S \rangle| \ \sim \ 10\ \mu_N \sqrt{\frac{v_{nuc}}{v_{mol}}} e^{-G} \tag{490}$$

For an external magnetic field from a magnet, we take the expectation over the associated oscillators (associated with the spin wave modes) and write

$$\langle \Psi_{osc} | \hat{\mathbf{B}}\hat{\mathbf{B}} | \Psi_{osc} \rangle \ = \ B^2 \tag{491}$$

This results in a value of:

$$U^2_{magnetic} \ = \ \left( 10 \sqrt{\frac{v_{nuc}}{v_{mol}}} e^{-G} \right)^2 (\mu_N B)^2 \tag{492}$$

## 6.14 Parameters for electric dipole coupling in excitation transfer from $D_2$ donors to $^4$He receivers

The interaction term for electric coupling is

$$U^2_{electric} = \langle \Phi_{\mathrm{D}_2}|\mathbf{d}|\Phi_{^4\mathrm{He}}\rangle \cdot \langle \hat{\mathbf{E}}\hat{\mathbf{E}}\rangle_{osc} \cdot \langle \Phi_{\mathrm{D}_2}|\mathbf{d}|\Phi_{^4\mathrm{He}}\rangle \tag{493}$$

In what follows in this section we develop a rough estimate for this expression.

The approach that we will take involves relating the dipole moment to the radiative decay rate. From deuteron-deuteron collisions at low energy, we know that gammas can be emitted occasionally (Wilkinson et al, 1985 [92]), and that this gamma radiation occurs through electric quadrupole (E2) coupling from deuterons in the $^5$S channel. We are interested instead in electric dipole (E1) coupling from the $^3$P $J = 1$ channel. We know from Langenbrunner 1990 [91] that the fusion to gamma astrophysical S-factor is not very different at 100 keV than at lower energies, and that at 100 keV the $^3$P $J = 1$ channel makes a dominant contribution. This suggests that the E1 radiative decay is probably not different than the E2 radiative decay rate (which could only be true if the electric dipole coupling were more forbidden). This suggest that we might develop a rough estimate starting from the E2 radiative decay rate at zero energy.

The radiative decay rate for an electric dipole (E1) transtion can be expressed in terms of the dipole moment according to

$$\gamma_{rad} = \frac{4}{3}\frac{1}{4\pi\epsilon_0}\frac{\omega^3}{\hbar c^3}|\langle\phi_f|\mathbf{d}|\phi_i\rangle|^2 = \frac{4}{3}\alpha\frac{\omega^3}{c^2}\left|\left\langle\phi_f\left|\frac{\mathbf{d}}{e}\right|\phi_i\right\rangle\right|^2 \tag{494}$$

where $\hbar\omega = \Delta Mc^2$ and where the fine structure constant is

$$\alpha = \frac{e^2}{4\pi\epsilon_0\hbar c} \tag{495}$$

We can write the square of the dipole moment in terms of the radiative decay rate according to

$$\left|\left\langle\phi_f\left|\frac{\mathbf{d}}{e}\right|\phi_i\right\rangle\right|^2 = \frac{3}{4}\frac{1}{\alpha}\frac{c^2}{\omega^3}\gamma_{rad} \tag{496}$$

For the coupling term for electric dipole interaction we have

$$\begin{aligned} U^2_{electric} &= \langle\phi_i|\mathbf{d}|\phi_f\rangle \cdot \langle\hat{\mathbf{E}}\hat{\mathbf{E}}\rangle_{osc} \cdot \langle\phi_f|\mathbf{d}|\phi_i\rangle \\ &\to |\langle\phi_f|d_z|\phi_i\rangle|^2\langle E_z^2\rangle_{osc} \end{aligned} \tag{497}$$

in the event that the ground state is a $J = 0, M_J = 0$ state, and the upper state is a $J = 1, M_J = 0$ state. We recall that the electric field responsible for accelerating nuclei in the lattice due to vibrations is related to the energy in the vibrational model, and that we can write

$$\langle\hat{\mathbf{E}}_j\hat{\mathbf{E}}_k\rangle_{osc} \to \hat{\mathbf{i}}_z\hat{\mathbf{i}}_z\frac{\sqrt{M_jM_k}\omega_A^2}{NZ_jZ_ke^2}(P_{diss}\tau) \tag{498}$$

where we assume the acoustic mode vibrations are z-directed and $\hbar\omega_A$ is the acoustic phonon mode energy. We have assumed that the unit vector associated with the phonon mode $\hat{\mathbf{e}} \to \hat{\mathbf{i}}_z$ which assumes z-directed vibrations and also assumes a simple unit cell. This is the result we would get making use of the phonon based electric field operators that we defined in Eq. 25. We can use this to write

$$\begin{aligned} U^2_{electric} &= \frac{3}{4}\frac{1}{\alpha}\frac{c^2}{\omega^3}\gamma_{rad}\frac{M_j\omega_A^2}{NZ^2}(P_{diss}\tau) \\ &= \frac{3}{4\alpha}\frac{\gamma_{rad}}{\omega}\frac{\omega_A^2}{\omega^2}\frac{M_jc^2}{NZ^2}(P_{diss}\tau) \\ &= \frac{3}{4\alpha}\frac{\hbar\gamma_{rad}}{\Delta Mc^2}\frac{(\hbar\omega_A)^2}{(\Delta Mc^2)^2}\frac{M_jc^2}{NZ^2}(P_{diss}\tau) \end{aligned} \tag{499}$$

In this case $\gamma_{rad}$ is given in terms of the fusion rate according to

$$\gamma_{rad} = 6 \times 10^{-8}\gamma_{DD} \tag{500}$$

where the ratio $\gamma_{rad}/\gamma_{DD}$ (portion of the DD fusion reactions that result in gamma emission) can be taken from [92] and where the spontaneous fusion rate per $D_2$, $\gamma_{DD}$, given by Koonin and Nauenberg 1989 [20] as

$$\gamma_{DD} \approx 3 \times 10^{-64}\ \mathrm{s}^{-1} \tag{501}$$

## 6.15 Parameters for relativistic coupling in excitation transfer from $D_2$ donors to $^4$He receivers

For relativistic coupling, the interaction strength is

$$U^2_{relativistic} = \langle \Phi_{\mathrm{D}_2} | \mathbf{a} | \Phi_{^4\mathrm{He}} \rangle \cdot c^2 \langle \hat{\mathbf{P}}_{^4\mathrm{He}} \hat{\mathbf{P}}_{^4\mathrm{He}} \rangle \cdot \langle \Phi_{\mathrm{D2}} | \mathbf{a} | \Phi_{^4\mathrm{He}} \rangle \tag{502}$$

In what follows in this section we develop a rough estimate for this expression.

Working with the **a** operator in the form given above is labor-intensive since it involves two-body interactions (and potentially three-body interactions in a nuclear model where three-body potentials are used). In [93] an identity is given which allows the operator to be expressed in terms of one-body terms, which provides for a major reduction in the effort needed for detailed calculations. The nonrelativistic reduction of the **a** operator based on the identity is given by

$$\mathbf{a} = \frac{1}{2i} \frac{\Delta E}{Mc^2} \left( \sum_j \boldsymbol{\sigma}_j \times \frac{\boldsymbol{\pi}_j}{mc} \right) \tag{503}$$

We can get a rough estimate for the **a** matrix element for a $D_2$ $^3$P (J=1) to $^4$He transition according to

$$\begin{aligned} |\mathbf{a}| &\sim \frac{1}{2} \frac{24\text{ MeV}}{4 \times 931.5\text{ MeV}} \frac{\hbar\,(1\text{ fm}^{-1})}{mc} e^{-G} \sqrt{\frac{v_{nuc}}{v_{mol}}} \\ &= 1.7 \times 10^{-9} e^{-G} \end{aligned} \tag{504}$$

where $\frac{v_{nuc}}{v_{mol}}$ is the ratio of the nuclear volume to molecular volume, which we take to be $2.6 \times 10^{-12}$.

We then have an estimate for the matrix element associated with the $D_2$/$^4$He transition

$$\begin{aligned} \langle \Phi_{D2} | \mathbf{a} | \Phi_{^4\mathrm{He}} \rangle &= \frac{1}{2} \frac{24\text{ MeV}}{4 \times 931.5\text{ MeV}} \frac{\hbar\,(1\text{ fm}^{-1})}{mc} \sqrt{\frac{v_{nuc}}{v_{mol}}} e^{-G} \\ &= 6.7 \times 10^{-4} \sqrt{\frac{v_{nuc}}{v_{mol}}} e^{-G} \end{aligned} \tag{505}$$

In section 6.9, we provide a more detailed calculation of the matrix element, which yields:

$$\langle \Phi_{D2} | \mathbf{a} | \Phi_{^4\mathrm{He}} \rangle = 0.0362 \sqrt{\frac{v_{nuc}}{v_{mol}}} e^{-G} \tag{506}$$

The oscillator expectation value in the case of acoustic phonons is

$$\langle \hat{\mathbf{P}}_i \hat{\mathbf{P}}_j \rangle_{osc} = \hat{\mathbf{i}}_z \hat{\mathbf{i}}_z \hbar\omega_A \frac{\sqrt{M_j M_k}}{2N} \left( 2\langle \hat{n}_A \rangle + 1 \right) \tag{507}$$

We can write this in terms of the dissipated power according to

$$\langle \hat{\mathbf{P}}_j \hat{\mathbf{P}}_k \rangle_{osc} \to \hat{\mathbf{i}}_z \hat{\mathbf{i}}_z \frac{\hbar\omega_A \sqrt{M_j M_k}}{N} \frac{P_{diss}\tau}{\hbar\omega_A} = \hat{\mathbf{i}}_z \hat{\mathbf{i}}_z \frac{\sqrt{M_j M_k}}{N} P_{diss}\tau \tag{508}$$

where j is the ground state nucleus on the donor side and k is the ground state nucleus on the receiver side. The two masses ($M_{^4He}$ and $M_{D_2}$) are essentially the same as far as the phonons are concerned. In the case of transfer from $D_2$ donors to $^4$He receivers, we end up with

$$U^2_{relativistic} = \left( 0.0362 \sqrt{\frac{v_{nuc}}{v_{mol}}} e^{-G} \right)^2 c^2 \frac{M_j}{N} P_{diss} \tau \tag{509}$$

## 6.16 Parameters for relativistic coupling in excitation transfer from $D_2$ donors to Pd receivers

For relativistic coupling from a DD fusion transition to a heavy nucleus excitation the interaction strength is

$$(UV)_{relativistic} = \langle \Phi_{\mathrm{D}_2} | \mathbf{a} | \Phi_{^4\mathrm{He}} \rangle \cdot c^2 \langle \hat{\mathbf{P}}_{^4\mathrm{He}} \hat{\mathbf{P}}_{\mathrm{Pd}} \rangle \cdot \langle \Phi_{\mathrm{Pd}*} | \mathbf{a} \cdot \hat{\mathbf{i}}_z | \Phi_{\mathrm{Pd}} \rangle \tag{510}$$

In what follows in this section we develop a rough estimate for this expression.

For the first term, the same value applies as estimated in the previous section 6.15. For the last term, a quantitative estimate is developed in section 6.11. This results in

$$|\langle \Phi_{\mathrm{Pd}*} | \mathbf{a} \cdot \hat{\mathbf{i}}_z | \Phi_{\mathrm{Pd}} \rangle| \rightarrow 2.6 \times 10^{-5}\, O \tag{511}$$

This results in an interaction strength of:

$$\begin{aligned} (UV)_{relativistic} &= \langle \Phi_{\mathrm{D}_2} | \mathbf{a} | \Phi_{^4\mathrm{He}} \rangle \cdot c^2 \langle \hat{\mathbf{P}}_{^4\mathrm{He}} \hat{\mathbf{P}}_{\mathrm{Pd}} \rangle \cdot \langle \Phi_{\mathrm{Pd}*} | \mathbf{a} \cdot \hat{\mathbf{i}}_z | \Phi_{\mathrm{Pd}} \rangle \\ &= \left( 0.0362 \sqrt{\frac{v_{nuc}}{v_{mol}}} e^{-G} \right) c^2 \frac{\sqrt{M_{^4He} M_{Pd}}}{N} P_{diss} \tau \left( 2.6 \times 10^{-5}\, O \right) \end{aligned} \tag{512}$$

## 6.17 Proton, neutron and alpha removal energies of the stable Pd isotopes

The stability of the Pd bound states is limited by proton, neutron and alpha removal energies. These removal energies for the stable Pd isotopes are given in Table 10:

**Table 10:** Proton, neutron and alpha removal energies.

| Isotope | $I_p$ (MeV) | $I_n$ (MeV) | $I_\alpha$ (MeV) |
|---|---|---|---|
| Pd-102 | 7.732 | 10.467 | 2.131 |
| Pd-104 | 8.572 | 9.884 | 2.596 |
| Pd-105 | 8.667 | 7.026 | 2.888 |
| Pd-106 | 9.257 | 9.470 | 3.229 |
| Pd-108 | 9.857 | 9.137 | 3.856 |
| Pd-110 | 10.523 | 8.716 | 4.434 |

We can see that it takes the least energy to remove an alpha particle for all of the stable Pd isotopes. We expect states above the alpha removal energy to be unstable against alpha decay.

## 6.18 Estimating nuclear density of states for Pd nuclei

The density of nuclear states allows us to estimate how near the resonance condition the closest available nuclear excited state is. It can be expressed as a summation over the different nuclear transitions

$$\rho_N(E) \;=\; \sum_f \delta(E - E_f) \tag{513}$$

Given the model for the levels of $^{120}$Sn in section 5.4, we can evaluate this expression.

No data sets for the nuclear density of states of Pd nuclei are available at this point. However, since Pd and Sn nuclei are similarly sized, we use the $^{120}$Sn data as a proxy for our Pd estimates.

### Gaussian fits for the nuclear density of states

The resulting density of states, along with a Gaussian fit, is shown in Figure 57.

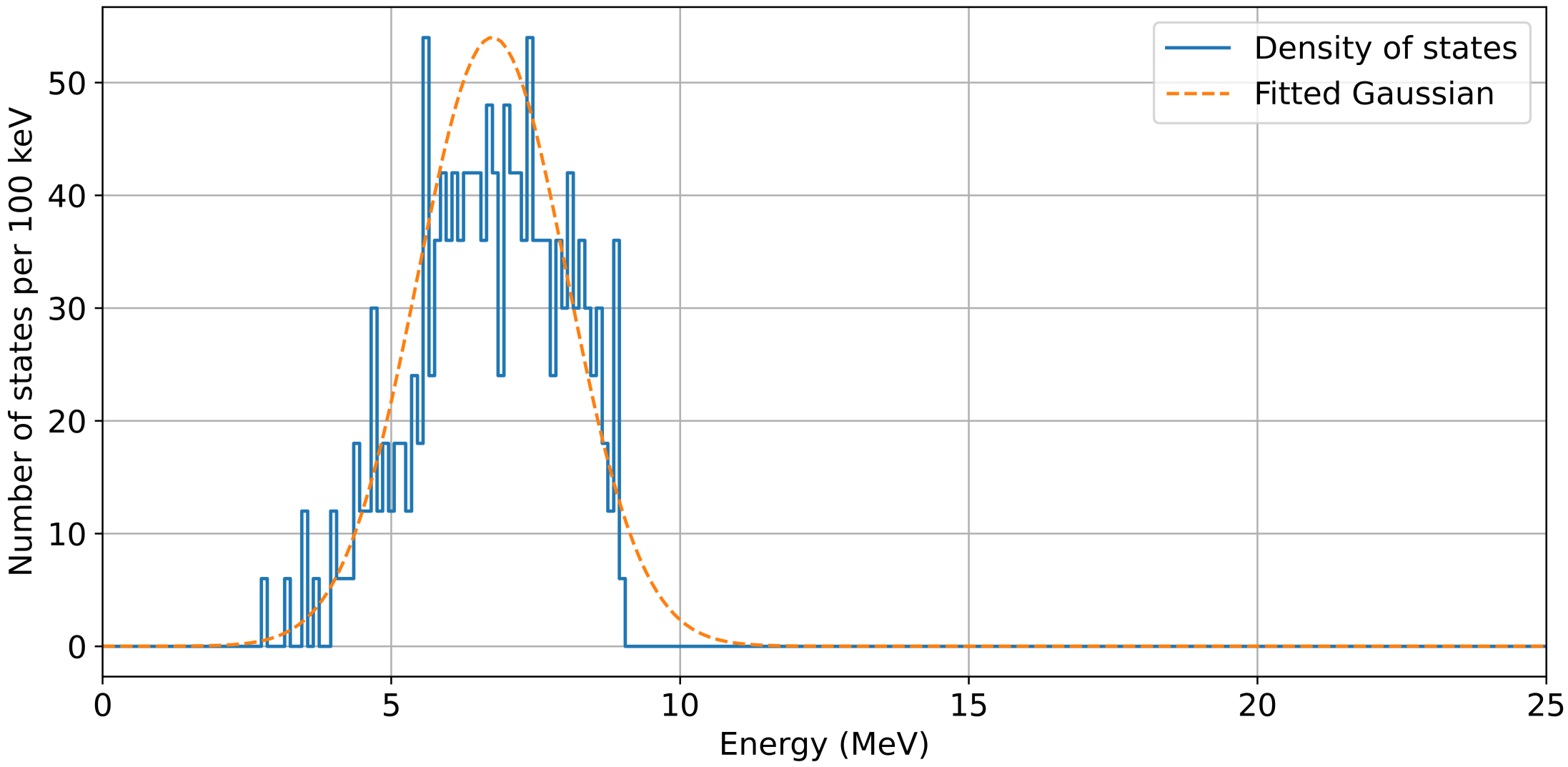

**Figure 57:** Result of the calculation of $\sum_i \delta(E - E_i)$ evaluated with the $^{120}$Sn data set (and assumed to be relevant for all stable Pd isotopes).

Due to the disintegration threshold <9 MeV, no single state is available near 23.85 MeV. However, in section 5.9 and onward we consider excitation transfer to a combination of lower-energy transitions that collectively meet the resonance condition.

In the remainder of this section, we will estimate the density of states near 23.85 MeV that results from combinations of lower-energy of states. We will start with considering combinations of two states (second order) and then combinations of three states (third order).

We find that the distributions associated with the different orders are roughly Gaussian in shape. After evaluating the summations at different orders, we fit them to Gaussian distributions.

For the lowest order contribution we evaluated

$$\sum_i \delta(E - E_i) \;\approx\; A_1 \frac{1}{\sqrt{2\pi\sigma_1^2}} e^{-(E-\mu_1)^2/2\sigma_1^2} \tag{514}$$

where

$$\mu_1 \;\approx\; 6.75\,\mathrm{MeV} \qquad \sigma_1 \;\approx\; 1.28\,\mathrm{MeV} \qquad A_1 \;\approx\; 1.76\times 10^5 \tag{515}$$

**Nuclear density of states when considering combinations of two Pd transitions**

At next order we evaluated the different terms, and once again fit them to a Gaussian

$$\sum_i \sum_j \delta(E - E_i - E_j) \;\approx\; A_2 \frac{1}{\sqrt{2\pi\sigma_2^2}} e^{-(E-\mu_2)^2/2\sigma_2^2} \tag{516}$$

with

$$\mu_2 \;\approx\; 2\times 6.75\,\mathrm{MeV} \qquad \sigma_2 \;\approx\; \sqrt{2}\times 1.28\,\mathrm{MeV} \qquad A_2 \;\approx\; 2.15\times 10^8 \tag{517}$$

The density of states from combinations of two transitions is shown in Figure 58.

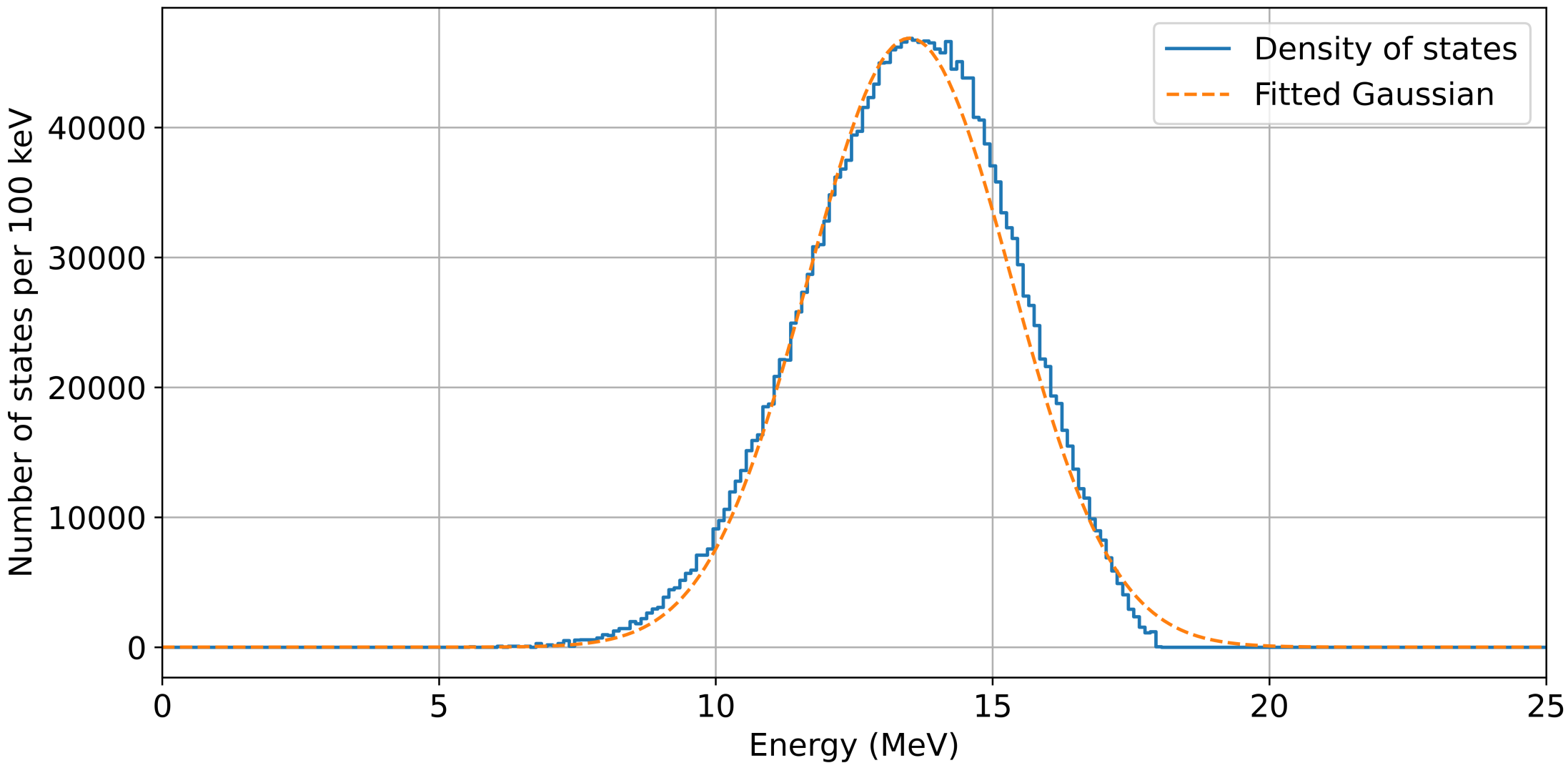

**Figure 58:** Result of the calculation of $\sum_i \sum_j \delta(E - E_i - E_j)$ evaluated with the $^{120}$Sn data set (and assumed to be relevant for all stable Pd isotopes).

**Nuclear density of states when considering combinations of three Pd transitions**

For terms at the next order we obtained

$$\sum_i \sum_j \sum_k \delta(E - E_i - E_j - E_k) \;\approx\; A_3 \frac{1}{\sqrt{2\pi\sigma_3^2}} e^{-(E-\mu_3)^2/2\sigma_3^2} \tag{518}$$

with

$$\mu_3 \;\approx\; 3 \times 6.75\,\text{MeV} \qquad \sigma_3 \;\approx\; \sqrt{3} \times 1.28\,\text{MeV} \qquad A_3 \;\approx\; 3.28 \times 10^{11} \tag{519}$$

The density of states from combinations of three transitions is shown in Figure 59.

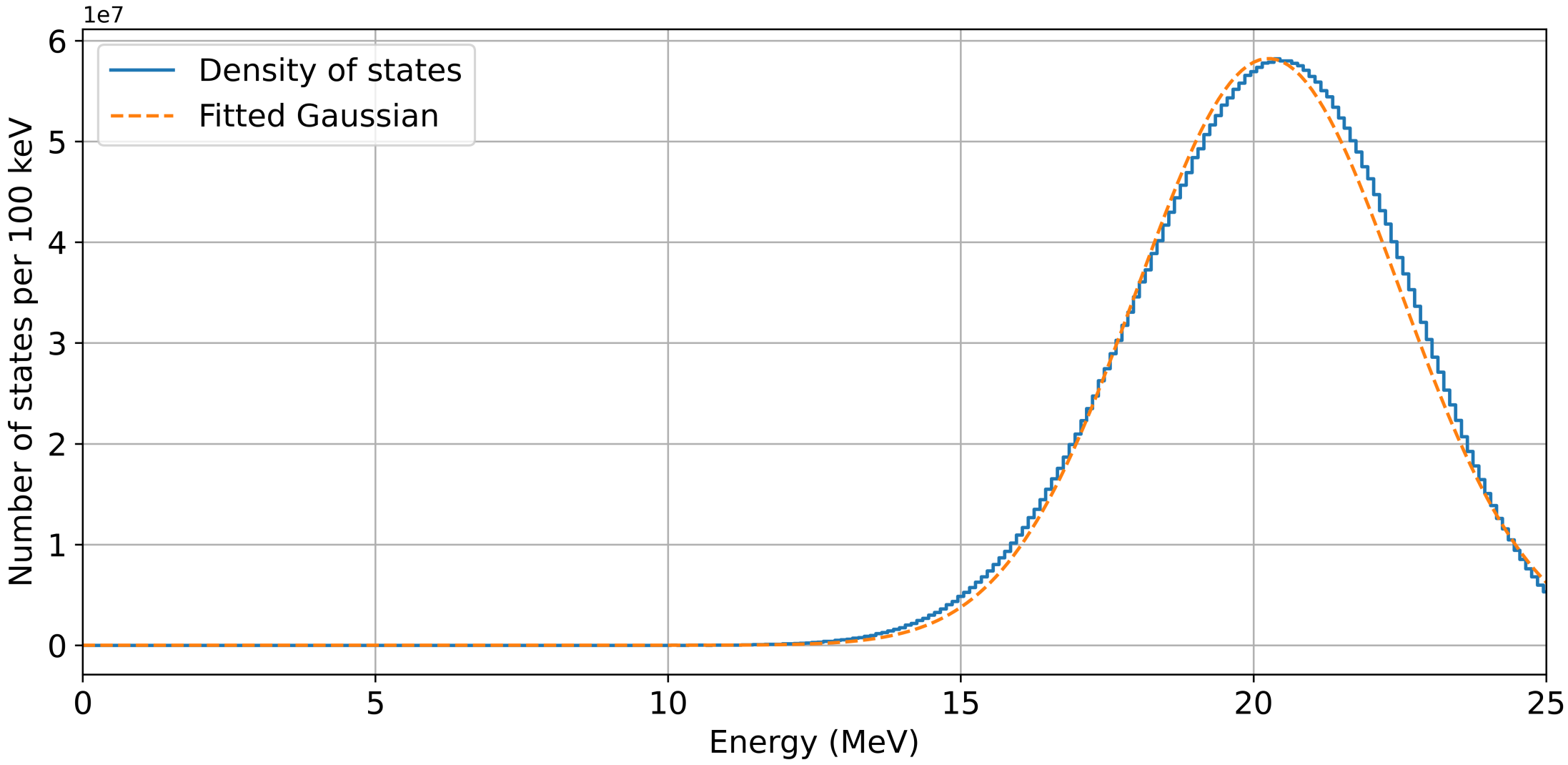

**Figure 59:** Result of the calculation of $\sum_i \sum_j \sum_k \delta(E - E_i - E_j - E_k)$ evaluated with the $^{120}$Sn data set (and assumed to be relevant for all stable Pd isotopes).

## Conclusions

Figure 60 shows the sum total density of states for one, two, and three transitions.

In the region of the resonance condition (23848109 eV), the resulting density of states is estimated to be near:

$$\rho_N^{(0)} \;=\; \rho_N(\Delta M c^2) \;\rightarrow\; 178\,\text{eV}^{-1} \tag{520}$$

Note that by including the possibility of more than three transitions in this estimation, the density of states is expected to be even higher.

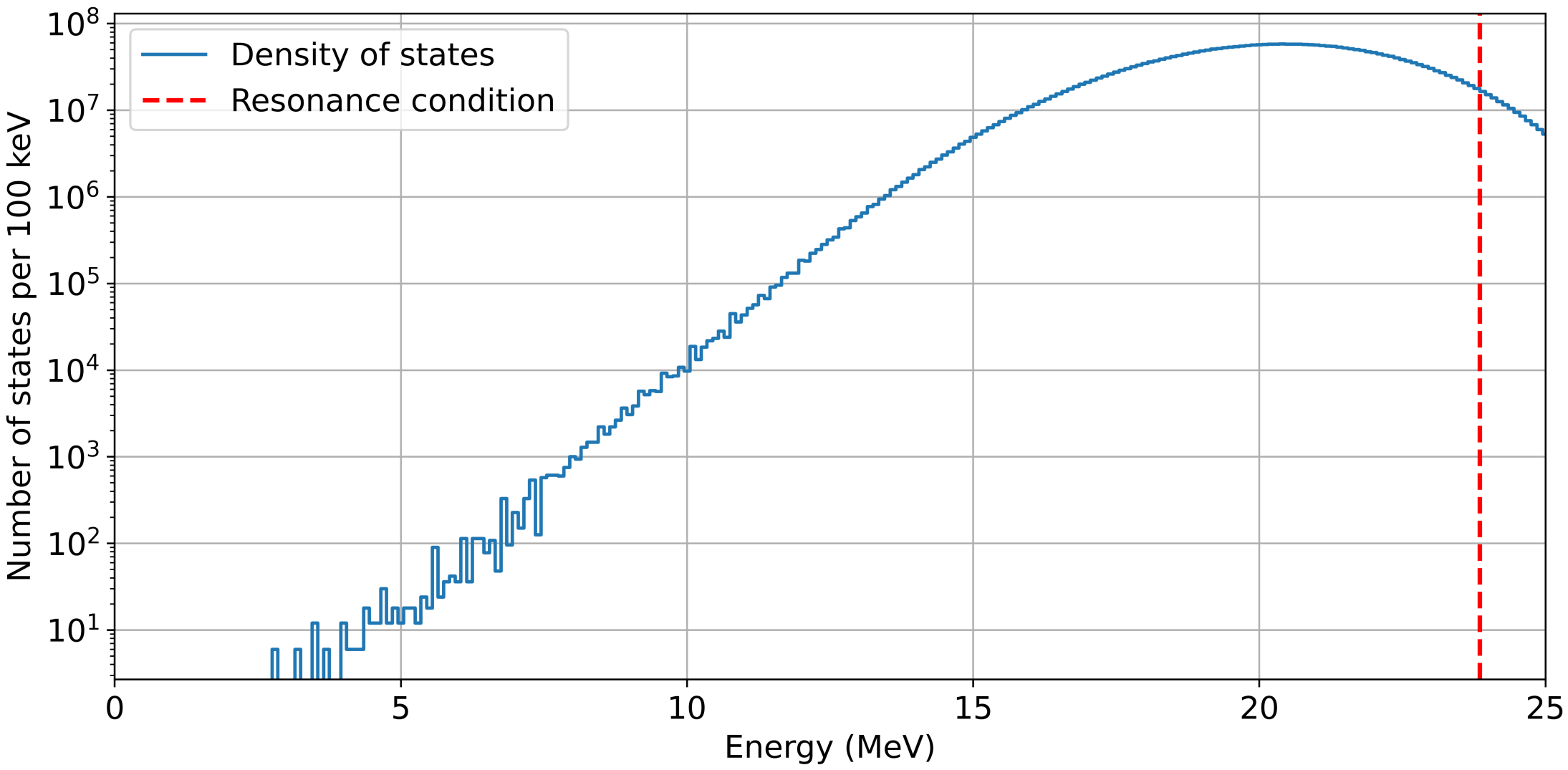

**Figure 60:** Result of the summing the density of states for combinations of one, two and three nuclear transitions, evaluated with the $^{120}$Sn data set (and assumed to be relevant for all stable Pd isotopes)

### 6.19 Estimating generalized nuclear density of states for Pd nuclei

In the expression for the Golden Rule in Eq. 209, there appears a summation over the different nuclear transitions which we identify as a *generalized density of states*

$$\tilde{\rho}_N(E) \ = \ \sum_f \left( \prod F(g) \right)^2_f \delta(E - E_f) \tag{521}$$

With this definition, the excitation transfer rate for transfer to multiple resonant receiver states can be expressed as

$$\Gamma_{transfer}(E) \ = \ \left( \frac{2\pi}{\hbar} |\Delta \mathcal{U}|^2 \right) (\tilde{\rho}_N * \rho_A * \rho_O * \rho_P)(E) \tag{522}$$

where the $*$ notation indicates that on the RHS we are convolving three densities of states $\rho_A * \rho_O * \rho_P$ with this new generalized density of states $\tilde{\rho}_N$, and evaluating the result at energy $E$.

#### Gaussian fits for the generalized nuclear density of states

Given the model for the levels and $O$-values from $^{120}$Sn in section 5.4, it is in principle straightforward to evaluate the terms that appear in the summation. We put together some code to do exactly that, and after plotting the results, we found that the distributions associated with the different orders were roughly Gaussian in shape. After evaluating the summations at different orders, we fit them to Gaussian distributions, with parameters that depend on $E_A$.

We recall (Eq. 199) that the dimensionless coupling constant for a Pd*/Pd transition is

$$g \ = \ 865 \, O_{\text{Pd}} \sqrt{\frac{P_{diss}\tau_A}{1 \text{ J}}} \sqrt{\frac{N_{^A\text{Pd}}}{N}} \tag{523}$$

where

$$E_A \ = \ P_D^{(A)} \tau_A \tag{524}$$

is the energy in the acoustic phonon mode. We identify the ratio $N_{PdA}/N$ as the isotopic fraction of the Pd isotope of the transition. From the discussion above we know that the center of mass momentum operator includes contributions from the acoustic phonon mode, the optical phonon mode and the plasmon mode. However, for the primary nuclear transitions under discussion here, the contribution from the acoustic dominates under most conditions.

For the lowest order contribution we evaluated

$$\sum_i F^2(g_i) \delta(E - E_i) \ \approx \ A_1 \frac{1}{\sqrt{2\pi\sigma_1^2}} e^{-(E-\mu_1)^2/2\sigma_1^2} \tag{525}$$

where

$$\mu_1 \ \approx \ 6.75 \text{ MeV} \qquad \sigma_1 \ \approx \ 1.28 \text{ MeV} \tag{526}$$

and

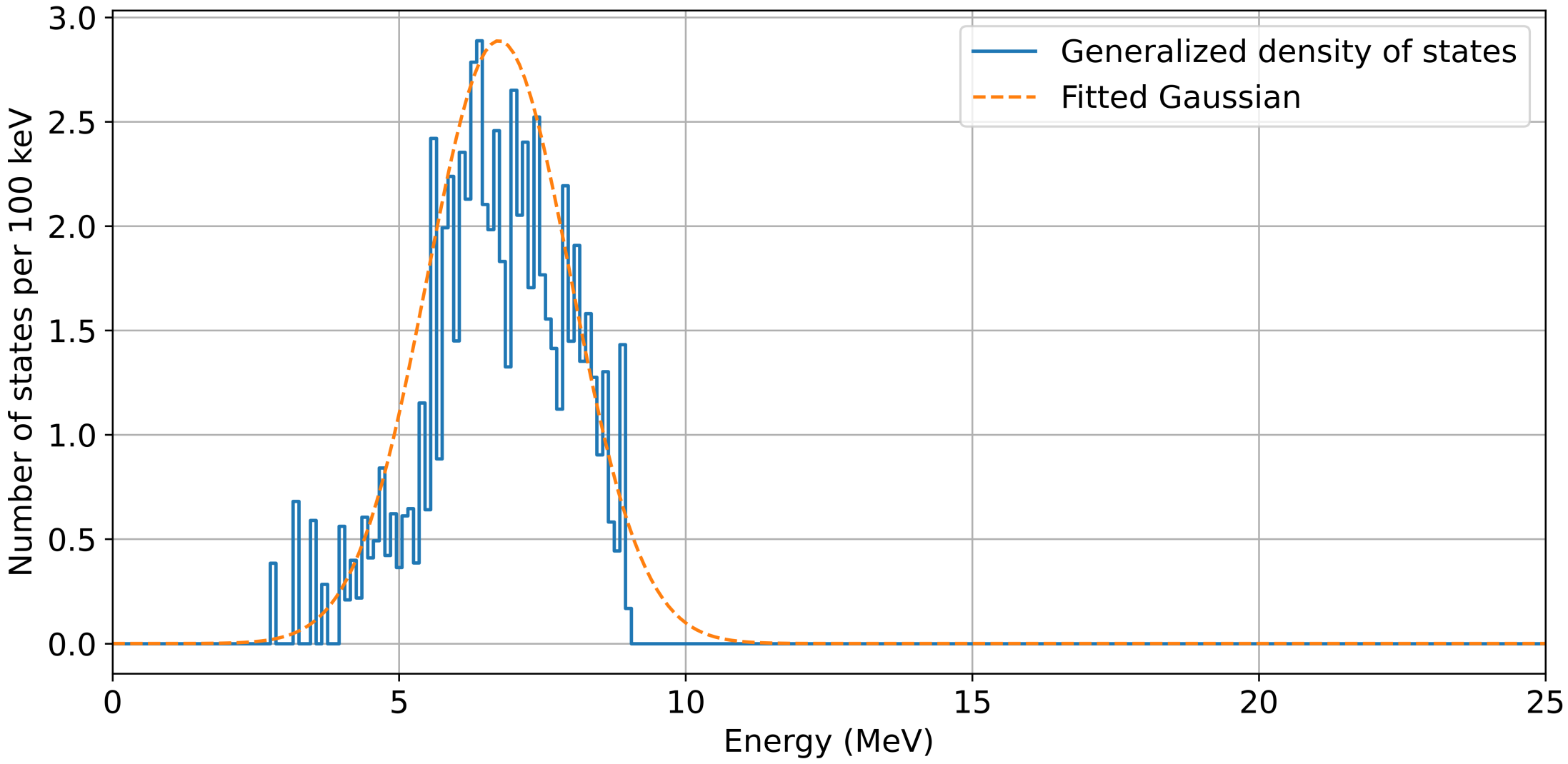

**Figure 61:** Result of the calculation of $\sum_i F^2(g_i)\delta(E-E_i)$ evaluated with the $^{120}$Sn data set (and assumed to be relevant for all stable Pd isotopes) at $E_A = 10^{-3}$ J.

$$A_1 \;=\; \alpha \frac{E_A}{(E_0^s + E_A^s)^{1/s}} \tag{527}$$

with

$$\alpha \;=\; 365.04 \qquad E_0 \;=\; 1.2132 \times 10^{-3} \text{ J} \qquad s \;=\; 0.45300 \tag{528}$$

The sum evaluated an an acoustic energy of $E_A = 1$ mJ is shown in Figure 61.

At next order we evaluated the different terms, and once again fit them to a Gaussian

$$\sum_i \sum_j F^2(g_i)F^2(g_j)\delta(E-E_i-E_j) \;\approx\; A_2 \frac{1}{\sqrt{2\pi\sigma_2^2}} e^{-(E-\mu_2)^2/2\sigma_2^2} \tag{529}$$

with

$$\mu_2 \;\approx\; 2 \times 6.75 \text{ MeV} \qquad \sigma_2 \;\approx\; \sqrt{2} \times 1.28 \text{ MeV} \tag{530}$$

and

$$A_2 \;=\; \alpha \frac{E_A^2}{(E_0^s + E_A^s)^{2/s}} \tag{531}$$

with

$$\alpha \;=\; 133256 \qquad E_0 \;=\; 1.2132 \times 10^{-3} \text{ J} \qquad s \;=\; 0.453003 \tag{532}$$

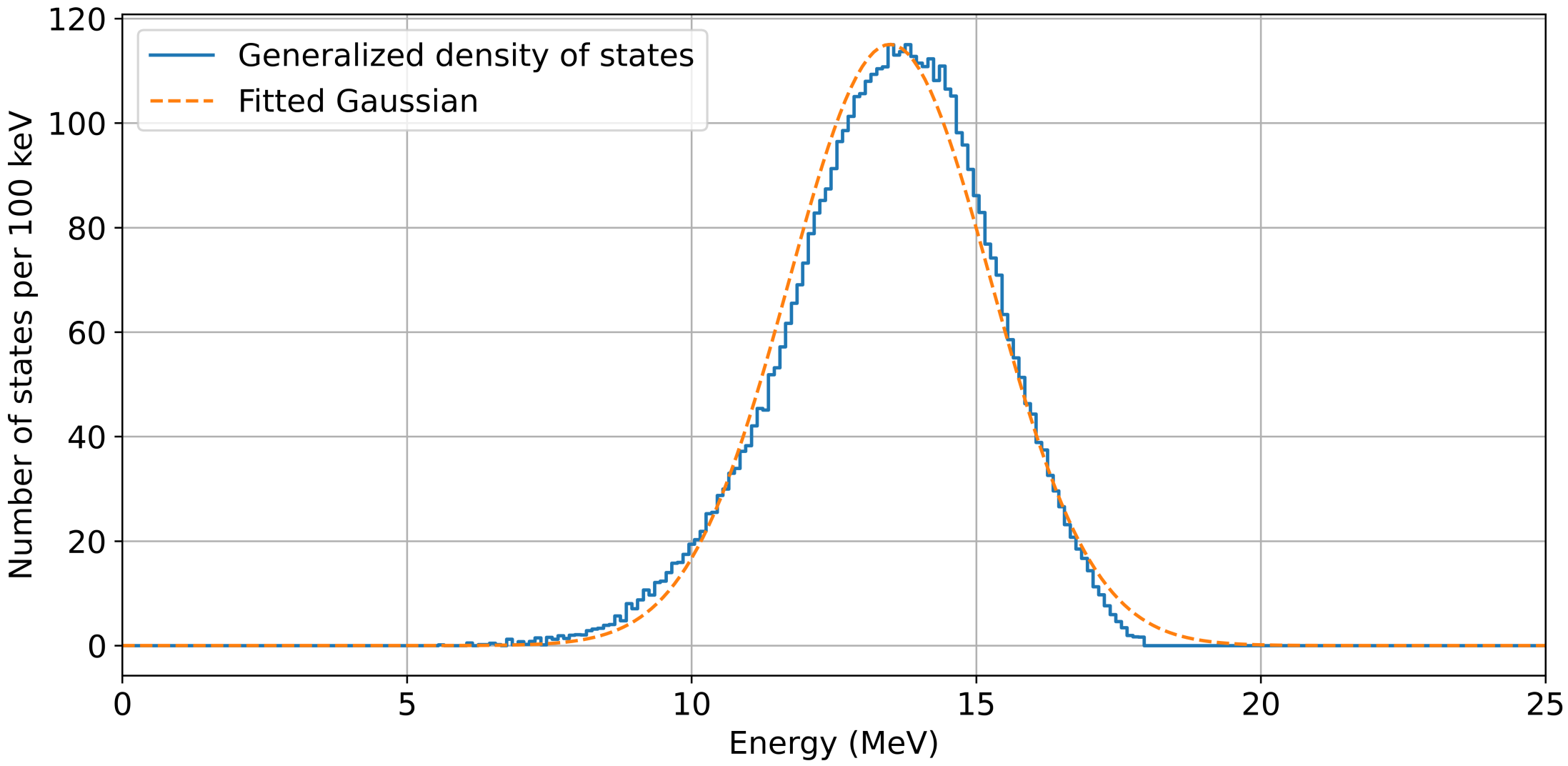

**Figure 62:** Result of the calculation of $\sum_i \sum_j F^2(g_i)F^2(g_j)\delta(E - E_i - E_j)$ evaluated with the $^{120}$Sn data set (and assumed to be relevant for all stable Pd isotopes) at $E_A = 10^{-3}$ J.

The sum evaluated an an acoustic energy of $E_A = 1$ mJ is shown in Figure 62.

For terms at the next order we obtained

$$\sum_i \sum_j \sum_k F^2(g_i)F^2(g_j)F^2(g_k)\delta(E - E_i - E_j - E_k) \approx A_3 \frac{1}{\sqrt{2\pi\sigma_3^2}} e^{-(E-\mu_3)^2/2\sigma_3^2} \tag{533}$$

with

$$\mu_3 \approx 3 \times 6.75 \text{ MeV} \qquad \sigma_3 \approx \sqrt{3} \times 1.28 \text{ MeV} \tag{534}$$

and

For the sum we have

$$A_3 = \alpha \frac{E_A^3}{(E_0^s + E_A^s)^{3/s}} \tag{535}$$

with

$$\alpha = 4.864 \times 10^7 \qquad E_0 = 1.2132 \times 10^{-3} \text{ J} \qquad s = 0.453003 \tag{536}$$

The sum evaluated an an acoustic energy of $E_A = 1$ mJ is shown in Figure 63.

These results suggest that we might develop a model for all transitions based on

$$\tilde{\rho}_N(E) = \sum_m A_m \frac{1}{\sqrt{2\pi\sigma_m^2}} e^{-(E-\mu_m)^2/2\sigma_m^2} \tag{537}$$

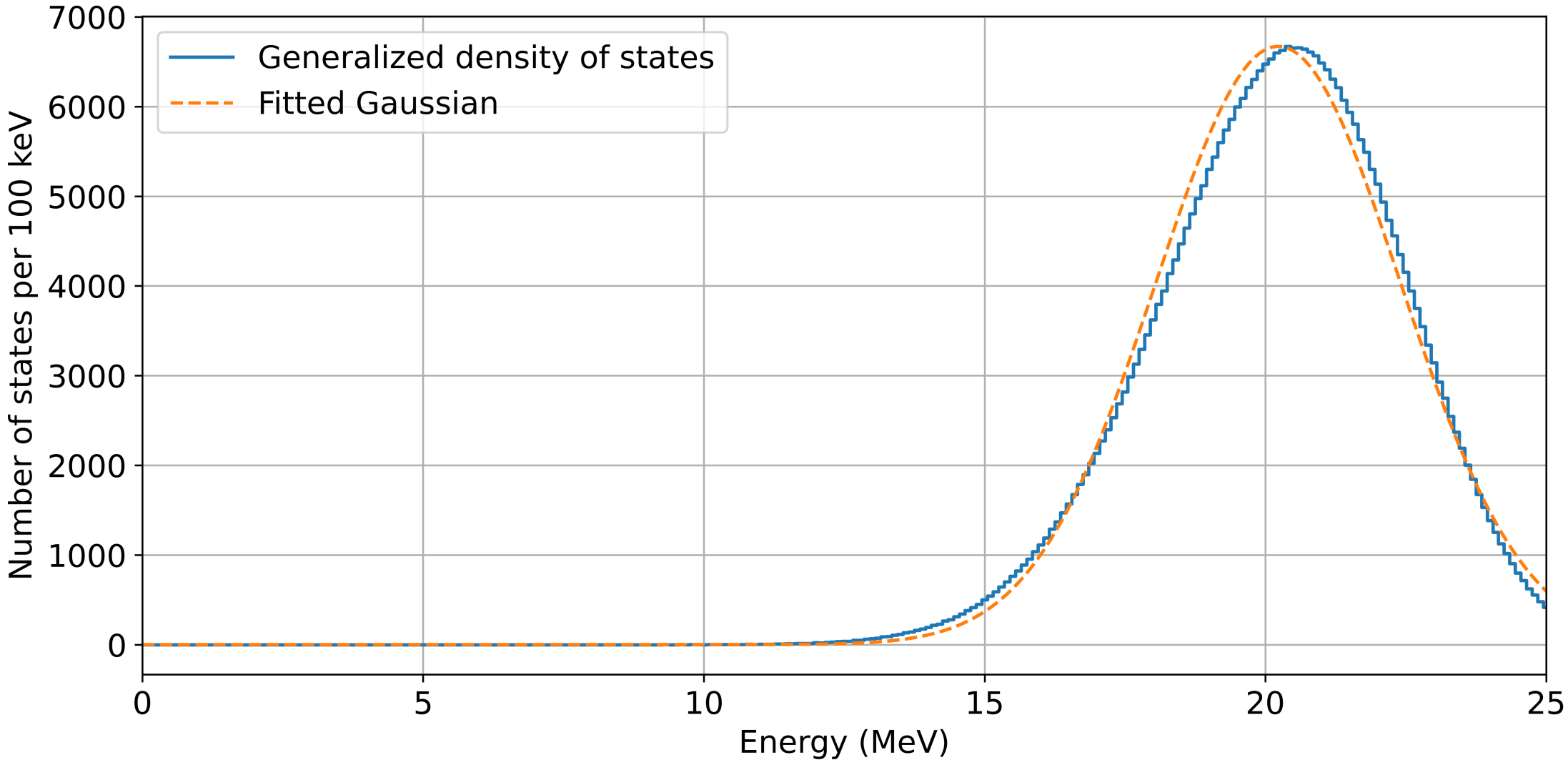

**Figure 63:** Result of the calculation of $\sum_i \sum_j \sum_k F^2(g_i)F^2(g_j)F^2(g_k)\delta(E-E_i-E_j-E_k)$ evaluated with the $^{120}$Sn data set (and assumed to be relevant for all stable Pd isotopes) at $E_A = 10^{-3}$ J.

$$\mu_m \;\approx\; m \times 6.75 \text{ MeV} \qquad \sigma_m \;\approx\; \sqrt{m} \times 1.28 \text{ MeV} \tag{538}$$

and

$$A_m \;=\; \alpha^m \frac{E_A^m}{(E_0^s + E_A^s)^{m/s}} \tag{539}$$

with

$$\alpha \;=\; 365.03 \qquad E_0 \;=\; 1.2132 \times 10^{-3} \text{ J} \qquad s \;=\; 0.45300 \tag{540}$$

**Generalized nuclear density of states at the fusion energy**

It will be useful to have a parameterization of the generalized nuclear density of states

$$\tilde{\rho}_N(E) \;=\; \sum_f \left( \prod F(g) \right)_f^2 \delta(E - E_f) \tag{541}$$

evaluated specifically at the fusion transition energy

$$E \;=\; \Delta M c^2 \;=\; 23.85 \text{ MeV} \tag{542}$$

Results are shown in Figure 64.

We have fit this function over the 6 orders of magnitude shown according to

$$\tilde{\rho}_N(\Delta Mc^2) \;=\; a\frac{E_A^n}{(E_0^s + E_A^s)^{m/s}} \tag{543}$$

with fitting parameters

$$\begin{aligned} a \;&=\; 3041.48\,\text{eV}^{-1} \qquad E_0 \;=\; 2.50643\times 10^{-3}\ \text{J} \\ s \;&=\; 0.55385 \qquad n \;=\; 3.51605 \qquad m \;=\; 3.35353 \end{aligned} \tag{544}$$

with $E_A$ in Joules. The resulting fit is shown against the Gaussian model in Figure 64. It will be convenient to have a reference value against which to normalize. At $E_A = 1$ J we can write

$$\tilde{\rho}_N^{(0)} \;=\; \tilde{\rho}_N(\Delta Mc^2)\bigg|_{1\ \text{J}} \;\rightarrow\; 2451\ \text{eV}^{-1} \tag{545}$$

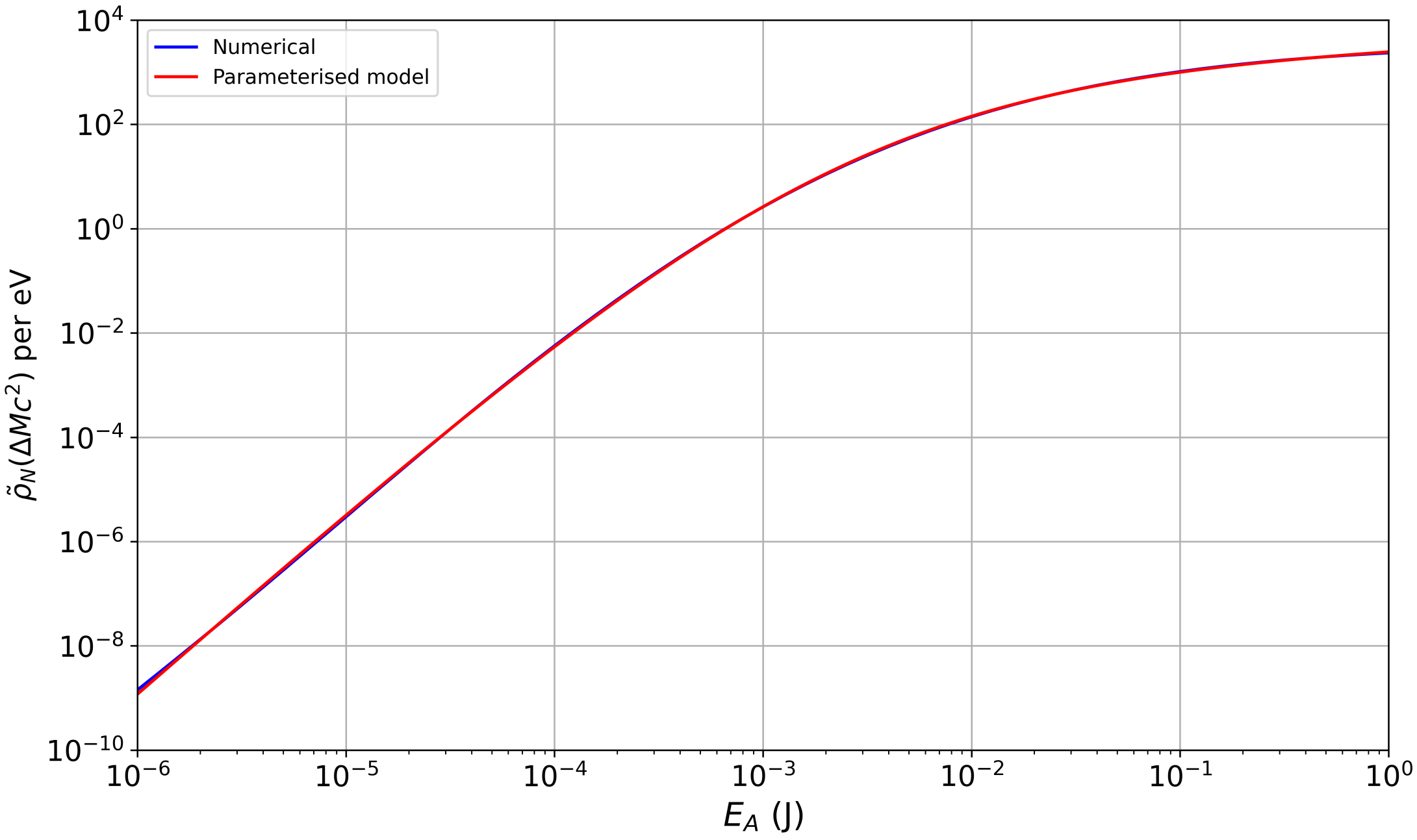

**Figure 64:** Generalized nuclear density of states $\tilde{\rho}_N(E_A)$ (blue); result of least squares fitting to Eq 543 (red).

## 6.20 Nuclear molecule cluster states

The modeling of atomic nuclei has long stood as a cornerstone in the field of nuclear physics, providing significant insights into the structure, stability, and dynamics of nuclei. Traditional approaches, rooted in the Liquid Drop Model (LDM) and the Shell Model, have offered a macroscopic and microscopic understanding of nuclear forces, energy distributions, and the magic numbers that denote extra stability for certain nucleon configurations. These models, while foundational, primarily treat the nucleus as a homogeneous collection of nucleons or as shells of particles without explicit consideration of cluster formations.

Over time, the notion emerged that nucleons may form clusters associated with different nuclear states. Many different cluster configurations are conceivable. Nuclear states have also been identified where nucleons are concentrated in two (or more) separate regions rather than in a single nucleus that may be highly deformed. Such nuclei have been referred to as *nuclear molecules*. Such considerations were motivated by the observation of resonances in collisions between $^{12}$C nuclei [133]. Two nuclei close together at the fermi scale have the possibility of forming a nuclear molecule, by analogy at the atomic scale of molecules formed through electron bonding [134, 135, 136]. Earlier work on nuclear fission proposed essentially the same notion [137].

In [122] we consider variations of the liquid drop model for nuclear molecules from $^{106}$Pd. See Figure 65 as an example for a clustered state of Pd modeled with a basic (Bohr-Wheeler) liquid drop model approach. An important conclusion is that reasonably stable binary nuclear molecules are not expected from liquid drop models as normally used for fission calculations. However, this is likely a consequence of the modeling approach and not because long-lived nuclear molecule states cannot exist. Perhaps the most significant drawback of the Bohr-Wheeler liquid drop model for describing fission and fusion processes is that the strong force interaction comes in through volume and surface energy terms, so that there is no residual strong force interaction when the daughters are separated. More sophisticated model that we explored such as the folded Yukawa model and the finite range liquid drop model (FRLDM) remedy this as the strong force interaction is modeled through a scalar Yukawa potential between different elements of the liquid drop mass density, which leads to a short range daughter-daughter attraction as might be expected intuitively. However, what these models have in common is that they consider highly deformed single nuclei rather than a molecule-like state consisting of two separate clusters and their lifetimes are generally very short.

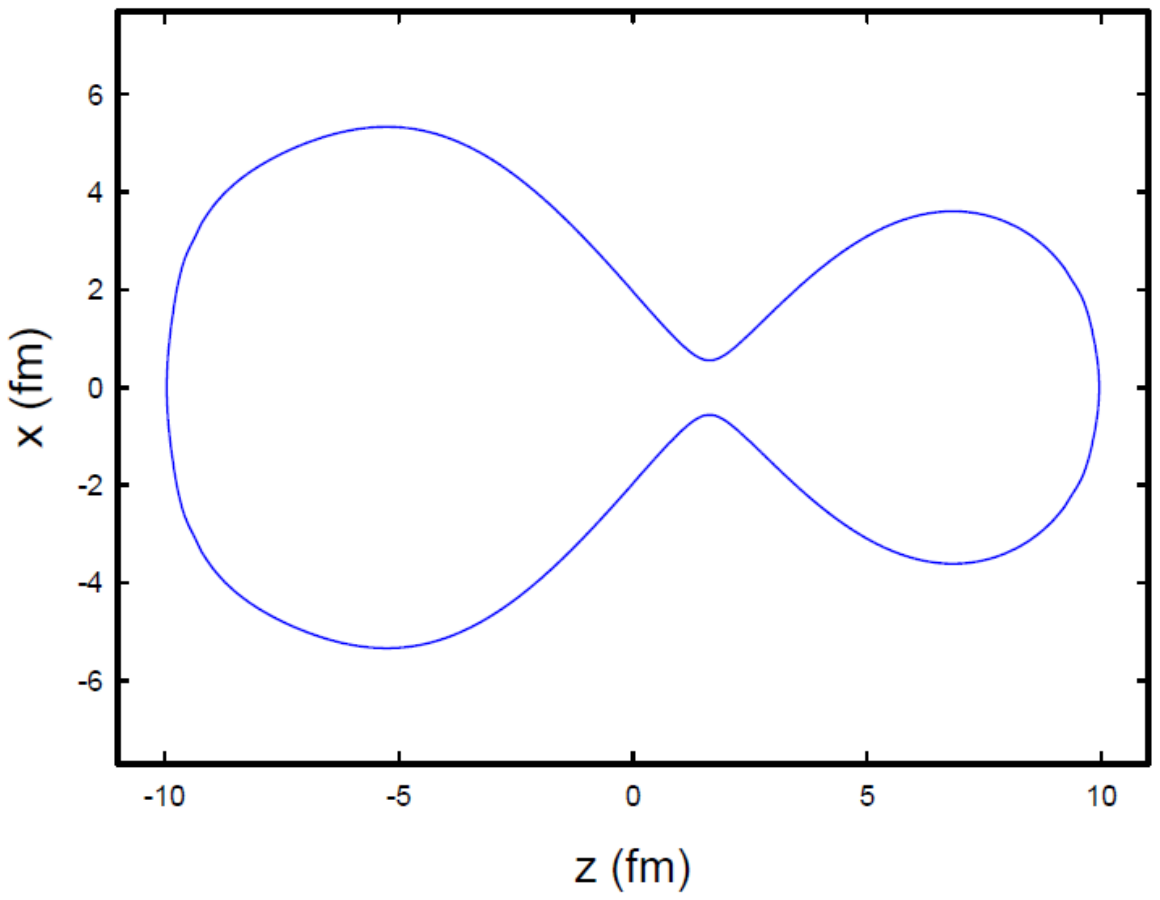

**Figure 65:** Liquid drop model surface for an asymmetric binary nuclear molecule, where the parent A = 106 nucleus is split into two daughter nuclei with Ab = 80, Ac = 26. Reproduced with permission from [122].

However, calculations done for C-C and for O-O suggest the existence of reasonably stable nuclear molecule states which involve separated clusters rather than extreme nuclear deformation. The idea is that the daughters remain localized as separate units with essentially no net relative kinetic energy. In [122], we developed a simple model based on this picture, which draws on the FRLDM approach not applied to the nucleus as a whole but rather to the two daughter clusters individually. We assume that the daughters spend a substantial fraction of their time apart. Then we can estimate the associated mass energy from the isotope mass tables. And we can use the FRLDM model to get a rough estimate for the binding energy, and also for the tunneling barrier potential. Similar models have been described in Greiner as two-core models (see Figure 66).

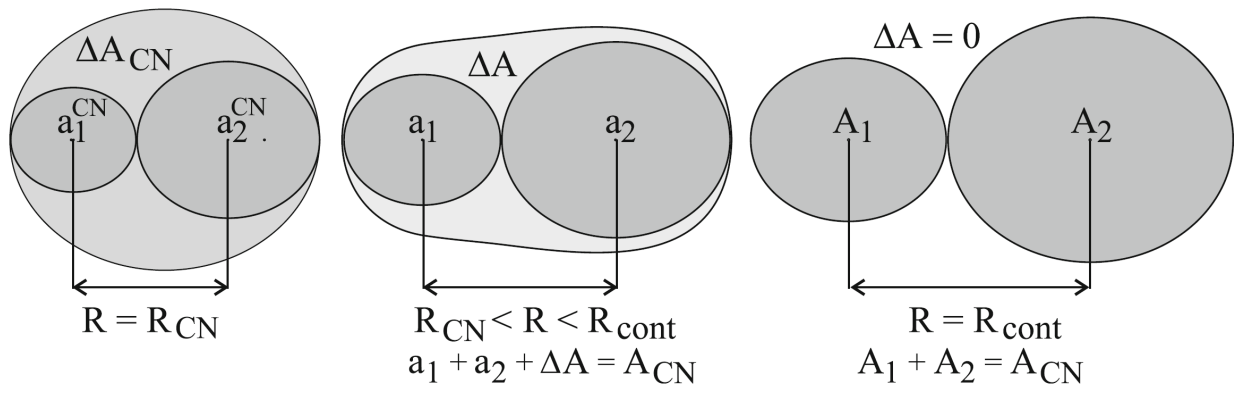

**Figure 66:** Schematic picture of a nuclear system in the two-core approximation. Reproduced with permission from [138].

The model that we have studied estimates the nuclear molecule binding energy simply based on the difference between the Coulomb and YPE energies for the nuclear molecule configuration and for infinite separation. Assuming that the surface separation is fixed at 0.5 fm, then we can write

$$-B \;=\; E_C(R_{min}) - E_C(\infty) + E_{YPE}(R_{min}) - E_{YPE}(\infty) + E_D(R_{min}) - E_D(\infty) \tag{546}$$

where $B$ is the binding energy of the nuclear molecule treated as two clusters, and where $E_C$, $E_{YPE}$ and $E_D$ are the Coulomb energy, Yukawa plus exponential energy, and density correction to the Coulomb energy. From this we can find the excitation energy

$$E_{bc} \;=\; M_B c^2 + M_c^2 - M_a c^2 - B \tag{547}$$

To estimate the tunneling rate we use

$$\gamma(\text{tunnel}) \;=\; \Gamma_0 e^{-2G} \tag{548}$$

with the Gamow factor estimated using

$$G \;=\; \int_{R_{min}}^{R_{max}} \sqrt{\frac{2\mu(V(r)-E)}{\hbar^2}}dr \tag{549}$$

and

$$\Omega_0 \;=\; \frac{1}{2}\omega_0 \tag{550}$$

where the potential is approximated by

$$V(r) \;\to\; V_0 + \frac{1}{2}\mu\omega_0^2(r-R_{min})^2 \tag{551}$$

with the SHO potential energy taken arbitrarily to be 3 MeV at 1 fm distance from the bottom of the well.

The results of this exercise are seen in Figs. 20 and 21, which contain first estimates for energy levels and lifetimes of nuclear molecule states of stable Pd isotopes.

If we restrict attention to ground states of binary nuclear molecules in a naive picture where total angular momentum is ignored, then the maximum density of nuclear molecule states is near 80/MeV at an excitation energy near 50 MeV, which if they could all be reached would require an energy exchange of 12 keV per excitation transfer. The total density of states including deformation, total angular momentum, rotations, vibrations, and excitation of individual clusters to reasonably stable excited states would be expected to increase the density of states by more than ten-fold (there remains issues of accessibility and suitability). If we include the same generalizations for ternary and quaternary nuclear molecules (based on clusters), then we would expect another order of magnitude or more increase in the density of states (again with issues of reachability and suitability based on stability).

## 6.21 $D_2$ molecules in Pd and $N_{D_2}$ per unit cell

Through much of this document, we presuppose the existence of comparatively large numbers of $D_2$ molecules in a PdD sample (and in section 6.30 a large number of HD molecules in a PdDH sample). The number of number of $D_2$ molecules per unit cell ($N_{D_2}/N$) is a direct input to the excitation transfer rate expressions (*e.g.*, Eq. 154). This section will review in more detail the relevant literature and motivate the specific values used in the rate estimates.

Since this section deals primarily with questions of nanostructure and atomic diffusion, which are governed by the electronic configuration of the involved atoms, most discussions apply to different isotopes of hydrogen equally. Some paragraphs explicitly refer to deuterium and some explicitly refer to hydrogen, depending on what related literature is drawn on.

#### Di-deuterium

In diagrams of hydrogen bonding in $H_2O$, the angle between the two hydrogen atoms is 104.5°. In this case, the two hydrogen atoms bond individually to the oxygen atom, and are well separated (1.515 Å). Oxygen does not bind to $H_2$ as a molecule, but instead the $H_2$ is split into two separate H atoms. It was widely believed to be impossible for there to be binding to an $H_2$ molecule as a molecule up until about 1980, when di-hydrogen complexes were discovered in experiments and studied theoretically (see the review Kubas 2014 [139]). Palladium is special in that a Pd atom can bind with molecular $H_2$, so that the ground state of $PdH_2$ is a di-hydrogen complex [140]. The di-hydrogen complex $PdH_2$ (and $PdD_2$) has been observed in experiments [141] [142] [143]. The H-H separation in ground state $PdH_2$ is 0.854 Å, which is somewhat larger than the 0.741 Å separation in molecular $H_2$ in vacuum (the numbers are very similar for $PdD_2$ and $D_2$).

Note that di-hydrogen formation has been predicted for Pd clusters [144], which indicates that di-hydrogen formation is not limited to individual Pd atoms.

#### D in monovacancies

While $D_2$ is stable in vacuum, the conjectured formation of $D_2$ in a metal deuteride is nontrivial. Earlier literature considered the possibility of D double occupation of octahedral sites (O-sites) in bulk Pd but concluded that the local electron density was too high, resulting in the occupation of anti-bonding sites [145] [146] [147]. In contrast, Pd monovacancies have been suggested to allow for $D_2$ molecule formation [148].

Monovacancies are traps for atomic D in $PdD_x$. Moreover, octahedral sites and tetrahedral sites (T-sites) in a Pd monovacancy are nearly degenerate [149]. There are 6 O-sites and 8 T-sites in a Pd monovacancy, which means that in principle there can be up to 14 deuterium atoms within a single monovacancy at high D/Pd loading. As the loading of a monovacancy increases, the repulsion energy between nearby O-site and T-site hydrogen atoms increases [150].

#### Surface chemisorbed $H_2$

A computational paper suggests that an $H_2$ molecule at the center of a Pd monovacancy is stable [151]. More relevant to us would be a computational study of a dideuterium complex at the surface of a monovacancy, but this particular configuration does not seem to have been studied in the literature to date.

However, the closely related problem of di-hydrogen formation on a clean Pd surface has been studied in connection with catalytic activity. Metallic palladium is a catalyst for hydrogen dissociation, and when $H_2$ approaches a clean Pd surface, it breaks apart to form adsorbed hydrogen atoms on the surface (which can subsequently diffuse into the lattice to become absorbed hydrogen). Because of this, the interaction of $H_2$ with a clean Pd surface has been studied both experimentally and theoretically [152]. Pathways have been identified in which the relative H-H separation increases as the molecular center of mass approaches the surface.

It was unexpected that a weakly bound (chemisorbed) $H_2$ feature showed up at low temperature in experiments, where adsorbed H is present on a $PdH_x$ surface [153]. This has been studied theoretically in [154]. The chemisorbed $H_2$ was observed to have a binding energy of 0.25 eV, which is lower than the 0.78 eV binding energy of $H_2$ in $PdH_2$.

Due to the similarity between the clean surface of PdHx and the interior surface of a monovacancy in PdDx, we expect chemisorbed $D_2$ to be present in monovacancies. Based on the cited calculations of di-hydrogen formation on Pd clusters [144], we conjecture that di-hydrogen (and dideuterium) can also form in Pd monovacancies.

### Superabundant vacancy formation

As mentioned above, monovacancies are traps for H in Pd (and also in other metals). The H binding energy at an O-site is increased if the Pd neighbor is removed [148], which is equivalent to a reduction of the Pd removal energy in fully loaded PdD relative to unloaded Pd. In essence, vacancy formation is made easier in a highly-loaded metal hydride or deuteride. If the loading is sufficiently high, then we expect spontaneous vacancy formation to occur as the vacancies are thermodynamically favored. At room temperature the atomic self-diffusion coefficient of Pd is very low, which means that in the absence of stress, one would not expect much vacancy formation to occur in an experiment at experimental timescales in a highly loaded sample.

At high temperature, the situation is very different. Fukai and Okuma worked with highly-loaded PdH and NiH at high temperature near 700 C (in a diamond anvil cell to provide a high-pressure environment) where atomic self-diffusion is much faster, and reported a change in the lattice constant, which they attributed to superabundant vacancy formation [155] [156]. A superabundant vacancy phase was proposed, in which 25% of the Pd atoms are absent in an ordered $Pd_3vacH_4$ lattice. An x-ray diffraction study in $Ni_3vacH_4$ was reported in support of this interpretation [157]. Statistical mechanics models have been developed for the vacancy concentration in thermal equilibrium, for example as reported in [158].

We expect Pd samples made this way with superabundant vacancies to be particularly relevant to the models presented in this document, which assume high concentrations of molecule-like deuteron pairs per unit cell.

The arguments above refer to pre-existing metal hydride samples. Another line of inquiry involves samples that are built up via the co-deposition of metal and hydrogen atoms. For instance, newly formed PdD in a co-deposition experiment has been proposed to include regions in the sample with superabundant vacancies (at least as long as the co-deposition occurred slowly, and if the D/Pd ratio was high) [159] [160]. Thermal desorption measurements have supported the conjecture that large numbers of vacancies can be present in electrodeposited Ni and Cu samples [161].

We draw attention to recent work by Staker, who has proposed additional routes to vacancy formation in PdD that do not involve co-deposition or heating [162].

### Assumptions concerning vacancies in the models of this document

The discussion above is relevant to the quantitative estimates used in the models in this paper. If one were to somehow work with an ideal single crystal PdD sample with no vacancies, then we would expect no dideuterium or chemisorbed deuterium to be present, in which case $N_{\mathrm{D}_2} \to 0$.

As a result, no obserable fusion rates would be expected in such a sample, independent of how the sample is stimulated. For the models developed in this document to apply, we require molecular $D_2$ to be present in the lattice (in the PdD lattice, in the given case), which means that loaded vacancies are required. For the quantitative estimates that we report in this work, we use

$$\frac{N_{\mathrm{D}_2}}{N} \ \to \ 0.25 \tag{552}$$

(including all nuclear spin states) which is a rough estimate that corresponds to the proposed superabundant vacancy phase $Pd_3vacD_4$.

Note that there is no difficulty loading up molecular $D_2$ in microvoids in $PdD_x$, which could be argued would result in a large $N_{D_2}$ value. However, for such deuteron pairs we would expect little screening from the comparably distant Pd electrons. We would also expect the coupling with optical phonon and plasmon modes to be minimal. Chemisorbed $D_2$ appears more relevant due to the greater chance for screening, and for optical phonon exchange. Most interesting are dideuterium complexes, which expect to experience comparably strong screening and coupling to lattice modes.

#### Relevant $D_2$ molecule spin states

Among all available $D_2$ molecules, there are only three $J = 1$ $^3$P states which have the potential to be coupled to via the relativistic coupling, which we identified to be the strongest (by far) and therefore the most relevant of the available couplings (5.3).

If we consider a highly-excited acoustic phonon mode with linear polarization, then only one of the nine $J = 1$ states will participate in the excitation transfer process.

This reduces the ratio by another factor of $\frac{1}{9}$, resulting in an overall

$$\frac{N_{D_2}}{N} = 0.25 \times \frac{1}{9} \tag{553}$$

This is the value used in many of the transfer rate estimates in section 5.

## 6.22 Decoherence time and Dicke state fragility

The described excitation transfer dynamics require the development of cooperative (Dicke) enhancement factors to increase the excitation transfer rate from the minimum rate $\Gamma_{min}$ toward the maximum rate $\Gamma_{max}$. In order for the system to develop a cooperative (Dicke) enhancement there needs to be a commonality of phase between the different constituent configurations.

Specifically, we have considered Dicke states on the donor side consisting of deuteron pairs and Dicke states on the receiver side consisting of Pd nuclei.

If all of the dideuterium molecules are in the upper state (*i.e.*, no transition to $^4$He has yet occured), then we can think of having only a single configuration in connection with the $\mathbf{a} \cdot c\mathbf{P}$ transitions. When one excitation transfer occurs (with uniform probability over all participating dideuterium molecules within the coherent domain of the highly-excited phonon mode), then a very large number of configurations are generated, all initially with a common phase (a Dicke state).

However, Dicke states are fragile, since the accumulation of phase differences among the different configurations will destroy the associated cooperative enhancement factors. A critical parameter concerns how long it takes for this to happen.

For cooperative (Dicke) enhancement factors to build up, we need the first excitation transfer from the $D_2$/$^4$He fusion transition to occur in a time on the order of, or shorter than, this dephasing time.

### Estimation of the $D_2$ dissociation time

In order to develop an estimate for the dephasing time, we consider how long it might take for a $D_2$ molecule to dissociate in a monovacancy as the primary location of dideuterium molecules.

We first consider the problem of H diffusion in bulk palladium. The diffusion coefficient $D$ can be modeled as arising from jumps from one O-site to another according to [163]

$$D \ = \ \frac{a^2}{12\tau} \tag{554}$$

where $a$ is the lattice parameter and $\tau$ is the time between jumps. At 20 C this time constant in beta phase PdH$_x$ is reported to be [164] (where H/Pd is 0.65):

$$\tau \ = \ 44 \text{ ps} \tag{555}$$

The conclusion so far is that if $D_2$ molecular dissociation works like diffusion, then the molecule cannot be expected to survive very long near room temperature. Note that at low temperature, a longer dissociation time can be expected.

If neighboring O-sites are occupied, then jumps to those sites are blocked, and the diffusion rate is reduced by a blocking factor $(1-\theta)$ where $\theta$ is the D/Pd ratio [165].

We expect a deuterium atom in an O-site of a monovacancy that is partially loaded to make jumps to neighboring unoccupied sites on a commensurate timescale. When a dideuterium molecule forms, we expect it to dissociate in tens of picoseconds.

Since a monovacancy in Pd acts a trap for H and D, we would expect the monovacancy to be highly loaded, even when the bulk has only modest loading. Besenbacher et al. (1990) [149] estimated the trapping energy for D in a monovacancy to be near 0.10 eV, which is comparable to the defect trapping energy of 0.13 eV calculated in [166].

Consequently, at high D/Pd loading, we expect neighboring sites in a monovacancy to be nearly completely blocked. For the hopping time, we obtain a broad estimate from

$$\tau = \left(\frac{1-0.65}{1-\theta_t}\right) 44 \text{ ps} \tag{556}$$

where $\theta_t$ is the ratio of D occupation to available O-sites and T-sites. If all of the neighboring sites are blocked, then the hopping time diverges in this model. Based on these considerations and assuming a $\theta_t = 0.985$, we estimate the dissociation time from dissociation as

$$\tau \sim 1 \text{ ns} \tag{557}$$

This number is consistent with a picture, in which the Pd monovacancies are fully packed with D, with essentially no place for them to hop.

Note that $D_2$ molecule occupation of the center of the monovacancy would be expected at high D/Pd loading if all of the sites near the surface are occupied. Such states would have much greater stability, but also reduced screening and weaker coupling to lattice vibrations.

When cooperative (Dicke) enhancement factors build up and a significant admixture of $^4$He occupation develops, the lower average energy of the mixed state may help stabilize the dideuterium component.

**Angular momentum changes**

Angular momentum transitions are likely to dominate dephasing (likely to be faster than 1 ns but on a similar order). At present, we do not have an estimate for this number based on experimental data. A model will have to be developed for providing a precise estimate.

**Fast phonon exchange and Dicke states**

If there is substantial energy in the uniform acoustic phonon mode in the model, then we would expect fast acoustic phonon exchange, as has been discussed in 5.11. This is significant in the context of Dicke state fragility in that we would expect such transitions to help "refresh" the phase coherence between the different configurations that make up the Dicke states, especially in the case of the Pd*/Pd transitions. For this to be a factor in the case of $D_2$/$^4$He Dicke states, the rate of acoustic phonon exchange associated with the transition needs to exceed the corresponding dephasing rate.

## 6.23 Excitation transfer to multiple transitions

We will show in this section that excitation transfer can take place, in principle unmitigated to a combination of multiple receiver states as long as the sum of those states meets the resonance condition.

We recall that destructive interference hinders excitatidon transfer, and we require there to be differences in the pathways associated with loss in order for a finite contribution to arise at low order in perturbation theory. At present we have an evaluation of the **a**-matrix element both with loss and without loss (sections 6.5 and 6.6), which means that we can make use of these estimates to evaluate indirect coupling matrix elements at lowest order in perturbation theory.

Even though we are interested in excitation transfer from one transition to multiple transitions, since we are not used to non-zero contributions showing up at lowest order, it makes sense to see how this works for the simplest case of excitation transfer from one transition to another first.

### Transfer to one transition

A schematic of the excitation transfer scheme is shown in Figure 67, where we see the usual diamond type of diagram that we associated with second-order excitation transfer. What is interesting about this scheme is that the $D_2$/$^4$He matrix element has one value $U'$ when loss is present (from state 1), and a different value $U$ when there is no loss (from state 3).

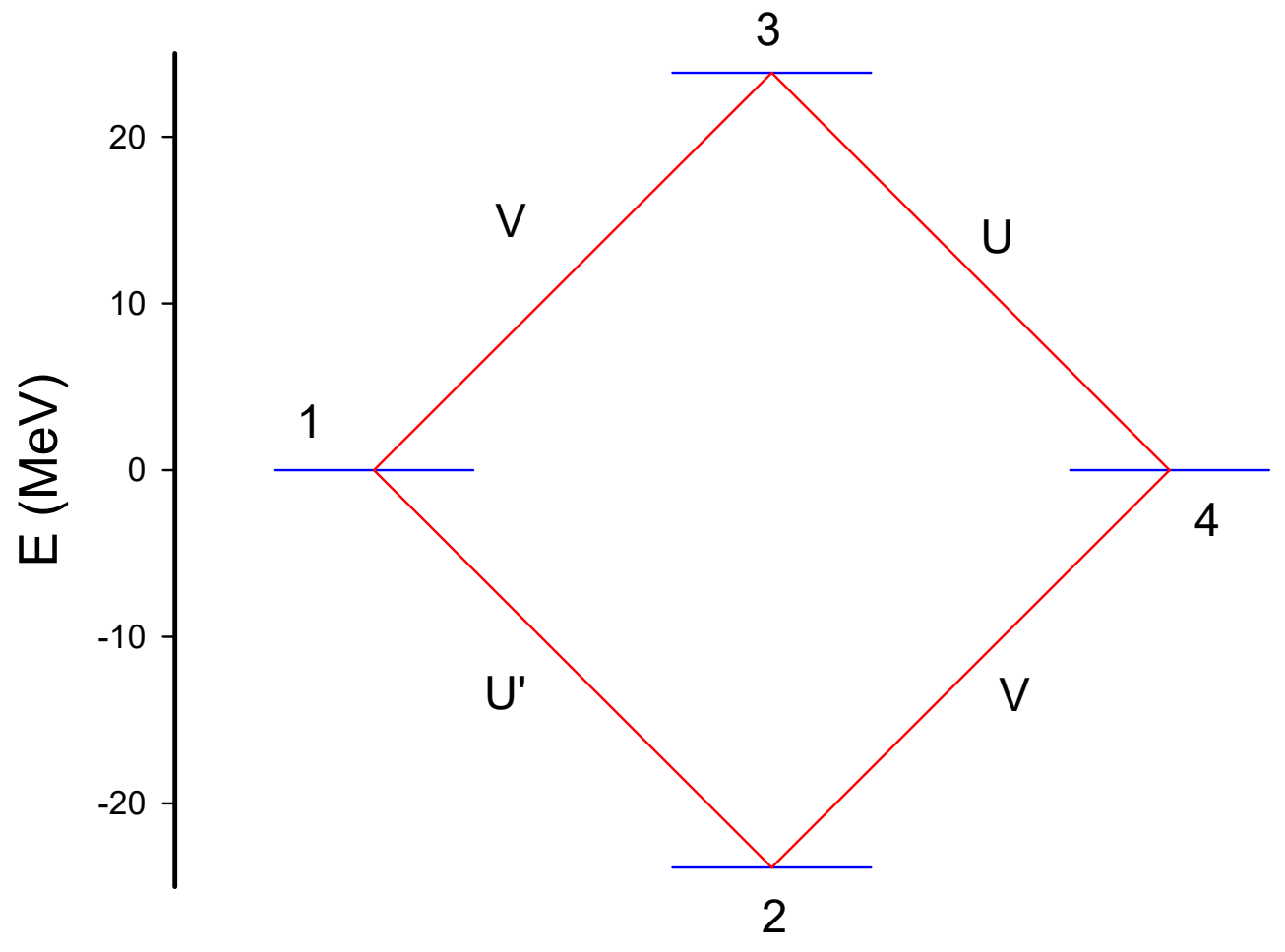

**Figure 67:** Schematic of excitation transfer from the $D_2$/$^4$He transition to one Pd*/Pd transition.

For the different states in this model we define

$$\Phi_1 = \Big|D_2, \text{ Pd}\Big\rangle \qquad \Phi_2 = \Big|{}^4\text{He}, \text{ Pd}\Big\rangle \qquad \Phi_3 = \Big|D_2, \text{ Pd}^*\Big\rangle \qquad \Phi_4 = \Big|{}^4\text{He}, \text{ Pd}^*\Big\rangle \tag{558}$$

We recall that the $\mathbf{a} \cdot c\mathbf{P}$ interaction preserves isospin, so that the $D_2$ and $^4$He states are $T = 0$ states. The $D_2$ state of interest is the $^3$P $J^\pi = 1^-$ state, and the ground state is $^1$S $J^\pi = 0^+$. If the phonon (or plasmon) mode has linear polarization in $z$, then the molecular $D_2$ state would have $M_J = 0$.

The lowest-order contribution to the indirect matrix element $H_{14}$ for this scheme is

$$H_{14} \;\rightarrow\; \frac{H_{12}H_{24}}{E-E_2} + \frac{H_{13}H_{34}}{E-E_3} \tag{559}$$

For the state energies we take

$$E_1 \;=\; E_{\mathrm{D}_2} + E_{\mathrm{Pd}} \qquad E_2 \;=\; E_{^4\mathrm{He}} + E_{\mathrm{Pd}} \qquad E_3 \;=\; E_{\mathrm{D}_2} + E_{\mathrm{Pd}*} \qquad E_4 \;=\; E_{^4\mathrm{He}} + E_{\mathrm{Pd}*} \tag{560}$$

We assume that

$$E_{\mathrm{Pd}*} - E_{\mathrm{Pd}} \;=\; E_{\mathrm{D}_2} - E_{^4\mathrm{He}} \;=\; \Delta Mc^2 \tag{561}$$

and take

$$E \;=\; E_1 \tag{562}$$

For the matrix elements we parameterize according to

$$H_{12} \;=\; U' \qquad H_{13} \;=\; V \qquad H_{24} \;=\; V \qquad H_{34} \;=\; U \tag{563}$$

Because fusion loss occurs in state 1 and fusion loss does not occur in state 3, the $\mathbf{D_2}$ wave function is different, and $U' \neq U$. It is useful to define

$$\Delta U \;=\; U' - U \tag{564}$$

The indirect coupling matrix element in terms of these parameters is

$$H_{14} \;\rightarrow\; \Delta U \frac{V}{\Delta Mc^2} \tag{565}$$

Note that we are getting a non-zero result for excitation transfer at second order in this model, since loss impact the wave functions, which impacts the matrix elements, and removes some of the destructive interference.

If we define

$$g \;=\; \frac{|V|}{\Delta Mc^2} \tag{566}$$

then we can write

$$H_{14} \;=\; \Delta U g \tag{567}$$

and

$$\Gamma_{transfer} \;=\; \frac{2}{\hbar}|H_{14}| \;=\; \frac{2}{\hbar}|\Delta U| g \tag{568}$$

We expect perturbation theory to give good results for

$$g \;\ll\; 1 \tag{569}$$

For $g \geq 1$ we will need to implement a non-perturbative analysis to get reliable results.

**Transfer to two transitions**

Now we would like to extend this calculation to model the lowest-order contribution to indirect coupling in the case of excitation transfer to two (different) transitions. The levels and coupling are illustrated in Figure 68. In the absence of loss we would expect no indirect coupling at fourth-order in perturbation theory; however, if the $\mathbf{D_2}$ wave function in the initial state (state 1) is modified by fusion decay, then the associated **a**-matrix element will be different than if there were no loss. For intermediate states at higher energy (states 3,4, and 7) fusion decay is not allowed since the basis state energies are well above the energy eigenvalue $E$. This is captured in the figure through the different coupling matrix element between states 1 and 2 ($U'$), where the other downward transitions are associated with $U$. It is this difference which gives rise to indirect coupling at lowest order.

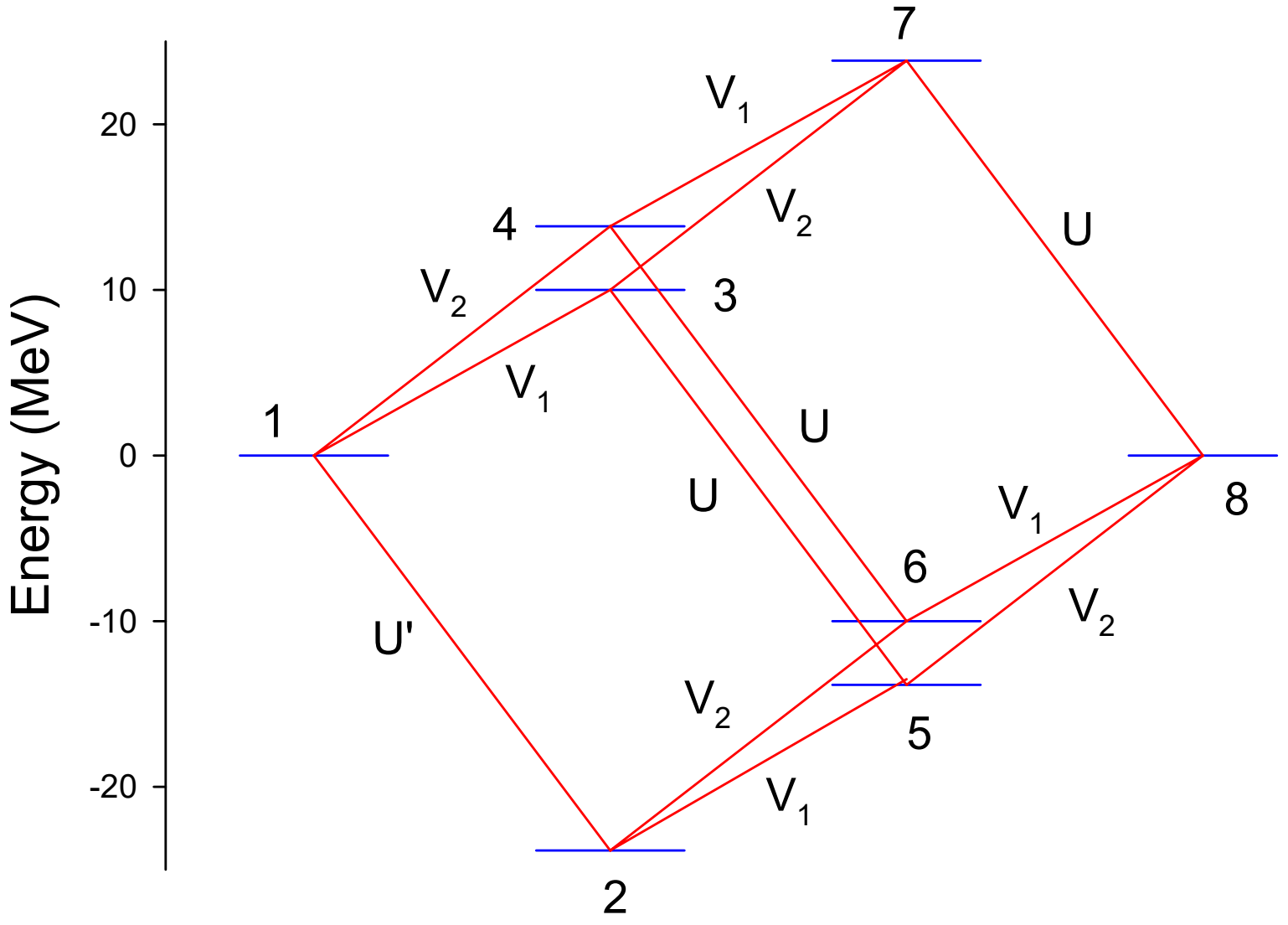

**Figure 68:** Schematic of excitation transfer from the $D_2$/$^4$He transition to two Pd*/Pd transitions.

The state definitions in this case are

$$
\begin{aligned}
\Phi_1 &= \left| \mathrm{D}_2,\ \mathrm{Pd}_a,\ \mathrm{Pd}_b \right\rangle & \Phi_2 &= \left| {}^4\mathrm{He},\ \mathrm{Pd}_a,\ \mathrm{Pd}_b \right\rangle & \Phi_3 &= \left| \mathrm{D}_2,\ \mathrm{Pd}_a^*,\ \mathrm{Pd}_b \right\rangle \\
\Phi_4 &= \left| \mathrm{D}_2,\ \mathrm{Pd}_a,\ \mathrm{Pd}_b^* \right\rangle & \Phi_5 &= \left| {}^4\mathrm{He},\ \mathrm{Pd}_a^*,\ \mathrm{Pd}_b \right\rangle & \Phi_6 &= \left| {}^4\mathrm{He},\ \mathrm{Pd}_a,\ \mathrm{Pd}_b^* \right\rangle \\
& & \Phi_7 &= \left| \mathrm{D}_2,\ \mathrm{Pd}_a^*,\ \mathrm{Pd}_b^* \right\rangle & \Phi_8 &= \left| {}^4\mathrm{He},\ \mathrm{Pd}_a^*,\ \mathrm{Pd}_b^* \right\rangle
\end{aligned}
\tag{570}
$$

The lowest-order contribution to the indirect matrix element is

$$
\begin{aligned}
H_{18} \to\ & \frac{H_{12}H_{25}H_{58}}{(E-E_2)(E-E_5)} + \frac{H_{12}H_{26}H_{68}}{(E-E_2)(E-E_6)} + \frac{H_{13}H_{35}H_{58}}{(E-E_3)(E-E_5)} \\
& + \frac{H_{13}H_{37}H_{78}}{(E-E_3)(E-E_7)} + \frac{H_{14}H_{46}H_{68}}{(E-E_4)(E-E_6)} + \frac{H_{14}H_{47}H_{78}}{(E-E_4)(E-E_7)}
\end{aligned}
\tag{571}
$$

For the energies we take

$$\begin{aligned} E_1 &= E_{\mathrm{D}_2} + E_{Pda} + E_{Pdb} \qquad E_2 = E_{^4\mathrm{He}} + E_{Pda} + E_{Pdb} \qquad E_3 = E_{\mathrm{D}_2} + E_{Pda*} + E_{Pdb} \\ E_4 &= E_{\mathrm{D}_2} + E_{Pda} + E_{Pdb*} \qquad E_5 = E_{^4\mathrm{He}} + E_{Pda*} + E_{Pdb} \qquad E_6 = E_{^4\mathrm{He}} + E_{Pda} + E_{Pdb*} \\ E_7 &= E_{\mathrm{D}_2} + E_{Pda*} + E_{Pdb*} \qquad E_8 = E_{^4\mathrm{He}} + E_{Pda*} + E_{Pdb*} \end{aligned} \tag{572}$$

with

$$E_{\mathrm{D}_2} - E_{^4\mathrm{He}} = \Delta Mc^2 = E_{Pda*} + E_{Pdb*} - E_{Pda} - E_{Pdb} \tag{573}$$

and

$$E = E_1 \tag{574}$$

For the matrix elements we parameterize according to

$$\begin{aligned} H_{12} &= U' \qquad H_{13} = V_1 \qquad H_{14} = V_2 \\ H_{25} &= V_1 \qquad H_{26} = V_2 \\ H_{35} &= U \qquad H_{37} = V_2 \\ H_{46} &= U \qquad H_{47} = V_1 \\ H_{58} &= V_2 \qquad H_{68} = V_1 \qquad H_{78} = U \end{aligned} \tag{575}$$

We can evaluate and obtain

$$H_{18} = \Delta U \frac{V_1}{\epsilon_1} \frac{V_2}{\epsilon_2} \tag{576}$$

where

$$\Delta U = U' - U \qquad \epsilon_1 = E_{Pda*} - E_{Pda} \qquad \epsilon_2 = E_{Pdb*} - E_{Pdb} \tag{577}$$

As was the case above for excitation transfer to a single transition, the indirect coupling is proportional to the difference in $D_2$/$^4$He matrix element, which is a result of the fusion loss being different on and off of resonance.

It is convenient to define dimensionless coupling constants according to

$$g_1 = \frac{|V_1|}{\epsilon_1} \qquad g_2 = \frac{|V_2|}{\epsilon_2} \tag{578}$$

In terms of these dimensionless coupling constants we have

$$H_{18} = \Delta U g_1 g_2 \tag{579}$$

and

$$\Gamma_{transfer} = \frac{2}{\hbar}|H_{18}| = \frac{2}{\hbar}|\Delta U|g_1 g_2 \tag{580}$$

Once again we expect that when the dimensionless coupling constants are much less than unity that perturbation theory will give us a good answer, but that a much bigger calculation will be required to get accurate answers in the strong coupling limit.

**Transfer to three transitions**

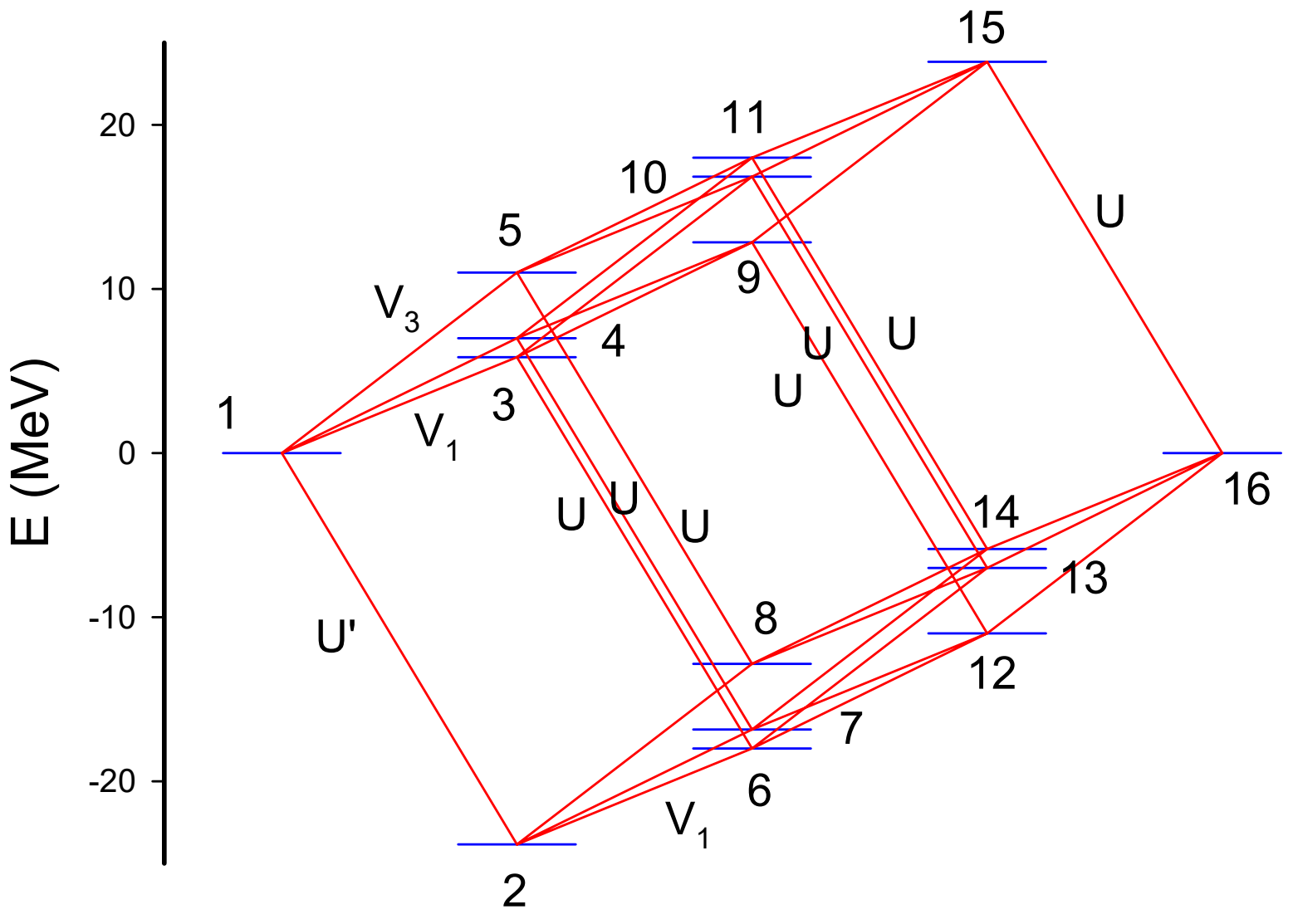

**Figure 69:** Schematic of excitation transfer from the $D_2/^4He$ transition to three Pd*/Pd transitions.

The generalization of the model to excitation transfer from one transition to three transitions is straightforward conceptually, but begins to take some work since more states and transitions are involved. A schematic of the levels and transitions that go into the lowest-order contribution to the indirect coupling coefficient in perturbation theory are illustrated in Figure 69. Due to the complexity of the diagram, only a few of the transitions are labeled. As was the case in the examples above, it is the difference in the coupling matrix elements $U$ and $U'$ for the downward transitions that gives rise to the removal of part of the destructive interference.

For the state definitions we can write

$$
\begin{aligned}
\Phi_1 &= \left|\mathrm{D}_2,\ \mathrm{Pd}_a,\ \mathrm{Pd}_b,\ \mathrm{Pd}_c\right\rangle & \Phi_2 &= \left|{}^4\mathrm{He},\ \mathrm{Pd}_a,\ \mathrm{Pd}_b,\ \mathrm{Pd}_c\right\rangle & \Phi_3 &= \left|\mathrm{D}_2,\ \mathrm{Pd}_a^*,\ \mathrm{Pd}_b,\ \mathrm{Pd}_c\right\rangle \\
\Phi_4 &= \left|\mathrm{D}_2,\ \mathrm{Pd}_a,\ \mathrm{Pd}_b^*,\ \mathrm{Pd}_c\right\rangle & \Phi_5 &= \left|\mathrm{D}_2,\ \mathrm{Pd}_a,\ \mathrm{Pd}_b,\ \mathrm{Pd}_c^*\right\rangle & \Phi_6 &= \left|{}^4\mathrm{He},\ \mathrm{Pd}_a^*,\ \mathrm{Pd}_b,\ \mathrm{Pd}_c\right\rangle \\
\Phi_7 &= \left|{}^4\mathrm{He},\ \mathrm{Pd}_a,\ \mathrm{Pd}_b^*,\ \mathrm{Pd}_c\right\rangle & \Phi_8 &= \left|{}^4\mathrm{He},\ \mathrm{Pd}_a,\ \mathrm{Pd}_b,\ \mathrm{Pd}_c^*\right\rangle & \Phi_9 &= \left|\mathrm{D}_2,\ \mathrm{Pd}_a^*,\ \mathrm{Pd}_b^*,\ \mathrm{Pd}_c\right\rangle \\
\Phi_{10} &= \left|\mathrm{D}_2,\ \mathrm{Pd}_a^*,\ \mathrm{Pd}_b,\ \mathrm{Pd}_c^*\right\rangle & \Phi_{11} &= \left|\mathrm{D}_2,\ \mathrm{Pd}_a,\ \mathrm{Pd}_b^*,\ \mathrm{Pd}_c^*\right\rangle & \Phi_{12} &= \left|{}^4\mathrm{He},\ \mathrm{Pd}_a^*,\ \mathrm{Pd}_b^*,\ \mathrm{Pd}_c\right\rangle \\
\Phi_{13} &= \left|{}^4\mathrm{He},\ \mathrm{Pd}_a^*,\ \mathrm{Pd}_b,\ \mathrm{Pd}_c^*\right\rangle & \Phi_{14} &= \left|{}^4\mathrm{He},\ \mathrm{Pd}_a,\ \mathrm{Pd}_b^*,\ \mathrm{Pd}_c^*\right\rangle & \Phi_{15} &= \left|\mathrm{D}_2,\ \mathrm{Pd}_a^*,\ \mathrm{Pd}_b^*,\ \mathrm{Pd}_c^*\right\rangle \\
& & & & \Phi_{16} &= \left|{}^4\mathrm{He},\ \mathrm{Pd}_a^*,\ \mathrm{Pd}_b^*,\ \mathrm{Pd}_c^*\right\rangle
\end{aligned}
\tag{581}
$$

For this scheme we can write for the indirect matrix element

$$
\begin{aligned}
\hat{H}_{1,16} \rightarrow\ & \frac{H_{1,2}H_{2,6}H_{6,12}H_{12,16}}{(E-E_2)(E-E_6)(E-E_{12})} + \frac{H_{1,2}H_{2,6}H_{6,13}H_{13,16}}{(E-E_2)(E-E_6)(E-E_{13})} + \frac{H_{1,2}H_{2,7}H_{7,12}H_{12,16}}{(E-E_2)(E-E_7)(E-E_{12})} \\
& + \frac{H_{1,2}H_{2,7}H_{7,14}H_{14,16}}{(E-E_2)(E-E_7)(E-E_{14})} + \frac{H_{1,2}H_{2,8}H_{8,13}H_{13,16}}{(E-E_2)(E-E_8)(E-E_{13})} + \frac{H_{1,2}H_{2,8}H_{8,14}H_{14,16}}{(E-E_2)(E-E_8)(E-E_{14})} \\
& + \frac{H_{1,3}H_{3,6}H_{6,12}H_{12,16}}{(E-E_3)(E-E_6)(E-E_{12})} + \frac{H_{1,3}H_{3,6}H_{6,13}H_{13,16}}{(E-E_3)(E-E_6)(E-E_{13})} + \frac{H_{1,3}H_{3,9}H_{9,12}H_{12,16}}{(E-E_3)(E-E_9)(E-E_{12})} \\
& + \frac{H_{1,3}H_{3,9}H_{9,15}H_{15,16}}{(E-E_3)(E-E_9)(E-E_{15})} + \frac{H_{1,3}H_{3,10}H_{10,13}H_{13,16}}{(E-E_3)(E-E_{10})(E-E_{13})} + \frac{H_{1,3}H_{3,10}H_{10,15}H_{15,16}}{(E-E_3)(E-E_{10})(E-E_{15})} \\
& + \frac{H_{1,4}H_{4,7}H_{7,12}H_{12,16}}{(E-E_4)(E-E_7)(E-E_{12})} + \frac{H_{1,4}H_{4,7}H_{7,14}H_{14,16}}{(E-E_4)(E-E_7)(E-E_{14})} + \frac{H_{1,4}H_{4,9}H_{9,12}H_{12,16}}{(E-E_4)(E-E_9)(E-E_{12})} \\
& + \frac{H_{1,4}H_{4,9}H_{9,15}H_{15,16}}{(E-E_4)(E-E_9)(E-E_{15})} + \frac{H_{1,4}H_{4,11}H_{11,14}H_{14,16}}{(E-E_4)(E-E_{11})(E-E_{14})} + \frac{H_{1,4}H_{4,11}H_{11,15}H_{15,16}}{(E-E_4)(E-E_{11})(E-E_{15})} \\
& + \frac{H_{1,5}H_{5,8}H_{8,13}H_{13,16}}{(E-E_5)(E-E_8)(E-E_{13})} + \frac{H_{1,5}H_{5,8}H_{8,14}H_{14,16}}{(E-E_5)(E-E_8)(E-E_{14})} + \frac{H_{1,5}H_{5,10}H_{10,13}H_{13,16}}{(E-E_5)(E-E_{10})(E-E_{13})} \\
& + \frac{H_{1,5}H_{5,10}H_{10,15}H_{15,16}}{(E-E_5)(E-E_{10})(E-E_{15})} + \frac{H_{1,5}H_{5,11}H_{11,14}H_{14,16}}{(E-E_5)(E-E_{11})(E-E_{14})} + \frac{H_{1,5}H_{5,11}H_{11,15}H_{15,16}}{(E-E_5)(E-E_{11})(E-E_{15})}
\end{aligned}
\tag{582}
$$

We used energy definitions similar to what is described above

$$
\begin{aligned}
E_1 &= E_{\mathrm{D}_2} + E_{Pda} + E_{Pdb} + E_{Pdc} \\
&\vdots \\
E_{16} &= E_{^4\mathrm{He}} + E_{Pda*} + E_{Pdb*} + E_{Pdc*}
\end{aligned}
\tag{583}
$$

We took

$$
E = E_1 \tag{584}
$$

For the fusion transition matrix elements we parameterized according to

$$\begin{aligned} H_{1,2} &= U' \quad H_{3,6} = U \quad H_{4,7} = U \quad H_{5,8} = U \\ H_{9,12} &= U \quad H_{10,13} = U \quad H_{11,14} = U \quad H_{15,16} = U \end{aligned} \tag{585}$$

For the Pda*/Pda transition matrix elements we have

$$\begin{aligned} H_{1,3} &= V_1 \quad H_{2,6} = V_1 \quad H_{4,9} = V_1 \quad H_{5,10} = V_1 \\ H_{7,12} &= V_1 \quad H_{8,13} = V_1 \quad H_{11,15} = V_1 \quad H_{14,16} = V_1 \end{aligned} \tag{586}$$

For the Pdb*/Pdb transition matrix elements we have

$$\begin{aligned} H_{1,4} &= V_2 \quad H_{2,7} = V_2 \quad H_{3,9} = V_2 \quad H_{5,11} = V_2 \\ H_{6,12} &= V_2 \quad H_{8,14} = V_2 \quad H_{10,15} = V_2 \quad H_{13,16} = V_2 \end{aligned} \tag{587}$$

For the Pdc*/Pdc transition matrix elements we have

$$\begin{aligned} H_{1,5} &= V_3 \quad H_{2,8} = V_3 \quad H_{3,10} = V_3 \quad H_{4,11} = V_3 \\ H_{6,13} &= V_3 \quad H_{7,14} = V_3 \quad H_{9,15} = V_3 \quad H_{12,16} = V_3 \end{aligned} \tag{588}$$

We evaluated the indirect matrix element using Mathematica, with the result

$$H_{1,16} \rightarrow \Delta U \frac{V_1}{\epsilon_1} \frac{V_2}{\epsilon_2} \frac{V_3}{\epsilon_3} \tag{589}$$

where

$$\epsilon_1 = E_{Pda*} - E_{Pda} \qquad \epsilon_2 = E_{Pdb*} - E_{Pdb} \qquad \epsilon_3 = E_{Pdc*} - E_{Pdc} \tag{590}$$

and

$$E_{\mathrm{D}_2} - E_{^4\mathrm{He}} = \Delta M c^2 = E_{Pda*} + E_{Pdb*} + E_{Pdc*} - E_{Pda} - E_{Pdb} - E_{Pdc} \tag{591}$$

If we define dimensionless coupling coefficients according to

$$g_1 = \frac{|V_1|}{\epsilon_1} \qquad g_2 = \frac{|V_2|}{\epsilon_2} \qquad g_3 = \frac{|V_3|}{\epsilon_3} \tag{592}$$

then we can write

$$H_{1,16} = \Delta U g_1 g_2 g_3 \tag{593}$$

$$\Gamma_{transfer} = \frac{2}{\hbar}|H_{1,16}| = \frac{2}{\hbar}|\Delta U| g_1 g_2 g_3 \tag{594}$$

Once again we expect these expressions to be accurate when all of the dimensionless coupling coefficients are much less than unity. A much larger calculation will be required to get accurate indirect coupling coefficients are rates in the strong coupling regime.

## 6.24 Model development for excitation transfer from $D_2$ to multiple receiver states with oscillator energy exchange

In section 5.11, the goal is to estimate the transfer rate for excitation transfer from the fusion transition to multiple receiver states, where the receiver states comprise all possible (energetically allowed) combinations of Pd*/Pd transitions under conditions where energy is exchanged from phonons and plasmons. Building a model to describe excitation transfer in such a regime is complicated not least because, as we have seen in section 5.11, we are in the strong coupling limit.

In this section we will draw upon non-perturbative techniques in particular the Lippmann-Schwinger formalism. After some adjustment to the problem at hand, we will arrive at an expression for the transfer rate that is comparably simple and takes the form of the convolution expression seen in Eq. 236 in section 5.11.

**Starting point for a model for nuclear transitions and oscillators**

Similar to what was done in section 5.1, we start out with a comprehensive Hamiltonian analogous to Eq. 17 (in this case not including magnons as we also did later in section 5.1 for simplicity):

$$\hat{H} \;=\; \hat{H}_{nuclei} + \hat{H}_{phonons} + \hat{H}_{plasmons} + \hat{V}_{nuclei,phonons} + \hat{V}_{nuclei,plasmons} \tag{595}$$

We include selective loss in the Hamiltonian for the nuclei, per section 5.2 and for the interaction part, we focus on the relativistic interaction as it has been shown to be the strongest among the relevant interactions (5.3).

This results in the Hamiltonian

$$\begin{aligned}\hat{H} \;=\; & \sum_j \sum_k \left\{ |\phi_j\rangle \left( M_j c^2 - i\frac{\hbar}{2}\gamma_j(E) \right) \langle\phi_j| \right\}_k + \hbar\omega_A \hat{a}_A^\dagger \hat{a}_A + \hbar\omega_O \hat{a}_O^\dagger \hat{a}_O + \hbar\omega_P \hat{a}_P^\dagger \hat{a}_P \\ & + \sum_{j,j'} \sum_k \left\{ |\phi_{j'}\rangle\langle\phi_{j'}| \mathbf{a}\cdot c\hat{\mathbf{P}}_j |\phi_j\rangle\langle\phi_j| \right\}_k\end{aligned} \tag{596}$$

where the center of mass momentum operators for a uniform acoustic phonon mode, a uniform optical phonon mode, and a uniform plasmon mode are given by

$$\hat{\mathbf{P}}_j \;\rightarrow\; \mathbf{e}_j^{(A)} \sqrt{\frac{\hbar M_j \omega_A}{2N}} \left( \frac{\hat{a}_A - \hat{a}_A^\dagger}{i} \right) + \mathbf{e}_j^{(O)} \sqrt{\frac{\hbar M_j \omega_O}{2N}} \left( \frac{\hat{a}_O - \hat{a}_O^\dagger}{i} \right) + \mathbf{e}_j^{(P)} \sqrt{\frac{Z_j^2 \hbar M_j \omega_P}{2N_e}} \left( \frac{\hat{a}_P - \hat{a}_P^\dagger}{i} \right) \tag{597}$$

**Pseudo-spin model for nuclear transitions and oscillators**

Even though the model in Eq. 596 has already been drastically reduced from where we started in section 5.1 (where initially all nuclear transitions, all possible low-order interactions with the lattice, and coupling with all phonon modes, and potentially all plasmon modes, was considered), it is still quite complicated due to the many-level formalism associated with the nuclear states. The technical issue here is that the ground state(s) of a stable Pd nucleus (any isotope) will have $\mathbf{a} \cdot c\mathbf{P}$ transitions to many different excited states, and in the strong coupling regime this will impact the model.

It is possible to develop an approximate version of the model, where we work instead with many two-level systems. With such an approximation we are able to develop first quantitative estimates for excitation transfer rates (which can later be updated based on future versions of such models).

We start by writing the model that results as

$$\hat{H} = \hat{H}_{fus} + \hat{H}_{Pd} + \hat{H}_{osc} + \hat{V} + \hat{U} \tag{598}$$

with

$$\begin{aligned}
\hat{H}_{fus} &= \Delta M c^2 \frac{\hat{S}_z^{(fus)}}{\hbar} \\
\hat{H}_{Pd} &= \sum_j \epsilon_j \frac{\hat{S}_z^{(j)}}{\hbar} \\
\hat{H}_{osc} &= \hbar\omega_A \hat{a}_A^\dagger \hat{a}_A + \hbar\omega_O \hat{a}_O^\dagger \hat{a}_O + \hbar\omega_P \hat{a}_P^\dagger \hat{a}_P \\
\hat{V} &= \langle \mathrm{D}_2 | \mathbf{a} \cdot c\hat{\mathbf{P}}_{^4He} | ^4\mathrm{He} \rangle \left( \frac{\hat{S}_+^{(fus)}}{\hbar} + \frac{\hat{S}_-^{(fus)}}{\hbar} \right) \\
\hat{U} &= \sum_j \langle \mathrm{Pd}_j^* | \mathbf{a} \cdot c\hat{\mathbf{P}}_{Pd_j} | \mathrm{Pd}_j \rangle \left( \frac{\hat{S}_+^{(j)}}{\hbar} + \frac{\hat{S}_-^{(j)}}{\hbar} \right)
\end{aligned} \tag{599}$$

Note that in the model above (Eq. 596), we included nuclear decay channels (such as radiative decay, alpha decay, and so forth), while in the pseudo-spin version of the model here, these terms are not included. The issue here is that we anticipate fast transitions due to exchange with the oscillators, which has the potential to slow down such decay processes. This effect is discussed in more details in section 5.13 and is related to the quantum Zeno effect.

**A formulation for excitation transfer to multiple receiver states**

As can be see from sections 6.18 and 6.19 and as was discussed in section 5.10, the density of states and therefore the achievable transfer rates are much higher at higher energies. At first glance, this is not helpful, since the energy available from the fusion transition is fixed. However, if substantial energy (here we consider tens of MeV and more) were made available to the nuclear system from the oscillator modes, then it would be as if the fusion transition energy were boosted to much higher energy, where the number of combinations of multiple Pd*/Pd transitions is exponentially greater, and a much faster excitation transfer rate is predicted. Since the exchange of phonons and plasmons with the nuclear system can be fast (see sections 5.11 and 6.25), we consider such a large transfer of energy to be a possibility.

However, it takes time for a large amount of energy to be exchanged between lattice oscillator modes and nuclear states. This timing requirement impacts the associated rate calculation. For example, at insufficiently fast transfer rates associated with the comparatively low density of states region of the energy spectrum, it is not possible for an excitation transfer to be completed until energy from the oscillators has been exchanged. This means that we cannot expect a simple exponential decay rate. Also, if more energy is exchanged, then the number of combinations of multiple Pd*/Pd transitions is even larger (exponentially), which means that after an initial delay, the system is much more likely to make a transition. Once again, this kind of process is not going to be described by an exponential random process.

The above considerations imply a rather complicated quantum dynamics model, which goes beyond what we recognize to exist in the literature from other contexts.

Here, we will outline how we are thinking about the problem. We make use of a sector decomposition, or dynamic resonating group method, and write coupled equations according to

$$i\hbar \frac{\partial}{\partial t} \Psi_0 = \left( \hat{H}_{fus} + \hat{H}_{Pd} + \hat{H}_{osc} + \hat{U} \right) \Psi_0 + \hat{V}_- \Psi_1$$

$$
\begin{aligned}
i\hbar\frac{\partial}{\partial t}\Psi_1 &= \left(\hat{H}_{fus} + \hat{H}_{Pd} + \hat{H}_{osc} + \hat{U}\right)\Psi_1 + \hat{V}_-\Psi_2 + \hat{V}_+\Psi_0 \\
i\hbar\frac{\partial}{\partial t}\Psi_2 &= \left(\hat{H}_{fus} + \hat{H}_{Pd} + \hat{H}_{osc} + \hat{U}\right)\Psi_2 + \hat{V}_-\Psi_3 + \hat{V}_+\Psi_1 \\
&\vdots
\end{aligned}
\tag{600}
$$

where $\Psi_0$ includes states where no fusion transitions have occurred, where $\Psi_1$ includes states where one fusion transition has occurred, where $\Psi_2$ includes states where two fusion transitions have occurred; and so forth. The part of the $\hat{V}$ operator that increases the number of fusion transitions that have occurred is $\hat{V}_+$, and the part of $\hat{V}$ that reduces the number of fusion transitions that have occurred is $\hat{V}_-$. We would expect the system to evolve from the initial $\Psi_0$ sector to the $\Psi_1$ sector, then to the $\Psi_2$ sector, and so forth.

Suppose that the system starts in the $\Psi_0$ sector, and we were to "turn on" interactions favorable to excitation transfer from the fusion transition. Because of the time needed for energy exchange with the oscillators, we would expect no initial build up of probability amplitude in the $\Psi_1$ sector. A dynamical Golden Rule calculation would lead to an initially slow rate of moving from sector 0 to sector 1, *i.e.*, a slow $\gamma_{1,0}(t)$. As time goes on, many fast Pd*/Pd transitions occur, which result in off-resonant occupation of excited Pd* states, along with the fast exchange of oscillator quanta. Eventually there is sufficient energy exchange with the oscillators such that the dynamic Golden Rule rate estimate $\gamma_{1,0}(t)$ increases to the point where we would expect the $\Psi_1$ sector to begin accumulating occupation probability. As $t$ continues to increase, $\gamma_{1,0}(t)$ gets exponentially faster (since more energy from the oscillators is available, allowing for more Pd*/Pd transitions, when combined with the fusion mass energy difference), and we expect the probability that the system remains in the $\Psi_0$ sector to decrease rapidly.

We anticipate dynamics in the other sectors that is qualitatively similar. During an excitation transfer event, the new sector is occupied over a relatively short duration, then energy exchange with oscillators begins. The dynamical excitation transfer rate to the next sector is initially very slow, but builds up roughly exponentially as more energy exchange with phonons occurs, and then excitation is transferred to the next sector.

**Decay rate in terms of the Lippmann-Schwinger transition operator**

Given the discussion above, we are interested in developing a relevant approximation scheme for the corresponding excitation transfer rate. Based on the process outlined above, we are thinking of a dynamical Golden Rule transition rate, where ultimately we will require self-consistency between the estimated dynamical Golden Rule rate and the time spent in a sector. This suggests the use of a finite time formalism related to the Lippmann-Schwinger formalism.

The Lippmann-Schwinger formalism was developed for scattering problems, where the particles are initially separate, then approach and interact, and then separate. Since the interaction in this kind of model is localized to where the particles are close to each other, the calculation is "clean" in the sense that it is possible in principle to include all effects that are connected to the interaction between the particles. The Golden Rule rate for transitions in this model can be expressed in terms of matrix elements of the transition operator $\hat{T}$ according to

$$
\gamma_{fi} = \frac{2\pi}{\hbar}|\langle f|\hat{T}|i\rangle|^2\rho \tag{601}
$$

This formalism is attractive for the above reasons. However, we need to adapt it to our problem since it is qualitatively different from a scattering problem. In order to do this, we need to first recall how the transition operator is calculated in the Lippmann-Schwinger formalism.

We recall that the Lippmann-Schwinger equation is concerned with a scattering problem, in which an initial incoming free state $|\phi\rangle$ interacts with a potential $\hat{V}$, resulting in an outgoing state $|\psi^{(+)}\rangle$ according to

$$|\psi^{(+)}\rangle \; = \; |\phi\rangle + \frac{1}{E - \hat{H}_0 + i\epsilon}\hat{V}|\psi^{(+)}\rangle \tag{602}$$

The transition operator satisfies

$$\hat{V}|\psi^{(+)}\rangle \; = \; \hat{T}|\phi\rangle \tag{603}$$

This leads to

$$\hat{T}|\phi\rangle \; = \; \hat{V}|\phi\rangle + \hat{V}\frac{1}{E - \hat{H}_0 + i\epsilon}\hat{T}|\phi\rangle \tag{604}$$

The transition operator $\hat{T}$ satisfies

$$\hat{T} \; = \; \hat{V} + \hat{V}\frac{1}{E - \hat{H}_0 + i\epsilon}\hat{T} \tag{605}$$

Scholes and Ghiggino (1995) used this kind of formalism in the case of electronic energy transfer in multichromophoric assemblies [167].

**Finite duration Lippmann-Schwinger transition operator**

We consider a modification of the Lippmann-Schwinger formulation, by working with a finite time duration version of the model, which will provide for a rough approximation for the excitation transfer rate from the fusion transition We can write the wave function in the Schrödinger picture at time $t = \tau$ in terms of an initial state at $t = 0$ according to

$$\psi(\tau) \; = \; e^{-i\hat{H}_0\tau/\hbar}\psi(0) + \frac{1}{i\hbar}\int_0^\tau e^{-i\hat{H}_0(\tau - t')/\hbar}\hat{V}\psi(t')dt' \tag{606}$$

Recall that the Lipmann-Schwinger formalism is expressed in the interaction representation. Here we are obliged to work in the Schrödinger picture due to the difficulties in developing eigenfunctions of the coupled oscillator and nuclear problem. Because of this, in what follows we pursue the development of a crude model in the Schrödinger picture.

A natural generalization of the Lippmann-Schwinger transition operator for this model is given by

$$\hat{V}\psi(\tau) \; = \; \hat{T}_S(\tau)\psi(0) \tag{607}$$

where $\hat{T}_S(\tau)$ is a Schrodinger picture finite time generalization of the Lipmann-Schwinger $\hat{T}$ operator. It follows that the finite time transition operator satisfies

$$\hat{T}_S(\tau)\psi(0) \; = \; \hat{V}e^{-i\hat{H}_0\tau/\hbar}\psi(0) + \hat{V}\frac{1}{i\hbar}\int_0^\tau e^{-i\hat{H}_0(\tau - t')/\hbar}\hat{T}_S(t')\psi(0)dt' \tag{608}$$

or

$$\hat{T}_S(\tau) \; = \; \hat{V}e^{-i\hat{H}_0\tau/\hbar} + \hat{V}\frac{1}{i\hbar}\int_0^\tau e^{-i\hat{H}_0(\tau - t')/\hbar}\hat{T}_S(t')dt' \tag{609}$$

We note that this Schrödinger picture $\hat{T}_S(\tau)$ operator cannot be used directly in connection with the Golden Rule (in contrast to the Lipmann-Schwinger formalism), since we do not have a clean isolation of the interaction responsible for a transition (again, in contrast to the Lipmann-Schwinger formalism). For excitation transfer from the fusion transition, we need for there to be a fusion transition as well as a large number of Pd*/Pd transitions to exchange oscillator quanta. While the Schrödinger picture $\hat{T}_S(\tau)$ operator includes these Pd*/Pd transitions related to the excitation transfer process of interest to us, it also includes many Pd*/Pd transitions that are unrelated to the excitation transfer process of interest. If were were able to separate the Pd* transitions involved in the excitation transfer process from all of the others, then we could use the part of $\hat{T}_S(\tau)$ that is connected to the excitation transfer. We can implement this according to

$$\gamma_{fi} \approx \frac{2\pi}{\hbar} |\langle f|\hat{T}'_S(\tau)|i\rangle|^2 \rho \tag{610}$$

subject to the constraint that

$$\tau = \frac{1}{\gamma_{fi}} \tag{611}$$

At present we do not have a way to develop the clean separation that we need for the construction of $\hat{T}'_S(\tau)$. On the other hand, we do have a way to develop a crude approximation for $\hat{T}'_S(\tau)$, which will lead to a rough estimate for the decay rate associated with excitation transfer from the fusion transition.

**Notion of "free" energy exchange between nuclei and oscillators**

To estimate the excitation transfer rate, we need to evaluate matrix elements of the finite time transition operator, which in general is a challenging calculation. From a mathematical point of view, we would want to work in terms of eigenstates of the strongly coupled Pd and oscillator systems. At present, such an approach appears formidable, and it is not obvious how to implement it practically.

The key feature of the model that we would like to address is the fast energy exchange between the Pd*/Pd transitions and the oscillators. When sufficient energy has been exchanged, the dynamical Golden Rule excitation transfer rate becomes sufficiently fast that the system can go from one sector to another. A complicating issue is that every ground state nucleus in the lattice makes transitions driven by the strongly excited oscillators. Since the oscillators are at much lower frequency (*i.e.*, energy) than the nuclear transitions, in general this involves a polarization effect. If we were able to diagonalize the overall Pd and oscillator parts of the Hamiltonian, we would be able to identify which part of the energy exchange can be associated with the formation of "dressed" states of the coupled system, and which part of the energy exchange involves the conversion of oscillator energy to nuclear energy available for additional excitation transfers.

To develop an approximation, we make use of the notion of "free" energy exchange between Pd nuclei and oscillators. The Pd*/Pd transitions that are involved in the excitation transfer process create and destroy oscillator quanta rapidly, and we seek to model these dynamics. We expect that only the transitions directly involved in the nuclear part of the excitation transfer should be included in an estimation of the energy exchange for that excitation transfer.

In principle, we might be able to make use of some kind of clever summation of Feynman diagrams in order to determine the energy exchange along with the nuclear transitions directly associated with excitation transfer. However, this is complicated by the very large number of them, by the divergences expected in the strong coupling regime, and also by the finite time limitation. Presumably, all of these issues could in the future be addressed. At the moment, this kind of detailed approach goes beyond the scope of this document.

In the approach presented here, we consider as an idealization a $t = 0$ situation, both for the initial excitation transfer (from the fusion transition) as well as for subsequent excitation transfers, where there has been no interaction between the oscillators and Pd*/Pd transitions involved in the excitation transfer of interest. We would expect that the few Pd*/Pd transitions that can be excited with the fusion energy $\Delta Mc^2$ will be excited essentially immediately, allowing there to be mixing (due to these transitions) with the oscillators. At this point we would expect coherent energy exchange with the oscillators to get started.

As more energy is exchanged, there is a probability distribution associated with energy exchange between oscillators and nuclear states. Within this probability distribution is the possibility of energy beyond what is supplied by the fusion transition and more Pd*/Pd transitions can be part of the excitation transfer from the fusion transition. This makes energy exchange with the oscillator faster. These dynamics continue up until the point where eventually probability amplitude couples into the next sector.

When excitation transfer from the fusion transition is complete, we expect that energy will be conserved; the fusion energy and energy from the oscillators go into promoting Pd* excitation. However, between $t = 0$ and $t = \tau$ there will be many states with some occupation that are off of resonance, and many of these will have more Pd* excitation. The question is whether we should think of the dynamics as being mostly on resonance, in which case the number of Pd*/Pd transitions that couple to the oscillators will be limited, and the energy exchange that we should model will be correspondingly limited. Alternatively, if what matters is the number of Pd*/Pd transitions that are promoted in the end, then we should use the larger number throughout.

Were we to imagine a calculation in which all possible Feynman diagrams were included consistent with energy conservation at the end, we would expect the majority of basis states contributing to not be so constrained by energy conservation. Energy conservation comes into the problem at the end; however, since many of the transitions are in the strong coupling regime, we expect contributions from basis states that are far off of resonance. This argument suggests that it would be reasonable to assume that (at least) the number of transitions that are promoted as a result of excitation from the fusion transition should contribute to energy exchange with the oscillator.

Keeping our focus on energy exchange between the Pd*/Pd transitions directly involved in the excitation transfer process with the oscillators, we note that there is not much difference between the energy exchange with one configuration of Pd*/Pd transitions and another. This suggests that we can work with an estimate of the average energy exchange, based on an average number of final state transitions involved in the excitation transfer, and an average strength of an individual transition (in terms of the transitions' $O$-values, see section 5.4). Next, we will develop an implementation of this idea, consistent with the finite time transition operator introduced above.

### The finite time transition operator and energy exchange

The finite time transition operator $\hat{T}(\tau)$ in this expression includes the dynamics described by

$$\hat{T}_S(\tau)\Psi(0) \ = \ \hat{V}e^{-i\hat{H}_0\tau/\hbar}\Psi(0) + \hat{V}\frac{1}{i\hbar}\int_0^\tau e^{-i\hat{H}_0(\tau-t')/\hbar}\hat{T}_S(t')\Psi(0)dt' \tag{612}$$

where

$$\hat{H}_0 \ = \ \hat{H}_{fus} + \hat{H}_{Pd} + \hat{H}_{osc} + \hat{U} \tag{613}$$

This connects the formalism under discussion to the conceptual model laid out above. The finite time transition operator includes the fusion transition. Based on the discussion above, our focus will be on the dynamics connected to a single fusion transition. Of key interest is the nuclear part of the problem, where excitation is transferred from a fusion transition to many Pd*/Pd transitions, and the energy exchange part of the problem where oscillator energy is made available to the nuclei. However, because nuclear and oscillator degrees of freedom are mixed in the absence of a fusion transition, we need to separate unwanted interactions from the ones that are critical.

### Approximation for the finite time transition operator

To proceed we need to make use of our intuition about the associated dynamics. If we focus on energy exchange with the oscillator, then we consider a plausible approximation to be

$$\hat{T}_S'(\tau) \ \sim \ e^{-i\hat{H}_{free}\tau/\hbar} \tag{614}$$

where $\hat{H}_{free}$ is an average model for oscillator exchange that will be described shortly (and which we can carry out calculations with).

However, there is a challenge in separating the nuclear and oscillator degrees of freedom in the finite time transition operator. Energy exchange with the oscillator will provide for coupling to oscillator states, where a very large number of oscillator quanta have been gained or lost, so that the oscillator energy will be different. In weak coupling, we would expect the Pd plus oscillator energy not to change much (except due to energy input from the fusion transition). Since there must be overall energy conservation at the end, if we couple to an oscillator state that has less energy, then whatever nuclear states that we couple to have more energy.

What is lacking here is a useful notation. For this, we use

$$\hat{T}_S'(\tau) \ \rightarrow \ \left( \hat{T}_{fus,Pd}'(\tau) e^{-i\hat{H}_{free}\tau/\hbar} \right)_{\Delta Mc^2} \tag{615}$$

where the idea is to implement a separation of the nuclear and oscillator degrees of freedom, but in doing so assuring that energy conservation is maintained in the process. The transition operator associated with the fusion transition and Pd system transitions $\hat{T}_{fus,Pd}'(\tau)$ in this case is equivalent to what we would get in the absence of energy exchange with the oscillators, but with a different fusion mass difference energy.

### Initial and final states

In the excitation transfer process, a $D_2$ molecule is converted into a $^4$He atom; many ground state Pd nuclei are promoted to excited states and a great many phonons and plasmons are exchanged. If there are initially excited Pd$^*$ states (due to off-resonance effects in connection with strong coupling), then we expect that some of these will be de-excited during the excitation transfer process. To model this, we can write for the initial state

$$\Psi_i \ = \ |S, M\rangle_{fus} \left( \prod_j |S_j, M_j\rangle \right)_{Pd} \Big| n_A, n_O, n_P \Big\rangle_{osc} \tag{616}$$

Similarly, for the final state we have

$$\Psi_f \ = \ |S, M-1\rangle_{fus} \left( \prod_j |S_j, M_j'\rangle \right)_{Pd} \Big| n_A', n_O', n_P' \Big\rangle_{osc} \tag{617}$$

Energy conservation leads to the requirement that

$$\Delta Mc^2 + \sum_j \epsilon_j M_j + n_A \hbar\omega_A + n_O \hbar\omega_O + n_P \hbar\omega_P \ = \ \sum_j \epsilon_j M_j' + n_A' \hbar\omega_A + n_O' \hbar\omega_O + n_P' \hbar\omega_P \tag{618}$$

**Estimate for the excitation transfer rate**

For the excitation transfer rate $\Gamma_{transfer}$ we can write

$$\Gamma_{transfer} \ = \ \sum_f |\langle \Psi_f | \hat{T}'_S(\tau) | \Psi_i \rangle|^2 \delta(E_i - E_f) \tag{619}$$

We can make use of the crude approximation for the transition operator and write

$$\begin{aligned} \langle \Psi_f | \hat{T}'_S(\tau) | \Psi_i \rangle \ &= \ \Big\langle |S, M-1\rangle_{fus} \Big( \prod_j |S_j, M'_j\rangle \Big)_{Pd} |n'_A, n'_O, n'_P\rangle_{osc} \Big| \\ &\quad \hat{T}'_{fus,Pd}(\tau) e^{-i\hat{H}_{free}\tau/\hbar} \Big| |S, M\rangle_{fus} \Big( \prod_j |S_j, M_j\rangle \Big)_{Pd} |n_A, n_O, n_P\rangle_{osc} \Big\rangle \\ &= \ \Big\langle |S, M-1\rangle_{fus} \Big( \prod_j |S_j, M'_j\rangle \Big)_{Pd} \Big| \hat{T}'_{fus,Pd}(\tau) \Big| |S, M\rangle_{fus} \Big( \prod_j |S_j, M_j\rangle \Big)_{Pd} \Big\rangle \\ &\quad \Big\langle n'_A, n'_O, n'_P \Big| e^{-i\hat{H}_{free}\tau/\hbar} \Big| n_A, n_O, n_P \Big\rangle_{osc} \end{aligned} \tag{620}$$

where we keep in mind that we need to enforce energy conservation. We can use this to write for the excitation transfer rate

$$\begin{aligned} \Gamma_{transfer} \ &= \ \sum_{\{M'_j\}} \sum_{n'_A} \sum_{n'_O} \sum_{n'_P} \left| \Big\langle |S, M-1\rangle_{fus} \Big( \prod_j |S_j, M'_j\rangle \Big)_{Pd} \Big| \hat{T}'_{fus,Pd}(\tau) \Big| |S, M\rangle_{fus} \Big( \prod_j |S_j, M_j\rangle \Big)_{Pd} \Big\rangle \right|^2 \\ &\quad \left| \Big\langle n'_A, n'_O, n'_P \Big| e^{-i\hat{H}_{free}\tau/\hbar} \Big| n_A, n_O, n_P \Big\rangle_{osc} \right|^2 \\ &\quad \delta\Big( \Delta M c^2 - \sum_j \epsilon_j (M'_j - M_j) - (n'_A - n_A)\hbar\omega_A - (n'_O - n_O)\hbar\omega_O - (n'_P - n_P)\hbar\omega_P \Big) \end{aligned} \tag{621}$$

Eq. 621 represents a major goal of this section and is equivalent to the simplified transfer rate expression (Eq. 236) used in section 5.11.

In the limit of a continuum density of states the sums becomes essentially the familiar integral from Eq. 236.

Note that there are many more terms associated with the approximation far off of resonance contained within the summation than would have been obtained were we able to work with the exact finite time transition operator; however, by enforcing energy conservation, these non-physical terms are eliminated.

This excitation transfer rate depends explicitly on $\tau$, and we can work with this value in order to estimate the time-dependent Golden Rule rate associated with a completed excitation transfer. To that end we set

$$\tau \ = \ \frac{1}{\Gamma_{transfer}} \tag{622}$$

for self-consistency. We would like for the excitation transfer rate to be on the order of the time $\tau$ required to reach conditions where the excitation transfer can happen on a relevant timescale.

The goal now is to relate the expressions in Eq. 621 to familiar physical parameters such as the power in a phonon mode P. The remainder of this section will focus on deriving an expression for

$$\left| \left\langle n'_A, n'_O, n'_P \middle| e^{-i\hat{H}_{free}\tau/\hbar} \middle| n_A, n_O, n_P \right\rangle_{osc} \right|^2 \tag{623}$$

as an integral part of Eq. 621. As a part of this process $\hat{H}_{free}$ needs to be determined, which is done next.

**Time-dependent Hartree approximation**

It is possible to gain some understanding of how the three systems interact through the use of a time-dependent Hartree approximation. We can approximate the overall wave function as a product of individual fusion transition, Pd, and oscillator components according to

$$\Psi \ = \ \psi_{fus}\psi_{Pd}\psi_{osc} \tag{624}$$

where the associated time-dependent Hartree equations are

$$i\hbar\frac{\partial}{\partial t}\psi_{fus} \ = \ \hat{H}_{fus}\psi_{fus} + \langle\psi_{osc}|\hat{V}|\psi_{osc}\rangle_{osc}\psi_{fus} \tag{625}$$

$$i\hbar\frac{\partial}{\partial t}\psi_{Pd} \ = \ \hat{H}_{Pd}\psi_{Pd} + \langle\psi_{osc}|\hat{U}|\psi_{osc}\rangle_{osc}\psi_{Pd} \tag{626}$$

$$i\hbar\frac{\partial}{\partial t}\psi_{osc} \ = \ \hat{H}_{osc}\psi_{osc} + \langle\psi_{fus}|\hat{V}|\psi_{fus}\rangle_{fus}\psi_{osc} + \langle\psi_{Pd}|\hat{U}|\psi_{Pd}\rangle_{Pd}\psi_{osc} \tag{627}$$

By separating the different degrees of freedom, it becomes much easier to solve for the dynamics and distributions associated with each degree of freedom individually. It may be that in the future an improved model for excitation transfer could be developed based on this kind of approach. However, for this to work, there may need to be a separate calculation of the part of the oscillator distribution that can be associated directly with excitation transfer transitions in connection with energy exchange as mentioned above.

To estimate the part of the energy exchange with the oscillators that is associated with the Pd*/Pd transitions involved in the excitation transfer process, we write

$$\hat{H}_{free} \ \to \ \hat{H}_{osc} + \bigg(\langle\psi_{Pd}|\hat{U}|\psi_{Pd}\rangle_{Pd}\bigg)_{transfer} \tag{628}$$

It is unlikely that this can be evaluated accurately in the framework of a time-dependent Hartree model. More practical is to make use of

$$\hat{H}_{free} \ \approx \ \hat{H}_{osc} + \overline{\bigg(\langle\psi_{Pd}|\hat{U}|\psi_{Pd}\rangle_{Pd}\bigg)_{transfer}} \tag{629}$$

which involves the associated average (or an estimate for the average) of $\hat{U}$ defined in Eq. 6.24.

For simplicity, we will focus on energy exchange with highly-excited acoustic phonons, so that

$$\hat{H}_{osc} \;\rightarrow\; \hbar\omega_A \hat{a}_A^\dagger \hat{a}_A \tag{630}$$

For the interaction with the Pd*/Pd transitions involved in excitation transfer from the fusion transition, we take

$$\begin{aligned}
\overline{\left(\langle \psi_{Pd}|\hat{U}|\psi_{Pd}\rangle_{Pd}\right)}_{transfer} &= \overline{\left(\left\langle \psi_{Pd}\middle| \sum_j \langle \text{Pd}_j^*|\mathbf{a}\cdot c\hat{\mathbf{P}}_{Pd_j}|\text{Pd}_j\rangle \left(\frac{\hat{S}_+^{(j)}}{\hbar} + \frac{\hat{S}_-^{(j)}}{\hbar}\right)\middle|\psi_{Pd}\right\rangle_{Pd}\right)}_{transfer} \\
&= \overline{\left(\sum_j \left\langle \psi_{Pd}\middle| \left(\frac{\hat{S}_+^{(j)}}{\hbar} + \frac{\hat{S}_-^{(j)}}{\hbar}\right)\middle|\psi_{Pd}\right\rangle_{Pd} \langle \text{Pd}_j^*|\mathbf{a}|\text{Pd}_j\rangle \cdot c\hat{\mathbf{P}}_{Pd_j}\right)}_{transfer} \\
&= \overline{\left(\sum_j \left\langle \frac{\hat{S}_+^{(j)}}{\hbar} + \frac{\hat{S}_-^{(j)}}{\hbar}\right\rangle_{Pd} \langle \text{Pd}_j^*|\mathbf{a}|\text{Pd}_j\rangle \cdot \mathbf{e}_j^{(A)} \sqrt{\frac{M_j c^2 \hbar\omega_A}{2N}}\right)}_{transfer} \left(\frac{\hat{a}_A - \hat{a}_A^\dagger}{i}\right)
\end{aligned} \tag{631}$$

We can now relate this to our estimate of the acoustic phonon exchange rate $\Gamma_A(j)$ which we first introduced in section 5.11 and expanded on in section 6.25, namely:

$$\Gamma_A(j) \;\approx\; \frac{2}{\hbar}|\langle \text{Pd}^*(j)|\mathbf{a}|\text{Pd}\rangle \cdot \mathbf{e}_j^{(A)}|\sqrt{M_j c^2 n_A \hbar\omega_A}\sqrt{\frac{N_{Pd_j}}{N}} \tag{632}$$

In terms of this rate we have

$$\overline{\left(\langle \psi_{Pd}|\hat{U}|\psi_{Pd}\rangle_{Pd}\right)}_{transfer} = \frac{\hbar}{2}\overline{\left(\sum_j \left\langle \frac{\hat{S}_+^{(j)}}{\hbar} + \frac{\hat{S}_-^{(j)}}{\hbar}\right\rangle_{Pd} \sqrt{\frac{1}{N_{Pd_j}}}\Gamma_A(j)\right)}_{transfer} \sqrt{\frac{1}{2n_A}}\left(\frac{\hat{a}_A - \hat{a}_A^\dagger}{i}\right) \tag{633}$$

assuming that

$$\langle \text{Pd}_j^*|\mathbf{a}|\text{Pd}_j\rangle \cdot \mathbf{e}_j^{(A)} \;=\; |\langle \text{Pd}_j^*|\mathbf{a}|\text{Pd}_j\rangle \cdot \mathbf{e}_j^{(A)}| \tag{634}$$

which involves defining the phases of the excited Pd* states relative to the ground states appropriately. We estimate the number of transitions involved in the excitation transfer according to

$$\overline{\left(\sum_j \left\langle \frac{\hat{S}_+^{(j)}}{\hbar} + \frac{\hat{S}_-^{(j)}}{\hbar}\right\rangle_{Pd} \sqrt{\frac{1}{N_{Pd_j}}}\Gamma_A(j)\right)}_{transfer} \;\rightarrow\; \overline{n_j}\Gamma_A \tag{635}$$

where $\Gamma_A$ is the average of $\Gamma_A(j)$, and where $\overline{n_j}$ is the estimated number of transitions involved.

This leads to

$$\overline{\left(\langle \psi_{Pd}|\hat{U}|\psi_{Pd}\rangle_{Pd}\right)}_{transfer} \;\rightarrow\; \frac{\hbar}{2}(\overline{n_j}\Gamma_A)\sqrt{\frac{1}{2n_A}}\left(\frac{\hat{a}_A - \hat{a}_A^\dagger}{i}\right) \tag{636}$$

where it should be noted that we've neglected Dicke enhancement factors for the Pd*/Pd transitions (which are not straightforward to estimate in this kind of model), and assumed that the interaction between the acoustic phonons and nuclei can be modeled as a polarization [168].

In the end, the Hamiltonian for "free" exchange in the case of acoustic phonons is

$$\hat{H}_{free} \rightarrow \hbar\omega_A \hat{a}_A^\dagger \hat{a}_A + \frac{\hbar}{2}(\overline{n_j}\Gamma_A)\sqrt{\frac{1}{2n_A}}\left(\frac{\hat{a}_A - \hat{a}_A^\dagger}{i}\right) \tag{637}$$

**Finite basis expansion and evolution equation for the expansion coefficients**

Assuming only acoustic phonons in Eq. 627, then $\psi_{osc} \rightarrow \psi_A$. We make a finite basis expansion for the acoustic phonons according to

$$\psi_A = \sum_n i^n c_n(t)\phi_n^{(A)} \tag{638}$$

In addition, we can drop the fusion part from Eq. 627 because it is small due to the Coulomb barrier. The associated evolution equation for the expansion coefficients is then

$$i\hbar\frac{d}{dt}c_n(t) = \hbar\omega_A n c_n(t) - \frac{\hbar}{2}(\overline{n_j}\Gamma_A)\left(\frac{\sqrt{n}c_{n-1}(t) + \sqrt{n+1}c_{n+1}(t)}{\sqrt{2n_A^{(0)}}}\right) \tag{639}$$

where $n_A^{(0)}$ is the number of acoustic phonons used for the estimate of $\Gamma_A$. In the event that the fusion energy $\Delta Mc^2$ is much smaller than the energy in the acoustic phonon mode

$$\Delta Mc^2 \ll n_A\hbar\omega_A = E_A \tag{640}$$

then we can simplify to

$$i\hbar\frac{d}{dt}c_n(t) = n\hbar\omega_A c_n(t) - \frac{\hbar}{2}(\overline{n_j}\Gamma_A)\left(\frac{c_{n-1}(t) + c_{n+1}(t)}{\sqrt{2}}\right) \tag{641}$$

In the event that acoustic phonon exchange is fast such that

$$\frac{\hbar}{2}(\overline{n_j}\Gamma_A) \gg 10\,\Delta Mc^2 \ll n\hbar\omega_A \tag{642}$$

then $\hbar\omega_A n c_n(t)$ can be considered as a constant and dropped form the equation, giving:

$$i\hbar\frac{d}{dt}c_n(t) = -\frac{\hbar}{2}(\overline{n_j}\Gamma_A)\left(\frac{c_{n-1}(t) + c_{n+1}(t)}{\sqrt{2}}\right) \tag{643}$$

Next, we solve the evolution equation for the expansion coefficients as an initial value problem with

$$c_n(0) = \begin{cases} 1 & n = n_A^{(0)} \\ 0 & n \neq n_A^{(0)} \end{cases} \tag{644}$$

The solution for large time has an associated probability distribution that is Gaussian according to

$$|c_n(\tau)|^2 \;=\; \frac{1}{\sqrt{2\pi\sigma^2_{n_A}(\tau)}} \,\mathrm{exp}\left\{ -\frac{(n-n_A^{(0)})^2}{2\sigma^2_{n_A}(\tau)} \right\} \tag{645}$$

with

$$\sigma_{n_A}(T) \;=\; \frac{1}{2}\overline{n_j}\,\Gamma_A \tau \tag{646}$$

For the associated matrix element that appears in the reduction of the Golden Rule rate estimate in Eq. 621 earlier in this section we can then write

$$\left| \left\langle n'_A, n_O, n_P \middle| e^{-i\hat{H}_{free}\tau/\hbar} \middle| n_A, n_O, n_P \right\rangle_{osc} \right|^2 \;\approx\; \frac{1}{\sqrt{2\pi\sigma^2_{n_A}(\tau)}} \,\mathrm{exp}\left\{ -\frac{(n'_A - n_A)^2}{2\sigma^2_{n_A}(\tau)} \right\} \tag{647}$$

This expression effectively represents the phonon distribution and it enters into the simplified transfer rate expression (Eq. 236) in section 5.11.

## 6.25 Energy exchange rates between Pd and oscillator modes

Of interest in sections 5.11 and 6.24 is the question of how fast energy exchange occurs with the different oscillator modes in the lattice. We have focused on phonon (acoustic and optical) and plasmon modes in this document. Characteristics of these oscillators result in different behaviors with respect to energy exchange. For example, the acoustic phonon lifetime is long, making it much easier in practice for the acoustic phonons to be driven so that there is substantial energy in the mode. On the other hand, the energy associated with a single optical phonon or plasmon quantum is much larger. In this section we derive rates of energy exchange between Pd and highly excited uniform modes of different kinds.

### Acoustic phonon mode

We can define a coherent exchange rate for acoustic phonons associated with a single transition:

$$\Gamma_A(j) \;=\; \frac{2}{\hbar}|\langle \text{Pd}^*(j)|\mathbf{a}\cdot c\mathbf{P}(j)|\text{Pd}\rangle_A|\sqrt{N_{Pd_j}} \tag{648}$$

assuming a minimal cooperative (Dicke) factor.

$$\begin{aligned}\Gamma_A(j) \;&\approx\; \frac{2}{\hbar}|\mathbf{e}_j^{(A)}||\langle \text{Pd}^*(j)|a_z|\text{Pd}\rangle|c\sqrt{\langle|P(j)|^2\rangle_A}\sqrt{N_{Pd_j}} \\ &\approx\; \frac{2}{\hbar}|\mathbf{e}_j^{(A)}||\langle \text{Pd}^*(j)|a_z|\text{Pd}\rangle|\sqrt{M_jc^2P_D^{(A)}\tau_A}\sqrt{\frac{N_{Pd_j}}{N}}\end{aligned} \tag{649}$$

This evaluates numerically to

$$\Gamma_A(j) \;\approx\; 6.21\times 10^{25}\,O_j\,|\mathbf{e}_j^{(A)}|\sqrt{\frac{P_D^{(A)}}{1\text{ watt}}}\left(\frac{1\text{ MHz}}{f_A}\right)^{3/4}\sqrt{\frac{N_{Pdj}}{N}}\ \text{s}^{-1} \tag{650}$$

We see that the associated energy exchange rate can be very fast.

The associated energy exchange rate is

$$\begin{aligned}\hbar\omega_A\Gamma_A(j) \;&\approx\; 2\omega_A|\mathbf{e}_j^{(A)}||\langle \text{Pd}^*(j)|a_z|\text{Pd}\rangle|c\sqrt{\langle|P(j)|^2\rangle_A}\sqrt{N_{Pd_j}} \\ &\approx\; 2\omega_A|\mathbf{e}_j^{(A)}||\langle \text{Pd}^*(j)|a_z|\text{Pd}\rangle|\sqrt{M_jc^2P_D^{(A)}\tau_A}\sqrt{\frac{N_{Pd_j}}{N}}\end{aligned} \tag{651}$$

We can develop a numerical estimate for this of

$$\hbar\omega_A\Gamma_A(j) \;\approx\; 0.0073\,O_j\,|\mathbf{e}_j^{(A)}|\sqrt{\frac{P_D^{(A)}}{1\text{ watt}}}\left(\frac{f_A}{1\text{ MHz}}\right)^{1/4}\sqrt{\frac{N_{Pdj}}{N}}\ \text{watts} \tag{652}$$

Given that the rate in Eq. 650 is so fast, there is the potential for considerable energy exchange from acoustic phonons.

**Optical phonon mode**

We can use a similar approach to develop an estimate in the case of optical phonon exchange; for the excitation transfer rate we obtain

$$\Gamma_O(j) \ \approx \ 6.2 \times 10^{19} \, O_j \, |\mathbf{e}_j^{(O)}| \sqrt{\frac{P_D^{(O)}}{1 \text{ watt}}} \left( \frac{10 \text{ THz}}{f_O} \right)^{3/4} \sqrt{\frac{N_{Pdj}}{N}} \ \text{s}^{-1} \tag{653}$$

For the energy exchange rate we have

$$\hbar\omega_O \Gamma_O(j) \ \approx \ 0.41 \, O_j \, |\mathbf{e}_j^{(O)}| \sqrt{\frac{P_D^{(O)}}{1 \text{ watt}}} \left( \frac{f_O}{10 \text{ THz}} \right)^{1/4} \sqrt{\frac{N_{Pdj}}{N}} \ \text{watts} \tag{654}$$

The rate of exchange is much slower per unit dissipated power than what we found for the acoustic phonons, but since the quantum is so much larger, there can be a comparable if not larger rate for energy exchange.

**Plasmon mode**

In the case of plasmon exchange, we can write

$$\begin{aligned} \Gamma_P(j) \ &\approx \ \frac{2}{\hbar} |\mathbf{e}_j^{(P)}| |\langle \text{Pd}^*(j)| a_z |\text{Pd}\rangle| c \sqrt{\langle |P(j)|^2 \rangle_P} \sqrt{N_{Pdj}} \\ &\approx \ \frac{2}{\hbar} |\mathbf{e}_j^{(P)}| |\langle \text{Pd}^*(j)| a_z |\text{Pd}\rangle| Z_j \sqrt{m_e c^2 P_D^{(P)} \tau_P} \sqrt{\frac{N_{Pdj}}{N_e}} \end{aligned} \tag{655}$$

This can be evaluated to give

$$\Gamma_P(j) \ \approx \ 2.05 \times 10^{17} \, O_j \, |\mathbf{e}_j^{(P)}| \sqrt{\frac{P_D^{(P)}}{1 \text{ watt}}} \sqrt{\frac{N_{Pdj}}{N_e}} \ \text{s}^{-1} \tag{656}$$

The associated rate for energy exchange is

$$\begin{aligned} \hbar\omega_P \Gamma_P(j) \ &\approx \ 2\omega_P |\mathbf{e}_j^{(P)}| |\langle \text{Pd}^*(j)| a_z |\text{Pd}\rangle| c \sqrt{\langle |P(j)|^2 \rangle_P} \sqrt{N_{Pdj}} \\ &\approx \ 0.156 \, O_j \, |\mathbf{e}_j^{(P)}| \sqrt{\frac{P_D^{(P)}}{1 \text{ watt}}} \left( \frac{\hbar\omega_P}{4.75 \text{ eV}} \right) \sqrt{\frac{N_{Pdj}}{N_e}} \ \text{watts} \end{aligned} \tag{657}$$

All of these energy exchange rates can be quite fast in connection with the model under discussion if uniform modes are driven efficiently at the Watt level.

## 6.26 Analytic approximations for excitation transfer from $D_2$ to multiple receiver states with oscillator energy exchange

Section 5.11 presented a model for describing excitation transfer from fusion transitions to multiple receiver states under conditions where energy exchange between nuclei and lattice oscillators can take place. The model was rationalized by the considerations laid out on section 6.24.

At the center of the resulting transfer rate expressions are estimates for the generalized nuclear density of states $\tilde{\rho}_n$ (as discussed in section 6.19) and Gaussian probability distributions for the energy exchange between nuclear states and lattice oscillators such as acoustic phonons.

Here we present analytic approximations for these expressions. At the end of the section, we also considering some key scaling laws.

Recall the transfer rate expression from Eq. 5.11:

$$\begin{aligned}\Gamma_{transfer}(E) &= \left(\Gamma^{(0)}_{transfer} * f_{\epsilon_A}\right)(E) \\ &= \frac{2\pi}{\hbar}|\Delta\mathcal{U}|^2\left(\tilde{\rho}_N * f_{\epsilon_A}\right)(E) \\ &= \frac{2\pi}{\hbar}|\Delta\mathcal{U}|^2\int_{-\infty}^{\infty}\tilde{\rho}_N(E-\epsilon_A)f_{\epsilon_A}(\epsilon_A,\tau)d\epsilon\end{aligned} \tag{658}$$

**Fitting the generalized nuclear density of states**

We have developed an empirical model according to

$$\tilde{\rho}_n(E) = \alpha(E_A)e^{\beta(E_A)E} \tag{659}$$

with

$$\ln\alpha(E_A) = -15.4589 - 0.00238164\ \ln\left(\frac{E_A}{1\ \mathrm{J}}\right) - 0.00207085\left\{\ln\left(\frac{E_A}{1\ \mathrm{J}}\right)\right\}^2 \tag{660}$$

with $\alpha(E_A)$ in $\mathrm{eV}^{-1}$.

and

$$\Big(\beta(E_A)\Big)^{-1} = \frac{0.0383705}{\left(\frac{E_A}{1\ \mathrm{J}}\right)^{1/3} - \left(3.8901\times10^{-6}\right)^{1/3}} + 1.00497\ \mathrm{MeV} \tag{661}$$

The fit for $\alpha(E_A)$ is shown in figure 70, and for $\beta^{-1}$ is shown in Figure 71. When the acoustic mode energy $E_A$ is less than 3.8901 $\mu$J in this model, the excitation transfer rate is not increased by energy exchange from the highly-excited acoustic phonon mode.

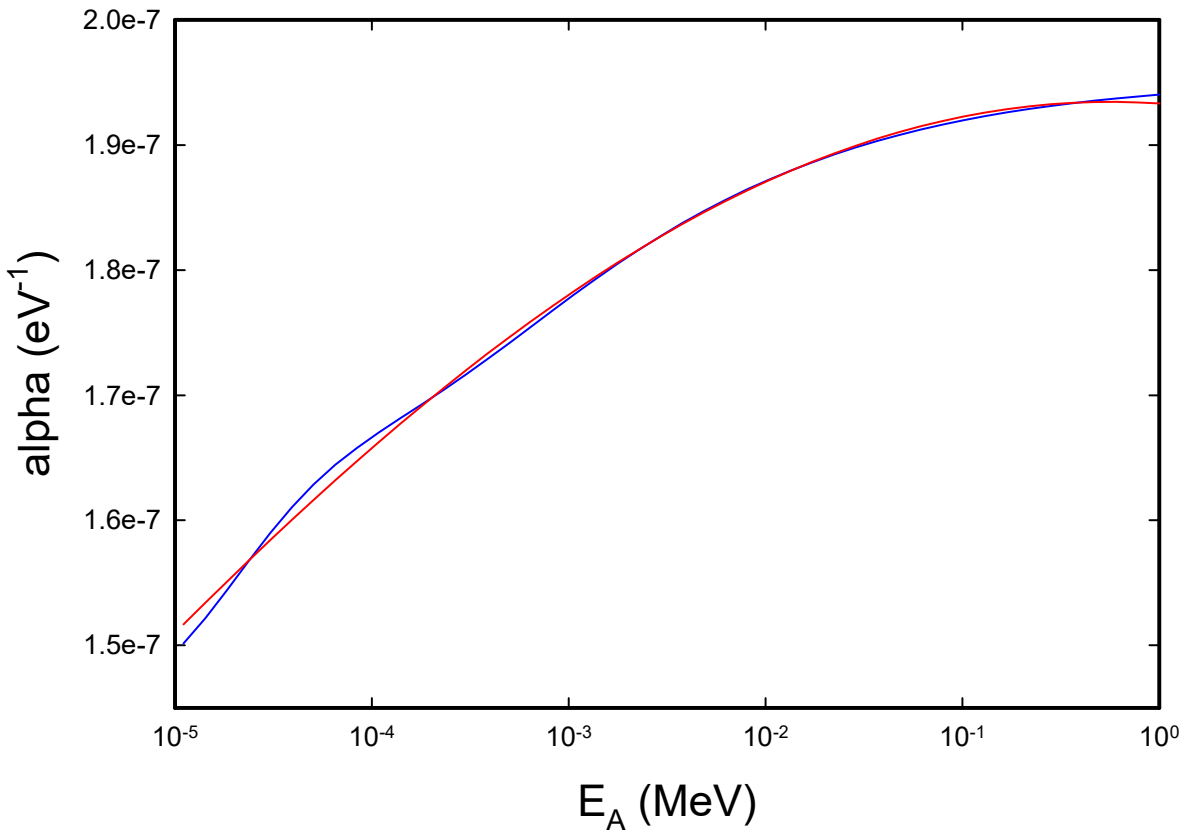

**Figure 70:** Results for $\alpha$ (blue) and the associated fit (red).

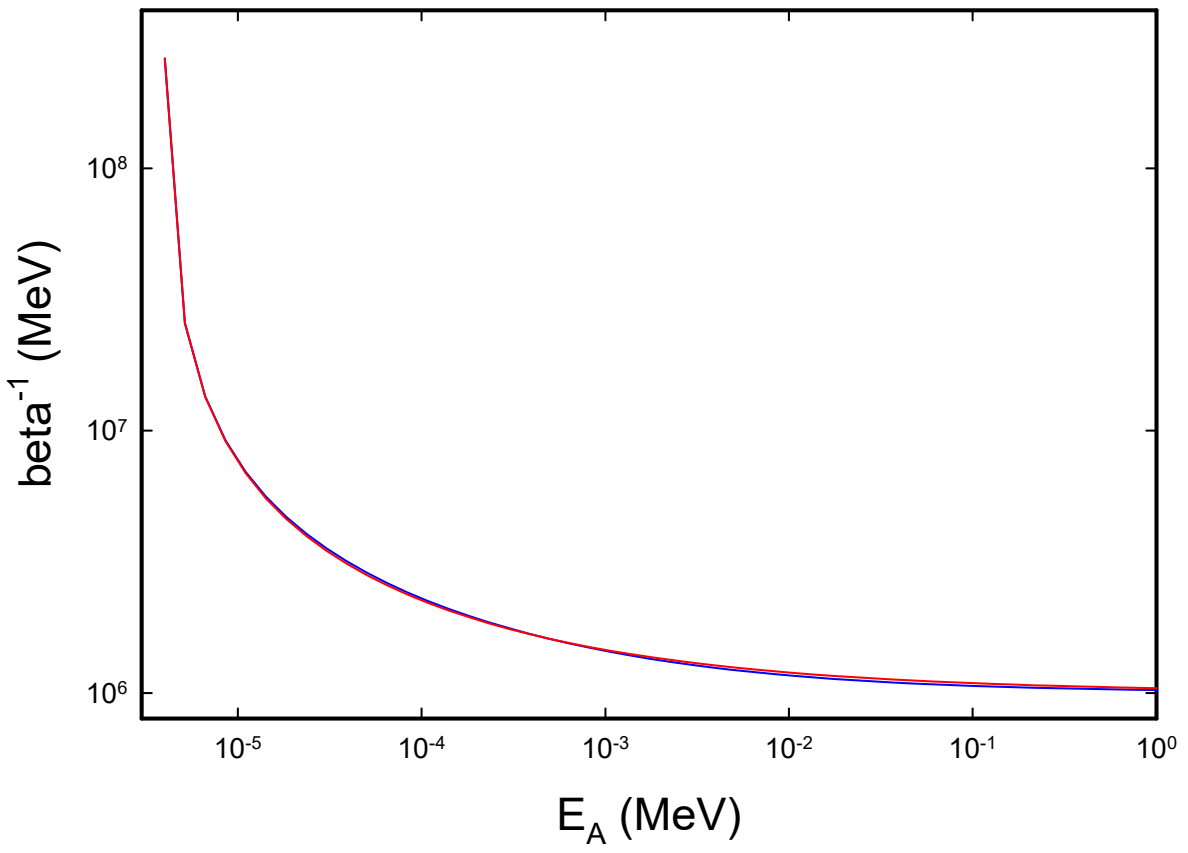

**Figure 71:** Results for $\beta^{-1}$ (blue) and the associated fit (red).

## Convolution

$\tilde{\rho}_n$ represents the first part of the convolution in 5.11. For the second part, we take

$$f_{\epsilon_A}(\epsilon) \;=\; \frac{1}{\sqrt{2\pi}\sigma_A} e^{-\epsilon^2/2\sigma_A^2} \tag{662}$$

For the convolution we can write

$$\begin{aligned}\left(\tilde{\rho}_N * f_{\epsilon_A}\right)(E) \;&=\; \int_{-\infty}^{\infty} \alpha(E_A) e^{\beta(E_A)(E-\epsilon)} \frac{1}{\sqrt{2\pi}\sigma_A} e^{-\epsilon^2/2\sigma_A^2} d\epsilon \\ &=\; \alpha(E_A) e^{\beta(E_A)E} e^{\beta^2(E_A)\sigma_A^2/2} \\ &=\; \tilde{\rho}_n(E) e^{\beta^2(E_A)\sigma_A^2/2}\end{aligned} \tag{663}$$

**Average energy exchange with the phonons**

The mean energy gain of the acoustic mode is

$$\begin{aligned}\bar{\epsilon} \ &= \ \frac{\int_{-\infty}^{\infty} \epsilon \alpha(E_A) e^{\beta(E_A)(E-\epsilon)} \frac{1}{\sqrt{2\pi}\sigma_A} e^{-\epsilon^2/2\sigma_A^2} d\epsilon}{\int_{-\infty}^{\infty} \alpha(E_A) e^{\beta(E_A)(E-\epsilon)} \frac{1}{\sqrt{2\pi}\sigma_A} e^{-\epsilon^2/2\sigma_A^2} d\epsilon} \\ &= \ -\beta(E_A)\sigma_A^2 \end{aligned} \tag{664}$$

where we used Mathematica to evaluate the integrals.

Note that $E$ in the formula above is $\Delta Mc^2$. The mean energy associated with the acoustic phonon mode is relative to $n_A \hbar \omega_A$. Since it is negative, we know that energy is being given to the nuclei.

**Rough estimate for the number of transitions participating**

The total energy available to the nuclei for Pd*/Pd excitation in this model is

$$\Delta Mc^2 - \bar{\epsilon} \ = \ \Delta Mc^2 + \beta(E_A)\sigma_A^2 \tag{665}$$

We can use this to develop an initial estimate $\overline{n_j}$. We get

$$\overline{n_j} \ \approx \ \frac{\Delta Mc^2 - \bar{\epsilon}}{\delta E} \ = \ \frac{\Delta Mc^2 + \beta(E_A)\sigma_A^2}{\delta E} \tag{666}$$

where

$$\delta E \ = \ 6.75 \text{ MeV} \tag{667}$$

**Improved estimate for the number of transitions participating**

It will be useful to improve this estimate. To understand why there should be a correction, recall the expansion of the generalized nuclear density of states in the absence of energy exchange in terms of Gaussians

$$\tilde{\rho}_n(E) \ = \ \sum_m A_m \frac{1}{\sqrt{2\pi\sigma_m^2}} e^{-(E-\mu_m)^2/2\sigma_m^2} \tag{668}$$

For the convolution we can write

$$\begin{aligned} &(\tilde{\rho}_n * f_A)(E) \\ = \ &\sum_m A_m \int_{-\infty}^{\infty} \frac{1}{\sqrt{2\pi\sigma_m^2}} e^{-(E-\mu_m-\epsilon)^2/2\sigma_m^2} \frac{1}{\sqrt{2\pi}\sigma_A} e^{-\epsilon^2/2\sigma_A^2} d\epsilon \\ = \ &\sum_m A_m \int_{-\infty}^{\infty} \frac{1}{\sqrt{2\pi\sigma_{AN}^2}} e^{-(E-\mu_m)^2/2\sigma_{AN}^2} \end{aligned} \tag{669}$$

where

$$\sigma_{AN}^2 \ = \ \sigma_A^2 + \sigma_m^2 \tag{670}$$

We can develop an expression for the average energy exchange with the acoustic phonons based on

$$\begin{aligned}\bar{\epsilon} \ &= \ \frac{\int_{-\infty}^{\infty} \epsilon \sum_m A_m \frac{1}{\sqrt{2\pi\sigma_m^2}} e^{-(E-\epsilon-\mu_m)^2/2\sigma_m^2} \frac{1}{\sqrt{2\pi}\sigma_A} e^{-\epsilon^2/2\sigma_A^2} d\epsilon}{\int_{-\infty}^{\infty} \sum_m A_m \frac{1}{\sqrt{2\pi\sigma_m^2}} e^{-(E-\epsilon-\mu_m)^2/2\sigma_m^2} \frac{1}{\sqrt{2\pi}\sigma_A} e^{-\epsilon^2/2\sigma_A^2} d\epsilon} \\ &= \ \frac{\sum_m A_m \frac{\sigma_A^2}{\sigma_A^2+\sigma_m^2}(E-\mu_m) e^{-(E-\mu_m)^2/2\sigma^2} \frac{1}{\sqrt{2\pi\sigma^2}}}{\sum_m A_m e^{-(E-\mu_m)^2/2\sigma^2} \frac{1}{\sqrt{2\pi\sigma^2}}}\end{aligned} \tag{671}$$

In this expression we see a somewhat complicated expression that acts as the difference energy associated with the acoustic phonons.

$$\epsilon \ \rightarrow \ \frac{\sigma_A^2}{\sigma_A^2 + \sigma_m^2}(E - \mu_m) \ = \ \frac{\sigma_A^2}{\sigma_A^2 + m\sigma_0^2}(E - m\delta E) \tag{672}$$

where $\delta E = 6.75$ MeV and where $\sigma_0 = 1.28$ MeV. The summation over $m$ is an average over order, where $m$ is the number of transitions involved. We can get an estimate for the average number of transitions $\overline{n_j}$ by solving this as an algebraic relation for the means according to

$$\bar{\epsilon} \ \approx \ \frac{\sigma_A^2}{\sigma_A^2 + \overline{n_j}\sigma_0^2}(E - \overline{n_j}\delta E) \tag{673}$$

which assumes that contributions in the summation and integral are well localized around the maximum. We can solve this and obtain

$$\overline{n_j} \ \approx \ \frac{\sigma_A^2(\Delta M c^2 - \bar{\epsilon})}{\sigma_A^2 \delta E + \bar{\epsilon}\sigma_0^2} \ = \ \frac{\Delta M c^2 + \beta\sigma_A^2}{\delta E - \beta\sigma_0^2} \tag{674}$$

**Self-consistent model**

We can put everything together for a self-consistent model. We need to solve

$$\begin{aligned}\Gamma_{transfer} \ &= \ \Gamma_{transfer}(\Delta M c^2) \ = \ \frac{2\pi}{\hbar}|\Delta\mathcal{U}|^2 \alpha(E_A) e^{\beta(E_A)\Delta M c^2} e^{\beta^2(E_A)\sigma_A^2/2} \\ \sigma_A \ &= \ \frac{1}{2}(\hbar\omega_A)\overline{n_j}\Gamma_A \tau \\ \overline{n_j} \ &= \ \frac{\Delta M c^2 + \beta\sigma_A^2}{\delta E - \beta\sigma_0^2} \\ \tau \ &= \ \frac{1}{\Gamma_{transfer}}\end{aligned} \tag{675}$$

### Results

Self-consistent solutions have been obtained from this model. The excitation transfer rate from the fusion transition is shown in Figure 72. We see that there is generally good agreement between the numerical results and the analytic model.

The standard deviation for the numerically exact and the analytic model is shown in Figure 73, where there is essentially no disagreement.

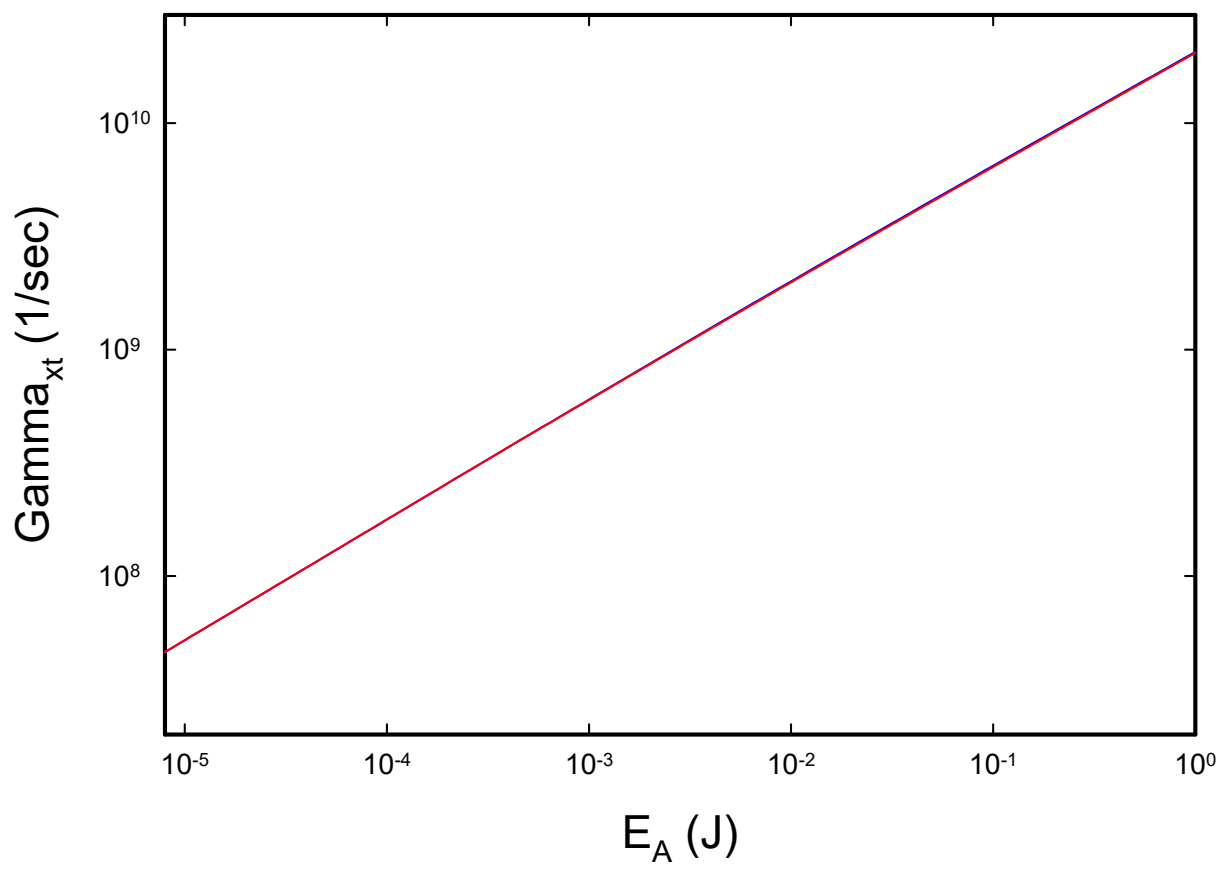

**Figure 72:** Results for the excitation transfer rate from the fusion transition $\Gamma_{transfer}$ (blue) and the analytic model (red).

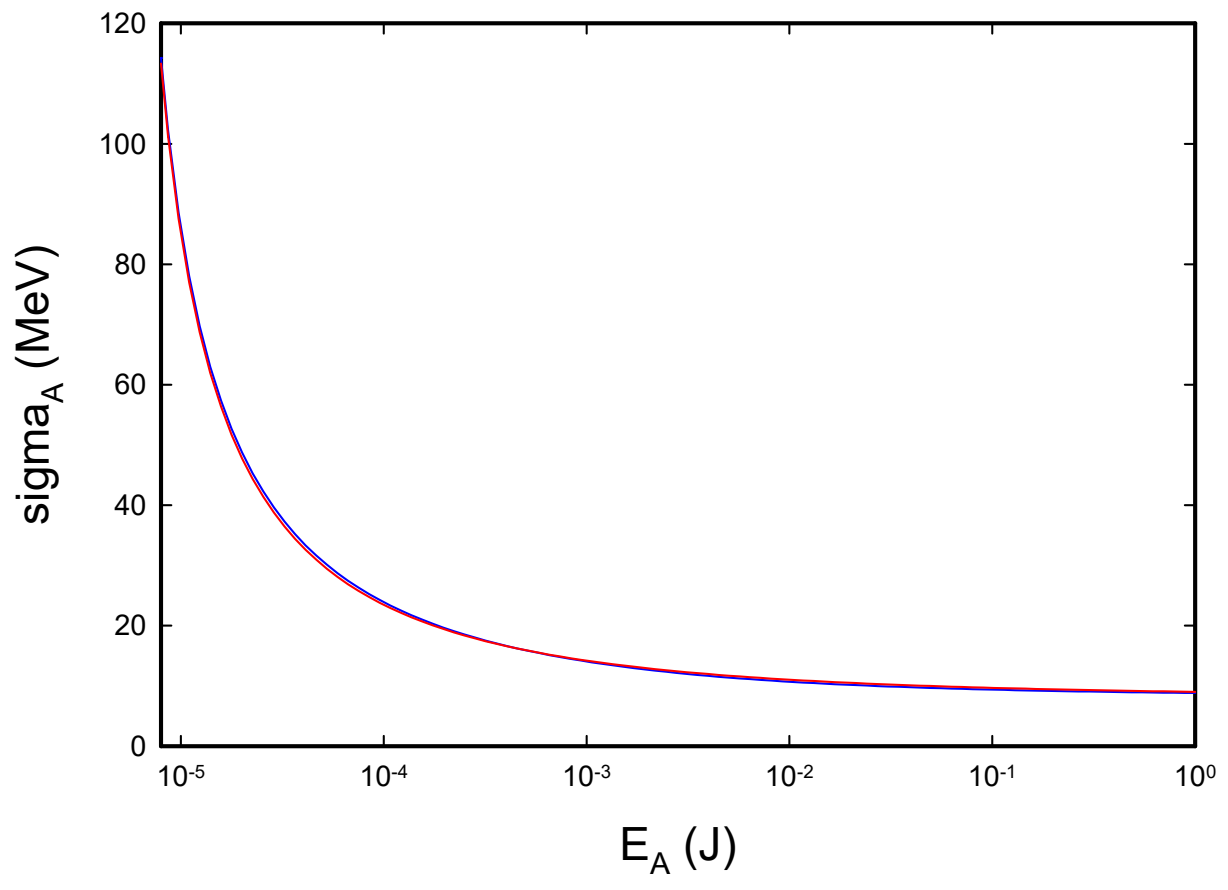

**Figure 73:** Results for $\sigma_A$ (blue) and the analytic model (red).

The energy exchange from the acoustic phonons are the same as shown in Figure 74. Once again we see good agreement between the numerical calculation and the analytic model.

The mean number of transitions involved is shown in Figure 75. There is good agreement between the numerical calculation and the analytic model.

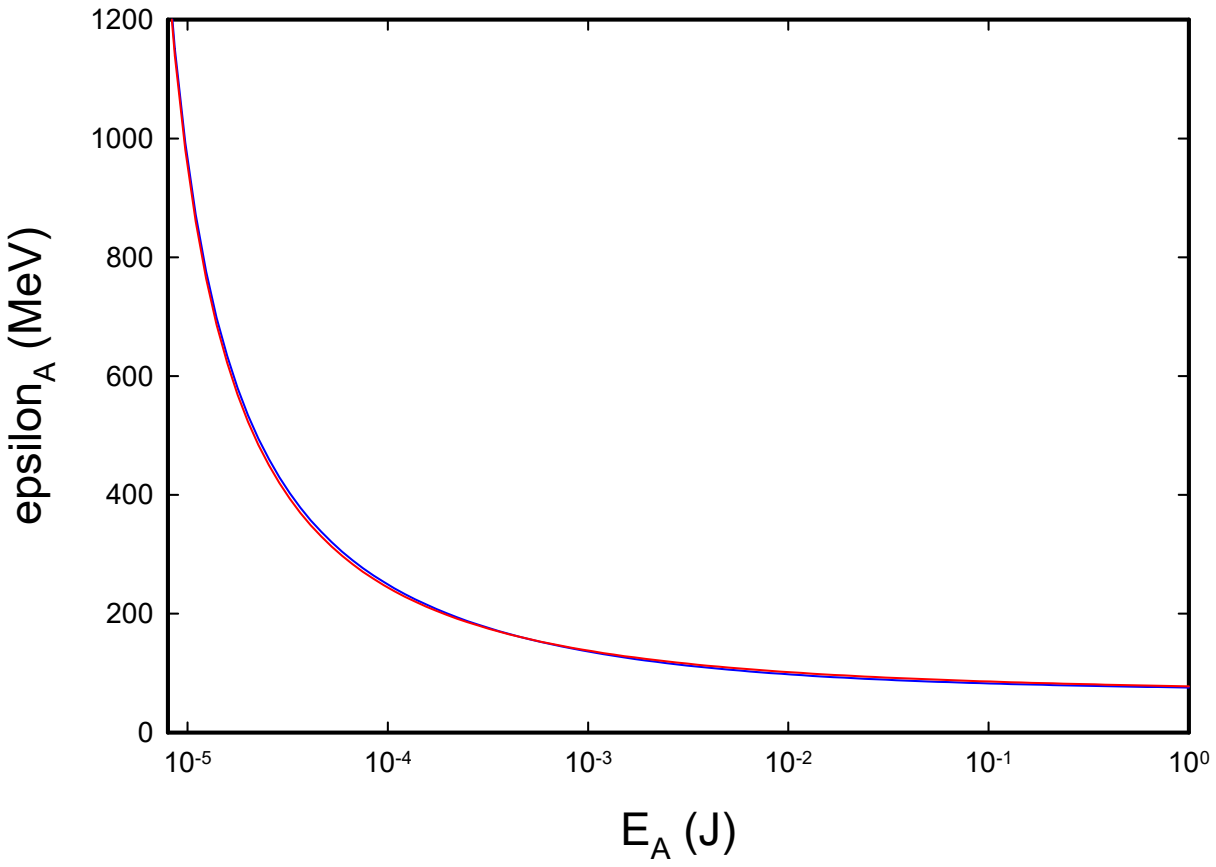

**Figure 74:** Results for the mean energy exchange from the acoustic phonons (blue) and the analytic model (red).

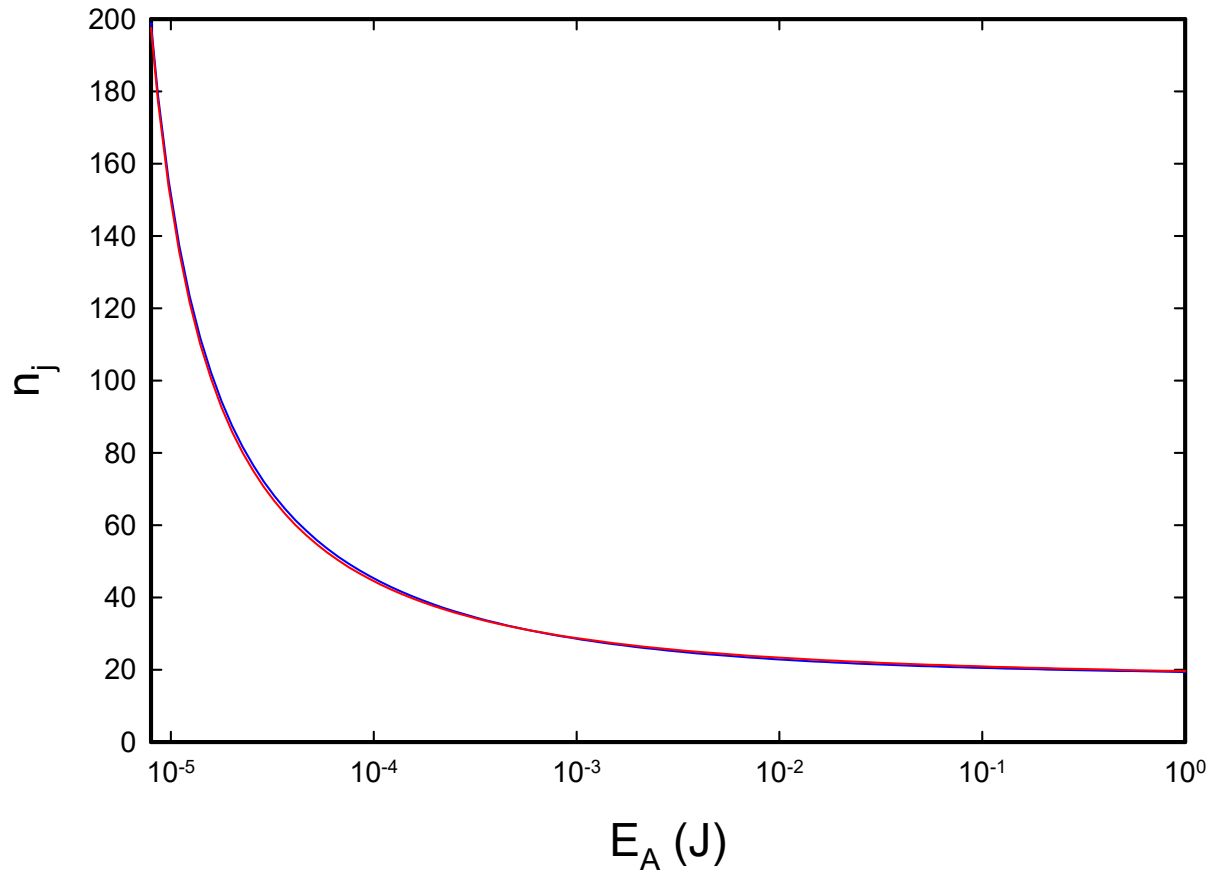

**Figure 75:** Results for $\overline{n_j}$ (blue) and the analytic model (red).

### Scaling with the acoustic phonon energy

We have made use of the analytic model to see how the excitation transfer rate from the fusion transition at start-up depends on the acoustic phonon frequency. For this study, we used the phonon frequencies

$$\omega_A \ = \ 2\pi 10^6, \ 2\pi 10^7, \ 2\pi 10^8 \ \frac{\text{rad}}{\text{sec}} \tag{676}$$

while keeping all of the other model parameters the same (as in the calculations above). Results for the excitation transfer rate are shown in Figure 76. We see that the excitation transfer rate is roughly linear in the acoustic phonon energy. Results for the mean energy exchanged from the acoustic phonon modes is shown in Figure 77. There is little difference in this model parameter for the three different phonon frequencies, which is similar for $\sigma_A$ and $\overline{n_j}$ (the results are very close to those plotted above, so we have not included figures).

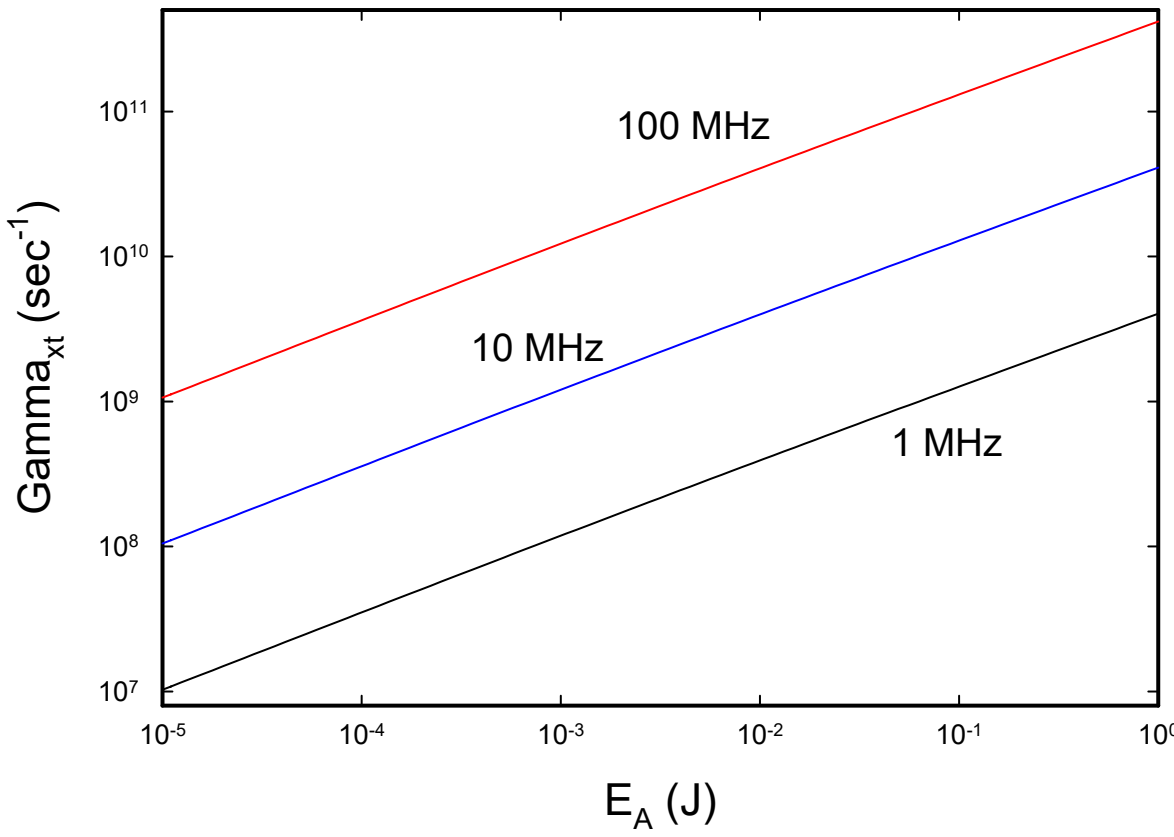

**Figure 76:** Results for $\Gamma_{transfer}$ for 1 MHz (black), 10 MHz (blue), and 100 MHz phonons (red).

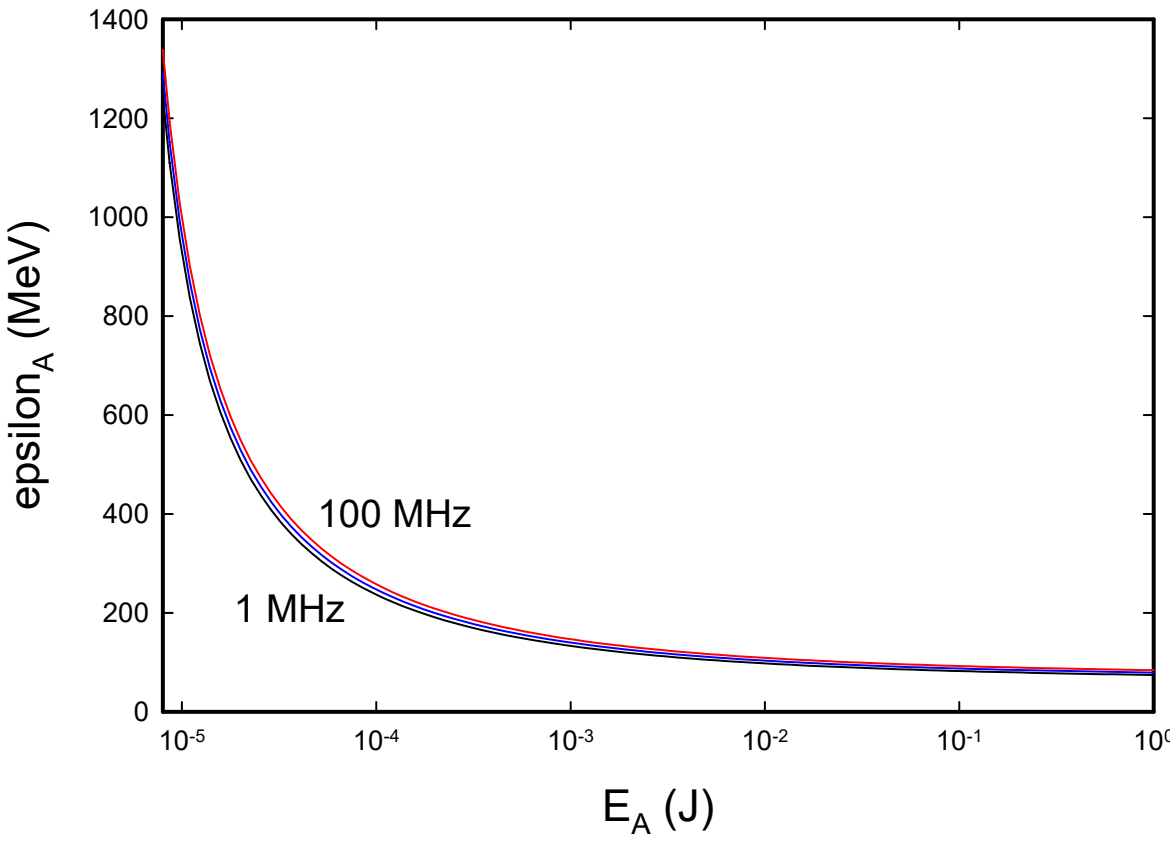

**Figure 77:** Mean energy exchanged from acoustic phonons for 1 MHz (black), 10 MHz (blue), 100 MHz phonons (red).

### Scaling with the screening energy $U_e$

We have argued that the Coulomb barrier hinders $D_2/^4He$ transitions due to tunneling, and that the large screening that has been seen in low energy ion beam experiments suggests that we might expect screening to be important in our model. We have made use of the analytic model to study the effect of screening on the excitation transfer rate, and on the other model parameters.

Results for the excitation transfer rate for different values of the screening energy including

$$U_e \ = \ 0, \ 50, \ 100, \ 150, \ 200, \ 250, \ 300, \ 350, \ 400 \text{ eV} \tag{677}$$

are shown in Figure 78.

We see that the excitation transfer rate decreases a minor amount as the screening increases, a result which is counter intuitive – we would expect that excitation transfer from the fusion transition should be faster if there is more screening. Note that the $\mathbf{a} \cdot c\mathbf{P}$ matrix element for the $D_2/^4He$ transition increases by more than 22 orders of magnitude between

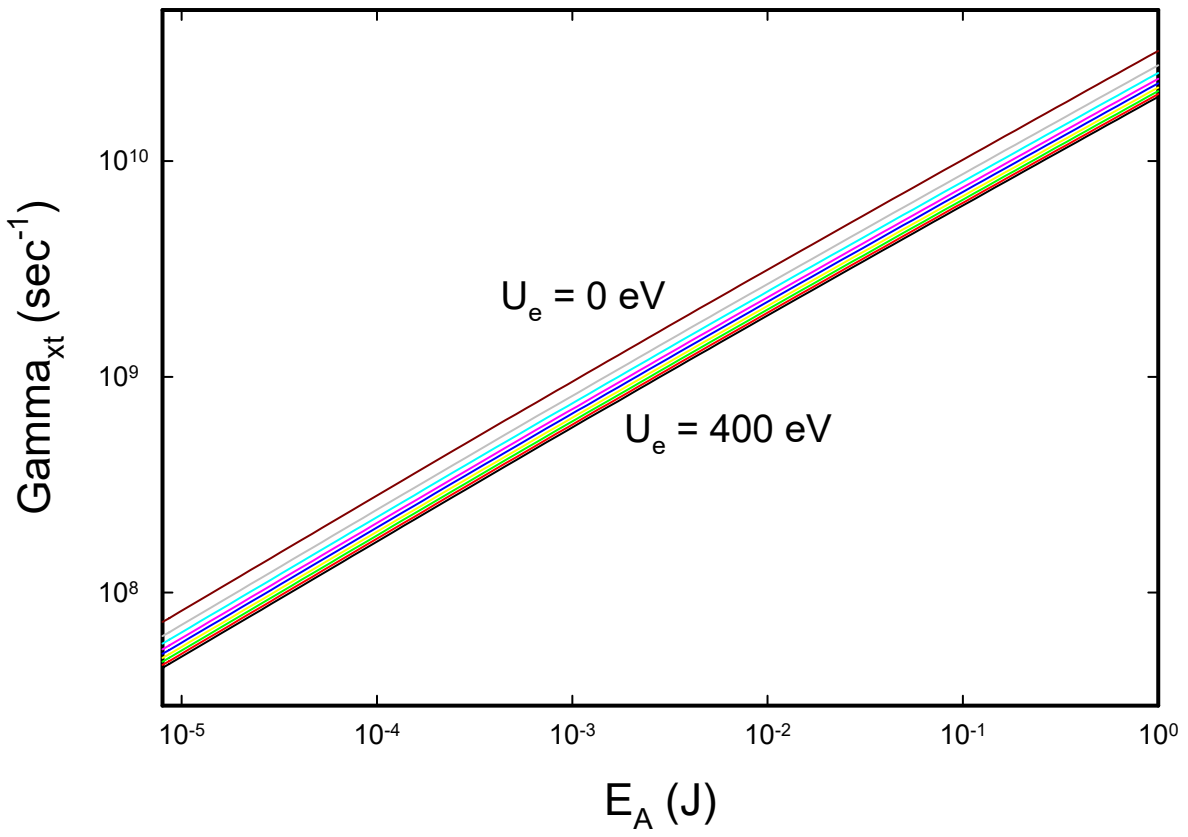

**Figure 78:** Results for the excitation transfer rate for screening energies between 0 eV and 400 eV.

$U_e = 0$ and $U_e = 400$ eV, which the excitation transfer rate changes by on the order of a factor of 2. This tells us that the excitation transfer rate has a very weak dependence on the screening energy. From the analytic model described above, and also from the approximate solution discussed below, the tunneling factor comes into the model through the square root of the logarithm of $e^{-2G}$. This is consistent with the weak dependence observe in this parameter study.

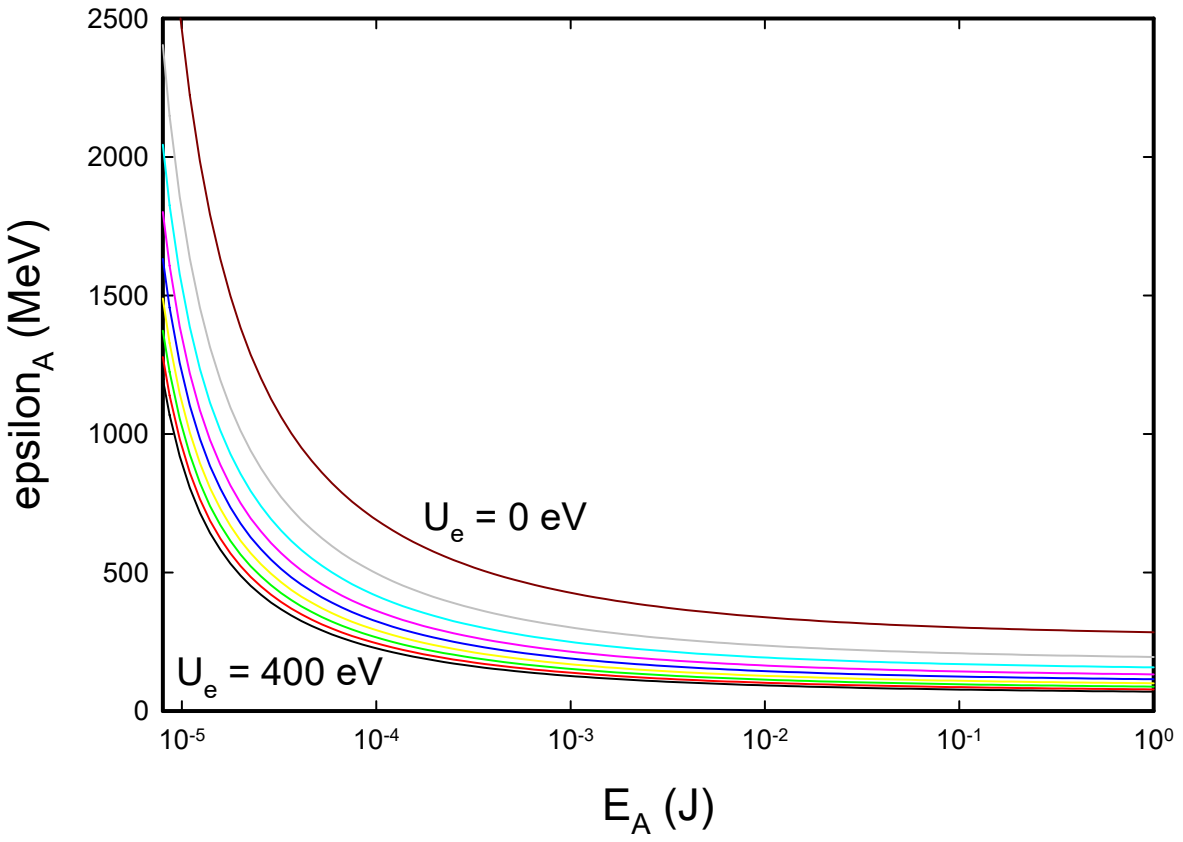

**Figure 79:** Mean energy exchanged from the acoustic phonons for screening energies between 0 eV and 400 eV.

If we pursue the issue further, it becomes clear what the origin of this very counter intuitive result is. When little screening occurs, more energy is needed from the acoustic phonons, leading to more Pd*/Pd transitions being involved, and this increases the number of transitions involved in exchanging energy with the acoustic phonons, so that the rate at which acoustic phonons are exchanged is larger. In Figure 79 we show the mean energy from the acoustic phonons as a function of the acoustic mode energy for the same values of the screening energy. Results for $\overline{n_j}$ for the different screening energies are shown in Figure 80.

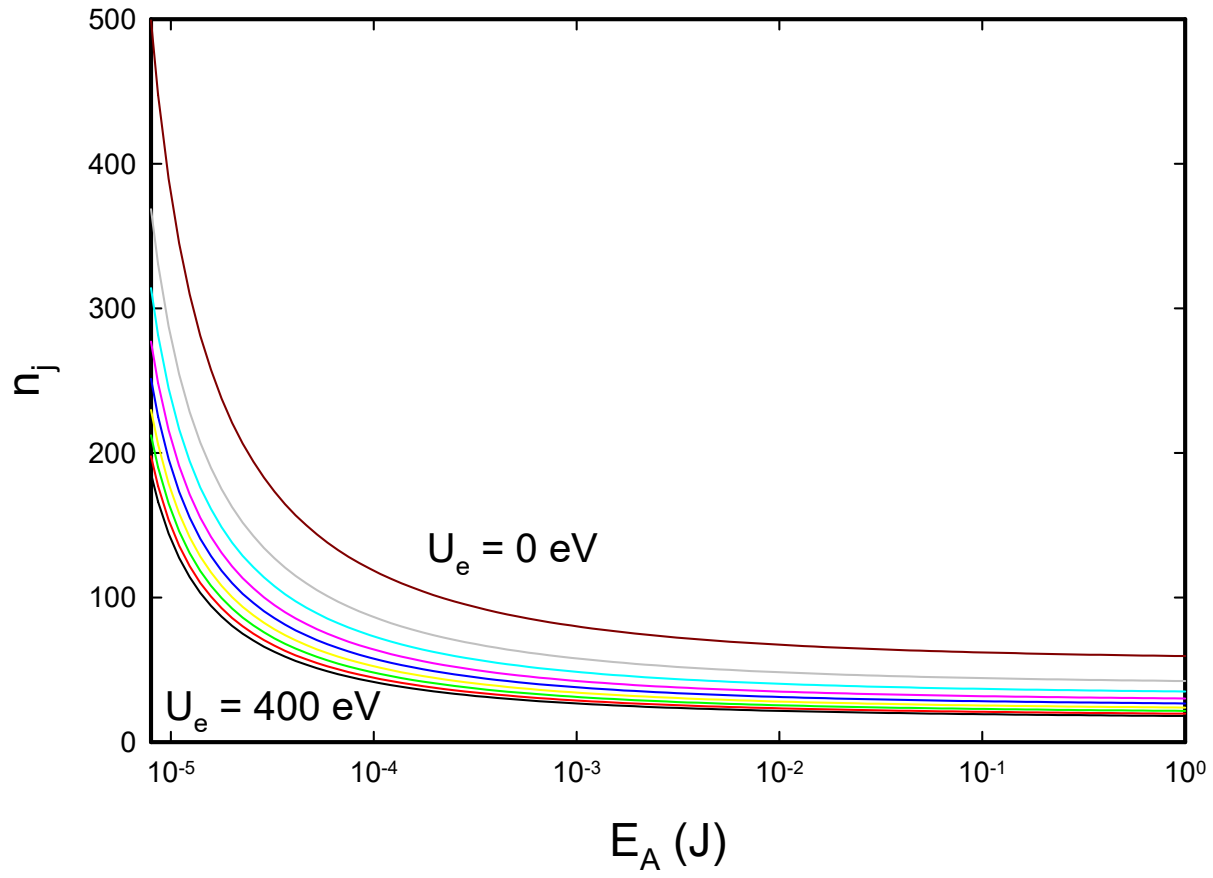

**Figure 80:** Mean number of Pd*/Pd transitions $\overline{n_j}$ for screening energies between 0 eV and 400 eV.

Based on this result, one might naively assume that the model somehow favors less screening and a more hindered $D_2/^4He$ matrix element since the excitation transfer rate from the fusion transition to many Pd*/Pd transitions is a little faster. After some thought, this is not the case. At start-up the Pd*/Pd transitions needs more energy from the acoustic mode when there is less screening, which makes it more difficult for the system to get started and to evolve to faster excitation transfers. Then, if the system manages to evolve past start up, it will require a much larger Dicke enhancement factor on the $D_2/^4He$ transition in order to make it into a regime where excess energy can be produced efficiently.

**Approximate solution for the excitation transfer rate**

It would be nice to have an analytic approximation for the excitation transfer rate including energy exchange with the highly-excited acoustic phonon mode. The analytic version of the model allows us to develop such an approximation.

We start by writing the excitation transfer rate including energy exchange in terms of the excitation transfer rate with no energy exchange

$$\Gamma_{transfer} = \Gamma_{transfer}^{(0)} e^{\beta^2 \sigma_A^2/2} \tag{678}$$

where

$$\Gamma_{transfer}^{(0)} = \frac{2\pi}{\hbar} |\Delta \mathcal{U}|^2 \tilde{\rho}_N \tag{679}$$

We can use this to write $\sigma_A$ in terms of $\Gamma_{transfer}$ according to

$$\sigma_A = \frac{\sqrt{2}}{\beta} \sqrt{\ln \frac{\Gamma_{transfer}}{\Gamma_{transfer}^{(0)}}} \tag{680}$$

We recall that the standard deviation $\sigma_A$ is related to the model parameters through

$$\sigma_A = \frac{1}{2}(\hbar\omega_A)\overline{n_j}\Gamma_A \tau \tag{681}$$

Recalling that $\tau = 1/\Gamma_{transfer}$, this leads to the constraint

$$\Gamma_{transfer} = \frac{1}{\tau} = \frac{\frac{1}{2}(\hbar\omega_A)\overline{n_j}\Gamma_A}{\frac{\sqrt{2}}{\beta}\sqrt{\ln \frac{\Gamma_{transfer}}{\Gamma^{(0)}_{transfer}}}} \tag{682}$$

It will be useful to define a new parameter $\kappa$ according to

$$\kappa = \sqrt{\ln \frac{\Gamma_{transfer}}{\Gamma^{(0)}_{transfer}}} \tag{683}$$

which we can use to write

$$\Gamma_{transfer} = \frac{\frac{1}{2}(\hbar\omega_A)\overline{n_j}\Gamma_A}{\frac{\sqrt{2}}{\beta}\kappa} \tag{684}$$

For the average number of Pd*/Pd transitions we can write

$$\overline{n_j} = \frac{\Delta M c^2}{\delta E - \beta\sigma_0^2} + \frac{\beta}{\delta E - \beta\sigma_0^2}\frac{2}{\beta^2}\kappa^2 \tag{685}$$

We can use this to write the analytic model according to

$$\begin{aligned} \Gamma_{transfer} &= \frac{\Gamma_A}{2\sqrt{2}}\frac{\hbar\omega_A}{(\delta E - \beta\sigma_0^2)}\left(\frac{\beta\Delta M c^2}{\kappa} + 2\kappa\right) \\ \kappa &= \sqrt{\ln \frac{\Gamma_{transfer}}{\Gamma^{(0)}_{transfer}}} \end{aligned} \tag{686}$$

An analytic self-consistent solution does not seem to be possible. However, it is possible to develop an iteration scheme based on

$$\begin{aligned} \Gamma_{transfer}[n] &= \frac{\Gamma_A}{2\sqrt{2}}\frac{\hbar\omega_A}{(\delta E - \beta\sigma_0^2)}\left(\frac{\beta\Delta M c^2}{\kappa[n]} + 2\kappa[n]\right) \\ \kappa[n+1] &= \sqrt{\ln \frac{\Gamma_{transfer}[n]}{\Gamma^{(0)}_{transfer}}} \end{aligned} \tag{687}$$

To start the iterations we might approximate

$$\Gamma[0] = 10^9 \text{ s}^{-1} \tag{688}$$

The first-order approximation for the rate that results is

$$\Gamma_{transfer} = \frac{1}{2\sqrt{2}}\beta(\hbar\omega_A)\Gamma_A \frac{1}{(\delta E - \beta\sigma_0^2)} \left( \Delta Mc^2 \sqrt{\ln \frac{\Gamma_{transfer}[0]}{\Gamma^{(0)}_{transfer}}}^{-1} + \frac{2}{\beta}\sqrt{\ln \frac{\Gamma_{transfer}[0]}{\Gamma^{(0)}_{transfer}}} \right) \tag{689}$$

We recall that the results of the Gaussian fits for the $^{120}$Sn data resulted in

$$\delta E = 6.75 \text{ MeV} \qquad \sigma_0 = 1.28 \text{ MeV} \tag{690}$$

A comparison of this lowest-order analytic approximation with the analytic result for the excitation transfer rate is shown in Figure 81. The model parameters for this calculation are

$$|\Delta \mathcal{U}| = \frac{\hbar}{2} 5.48 \times 10^3 \sqrt{\frac{E_A}{1 \text{ J}}}\sqrt{\frac{N_{D_2}}{N}} \text{ eV} \qquad \frac{N_{D_2}}{N} = 0.25 \times \frac{1}{9} \tag{691}$$

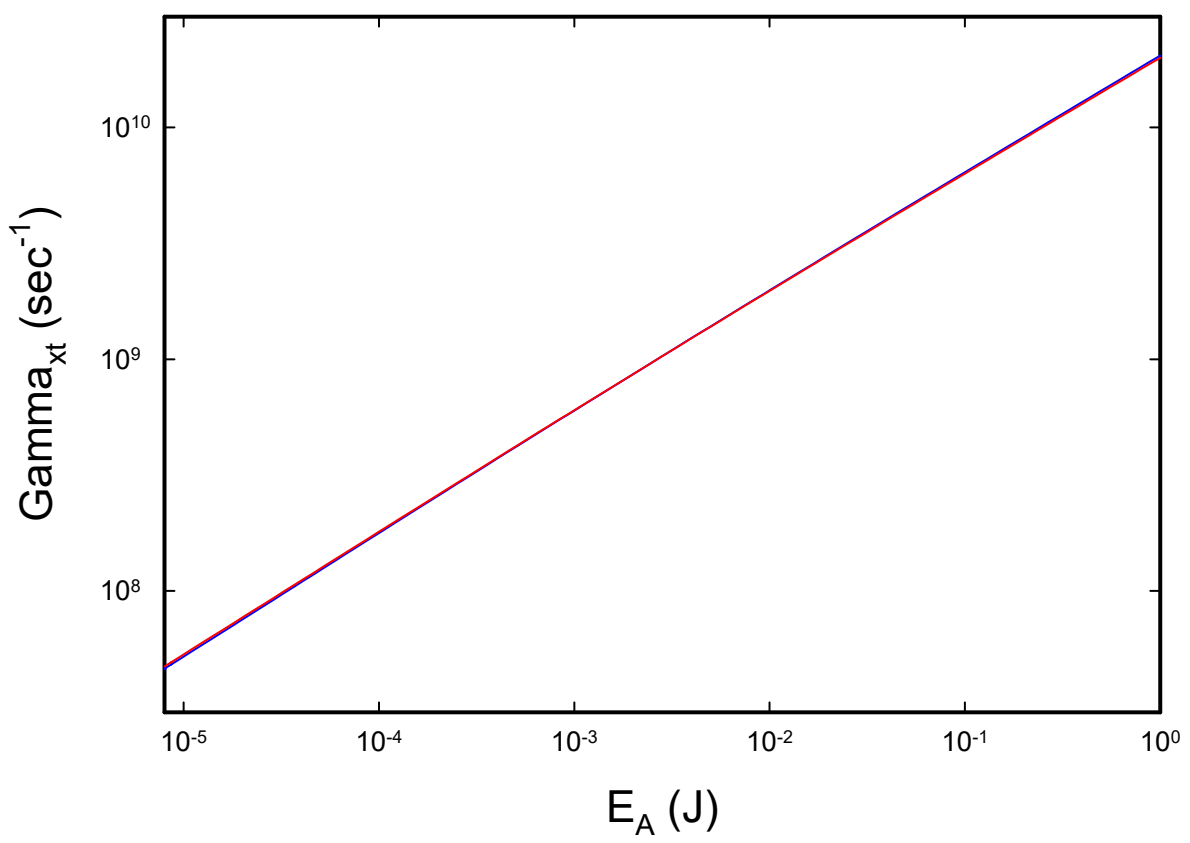

**Figure 81:** Results for $\Gamma_{transfer}$ from the analytic model (blue), and from the lowest order approximation (red).

We see that the lowest-order approximation works quite well. From the lowest-order approximation we see that the excitation transfer rate is roughly linear in $\hbar\omega_A$, which is understandable in that if the phonon energy quantum is larger then it requires fewer phonons to be exchanged for a given $\epsilon_A$, which means that it takes less time for the phonons to be exchanged, which means that the excitation transfer rate is faster.

We see that the excitation transfer rate is roughly proportional to the acoustic mode phonon exchange rate $\Gamma_A$, since it takes less time to transfer the large number of phonons if this transfer rate is faster. Since $\Gamma_A$ is proportional to the square root of the acoustic mode energy $E_A$, the excitation transfer rate is roughly proportional to the square root of $E_A$. The dependence on the number of $D_2$ molecules is weak, since $\Gamma^{(0)}_{transfer}$ is inside of a logarithm which is inside of a square root. Since $\Gamma^{(0)}_{transfer}$ is proportional to $e^{-2G}$, the Gamow factor comes into the excitation transfer rate as $G$ (and not through an exponential). This is a consequence of the very large generalized density of states that can be reached if more energy is transferred from the acoustic phonon mode.

## 6.27 Optical phonon and plasmon exchange

The generalized excitation transfer rate including optical phonon and plasmon exchange can be written as

$$\Gamma_{transfer}(E) \ = \ \frac{2\pi}{\hbar}|\Delta \mathcal{U}|^2 \biggl( \tilde{\rho}_N * f_{\epsilon_A} * f_{\epsilon_O} * f_{\epsilon_P} \biggr)(E) \tag{692}$$

The use of the generalized excitation transfer rate is convenient here in connection with the convolutions, but ultimately we are interested in the rate evaluated at the fusion energy

$$\Gamma_{transfer} \ = \ \Gamma_{transfer}(\Delta M c^2) \tag{693}$$

The simple quantum diffusion model that we used for acoustic phonon exchange is appropriate with the amount of energy exchange is much less than the total energy in the mode. This may not be the case for optical phonons or plasmons, which means that a more complicated distribution would be needed. Since the plasmon lifetime is so short (on the order of 1 fs), and since the associated power level is high

$$\frac{\Delta M c^2}{\tau_P} \ \approx \ \frac{23.85 \text{ MeV}}{1 \text{ fs}} \ = \ 3821 \text{ watts} \tag{694}$$

it will require a high-power system with where the plasmons are driven efficiently in order for there to be sufficient energy in the uniform plasmon mode to exchange with the fusion and Pd system to make a difference. The most promising approach would be in a transient experiment driven by a high-power short laser pulse tuned to the plasmon energy.

We would expect that under less extreme circumstances we would be able to neglect energy exchange with plasmons and make use of

$$\frac{2\pi}{\hbar}|\Delta \mathcal{U}|^2 \biggl( \tilde{\rho}_N * f_{\epsilon_A} * f_{\epsilon_O} * f_{\epsilon_P} \biggr)(E) \ \rightarrow \ \frac{2\pi}{\hbar}|\Delta \mathcal{U}|^2 \biggl( \tilde{\rho}_N * f_{\epsilon_A} * f_{\epsilon_O} \biggr)(E) \tag{695}$$

An analogous argument could be made in the case of optical phonons, where the optical phonon lifetime can be on the order of 1 ps. The dissipated power that corresponds to a highly-excited optical phonon mode with and energy of $\Delta M c^2$ would be a few watts. In general it is not easy to arrange for significant energy in a uniform optical phonon mode. It can be done using a THz laser driving the surface of a thin film, or taking advantage of the beat frequency between two optical lasers, where the drive or beat frequency is tuned to an optical phonon mode at the $\Gamma$-point.

## 6.28 Average oscillator energy exchange during excitation transfer

The self-consistent solution to the excitation transfer rate (Eq. 236) was calculated in section 5.12. We reproduce the equation and the corresponding figure (Figure 37, here: Figure 82) for convenience:

$$\Gamma_{transfer} = \Gamma_{transfer}(\Delta Mc^2) = \frac{2\pi}{\hbar}|\Delta\mathcal{U}|^2 \int_{-\infty}^{\infty} \tilde{\rho}_N(\Delta Mc^2 - \epsilon) f_{\epsilon_A}(\epsilon_A, \tau) d\epsilon \tag{696}$$

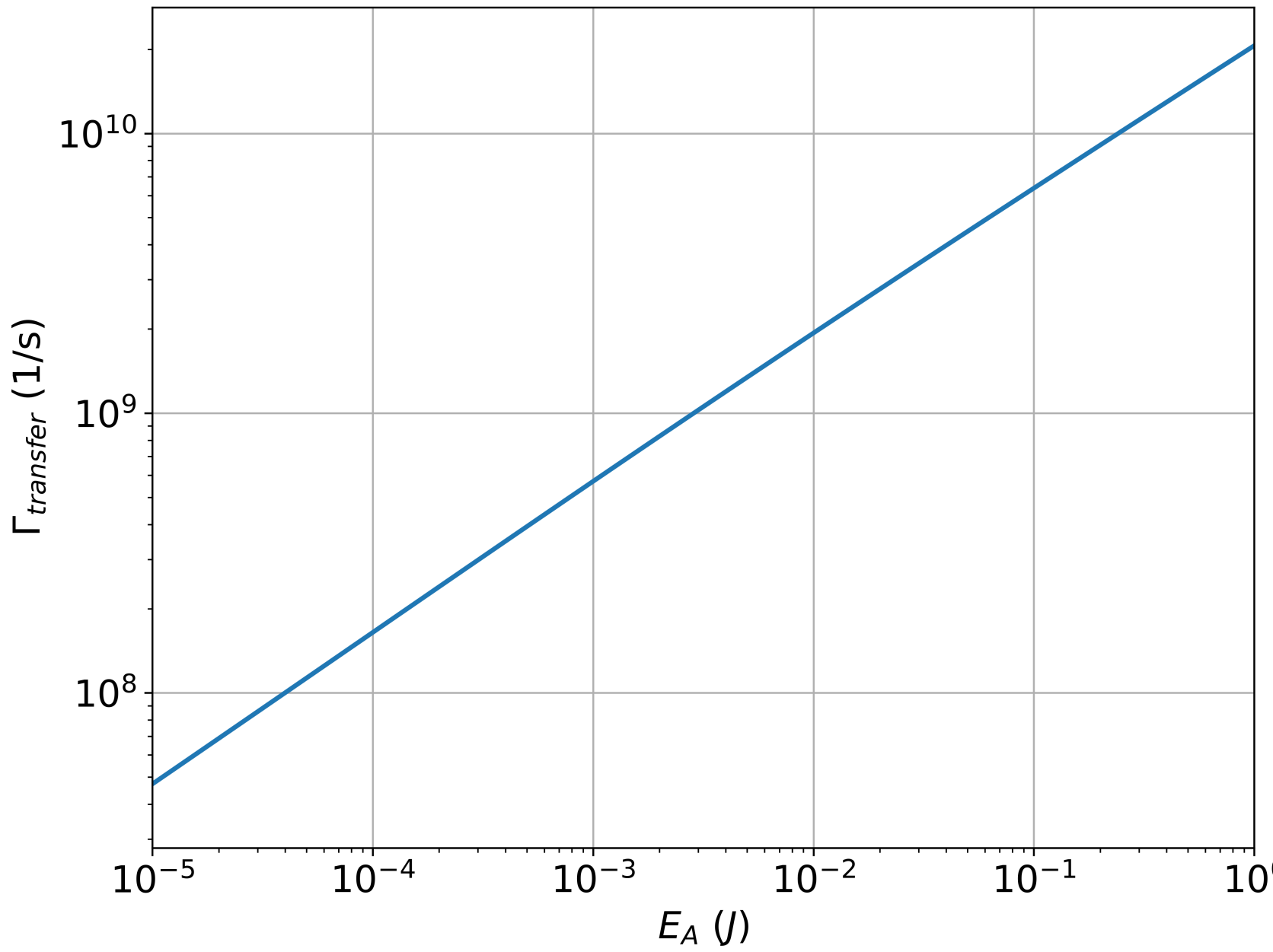

**Figure 82:** Excitation transfer rate as a function of $E_A$.

The rate above allows us to study the characteristics of the phonon energy distribution defined in Eq 228. Specifically, the spread in phonon energy (measured by the standard deviation $\sigma_{\epsilon_A}$) and the average energy exchanged (see Eq. 6.26)

In Figure 83 we show the standard deviation $\sigma_{\epsilon_A}$ in MeV associated with acoustic phonon exchange. We see that at all values of $E_A$ shown the standard deviation is substantial. Substantial energy exchange with the acoustic phonons is theefore required in order for excitation transfer from the fusion transition to occur, even when there is significant energy in the acoustic mode.

In Figure 84 we show the mean energy in MeV transferred from the acoustic phonons to the nuclear transition as a function of $E_A$. We see that substantial energy exchange is needed for excitation transfer from the fusion transition, with on the order of 1000 MeV at the lowest $E_A$ value (10 $\mu$J) to about 77 MeV for $E_A$ near 1 J.

From this result we can gain some intuition about how the model works. When the energy in the acoustic mode $E_A$ is low, then the excitation transfer rate is very slow if only the fusion energy $\Delta Mc^2$ were available to promote Pd*/Pd transitions. But if the excitation transfer rate is slow, there is more time for energy exchange with the acoustic phonon mode. With energy from the acoustic mode, the excitation transfer rate can be faster. In this model, the rate near $E_A = 10$ $\mu$J is sufficiently slow that there is time for on the order of 1000 MeV to be transferred.

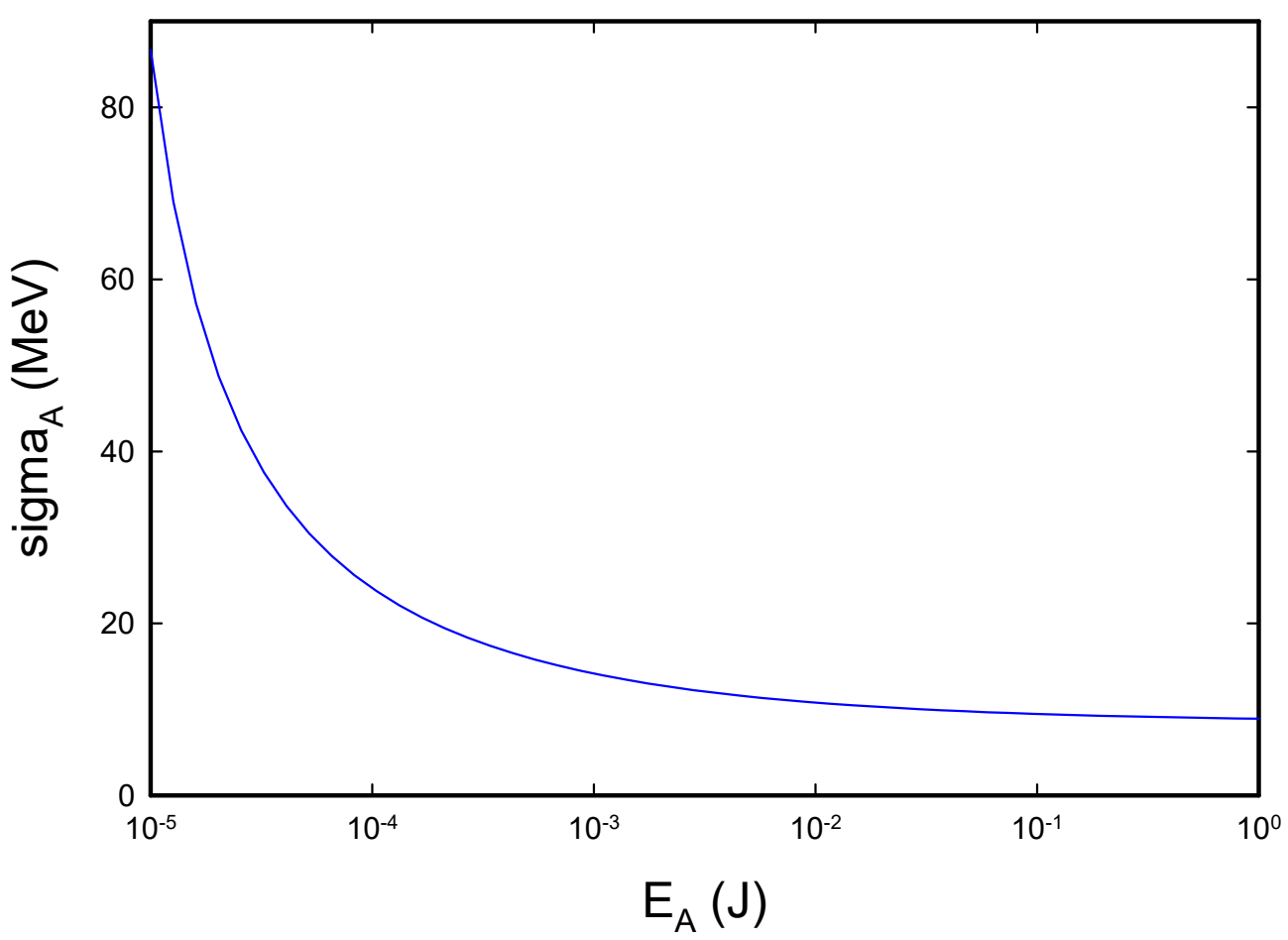

**Figure 83:** Standard deviation $\sigma_{\epsilon_A}$ as a function of $E_A$.

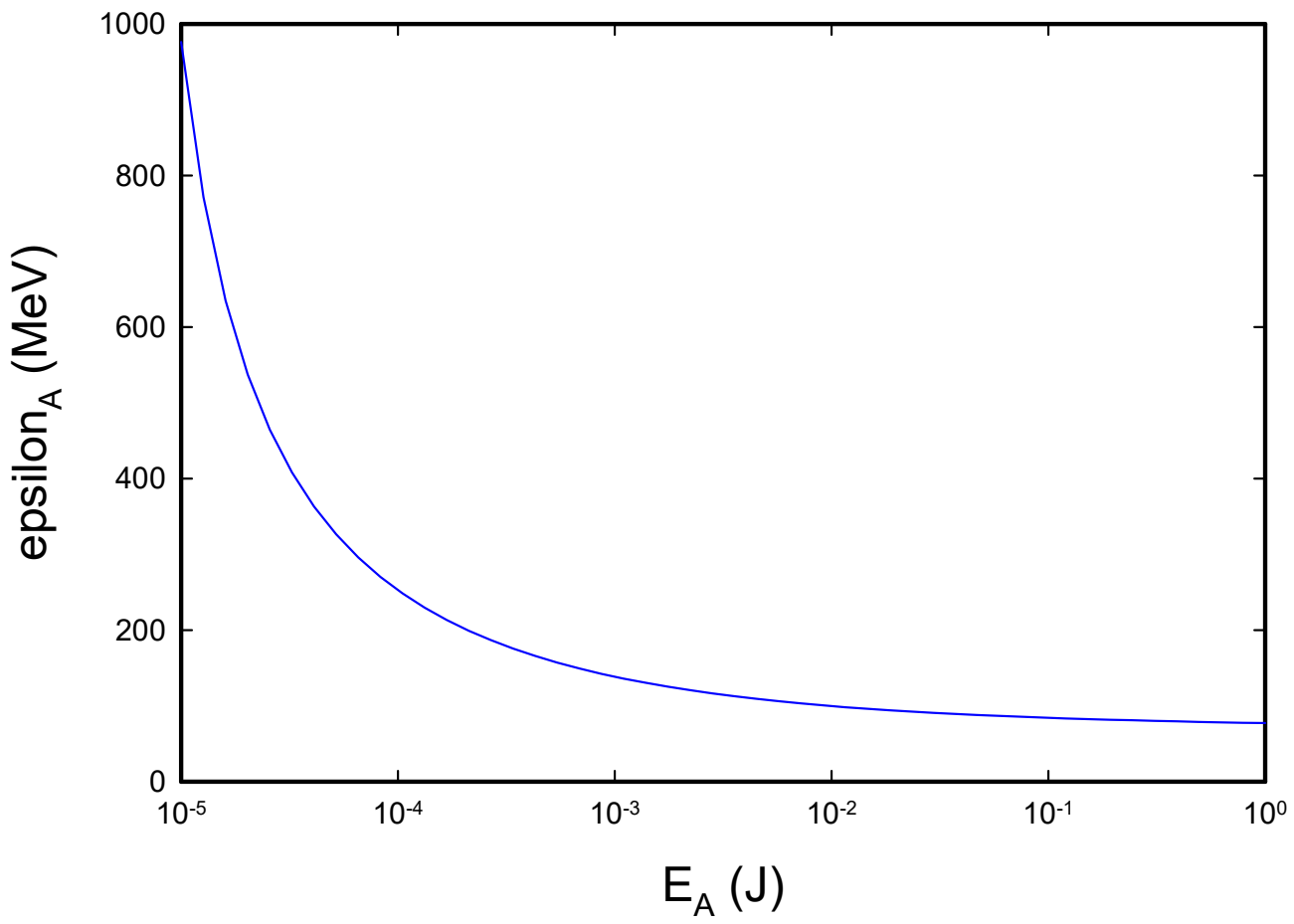

**Figure 84:** Mean energy transferred from the acoustic phonons $\epsilon_A$ as a function of $E_A$.

## 6.29 Dicke enhancement with excitation transfer to multiple receiver states

According to the discussion above, the system can only start up by taking considerable energy from the acoustic vibrations. If sufficient energy can be provided by the uniform acoustic phonon mode to sustain many such transitions, and if the rate is sufficiently high to exceed the molecular $D_2$ decoherence rate (which we have estimated might be as low as $10^9$ s$^{-1}$ in monovacancies where available sites to hop to are filled), then it is possible for the Dicke factor associated with the $D_2/^4He$ fusion transition to increase. In order to show results under these conditions, we make use of $N_{^4He}^{(Dicke)}$ for the number of $^4$He atoms associated with the Dicke state of the $D_2/^4He$ fusion transition according to

$$\left\langle S, M+1 \left| \frac{\hat{S}_+}{\hbar} \right| S, M \right\rangle = \sqrt{(S+M+1)(S-M)} = \sqrt{(N_{D_2}^{(Dicke)}+1)N_{^4He}^{(Dicke)}} \tag{697}$$

where

$$\begin{aligned} N_{D2}^{(Dicke)} &= S+M \\ N_{^4He}^{(Dicke)} &= S-M \end{aligned} \tag{698}$$

This factor comes into the model in connection with $|\Delta\mathcal{U}|$, which after start-up (still assuming $U_e = 350$ eV) would become

$$\frac{2}{\hbar}|\Delta\mathcal{U}| \rightarrow 5.48 \times 10^3 \sqrt{N_{^4He}^{(Dicke)}} \sqrt{\frac{E_A}{1 \text{ J}}} \sqrt{\frac{N_{D2}^{(Dicke)}}{N}} \text{ s}^{-1} \tag{699}$$

For this discussion we assume that

$$N_{^4He}^{(Dicke)} \ll N_{D_2} \tag{700}$$

so that

$$N_{D_2}^{(Dicke)} \approx N_{D_2} \tag{701}$$

As mentioned above, we have not included Dicke enhancement factors for the Pd*/Pd transitions in the model. We might expect a baseline level of Pd* admixture due to the polarization of the Pd*/Pd transitions by the highly-excited acoustic phonon mode, unconnected with the excitation transfer from the fusion transition. Because of this, additional excitation of the Pd*/Pd transitions will result in Dicke enhancement factors contributing even with only a few transitions participating at start-up. If the energy exchange with the oscillators is significant, then the number of Pd*/Pd transitions promoted in the excitation transfer process can be on the order of 50-100, which means that there will be many transitions that are multiply excited which will come with Dicke enhancement factors. After starting up, if the system has transferred some excitation to the Pd*/Pd transitions, then there can be residual excitation which would lead to additional Dicke enhancement factors. Because of the complications involved, we have chosen to carry out exploratory modeling neglecting all Pd*/Pd Dicke enhancement above the ground states, in order to begin understanding how the model works and to obtain "worst case" excitation transfer rates.

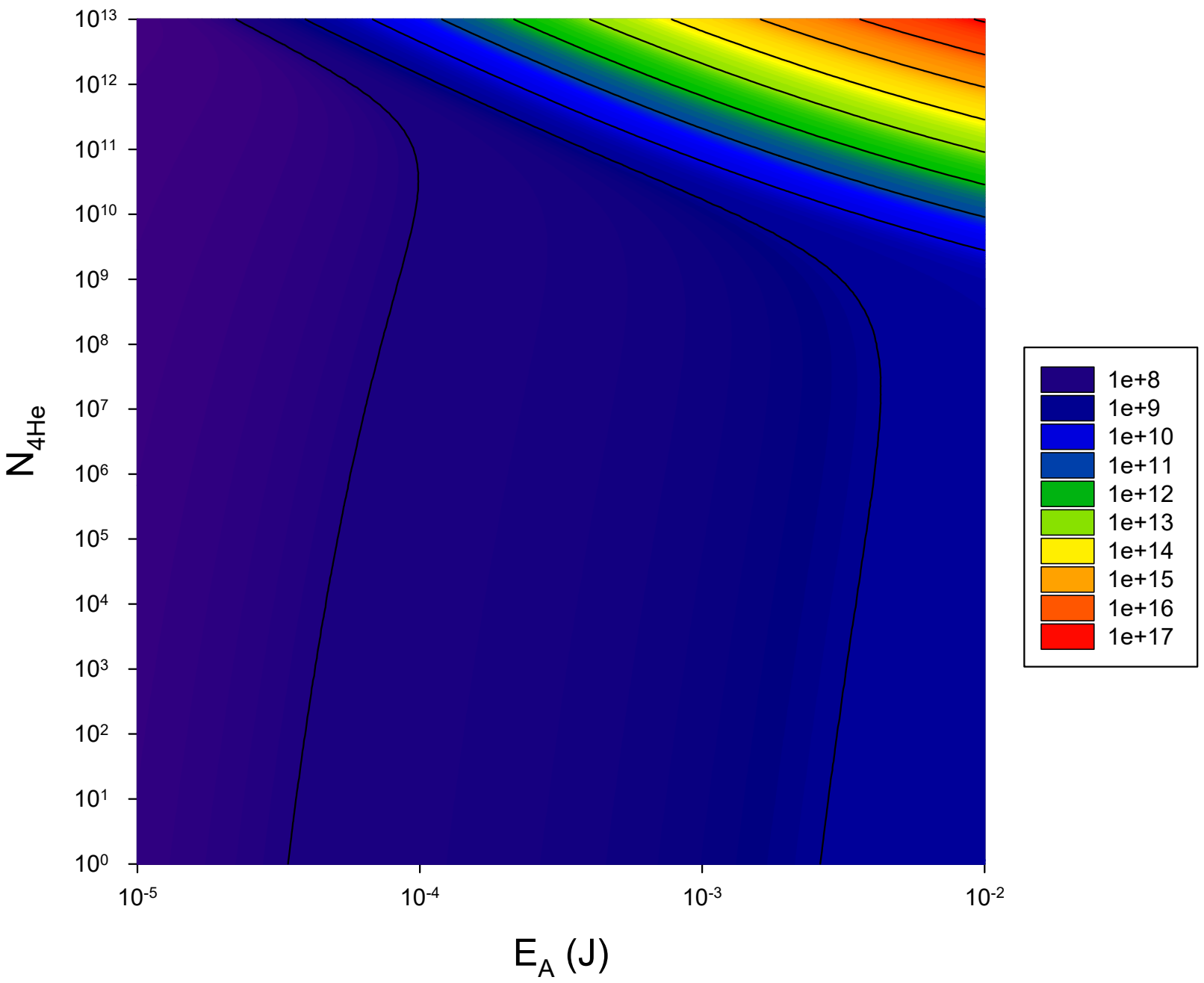

**Figure 85:** Excitation transfer rate $\Gamma_{transfer}$ in s$^{-1}$ as a function of $E_A$ and $N^{(Dicke)}_{^4He}$.

**Excitation transfer rate with Dicke enhancement of the $D_2$/$^4$He fusion transition**

Self-consistent results for the excitation transfer rate including the Dicke enhancement factor for the $D_2$/$^4$He transition from Equation (236), with the same model parameters as above, are shown in Figure 85 as a function of $E_A$ and $N^{(Dicke)}_{^4He}$.

We see that in this model the excitation transfer rate is almost independent of the number of $^4$He atoms which are part of the Dicke state over many orders of magnitude. In this regime, excitation transfers occur only because substantial energy is coming from the acoustic phonons, which results in a limitation on the associated rate. When the Dicke enhancement factor gets sufficiently large, then the nuclear system no longer requires energy input from the acoustic phonons, and the excitation transfer rate increases with the Dicke enhancement factor. This is seen to occur in this model at $N_{^4He}$ above about $10^{11}$ at $E_A = 10$ mJ in Figure 85, and at higher $N_{^4He}$ at lower energy.

We see that quite high excitation transfer rates from the fusion transition are predicted at large $N^{(Dicke)}_{^4He}$ values. As more excitation transfer from the fusion transition occurs, the Dicke factor increases, and the rate increases. This effect would be enhanced further if Dicke factors for the Pd*/Pd transition were included. This kind of "runaway" effect is a basic feature of the model, and would naturally be associated with an optimized "large" experiment. In this mode of operation, we would expect the supply of molecular $D_2$ to be depleted, in part from being used, and in part from being lost from the $PdD_x$ as the sample heats.

There have been a some claims of an excess heat runaway effect observed in certain metal-hydrogen experiments. We surmise that this effect is connected with operation in the unstable regime at large $N^{(Dicke)}_{^4He}$ as predicted in this kind of model. Clearly, if the principles presented here are harnessed in future energy technologies (see 5.16), operating in a stable regime rather than a runaway regime is important.

Finally, in the calculation above, the focus was on the impact of the Dicke factor for the fusion transition on the donor side, neglecting Dicke factors associated with the Pd*/Pd transitions on the receiver side. The results suggest that a (unrealistically) large Dicke enhancement of the fusion transition is required for runaway to occur. In a more complete calculation, we expect contributions to the transfer rate due to Dicke enhancement from both the fusion transition and Pd*/Pd transitions, as well as contributions from many additional Pd*/Pd transitions with lower $O_{Pd}$ values. Consequently, the threshold to cross into the runaway regime is predicted to be lower in such a model (*i.e.*, occurring already with more modest Dicke enhancement of the fusion transition). A rough scaling argument suggests that with this model, the runaway regime is accessible at a transfer rate $\Gamma_{transfer}$ corresponding to on the order of 100 watts for $E_A = 1$ J. These are aspects that need to be taken into account in the future design of experiments and, possibly, technology based on these principles.

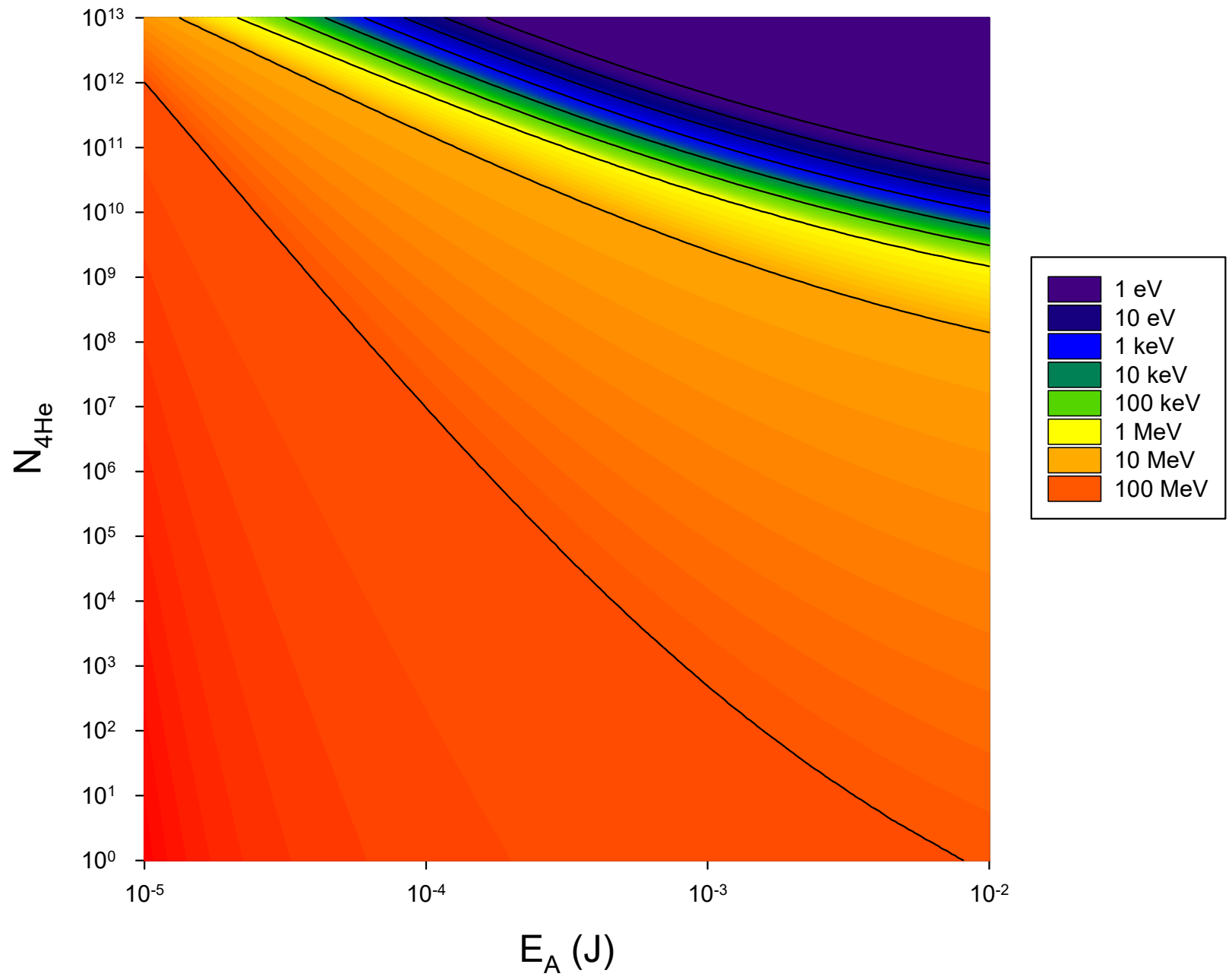

**Figure 86:** Acoustic phonon energy exchanged as a function of $E_A$ and $N_{He}$.

### Energy exchange with Dicke enhancement of the $D_2$/$^4$He fusion transition

The energy exchanged with the acoustic phonons as a function of $E_A$ and $N_{^4He}^{(Dicke)}$ is shown in Figure 86. The result from Figure 84 at $N_{^4He}^{(Dicke)} = 1$ is the same as at the bottom if this figure, with the exchanged energy approaching 1000 MeV in the lower left corner. We see that as the Dicke factor of the $D_2$/$^4$He fusion transition (as represented here through $N_{^4He}^{(Dicke)}$) increases, the energy exchanged from the acoustic phonon mode decreases. In this model a large Dicke enhancement is required in order for the energy exchange to be suppressed by the fast excitation transfer rate without energy exchange, an effect that results from the excitation transfer being completed before there is time for much energy exchange to occur.

### Internal coefficient of performance

Under conditions where a substantial amount of energy from vibrations is required for excitation transfer to occur, there can be net energy gain, but the associated internal coefficient of performance

$$\text{CoP}_{int} = \frac{\Delta M c^2 + \epsilon_A}{\epsilon_A} \tag{702}$$

will be poor. Once the Dicke enhanced excitation transfer rate without energy exchange becomes sufficiently fast, then the internal coefficient of performance can be high. This regime is of interest for efficient energy generation.

## 6.30 Excitation transfer from the HD/$^3$He fusion transition

The large majority of this document is focused on deuterium pairs as donor systems in nuclear excitation transfer dynamics via the $D_2$ to 4He transition that releases 23.85 MeV of nuclear binding energy. However, the arguments developed can be readily transferred to other materials. On the donor systems side, an obvious candidate is the HD to $^3$He transition that releases 5.49 MeV of nuclear binding energy. Peculiarities of this transition will be briefly discussed in this section. For a calculation of the HD to $^3$He matrix element with respect to the relativistic coupling discussed in section 6.7, see section 6.10.

In the literature, anomalous excess heat production has been particularly associated with light water and nickel systems. This provides motivation to speculate about the possibility of excitation transfer from the HD/$^3$He fusion transition to excited Ni$^*$ states, by analogy with the mechanisms discussed in this document. Note that when using light water in an experiment, deuterium is still present, at a natural abundance ratio of roughly 1 to 6500 D to H atoms.

Molecular HD is conjectured to form in different lattices where defects provide for low enough electron density. In the case of Pd, we have been interested in the possibility of monovacancy formation in PdD$_x$, as the removal energy of Pd is reduced roughly in proportion to the occupation of nearby O sites with H or D; the situation is similar in NiH$_x$, but there is a larger reduction of the removal energy. Pd attracted our attention in connection with sigma-bonded di-hydrogen complexes, as $H_2$ bonds with atomic Pd resulting in Pd($H_2$), where the atom bonds to the molecule and not individually to atoms in the ground state. A qualitatively similar effect occurs for nickel, where there is a low-lying Ni($H_2$) complex [169]. This suggests that it may be possible to arrange for substantial molecular HD occupation near Ni atoms in monovacancies in a dedicated experiment.

Note that the HD system exhibits a lower reduced mass than the $D_2$ system, which means that tunneling through the Coulomb barrier would be orders of magnitude faster. Consequently, the Gamow factor for molecular HD is significantly smaller than for molecular $D_2$, which is encouraging for excitation transfer from the HD/$^3$He fusion transition. Also encouraging is that we would expect a strong screening effect for molecular HD, similar to the screening effect seen for molecular $D_2$ based on deuteron-deuteron ion beam collision experiments. A screening effect has been reported for deuteron-deuteron collisions in Ni [54].

We have commented above on the destructive interference associated with excitation transfer, and the need for a mechanism to eliminate some of the destructive interference to accelerate the rate. For the $D_2$/$^4$He transition, we noted that loss due to fusion to the 3+1 channel can remove some of the destructive interference, and there is a bigger effect in the $D_2$/3+1/$^4$He pathway due to the fast tunnel decay of the 3+1 state. There is no equivalent intermediate state with a fast tunneling decay in the HD/$^3$He transition, which means that this transition will not be as effective in removing the destructive interference. There is a radiative decay mechanism that comes into the model the same way as fusion loss comes into the model for the $D_2$/$^4$He transition. Since the radiative decay rate is orders of magnitude slower, the associated $1 - \eta|$ factor due to the this loss is a comparable number of orders of magnitude slower. If there were no other mechanisms available to eliminate the destructive interference, then we would conclude that the HD/$^3$He fusion transition would not work very well (in comparison to the $D_2$/$^4$He transition and in comparison to the $D_2$/3+1/$^4$He pathway). We note that it is possible that fast loss associated with a Pd$^*$/Pd transition or Ni$^*$/Ni transition could reduce the destructive interference, and accelerate excitation transfer.

Excitation transfer from the HD/$^3$He fusion transition to many Ni$^*$/Ni transitions with energy input from the highly-excited acoustic phonon mode should be able to work in much the same way as for excitation transfer from the $D_2$/$^4$He transition. If fast coherent transitions can stabilize excited Pd$^*$ states in the PdD$_x$ system, then there seems to be no reason that a similar stabilization should not occur for the excited Ni$^*$ states. We would expect energy exchange from the Ni$^*$ states to uniform optical phonon modes and uniform plasmon modes to be similar.